\documentclass{pasj01}

\usepackage[OT2,T1]{fontenc}

\def\commenta{$^*$}
\def\commentb{$^\dagger$}
\def\commentc{$^\ddagger$}
\def\commentd{$^\S$}
\def\commente{$^\|$}
\def\commentf{$^\#$}

\newcounter{author}
\def\authorcount#1#2{\refstepcounter{author}\label{#1}
                     \altaffiltext{\ref{#1}}{#2}}

\def\Ohtprep{T. Ohshima et al. in preparation}
\def\Nakataprep{C. Nakata et al. in preparation}

\def\Namekataprep{K. Namekata et al. in preparation}

\begin{document}
\SetRunningHead{T. Kato et al.}{Period Variations in SU UMa-Type Dwarf Novae VIII}

\Received{201X/XX/XX}%{yyyy/mm/dd}
\Accepted{201X/XX/XX}%{yyyy/mm/dd}

\title{Survey of Period Variations of Superhumps in SU UMa-Type Dwarf Novae.
    VIII: The Eighth Year (2015--2016)}

\author{Taichi~\textsc{Kato},\altaffilmark{\ref{affil:Kyoto}*}
%% 53
        Franz-Josef~\textsc{Hambsch},\altaffilmark{\ref{affil:GEOS}}$^,$\altaffilmark{\ref{affil:BAV}}$^,$\altaffilmark{\ref{affil:Hambsch}}
%% 29
        Berto~\textsc{Monard},\altaffilmark{\ref{affil:Monard}}$^,$\altaffilmark{\ref{affil:Monard2}}
%% 27
        Tonny~\textsc{Vanmunster},\altaffilmark{\ref{affil:Vanmunster}}
%% 26
        Yutaka~\textsc{Maeda},\altaffilmark{\ref{affil:Mdy}}
        Ian~\textsc{Miller},\altaffilmark{\ref{affil:Miller}}
%% 24
        Hiroshi~\textsc{Itoh},\altaffilmark{\ref{affil:Ioh}}
        Seiichiro~\textsc{Kiyota},\altaffilmark{\ref{affil:Kis}}
%% 9+alpha
        Keisuke~\textsc{Isogai},\altaffilmark{\ref{affil:Kyoto}}
        Mariko~\textsc{Kimura},\altaffilmark{\ref{affil:Kyoto}}
        Akira~\textsc{Imada},\altaffilmark{\ref{affil:HidaKwasan}}
%% 18
        Tam\'as~\textsc{Tordai},\altaffilmark{\ref{affil:Polaris}}
%% 13
        Hidehiko~\textsc{Akazawa},\altaffilmark{\ref{affil:OUS}}
        Kenji~\textsc{Tanabe},\altaffilmark{\ref{affil:OUS}}
        Noritoshi~\textsc{Otani},\altaffilmark{\ref{affil:OUS}}
        Minako~\textsc{Ogi},\altaffilmark{\ref{affil:OUS}}
        Kazuko~\textsc{Ando},\altaffilmark{\ref{affil:OUS}}
        Naoki~\textsc{Takigawa},\altaffilmark{\ref{affil:OUS}}
        Pavol~A.~\textsc{Dubovsky},\altaffilmark{\ref{affil:Dubovsky}}
        Igor~\textsc{Kudzej},\altaffilmark{\ref{affil:Dubovsky}}
        Sergey~Yu.~\textsc{Shugarov},\altaffilmark{\ref{affil:Sternberg}}$^,$\altaffilmark{\ref{affil:Slovak}}
        Natalia~\textsc{Katysheva},\altaffilmark{\ref{affil:Sternberg}}
        Polina~\textsc{Golysheva},\altaffilmark{\ref{affil:Sternberg}}
        Natalia~\textsc{Gladilina},\altaffilmark{\ref{affil:RASInstAst}}
        Drahomir~\textsc{Chochol},\altaffilmark{\ref{affil:Slovak}}
%% 11
        Peter~\textsc{Starr},\altaffilmark{\ref{affil:Starr}}
%% 10
        Kiyoshi~\textsc{Kasai},\altaffilmark{\ref{affil:Kai}}
        Roger~D.~\textsc{Pickard},\altaffilmark{\ref{affil:BAAVSS}}$^,$\altaffilmark{\ref{affil:Pickard}}
%% 9
        Enrique~de~\textsc{Miguel},\altaffilmark{\ref{affil:Miguel}}$^,$\altaffilmark{\ref{affil:Miguel2}}
%% 8
        Naoto~\textsc{Kojiguchi},\altaffilmark{\ref{affil:OKU}}
        Yuki~\textsc{Sugiura},\altaffilmark{\ref{affil:OKU}}
        Daiki~\textsc{Fukushima},\altaffilmark{\ref{affil:OKU}}
        Eiji~\textsc{Yamada},\altaffilmark{\ref{affil:OKU}}
        Yusuke~\textsc{Uto},\altaffilmark{\ref{affil:OKU}}
        Taku~\textsc{Kamibetsunawa},\altaffilmark{\ref{affil:OKU}}
        Taiki~\textsc{Tatsumi},\altaffilmark{\ref{affil:OKU}}
        Nao~\textsc{Takeda},\altaffilmark{\ref{affil:OKU}}
        Katsura~\textsc{Matsumoto},\altaffilmark{\ref{affil:OKU}}
        Lewis~M.~\textsc{Cook},\altaffilmark{\ref{affil:LewCook}}
%% 7
        Elena~P.~\textsc{Pavlenko},\altaffilmark{\ref{affil:CrAO}}
        Julia~V.~\textsc{Babina},\altaffilmark{\ref{affil:CrAO}}
        Nikolaj~V.~\textsc{Pit},\altaffilmark{\ref{affil:CrAO}}
        Oksana~I.~\textsc{Antonyuk},\altaffilmark{\ref{affil:CrAO}}
        Kirill~A.~\textsc{Antonyuk},\altaffilmark{\ref{affil:CrAO}}
        Aleksei~A.~\textsc{Sosnovskij},\altaffilmark{\ref{affil:CrAO}}
        Aleksei~V.~\textsc{Baklanov},\altaffilmark{\ref{affil:CrAO}}
        Stella~\textsc{Kafka},\altaffilmark{\ref{affil:AAVSO}}
%% 5
        William~\textsc{Stein},\altaffilmark{\ref{affil:Stein}}
%% 4
        Irina~B.~\textsc{Voloshina},\altaffilmark{\ref{affil:Sternberg}}
%% 3
        Javier~\textsc{Ruiz},\altaffilmark{\ref{affil:Ruiz1}}$^,$\altaffilmark{\ref{affil:Ruiz2}}$^,$\altaffilmark{\ref{affil:Ruiz3}}
        Richard~\textsc{Sabo},\altaffilmark{\ref{affil:Sabo}}
        Shawn~\textsc{Dvorak},\altaffilmark{\ref{affil:Dvorak}}
        Geoff~\textsc{Stone},\altaffilmark{\ref{affil:AAVSO}}
%% 2
        Maksim~V.~\textsc{Andreev},\altaffilmark{\ref{affil:Terskol}}$^,$\altaffilmark{\ref{affil:ICUkraine}}
        Sergey~V.~\textsc{Antipin},\altaffilmark{\ref{affil:Sternberg}}$^,$\altaffilmark{\ref{affil:RASInstAst}}
        Alexandra~M.~\textsc{Zubareva},\altaffilmark{\ref{affil:RASInstAst}}$^,$\altaffilmark{\ref{affil:Sternberg}}
        Anna~M.~\textsc{Zaostrojnykh},\altaffilmark{\ref{affil:Kazan}}
        Michael~\textsc{Richmond},\altaffilmark{\ref{affil:RIT}}
        Jeremy~\textsc{Shears},\altaffilmark{\ref{affil:Shears}}$,$\altaffilmark{\ref{affil:BAAVSS}}
        Franky~\textsc{Dubois},\altaffilmark{\ref{affil:Dubois}}
        Ludwig~\textsc{Logie},\altaffilmark{\ref{affil:Dubois}}
        Steve~\textsc{Rau},\altaffilmark{\ref{affil:Dubois}}
        Siegfried~\textsc{Vanaverbeke},\altaffilmark{\ref{affil:Dubois}}
%% 1
        Andrei~\textsc{Simon},\altaffilmark{\ref{affil:Lesniki}}
        Arto~\textsc{Oksanen},\altaffilmark{\ref{affil:Nyrola}}
        William~N.~\textsc{Goff},\altaffilmark{\ref{affil:Goff}}
        Greg~\textsc{Bolt},\altaffilmark{\ref{affil:Bolt}}
        Bart{\l}omiej~\textsc{D\k{e}bski},\altaffilmark{\ref{affil:Debski}}
%% discoveries and related
        Christopher~S.~\textsc{Kochanek},\altaffilmark{\ref{affil:Ohio}}
        Benjamin~\textsc{Shappee},\altaffilmark{\ref{affil:Ohio}}
        Krzysztof~Z.~\textsc{Stanek},\altaffilmark{\ref{affil:Ohio}}
        Jos\'e~L.~\textsc{Prieto},\altaffilmark{\ref{affil:DiegoPortales}}$^,$\altaffilmark{
\ref{affil:Princeton}}
%% outbursts detection
        Rod~\textsc{Stubbings},\altaffilmark{\ref{affil:Stubbings}}
        Eddy~\textsc{Muyllaert},\altaffilmark{\ref{affil:VVSBelgium}}
        Mitsutaka~\textsc{Hiraga},\altaffilmark{\ref{affil:Hrm}}
        Tsuneo~\textsc{Horie},\altaffilmark{\ref{affil:Heo}}
        Patrick~\textsc{Schmeer},\altaffilmark{\ref{affil:Schmeer}}
        Kenji~\textsc{Hirosawa},\altaffilmark{\ref{affil:Hsk}}
}

\authorcount{affil:Kyoto}{
     Department of Astronomy, Kyoto University, Kyoto 606-8502, Japan}
\email{$^*$tkato@kusastro.kyoto-u.ac.jp}

\authorcount{affil:GEOS}{
     Groupe Europ\'een d'Observations Stellaires (GEOS),
     23 Parc de Levesville, 28300 Bailleau l'Ev\^eque, France}

\authorcount{affil:BAV}{
     Bundesdeutsche Arbeitsgemeinschaft f\"ur Ver\"anderliche Sterne
     (BAV), Munsterdamm 90, 12169 Berlin, Germany}

\authorcount{affil:Hambsch}{
     Vereniging Voor Sterrenkunde (VVS), Oude Bleken 12, 2400 Mol, Belgium}

\authorcount{affil:Monard}{
     Bronberg Observatory, Center for Backyard Astrophysics Pretoria,
     PO Box 11426, Tiegerpoort 0056, South Africa}

\authorcount{affil:Monard2}{
     Kleinkaroo Observatory, Center for Backyard Astrophysics Kleinkaroo,
     Sint Helena 1B, PO Box 281, Calitzdorp 6660, South Africa}

\authorcount{affil:Vanmunster}{
     Center for Backyard Astrophysics Belgium, Walhostraat 1A,
     B-3401 Landen, Belgium}

\authorcount{affil:Mdy}{
     Kaminishiyamamachi 12-14, Nagasaki, Nagasaki 850-0006, Japan}

\authorcount{affil:Miller}{
     Furzehill House, Ilston, Swansea, SA2 7LE, UK}

\authorcount{affil:Ioh}{
     Variable Star Observers League in Japan (VSOLJ),
     1001-105 Nishiterakata, Hachioji, Tokyo 192-0153, Japan}

\authorcount{affil:Kis}{
     VSOLJ, 7-1 Kitahatsutomi, Kamagaya, Chiba 273-0126, Japan}

\authorcount{affil:HidaKwasan}{
     Kwasan and Hida Observatories, Kyoto University, Yamashina,
     Kyoto 607-8471, Japan}

\authorcount{affil:Polaris}{
     Polaris Observatory, Hungarian Astronomical Association,
     Laborc utca 2/c, 1037 Budapest, Hungary}

\authorcount{affil:OUS}{
     Department of Biosphere-Geosphere System Science, Faculty of Informatics,
     Okayama University of Science, 1-1 Ridai-cho, Okayama,
     Okayama 700-0005, Japan}

\authorcount{affil:Dubovsky}{
     Vihorlat Observatory, Mierova 4, 06601 Humenne, Slovakia}

\authorcount{affil:Sternberg}{
     Sternberg Astronomical Institute, Lomonosov Moscow State University, 
     Universitetsky Ave., 13, Moscow 119992, Russia}

\authorcount{affil:RASInstAst}{
     Institute of Astronomy, Russian Academy of Sciences,
     Moscow 119017, Russia}

\authorcount{affil:Slovak}{
     Astronomical Institute of the Slovak Academy of Sciences,
     05960 Tatranska Lomnica, Slovakia}

\authorcount{affil:Starr}{
     Warrumbungle Observatory, Tenby, 841 Timor Rd,
     Coonabarabran NSW 2357, Australia}

\authorcount{affil:Kai}{
     Baselstrasse 133D, CH-4132 Muttenz, Switzerland}

\authorcount{affil:BAAVSS}{
     The British Astronomical Association, Variable Star Section (BAA VSS),
     Burlington House, Piccadilly, London, W1J 0DU, UK}

\authorcount{affil:Pickard}{
     3 The Birches, Shobdon, Leominster, Herefordshire, HR6 9NG, UK}

\authorcount{affil:Miguel}{
     Departamento de F\'isica Aplicada, Facultad de Ciencias
     Experimentales, Universidad de Huelva,
     21071 Huelva, Spain}

\authorcount{affil:Miguel2}{
     Center for Backyard Astrophysics, Observatorio del CIECEM,
     Parque Dunar, Matalasca\~nas, 21760 Almonte, Huelva, Spain}

\authorcount{affil:OKU}{
     Osaka Kyoiku University, 4-698-1 Asahigaoka, Osaka 582-8582, Japan}

\authorcount{affil:LewCook}{
     Center for Backyard Astrophysics Concord, 1730 Helix Ct. Concord,
     California 94518, USA}

\authorcount{affil:CrAO}{
     Federal State Budget Scientific Institution ``Crimean Astrophysical
     Observatory of RAS'', Nauchny, 298409, Republic of Crimea}

\authorcount{affil:AAVSO}{
     American Association of Variable Star Observers, 49 Bay State Rd.,
     Cambridge, MA 02138, USA}

\authorcount{affil:Stein}{
     6025 Calle Paraiso, Las Cruces, New Mexico 88012, USA}

\authorcount{affil:Ruiz1}{
     Observatorio de C\'antabria, Ctra. de Rocamundo s/n, Valderredible, 
     39220 Cantabria, Spain}

\authorcount{affil:Ruiz2}{
     Instituto de F\'{\i}sica de Cantabria (CSIC-UC), Avenida Los Castros s/n, 
     E-39005 Santander, Cantabria, Spain}

\authorcount{affil:Ruiz3}{
     Agrupaci\'on Astron\'omica C\'antabria, Apartado 573,
     39080, Santander, Spain}

\authorcount{affil:Sabo}{
     2336 Trailcrest Dr., Bozeman, Montana 59718, USA}

\authorcount{affil:Dvorak}{
     Rolling Hills Observatory, 1643 Nightfall Drive,
     Clermont, Florida 34711, USA}

\authorcount{affil:Terskol}{
     Terskol Branch of Institute of Astronomy, Russian Academy of Sciences,
     361605, Peak Terskol, Kabardino-Balkaria Republic, Russia}

\authorcount{affil:ICUkraine}{
     International Center for Astronomical, Medical and Ecological Research
     of NASU, Ukraine 27 Akademika Zabolotnoho Str. 03680 Kyiv,
     Ukraine}

\authorcount{affil:Kazan}{
     Institute of Physics, Kazan Federal University,
     Ulitsa Kremlevskaya 16a, Kazan 420008, Russia}

\authorcount{affil:RIT}{
     Physics Department, Rochester Institute of Technology, Rochester,
     New York 14623, USA}

\authorcount{affil:Shears}{
     ``Pemberton'', School Lane, Bunbury, Tarporley, Cheshire, CW6 9NR, UK}

\authorcount{affil:Dubois}{
     Public observatory Astrolab Iris, Verbrandemolenstraat 5,
     B 8901 Zillebeke, Belgium}

\authorcount{affil:Lesniki}{
     Taras Shevchenko National University of Kyiv, the Faculty of Physics,
     Astronomy and Space Physics Department, 64/13, Volodymyrska Street, 
     Kyiv 01601, Ukraine}

\authorcount{affil:Nyrola}{
     Hankasalmi observatory, Jyvaskylan Sirius ry, Verkkoniementie 30,
     FI-40950 Muurame, Finland}

\authorcount{affil:Goff}{
     13508 Monitor Ln., Sutter Creek, California 95685, USA}

\authorcount{affil:Bolt}{
     Camberwarra Drive, Craigie, Western Australia 6025, Australia}

\authorcount{affil:Debski}{
     Astronomical Observatory, Jagiellonian University,
     ul. Orla 171 30-244 Krak\'ow, Poland}

\authorcount{affil:Ohio}{
     Department of Astronomy, the Ohio State University, Columbia,
     OH 43210, USA}

\authorcount{affil:DiegoPortales}{
     N\'ucleo de Astronom\'ia de la Facultad de Ingenier\'ia, Universidad
     Diego Portales, Av. Ej\'ercito 441, Santiago, Chile}

\authorcount{affil:Princeton}{
     Department of Astrophysical Sciences, Princeton University,
     NJ 08544, USA}

\authorcount{affil:Stubbings}{
     Tetoora Observatory, 2643 Warragul-Korumburra Road, Tetoora Road,
     Victoria 3821, Australia}

\authorcount{affil:VVSBelgium}{
     Vereniging Voor Sterrenkunde (VVS), Moffelstraat 13 3370
     Boutersem, Belgium}

\authorcount{affil:Hrm}{
     19-27 Higashikannon-cho, Shimonoseki, Yamaguchi 752-0906, Japan}

\authorcount{affil:Heo}{
     759-10 Tokawa, Hadano-shi, Kanagawa 259-1306, Japan}

\authorcount{affil:Schmeer}{
     Bischmisheim, Am Probstbaum 10, 66132 Saarbr\"{u}cken, Germany}

\authorcount{affil:Hsk}{
     216-4 Maeda, Inazawa-cho, Inazawa-shi, Aichi 492-8217, Japan}

%%% end:list of authors


\maketitle

\begin{abstract}
Continuing the project described by \citet{Pdot}, we collected
times of superhump maxima for 128 SU UMa-type dwarf novae 
observed mainly during the 2015--2016 season and characterized
these objects.  The data have improved the distribution of
orbital periods, the relation between the orbital period
and the variation of superhumps, the relation between
period variations and the rebrightening type in WZ Sge-type
objects.  Coupled with new measurements of mass ratios
using growing stages of superhumps, we now have a clearer
and statistically greatly improved evolutionary path
near the terminal stage of evolution of cataclysmic variables.
Three objects (V452 Cas, KK Tel, ASASSN-15cl) appear
to have slowly growing superhumps, which is proposed to
reflect the slow growth of the 3:1 resonance near
the stability border.  ASASSN-15sl, ASASSN-15ux,
SDSS J074859.55$+$312512.6 and CRTS J200331.3$-$284941
are newly identified eclipsing SU UMa-type (or WZ Sge-type)
dwarf novae.  ASASSN-15cy has a short ($\sim$0.050~d)
superhump period and appears to belong to EI Psc-type
objects with compact secondaries having an evolved core.
ASASSN-15gn, ASASSN-15hn, ASASSN-15kh and ASASSN-16bu
are candidate period bouncers with superhump periods
longer than 0.06~d.  We have newly obtained superhump
periods for 79 objects and 13 orbital periods, including
periods from early superhumps.
In order that the future observations will be
more astrophysically beneficial and rewarding
to observers, we propose guidelines how to organize
observations of various superoutbursts.
\end{abstract}

\section{Introduction}

   This paper is one of series of papers \citet{Pdot},
\citet{Pdot2}, \citet{Pdot3}, \citet{Pdot4}, \citet{Pdot5},
\citet{Pdot6} and \citet{Pdot7} dealing with
superhumps in SU UMa-type dwarf novae (DNe).
SU UMa-type dwarf novae are a class of cataclysmic
variables (CVs) which are close binary systems
transferring matter from a low-mass dwarf secondary to
a white dwarf, forming an accretion disk.
In SU UMa-type dwarf novae, two types of outbursts
are seen: normal outbursts and superoutbursts.
During superoutbursts, small-amplitude variations
with period a few percent longer than
the orbital period ($P_{\rm orb}$) called superhumps
are observed.  These superhumps are considered to
be a result of the precession of the eccentric
(or flexing) disk deformed by the tidal instability
at the 3:1 resonance [see e.g. \citet{whi88tidal};
\citet{hir90SHexcess}; \citet{lub91SHa}; \citet{woo11v344lyr};
for general information of CVs, DNe, SU UMa-type 
dwarf novae and superhumps, see e.g. \citet{war95book}].

\begin{figure}
  \begin{center}
%    \FigureFile(80mm,110mm){stagerev.eps}
    \FigureFile(80mm,110mm){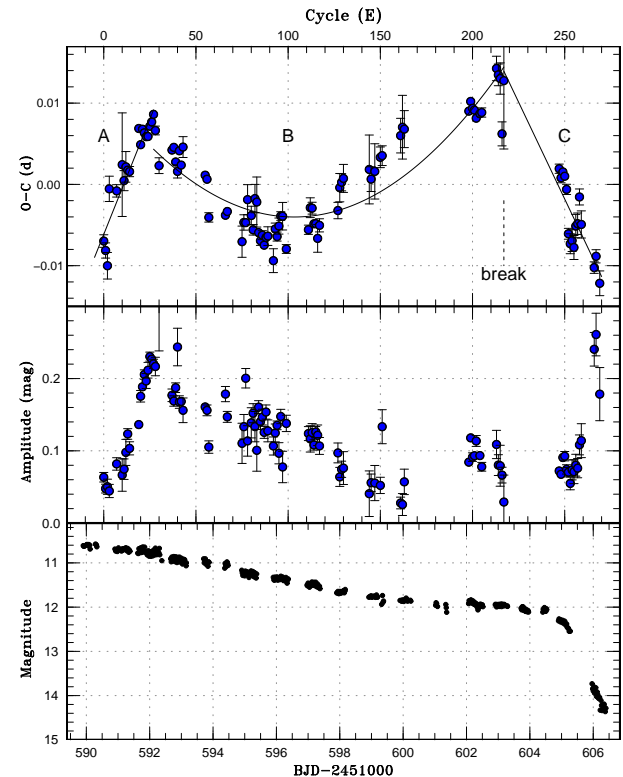}
  \end{center}
  \caption{Representative $O-C$ diagram showing three stages (A--C)
  of $O-C$ variation.  The data were taken from the 2000 superoutburst
  of SW UMa.  (Upper:) $O-C$ diagram.  Three distinct stages
  (A -- evolutionary stage with a longer superhump period, 
  B -- middle stage, and C -- stage after
  transition to a shorter period) and the location of the period break
  between stages B and C are shown. (Middle): Amplitude of superhumps.
  During stage A, the amplitude of the superhumps grew.
  (Lower:) Light curve.
  (Reproduction of figure 1 in \cite{kat13qfromstageA})}
  \label{fig:stagerev}
\end{figure}

   In recent years, it has been demonstrated that
the periods of superhumps systematically vary
during superoutburst and \citet{Pdot} introduced
superhump stages (stages A, B and C):
initial growing stage with a long period (stage A) and
fully developed stage with a systematically
varying period (stage B) and later stage C with a shorter,
almost constant period (see figure \ref{fig:stagerev}).
Although the origin of these stages was unknown 
at the time of \citet{Pdot},
the phenomenon has been repeatedly confirmed
by observations reported in \citet{Pdot2}--\citet{Pdot7}.
Quite recently, partly with the help of Kepler \citep{Kepler}
observations, \citet{osa13v344lyrv1504cyg} proposed
that stage A superhumps reflect the dynamical precession rate
at the 3:1 resonance radius and that the rapid decrease
of the period (stage B) reflects the pressure effect
which has an effect of retrograde precession
(\cite{lub92SH}; \cite{hir93SHperiod};
\cite{mur98SH}; \cite{mon01SH}; \cite{pea06SH}).
\citet{kat13qfromstageA} further extended this
interpretation and confirmed that stage A superhumps
indeed reflect the dynamical precession rate
at the 3:1 resonance radius by using objects with
mass ratios ($q$) established by eclipse observations.
After this physical identification of the superhump stages,
observations of superhumps during superoutbursts
became an important tool not only for diagnosing
the accretion disk but also for obtaining $q$ values,
which are most essential in understanding the nature
of binaries and their evolutions.
Applications of the stage A superhump method
have been numerous: e.g. \citet{kat13j1222}; \citet{nak13j2112j2037};
\citet{ohs14eruma}; \citet{kat14j0902}; \citet{nak14j0754j2304}.

   Outbursts and superoutbursts in SU UMa-type dwarf novae
are considered to be a result of the combination of
thermal and tidal instabilities [thermal-tidal instability (TTI)
model by \citet{osa89suuma}; \citet{osa96review}].
Although there have been claims of other mechanisms
[the enhanced mass-transfer model
\citep{sma91suumamodel} and pure thermal instability model
\citep{can10v344lyr}], it has been demonstrated using
Kepler observations of V1504 Cyg and V344 Lyr that
the TTI model is the best one to explain the observations
\citep{osa13v1504cygKepler}
(see also \cite{osa14v1504cygv344lyrpaper3}).

   In this paper, we report observations of
superhumps and associated phenomena in SU UMa-type
dwarf novae whose superoutbursts were observed mainly
in 2015--2016.
We present basic observational materials
and discussions in relation to individual objects.
Starting from \citet{Pdot6}, we have been intending
these series of papers to be also a source of compiled
information, including historical, of individual dwarf novae.

   The material and methods of analysis are given in
section \ref{sec:obs}, observations and analysis of
individual objects are given in section \ref{sec:individual},
including discussions particular to the objects,
the general discussion is given in section
\ref{sec:discuss} and the summary is given in section
\ref{sec:summary}.  Some tables and figures are available
online only.

\section{Observation and Analysis}\label{sec:obs}

   The data were obtained under campaigns led by 
the VSNET Collaboration \citep{VSNET}.
We also used the public data for some objects from 
the AAVSO International Database\footnote{
   $<$http://www.aavso.org/data-download$>$.
}.
Outburst detections heavily relied on
the ASAS-SN CV patrol \citep{dav15ASASSNCVAAS}\footnote{
   $<$http://cv.asassn.astronomy.ohio-state.edu/$>$.
}, 
Catalina Real-time Transient Survey
(CRTS; \cite{CRTS})\footnote{
   $<$http://nesssi.cacr.caltech.edu/catalina/$>$.
   For the information of the individual Catalina CVs, see
   $<$http://nesssi.cacr.caltech.edu/catalina/AllCV.html$>$.
} in addition to outburst detections reported to
VSNET, AAVSO\footnote{
  $<$https://www.aavso.org/$>$.
}, BAAVSS alert\footnote{
  $<$https://groups.yahoo.com/group/baavss-alert/$>$.
} and cvnet-outburst.\footnote{
  $<$https://groups.yahoo.com/neo/groups/cvnet-outburst/$>$.
}
There were some detections by the MASTER network
\citep{MASTER}.

   The majority of the data were acquired
by time-resolved CCD photometry by using 30cm-class telescopes
located world-wide.  The details of these observations
will be presented in future papers dealing with analysis
and discussion on individual objects of interest.
The list of outbursts and observers is summarized in 
table \ref{tab:outobs}.
The data analysis was performed just in the same way described
in \citet{Pdot} and \citet{Pdot6} and we mainly used
R software\footnote{
   The R Foundation for Statistical Computing:\\
   $<$http://cran.r-project.org/$>$.
} for data analysis.
In de-trending the data, we used both lower (1--3rd order)
polynomial fitting and locally-weighted polynomial regression 
(LOWESS: \cite{LOWESS}).
The times of superhumps maxima were determined by
the template fitting method as described in \citet{Pdot}.
The times of all observations are expressed in 
barycentric Julian days (BJD).

   The abbreviations used in this paper are the same
as in \citet{Pdot6}: $P_{\rm orb}$ means
the orbital period and $\epsilon \equiv P_{\rm SH}/P_{\rm orb}-1$ for 
the fractional superhump excess.   Following \citet{osa13v1504cygKepler},
the alternative fractional superhump excess in the frequency unit
$\epsilon^* \equiv 1-P_{\rm orb}/P_{\rm SH}-1 = \epsilon/(1+\epsilon)$
has been introduced because this fractional superhump excess
is a direct measure of the precession rate.  We therefore
used $\epsilon^*$ in discussing the precession rate.

   We used phase dispersion minimization (PDM; \cite{PDM})
for period analysis and 1$\sigma$ errors for the PDM analysis
was estimated by the methods of \citet{fer89error} and \citet{Pdot2}.
We introduced a variety of bootstrapping in
estimating the robustness of the result of the PDM analysis
since \citet{Pdot3}.
We typically analyzed 100 samples which randomly contain 50\% of
observations, and performed PDM analysis for these samples.
The bootstrap result is shown as a form of 90\% confidence intervals
in the resultant PDM $\theta$ statistics.

   If this paper provides the first solid presentation of
a new SU UMa-type classification, we provide the result
of PDM period analysis and averaged superhump profile.

   The resultant $P_{\rm SH}$, $P_{\rm dot}$ and other parameters
are listed in table \ref{tab:perlist} in same format as in
\citet{Pdot}.  The definitions of parameters $P_1, P_2, E_1, E_2$
and $P_{\rm dot}$ are the same as in \citet{Pdot}:
$P_1$ and $P_2$ represent periods in stage B and C, respectively,
and $E_1$ and $E_2$ represent intervals (in cycle numbers)
to determine $P_1$ and $P_2$, respectively.\footnote{
   The intervals ($E_1$ and $E_2$) for the stages B and C given in the table
   sometimes overlap because there is sometimes observational
   ambiguity (usually due to the lack of observations)
   in determining the stages.
}
Comparisons of $O-C$ diagrams between different
superoutbursts are also presented whenever available,
since this comparison was one of the main motivations
in of these series papers.
In drawing combined $O-C$ diagrams, we usually used
$E=$0 for the start of the superoutburst, which usually
refers to the first positive detection of the outburst.
This epoch usually has an accuracy of $\sim$1 d for
well-observed objects, and if the outburst was not sufficiently
observed, we mentioned in the figure caption how to estimate
$E$ in such an outburst.
In some cases, this $E=$0 is defined as the appearance
of superhumps.  This treatment is necessary since
some objects have a long waiting time before
appearance of superhumps.
Combined $O-C$ diagrams also help identifying
superhump stages particularly when observations are
insufficient.
We also note that there is sometimes an ambiguity in
selecting the true period among aliases.  In some
cases, this can be resolved by the help of
the $O-C$ analysis.  The procedure and example
are shown in subsection 2.2 in \citet{Pdot7}.

   We also present $O-C$ diagrams and 
light curves especially for WZ Sge-type dwarf novae.
WZ Sge-type dwarf novae are a subclass of SU UMa-type
dwarf novae characterized by the presence of early superhumps.
They are seen during the early stages of superoutburst,
and have period close to the orbital
periods (\cite{kat96alcom}; \cite{kat02wzsgeESH}; 
\cite{osa02wzsgehump}; \cite{kat15wzsge}).
These early superhumps are considered to be a result
of the 2:1 resonance \citep{osa02wzsgehump}.
These objects usually show very rare outbursts
(once in several years to decades) and often have
complex light curves \citep{kat15wzsge} and are
of special astrophysical interest since the origin
of the complex light curves, including repetitive
rebrightenings, is not well understood.
They receive special attention since they are considered
to represent the terminal stage of CV evolution and
they may have brown-dwarf secondaries.
We used the period of early superhumps as the approximate
orbital period (\cite{Pdot6}; \cite{kat15wzsge}).

   In figures, the points are accompanied by 1$\sigma$
error bars whenever available, which are omitted
when the error is smaller than the plot mark.

   We used the same terminology of superhumps summarized in
\citet{Pdot3}.  We especially call attention to
the term ``late superhumps''.  Although this term has been
used to express various phenomena, we only used
the concept of ``traditional'' late superhumps when
there is an $\sim$0.5 phase shift
[\citet{vog83lateSH}; see also table 1 in \citet{Pdot3} 
for various types of superhumps], 
since we suspect that many of the past
claims of detections of ``late superhumps'' were likely
stage C superhumps before it became apparent that
there are complex structures in the $O-C$ diagrams
of superhumps (see discussion in \cite{Pdot}).

   For objects detected in CRTS, we preferably used the names 
provided in \citet{dra14CRTSCVs}.
If these names are not yet available,
we used the International Astronomical
Union (IAU)-format names provided by the CRTS team 
in the public data release\footnote{
  $<$http://nesssi.cacr.caltech.edu/DataRelease/$>$.
}
As in \citet{Pdot}, we have used coordinate-based 
optical transient (OT) designations for some objects, such as
apparent dwarf nova candidates reported in
the Transient Objects Confirmation Page of
the Central Bureau for Astronomical Telegrams\footnote{
   $<$http://www.cbat.eps.harvard.edu/unconf/tocp.html$>$.
} and CRTS objects without registered designations
in \citet{dra14CRTSCVs} or in the CRTS public data release
and listed the original identifiers in table \ref{tab:outobs}.

   We provided coordinates from astrometric catalogs
for ASAS-SN \citep{ASASSN} CVs and two objects
without coordinate-based names other than listed
in the General Catalog of Variable Stars \citep{GCVS}
in table \ref{tab:coord}.
We used Sloan Digital Sky Survey (SDSS, \cite{SDSS9}),
the Initial Gaia Source List (IGSL, \cite{IGSL}) and 
Guide Star Catalog 2.3.2 (GSC 2.3.2, \cite{GSC232})
and some other catalogs.
The coordinates used in this paper are J2000.0.
We also supplied SDSS $g$ magnitudes and GALEX
NUV magnitudes when counterparts are present.

\begin{table*}
\caption{List of Superoutbursts.}\label{tab:outobs}
\begin{center}
\begin{tabular}{ccccl}
\hline
Subsection & Object & Year & Observers or references\commenta & ID\commentb \\
\hline
\ref{obj:kvand}      & KV And          & 2015 & DPV, RPc, Rui & \\
\ref{obj:egaqr}      & EG Aqr          & 2015 & Aka, Kis & \\
\ref{obj:nncam}      & NN Cam          & 2015 & DPV, NKa, Aka, Trt & \\
\ref{obj:v452cas}    & V452 Cas        & 2016 & RPc, IMi, Trt & \\
\ref{obj:v1040cen}   & V1040 Cen       & 2015 & HaC & \\
\ref{obj:pucma}      & PU CMa          & 2016 & GBo, Aka, Kis, SPE & \\
\ref{obj:alcom}      & AL Com          & 2015 & \citet{kim16alcom} & \\
\ref{obj:vwcrb}      & VW CrB          & 2015 & Kis, Dub & \\
\ref{obj:v550cyg}    & V550 Cyg        & 2015 & Mdy & \\
--                   & V1006 Cyg       & 2015 & \citet{kat16v1006cyg} & \\
\ref{obj:v1028cyg}   & V1028 Cyg       & 2016 & Aka, RPc, Trt & \\
\ref{obj:v1113cyg}   & V1113 Cyg       & 2015 & Trt, Ter, Kai, Shu, COO, & \\
                     &                 &      & DPV, IMi, Kis, Rui & \\
\ref{obj:hodel}      & HO Del          & 2015 & Trt & \\
\ref{obj:aqeri}      & AQ Eri          & 2016 & Aka, Kis, Mdy, AAVSO & \\
\ref{obj:axfor}      & AX For          & 2015 & HaC, Aka & \\
\ref{obj:v844her}    & V844 Her        & 2015 & OKU, DPV, IMi & \\
\ref{obj:rzleo}      & RZ Leo          & 2016 & Kis, deM, HaC, Aka, AAVSO, & \\
                     &                 &      & CRI, Mdy, IMi, SWI, DKS, & \\
                     &                 &      & SRI, SGE, Trt, RPc & \\
\ref{obj:mmhya}      & MM Hya          & 2015 & COO & \\
\ref{obj:v585lyr}    & V585 Lyr        & 2015 & RPc, DPV, OKU & \\
\ref{obj:v2051oph}   & V2051 Oph       & 2015 & HaC, Kis, Aka & \\
\hline
  \multicolumn{5}{l}{\parbox{500pt}{\commenta Key to observers:
Aka (H. Akazawa, OUS),
Ant (S. Antipin and A. Zubareva team),
COO (L. Cook),
CRI (Crimean Astrophys. Obs.),
Deb (B. Debski),
deM (E. de Miguel),
DKS\commentc (S. Dvorak),
DPV (P. Dubovsky),
Dub (F. Dubois team),
GBo (G. Bolt),
GFB\commentc (W. Goff),
HMB (F.-J. Hambsch),
HaC (F.-J. Hambsch, remote obs. in Chile),
Han (Hankasalmi Obs., by A. Oksanen),
IMi\commentc (I. Miller),
Ioh (H. Itoh),
JSh\commentc (J. Shears),
KU (Kyoto U., campus obs.),
Kai (K. Kasai),
Kis (S. Kiyota),
Les (Lesniki Obs.),
MLF (B. Monard),
Mdy (Y. Maeda),
NKa (N. Katysheva and S. Shugarov),
OKU (Osaya Kyoiku U.),
OUS (Okayama U. of Science),
RIT (M. Richmond),
RPc\commentc (R. Pickard),
Rui (J. Ruiz),
SGE\commentc (G. Stone),
SPE\commentc (P. Starr),
SRI\commentc (R. Sabo),
SWI\commentc (W. Stein),
Shu (S. Shugarov team),
Ter (Terskol Obs.),
Trt (T. Tordai),
Van (T. Vanmunster),
Vol (I. Voloshina),
AAVSO (AAVSO database)
}} \\
  \multicolumn{5}{l}{\commentb Original identifications, discoverers or data source.} \\
  \multicolumn{5}{l}{\commentc Inclusive of observations from the AAVSO database.} \\
\end{tabular}
\end{center}
\end{table*}

\addtocounter{table}{-1}
\begin{table*}
\caption{List of Superoutbursts (continued).}
\begin{center}
\begin{tabular}{ccccl}
\hline
Subsection & Object & Year & Observers or references\commenta & ID\commentb \\
\hline
\ref{obj:v368peg}    & V368 Peg        & 2015 & Trt & \\
\ref{obj:v650peg}    & V650 Peg        & 2015 & HaC, Shu, Mdy, Aka, Kai, & \\
                     &                 &      & Trt, Ioh & \\
\ref{obj:typsc}      & TY Psc          & 2015 & Trt, Deb, Kis & \\
\ref{obj:puper}      & PU Per          & 2015 & KU, Mdy, Vol, IMi, Kis & \\
\ref{obj:puper}      & QY Per          & 2015 & Trt, Shu & \\
\ref{obj:v493ser}    & V493 Ser        & 2015 & DPV, HaC, IMi & \\
\ref{obj:v1212tau}   & V1212 Tau       & 2016 & Mdy & \\
\ref{obj:kktel}      & KK Tel          & 2015 & HaC, SPE & \\
\ref{obj:ciuma}      & CI UMa          & 2016 & SGE, Trt & \\
\ref{obj:ksuma}      & KS UMa          & 2015 & Kis, Ioh, AAVSO, Aka, Kai & \\
\ref{obj:mruma}      & MR UMa          & 2015 & DPV & \\
\ref{obj:nsv2026}    & NSV 2026        & 2015 & AAVSO, IMi & \\
                     &                 & 2016 & JSh, RPc, IMi, KU, Dub, AAVSO & \\
\ref{obj:asassn13ah} & ASASSN-13ah     & 2016 & Van, IMi & \\
\ref{obj:asassn13ak} & ASASSN-13ak     & 2015 & deM, KU, RPc, IMi & \\
\ref{obj:asassn13az} & ASASSN-13az     & 2016 & Van & \\
\ref{obj:asassn14ca} & ASASSN-14ca     & 2015 & IMi & \\
--                   & ASASSN-14cc     & 2014 & \citet{kat15asassn14cc} & \\
\ref{obj:asassn14dh} & ASASSN-14dh     & 2015 & HaC, Mdy, Ioh & \\
\ref{obj:asassn14fz} & ASASSN-14fz     & 2015 & HaC & \\
\ref{obj:asassn14le} & ASASSN-14le     & 2014 & KU & \\
\ref{obj:asassn15cl} & ASASSN-15cl     & 2016 & Kis, HaC, Ioh & \\
\ref{obj:asassn15cy} & ASASSN-15cy     & 2015 & MLF, HaC, deM, Kis & \\
\ref{obj:asassn15dh} & ASASSN-15dh     & 2015 & Van, SWI & \\
\ref{obj:asassn15dp} & ASASSN-15dp     & 2015 & DPV, Van, Trt, IMi, SWI & \\
\ref{obj:asassn15dr} & ASASSN-15dr     & 2015 & HaC, MLF & \\
\ref{obj:asassn15ea} & ASASSN-15ea     & 2015 & deM, Van & \\
\ref{obj:asassn15ee} & ASASSN-15ee     & 2015 & MLF, HaC, SPE & \\
\ref{obj:asassn15eh} & ASASSN-15eh     & 2015 & MLF & \\
\ref{obj:asassn15ev} & ASASSN-15ev     & 2015 & MLF, HaC & \\
\ref{obj:asassn15fo} & ASASSN-15fo     & 2015 & MLF, HaC & \\
\ref{obj:asassn15fu} & ASASSN-15fu     & 2015 & MLF, HaC & \\
\ref{obj:asassn15gf} & ASASSN-15gf     & 2015 & Van, Kai & \\
\hline
\end{tabular}
\end{center}
\end{table*}

\addtocounter{table}{-1}
\begin{table*}
\caption{List of Superoutbursts (continued).}
\begin{center}
\begin{tabular}{ccccl}
\hline
Subsection & Object & Year & Observers or references\commenta & ID\commentb \\
\hline
\ref{obj:asassn15gh} & ASASSN-15gh     & 2015 & HaC, MLF & \\
\ref{obj:asassn15gi} & ASASSN-15gi     & 2015 & MLF, HaC & \\
\ref{obj:asassn15gn} & ASASSN-15gn     & 2015 & MLF, HaC, COO & \\
\ref{obj:asassn15gq} & ASASSN-15gq     & 2015 & Van, IMi, COO & \\
\ref{obj:asassn15gs} & ASASSN-15gs     & 2015 & MLF & \\
\ref{obj:asassn15hd} & ASASSN-15hd     & 2015 & Van, OKU, RIT, DPV, IMi, & \\
                     &                 &      & Kai, Ioh, deM, CRI, Trt, & \\
                     &                 &      & GFB, HMB, SRI, COO, DKS, & \\
                     &                 &      & HaC & \\
\ref{obj:asassn15hl} & ASASSN-15hl     & 2015 & MLF & \\
\ref{obj:asassn15hm} & ASASSN-15hm     & 2015 & HaC, Kis, COO & \\
\ref{obj:asassn15hn} & ASASSN-15hn     & 2015 & MLF, HaC, Ioh, COO, Kis, & \\
                     &                 &      & KU, deM & \\
\ref{obj:asassn15ia} & ASASSN-15ia     & 2015 & HaC & \\
\ref{obj:asassn15ie} & ASASSN-15ie     & 2015 & HaC & \\
\ref{obj:asassn15iv} & ASASSN-15iv     & 2015 & HaC & \\
\ref{obj:asassn15iz} & ASASSN-15iz     & 2015 & HaC & \\
--                   & ASASSN-15jd     & 2015 & \citet{kim16asassn15jd} & \\
\ref{obj:asassn15jj} & ASASSN-15jj     & 2015 & HaC, MLF & \\
\ref{obj:asassn15kf} & ASASSN-15kf     & 2015 & MLF, HaC & \\
\ref{obj:asassn15kh} & ASASSN-15kh     & 2015 & MLF, HaC, SPE & \\
\ref{obj:asassn15le} & ASASSN-15le     & 2015 & Van, HaC & \\
\ref{obj:asassn15lt} & ASASSN-15lt     & 2015 & HaC, SPE & \\
\ref{obj:asassn15mb} & ASASSN-15mb     & 2015 & HaC & \\
\ref{obj:asassn15mt} & ASASSN-15mt     & 2015 & Kis, Shu, Ant, Ioh, COO, IMi & \\
\ref{obj:asassn15na} & ASASSN-15na     & 2015 & MLF, HaC & \\
\ref{obj:asassn15ni} & ASASSN-15ni     & 2015 & OKU, DPV, IMi, Van, Kis, & \\
                     &                 &      & Shu, SPE, Ioh, Trt, AAVSO & \\
\ref{obj:asassn15nl} & ASASSN-15nl     & 2015 & Shu, Van, Trt & \\
\ref{obj:asassn15ob} & ASASSN-15ob     & 2015 & CRI, HaC & \\
\ref{obj:asassn15oj} & ASASSN-15oj     & 2015 & MLF & \\
\ref{obj:asassn15ok} & ASASSN-15ok     & 2015 & MLF, HaC & \\
\hline
\end{tabular}
\end{center}
\end{table*}

\addtocounter{table}{-1}
\begin{table*}
\caption{List of Superoutbursts (continued).}
\begin{center}
\begin{tabular}{ccccl}
\hline
Subsection & Object & Year & Observers or references\commenta & ID\commentb \\
\hline
\ref{obj:asassn15pi} & ASASSN-15pi     & 2015 & Van & \\
--                   & ASASSN-15po     & 2015 & \Namekataprep & \\
\ref{obj:asassn15pu} & ASASSN-15pu     & 2015 & MLF, SPE, HaC & \\
\ref{obj:asassn15qe} & ASASSN-15qe     & 2015 & KU, Mdy, Van, Ioh & \\
\ref{obj:asassn15ql} & ASASSN-15ql     & 2015 & HaC & \\
\ref{obj:asassn15qo} & ASASSN-15qo     & 2015 & IMi, Van, Ioh & \\
\ref{obj:asassn15qq} & ASASSN-15qq     & 2015 & MLF, HaC & \\
\ref{obj:asassn15rf} & ASASSN-15rf     & 2015 & HaC & \\
\ref{obj:asassn15rj} & ASASSN-15rj     & 2015 & Mdy, KU, CRI & \\
\ref{obj:asassn15ro} & ASASSN-15ro     & 2015 & Mdy & \\
\ref{obj:asassn15rr} & ASASSN-15rr     & 2015 & MLF & \\
\ref{obj:asassn15rs} & ASASSN-15rs     & 2015 & Mdy, Ioh, Trt, Kis & \\
\ref{obj:asassn15ry} & ASASSN-15ry     & 2015 & Kis, OUS, Ioh & \\
\ref{obj:asassn15sc} & ASASSN-15sc     & 2015 & Ioh, Rui, Mdy, SWI, DPV, & \\
                     &                 &      & Van, CRI, Kis, OUS, Les, & \\
                     &                 &      & Kai, IMi & \\
\ref{obj:asassn15sd} & ASASSN-15sd     & 2015 & HaC, MLF & \\
\ref{obj:asassn15se} & ASASSN-15se     & 2015 & OKU, Van, Kis, Rui, IMi, & \\
                     &                 &      & HaC & \\
\ref{obj:asassn15sl} & ASASSN-15sl     & 2015 & Shu, Ioh, Trt, Van, IMi, & \\
                     &                 &      & CRI & \\
\ref{obj:asassn15sn} & ASASSN-15sn     & 2015 & Kai, Mdy & \\
\ref{obj:asassn15sp} & ASASSN-15sp     & 2015 & HaC & \\
\ref{obj:asassn15su} & ASASSN-15su     & 2015 & Trt & \\
\ref{obj:asassn15sv} & ASASSN-15sv     & 2015 & Kai, Shu & \\
\ref{obj:asassn15ud} & ASASSN-15ud     & 2015 & Mdy & \\
\ref{obj:asassn15uj} & ASASSN-15uj     & 2015 & MLF, HaC, SPE & \\
\ref{obj:asassn15ux} & ASASSN-15ux     & 2015 & deM, Ioh, KU, Mdy, Van, & \\
                     &                 &      & Kis, Shu, IMi & \\
\ref{obj:asassn16af} & ASASSN-16af     & 2016 & Mdy, Van, HaC, Ioh & \\
\ref{obj:asassn16ag} & ASASSN-16ag     & 2016 & Mdy, IMi, Van, Ioh & \\
\ref{obj:asassn16ao} & ASASSN-16ao     & 2016 & MLF & \\
\ref{obj:asassn16aq} & ASASSN-16aq     & 2016 & IMi & \\
\hline
\end{tabular}
\end{center}
\end{table*}

\addtocounter{table}{-1}
\begin{table*}
\caption{List of Superoutbursts (continued).}
\begin{center}
\begin{tabular}{ccccl}
\hline
Subsection & Object & Year & Observers or references\commenta & ID\commentb \\
\hline
\ref{obj:asassn16bh} & ASASSN-16bh     & 2016 & MLF, SPE, DKS, Kis, Mdy, & \\
                     &                 &      & HaC, Ioh & \\
\ref{obj:asassn16bi} & ASASSN-16bi     & 2016 & MLF, SPE, HaC & \\
\ref{obj:asassn16bu} & ASASSN-16bu     & 2016 & OKU, Ioh, Kis, IMi, Van, & \\
                     &                 &      & Mdy, deM & \\
\ref{obj:asassn16de} & ASASSN-16de     & 2016 & Van, HaC & \\
\ref{obj:j0819}      & CRTS J081936    & 2015 & Kai & CRTS J081936.1$+$191540 \\
\ref{obj:j0959}      & CRTS J095926    & 2015 & MLF, HaC & CRTS J095926.4$-$160147 \\
\ref{obj:j1200}      & CRTS J120052    & 2011 & \citet{Pdot3} & CRTS J120052.9$-$152620 \\
                     &                 & 2016 & OKU, HaC & \\
\ref{obj:j1631}      & CRTS J163120    & 2015 & RIT & CRTS J163120.9$+$103134 \\
\ref{obj:j2003}      & CRTS J200331    & 2015 & HaC & CRTS J200331.3$-$284941 \\
\ref{obj:j2125}      & CRTS J212521    & 2015 & Shu, Ter & CRTS J212521.8$-$102627 \\
\ref{obj:j2147}      & CRTS J214738    & 2015 & Ioh & CRTS J214738.4$+$244554 \\
\ref{obj:j2218}      & CSS J221822     & 2015 & Mdy, Van, Vol & CSS120812:221823$+$344509 \\
\ref{obj:dde26}      & DDE 26          & 2015 & Kai, Ioh, Mdy & \\
\ref{obj:j2305}      & IPHAS J230538   & 2015 & CRI, IMi, Van & IPHAS J230538.39$+$652158.7 \\
\ref{obj:j0038}      & MASTER J003831  & 2016 & HaC & MASTER OT J003831.10$-$640313.7 \\
\ref{obj:j0733}      & MASTER J073325  & 2016 & Mdy, Van, DPV, RPc, NKa & MASTER OT J073325.52$+$373744.9 \\
\ref{obj:j1202}      & MASTER J120251  & 2015 & MLF, HaC, SPE & MASTER OT J120251.56$-$454116.7 \\
\ref{obj:j1313}      & MASTER J131320  & 2016 & Van, SGE & MASTER OT J131320.24$+$692649.1 \\
\ref{obj:j1815}      & MASTER J181523  & 2015 & Van & MASTER OT J181523.78$+$692037.4 \\
\ref{obj:j2126}      & MASTER J212624  & 2015 & Shu, Ioh & MASTER OT J212624.16$+$253827.2 \\
\ref{obj:n080829a}   & N080829A        & 2015 & KU, Mdy, OUS, SWI, Vol, & \\
                     &                 &      & Ioh & \\
\ref{obj:j1914}      & OT J191443      & 2015 & Mdy, OKU & OT J191443.6$+$605214 \\
--                   & PM J03338$+$3320 & 2015 & \citet{kat16j0333} & \\
\ref{obj:j0748}      & SDSS J074859    & 2015 & Mdy, RPc, Kis & SDSS J074859.55$+$312512.6 \\
\ref{obj:j1457}      & SDSS J145758    & 2015 & Vol, JSh, AAVSO, Shu, IMi, & SDSS J145758.21$+$514807.9 \\
                     &                 &      & SRI, Han, deM, Mdy, RPc, Ioh & \\
\ref{obj:j1642}      & SDSS J164248    & 2016 & HaC, Van & SDSS J164248.52$+$134751.4 \\
\hline
\end{tabular}
\end{center}
\end{table*}

\begin{table*}
\caption{Coordinates of objects without coordinate-based names.}\label{tab:coord}
\begin{center}
\begin{tabular}{cccccc}
\hline
Object & Right Ascention & Declination & Source\commenta & SDSS $g$ & GALEX NUV \\
\hline
ASASSN-13ah & \timeform{18h 32m 11.37s} & \timeform{+61D 55' 05.6''} & IGSL & -- & 20.6(2) \\
ASASSN-13ak & \timeform{17h 48m 27.88s} & \timeform{+50D 50' 39.7''} & SDSS & 19.9 & -- \\
ASASSN-13az & \timeform{18h 42m 58.18s} & \timeform{+73D 42' 28.4''} & ASAS-SN\commentb & -- & 20.9(3) \\
ASASSN-14ca & \timeform{23h 53m 13.22s} & \timeform{+27D 42' 01.8''} & SDSS\commentb & 20.6 & -- \\
ASASSN-14dh & \timeform{21h 23m 25.65s} & \timeform{-15D 39' 54.3''} & IGSL & -- & 19.7(1) \\
ASASSN-14fz & \timeform{19h 00m 05.25s} & \timeform{-49D 30' 34.4''} & IGSL & -- & -- \\
ASASSN-14le & \timeform{21h 43m 21.84s} & \timeform{+73D 41' 12.9''} & SDSS & 22.4 & -- \\
ASASSN-15cl & \timeform{07h 39m 51.35s} & \timeform{-15D 41' 00.3''} & IGSL & -- & -- \\
ASASSN-15cy & \timeform{08h 11m 50.53s} & \timeform{-12D 27' 51.5''} & Kis\commentb & -- & -- \\
ASASSN-15dh & \timeform{04h 59m 55.75s} & \timeform{+77D 11' 17.9''} & IGSL & -- & 17.44(3) \\
ASASSN-15dp & \timeform{04h 49m 32.28s} & \timeform{+36D 05' 14.0''} & GSC2.3.2 & -- & -- \\
ASASSN-15dr & \timeform{11h 05m 49.38s} & \timeform{-42D 41' 15.9''} & IGSL & -- & -- \\
ASASSN-15ea & \timeform{18h 50m 50.59s} & \timeform{+40D 44' 06.0''} & deM & -- & -- \\
ASASSN-15ee & \timeform{06h 36m 17.49s} & \timeform{-31D 14' 43.6''} & IGSL & -- & -- \\
ASASSN-15eh & \timeform{17h 53m 51.25s} & \timeform{-64D 17' 40.2''} & GSC2.3.2 & -- & -- \\
ASASSN-15ev & \timeform{07h 38m 19.36s} & \timeform{-82D 50' 40.2''} & IGSL & -- & -- \\
ASASSN-15fo & \timeform{12h 39m 56.28s} & \timeform{-47D 16' 23.9''} & IGSL & -- & -- \\
ASASSN-15fu & \timeform{11h 10m 46.21s} & \timeform{-23D 37' 44.2''} & IGSL & -- & 21.6(2) \\
ASASSN-15gf & \timeform{06h 10m 04.01s} & \timeform{+12D 40' 08.5''} & IPHAS DR2 & -- & -- \\
ASASSN-15gh & \timeform{17h 48m 16.13s} & \timeform{-57D 33' 16.3''} & ASAS-SN & -- & -- \\
ASASSN-15gi & \timeform{09h 11m 51.74s} & \timeform{-77D 56' 33.8''} & IGSL & -- & -- \\
ASASSN-15gn & \timeform{15h 19m 29.60s} & \timeform{-24D 40' 00.7''} & GSC2.3.2 & -- & -- \\
ASASSN-15gq & \timeform{10h 15m 11.32s} & \timeform{+81D 24' 17.5''} & ASAS-SN\commentb & 21.6? & -- \\
ASASSN-15gs & \timeform{13h 59m 17.48s} & \timeform{-37D 52' 42.2''} & GSC2.3.2 & -- & -- \\
ASASSN-15hd & \timeform{17h 31m 25.45s} & \timeform{+27D 54' 28.5''} & SDSS & 21.4--21.9 & -- \\
ASASSN-15hl & \timeform{05h 53m 30.40s} & \timeform{-48D 06' 23.2''} & GSC2.3.2 & -- & -- \\
\hline
  \multicolumn{6}{l}{\parbox{400pt}{\commenta source of the coordinates:
2MASS (2MASS All-Sky Catalog of Point Sources; \cite{2MASS}),
ASAS-SN (ASAS-SN measurements),
Gaia (Gaia measurements),
GSC2.3.2 (The Guide Star Catalog, Version 2.3.2, \cite{GSC232}),
IGSL (The Initial Gaia Source List 3, \cite{IGSL}),
IPHAS DR2 (INT/WFC Photometric H$\alpha$ Survey, \cite{wit08IPHAS}),
KIC (The Kepler Input Catalog, \cite{KeplerInputCatalog}),
SDSS (The SDSS Photometric Catalog, Release 9, \cite{SDSS9}),
USNO-A2.0 (The USNO-A2.0 Catalogue, \cite{USNOA20}),
the others are observers' symbols (see table \ref{tab:outobs}).
}} \\
  \multicolumn{5}{l}{\commentb See text for more details.} \\
\end{tabular}
\end{center}
\end{table*}

\addtocounter{table}{-1}
\begin{table*}
\caption{Coordinates of objects without coordinate-based names (continued).}
\begin{center}
\begin{tabular}{cccccc}
\hline
Object & Right Ascention & Declination & Source\commenta & SDSS $g$ & GALEX NUV \\
\hline
ASASSN-15hm & \timeform{11h 00m 38.05s} & \timeform{-11D 56' 46.4''} & IGSL & -- & 21.7(3) \\
ASASSN-15hn & \timeform{09h 07m 05.42s} & \timeform{-10D 42' 45.4''} & GSC2.3.2 & -- & -- \\
ASASSN-15ia & \timeform{18h 10m 33.83s} & \timeform{-53D 49' 45.9''} & IGSL & -- & 21.1(3) \\
ASASSN-15ie & \timeform{20h 36m 44.20s} & \timeform{-13D 11' 56.5''} & IGSL & -- & 21.4(2) \\
ASASSN-15iv & \timeform{14h 36m 15.50s} & \timeform{-36D 04' 17.0''} & IGSL & -- & -- \\
ASASSN-15iz & \timeform{13h 12m 13.26s} & \timeform{-42D 18' 00.4''} & ASAS-SN & -- & -- \\
ASASSN-15jj & \timeform{19h 17m 56.80s} & \timeform{-56D 57' 58.1''} & IGSL & -- & -- \\
ASASSN-15kf & \timeform{15h 38m 38.23s} & \timeform{-30D 35' 49.4''} & IGSL & -- & -- \\
ASASSN-15kh & \timeform{10h 38m 59.93s} & \timeform{-36D 54' 41.5''} & ASAS-SN & -- & -- \\
ASASSN-15le & \timeform{18h 21m 13.93s} & \timeform{+17D 09' 16.2''} & IGSL & -- & 19.5(1) \\
ASASSN-15lt & \timeform{20h 03m 59.59s} & \timeform{-39D 28' 30.7''} & IGSL & -- & 20.6(2) \\
ASASSN-15mb & \timeform{02h 52m 48.21s} & \timeform{-39D 59' 11.3''} & IGSL & -- & -- \\
ASASSN-15mt & \timeform{19h 12m 35.54s} & \timeform{+50D 34' 31.3''} & IGSL & -- & -- \\
ASASSN-15na & \timeform{19h 19m 08.77s} & \timeform{-49D 45' 41.7''} & IGSL & -- & 21.0(2) \\
ASASSN-15ni & \timeform{18h 39m 57.96s} & \timeform{+22D 21' 18.7''} & ASAS-SN & -- & -- \\
ASASSN-15nl & \timeform{14h 14m 59.70s} & \timeform{+38D 35' 47.8''} & SDSS & 19.3 & 20.1(1) \\
ASASSN-15ob & \timeform{16h 50m 59.48s} & \timeform{+01D 20' 06.5''} & SDSS & 21.1 & -- \\
ASASSN-15oj & \timeform{13h 45m 18.90s} & \timeform{-36D 30' 15.0''} & IGSL & -- & -- \\
ASASSN-15ok & \timeform{00h 24m 30.76s} & \timeform{-66D 35' 52.7''} & 2MASS & -- & 20.7(2) \\
ASASSN-15pi & \timeform{18h 50m 22.29s} & \timeform{+74D 56' 03.3''} & ASAS-SN & -- & -- \\
ASASSN-15pu & \timeform{21h 11m 04.70s} & \timeform{-39D 56' 33.9''} & GSC2.3.2 & -- & -- \\
ASASSN-15qe & \timeform{22h 53m 44.43s} & \timeform{+35D 53' 54.4''} & GSC2.3.2 & -- & -- \\
ASASSN-15ql & \timeform{20h 21m 56.11s} & \timeform{-85D 59' 09.5''} & IGSL & -- & -- \\
ASASSN-15qo & \timeform{18h 11m 22.86s} & \timeform{+22D 42' 06.7''} & SDSS & 22.0 & -- \\
ASASSN-15qq & \timeform{22h 54m 35.69s} & \timeform{-36D 12' 27.6''} & IGSL & -- & 21.0(2) \\
ASASSN-15rf & \timeform{21h 57m 23.06s} & \timeform{+10D 49' 59.3''} & SDSS & 20.7--20.8 & 21.6(2) \\
ASASSN-15rj & \timeform{02h 59m 38.33s} & \timeform{+44D 57' 04.7''} & SDSS & 21.3 & -- \\
ASASSN-15ro & \timeform{01h 33m 07.56s} & \timeform{+41D 07' 18.6''} & SDSS & 21.6 & -- \\
ASASSN-15rr & \timeform{19h 08m 47.24s} & \timeform{-58D 31' 06.8''} & GSC2.3.2 & -- & -- \\
ASASSN-15rs & \timeform{04h 46m 33.68s} & \timeform{+48D 57' 55.6''} & 2MASS & -- & 17.89(4) \\
ASASSN-15ry & \timeform{05h 28m 55.66s} & \timeform{+36D 18' 38.9''} & IGSL & -- & -- \\
\hline
\end{tabular}
\end{center}
\end{table*}

\addtocounter{table}{-1}
\begin{table*}
\caption{Coordinates of objects without coordinate-based names (continued).}
\begin{center}
\begin{tabular}{cccccc}
\hline
Object & Right Ascention & Declination & Source\commenta & SDSS $g$ & GALEX NUV \\
\hline
ASASSN-15sc & \timeform{02h 21m 11.94s} & \timeform{+60D 19' 49.2''} & ASAS-SN\commentb & -- & -- \\
ASASSN-15sd & \timeform{23h 18m 33.27s} & \timeform{-35D 37' 22.7''} & IGSL & -- & 18.75(5) \\
ASASSN-15se & \timeform{09h 33m 09.37s} & \timeform{+10D 28' 02.1''} & SDSS & 20.5--20.6 & -- \\
ASASSN-15sl & \timeform{07h 23m 12.73s} & \timeform{+50D 50' 07.7''} & IGSL & -- & 21.6(4) \\
ASASSN-15sn & \timeform{20h 04m 23.09s} & \timeform{+44D 20' 30.2''} & KIC & -- & -- \\
ASASSN-15sp & \timeform{07h 58m 08.47s} & \timeform{-57D 22' 41.9''} & IGSL & -- & -- \\
ASASSN-15su & \timeform{05h 05m 03.08s} & \timeform{+22D 25' 30.3''} & IGSL & -- & -- \\
ASASSN-15sv & \timeform{00h 39m 00.55s} & \timeform{+27D 13' 45.6''} & SDSS & 22.1--22.6 & -- \\
ASASSN-15ud & \timeform{08h 52m 28.59s} & \timeform{-08D 44' 11.4''} & GSC2.3.2 & -- & -- \\
ASASSN-15uj & \timeform{04h 36m 21.63s} & \timeform{-55D 25' 07.4''} & IGSL & -- & -- \\
ASASSN-15ux & \timeform{06h 52m 26.66s} & \timeform{+47D 10' 56.5''} & ASAS-SN & -- & -- \\
ASASSN-16af & \timeform{09h 06m 06.44s} & \timeform{+00D 04' 34.7''} & SDSS & 21.1--21.9 & -- \\
ASASSN-16ag & \timeform{01h 34m 38.24s} & \timeform{+52D 06' 16.5''} & SDSS & 19.7 & 21.0(3) \\
ASASSN-16ao & \timeform{04h 40m 47.12s} & \timeform{-58D 07' 28.5''} & GSC2.3.2 & -- & 21.8(3) \\
ASASSN-16aq & \timeform{16h 55m 25.06s} & \timeform{+37D 21' 36.8''} & SDSS & 22.2 & 22.1(4) \\
ASASSN-16bh & \timeform{13h 24m 57.23s} & \timeform{-27D 56' 10.6''} & GSC2.3.2 & -- & 21.5(4) \\
ASASSN-16bi & \timeform{07h 46m 22.50s} & \timeform{-77D 47' 16.7''} & Gaia & -- & -- \\
ASASSN-16bu & \timeform{07h 27m 31.65s} & \timeform{+33D 46' 35.1''} & SDSS & 22.1 & 22.8(5) \\
ASASSN-16de & \timeform{18h 42m 34.00s} & \timeform{+17D 41' 22.7''} & USNO-A2.0 & -- & -- \\
DDE 26      & \timeform{22h 03m 28.22s} & \timeform{+30D 56' 36.4''} & SDSS & 19.6 & -- \\
N080829A    & \timeform{21h 42m 54.30s} & \timeform{+15D 36' 42.3''} & SDSS & 22.6 & -- \\
\hline
\end{tabular}
\end{center}
\end{table*}

\begin{table*}
\caption{Superhump Periods and Period Derivatives}\label{tab:perlist}
\begin{center}
\begin{tabular}{c@{\hspace{7pt}}c@{\hspace{7pt}}c@{\hspace{7pt}}c@{\hspace{7pt}}c@{\hspace{7pt}}c@{\hspace{7pt}}c@{\hspace{7pt}}c@{\hspace{7pt}}c@{\hspace{7pt}}c@{\hspace{7pt}}c@{\hspace{7pt}}c@{\hspace{7pt}}c@{\hspace{7pt}}c}
\hline
Object & Year & $P_1$ (d)\commenta & err & \multicolumn{2}{c}{$E_1$\commentb} & $P_{\rm dot}$\commentc & err\commentc & $P_2$ (d)\commenta & err & \multicolumn{2}{c}{$E_2$\commentb} & $P_{\rm orb}$ (d)\commentd & Q\commente \\
\hline
KV And & 2015 & 0.074283 & 0.000011 & 15 & 56 & 0.0 & 2.6 & 0.074108 & 0.000054 & 66 & 96 & -- & B \\
EG Aqr & 2015 & 0.078688 & 0.000122 & 51 & 65 & -- & -- & -- & -- & -- & -- & -- & C \\
NN Cam & 2015 & 0.074226 & 0.000037 & 0 & 43 & 11.6 & 7.8 & 0.073768 & 0.000035 & 67 & 150 & 0.0717 & B \\
V452 Cas & 2016 & 0.088828 & 0.000088 & 11 & 57 & -- & -- & -- & -- & -- & -- & -- & C \\
V1040 Cen & 2015 & 0.062244 & 0.000024 & 0 & 67 & -- & -- & -- & -- & -- & -- & 0.06049 & CG2 \\
PU CMa & 2016 & 0.057968 & 0.000018 & 0 & 107 & 6.9 & 1.7 & -- & -- & -- & -- & 0.056694 & B \\
AL Com & 2015 & 0.057293 & 0.000010 & 0 & 128 & 1.6 & 0.8 & -- & -- & -- & -- & 0.056669 & B \\
VW CrB & 2015 & 0.072820 & 0.000038 & 0 & 112 & 3.6 & 3.1 & -- & -- & -- & -- & -- & C \\
V1006 Cyg & 2015 & 0.105407 & 0.000044 & 36 & 94 & 20.8 & 2.0 & 0.104437 & 0.000050 & 102 & 200 & 0.09903 & B \\
V1028 Cyg & 2016 & 0.062009 & 0.000055 & 0 & 58 & 11.8 & 8.3 & -- & -- & -- & -- & -- & C \\
V1113 Cyg & 2015 & 0.078937 & 0.000018 & 0 & 107 & $-$4.3 & 1.3 & -- & -- & -- & -- & -- & BG \\
HO Del & 2015 & 0.064326 & 0.000018 & 0 & 33 & -- & -- & -- & -- & -- & -- & 0.06266 & C \\
AQ Eri & 2016 & 0.062432 & 0.000030 & 0 & 83 & 10.8 & 2.5 & -- & -- & -- & -- & 0.06094 & C \\
AX For & 2015 & -- & -- & -- & -- & -- & -- & 0.081041 & 0.000073 & 0 & 26 & -- & C \\
V844 Her & 2015 & 0.055902 & 0.000025 & 32 & 120 & 10.4 & 2.6 & 0.055819 & 0.000047 & 174 & 228 & 0.054643 & C \\
RZ Leo & 2016 & 0.078675 & 0.000035 & 0 & 48 & 15.6 & 5.9 & 0.078229 & 0.000022 & 53 & 138 & 0.076030 & A \\
V585 Lyr & 2015 & 0.060360 & 0.000018 & 11 & 84 & 9.9 & 2.4 & -- & -- & -- & -- & -- & C \\
V2051 Oph & 2015 & 0.064708 & 0.000088 & 0 & 16 & -- & -- & 0.064144 & 0.000044 & 40 & 72 & 0.062428 & Ce \\
V650 Peg & 2015 & 0.069777 & 0.000020 & 0 & 125 & 6.7 & 0.9 & 0.069271 & 0.000029 & 135 & 210 & -- & C \\
\hline
  \multicolumn{14}{l}{\commenta $P_1$ and $P_2$ are mean periods of stage B and C superhumps, respectively.} \\
  \multicolumn{14}{l}{\commentb Interval used for calculating the period (corresponding to $E$ in the individual tables in section \ref{sec:individual}).} \\
  \multicolumn{14}{l}{\commentc $P_{\rm dot} = \dot{P}/P$ for stage B superhumps, unit $10^{-5}$.} \\
  \multicolumn{14}{l}{\parbox{460pt}{\commentd References:
NN Cam (Denisenko, D. 2007, vsnet-alert 9557),
V1040 Cen (Longa-Pe{\~n}a et al., see text),
PU CMa \citep{tho03kxaqlftcampucmav660herdmlyr},
AL Com \citep{Pdot6},
V1006 Cyg \citep{she07CVspec},
HO Del \citep{pat03suumas},
AQ Eri \citep{tho96Porb},
V844 Her \citep{tho02gwlibv844herdiuma},
RZ Leo (\cite{dai16KepCVs}; improved by this work),
V2051 Oph (this work),
TY Psc \citep{tho96Porb},
V493 Ser \citep{Pdot},
KS UMa \citep{pat03suumas},
PM J03338 \citep{ski14SuperblinkCVs},
SDSS J145758 \citep{uth11CVthesis},
ASASSN-15gq, ASASSN-15hd, ASASSN-15na, ASASSN-15ni, ASASSN-15pu,
ASASSN-15sc, ASASSN-15uj, ASASSN-15ux, ASASSN-16bh, ASASSN-16bi,
ASASSN-16bu, CRTS J095926, CRTS J200331 (this work),
ASASSN-15po (\Namekataprep)
}} \\
  \multicolumn{14}{l}{\parbox{460pt}{\commente Data quality and comments. A: excellent, B: partial coverage or slightly low quality, C: insufficient coverage or observations with large scatter, G: $P_{\rm dot}$ denotes global $P_{\rm dot}$, M: observational gap in middle stage, U: uncertainty in alias selection, 2: late-stage coverage, the listed period may refer to $P_2$, E: $P_{\rm orb}$ refers to the period of early superhumps, e: eclipsing system.}} \\
\end{tabular}
\end{center}
\end{table*}

\addtocounter{table}{-1}
\begin{table*}
\caption{Superhump Periods and Period Derivatives (continued)}
\begin{center}
\begin{tabular}{c@{\hspace{7pt}}c@{\hspace{7pt}}c@{\hspace{7pt}}c@{\hspace{7pt}}c@{\hspace{7pt}}c@{\hspace{7pt}}c@{\hspace{7pt}}c@{\hspace{7pt}}c@{\hspace{7pt}}c@{\hspace{7pt}}c@{\hspace{7pt}}c@{\hspace{7pt}}c@{\hspace{7pt}}c}
\hline
Object & Year & $P_1$ & err & \multicolumn{2}{c}{$E_1$} & $P_{\rm dot}$ & err & $P_2$ & err & \multicolumn{2}{c}{$E_2$} & $P_{\rm orb}$ & Q \\
\hline
PU Per & 2015 & -- & -- & -- & -- & -- & -- & 0.067975 & 0.000020 & 0 & 51 & -- & C \\
QY Per & 2015 & 0.078593 & 0.000032 & 0 & 40 & 14.7 & 2.8 & -- & -- & -- & -- & -- & C \\
TY Psc & 2015 & 0.070093 & 0.000330 & 0 & 10 & -- & -- & -- & -- & -- & -- & 0.068348 & C \\
V493 Ser & 2015 & 0.082793 & 0.000062 & 39 & 76 & -- & -- & 0.082576 & 0.000033 & 87 & 173 & 0.08001 & B \\
KK Tel & 2015 & 0.087606 & 0.000023 & 42 & 136 & 0.8 & 1.9 & -- & -- & -- & -- & -- & B \\
CI UMa & 2016 & 0.063283 & 0.000354 & 0 & 5 & -- & -- & -- & -- & -- & -- & -- & C \\
KS UMa & 2015 & 0.069948 & 0.000058 & 0 & 19 & -- & -- & 0.069763 & 0.000086 & 75 & 93 & 0.06796 & C \\
NSV 2026 & 2015 & 0.069829 & 0.000015 & 0 & 102 & 0.4 & 1.5 & -- & -- & -- & -- & -- & CG \\
NSV 2026 & 2016 & 0.069795 & 0.000013 & 0 & 130 & $-$0.4 & 1.1 & -- & -- & -- & -- & -- & CG \\
ASASSN-13ah & 2016 & 0.066141 & 0.000013 & 0 & 33 & -- & -- & -- & -- & -- & -- & -- & C \\
ASASSN-13ak & 2015 & 0.086655 & 0.000040 & 0 & 34 & -- & -- & -- & -- & -- & -- & -- & C \\
ASASSN-14cc & 2014 & 0.015610 & 0.000010 & -- & -- & -- & -- & -- & -- & -- & -- & -- & C \\
ASASSN-14dh & 2015 & -- & -- & -- & -- & -- & -- & 0.073629 & 0.000031 & 0 & 91 & -- & C \\
ASASSN-14fz & 2015 & 0.078028 & 0.000030 & 13 & 103 & $-$2.9 & 3.0 & -- & -- & -- & -- & -- & CG \\
ASASSN-15cl & 2016 & 0.094633 & 0.000104 & 22 & 33 & -- & -- & 0.093907 & 0.000071 & 43 & 96 & -- & B \\
ASASSN-15cy & 2015 & 0.049955 & 0.000009 & 0 & 122 & 3.4 & 1.2 & -- & -- & -- & -- & -- & C \\
ASASSN-15dh & 2015 & 0.088014 & 0.000087 & 0 & 19 & $-$103.2 & 14.1 & -- & -- & -- & -- & -- & CG \\
ASASSN-15dp & 2015 & 0.060005 & 0.000015 & 49 & 200 & 0.4 & 1.1 & -- & -- & -- & -- & -- & B \\
ASASSN-15dr & 2015 & 0.056387 & 0.000045 & 25 & 80 & -- & -- & -- & -- & -- & -- & -- & C \\
ASASSN-15ea & 2015 & 0.095225 & 0.000034 & -- & -- & -- & -- & -- & -- & -- & -- & -- & CU \\
ASASSN-15ee & 2015 & 0.057136 & 0.000018 & 15 & 120 & 8.1 & 1.2 & -- & -- & -- & -- & -- & B \\
ASASSN-15eh & 2015 & 0.085665 & 0.000033 & 0 & 60 & $-$10.3 & 2.4 & -- & -- & -- & -- & -- & CG \\
ASASSN-15ev & 2015 & 0.058015 & 0.000041 & 0 & 20 & -- & -- & -- & -- & -- & -- & -- & C \\
ASASSN-15fo & 2015 & 0.0630 & 0.0030 & 0 & 4 & -- & -- & -- & -- & -- & -- & -- & C \\
ASASSN-15fu & 2015 & 0.074770 & 0.000062 & 0 & 44 & -- & -- & 0.074001 & 0.000228 & 43 & 71 & -- & C \\
ASASSN-15gf & 2015 & 0.06690 & 0.00012 & 0 & 15 & -- & -- & -- & -- & -- & -- & -- & CU \\
ASASSN-15gh & 2015 & 0.05905 & 0.00030 & 0 & 68 & -- & -- & -- & -- & -- & -- & -- & CU \\
ASASSN-15gi & 2015 & 0.061270 & 0.000050 & 0 & 66 & 15.5 & 8.1 & 0.060928 & 0.000038 & 64 & 130 & -- & C \\
ASASSN-15gn & 2015 & 0.063641 & 0.000033 & 18 & 112 & $-$3.2 & 3.8 & -- & -- & -- & -- & -- & C \\
ASASSN-15gq & 2015 & 0.066726 & 0.000034 & 15 & 120 & 11.9 & 0.8 & -- & -- & -- & -- & 0.06490 & BE \\
ASASSN-15hd & 2015 & 0.056105 & 0.000007 & 22 & 273 & 1.5 & 0.3 & -- & -- & -- & -- & 0.05541 & BE \\
ASASSN-15hl & 2015 & 0.067947 & 0.000035 & 0 & 89 & $-$5.9 & 3.6 & -- & -- & -- & -- & -- & CG \\
\hline
\end{tabular}
\end{center}
\end{table*}

\addtocounter{table}{-1}
\begin{table*}
\caption{Superhump Periods and Period Derivatives (continued)}
\begin{center}
\begin{tabular}{c@{\hspace{7pt}}c@{\hspace{7pt}}c@{\hspace{7pt}}c@{\hspace{7pt}}c@{\hspace{7pt}}c@{\hspace{7pt}}c@{\hspace{7pt}}c@{\hspace{7pt}}c@{\hspace{7pt}}c@{\hspace{7pt}}c@{\hspace{7pt}}c@{\hspace{7pt}}c@{\hspace{7pt}}c}
\hline
Object & Year & $P_1$ & err & \multicolumn{2}{c}{$E_1$} & $P_{\rm dot}$ & err & $P_2$ & err & \multicolumn{2}{c}{$E_2$} & $P_{\rm orb}$ & Q \\
\hline
ASASSN-15hm & 2015 & 0.056165 & 0.000031 & 34 & 159 & 5.4 & 2.0 & -- & -- & -- & -- & -- & C \\
ASASSN-15hn & 2015 & 0.061831 & 0.000018 & 32 & 178 & $-$0.5 & 1.5 & -- & -- & -- & -- & -- & B \\
ASASSN-15ia & 2015 & 0.070281 & 0.000072 & 0 & 29 & -- & -- & 0.069882 & 0.000059 & 28 & 72 & -- & C \\
ASASSN-15ie & 2015 & 0.058616 & 0.000022 & 35 & 224 & 4.0 & 0.9 & -- & -- & -- & -- & -- & C \\
ASASSN-15iv & 2015 & 0.067435 & 0.000051 & 0 & 89 & 17.4 & 2.9 & 0.067099 & 0.000068 & 87 & 149 & -- & C \\
ASASSN-15iz & 2015 & 0.081434 & 0.000049 & 0 & 61 & -- & -- & -- & -- & -- & -- & -- & CU \\
ASASSN-15jd & 2015 & 0.064981 & 0.000008 & 24 & 90 & 9.9 & 4.2 & -- & -- & -- & -- & -- & C \\
ASASSN-15jj & 2015 & 0.062388 & 0.000025 & 0 & 146 & 8.1 & 0.6 & 0.062124 & 0.000067 & 161 & 209 & -- & B \\
ASASSN-15kf & 2015 & 0.019251 & 0.000097 & 0 & 3 & -- & -- & -- & -- & -- & -- & -- & C \\
ASASSN-15kh & 2015 & 0.060480 & 0.000017 & 43 & 175 & 1.2 & 1.6 & -- & -- & -- & -- & -- & B \\
ASASSN-15le & 2015 & 0.078000 & 0.000053 & 0 & 43 & -- & -- & -- & -- & -- & -- & -- & CU \\
ASASSN-15lt & 2015 & 0.059815 & 0.000024 & 49 & 189 & -- & -- & 0.059633 & 0.000053 & 188 & 266 & -- & CM \\
ASASSN-15mb & 2015 & -- & -- & -- & -- & -- & -- & 0.068838 & 0.000056 & 43 & 203 & -- & CU \\
ASASSN-15mt & 2015 & 0.076342 & 0.000020 & 0 & 31 & -- & -- & 0.076032 & 0.000041 & 42 & 97 & -- & B \\
ASASSN-15na & 2015 & 0.063720 & 0.000027 & 0 & 111 & 3.1 & 2.6 & -- & -- & -- & -- & 0.06297 & CE \\
ASASSN-15ni & 2015 & 0.055854 & 0.000009 & 32 & 205 & 3.4 & 0.6 & -- & -- & -- & -- & 0.05517 & BE \\
ASASSN-15nl & 2015 & 0.060095 & 0.000098 & 0 & 33 & -- & -- & -- & -- & -- & -- & -- & C \\
ASASSN-15ob & 2015 & 0.060525 & 0.000060 & 0 & 83 & 15.1 & 2.8 & -- & -- & -- & -- & -- & C \\
ASASSN-15ok & 2015 & 0.078931 & 0.000122 & 0 & 25 & -- & -- & 0.078476 & 0.000022 & 25 & 115 & -- & C \\
ASASSN-15pi & 2015 & 0.078307 & 0.000300 & 0 & 4 & -- & -- & -- & -- & -- & -- & -- & C \\
ASASSN-15po & 2015 & 0.050916 & 0.000002 & 49 & 329 & 1.1 & 0.1 & -- & -- & -- & -- & 0.050457 & AE \\
ASASSN-15pu & 2015 & 0.058254 & 0.000024 & 34 & 146 & 3.3 & 2.1 & -- & -- & -- & -- & 0.05757 & BE \\
ASASSN-15qe & 2015 & 0.061092 & 0.000017 & 0 & 62 & -- & -- & -- & -- & -- & -- & -- & C \\
ASASSN-15qq & 2015 & 0.077131 & 0.000026 & 0 & 81 & $-$1.0 & 2.6 & -- & -- & -- & -- & -- & CG2 \\
ASASSN-15rj & 2015 & 0.092463 & 0.000164 & 0 & 15 & -- & -- & -- & -- & -- & -- & -- & C \\
ASASSN-15ro & 2015 & 0.072909 & 0.000088 & 0 & 29 & -- & -- & -- & -- & -- & -- & -- & C \\
ASASSN-15rr & 2015 & 0.054938 & 0.000039 & 0 & 75 & -- & -- & -- & -- & -- & -- & -- & CU \\
ASASSN-15rs & 2015 & 0.097854 & 0.000207 & 0 & 22 & -- & -- & -- & -- & -- & -- & -- & C \\
ASASSN-15ry & 2015 & 0.060876 & 0.000270 & 0 & 3 & -- & -- & -- & -- & -- & -- & -- & C \\
ASASSN-15sc & 2015 & 0.057735 & 0.000015 & 21 & 208 & 5.8 & 0.3 & -- & -- & -- & -- & -- & A \\
ASASSN-15sd & 2015 & -- & -- & -- & -- & -- & -- & 0.068894 & 0.000028 & 0 & 93 & -- & C \\
ASASSN-15se & 2015 & 0.063312 & 0.000042 & 10 & 58 & -- & -- & -- & -- & -- & -- & -- & C \\
\hline
\end{tabular}
\end{center}
\end{table*}

\addtocounter{table}{-1}
\begin{table*}
\caption{Superhump Periods and Period Derivatives (continued)}
\begin{center}
\begin{tabular}{c@{\hspace{7pt}}c@{\hspace{7pt}}c@{\hspace{7pt}}c@{\hspace{7pt}}c@{\hspace{7pt}}c@{\hspace{7pt}}c@{\hspace{7pt}}c@{\hspace{7pt}}c@{\hspace{7pt}}c@{\hspace{7pt}}c@{\hspace{7pt}}c@{\hspace{7pt}}c@{\hspace{7pt}}c}
\hline
Object & Year & $P_1$ & err & \multicolumn{2}{c}{$E_1$} & $P_{\rm dot}$ & err & $P_2$ & err & \multicolumn{2}{c}{$E_2$} & $P_{\rm orb}$ & Q \\
\hline
ASASSN-15sl & 2015 & 0.091065 & 0.000066 & 0 & 97 & 9.1 & 2.6 & -- & -- & -- & -- & 0.087048 & CGe \\
ASASSN-15sn & 2015 & 0.064684 & 0.000090 & 0 & 48 & $-$50.5 & 11.5 & -- & -- & -- & -- & -- & C \\
ASASSN-15sp & 2015 & 0.058366 & 0.000018 & 33 & 138 & 7.7 & 0.9 & 0.058292 & 0.000040 & -- & -- & -- & B \\
ASASSN-15ud & 2015 & 0.05649 & 0.00023 & 0 & 3 & -- & -- & -- & -- & -- & -- & -- & C \\
ASASSN-15uj & 2015 & 0.055805 & 0.000012 & 35 & 129 & $-$1.1 & 1.6 & -- & -- & -- & -- & 0.055266 & BE \\
ASASSN-15ux & 2015 & 0.056857 & 0.000012 & 73 & 131 & -- & -- & -- & -- & -- & -- & 0.056109 & Ce \\
ASASSN-16af & 2016 & 0.064204 & 0.000025 & 0 & 75 & 13.0 & 2.8 & -- & -- & -- & -- & -- & B \\
ASASSN-16ag & 2016 & 0.058479 & 0.000071 & 0 & 96 & -- & -- & -- & -- & -- & -- & -- & C \\
ASASSN-16bh & 2016 & 0.054027 & 0.000006 & 32 & 206 & 3.7 & 0.3 & -- & -- & -- & -- & 0.05346 & AE \\
ASASSN-16bu & 2016 & 0.060513 & 0.000071 & 42 & 82 & -- & -- & -- & -- & -- & -- & 0.05934 & BE \\
CRTS J095926 & 2015 & 0.089428 & 0.000041 & 10 & 67 & $-$4.4 & 5.0 & -- & -- & -- & -- & -- & C \\
CRTS J120052 & 2016 & 0.088950 & 0.000038 & 0 & 85 & $-$5.5 & 3.2 & -- & -- & -- & -- & -- & CG2 \\
CRTS J163120 & 2015 & 0.0645 & 0.0005 & 0 & 3 & -- & -- & -- & -- & -- & -- & -- & C \\
CRTS J200331 & 2015 & 0.059720 & 0.000088 & 49 & 84 & -- & -- & -- & -- & -- & -- & 0.058705 & Ce \\
CRTS J212521 & 2015 & 0.079090 & 0.000119 & 0 & 26 & -- & -- & -- & -- & -- & -- & -- & C \\
CSS J221822 & 2015 & 0.069294 & 0.000035 & 0 & 50 & -- & -- & -- & -- & -- & -- & -- & C \\
DDE 26 & 2015 & 0.088617 & 0.000042 & 0 & 49 & 10.3 & 2.5 & -- & -- & -- & -- & -- & C \\
IPHAS J230538 & 2015 & 0.072772 & 0.000017 & 15 & 83 & 4.6 & 2.4 & 0.072493 & 0.000060 & 82 & 125 & -- & C \\
MASTER J003831 & 2016 & 0.061605 & 0.000038 & 0 & 114 & 11.5 & 0.9 & 0.061310 & 0.000038 & 114 & 179 & -- & B \\
MASTER J073325 & 2016 & 0.061218 & 0.000027 & 32 & 162 & 5.5 & 1.1 & -- & -- & -- & -- & -- & C \\
MASTER J120251 & 2015 & 0.063372 & 0.000018 & 0 & 131 & $-$1.7 & 3.1 & -- & -- & -- & -- & -- & CGM \\
MASTER J131320 & 2016 & 0.069709 & 0.000044 & 0 & 34 & -- & -- & -- & -- & -- & -- & -- & CU \\
MASTER J181523 & 2015 & 0.058512 & 0.000056 & 0 & 20 & -- & -- & -- & -- & -- & -- & -- & C \\
MASTER J212624 & 2015 & 0.091193 & 0.000148 & 0 & 31 & -- & -- & -- & -- & -- & -- & -- & C \\
N 080829A & 2015 & 0.064282 & 0.000027 & 0 & 100 & 11.4 & 2.4 & -- & -- & -- & -- & -- & B \\
PM J03338 & 2015 & 0.069013 & 0.000021 & 70 & 115 & $-$12.7 & 4.7 & 0.068629 & 0.000020 & 115 & 246 & 0.06663 & A \\
SDSS J074859 & 2015 & 0.05958 & 0.00030 & -- & -- & -- & -- & -- & -- & -- & -- & 0.058311 & Ce \\
SDSS J145758 & 2015 & 0.054912 & 0.000018 & 0 & 128 & 2.2 & 2.9 & -- & -- & -- & -- & 0.054087 & C \\
SDSS J164248 & 2016 & 0.079327 & 0.000093 & 0 & 54 & -- & -- & -- & -- & -- & -- & -- & CG \\
\hline
\end{tabular}
\end{center}
\end{table*}

\section{Individual Objects}\label{sec:individual}

\subsection{KV Andromedae}\label{obj:kvand}

   KV And was discovered as dwarf nova by \citet{kur77kvandkwand}.
Since \citet{kur77kvandkwand} reported very faint (22.5 mag)
quiescent magnitude, this object was suspected to be
a dwarf nova with a very large outburst amplitude.
Later it turned out that \citet{kur77kvandkwand} underestimated
the quiescent brightness since they used the paper print of
the Palomar Observatory Sky Survey, and the true quiescent
magnitude is now considered to be around 20.
\citet{kat94kvand} and \citet{kat95kvand} reported superhump
observations in 1993 and 1994, respectively.
Although this object was recognized as an SU UMa-type
dwarf nova relatively early, none of observations sufficiently
recorded the development of superhumps.  Yet another
superoutburst in 2002 \citep{Pdot} suffered from rather
low signal-to-noise quality.

\begin{figure}
  \begin{center}
%    \FigureFile(85mm,70mm){kvandcomp.eps}
    \FigureFile(85mm,70mm){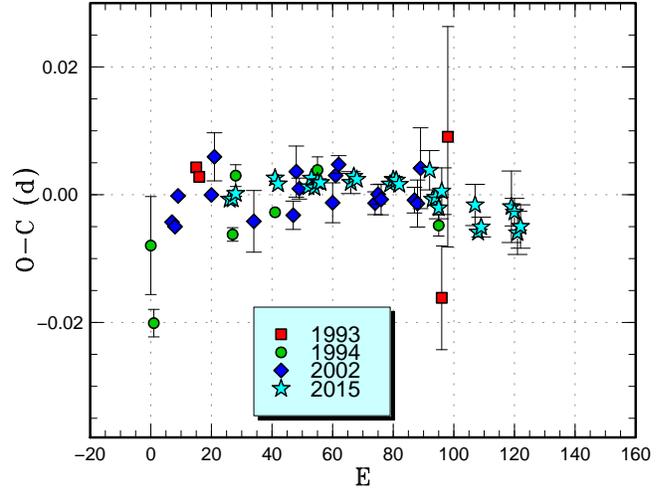}
  \end{center}
  \caption{Comparison of $O-C$ diagrams of KV And between different
  superoutbursts.  A period of 0.07428~d was used to draw this figure.
  Approximate cycle counts ($E$) after the outburst detections
  were used.  The actual starts of the outbursts were unknown.
  Since most of observations recorded large-amplitude
  superhumps, these observations probably recorded the early
  phases of the superoutbursts.
  }
  \label{fig:kvandcomp}
\end{figure}

   The 2015 superoutburst was visually detected by P. Dubovsky
on August 25 (vsnet-alert 19005).  Subsequent observations
detected superhumps (vsnet-alert 19011, 19020, 19022).
The times of superhump maxima are listed in
table \ref{tab:kvandoc2015}.  The $O-C$ data suggest that
The maxima for $E \le$2 probably recorded the final
part of stage A.  Although we give a comparison of
the $O-C$ diagrams in figure \ref{fig:kvandcomp},
the starts of these superoutbursts were not well determined
because this object has not been regularly monitored
by visual observers.  Since most of observations recorded large-amplitude
superhumps, these observations, however, probably recorded
the early phases of the superoutbursts and we treated these
$O-C$ data as if the initial outburst detection refers
to the start of the outburst.  It is likely that the 1994
observation partly recorded stage A, which is compatible
with low superhump amplitudes on the first two nights
(the data quality was very poor, though).

% SI

\begin{table}
\caption{Superhump maxima of KV And (2015)}\label{tab:kvandoc2015}
\begin{center}
\begin{tabular}{rp{55pt}p{40pt}r@{.}lr}
\hline
\multicolumn{1}{c}{$E$} & \multicolumn{1}{c}{max\commenta} & \multicolumn{1}{c}{error} & \multicolumn{2}{c}{$O-C$\commentb} & \multicolumn{1}{c}{$N$\commentc} \\
\hline
0 & 57261.4236 & 0.0006 & $-$0&0036 & 40 \\
1 & 57261.4979 & 0.0005 & $-$0&0035 & 40 \\
2 & 57261.5730 & 0.0006 & $-$0&0026 & 31 \\
15 & 57262.5411 & 0.0010 & 0&0006 & 41 \\
16 & 57262.6144 & 0.0012 & $-$0&0003 & 45 \\
26 & 57263.3568 & 0.0005 & $-$0&0002 & 60 \\
27 & 57263.4324 & 0.0005 & 0&0012 & 53 \\
28 & 57263.5051 & 0.0004 & $-$0&0003 & 108 \\
29 & 57263.5804 & 0.0004 & 0&0008 & 115 \\
30 & 57263.6546 & 0.0003 & 0&0007 & 35 \\
40 & 57264.3974 & 0.0017 & 0&0013 & 30 \\
41 & 57264.4726 & 0.0009 & 0&0023 & 37 \\
42 & 57264.5464 & 0.0004 & 0&0019 & 39 \\
53 & 57265.3628 & 0.0009 & 0&0018 & 38 \\
54 & 57265.4378 & 0.0008 & 0&0025 & 38 \\
55 & 57265.5119 & 0.0005 & 0&0025 & 39 \\
56 & 57265.5855 & 0.0006 & 0&0019 & 30 \\
66 & 57266.3306 & 0.0031 & 0&0047 & 31 \\
67 & 57266.4003 & 0.0009 & 0&0002 & 38 \\
68 & 57266.4749 & 0.0010 & 0&0005 & 40 \\
69 & 57266.5475 & 0.0017 & $-$0&0011 & 37 \\
70 & 57266.6244 & 0.0037 & 0&0016 & 9 \\
81 & 57267.4393 & 0.0032 & 0&0001 & 20 \\
82 & 57267.5093 & 0.0014 & $-$0&0041 & 37 \\
83 & 57267.5844 & 0.0015 & $-$0&0033 & 39 \\
93 & 57268.3304 & 0.0056 & 0&0005 & 18 \\
94 & 57268.4039 & 0.0022 & $-$0&0003 & 38 \\
95 & 57268.4749 & 0.0034 & $-$0&0035 & 25 \\
96 & 57268.5502 & 0.0034 & $-$0&0024 & 24 \\
\hline
  \multicolumn{6}{l}{\commenta BJD$-$2400000.} \\
  \multicolumn{6}{l}{\commentb Against max $= 2457261.4271 + 0.074224 E$.} \\
  \multicolumn{6}{l}{\commentc Number of points used to determine the maximum.} \\
\end{tabular}
\end{center}
\end{table}

\subsection{EG Aquarii}\label{obj:egaqr}

   EG Aqr was discovered as a blue eruptive object
(BV number 3 in \cite{luy59egaqrehaqr}) who reported
a photographic magnitude of 14.8 on 1951 August 6.
\citet{har60egaqrehaqr} reported full photographic
data, who gave a maximum of 14.0 mag (photographic) on
1958 November 5.  The date in \citet{luy59egaqrehaqr}
appears to have been a result of confusion.
\citet{vog82atlas} did not detect any further outburst.
\citet{szk92CVspec} obtained a K-type spectrum without
emission lines, which was probably due to mis-identification.

   The first well-confirmed outburst since the discovery
was detected by R. Stubbings on 2006 November 8 at
a visual magnitude of 12.4 (vsnet-alert 9103).
In vsnet-alert 9105, P. Schmeer reported a CCD detection
of the outburst earlier than R. Stubbings.
This 2006 superoutburst was well studied by \citet{ima08egaqr}.
The 2008 and 2011 superoutbursts were also observed
and analyzed in \citep{Pdot} and \citet{Pdot4},
respectively.  There was also a normal outburst
on 2013 October 1 at $V$=15.0 (ASAS-SN and R. Stubbings;
vsnet-obs 76596, vsnet-outburst 16054).
Although there was a bright outburst reaching a visual magnitude
of 13.0 (R. Stubbings) on 2014 July 21, the nature
of this outburst was unclear.

\begin{figure}
  \begin{center}
%    \FigureFile(85mm,70mm){egaqrcomp2.eps}
    \FigureFile(85mm,70mm){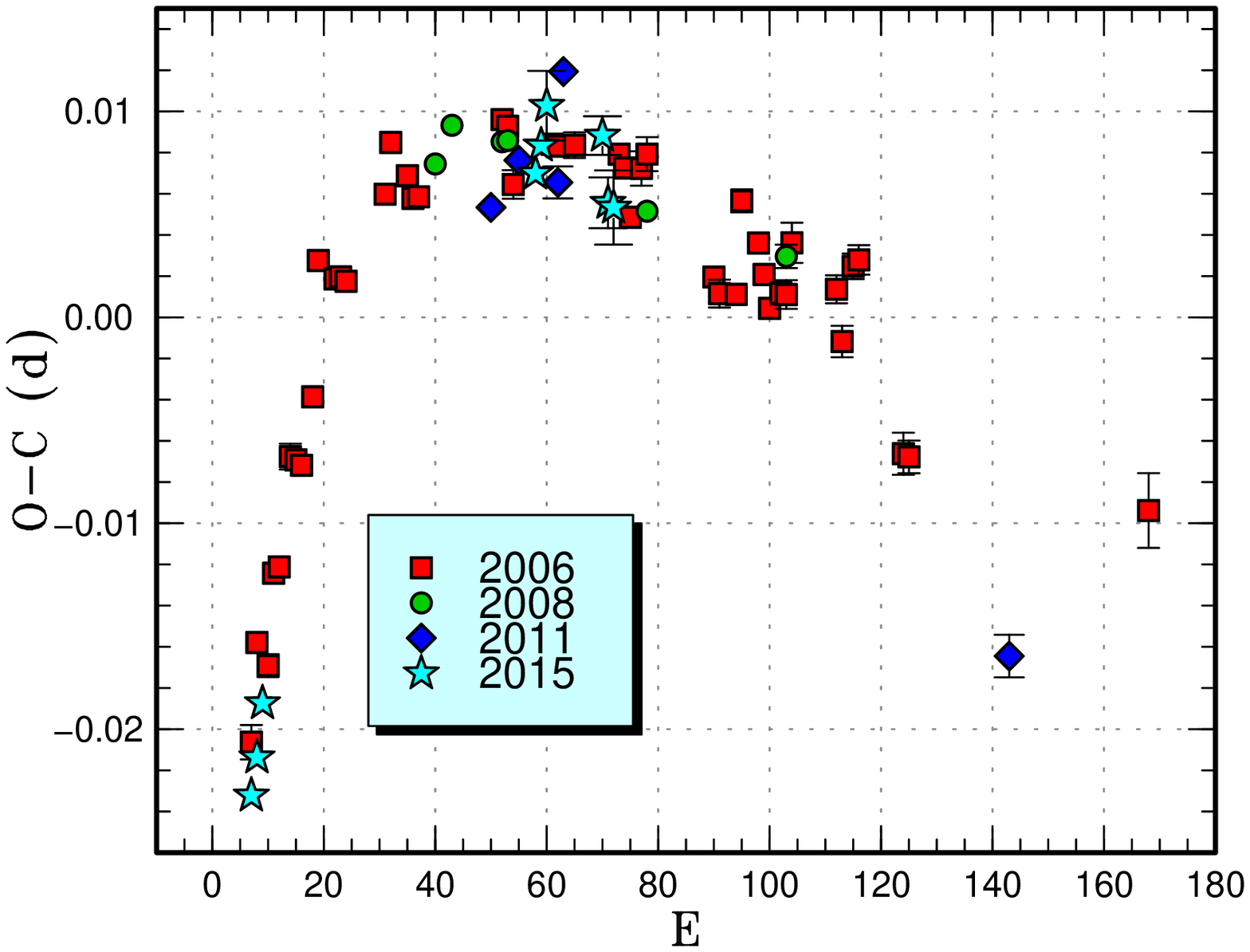}
  \end{center}
  \caption{Comparison of $O-C$ diagrams of EG Aqr between different
  superoutbursts.  A period of 0.07885~d was used to draw this figure.
  Approximate cycle counts ($E$) after the start of the superoutburst
  were used.  Since the starts of the 2008, 2011 and 2015
  superoutbursts were not well constrained, we shifted
  the $O-C$ diagrams to best fit the best-recorded 2006 one.
  }
  \label{fig:egaqrcomp2}
\end{figure}

   The 2015 superoutburst was visually detected by
R. Stubbings at a magnitude of 12.1 on August 21
(vsnet-alert 18998).  This outburst detection was
sufficiently early and growing phase of superhumps
was recorded (vsnet-alert 19003).  There was, however,
a long gap after the initial CCD observations and
the outburst was not well covered by observations.
The times of superhump maxima are listed in
table \ref{tab:egaqroc2015}.  The maxima for
$E \le$2 correspond to stage A superhumps
(figure \ref{fig:egaqrcomp2}).
The object showed a rebrightening on September 8
at $V$=16.37 (ASAS-SN detection, vsnet-alert 19038).

% SI

\begin{table}
\caption{Superhump maxima of EG Aqr (2015)}\label{tab:egaqroc2015}
\begin{center}
\begin{tabular}{rp{55pt}p{40pt}r@{.}lr}
\hline
\multicolumn{1}{c}{$E$} & \multicolumn{1}{c}{max\commenta} & \multicolumn{1}{c}{error} & \multicolumn{2}{c}{$O-C$\commentb} & \multicolumn{1}{c}{$N$\commentc} \\
\hline
0 & 57257.0963 & 0.0007 & $-$0&0024 & 137 \\
1 & 57257.1770 & 0.0006 & $-$0&0011 & 107 \\
2 & 57257.2585 & 0.0005 & 0&0011 & 131 \\
51 & 57261.1478 & 0.0005 & 0&0032 & 72 \\
52 & 57261.2281 & 0.0006 & 0&0041 & 81 \\
53 & 57261.3088 & 0.0017 & 0&0055 & 35 \\
63 & 57262.0959 & 0.0009 & $-$0&0007 & 52 \\
64 & 57262.1715 & 0.0012 & $-$0&0045 & 53 \\
65 & 57262.2501 & 0.0018 & $-$0&0052 & 42 \\
\hline
  \multicolumn{6}{l}{\commenta BJD$-$2400000.} \\
  \multicolumn{6}{l}{\commentb Against max $= 2457257.0987 + 0.079332 E$.} \\
  \multicolumn{6}{l}{\commentc Number of points used to determine the maximum.} \\
\end{tabular}
\end{center}
\end{table}

\subsection{NN Camelopardalis}\label{obj:nncam}

\begin{figure}
  \begin{center}
%    \FigureFile(85mm,70mm){nncamcomp3.eps}
    \FigureFile(85mm,70mm){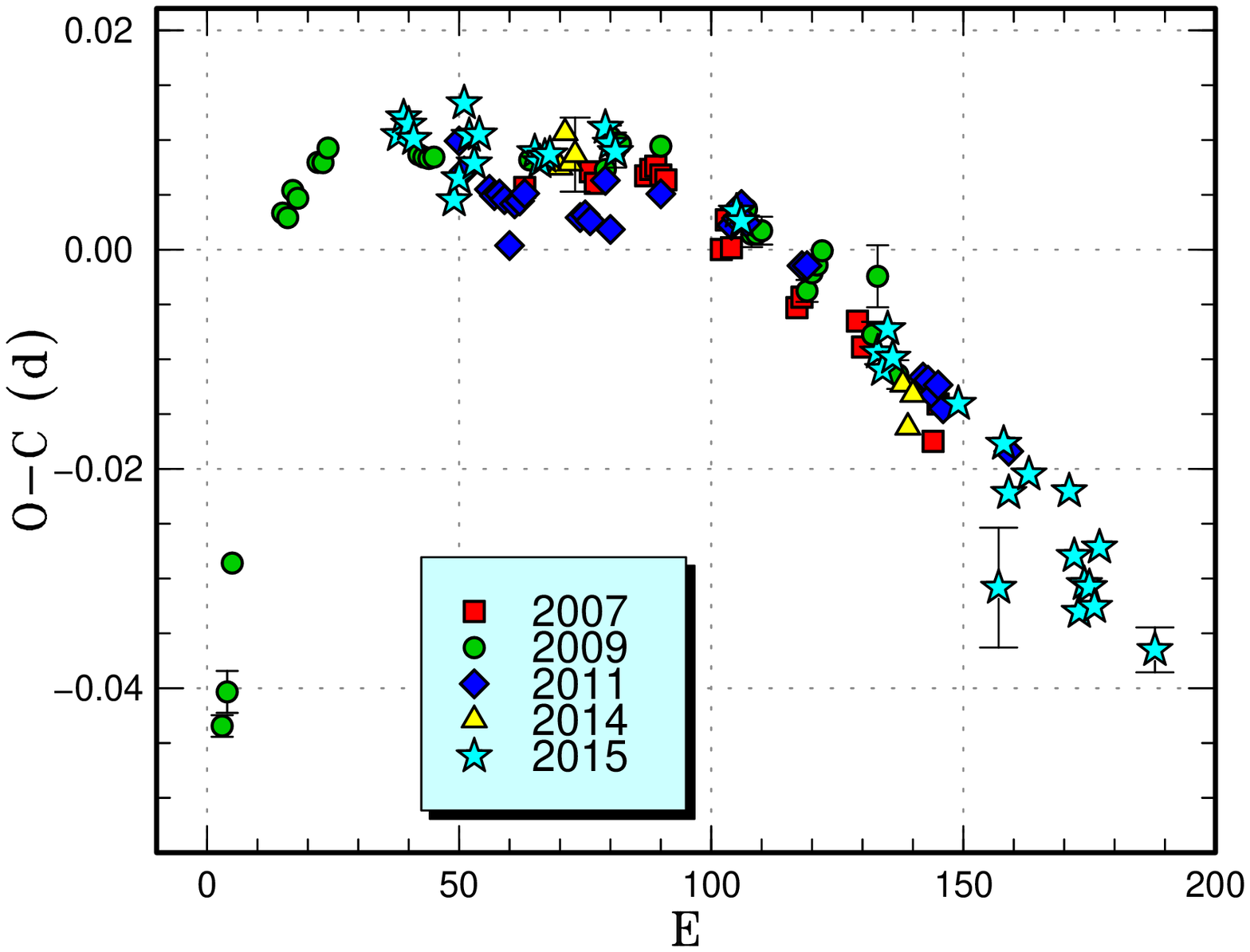}
  \end{center}
  \caption{Comparison of $O-C$ diagrams of NN Cam between different
  superoutbursts.  A period of 0.07425~d was used to draw this figure.
  Approximate cycle counts ($E$) after the start of the superoutburst
  were used.}
  \label{fig:nncamcomp3}
\end{figure}

   NN Cam = NSV 1485 was recognized as a dwarf nova
by \citet{khr05nsv1485}.  For more history, see \citet{Pdot7}.
The 2015 superoutburst was detected visually on August 11
by P. Dubovsky (vsnet-alert 18965).  Although there were possible
low-amplitude superhump-like modulations already on
the next night (vsnet-alert 18972), we could not
determine the period.  On August 13, the superhumps were
already stage B (vsnet-alert 18977, 18984).
The times of superhump maxima are listed in
table \ref{tab:nncamoc2015}.  The maxima for $E \ge$133
refer to rapidly fading part of the superoutburst.
The $O-C$ behavior was similar to that of past superoutbursts
(figure \ref{fig:nncamcomp3}).  There was no indication
of a phase jump as expected for traditional late superhumps.

% SI

\begin{table}
\caption{Superhump maxima of NN Cam (2015)}\label{tab:nncamoc2015}
\begin{center}
\begin{tabular}{rp{55pt}p{40pt}r@{.}lr}
\hline
\multicolumn{1}{c}{$E$} & \multicolumn{1}{c}{max\commenta} & \multicolumn{1}{c}{error} & \multicolumn{2}{c}{$O-C$\commentb} & \multicolumn{1}{c}{$N$\commentc} \\
\hline
0 & 57248.3330 & 0.0002 & $-$0&0055 & 67 \\
1 & 57248.4089 & 0.0002 & $-$0&0035 & 76 \\
2 & 57248.4825 & 0.0002 & $-$0&0039 & 76 \\
3 & 57248.5554 & 0.0003 & $-$0&0049 & 56 \\
11 & 57249.1438 & 0.0017 & $-$0&0080 & 40 \\
12 & 57249.2201 & 0.0011 & $-$0&0057 & 65 \\
13 & 57249.3012 & 0.0015 & 0&0015 & 79 \\
14 & 57249.3725 & 0.0004 & $-$0&0012 & 75 \\
15 & 57249.4442 & 0.0003 & $-$0&0034 & 77 \\
16 & 57249.5211 & 0.0004 & $-$0&0005 & 77 \\
27 & 57250.3362 & 0.0004 & 0&0013 & 76 \\
28 & 57250.4096 & 0.0003 & 0&0007 & 76 \\
29 & 57250.4843 & 0.0005 & 0&0016 & 75 \\
30 & 57250.5588 & 0.0004 & 0&0021 & 69 \\
41 & 57251.3780 & 0.0005 & 0&0079 & 76 \\
42 & 57251.4508 & 0.0005 & 0&0068 & 76 \\
43 & 57251.5242 & 0.0005 & 0&0063 & 64 \\
67 & 57253.3008 & 0.0009 & 0&0083 & 48 \\
68 & 57253.3742 & 0.0006 & 0&0077 & 42 \\
95 & 57255.3670 & 0.0013 & 0&0042 & 41 \\
96 & 57255.4396 & 0.0007 & 0&0029 & 103 \\
97 & 57255.5176 & 0.0009 & 0&0069 & 258 \\
98 & 57255.5892 & 0.0015 & 0&0046 & 63 \\
111 & 57256.5503 & 0.0005 & 0&0044 & 180 \\
119 & 57257.1275 & 0.0055 & $-$0&0099 & 27 \\
120 & 57257.2149 & 0.0009 & 0&0036 & 54 \\
121 & 57257.2847 & 0.0009 & $-$0&0006 & 55 \\
125 & 57257.5833 & 0.0006 & 0&0023 & 76 \\
133 & 57258.1758 & 0.0011 & 0&0033 & 52 \\
134 & 57258.2442 & 0.0006 & $-$0&0023 & 55 \\
135 & 57258.3133 & 0.0010 & $-$0&0071 & 109 \\
136 & 57258.3901 & 0.0006 & $-$0&0042 & 139 \\
137 & 57258.4641 & 0.0004 & $-$0&0042 & 104 \\
138 & 57258.5365 & 0.0008 & $-$0&0057 & 115 \\
139 & 57258.6162 & 0.0013 & 0&0000 & 42 \\
150 & 57259.4236 & 0.0021 & $-$0&0059 & 54 \\
\hline
  \multicolumn{6}{l}{\commenta BJD$-$2400000.} \\
  \multicolumn{6}{l}{\commentb Against max $= 2457248.3385 + 0.073940 E$.} \\
  \multicolumn{6}{l}{\commentc Number of points used to determine the maximum.} \\
\end{tabular}
\end{center}
\end{table}

\subsection{PU Canis Majoris}\label{obj:pucma}

   This object was originally selected as an ROSAT source
(RX J0640$-$24 = 1RXS J064047.8$-$242305: \cite{ROSATRXP}).
The object was classified as a dwarf nova based on
the detection at mag 11 on one ESO B plate
(cf. \cite{DownesCVatlas2}).  The dwarf nova-type nature
was established by monitoring by P. Schmeer with a CCD
attached to the 50-cm reflector at the Iowa Robotic Observatory.
The object underwent an outburst in 2000 January
and February.  The rapid fading recorded in the 2000 January
outburst suggested a normal outburst in an SU UMa-type
dwarf nova \citep{kat03v877arakktelpucma}.
\citet{tho03kxaqlftcampucmav660herdmlyr} obtained
an orbital period of 0.05669(4)~d by a radial-velocity study.
Using the data during superoutbursts in 2003, 2005 and
2008, \citet{Pdot} established the superhump period.
The 2008 superoutburst was preceded by a precursor
outburst during which a long superhump period (now known
as stage A) was recorded \citep{Pdot}.
The 2009 superoutburst was reported in \citet{Pdot2}.
\citet{kat13qfromstageA} estimated the mass ratio
$q$=0.110(11) by using stage A superhumps.

\begin{figure}
  \begin{center}
%    \FigureFile(85mm,70mm){pucmacomp2.eps}
    \FigureFile(85mm,70mm){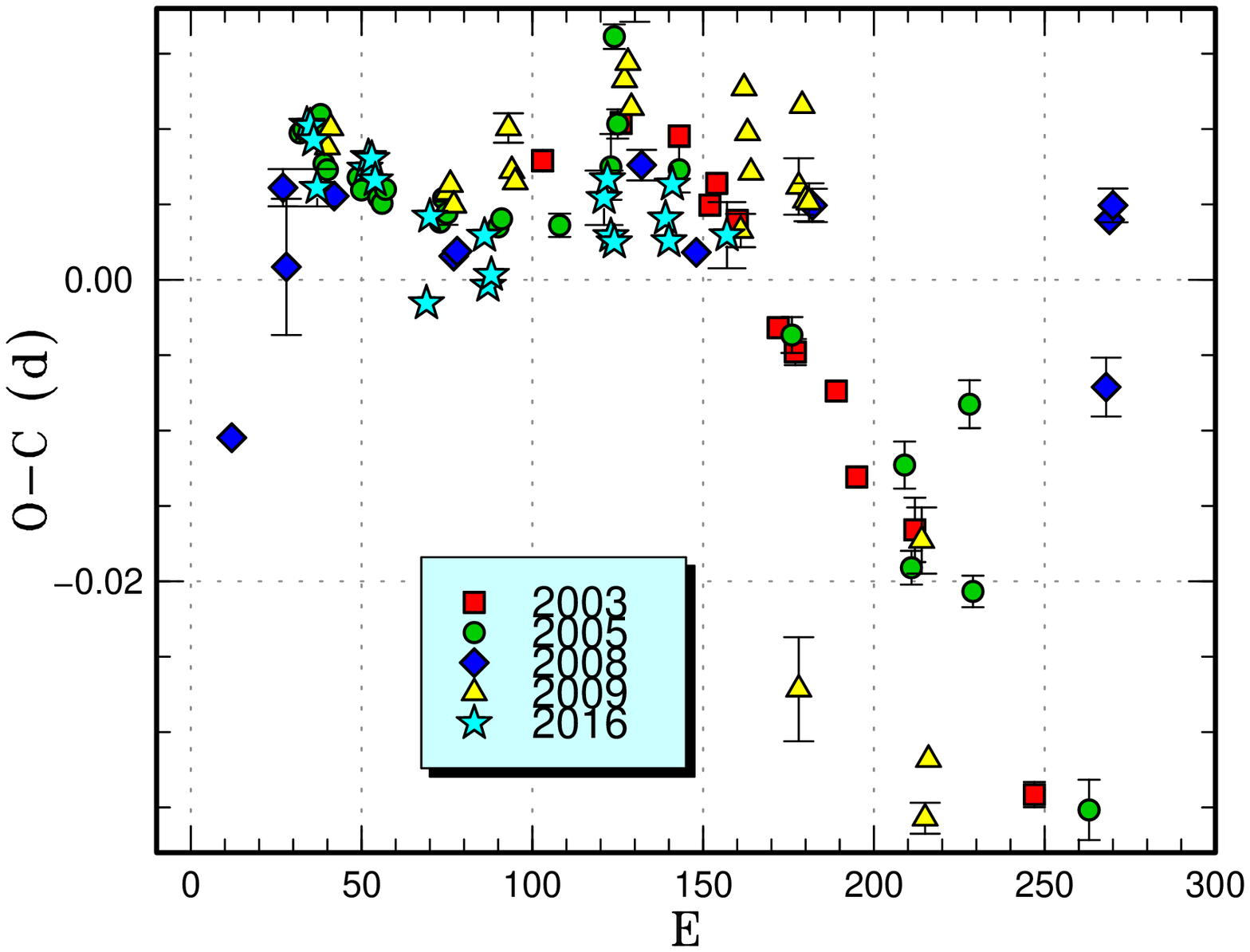}
  \end{center}
  \caption{Comparison of $O-C$ diagrams of PU CMa between different
  superoutbursts.  A period of 0.05801~d was used to draw this figure.
  Approximate cycle counts ($E$) after the start of the superoutburst
  were used.}
  \label{fig:pucmacomp2}
\end{figure}

   The 2016 superoutburst was visually detected by
T. Horie and R. Stubbings on February 29 (vsnet-alert 19543).
The object showed a precursor on March 1, as in
the 2008 superoutburst.  Due to the 1~d gap between
observations on March 1 and 2, we could not determine
the period of stage A superhumps.
The times of superhump maxima are listed in
table \ref{tab:pucma2016}.
The evolution of the $O-C$ variation was similar
to the past superoutbursts (figure \ref{fig:pucmacomp2}).

% SI

\begin{table}
\caption{Superhump maxima of PU CMa (2016)}\label{tab:pucma2016}
\begin{center}
\begin{tabular}{rp{55pt}p{40pt}r@{.}lr}
\hline
\multicolumn{1}{c}{$E$} & \multicolumn{1}{c}{max\commenta} & \multicolumn{1}{c}{error} & \multicolumn{2}{c}{$O-C$\commentb} & \multicolumn{1}{c}{$N$\commentc} \\
\hline
0 & 57449.9334 & 0.0005 & 0&0032 & 43 \\
1 & 57449.9913 & 0.0003 & 0&0031 & 48 \\
2 & 57450.0483 & 0.0003 & 0&0022 & 146 \\
3 & 57450.1032 & 0.0012 & $-$0&0009 & 64 \\
17 & 57450.9166 & 0.0008 & 0&0009 & 22 \\
18 & 57450.9754 & 0.0001 & 0&0018 & 139 \\
19 & 57451.0333 & 0.0002 & 0&0017 & 152 \\
20 & 57451.0899 & 0.0005 & 0&0003 & 113 \\
35 & 57451.9519 & 0.0010 & $-$0&0073 & 10 \\
36 & 57452.0156 & 0.0004 & $-$0&0015 & 30 \\
52 & 57452.9425 & 0.0007 & $-$0&0021 & 42 \\
53 & 57452.9972 & 0.0005 & $-$0&0054 & 38 \\
54 & 57453.0559 & 0.0010 & $-$0&0046 & 41 \\
87 & 57454.9754 & 0.0018 & 0&0018 & 43 \\
88 & 57455.0346 & 0.0011 & 0&0031 & 102 \\
89 & 57455.0889 & 0.0011 & $-$0&0006 & 134 \\
90 & 57455.1465 & 0.0005 & $-$0&0010 & 117 \\
105 & 57456.0182 & 0.0006 & 0&0012 & 117 \\
106 & 57456.0747 & 0.0005 & $-$0&0003 & 117 \\
107 & 57456.1365 & 0.0008 & 0&0035 & 115 \\
123 & 57457.0613 & 0.0022 & 0&0008 & 40 \\
\hline
  \multicolumn{6}{l}{\commenta BJD$-$2400000.} \\
  \multicolumn{6}{l}{\commentb Against max $= 2457449.9302 + 0.057970 E$.} \\
  \multicolumn{6}{l}{\commentc Number of points used to determine the maximum.} \\
\end{tabular}
\end{center}
\end{table}

\subsection{V452 Cassiopeiae}\label{obj:v452cas}

   V452 Cas was discovered as a dwarf nova (=S 10453)
with a photographic range of 14--17.5
by \citet{ric69v452cas}.  The SU UMa-type nature
was confirmed by T. Vanmunster in 2000
(vsnet-alert 3698, 3707).
\citet{she09v452cas} studied this object between 2005
and 2008, and obtained supercycle lengths of 146$\pm$16~d.
For more history, see \citet{Pdot6}.

\begin{figure}
  \begin{center}
%    \FigureFile(85mm,70mm){v452cascomp3.eps}
    \FigureFile(85mm,70mm){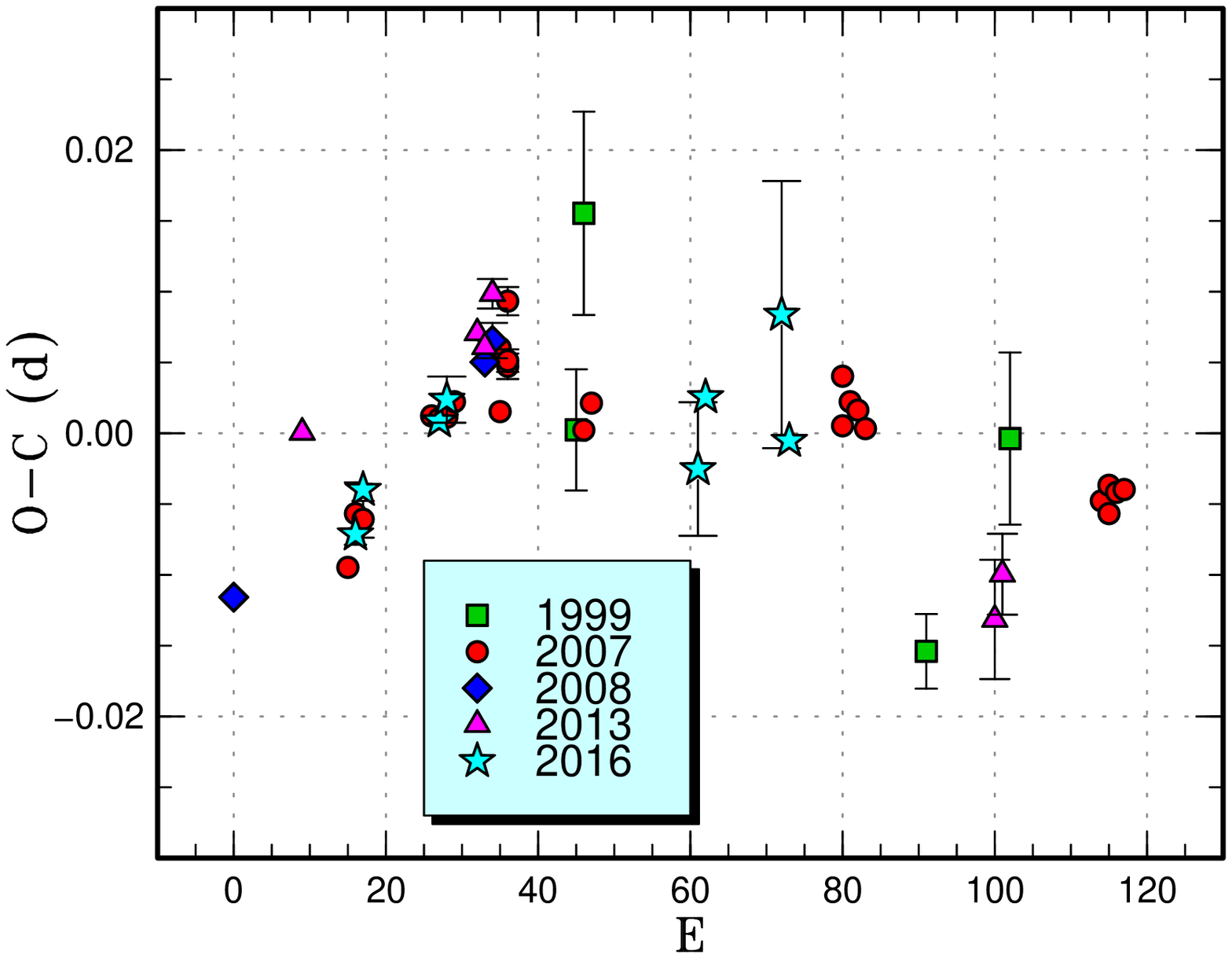}
  \end{center}
  \caption{Comparison of $O-C$ diagrams of V452 Cas between different
  superoutbursts.  A period of 0.08880~d was used to draw this figure.
  Approximate cycle counts ($E$) after the start of the superoutburst
  were used.  Since the start of the 2016 superoutburst was
  well observed, the 2007 $O-C$ diagram has been shifted
  by 15 cycles to match the 2016 one.}
  \label{fig:v452cascomp3}
\end{figure}

   The 2016 superoutburst was detected on February 9
by M. Hiraga at an unfiltered CCD magnitude of 15.5
(cf. vsnet-alert 19486).  This outburst was also
detected by I. Miller on February 10 and subsequent
observations detected superhumps (vsnet-alert 19486).
The times of superhump maxima are listed in
table \ref{tab:v452casoc2016}.  The initial part
($E <$11) probably refers to stage A.
A comparison of $O-C$ diagrams (figure \ref{fig:v452cascomp3})
suggests that V452 Cas has long-lasting stage A,
which has been recently established in long-$P_{\rm orb}$
systems such as V1006 Cyg (\cite{kat16v1006cyg};
\cite{Pdot6}; subsection \ref{sec:longstagea}).
Having a long superhumps period,
V452 Cas is an excellent candidate for this class
of objects.  Since stage A superhumps appears to be
easily observable, determination of the orbital period
and the period of stage A superhumps by systematic
observations will lead to an estimation of $q$
by the stage A superhump method \citep{kat13qfromstageA}.

% SI

\begin{table}
\caption{Superhump maxima of V452 Cas (2016)}\label{tab:v452casoc2016}
\begin{center}
\begin{tabular}{rp{55pt}p{40pt}r@{.}lr}
\hline
\multicolumn{1}{c}{$E$} & \multicolumn{1}{c}{max\commenta} & \multicolumn{1}{c}{error} & \multicolumn{2}{c}{$O-C$\commentb} & \multicolumn{1}{c}{$N$\commentc} \\
\hline
0 & 57429.2812 & 0.0013 & $-$0&0040 & 60 \\
1 & 57429.3731 & 0.0002 & $-$0&0009 & 231 \\
11 & 57430.2660 & 0.0012 & 0&0028 & 30 \\
12 & 57430.3563 & 0.0016 & 0&0042 & 34 \\
45 & 57433.2818 & 0.0047 & $-$0&0044 & 46 \\
46 & 57433.3757 & 0.0013 & 0&0006 & 71 \\
56 & 57434.2695 & 0.0094 & 0&0053 & 18 \\
57 & 57434.3494 & 0.0011 & $-$0&0037 & 89 \\
\hline
  \multicolumn{6}{l}{\commenta BJD$-$2400000.} \\
  \multicolumn{6}{l}{\commentb Against max $= 2457429.2851 + 0.088911 E$.} \\
  \multicolumn{6}{l}{\commentc Number of points used to determine the maximum.} \\
\end{tabular}
\end{center}
\end{table}

\subsection{V1040 Centauri}\label{obj:v1040cen}

   This object was originally selected as an X-ray source
(RX J1155.4$-$5641: \cite{mot98ROSATCV}).
After monitoring since 1999, B. Monard detected an outburst
on 2000 July 4 (vsnet-alert 5064).  The variable star
name V1040 Cen was given based on this observation
\citep{NameList77}.  The 2002 superoutburst was relatively
well observed.  \citet{pat03suumas} reported a superhump
period of 0.06215(10)~d and an orbital signal with a period
of 0.06028(6)~d.  Using the available data (part of the
data used in \cite{pat03suumas}), \citet{Pdot} studied
the evolution of superhumps and identified a period
of 0.060296(8)~d during the post-superoutburst phase.
This period has been listed as the orbital period
in \citet{RKCat}.  This period, however, was not confirmed
during the quiescent phase in 2008 \citep{Pdot}.
\citet{wou10CVperiod} studied this object on three nights
in 2008 when the object was returning to quiescence
from a normal outburst.  Although \citet{wou10CVperiod}
reported dwarf nova oscillations (DNOs) and quasi-periodic
oscillations (QPOs), there was no information about
the orbital variation.
Longa-Pe{\~n}a et al. found a spectroscopic orbital period
of 0.06049(10)~d\footnote{
  $<$http://www.noao.edu/meetings/wildstars2/posters/monday/p-longa-poster.png$>$.
}.
\citet{rut11v1040cen} studied this object
in 2009 in quiescence and two normal outbursts.
\citet{rut11v1040cen} derived an orbital period of
0.060458(80)~d using their photometric data.

\begin{figure}
  \begin{center}
%    \FigureFile(85mm,70mm){v1040cencomp.eps}
    \FigureFile(85mm,70mm){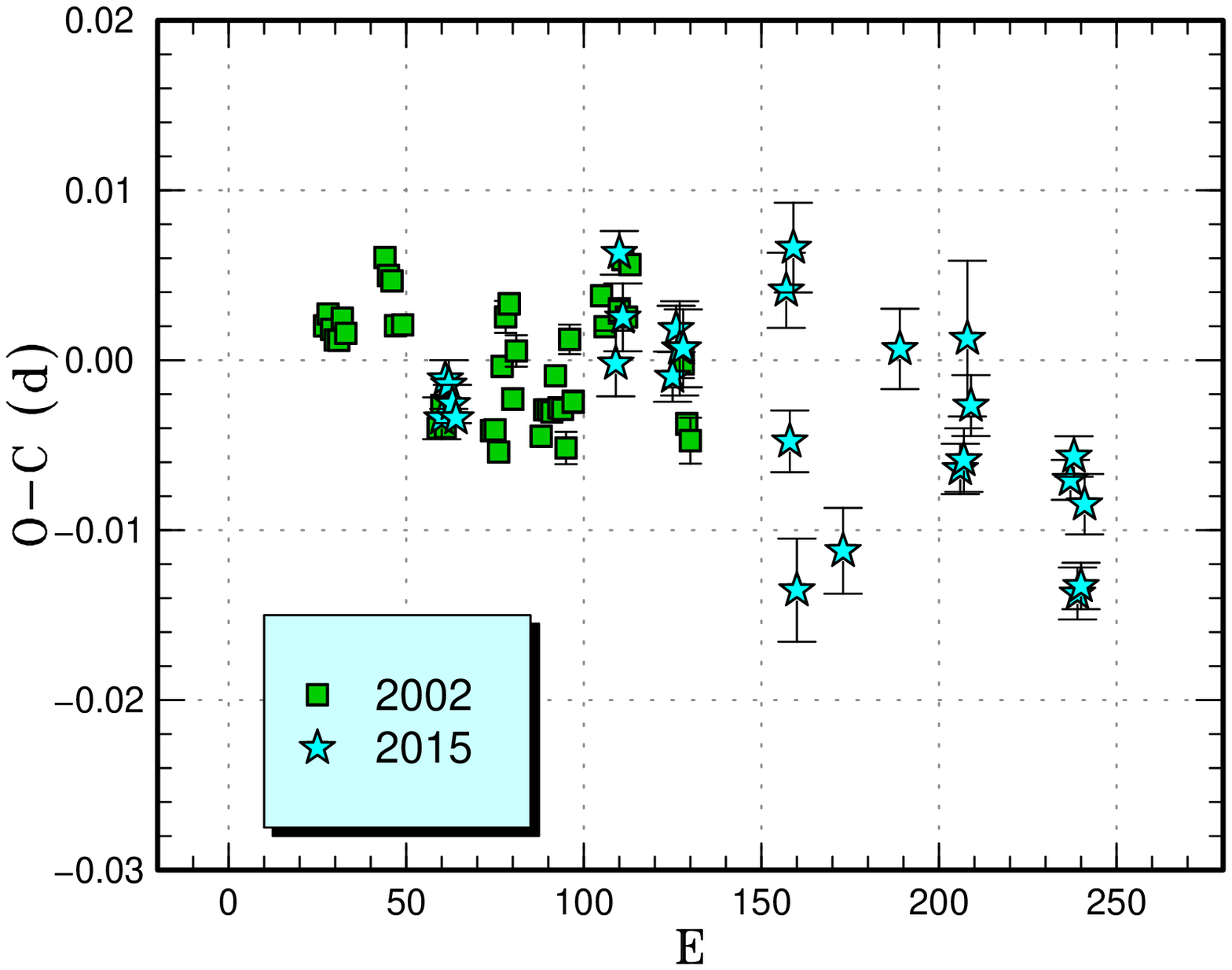}
  \end{center}
  \caption{Comparison of $O-C$ diagrams of V1040 Cen between different
  superoutbursts.  A period of 0.06218~d was used to draw this figure.
  Approximate cycle counts ($E$) after the start of the superoutburst
  were used.  Since there was a separate precursor outburst
  in 2015, we shifted the $O-C$ diagram so that it best
  matches the 2002 one.  The maxima for $E>$150 for the 2015
  superoutburst probably refer to the secondary maxima
  (see text for details).}
  \label{fig:v1040cencomp}
\end{figure}

   The 2015 superoutburst was detected by R. Stubbings
on April 4 at a visual magnitude of 12.0
(vsnet-outburst 18159).  This observation later turned
out to be a precursor outburst.  The object once faded
and the rising branch of the main superoutburst
was recorded in April 7 at $V$=13.49 by P. Starr.
The object took another 2~d to reach the maximum
around $V$=11.4.  Our time-resolved photometry started
1~d after this peak and immediately detected
superhumps (vsnet-alert 18532, 18560).
The times of superhump maxima are listed in
table \ref{tab:v1040cenoc2015}.  The maxima after
$E$=129 are post-superoutburst superhumps.
There was no phase jump around the termination
of the superoutburst.  A comparison of the $O-C$
diagrams (figure \ref{fig:v1040cencomp}) suggests that
the $O-C$ diagrams between the 2002 and 2015 do not
agree if we assume that superhumps immediately
started evolving around the precursor outburst.
This comparison suggests that superhumps started to
develop $\sim$40 cycles ($\sim$2.5~d) following
the peak of the precursor outburst.
This epoch corresponds to 0.5~d before the rising
phase to the main superoutburst.  This observation
gives a support to the suggestion that it can take
a long time to fully develop stage B superhumps
when the precursor occurred well before the main
superoutburst and the system stayed in low state
for a long time before the main superoutburst
starts [see subsection 5.4 in \citet{kat16j0333};
the case of CY UMa in \citet{Pdot7} might have been
similar].

\begin{figure}
  \begin{center}
%    \FigureFile(85mm,110mm){v1040censhprof.eps}
    \FigureFile(85mm,110mm){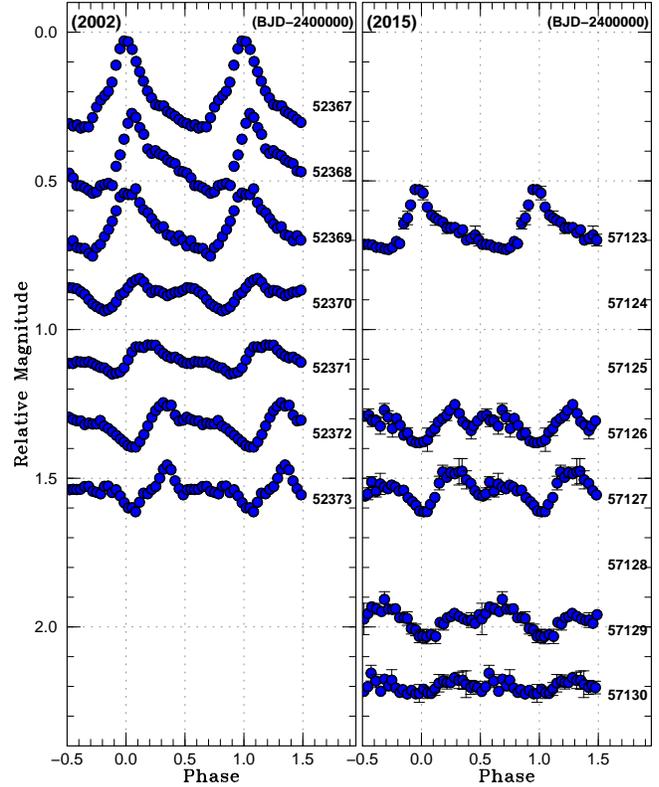}
  \end{center}
  \caption{Evolution of superhumps in V1040 Cen
     during the 2002 and 2015 superoutbursts.
     A period of 0.06190~d was used to draw this figure.
     The 2015 data were shifted by 2~d to reflect
     the shift in the cycle count in figure \ref{fig:v1040cencomp}.}
  \label{fig:v1040censhprof}
\end{figure}

   The evolutionary phase of superhumps also well match
between the 2002 and 2015 superoutburst by assuming
that superhumps started to develop $\sim$40 cycles
following the peak of the precursor outburst
(figure \ref{fig:v1040censhprof}).  Secondary maxima
of superhumps are relatively prominent in this system
and became stronger than the original maxima
during the later course of the superoutburst.
The same feature was recorded in the Kepler data
of V344 Lyr, which was considered to arise from
the accretion stream resulting a bright spot that 
sweeps around the rim of the non-axisymmetric disk
\citep{woo11v344lyr}.
The maxima for $E \ge$97 in table \ref{tab:v1040cenoc2015}
correspond to these secondary maxima, and are excluded
for obtaining the period in table \ref{tab:perlist}.
The interpretation in \citet{woo11v344lyr} would
suggest a high mass-transfer rate, and indeed
``traditional'' late superhumps with an $\sim$0.5
phase jump are seen only in systems with frequent
outbursts (e.g. \cite{Pdot4}).  V1040 Cen, however,
lacks frequent normal outbursts which are expected
for a high mass-transfer rate.  The small outburst
amplitude ($\sim$3.0 mag for superoutbursts)
of this systems suggests a bright disk in quiescence,
which may in turn suggest a high mass-transfer rate.
Normal outbursts in this system may be somehow
suppressed.  The duration of superoutbursts are
somewhat short for a short $P_{\rm orb}$ object
(the duration of the plateau phase were $\sim$8~d
in 2015).

% SI

\begin{table}
\caption{Superhump maxima of V1040 Cen (2015)}\label{tab:v1040cenoc2015}
\begin{center}
\begin{tabular}{rp{55pt}p{40pt}r@{.}lr}
\hline
\multicolumn{1}{c}{$E$} & \multicolumn{1}{c}{max\commenta} & \multicolumn{1}{c}{error} & \multicolumn{2}{c}{$O-C$\commentb} & \multicolumn{1}{c}{$N$\commentc} \\
\hline
0 & 57122.5102 & 0.0012 & $-$0&0045 & 21 \\
1 & 57122.5747 & 0.0009 & $-$0&0022 & 25 \\
2 & 57122.6365 & 0.0014 & $-$0&0025 & 13 \\
3 & 57122.6976 & 0.0011 & $-$0&0036 & 16 \\
4 & 57122.7589 & 0.0007 & $-$0&0044 & 17 \\
49 & 57125.5602 & 0.0019 & 0&0008 & 21 \\
50 & 57125.6289 & 0.0013 & 0&0074 & 13 \\
51 & 57125.6873 & 0.0020 & 0&0036 & 16 \\
65 & 57126.5543 & 0.0015 & 0&0007 & 24 \\
66 & 57126.6193 & 0.0014 & 0&0036 & 13 \\
67 & 57126.6803 & 0.0028 & 0&0025 & 17 \\
68 & 57126.7425 & 0.0023 & 0&0026 & 17 \\
97 & 57128.5492 & 0.0022 & 0&0072 & 24 \\
98 & 57128.6025 & 0.0018 & $-$0&0016 & 16 \\
99 & 57128.6760 & 0.0026 & 0&0098 & 17 \\
100 & 57128.7181 & 0.0030 & $-$0&0103 & 16 \\
113 & 57129.5287 & 0.0025 & $-$0&0074 & 25 \\
129 & 57130.5355 & 0.0024 & 0&0052 & 24 \\
146 & 57131.5855 & 0.0015 & $-$0&0011 & 18 \\
147 & 57131.6482 & 0.0019 & $-$0&0006 & 17 \\
148 & 57131.7175 & 0.0046 & 0&0066 & 16 \\
149 & 57131.7757 & 0.0018 & 0&0027 & 10 \\
177 & 57133.5124 & 0.0012 & $-$0&0004 & 30 \\
178 & 57133.5760 & 0.0012 & 0&0010 & 28 \\
179 & 57133.6301 & 0.0015 & $-$0&0070 & 15 \\
180 & 57133.6927 & 0.0014 & $-$0&0065 & 16 \\
181 & 57133.7597 & 0.0018 & $-$0&0017 & 13 \\
\hline
  \multicolumn{6}{l}{\commenta BJD$-$2400000.} \\
  \multicolumn{6}{l}{\commentb Against max $= 2457122.5147 + 0.062136 E$.} \\
  \multicolumn{6}{l}{\commentc Number of points used to determine the maximum.} \\
\end{tabular}
\end{center}
\end{table}

\subsection{AL Comae Berenices}\label{obj:alcom}

   We provide the table of superhumps maxima during
the 2015 superoutburst which was not presented in
\citet{kim16alcom} as a form of table \ref{tab:alcomoc2015}.
The maxima for $E \le$20
were not included in \citet{kim16alcom} and some maxima
with poor statistics have been removed.  The resultant
updated $P_{\rm dot}$ was $+1.6(0.8) \times 10^{-5}$.
The main conclusions in \citet{kim16alcom} are unchanged.

\begin{figure}
  \begin{center}
%    \FigureFile(85mm,70mm){alcomcomp3.eps}
    \FigureFile(85mm,70mm){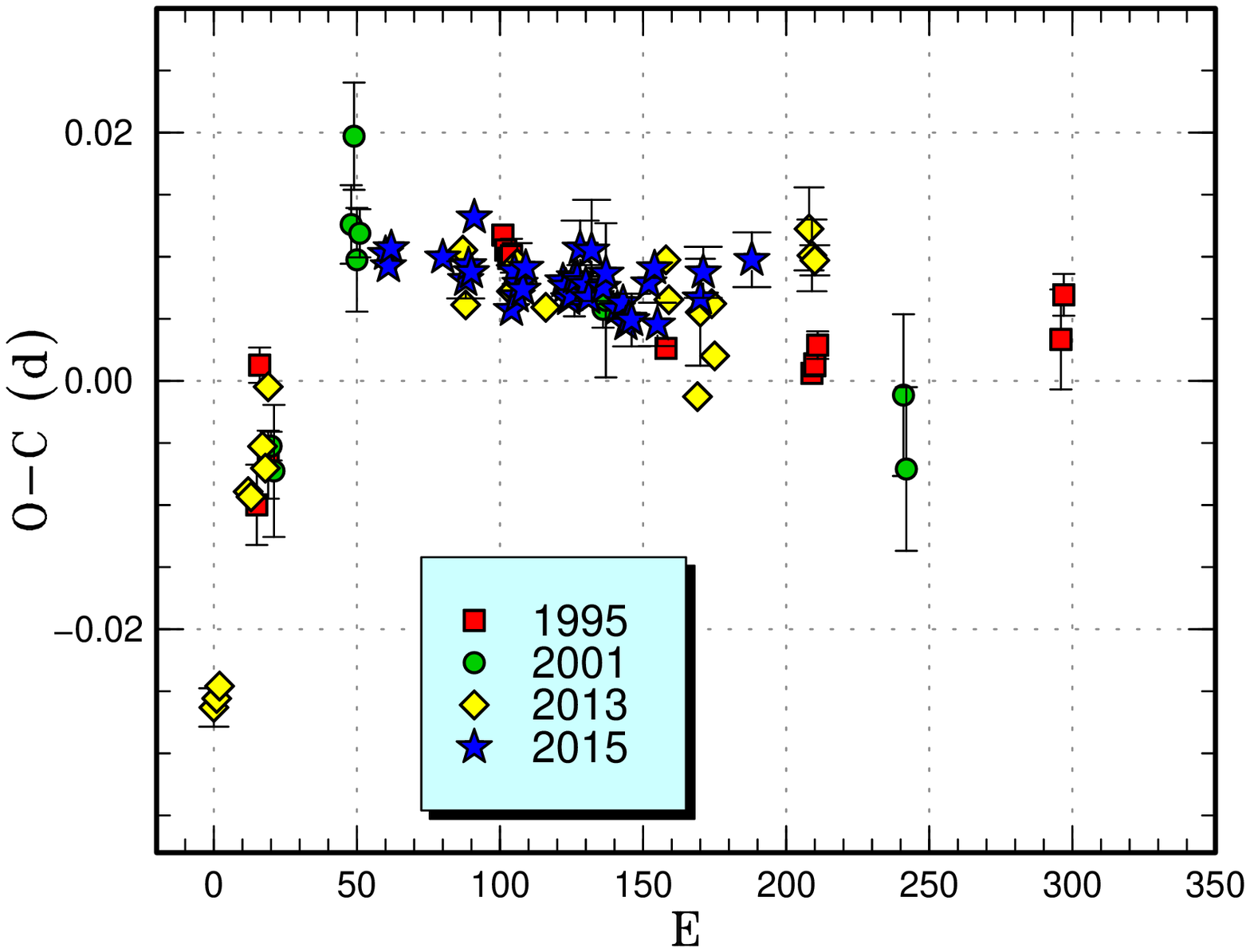}
  \end{center}
  \caption{Comparison of $O-C$ diagrams of AL Com between different
  superoutbursts.  A period of 0.05732~d was used to draw this figure.
  Approximate cycle counts ($E$) after the emergence of superhump
  were used.  Assuming that the stage A was best observed in
  2013, the 1995 and 2001 $O-C$ diagrams were shifted within
  20 cycles to best match the stage A-B transition in 2013.
  The 2015 $O-C$ diagram was shifted by 60 cycles to best
  match the others.}
  \label{fig:alcomcomp3}
\end{figure}

   A comparison of $O-C$ diagrams between different
superoutbursts is shown in figure \ref{fig:alcomcomp3}.
In order to match the other $O-C$ diagrams, the 2015
one had to be shifted by 60 cycles.  This implies
that stage A superhumps stared to appear $\sim$3~d
before the initial superhump observation on BJD 2457087
(2015 March 6).  Since the object was detected
on the rise on March 4, superhumps should have already
started to appear when the object was on the rise to
the superoutburst maximum.  This is consistent with
the lack of stage A superhumps in \citet{kim16alcom},
in which the earliest part of the superoutburst
was not well observed.

% SI

\begin{table}
\caption{Superhump maxima of AL Com (2015)}\label{tab:alcomoc2015}
\begin{center}
\begin{tabular}{rp{55pt}p{40pt}r@{.}lr}
\hline
\multicolumn{1}{c}{$E$} & \multicolumn{1}{c}{max\commenta} & \multicolumn{1}{c}{error} & \multicolumn{2}{c}{$O-C$\commentb} & \multicolumn{1}{c}{$N$\commentc} \\
\hline
0 & 57087.8281 & 0.0006 & 0&0006 & 22 \\
1 & 57087.8845 & 0.0006 & $-$0&0004 & 26 \\
2 & 57087.9432 & 0.0005 & 0&0011 & 26 \\
20 & 57088.9743 & 0.0008 & 0&0008 & 18 \\
28 & 57089.4310 & 0.0015 & $-$0&0008 & 30 \\
29 & 57089.4895 & 0.0004 & 0&0004 & 59 \\
30 & 57089.5463 & 0.0003 & $-$0&0001 & 56 \\
31 & 57089.6080 & 0.0010 & 0&0043 & 28 \\
44 & 57090.3457 & 0.0012 & $-$0&0028 & 43 \\
45 & 57090.4062 & 0.0005 & 0&0004 & 65 \\
46 & 57090.4615 & 0.0007 & $-$0&0016 & 78 \\
47 & 57090.5198 & 0.0008 & $-$0&0006 & 128 \\
48 & 57090.5766 & 0.0008 & $-$0&0010 & 90 \\
49 & 57090.6357 & 0.0007 & 0&0008 & 53 \\
62 & 57091.3798 & 0.0004 & 0&0000 & 41 \\
63 & 57091.4367 & 0.0003 & $-$0&0004 & 57 \\
64 & 57091.4930 & 0.0003 & $-$0&0013 & 57 \\
65 & 57091.5506 & 0.0005 & $-$0&0010 & 110 \\
66 & 57091.6094 & 0.0006 & 0&0005 & 85 \\
67 & 57091.6666 & 0.0007 & 0&0004 & 63 \\
68 & 57091.7263 & 0.0022 & 0&0028 & 21 \\
70 & 57091.8381 & 0.0016 & $-$0&0000 & 53 \\
71 & 57091.8943 & 0.0006 & $-$0&0011 & 81 \\
72 & 57091.9555 & 0.0041 & 0&0028 & 20 \\
76 & 57092.1817 & 0.0012 & $-$0&0001 & 112 \\
77 & 57092.2402 & 0.0015 & 0&0011 & 119 \\
81 & 57092.4666 & 0.0014 & $-$0&0017 & 30 \\
83 & 57092.5818 & 0.0013 & $-$0&0011 & 46 \\
84 & 57092.6374 & 0.0011 & $-$0&0028 & 55 \\
85 & 57092.6951 & 0.0007 & $-$0&0024 & 63 \\
86 & 57092.7523 & 0.0021 & $-$0&0024 & 26 \\
92 & 57093.0992 & 0.0010 & 0&0007 & 121 \\
94 & 57093.2151 & 0.0007 & 0&0019 & 121 \\
95 & 57093.2679 & 0.0017 & $-$0&0026 & 55 \\
110 & 57094.1298 & 0.0012 & $-$0&0001 & 119 \\
111 & 57094.1892 & 0.0021 & 0&0021 & 120 \\
128 & 57095.1646 & 0.0022 & 0&0035 & 104 \\
\hline
  \multicolumn{6}{l}{\commenta BJD$-$2400000.} \\
  \multicolumn{6}{l}{\commentb Against max $= 2457087.8276 + 0.057293 E$.} \\
  \multicolumn{6}{l}{\commentc Number of points used to determine the maximum.} \\
\end{tabular}
\end{center}
\end{table}

\subsection{VW Coronae Borealis}\label{obj:vwcrb}

   VW CrB was discovered as a dwarf nova (Antipin Var 21)
by \citet{ant96vwcrb}.  The observations by \citet{ant96vwcrb}
indicated the presence of two types of outbursts, which
was already suggestive of an SU UMa-type dwarf nova.
During the 1997 superoutburst, \citet{nov97vwcrb}
observed this object on one night and detected superhumps.
\citet{liu99CVspec1} obtained a spectrum in outburst
($B$=15.8).  This observation was made approximately
one month after the superoutburst observed in \citet{nov97vwcrb},
and was likely a normal outburst.
\citet{nog04vwcrb} reported observations of two superoutbursts
in 2001 and 2003, and detected a positive $P_{\rm dot}$
despite the relatively long superhump period.
These observations and the 2006 superoutburst were
analyzed in \citet{Pdot}.

\begin{figure}
  \begin{center}
%    \FigureFile(85mm,70mm){vwcrbcomp2.eps}
    \FigureFile(85mm,70mm){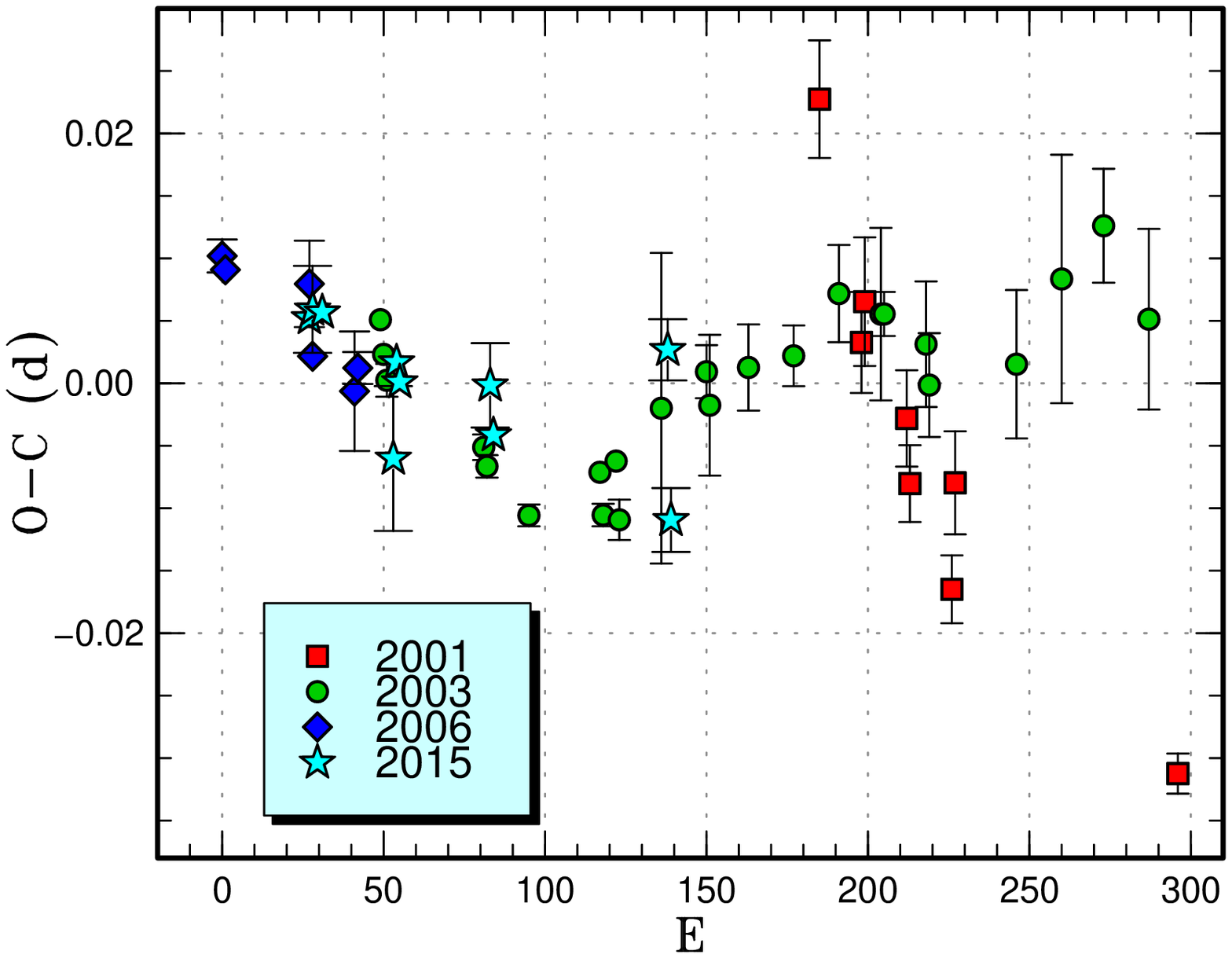}
  \end{center}
  \caption{Comparison of $O-C$ diagrams of VW CrB between different
  superoutbursts.  A period of 0.07290~d was used to draw this figure.
  Approximate cycle counts ($E$) after the start of the superoutburst
  were used.}
  \label{fig:vwcrbcomp2}
\end{figure}

   The 2015 superoutburst was detected by D. Denisenko
using the MASTER network (vsnet-alert 18577).
Subsequent observations detected superhumps
(vsnet-alert 18583).
The times of superhump maxima are listed in
table \ref{tab:vwcrboc2015}.
A comparison of $O-C$ diagrams between different
superoutbursts is shown in figure \ref{fig:vwcrbcomp2}.
The 2015 observation most likely covered stage B.

% SI

\begin{table}
\caption{Superhump maxima of VW CrB (2015)}\label{tab:vwcrboc2015}
\begin{center}
\begin{tabular}{rp{55pt}p{40pt}r@{.}lr}
\hline
\multicolumn{1}{c}{$E$} & \multicolumn{1}{c}{max\commenta} & \multicolumn{1}{c}{error} & \multicolumn{2}{c}{$O-C$\commentb} & \multicolumn{1}{c}{$N$\commentc} \\
\hline
0 & 57140.1102 & 0.0006 & 0&0018 & 155 \\
1 & 57140.1838 & 0.0035 & 0&0026 & 76 \\
4 & 57140.4022 & 0.0004 & 0&0026 & 41 \\
26 & 57141.9943 & 0.0058 & $-$0&0073 & 75 \\
27 & 57142.0749 & 0.0006 & 0&0005 & 157 \\
28 & 57142.1462 & 0.0006 & $-$0&0010 & 155 \\
56 & 57144.1872 & 0.0034 & 0&0009 & 51 \\
57 & 57144.2561 & 0.0011 & $-$0&0030 & 133 \\
111 & 57148.1995 & 0.0025 & 0&0082 & 126 \\
112 & 57148.2588 & 0.0026 & $-$0&0053 & 105 \\
\hline
  \multicolumn{6}{l}{\commenta BJD$-$2400000.} \\
  \multicolumn{6}{l}{\commentb Against max $= 2457140.1083 + 0.072820 E$.} \\
  \multicolumn{6}{l}{\commentc Number of points used to determine the maximum.} \\
\end{tabular}
\end{center}
\end{table}

\subsection{V550 Cygni}\label{obj:v550cyg}

   V550 Cyg was discovered by \citet{hof49newvar1}
as a dwarf nova (=S 3847) with a photographic range of 15
to fainter than 18 mag.  \citet{ahn52v550cyg} reported
a photographic range of 14.8 to fainter than 16.3.
Although the finding chart was provided by
\citet{hof57v550cyg} (Nr. 291), the scale was insufficient
to identify the object in quiescence.
\citet{pin72v1454cyg} recorded an outburst at a photographic
magnitude of 14.2 on 1961 September 20.
\citet{ski99VSID} was the first to identify the object
in 1999 and two outbursts were detected in 2000
(vsnet-alert 3993, 5191).
During the 2000 August outburst, superhumps were
detected \citep{Pdot}.

   The 2015 outburst was detected by E. Muyllaert
on October 11 at an unfiltered CCD magnitude of 15.12
(cf. vsnet-alert 19156).  Only single-night observations
(vsnet-alert 19173) yielded two superhumps maxima:
BJD 2457313.0659(5) ($N$=137) and 2457313.1256(18)
($N$=87).

\subsection{V1028 Cygni}\label{obj:v1028cyg}

   V1028 Cyg was discovered as a dwarf nova (=S 7854)
by \citet{hof63v1028cyg}.  The object has been famous
for the low frequency of outbursts (see e.g.
\cite{may68v1028cyg}; \cite{may70v1028cyg}).
Early observations \citep{tch63v1028cyg} were already
indicative of an SU UMa-type dwarf nova.
\citet{bru92CVspec2} reported a typical dwarf nova-type
spectrum in quiescence.  
The 1995 superoutburst was the best recorded
(cf. vsnet-alert 166, 168, 169, 172, 175, 177, 192,
193, 205).  This superoutburst was one of the first
examples showing positive $P_{\rm dot}$,
although the publication took some time \citep{bab00v1028cyg}.
Other superoutburst (not well observed) in 1996, 1999, 
2001, 2002, 2004 and 2008 were reported in \citet{Pdot}.

\begin{figure}
  \begin{center}
%    \FigureFile(85mm,70mm){v1028cygcomp2.eps}
    \FigureFile(85mm,70mm){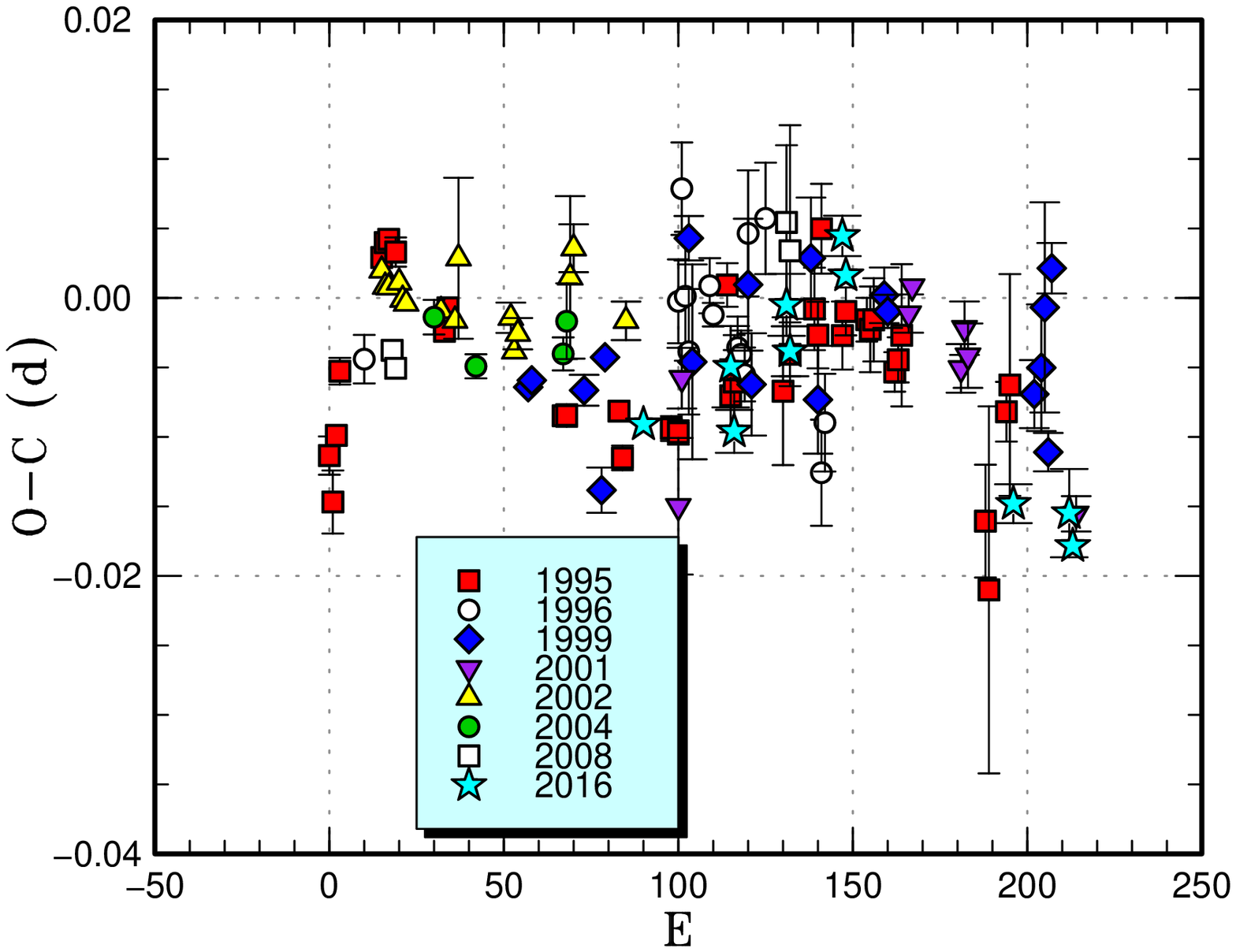}
  \end{center}
  \caption{Comparison of $O-C$ diagrams of V1028 Cyg between different
  superoutbursts.  A period of 0.06178~d was used to draw this figure.
  Approximate cycle counts ($E$) after the start of the superoutburst
  (the start of the main superoutburst when preceded by
  a precursor) were used.  The $E$ for the 2008 superoutburst was
  somewhat uncertain due to the lack of observations at the early stage.
  The 2016 superoutburst was shifted by 90 cycles
  to match the best observed 1995 one.}
  \label{fig:v1028cygcomp2}
\end{figure}

   The 2016 outburst was detected on March 14 probably
on the rising phase ($V$=14.8) by the ASAS-SN team and M. Hiraga
(vsnet-alert 19601).  Since the initial detection
magnitude was faint, the outburst did not receive
attention.  The initial time-resolved photometry started
on March 18 when the superoutburst state was confirmed.
Superhumps were soon recorded (vsnet-alert 19632).
The times of superhump maxima are listed in
table \ref{tab:v1028cygoc2016}.  The $O-C$ analysis
(figure \ref{fig:v1028cygcomp2}) clearly indicates
that the present observation recorded the later part
of stage B and stage C.  The period of stage C superhumps
was not determined due to the lack of data.
In figure \ref{fig:v1028cygcomp2}, we had to shift
90 cycles to match the $O-C$ curve to the 1995 one,
although initial superhump observations started
$\sim$60 cycles after the outburst detection.
It may have been that the 2016 superoutburst was
shorter than other ones, or it had a separate
precursor during which superhumps already started
to develop.

% SI

\begin{table}
\caption{Superhump maxima of V1028 Cyg (2016)}\label{tab:v1028cygoc2016}
\begin{center}
\begin{tabular}{rp{55pt}p{40pt}r@{.}lr}
\hline
\multicolumn{1}{c}{$E$} & \multicolumn{1}{c}{max\commenta} & \multicolumn{1}{c}{error} & \multicolumn{2}{c}{$O-C$\commentb} & \multicolumn{1}{c}{$N$\commentc} \\
\hline
0 & 57465.7060 & 0.0007 & $-$0&0081 & 51 \\
25 & 57467.2547 & 0.0029 & $-$0&0014 & 34 \\
26 & 57467.3118 & 0.0015 & $-$0&0060 & 46 \\
41 & 57468.2476 & 0.0022 & 0&0046 & 33 \\
42 & 57468.3061 & 0.0018 & 0&0014 & 47 \\
57 & 57469.2411 & 0.0015 & 0&0112 & 44 \\
58 & 57469.3000 & 0.0008 & 0&0084 & 47 \\
106 & 57472.2490 & 0.0014 & $-$0&0032 & 27 \\
122 & 57473.2368 & 0.0032 & $-$0&0023 & 34 \\
123 & 57473.2962 & 0.0010 & $-$0&0046 & 35 \\
\hline
  \multicolumn{6}{l}{\commenta BJD$-$2400000.} \\
  \multicolumn{6}{l}{\commentb Against max $= 2457465.7141 + 0.061680 E$.} \\
  \multicolumn{6}{l}{\commentc Number of points used to determine the maximum.} \\
\end{tabular}
\end{center}
\end{table}

\subsection{V1113 Cygni}\label{obj:v1113cyg}

   V1113 Cyg was discovered as a dwarf nova (=S 9382)
by \citet{hof66an289139}.
\citet{kat96v1113cyg} reported the detection of
superhumps.  \citet{kat01v1113cyg} reported a mean
supercycle of 189.8~d and that the number of normal outbursts
is too small for this short supercycle.
\citet{bak10v1113cyg} studied the 2003 and 2005
superoutbursts, and also confirmed the low frequency
of normal outbursts.  Although \citet{bak10v1113cyg}
reported large negative $P_{\rm dot}$, they probably
recorded stage B-C transition \citep{Pdot2}.

\begin{figure}
  \begin{center}
%    \FigureFile(85mm,70mm){v1113cygcomp2.eps}
    \FigureFile(85mm,70mm){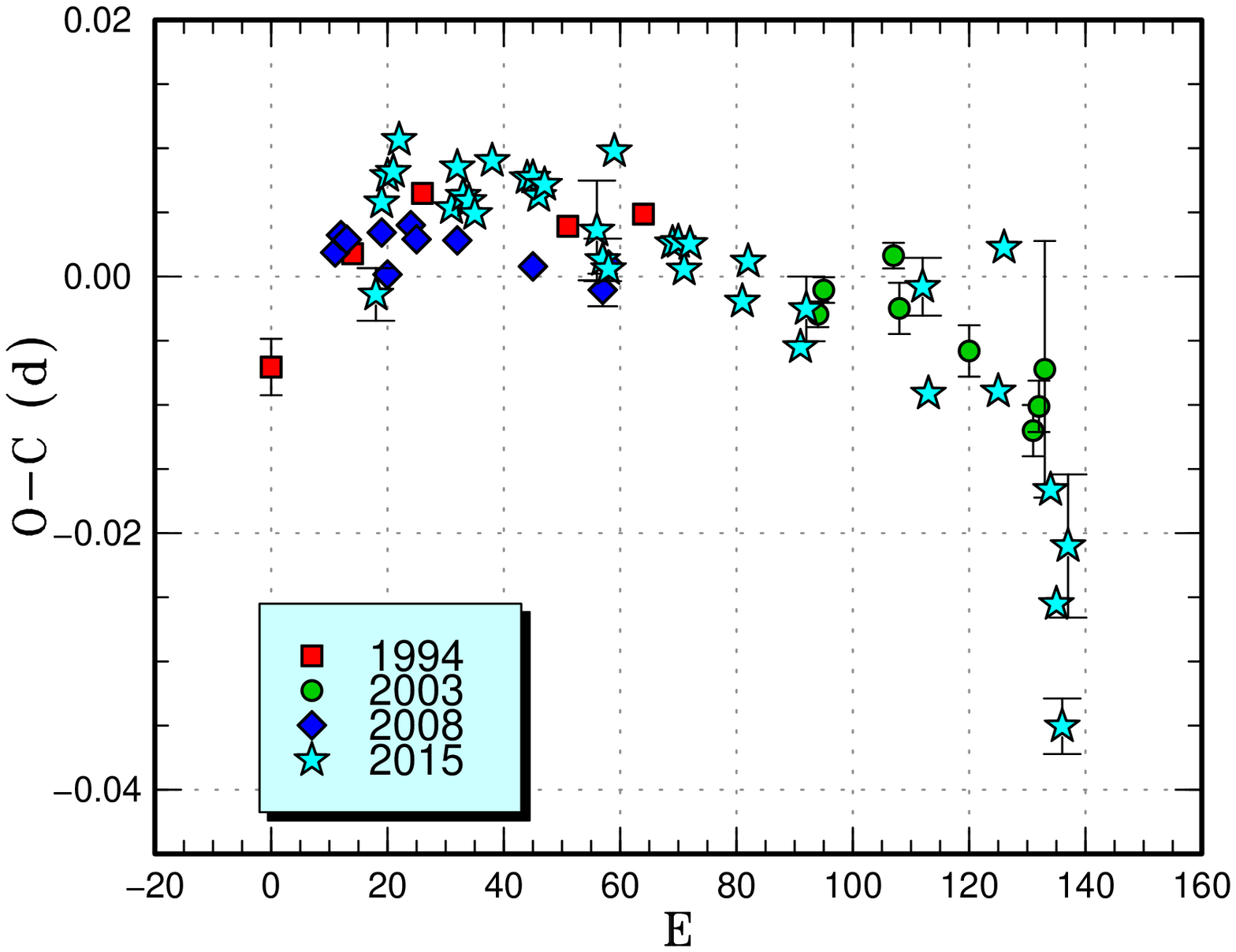}
  \end{center}
  \caption{Comparison of $O-C$ diagrams of V1113 Cyg between different
  superoutbursts.  A period of 0.07911~d was used to draw this figure.
  Approximate cycle counts ($E$) after the start of the superoutburst
  were used.}
  \label{fig:v1113cygcomp2}
\end{figure}

   The 2015 superoutburst was detected by the ASAS-SN team
on August 28 at $V$=13.95 (cf. vsnet-alert 19014).
Subsequent observations detected superhumps
(vsnet-alert 19018, 19019, 19023).
The times of superhump maxima are listed in
table \ref{tab:v1113cygoc2015}.  The stages are not
distinct (see also figure \ref{fig:v1113cygcomp2})
and we adopted a globally averaged period
except the rapidly fading part in table \ref{tab:perlist}.

% SI

\begin{table}
\caption{Superhump maxima of V1113 Cyg (2015)}\label{tab:v1113cygoc2015}
\begin{center}
\begin{tabular}{rp{55pt}p{40pt}r@{.}lr}
\hline
\multicolumn{1}{c}{$E$} & \multicolumn{1}{c}{max\commenta} & \multicolumn{1}{c}{error} & \multicolumn{2}{c}{$O-C$\commentb} & \multicolumn{1}{c}{$N$\commentc} \\
\hline
0 & 57264.2780 & 0.0021 & $-$0&0126 & 74 \\
1 & 57264.3643 & 0.0003 & $-$0&0052 & 308 \\
2 & 57264.4455 & 0.0003 & $-$0&0029 & 306 \\
3 & 57264.5249 & 0.0004 & $-$0&0023 & 223 \\
4 & 57264.6065 & 0.0009 & 0&0004 & 32 \\
13 & 57265.3132 & 0.0005 & $-$0&0028 & 106 \\
14 & 57265.3955 & 0.0003 & 0&0006 & 166 \\
15 & 57265.4724 & 0.0004 & $-$0&0014 & 166 \\
16 & 57265.5511 & 0.0004 & $-$0&0016 & 165 \\
17 & 57265.6292 & 0.0006 & $-$0&0024 & 161 \\
20 & 57265.8706 & 0.0011 & 0&0025 & 64 \\
26 & 57266.3440 & 0.0005 & 0&0026 & 103 \\
27 & 57266.4231 & 0.0004 & 0&0028 & 232 \\
28 & 57266.5008 & 0.0004 & 0&0016 & 282 \\
29 & 57266.5808 & 0.0007 & 0&0028 & 210 \\
38 & 57267.2891 & 0.0034 & 0&0012 & 23 \\
39 & 57267.3660 & 0.0017 & $-$0&0008 & 61 \\
40 & 57267.4444 & 0.0011 & $-$0&0013 & 42 \\
41 & 57267.5327 & 0.0009 & 0&0082 & 61 \\
51 & 57268.3166 & 0.0005 & 0&0033 & 40 \\
52 & 57268.3958 & 0.0002 & 0&0037 & 278 \\
53 & 57268.4728 & 0.0003 & 0&0018 & 187 \\
54 & 57268.5539 & 0.0003 & 0&0040 & 176 \\
63 & 57269.2614 & 0.0005 & 0&0017 & 146 \\
64 & 57269.3436 & 0.0002 & 0&0050 & 238 \\
73 & 57270.0489 & 0.0014 & 0&0004 & 138 \\
74 & 57270.1310 & 0.0025 & 0&0037 & 125 \\
94 & 57271.7150 & 0.0022 & 0&0101 & 113 \\
95 & 57271.7857 & 0.0006 & 0&0020 & 183 \\
103 & 57272.4141 & 0.0010 & $-$0&0006 & 82 \\
104 & 57272.4948 & 0.0006 & 0&0012 & 80 \\
105 & 57272.5739 & 0.0009 & 0&0015 & 82 \\
107 & 57272.7352 & 0.0012 & 0&0050 & 115 \\
116 & 57273.4396 & 0.0014 & $-$0&0005 & 73 \\
117 & 57273.5098 & 0.0012 & $-$0&0091 & 76 \\
118 & 57273.5795 & 0.0022 & $-$0&0183 & 79 \\
119 & 57273.6725 & 0.0056 & $-$0&0042 & 23 \\
\hline
  \multicolumn{6}{l}{\commenta BJD$-$2400000.} \\
  \multicolumn{6}{l}{\commentb Against max $= 2457264.2907 + 0.078874 E$.} \\
  \multicolumn{6}{l}{\commentc Number of points used to determine the maximum.} \\
\end{tabular}
\end{center}
\end{table}

\subsection{HO Delphini}\label{obj:hodel}

   HO Del (=S 10066) was discovered as a dwarf nova by \citet{hof67an29043}.
\citet{hof67an29043} recorded two outbursts in 1963 October and
1966 September.  The coordinates of this object
was wrongly given in \citet{hof67an29043} and it was only
corrected in the third volume of the fourth edition of
the GCVS \citep{GCVS} (the correct identification was found
by T.K. in 1990 while preparing charts by comparison with
the Palomar Sky Survey; the observations since 1990 by
the VSOLJ members referred to the correct object).
\citet{mun98CVspec5} confirmed the dwarf nova-type nature
by recording relatively strong Balmer and He{\sc i} emission lines.
Observations of superhumps during the 1994, 1996 and 2001
superoutbursts were analyzed in \citet{kat03hodel}.
\citet{pat03suumas} also reported the 1996 superoutburst
and the spectroscopic orbital period.
In \citet{kat03hodel}, HO Del was chosen as a prototypical
object showing the brightening trend near the end of
the plateau phase.  This phenomenon was later identified as
emergence of stage C superhumps \citep{Pdot}.
The 2008 superoutburst was also reported in \citet{Pdot}.

\begin{figure}
  \begin{center}
%    \FigureFile(85mm,70mm){hodelcomp2.eps}
    \FigureFile(85mm,70mm){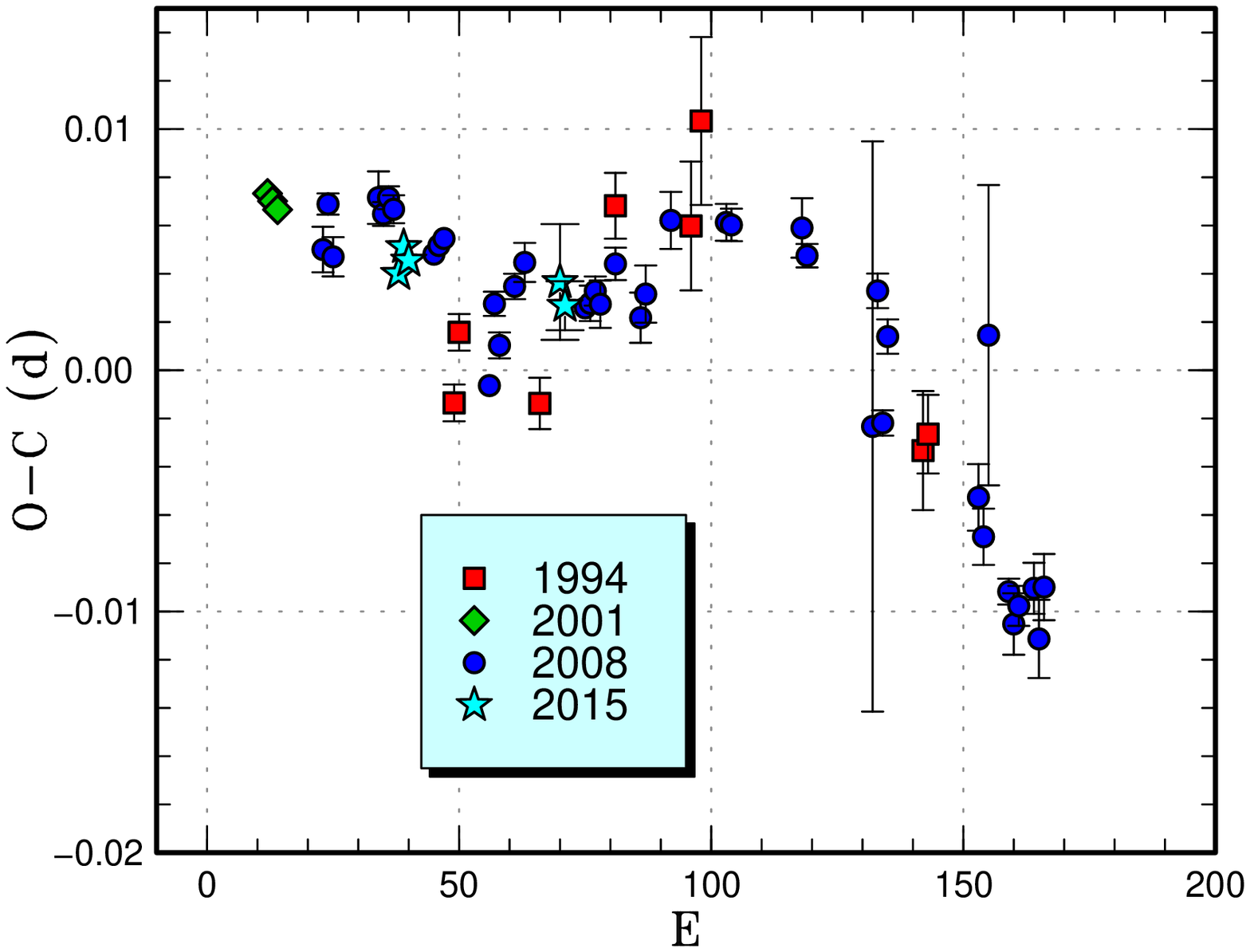}
  \end{center}
  \caption{Comparison of $O-C$ diagrams of HO Del between different
  superoutbursts.  A period of 0.06437~d was used to draw this figure.
  Approximate cycle counts ($E$) after the start of the superoutburst
  were used.  The 1994 superoutburst were artificially shifted
  to match the others.}
  \label{fig:hodelcomp2}
\end{figure}

   The 2015 outburst was detected by R. Stubbings and ASAS-SN
on July 18 (vsnet-alert 18865).  Superhump were detected
by observations which started 2~d later (vsnet-alert 18871).
The times of superhump maxima are listed in
table \ref{tab:hodeloc2015}.  The superhump period indicates
that these observations were already in stage B.
A comparison of $O-C$ diagrams between different
superoutbursts is shown in figure \ref{fig:hodelcomp2}.
It is likely that stage A is short in this system since stage B
superhumps already appeared 2~d after the outburst
detection despite that the outburst was detected sufficiently
early at least in 2001.

% SI

\begin{table}
\caption{Superhump maxima of HO Del (2015)}\label{tab:hodeloc2015}
\begin{center}
\begin{tabular}{rp{55pt}p{40pt}r@{.}lr}
\hline
\multicolumn{1}{c}{$E$} & \multicolumn{1}{c}{max\commenta} & \multicolumn{1}{c}{error} & \multicolumn{2}{c}{$O-C$\commentb} & \multicolumn{1}{c}{$N$\commentc} \\
\hline
0 & 57224.3507 & 0.0004 & $-$0&0006 & 59 \\
1 & 57224.4162 & 0.0005 & 0&0006 & 72 \\
2 & 57224.4800 & 0.0005 & 0&0000 & 59 \\
32 & 57226.4102 & 0.0024 & 0&0005 & 51 \\
33 & 57226.4736 & 0.0010 & $-$0&0005 & 66 \\
\hline
  \multicolumn{6}{l}{\commenta BJD$-$2400000.} \\
  \multicolumn{6}{l}{\commentb Against max $= 2457224.3513 + 0.064326 E$.} \\
  \multicolumn{6}{l}{\commentc Number of points used to determine the maximum.} \\
\end{tabular}
\end{center}
\end{table}

\subsection{AQ Eridani}\label{obj:aqeri}

   AQ Eri was discovered as a dwarf nova (=AN 431.1934) by
\citet{mor34an253441}.  \citet{hop35aqeri} studied
this object using 232 plates and found it unlikely
a Mira variable since it was invisible most of the time.
\citet{hop35aqeri} derived a possible period of 78~d
using three observed maxima (outbursts).
\citet{pet60DNe} listed
the object as a U Gem-type object with a cycle length
of 78~d. \citet{bon78bluevar2} obtained a spectrum
and recorded diffuse (broad) hydrogen emission lines
superposed on a blue continuum.  \citet{vog82atlas}
also listed the object as a U Gem-type variable.
\citet{GCVS} (printed version) listed the object as
a possible Z Cam-type dwarf nova based on \citet{bat81DNe1}.
Photometric observations in quiescence, however,
suggested a short orbital period \citet{szk87shortPCV}.

   Based on historical instances in which superoutbursts
were confused with Z Cam-type standstills, \citet{kat89aqeri}
studied this object during a long, bright outburst
in 1987 November both in photographic and visual
observations.  The detection of superhumps confirmed
the SU UMa-type nature.

   \citet{kat91aqeri} and \citet{kat01aqeri} reported
observations of superhumps using a CCD in 1991
and 1992, respectively.
\citet{kat99aqeri} also reported observations of
a normal outburst in 1998 December.
The spectroscopic orbital period was determined by
\citet{men93wxcetaqericuvel} and by \citet{tho96Porb}.
\citet{tap03DNelineprofile} reported a line-profile
analysis.  Further superoutbursts were observed
and reported in \citet{Pdot} (the 2006, 2008 superoutbursts),
\citet{Pdot2} (the 2010 superoutburst),
\citet{Pdot4} (the 2011 superoutburst) and
\citet{Pdot5} (the 2012 superoutburst).

\begin{figure}
  \begin{center}
%    \FigureFile(85mm,70mm){aqericomp4.eps}
    \FigureFile(85mm,70mm){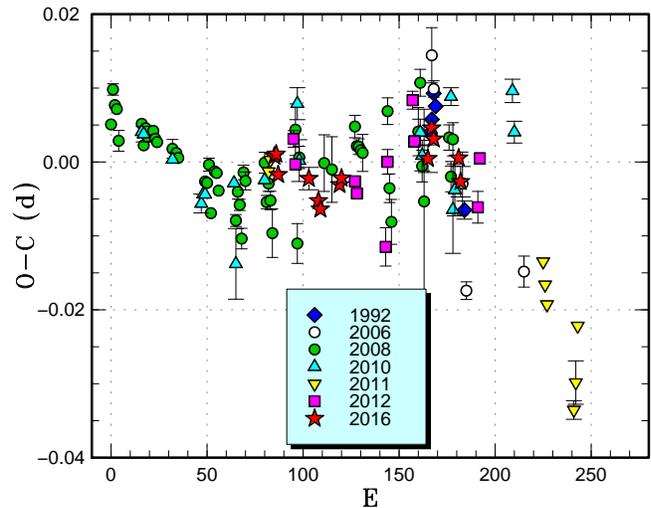}
  \end{center}
  \caption{Comparison of $O-C$ diagrams of AQ Eri between different
  superoutbursts.  A period of 0.06238~d was used to draw this figure.
  Approximate cycle counts ($E$) after the start of the superoutburst
  were used.  Since the starts of the 2012 and 2016 superoutbursts
  were not well constrained, we shifted the $O-C$ diagram
  to best fit the best-recorded 2008 one.
  }
  \label{fig:aqericomp4}
\end{figure}

   The 2016 superoutburst was visually detected
by R. Stubbings on January 24 (vsnet-alert 19438).
Subsequent observations detected superhumps
(vsnet-alert 19444).
The times of superhump maxima are listed in
table \ref{tab:aqerioc2016}.  We observed
stage B and initial part of stage C as inferred
from figure \ref{fig:aqericomp4}.

% SI

\begin{table}
\caption{Superhump maxima of AQ Eri (2016)}\label{tab:aqerioc2016}
\begin{center}
\begin{tabular}{rp{55pt}p{40pt}r@{.}lr}
\hline
\multicolumn{1}{c}{$E$} & \multicolumn{1}{c}{max\commenta} & \multicolumn{1}{c}{error} & \multicolumn{2}{c}{$O-C$\commentb} & \multicolumn{1}{c}{$N$\commentc} \\
\hline
0 & 57412.9419 & 0.0006 & 0&0032 & 69 \\
1 & 57413.0046 & 0.0006 & 0&0034 & 69 \\
2 & 57413.0642 & 0.0007 & 0&0006 & 69 \\
18 & 57414.0617 & 0.0015 & $-$0&0004 & 21 \\
23 & 57414.3706 & 0.0007 & $-$0&0036 & 50 \\
24 & 57414.4319 & 0.0009 & $-$0&0047 & 44 \\
34 & 57415.0590 & 0.0006 & $-$0&0018 & 162 \\
35 & 57415.1222 & 0.0005 & $-$0&0009 & 142 \\
80 & 57417.9320 & 0.0019 & 0&0003 & 97 \\
82 & 57418.0609 & 0.0008 & 0&0044 & 154 \\
83 & 57418.1218 & 0.0010 & 0&0029 & 18 \\
96 & 57418.9301 & 0.0012 & $-$0&0001 & 104 \\
97 & 57418.9893 & 0.0021 & $-$0&0033 & 34 \\
\hline
  \multicolumn{6}{l}{\commenta BJD$-$2400000.} \\
  \multicolumn{6}{l}{\commentb Against max $= 2457412.9392 + 0.062463 E$.} \\
  \multicolumn{6}{l}{\commentc Number of points used to determine the maximum.} \\
\end{tabular}
\end{center}
\end{table}

\subsection{AX Fornacis}\label{obj:axfor}

   This object was cataloged as 2QZ J021927.9$-$304545
in the 2dF QSO Redshift Survey \citep{boy002dFQSO}.
B. Monard monitored this object since 2005 January
and detected a bright (unfiltered CCD magnitude 11.9)
on 2005 July 2 (vsnet-alert 8521).  There were several
past outbursts in the ASAS-3 \citep{ASAS3} data
(vsnet-alert 8523).
The object was then established to be an SU UMa-type
dwarf nova by the detection of superhumps
\citep{ima06j0219}.  These two superoutbursts were
studied further in \citet{Pdot}.
The object was given a permanent
name of AX For in \citet{NameList80a}.

\begin{figure}
  \begin{center}
%    \FigureFile(85mm,70mm){axforcomp.eps}
    \FigureFile(85mm,70mm){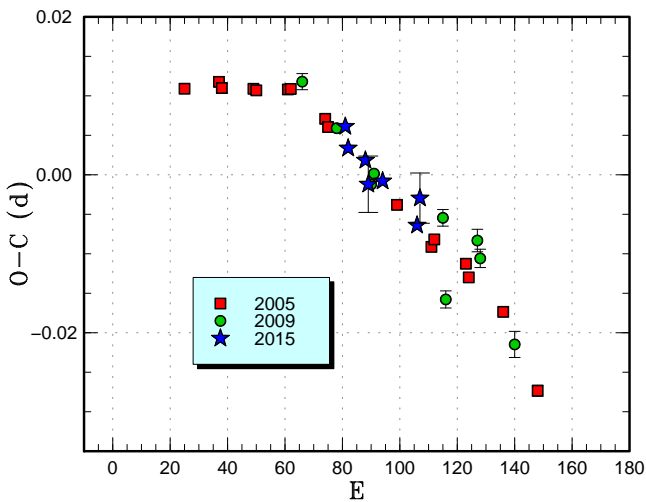}
  \end{center}
  \caption{Comparison of $O-C$ diagrams of AX For between different
  superoutbursts.  A period of 0.08140~d was used to draw this figure.
  Approximate cycle counts ($E$) after the start of the superoutburst
  were used.}
  \label{fig:axforcomp}
\end{figure}

   The 2015 superoutburst was visually detected by 
R. Stubbings at a magnitude of 12.0 on November 10
(cf. vsnet-alert 19255).  The times of superhump
maxima are listed in table \ref{tab:axforoc2015}.
A comparison of $O-C$ diagrams between different
superoutbursts (figure \ref{fig:axforcomp}) indicates
that we only observed stage C superhumps in 2015.
Although individual superhumps were not measured,
a PDM analysis of the post-superoutburst data
(BJD 2457346--2457350) yielded a strong signal
of 0.08109(6)~d, indicating that stage C superhumps
persisted after the rapid fading from the outburst
plateau.

% SI

\begin{table}
\caption{Superhump maxima of AX For (2015)}\label{tab:axforoc2015}
\begin{center}
\begin{tabular}{rp{55pt}p{40pt}r@{.}lr}
\hline
\multicolumn{1}{c}{$E$} & \multicolumn{1}{c}{max\commenta} & \multicolumn{1}{c}{error} & \multicolumn{2}{c}{$O-C$\commentb} & \multicolumn{1}{c}{$N$\commentc} \\
\hline
0 & 57341.5599 & 0.0008 & 0&0020 & 44 \\
1 & 57341.6386 & 0.0006 & $-$0&0004 & 36 \\
7 & 57342.1254 & 0.0007 & 0&0002 & 111 \\
8 & 57342.2038 & 0.0036 & $-$0&0024 & 76 \\
13 & 57342.6112 & 0.0010 & $-$0&0002 & 45 \\
25 & 57343.5824 & 0.0006 & $-$0&0015 & 44 \\
26 & 57343.6673 & 0.0032 & 0&0023 & 14 \\
\hline
  \multicolumn{6}{l}{\commenta BJD$-$2400000.} \\
  \multicolumn{6}{l}{\commentb Against max $= 2457341.5579 + 0.081041 E$.} \\
  \multicolumn{6}{l}{\commentc Number of points used to determine the maximum.} \\
\end{tabular}
\end{center}
\end{table}

\subsection{V844 Herculis}\label{sec:v844her}\label{obj:v844her}

   This object was discovered as a dwarf nova (Antipin Var 43)
by \citet{ant96newvar}.  The long outbursts were already
suggestive of an SU UMa-type dwarf nova.  The first
superhump detection was made by T. Vanmunster
during an outburst in 1996 October (vsnet-obs 4061, 4075).
The superhump period was first determined during 
the 1997 superoutburst by T. Vanmunster and L. Jensen
(Cataclysmic Variables Circular, No. 141, also
in vsnet-alert 935; vsnet-obs 5854).
\citet{pat98evolution} cited a superhump period of 0.05597(2)~d
determined from their observations.
\citet{kat00v844her} was the first solid publication
of superhumps in this system and reported a period
of 0.05592(2)~d.  \citet{tho02gwlibv844herdiuma}
obtained a spectroscopic orbital period of 0.054643(7)~d.
\citet{oiz07v844her} reported observations of
the 2002, 2003 and 2006 superoutbursts.
Positive $P_{\rm dot}$ was detected for the 2002 and
2006 superoutbursts.  \citet{oiz07v844her} also summarized
the known outbursts of this object.  Most of the outbursts
of this object were superoutbursts and only two out of
13 known outbursts were normal outbursts at the time
of \citet{oiz07v844her}.  The 2008 superoutburst was
reported in \citet{Pdot}.  The 2009 and 2010 superoutbursts
were reported in \citet{Pdot2}.  The second superoutburst
in 2010 (hereafter 2010b) was reported in \citet{Pdot3}.
Another superoutburst in 2012 was reported in \citet{Pdot4}.

   The 2015 superoutburst was detected by the ASAS-SN team
(cf. vsnet-alert 18617).  This detection was early enough
and stage A superhumps were partly observed
(vsnet-alert 18625, 18645).
The times of superhump maxima are listed in
table \ref{tab:v844heroc2015}.  The maxima for
$E \le$4 correspond to the growing stage of superhumps
and are stage A superhumps.  Stage B-C transition
probably fell in the observational gap between
$E$=119 and $E$=173 and the late phase of stage B
(with a long superhump period) was not properly
observed.

\begin{figure}
  \begin{center}
%    \FigureFile(85mm,70mm){v844hercomp5.eps}
    \FigureFile(85mm,70mm){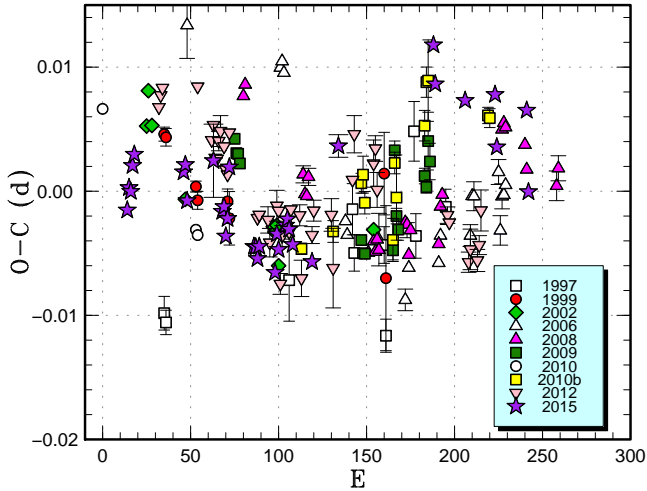}
  \end{center}
  \caption{Comparison of $O-C$ diagrams of V844 Her between different
  superoutbursts.  A period of 0.05595~d was used to draw this figure.
  Approximate cycle counts ($E$) after the start of the superoutburst
  were used.  The 2010b superoutburst were artificially shifted
  by 40 cycles to match the others.}
  \label{fig:v844hercomp5}
\end{figure}

   A comparison of $O-C$ diagrams between different
superoutbursts is shown in figure \ref{fig:v844hercomp5}.
Note that a different base period was used to draw
the figure compared to the earlier ones.
The 2010b superoutburst were artificially shifted
by 40 cycles to match the others.  This superoutburst
was either unusual (there was a normal outburst
$\sim$60~d preceding the superoutburst, and superhumps
may have started to develop before the superoutburst)
or the initial part of the superoutburst was missed
(there were only one negative observation with
a meaningful upper limit immediately before
this superoutbursts).

% SI

\begin{table}
\caption{Superhump maxima of V844 Her (2015)}\label{tab:v844heroc2015}
\begin{center}
\begin{tabular}{rp{55pt}p{40pt}r@{.}lr}
\hline
\multicolumn{1}{c}{$E$} & \multicolumn{1}{c}{max\commenta} & \multicolumn{1}{c}{error} & \multicolumn{2}{c}{$O-C$\commentb} & \multicolumn{1}{c}{$N$\commentc} \\
\hline
0 & 57157.3198 & 0.0003 & 0&0013 & 86 \\
1 & 57157.3776 & 0.0004 & 0&0031 & 104 \\
2 & 57157.4332 & 0.0002 & 0&0027 & 101 \\
3 & 57157.4912 & 0.0003 & 0&0048 & 100 \\
4 & 57157.5481 & 0.0002 & 0&0056 & 91 \\
32 & 57159.1133 & 0.0002 & 0&0033 & 51 \\
33 & 57159.1698 & 0.0002 & 0&0039 & 62 \\
34 & 57159.2229 & 0.0006 & 0&0009 & 23 \\
49 & 57160.0653 & 0.0029 & 0&0037 & 65 \\
54 & 57160.3410 & 0.0002 & $-$0&0005 & 103 \\
55 & 57160.3973 & 0.0002 & $-$0&0002 & 104 \\
56 & 57160.4509 & 0.0002 & $-$0&0026 & 104 \\
57 & 57160.5083 & 0.0001 & $-$0&0012 & 99 \\
58 & 57160.5684 & 0.0005 & 0&0029 & 33 \\
72 & 57161.3453 & 0.0003 & $-$0&0039 & 57 \\
73 & 57161.4012 & 0.0002 & $-$0&0040 & 57 \\
74 & 57161.4562 & 0.0003 & $-$0&0050 & 58 \\
75 & 57161.5131 & 0.0002 & $-$0&0040 & 57 \\
84 & 57162.0146 & 0.0004 & $-$0&0064 & 71 \\
85 & 57162.0736 & 0.0003 & $-$0&0034 & 119 \\
86 & 57162.1283 & 0.0004 & $-$0&0046 & 96 \\
90 & 57162.3541 & 0.0005 & $-$0&0028 & 57 \\
91 & 57162.4105 & 0.0003 & $-$0&0023 & 58 \\
92 & 57162.4657 & 0.0005 & $-$0&0032 & 53 \\
94 & 57162.5763 & 0.0005 & $-$0&0045 & 87 \\
105 & 57163.1904 & 0.0005 & $-$0&0062 & 64 \\
120 & 57164.0390 & 0.0009 & 0&0026 & 72 \\
174 & 57167.0684 & 0.0007 & 0&0091 & 98 \\
175 & 57167.1212 & 0.0004 & 0&0059 & 84 \\
192 & 57168.0710 & 0.0006 & 0&0040 & 57 \\
209 & 57169.0227 & 0.0005 & 0&0039 & 115 \\
210 & 57169.0744 & 0.0003 & $-$0&0003 & 120 \\
227 & 57170.0285 & 0.0004 & 0&0021 & 76 \\
228 & 57170.0779 & 0.0005 & $-$0&0045 & 119 \\
\hline
  \multicolumn{6}{l}{\commenta BJD$-$2400000.} \\
  \multicolumn{6}{l}{\commentb Against max $= 2457157.3185 + 0.055982 E$.} \\
  \multicolumn{6}{l}{\commentc Number of points used to determine the maximum.} \\
\end{tabular}
\end{center}
\end{table}

\subsection{MM Hydrae}\label{obj:mmhya}

   MM Hya was originally selected as a CV by
the Palomer-Green survey \citep{gre82PGsurveyCV}.
The SU UMa-type nature was confirmed by \citet{pat03suumas}.
See \citet{Pdot7} for more history.
The 2015 superoutburst was detected by R. Stubbings
on March 9 (vsnet-alert 18395).  Two superhumps were
recorded on March 19, when the object apparently
entered the rapid fading phase: 
BJD 2457096.9366(12) ($N$=36) and
2457096.9936(17) ($N$=43).

\subsection{RZ Leonis}\label{obj:rzleo}

   RZ Leo (=AN 30.1919) was discovered as a variable star
or a nova by \citet{wol19rzleo}.  The object was detected
at a photographic magnitude of 10--11 on 1918 March 13.
This magnitude scale was probably 1--2 mag
too bright compared to the modern scale (this tendency
is common to other Astron. Nach. papers in the 1910s,
see e.g. GR Ori in \cite{Pdot5}).
\citet{ber51Galnova} listed the object as a probable nova.
\citet{her58VSchart} provided an identification chart.
The identification by \citet{kha71novaID} was incorrect.
\citet{bru57DNatlas} and \citet{pet60DNe} listed this object
as a dwarf nova.  \citet{vog82atlas} listed this object
as a possible WZ Sge-type object (probably based on
the large outburst amplitude).

   Since the object had been suspected to be a dwarf nova
with rare outbursts, it had been sporadically monitored
by amateur observers since the 1970s.  Since 1982, it
had been monitored more systematically and the second
historical outburst was detected by R. Ducoty on 1984 December 29
at a visual magnitude of 12.9 (\cite{mat85rzleoiauc};
\cite{cri85rzleoiauc}; \cite{mcn85rzleoiauc}).
\citet{ric85rzleo} studied past photographic plates
and detected several (some of them were questionable)
outbursts only reaching 13 mag.  \citet{ric85rzleo}
suggested that the cycle length might be as short as 6~yr.
Although spectroscopic observation in outburst confirmed
the dwarf nova-type nature \citep{cri85rzleoiauc},
the object was listed as a recurrent nova in \citet{GCVS}.
\citet{szk87shortPCV} reported $JHK$ photometry on
1985 January 21, 23~d after the outburst detection.
\citet{szk87shortPCV} ascribed the magnitude $J$=14.0
to the intermediate (``Mid'') state.  In modern knowledge,
this observation probably reflected the ``red phase''
following a superoutburst (e.g. see subsection 4.6
in \cite{kat15wzsge}).

   On 1987 November 28, there was another outburst
reaching a visual magnitude of 12.3--12.5
detected by S. Lubbock
(\cite{hur87rzleoiauc}; \cite{mat87rzleoiauc}).
This outburst lasted at least for 12~d.
In the meantime, \citet{how88faintCV1} performed
time-resolved CCD photometry in quiescence and
detected 0.4 mag modulations with a period of 104~min
and suggested the SU UMa-type classification.
Since the object had been suspected to be a WZ Sge-type
dwarf nova \citep{vog82atlas}, it has been discussed
assuming this classification (\cite{dow90wxcet};
\cite{odo91wzsge}.  In \citet{odo91wzsge}, the presence
of a short (normal) outburst in 1989 \citep{nar89rzleoiauc}
was in particular discussed since various authors had claimed
the absence of short outbursts in WZ Sge.

   Despite these outburst detections, no secure outburst
had been detected before 2000 (there was a possible
outburst in 1990 October--November in the AAVSO data,
only detected by a single observer).
The 2000 outburst was detected by R. Stubbings on
December 20 at a visual magnitude of 12.1
(vsnet-alert 5437; \cite{mat00rzleoiauc}).
The detection of superhumps finally led to the identification
of an SU UMa-type dwarf nova (vsnet-alert 5446, 5448,
\cite{ish00rzleoiauc}; \cite{ish01rzleo}).
Since the orbital period had already been measured
to be 0.07651(26)~d (\cite{men01rzleo}; \cite{men99rzleo}),
\citet{ish01rzleo} identified the modulations
with a period of 0.07616(21)~d detected during
the early stage of the outburst to be early superhumps.
The orbital period has been updated to be 0.0760383(4)~d
by photometric observations in quiescence
\citep{pat03suumas}.  \citet{dai16KepCVs} further
determined the orbital period to be 0.07602997(4)~d
using the Kepler K2 mission data.
More analyses of superhumps
during the 2000 superoutburst were reported in
\citet{pat03suumas} and \citet{Pdot}.

   There was another superoutburst in 2006 detected
by S. Kerr on May 27 at a visual magnitude of 12.5
(vsnet-outburst 6885).  This outburst was not very
well observed due to the limited visibility in
the evening sky.  An analysis of superhumps was
reported in \citet{Pdot}.
There have been no confirmed normal outburst other than
the 1989 one.

   \citet{how10CVsecondarycarbon} and \citet{ham11CVsecondary}
reported the spectral type of the secondary to be
M3--M4V and M4$\pm$1, respectively, by infrared observations.
These results were consistent with the analysis of
spectral energy distribution by \citet{men02CVBD},
who suggested the spectral type M5 for the secondary.

\begin{figure}
  \begin{center}
%    \FigureFile(85mm,110mm){rzleo2016humpall.eps}
    \FigureFile(85mm,110mm){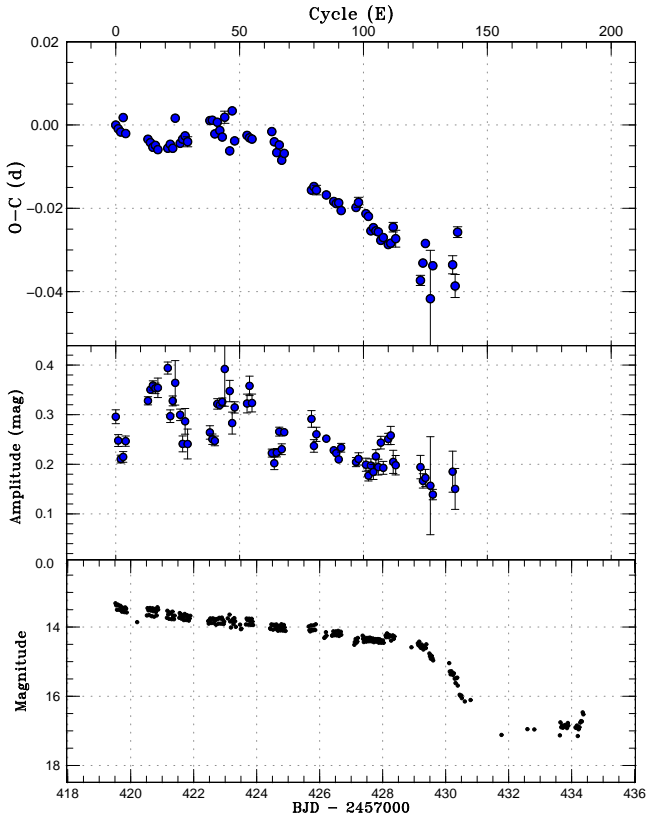}
  \end{center}
  \caption{$O-C$ diagram of superhumps in RZ Leo (2016).
     (Upper:) $O-C$ diagram.
     We used a period of 0.07865~d for calculating the $O-C$ residuals.
     (Middle:) Amplitudes of superhumps.  The modulations
     of the amplitudes with a period of $\sim$30 cycles
     in the initial part are the beat phenomenon between
     superhump period and the orbital period.
     (Lower:) Light curve.  The data were binned to 0.026~d.
  }
  \label{fig:rzleo2016humpall}
\end{figure}

   The 2016 outburst was detected by R. Stubbings
at a visual magnitude of 13.0 on January 31
(vsnet-alert 19448).  Subsequent observations already
recorded fully developed superhumps (vsnet-alert 19452,
19458, 19466).  The object rapidly faded on February
11--12 (vsnet-alert 19499).
There was also a post-superoutburst rebrightening
at a visual magnitude of 14.1 on February 15
(vsobs-share 12235).
The times of superhump maxima are listed in
table \ref{tab:rzleooc2016}.  There were clear stages
B and C (figure \ref{fig:rzleo2016humpall}).
In this figure, the amplitudes of superhumps
cyclically varied with a period of $\sim$30 cycles,
particularly in the early stage of the superoutburst.
This variation most likely reflects the beat phenomenon
between the superhump period and orbital period
(the estimated beat period is 2.3~d or 29 superhump
cycles).

\begin{figure}
  \begin{center}
%    \FigureFile(85mm,70mm){rzleocomp2.eps}
    \FigureFile(85mm,70mm){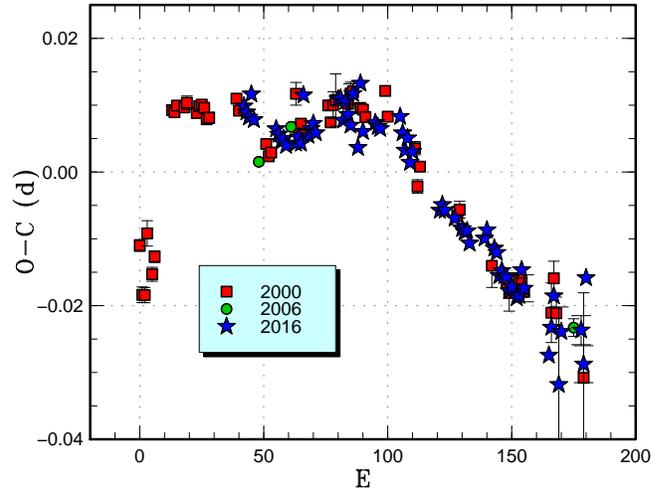}
  \end{center}
  \caption{Comparison of $O-C$ diagrams of RZ Leo between different
  superoutbursts.  A period of 0.07865~d was used to draw this figure.
  Approximate cycle counts ($E$) after the start of the appearance
  of ordinary superhumps.
  Since starts of the 2006 and 2016 outbursts were not constrained,
  we shifted the $O-C$ diagram of these outbursts
  to best fit the better-recorded 2000 one.
  We had to shift 48 and 42 cycles for the 2006 and 2016
  outbursts, respectively.
  }
  \label{fig:rzleocomp2}
\end{figure}

   A comparison of the $O-C$ diagrams (figure \ref{fig:rzleocomp2})
suggests that superhumps started $\sim$3~d before
the initial time-resolved observation in 2016.
It means that superhumps already started to develop
at the time of Stubbing's initial outburst detection.
The true start of the outburst was unknown due to
a 6~d gap in the observation.
The $O-C$ diagrams have stages B and C and
a large positive $P_{\rm dot}$ typical for
short-$P_{\rm orb}$ systems.  It would be worth noting
that stage C superhumps persisted long after
the main superoutburst and there was no phase
shift at the time of the rapid fading.

   Although the existence of early superhumps
was reported \citep{ish01rzleo}, these reported
early superhumps may have been different from
those of typical WZ Sge-type objects since
the phase of early superhumps was apparently
short.  The period of these modulations was not
sufficiently determined to make a secure comparison
with the orbital period.  Although \citet{Pdot}
concluded that these modulations could not be
considered as an extension of stage A superhumps,
the exact identification of modulations in
the earliest stage in RZ Leo still awaits
confirmation.  It was a pity that both the 2006
and 2016 superoutbursts were not detected sufficiently
early to confirm these modulations.  Future
intensive observations on the next occasion
are still strongly desired since RZ Leo is supposed
to be an atypical (long-$P_{\rm orb}$) WZ Sge-type
system (cf. \cite{kat15wzsge}) and confirmation
of early superhumps is very important to
verify this classification.

\begin{figure}
  \begin{center}
%    \FigureFile(85mm,110mm){rzleoporb.eps}
    \FigureFile(85mm,110mm){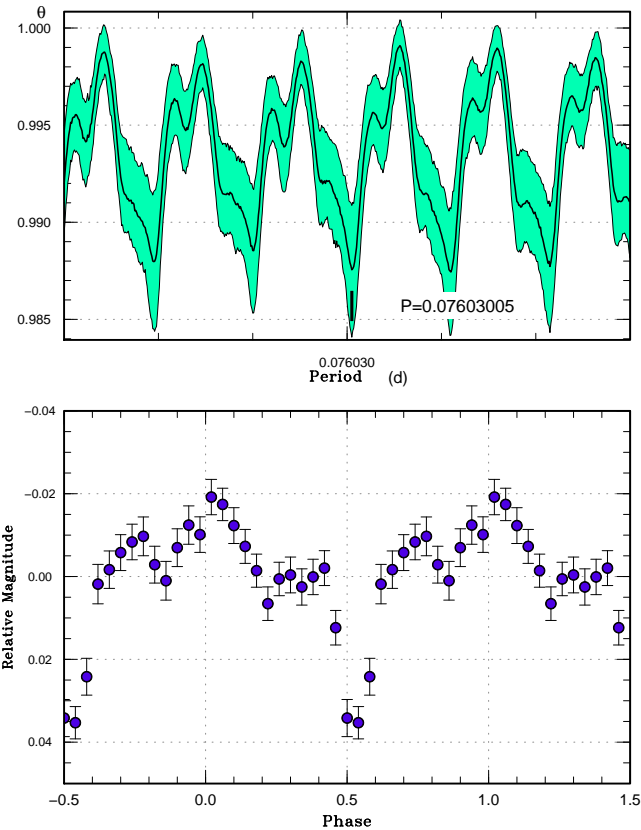}
  \end{center}
  \caption{Orbital variations in RZ Leo during superoutbursts (2000--2016).
     (Upper): PDM analysis.
     (Lower): Phase-averaged profile.}
  \label{fig:rzleoporb}
\end{figure}

   Since RZ Leo apparently has a high orbital inclination
(doubly peaked emission lines, ellipsoidal variations
in quiescence and beat phenomenon during superoutburst),
we attempted to detect the orbital variations during
the three superoutbursts (2000, 2006 and 2016).
All the combinations of these superoutbursts yielded
a consistent period (the alias was selected within the range
considering the error in \cite{dai16KepCVs})
within respective errors and we identified 0.07603005(2)~d
to be the updated orbital period (figure \ref{fig:rzleoporb}).

% SI

% SI

\begin{table*}
\caption{Superhump maxima of RZ Leo (2016)}\label{tab:rzleooc2016}
\begin{center}
\begin{tabular}{rp{55pt}p{40pt}r@{.}lrrp{55pt}p{40pt}r@{.}lr}
\hline
\multicolumn{1}{c}{$E$} & \multicolumn{1}{c}{max\commenta} & \multicolumn{1}{c}{error} & \multicolumn{2}{c}{$O-C$\commentb} & \multicolumn{1}{c}{$N$\commentc} & \multicolumn{1}{c}{$E$} & \multicolumn{1}{c}{max\commenta} & \multicolumn{1}{c}{error} & \multicolumn{2}{c}{$O-C$\commentb} & \multicolumn{1}{c}{$N$\commentc} \\
\hline
0 & 57419.5256 & 0.0004 & $-$0&0055 & 98 & 66 & 57424.7117 & 0.0003 & 0&0079 & 60 \\
1 & 57419.6033 & 0.0005 & $-$0&0061 & 121 & 67 & 57424.7867 & 0.0005 & 0&0045 & 74 \\
2 & 57419.6812 & 0.0003 & $-$0&0066 & 161 & 68 & 57424.8670 & 0.0003 & 0&0064 & 97 \\
3 & 57419.7633 & 0.0007 & $-$0&0028 & 46 & 79 & 57425.7233 & 0.0006 & 0&0006 & 33 \\
4 & 57419.8381 & 0.0004 & $-$0&0064 & 37 & 80 & 57425.8028 & 0.0005 & 0&0018 & 39 \\
13 & 57420.5446 & 0.0002 & $-$0&0053 & 144 & 81 & 57425.8806 & 0.0011 & 0&0012 & 18 \\
14 & 57420.6225 & 0.0002 & $-$0&0058 & 136 & 85 & 57426.1940 & 0.0003 & 0&0011 & 125 \\
15 & 57420.7000 & 0.0003 & $-$0&0067 & 177 & 88 & 57426.4284 & 0.0002 & 0&0004 & 159 \\
16 & 57420.7791 & 0.0003 & $-$0&0059 & 79 & 89 & 57426.5066 & 0.0002 & 0&0002 & 233 \\
17 & 57420.8567 & 0.0006 & $-$0&0067 & 31 & 90 & 57426.5854 & 0.0003 & 0&0006 & 176 \\
21 & 57421.1717 & 0.0002 & $-$0&0052 & 129 & 91 & 57426.6622 & 0.0003 & $-$0&0010 & 78 \\
22 & 57421.2513 & 0.0005 & $-$0&0040 & 106 & 97 & 57427.1348 & 0.0004 & 0&0014 & 145 \\
23 & 57421.3289 & 0.0003 & $-$0&0047 & 107 & 98 & 57427.2147 & 0.0012 & 0&0029 & 64 \\
24 & 57421.4148 & 0.0009 & 0&0028 & 47 & 101 & 57427.4479 & 0.0006 & 0&0010 & 67 \\
26 & 57421.5661 & 0.0003 & $-$0&0027 & 71 & 102 & 57427.5260 & 0.0006 & 0&0007 & 66 \\
27 & 57421.6457 & 0.0006 & $-$0&0014 & 85 & 103 & 57427.6011 & 0.0008 & $-$0&0025 & 60 \\
28 & 57421.7252 & 0.0009 & $-$0&0004 & 79 & 104 & 57427.6806 & 0.0008 & $-$0&0015 & 88 \\
29 & 57421.8024 & 0.0012 & $-$0&0015 & 91 & 105 & 57427.7584 & 0.0006 & $-$0&0020 & 34 \\
38 & 57422.5153 & 0.0005 & 0&0061 & 58 & 106 & 57427.8368 & 0.0008 & $-$0&0020 & 75 \\
39 & 57422.5941 & 0.0004 & 0&0064 & 67 & 107 & 57427.9135 & 0.0005 & $-$0&0037 & 140 \\
40 & 57422.6694 & 0.0004 & 0&0034 & 87 & 108 & 57427.9928 & 0.0007 & $-$0&0028 & 103 \\
41 & 57422.7509 & 0.0003 & 0&0065 & 101 & 110 & 57428.1484 & 0.0004 & $-$0&0039 & 185 \\
42 & 57422.8276 & 0.0002 & 0&0048 & 119 & 111 & 57428.2274 & 0.0007 & $-$0&0033 & 56 \\
43 & 57422.9046 & 0.0002 & 0&0035 & 85 & 112 & 57428.3099 & 0.0011 & 0&0008 & 55 \\
44 & 57422.9880 & 0.0015 & 0&0085 & 20 & 113 & 57428.3858 & 0.0020 & $-$0&0017 & 28 \\
46 & 57423.1373 & 0.0010 & 0&0010 & 99 & 123 & 57429.1622 & 0.0012 & $-$0&0089 & 44 \\
47 & 57423.2255 & 0.0008 & 0&0109 & 147 & 124 & 57429.2451 & 0.0009 & $-$0&0045 & 44 \\
48 & 57423.2970 & 0.0004 & 0&0039 & 146 & 125 & 57429.3284 & 0.0010 & 0&0005 & 45 \\
53 & 57423.6915 & 0.0006 & 0&0066 & 36 & 127 & 57429.4724 & 0.0116 & $-$0&0122 & 31 \\
54 & 57423.7696 & 0.0005 & 0&0063 & 34 & 128 & 57429.5590 & 0.0008 & $-$0&0040 & 94 \\
55 & 57423.8480 & 0.0005 & 0&0063 & 37 & 136 & 57430.1884 & 0.0022 & $-$0&0016 & 43 \\
63 & 57424.4790 & 0.0004 & 0&0103 & 71 & 137 & 57430.2620 & 0.0028 & $-$0&0064 & 44 \\
64 & 57424.5552 & 0.0007 & 0&0081 & 66 & 138 & 57430.3536 & 0.0013 & 0&0068 & 40 \\
65 & 57424.6312 & 0.0004 & 0&0058 & 80 & \multicolumn{1}{c}{--} & \multicolumn{1}{c}{--} & \multicolumn{1}{c}{--} & \multicolumn{2}{c}{--} & \multicolumn{1}{c}{--}\\
\hline
  \multicolumn{12}{l}{\commenta BJD$-$2400000.} \\
  \multicolumn{12}{l}{\commentb Against max $= 2457419.5310 + 0.078375 E$.} \\
  \multicolumn{12}{l}{\commentc Number of points used to determine the maximum.} \\
\end{tabular}
\end{center}
\end{table*}

\subsection{V585 Lyrae}\label{obj:v585lyr}

   V585 Lyr was discovered as a dwarf nova (TK 4)
by \citet{kry01v585lyrv587lyr}.  \citet{kry01v585lyrv587lyr}
already suggested the SU UMa-type classification based on
the presence of long and short outbursts.
The 2003 superoutburst was well observed and analyzed
in detail \citep{Pdot}.  The 2012 superoutburst was also
observed in \citet{Pdot4}.

   The object is located in the Kepler field and two 
superoutbursts (2010 January--February
and 2012 April--May) and one normal outburst (2013 January)
were recorded during the Kepler mission.  Although this
target was also observed in short-cadence mode in limited
epochs, all outbursts observed by Kepler were recorded
in long-cadence mode, making detailed analysis of
the superhump evolution difficult.  \citet{kat13j1939v585lyrv516lyr}
analyzed the Kepler long-cadence observations in 2010
using Markov-Chain Monte Carlo (MCMC)-based
modeling of long-exposure ($\sim$29 min)
sampling and derived an $O-C$ diagram.
This analysis confirmed the results of ground-based photometry
with higher time resolutions but frequent gaps \citep{Pdot}.
The particularly important point was that Kepler observation
confirmed the superhump stages (A--C) and the lack of
phase transition around the termination of the superoutburst.
V585 Lyr was also unique among Kepler observations of
dwarf novae: this object showed a rebrightening both in
the 2000 and 2012 superoutbursts, though such rebrightenings
are relatively common among other SU UMa/WZ Sge-type dwarf novae
(cf. \cite{Pdot}; \cite{kat15wzsge}).  There were
``mini-rebrightenings'' between the main superoutburst
and the rebrightening \citep{kat13j1939v585lyrv516lyr}.
\citet{mey15suumareb} interpreted these ``mini-rebrightenings''
as a result of reflection of cooling waves between
the lower branch and the intermediate branch
in the S-curve of the thermal equilibrium of the accretion disk. 
V585 Lyr was also the only object in the Kepler data
in which no precursor outburst was associated with
the superoutburst and a delay of development of superhumps
was recorded.  This was interpreted as a result of highly
accumulated mass in the disk before the outbursts started
\citep{kat13j1939v585lyrv516lyr}, corresponding to 
``case B'' superoutburst (the mass stored in the disk
before the superoutburst is large enough and the disk
can remain at radius of the 3:1 resonance or beyond 
for some time before superhumps start to develop) 
discussed by \citet{osa03DNoutburst}.

\begin{figure}
  \begin{center}
%    \FigureFile(85mm,70mm){v585lyrcomp.eps}
    \FigureFile(85mm,70mm){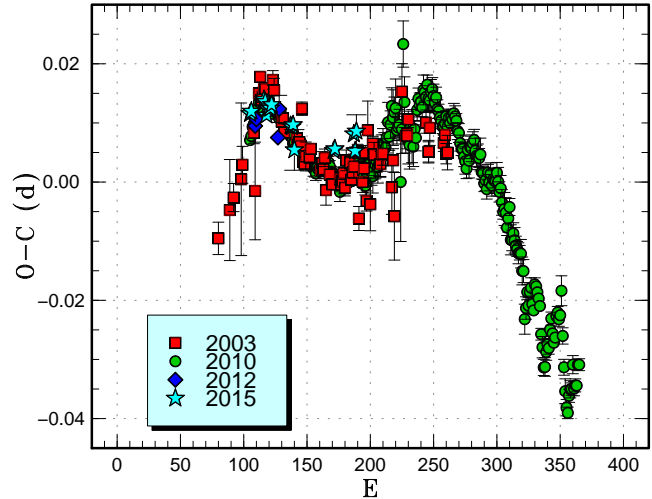}
  \end{center}
  \caption{Comparison of $O-C$ diagrams of V585 Lyr between different
  superoutbursts.  The 2010 data refer to Kepler observations
  analyzed in \citet{kat13j1939v585lyrv516lyr}.
  A period of 0.06045~d was used to draw this figure.
  Approximate cycle counts ($E$) after the start of the superoutburst
  were used.  The starts of the 2010 and 2012 superoutbursts
  were well determined by the Kepler data.
  We had to shift 49 and 80 cycles for the 2003 and 2015
  data, respectively, to obtain the best match with the 2010
  data.  These values suggest that the outbursts were missed
  for 3~d and 5~d for the 2003 and 2015 superoutbursts,
  respectively, or these superoutbursts had exceptionally
  smaller scales.  Considering the limited visual monitoring,
  the first possibility looks more likely.
  }
  \label{fig:v585lyrcomp}
\end{figure}

   The 2015 superoutburst was detected by the ASAS-SN team
(vsnet-alert 18688).  Subsequent observations detected
superhumps (vsnet-alert 18698, 18722).  The times of superhump
maxima are listed in table \ref{tab:v585lyroc2015}.
The final part of stage A and early part of
stage B were observed (figure \ref{fig:v585lyrcomp}).

% SI

\begin{table}
\caption{Superhump maxima of V585 Lyr (2015)}\label{tab:v585lyroc2015}
\begin{center}
\begin{tabular}{rp{55pt}p{40pt}r@{.}lr}
\hline
\multicolumn{1}{c}{$E$} & \multicolumn{1}{c}{max\commenta} & \multicolumn{1}{c}{error} & \multicolumn{2}{c}{$O-C$\commentb} & \multicolumn{1}{c}{$N$\commentc} \\
\hline
0 & 57179.5047 & 0.0005 & $-$0&0013 & 67 \\
1 & 57179.5655 & 0.0007 & $-$0&0008 & 60 \\
11 & 57180.1720 & 0.0002 & 0&0020 & 55 \\
12 & 57180.2321 & 0.0002 & 0&0017 & 67 \\
13 & 57180.2903 & 0.0011 & $-$0&0005 & 23 \\
14 & 57180.3518 & 0.0002 & 0&0008 & 45 \\
15 & 57180.4125 & 0.0004 & 0&0011 & 84 \\
16 & 57180.4728 & 0.0002 & 0&0009 & 124 \\
17 & 57180.5338 & 0.0003 & 0&0017 & 94 \\
32 & 57181.4370 & 0.0006 & $-$0&0006 & 52 \\
33 & 57181.4973 & 0.0005 & $-$0&0008 & 67 \\
34 & 57181.5581 & 0.0005 & $-$0&0003 & 67 \\
35 & 57181.6143 & 0.0034 & $-$0&0045 & 24 \\
66 & 57183.4881 & 0.0007 & $-$0&0020 & 67 \\
67 & 57183.5489 & 0.0006 & $-$0&0016 & 67 \\
82 & 57184.4581 & 0.0010 & 0&0021 & 66 \\
83 & 57184.5157 & 0.0010 & $-$0&0007 & 66 \\
84 & 57184.5795 & 0.0028 & 0&0028 & 36 \\
\hline
  \multicolumn{6}{l}{\commenta BJD$-$2400000.} \\
  \multicolumn{6}{l}{\commentb Against max $= 2457179.5060 + 0.060366 E$.} \\
  \multicolumn{6}{l}{\commentc Number of points used to determine the maximum.} \\
\end{tabular}
\end{center}
\end{table}

\subsection{V2051 Ophiuchi}\label{obj:v2051oph}

   V2051 Oph was discovered as an emission-line object
in outburst \citep{san72v2051oph}.  \citet{san72v2051oph}
suggested that this object to be a dwarf nova rather than
a nova.  \citet{bon77anumaiauc} listed this object as
a candidate polar, although this list included various
objects (EM Cyg: dwarf nova; V Sge: novalike variable;
CL Sco, HK Sco: symbiotic stars) which are not currently
considered to be related to polars.
\citet{ang77vvpupv2051ophiauc3065} reported that this
object is an eclipsing system with a period of 96 min.
\citet{ang77vvpupv2051ophiauc3065} also reported strong
Balmer, He\textsc{i} and He\textsc{ii} emission lines.
\citet{bon79CVspecproc} confirmed this spectroscopic finding.
After a suggestion of similarity with HT Cas
\citep{pat80htcasv2051ophaeaqr}, this object started to be
monitored by amateur astronomers in 1980.  Secure outburst
detections were not made in the 1980s.
F. Bateson, Royal Astronomical Society of New Zealand,
reported only limited number of outbursts up tp 1997
(only reaching 13.0--13.5 mag and without evidence
for superoutbursts: vsnet-chat 546; see also \cite{war83v2051oph}).
This object was considered to be too faint for amateur
telescopes (cf. citation in \cite{war87v2051oph}).

   The history of classification of this object had
long been confusing.
\citet{war83v2051oph} obtained high-speed photometry
and found eclipse profiles were highly variable.
\citet{war83v2051oph} suggested that this object is similar
to other SU UMa-type eclipsing dwarf novae Z Cha, OY Car
and HT Cas.  The analysis of flickering suggested a possibility
that the inner region of the disk may be truncated as in
polars \citep{war83v2051oph}.  A spectroscopic study
by \citet{coo83v2051oph} suggested
that the white dwarf is eclipsed.
\citet{wen84v2051oph} surveyed about 400 Sonneberg plates
with limiting magnitudes of 11.5--13 mag and found no major
outburst.  \citet{wen84v2051oph} considered this finding
to be compatible with a polar rather than a high-amplitude
dwarf nova.  \citet{wat86v2051ophspec} reported detailed
spectrophotometric analysis and suggested the similarity
of the disk structure to that of OY Car.
\citet{war87v2051oph} reported high-speed photometry and
the reconstructed eclipse map did not show evidence of
a well-established accretion disk.  \citet{war87v2051oph}
suggested that this object is a low-field polar based on
these observations.

   In 1997, several faint outbursts were recorded by
R. Stubbings: 13.9 mag on June 27 (vsnet-alert 1019),
13.6 mag on August 7 (vsnet-obs 6643, vsnet-alert 1129)
and 14.2 mag on September 24 (vsnet-obs 7540,
vsnet-alert 1239).  There was another outburst on
1998 March 27 at 13.6 mag (vsnet-alert 1600).
On 1998 March 18, there was a bright outburst reaching
11.9 mag (the outburst started one day before)
(vsnet-alert 1796).  CCD observations by L.~T. Jensen
revealed humps and eclipses (vsnet-alert 1814).
Observations by S. Kiyota identified these humps
as superhumps (vsnet-alert 1819).
Upon detection announcement of this superoutburst,
B. Warner wrote: The eclipse centered on a
``superhump'' shown on your Web page is just the enhanced
orbital hump that appears during outburst (vsnet-alert 1833).
T. Kato reported that superhump signatures were present
in observations by S. Kiyota and Osaka Kyoiku U. team
and pointed out that B. Warner actually recorded
the fading part of a superoutburst in \citet{war87v2051oph},
during which an apparent superhump was recorded only
one night (vsnet-alert 1835).
J. Patterson's team also reported the identification
of the observed humps as superhumps (vsnet-alert 1859).
The result by S. Kiyota was published in \citet{kiy98v2051oph}.

   Another superoutburst was recored in 1999 July--August.
(vsnet-alert 3284, 3308, 3315, 3330, 3347).
There was also a rebrightening (vsnet-alert 3354).
After this outburst, there was no doubt about
the SU UMa-type classification
of this object [see also \citet{vri00v2051oph};
\citet{vri03v2051oph}; \citet{pap08v2051oph}].
\citet{kat01v2051ophiyuma} reported a supercycle of
227~d and the recurrence time of normal outbursts
of 45~d.  In our series of papers, the 1999, 2003 and 2009
superoutbursts were analyzed in \citet{Pdot},
the 2010 one was reported in \citet{Pdot2}.
Despite that the recent observations suggest that this
object is an ordinary SU UMa-type dwarf nova,
there has been an argument using eclipse maps
that outbursts in this
system have properties different from what
the disk-instability model suggests
(\cite{bap07v2051oph}; \cite{and14v2051ophproc};
\cite{woj14v2051ophproc}).  \citet{bap12outburstmodel}
suggested that dwarf novae have two groups, the one
which can be understood in the framework of
the disk instability model and the other which
can only be explained in terms of the mass-transfer
instability model.  \citet{bap12outburstmodel}
claimed that V2051 Oph belongs to the latter group.

   In addition to above references, this object has been
thoroughly investigated since it is fairly close and
has received much attention from the early times: 
eclipse analysis (\cite{bap98v2051ophHST};
\cite{vri02v2051ophmapping}; \cite{pap08v2051oph}),
flickering (\cite{bru00htcasv2051ophippeguxumaflickering};
\cite{bap04v2051oph}),
spectroscopy (\cite{ste01v2051oph}; \cite{sai06v2051oph};
\cite{lon15ccsclv2051oph}),
and secular variation of the orbital period
(\cite{ech93v2051oph}; \cite{bap03v4140sgr}).

   The 2015 superoutburst was detected by R. Stubbings
(vsnet-alert 18650).  The outburst was detected on May 21
and reached a peak brightness of 11.9 mag on May 23.
The object faded rather quickly and it was already around
14 mag when observations of superhumps were performed
(vsnet-alert 18675).  During the past superoutbursts,
the object was mostly around 13 mag or fainter when
CCD time-resolved observations were performed despite
that the peak visual magnitude reached 12 or even brighter.
The short duration of the brightness peak may have
the reason why past plate collections and visual
monitoring failed to record relatively frequent
superoutbursts.  Such a short peak of the superoutburst
may be a result of a high inclination and would deserve
a further study.

\begin{figure}
  \begin{center}
%    \FigureFile(85mm,70mm){v2051ophcomp2.eps}
    \FigureFile(85mm,70mm){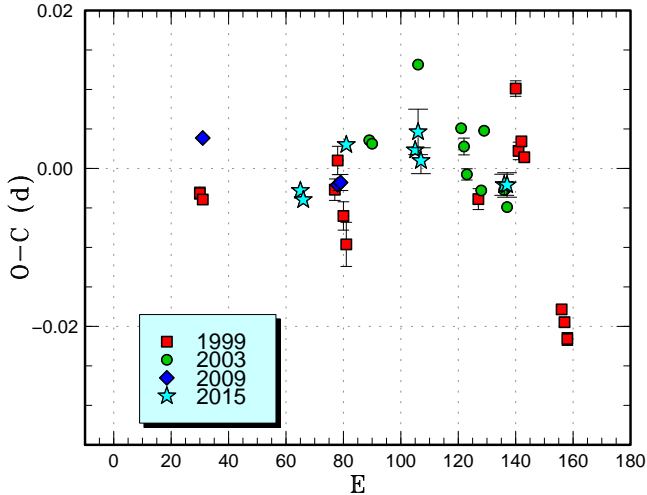}
  \end{center}
  \caption{Comparison of $O-C$ diagrams of V2051 Oph between different
  superoutbursts.  A period of 0.06430~d was used to draw this figure.
  Approximate cycle counts ($E$) after the start of the superoutburst
  were used.}
  \label{fig:v2051ophcomp2}
\end{figure}

   The times of superhump maxima in 2015 are listed in
table \ref{tab:v2051ophoc2015}.
We possibly detected stage B and initial part of
stage C by comparison with other superoutbursts of
this object (figure \ref{fig:v2051ophcomp2}).

   The object was observed after the superoutburst
and the superhump signal with a period of 0.06373(2)~d
was detected up to the next normal outburst
(BJD 2457186, June 13).
If we assume a disk radius of 0.35$\pm$0.04$a$, where
$a$ is the binary separation, for the post-superoutburst
state of an ordinary SU UMa-type dwarf nova
\citep{kat13qfromstageA}, the mass ratio is estimated
to be $q$=0.11(3).  Although this value is smaller than
$q$=0.19(3) \citep{bap98v2051ophHST}, who obtained
the value from eclipse observations, and $q$=0.18(5)
\citep{lon15ccsclv2051oph} by Doppler-tomography,
our smaller value for a short-$P_{\rm orb}$ object appears
to fit more comfortably on the evolutionary diagram
(e.g. figure 5 in \cite{kat13qfromstageA},
see also figure \ref{fig:qall5} in this paper).
Since eclipse light curves in this system were very
variable \citep{war87v2051oph} and ingress/egress features
of the hot spot are difficult to define
\citep{bap98v2051ophHST}, determination of $q$
from eclipse observations would suffer intrinsic
uncertainties.
Our new value using post-superoutburst
superhumps would be treated as a new measurement
of $q$ with comparable significance.

   We have also updated the eclipse ephemeris
by using MCMC analysis \citep{Pdot4}
of the eclipse observations of our 1999--2015 data:
\begin{equation}
{\rm Min(BJD)} = 2453189.48679(1) + 0.0624278552(2) E .
\label{equ:v2051ophecl}
\end{equation}
The epoch refers to the center of the entire observation.

% SI

\begin{table}
\caption{Superhump maxima of V2051 Oph (2015)}\label{tab:v2051ophoc2015}
\begin{center}
\begin{tabular}{rp{50pt}p{30pt}r@{.}lcr}
\hline
$E$ & max\commenta & error & \multicolumn{2}{c}{$O-C$\commentb} & phase\commentc & $N$\commentd \\
\hline
0 & 57168.1880 & 0.0006 & $-$0&0024 & 0.78 & 36 \\
1 & 57168.2511 & 0.0009 & $-$0&0035 & 0.79 & 59 \\
16 & 57169.2226 & 0.0009 & 0&0033 & 0.35 & 99 \\
40 & 57170.7651 & 0.0010 & 0&0023 & 0.06 & 18 \\
41 & 57170.8317 & 0.0029 & 0&0046 & 0.13 & 15 \\
42 & 57170.8924 & 0.0016 & 0&0009 & 0.10 & 10 \\
71 & 57172.7540 & 0.0013 & $-$0&0025 & 0.92 & 10 \\
72 & 57172.8183 & 0.0015 & $-$0&0026 & 0.95 & 11 \\
\hline
  \multicolumn{7}{l}{\commenta BJD$-$2400000.} \\
  \multicolumn{7}{l}{\commentb Against max $= 2457168.1904 + 0.064313 E$.} \\
  \multicolumn{7}{l}{\commentc Orbital phase.} \\
  \multicolumn{7}{l}{\commentd Number of points used to determine the maximum.} \\
\end{tabular}
\end{center}
\end{table}

\subsection{V368 Pegasi}\label{obj:v368peg}

   V368 Peg is a dwarf nova (Antipin Var 63) discovered by
\citet{ant99v368pegftcamv367pegv2209cyg}.
The SU UMa-type nature was identified by J. Pietz
during the 1999 superoutburst (vsnet-alert 3317).
The 2000, 2005 and 2006 superoutbursts were studied
in \citet{Pdot} and the 2009 superoutburst was reported
in \citet{Pdot2}.

   The 2015 superoutburst was visually
detected by P. Schmeer on September 14 (vsnet-alert 19063).
This superoutburst was also detected by C. Chiselbrook
(AAVSO) on the same night.  Single-night observations
on September 17 recorded two superhumps maxima:
BJD 2457283.3273(3) ($N$=76) and 2457283.3969(4)
($N$=64).

\subsection{V650 Pegasi}\label{obj:v650peg}

   This dwarf nova is an SU UMa-type dwarf nova selected
by P. Wils (cf. \cite{she11asas2243}).
The object was formerly referred to as ASAS J224349$+$0809.5.
For more history, see \citet{Pdot6}.

\begin{figure}
  \begin{center}
%    \FigureFile(85mm,70mm){asas2243comp2.eps}
    \FigureFile(85mm,70mm){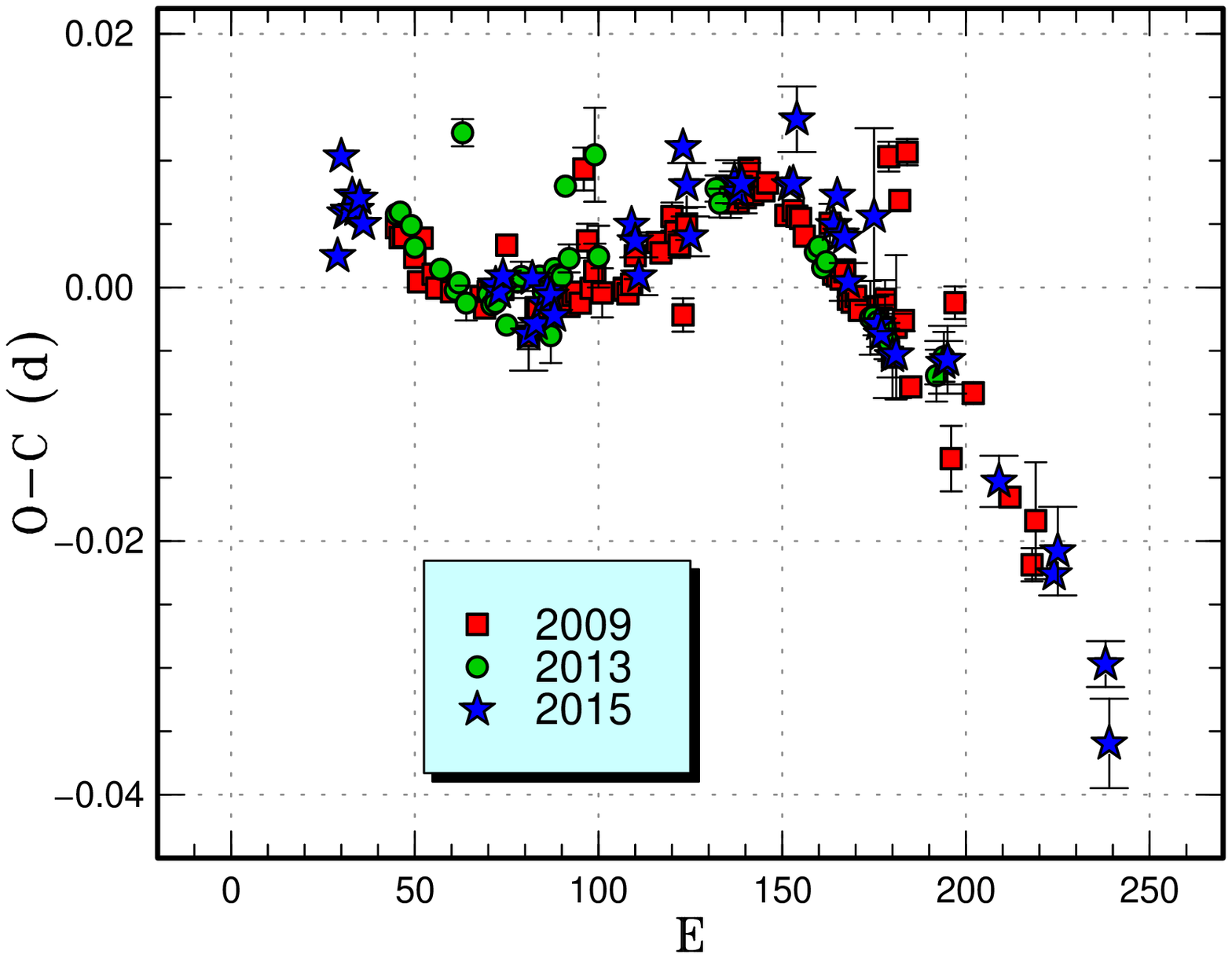}
  \end{center}
  \caption{Comparison of $O-C$ diagrams of V650 Peg between different
  superoutbursts.  A period of 0.06975~d was used to draw this figure.
  Approximate cycle counts ($E$) after the start of the superoutburst
  were used.  In \citet{Pdot6}, cycles were counted since
  the detection of the outbursts, on the contrary to the
  description in \citet{Pdot6}.  The starts of the outbursts
  were not known at the time of \citet{Pdot6}.  The start
  of the 2015 outburst is much better defined and we shifted
  the other outbursts to fit the 2015 one.
  }
  \label{fig:asas2243comp2}
\end{figure}

   The 2015 outburst was detected by the ASAS-SN team
on September 5 (cf. vsnet-alert 19032).  
Superhumps were soon detected (vsnet-alert 19034,
19059, 19066, 19069).
The times of superhumps maxima are listed in
table \ref{tab:asas2243oc2015}.
Although stage A was not recorded, stages B and C
were clearly detected (figure \ref{fig:asas2243comp2}).

% SI

\begin{table*}
\caption{Superhump maxima of V650 Peg (2015)}\label{tab:asas2243oc2015}
\begin{center}
\begin{tabular}{rp{55pt}p{40pt}r@{.}lrrp{55pt}p{40pt}r@{.}lr}
\hline
\multicolumn{1}{c}{$E$} & \multicolumn{1}{c}{max\commenta} & \multicolumn{1}{c}{error} & \multicolumn{2}{c}{$O-C$\commentb} & \multicolumn{1}{c}{$N$\commentc} & \multicolumn{1}{c}{$E$} & \multicolumn{1}{c}{max\commenta} & \multicolumn{1}{c}{error} & \multicolumn{2}{c}{$O-C$\commentb} & \multicolumn{1}{c}{$N$\commentc} \\
\hline
0 & 57273.0768 & 0.0004 & $-$0&0068 & 90 & 108 & 57280.6157 & 0.0017 & 0&0095 & 22 \\
1 & 57273.1544 & 0.0003 & 0&0012 & 140 & 109 & 57280.6846 & 0.0017 & 0&0087 & 26 \\
2 & 57273.2197 & 0.0003 & $-$0&0032 & 139 & 110 & 57280.7551 & 0.0016 & 0&0095 & 25 \\
3 & 57273.2897 & 0.0004 & $-$0&0029 & 144 & 123 & 57281.6617 & 0.0009 & 0&0106 & 26 \\
4 & 57273.3606 & 0.0001 & $-$0&0016 & 183 & 124 & 57281.7316 & 0.0010 & 0&0109 & 23 \\
5 & 57273.4292 & 0.0001 & $-$0&0027 & 289 & 125 & 57281.8064 & 0.0026 & 0&0160 & 10 \\
6 & 57273.4999 & 0.0002 & $-$0&0017 & 167 & 135 & 57282.4955 & 0.0005 & 0&0087 & 91 \\
7 & 57273.5676 & 0.0009 & $-$0&0036 & 43 & 136 & 57282.5676 & 0.0012 & 0&0111 & 34 \\
43 & 57276.0738 & 0.0004 & $-$0&0049 & 69 & 137 & 57282.6343 & 0.0012 & 0&0081 & 25 \\
44 & 57276.1430 & 0.0004 & $-$0&0054 & 77 & 138 & 57282.7038 & 0.0010 & 0&0079 & 24 \\
45 & 57276.2140 & 0.0004 & $-$0&0041 & 51 & 139 & 57282.7700 & 0.0015 & 0&0045 & 24 \\
52 & 57276.6977 & 0.0029 & $-$0&0080 & 11 & 146 & 57283.2634 & 0.0070 & 0&0104 & 104 \\
53 & 57276.7718 & 0.0009 & $-$0&0035 & 22 & 147 & 57283.3247 & 0.0004 & 0&0020 & 189 \\
54 & 57276.8380 & 0.0010 & $-$0&0069 & 22 & 148 & 57283.3936 & 0.0005 & 0&0013 & 125 \\
56 & 57276.9791 & 0.0005 & $-$0&0051 & 74 & 151 & 57283.6012 & 0.0033 & $-$0&0001 & 12 \\
57 & 57277.0498 & 0.0005 & $-$0&0041 & 77 & 152 & 57283.6711 & 0.0018 & 0&0001 & 26 \\
58 & 57277.1192 & 0.0004 & $-$0&0043 & 77 & 165 & 57284.5772 & 0.0017 & 0&0007 & 25 \\
59 & 57277.1874 & 0.0005 & $-$0&0058 & 77 & 166 & 57284.6471 & 0.0027 & 0&0010 & 50 \\
80 & 57278.6593 & 0.0012 & 0&0034 & 20 & 180 & 57285.6141 & 0.0020 & $-$0&0072 & 19 \\
81 & 57278.7278 & 0.0012 & 0&0022 & 29 & 195 & 57286.6530 & 0.0014 & $-$0&0131 & 26 \\
82 & 57278.7947 & 0.0015 & $-$0&0005 & 23 & 196 & 57286.7246 & 0.0035 & $-$0&0112 & 25 \\
94 & 57279.6419 & 0.0014 & 0&0109 & 25 & 209 & 57287.6224 & 0.0018 & $-$0&0189 & 24 \\
95 & 57279.7087 & 0.0017 & 0&0080 & 27 & 210 & 57287.6859 & 0.0035 & $-$0&0250 & 26 \\
96 & 57279.7744 & 0.0016 & 0&0040 & 29 & \multicolumn{1}{c}{--} & \multicolumn{1}{c}{--} & \multicolumn{1}{c}{--} & \multicolumn{2}{c}{--} & \multicolumn{1}{c}{--}\\
\hline
  \multicolumn{12}{l}{\commenta BJD$-$2400000.} \\
  \multicolumn{12}{l}{\commentb Against max $= 2457273.0836 + 0.069654 E$.} \\
  \multicolumn{12}{l}{\commentc Number of points used to determine the maximum.} \\
\end{tabular}
\end{center}
\end{table*}

\subsection{PU Persei}\label{obj:puper}

   PU Per was discovered as a dwarf nova (=S 9727)
by \citet{hof67an29043}, who detected two outbursts.
\citet{rom76DN} also detected two outbursts.
Both \citet{hof67an29043} and \citet{rom76DN}
recorded short and long outbursts.
\citet{bus79VS17} recorded further two long outbursts.
\citet{BruchCVatlas} detected another outburst.
The presence of two types of outbursts and
the large outburst amplitude made PU Per an excellent
candidate for an SU UMa-type dwarf nova.
\citet{kat95puper} observed a normal outburst in
1995 October and \citet{kat99puper} finally detected
superhumps during the 1998 September outburst.
The 2009 superoutburst was reported in \citet{Pdot}.

\begin{figure}
  \begin{center}
%    \FigureFile(85mm,70mm){pupercomp.eps}
    \FigureFile(85mm,70mm){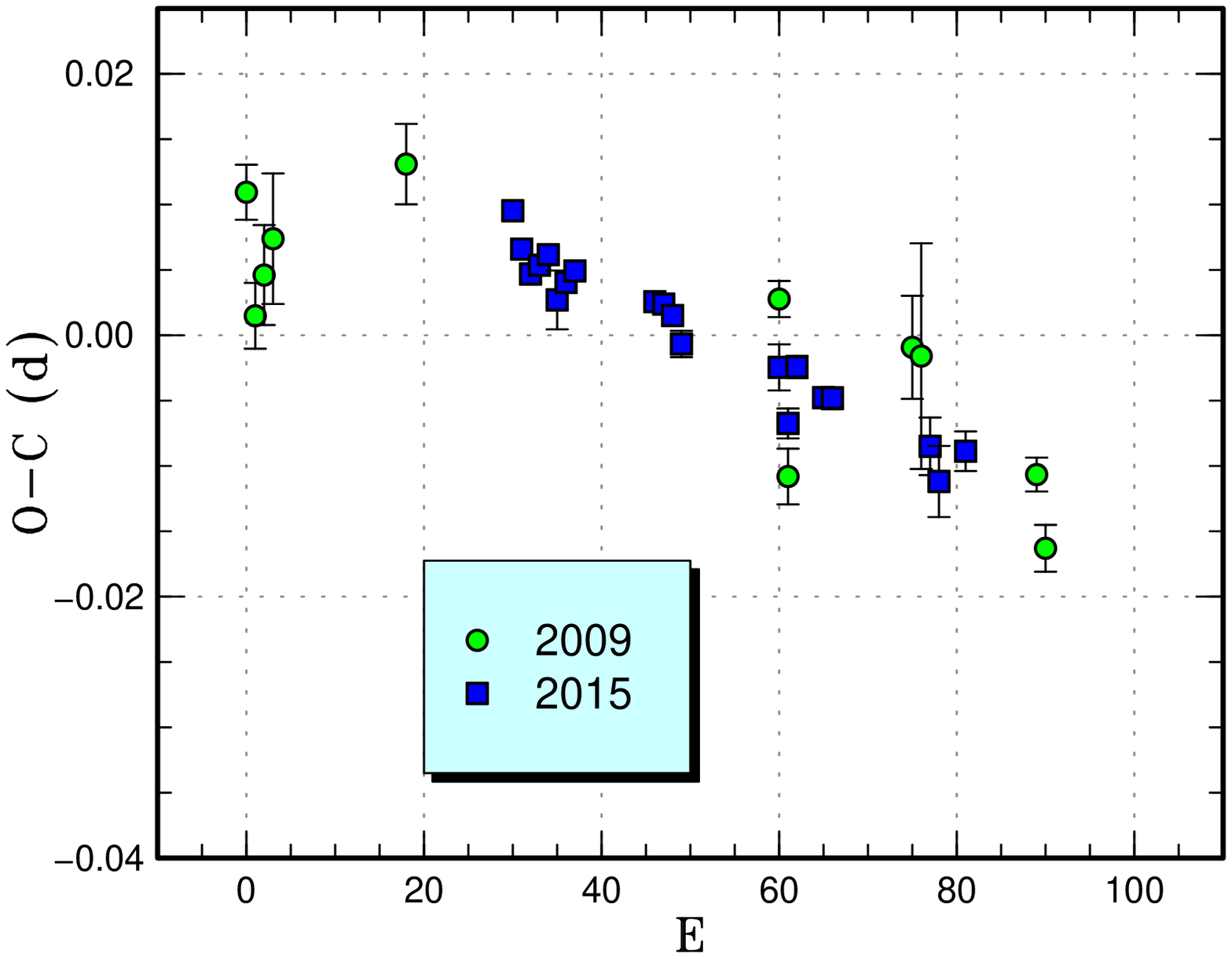}
  \end{center}
  \caption{Comparison of $O-C$ diagrams of PU Per between different
  superoutbursts.  A period of 0.06831~d was used to draw this figure.
  Approximate cycle counts ($E$) after the start of the observation
  were used.
  Since starts of neither outbursts were constrained, we shifted
  the $O-C$ diagram of the 2015 outburst
  to best fit the better-recorded 2009 one.
  }
  \label{fig:pupercomp}
\end{figure}

   The 2015 outburst was detected by E. Muyllaert
by CCD observations on October 3 (cf. vsnet-alert 19114).
Superhumps were detected (vsnet-alert 19124, 19130, 19144).
The times of superhump maxima are listed in
table \ref{tab:puperoc2015}.
A comparison with the 2009 data suggests that
we only recorded stage C superhumps
(figure \ref{fig:pupercomp}).

% SI

\begin{table}
\caption{Superhump maxima of PU Per (2015)}\label{tab:puperoc2015}
\begin{center}
\begin{tabular}{rp{55pt}p{40pt}r@{.}lr}
\hline
\multicolumn{1}{c}{$E$} & \multicolumn{1}{c}{max\commenta} & \multicolumn{1}{c}{error} & \multicolumn{2}{c}{$O-C$\commentb} & \multicolumn{1}{c}{$N$\commentc} \\
\hline
0 & 57301.0557 & 0.0007 & 0&0027 & 64 \\
1 & 57301.1211 & 0.0007 & 0&0001 & 73 \\
2 & 57301.1875 & 0.0007 & $-$0&0015 & 72 \\
3 & 57301.2565 & 0.0007 & $-$0&0005 & 69 \\
4 & 57301.3256 & 0.0008 & 0&0007 & 68 \\
5 & 57301.3904 & 0.0022 & $-$0&0025 & 71 \\
6 & 57301.4601 & 0.0005 & $-$0&0008 & 132 \\
7 & 57301.5293 & 0.0006 & 0&0004 & 132 \\
16 & 57302.1417 & 0.0006 & 0&0011 & 71 \\
17 & 57302.2098 & 0.0004 & 0&0012 & 218 \\
18 & 57302.2773 & 0.0004 & 0&0007 & 213 \\
19 & 57302.3434 & 0.0010 & $-$0&0011 & 148 \\
30 & 57303.0930 & 0.0018 & 0&0008 & 60 \\
31 & 57303.1570 & 0.0011 & $-$0&0032 & 75 \\
32 & 57303.2297 & 0.0009 & 0&0015 & 146 \\
35 & 57303.4323 & 0.0007 & 0&0001 & 178 \\
36 & 57303.5005 & 0.0008 & 0&0004 & 126 \\
47 & 57304.2482 & 0.0022 & 0&0004 & 127 \\
48 & 57304.3139 & 0.0027 & $-$0&0019 & 146 \\
51 & 57304.5211 & 0.0015 & 0&0014 & 59 \\
\hline
  \multicolumn{6}{l}{\commenta BJD$-$2400000.} \\
  \multicolumn{6}{l}{\commentb Against max $= 2457301.0530 + 0.067975 E$.} \\
  \multicolumn{6}{l}{\commentc Number of points used to determine the maximum.} \\
\end{tabular}
\end{center}
\end{table}

\subsection{QY Persei}\label{sec:qyper}\label{obj:qyper}

   QY Per was discovered as a dwarf nova (=S 9178)
by \citet{hof66an289139}.  \citet{pin76qyper} reported
another outburst at a photographic magnitude of 13.7
on 1971 September 21.  The object has been renowned
for its low frequency of outbursts and the 1989 October
outburst was reported by \citet{ros89qyperiauc}.
Although there was an outburst in 1994 October,
it faded rather quickly.  The next confirmed outburst
occurred in 1999 December and superhumps were detected
\citep{Pdot}.  Contrary to the expectation as a WZ Sge-type
dwarf nova, the superhump period was long ($\sim$0.0786~d).
The object was considered to be a long-period system
resembling WZ Sge-type dwarf novae, although no early
superhumps have been detected \citep{kat15wzsge}.
The next confirmed outburst (superoutburst) occurred
in 2005 September, which was not well observed
\citep{Pdot}.

\begin{figure}
  \begin{center}
%    \FigureFile(85mm,70mm){qypercomp2.eps}
    \FigureFile(85mm,70mm){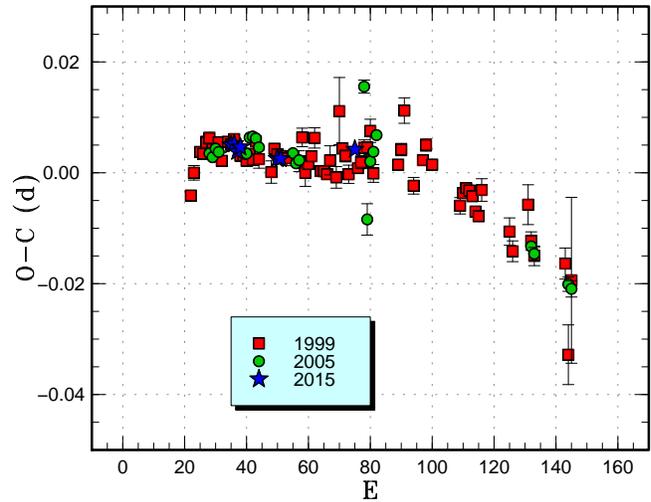}
  \end{center}
  \caption{Comparison of $O-C$ diagrams of QY Per between different
  superoutbursts.  A period of 0.07862~d was used to draw this figure.
  Approximate cycle counts ($E$) after the start of the superoutburst
  were used.  Since the start of the 2015 superoutburst
  was not well constrained, we shifted the $O-C$ diagram
  to best fit the others.
  }
  \label{fig:qypercomp2}
\end{figure}

   The 2015 outburst was detected by M. Hiraga at
an unfiltered CCD magnitude of 14.7 on November 14
(vsnet-alert 19263).  The object was fainter than
16.2 two days before.
Subsequent observations detected superhumps
(vsnet-alert 19267, 19281).
The times of superhump maxima are listed in
table \ref{tab:qyperoc2015}.
The early appearance of superhumps after the start
of the outbursts (figure \ref{fig:qypercomp2})
suggest that the object is less likely a WZ Sge-type
dwarf nova.  Since the 2015 outburst was much fainter
than the 1999 outburst, there remains a possibility
that the object shows superoutbursts of different
extent as in the WZ Sge-type object AL Com
\citep{kim16alcom}.

% SI

\begin{table}
\caption{Superhump maxima of QY Per (2015)}\label{tab:qyperoc2015}
\begin{center}
\begin{tabular}{rp{55pt}p{40pt}r@{.}lr}
\hline
\multicolumn{1}{c}{$E$} & \multicolumn{1}{c}{max\commenta} & \multicolumn{1}{c}{error} & \multicolumn{2}{c}{$O-C$\commentb} & \multicolumn{1}{c}{$N$\commentc} \\
\hline
0 & 57343.2396 & 0.0005 & 0&0007 & 70 \\
1 & 57343.3184 & 0.0004 & 0&0009 & 87 \\
2 & 57343.3956 & 0.0005 & $-$0&0004 & 87 \\
3 & 57343.4750 & 0.0005 & 0&0004 & 65 \\
15 & 57344.4164 & 0.0005 & $-$0&0013 & 58 \\
16 & 57344.4950 & 0.0008 & $-$0&0013 & 68 \\
40 & 57346.3836 & 0.0005 & 0&0010 & 57 \\
\hline
  \multicolumn{6}{l}{\commenta BJD$-$2400000.} \\
  \multicolumn{6}{l}{\commentb Against max $= 2457343.2389 + 0.078593 E$.} \\
  \multicolumn{6}{l}{\commentc Number of points used to determine the maximum.} \\
\end{tabular}
\end{center}
\end{table}

\subsection{TY Piscium}\label{obj:typsc}

\begin{figure}
  \begin{center}
%    \FigureFile(85mm,70mm){typsccomp3.eps}
    \FigureFile(85mm,70mm){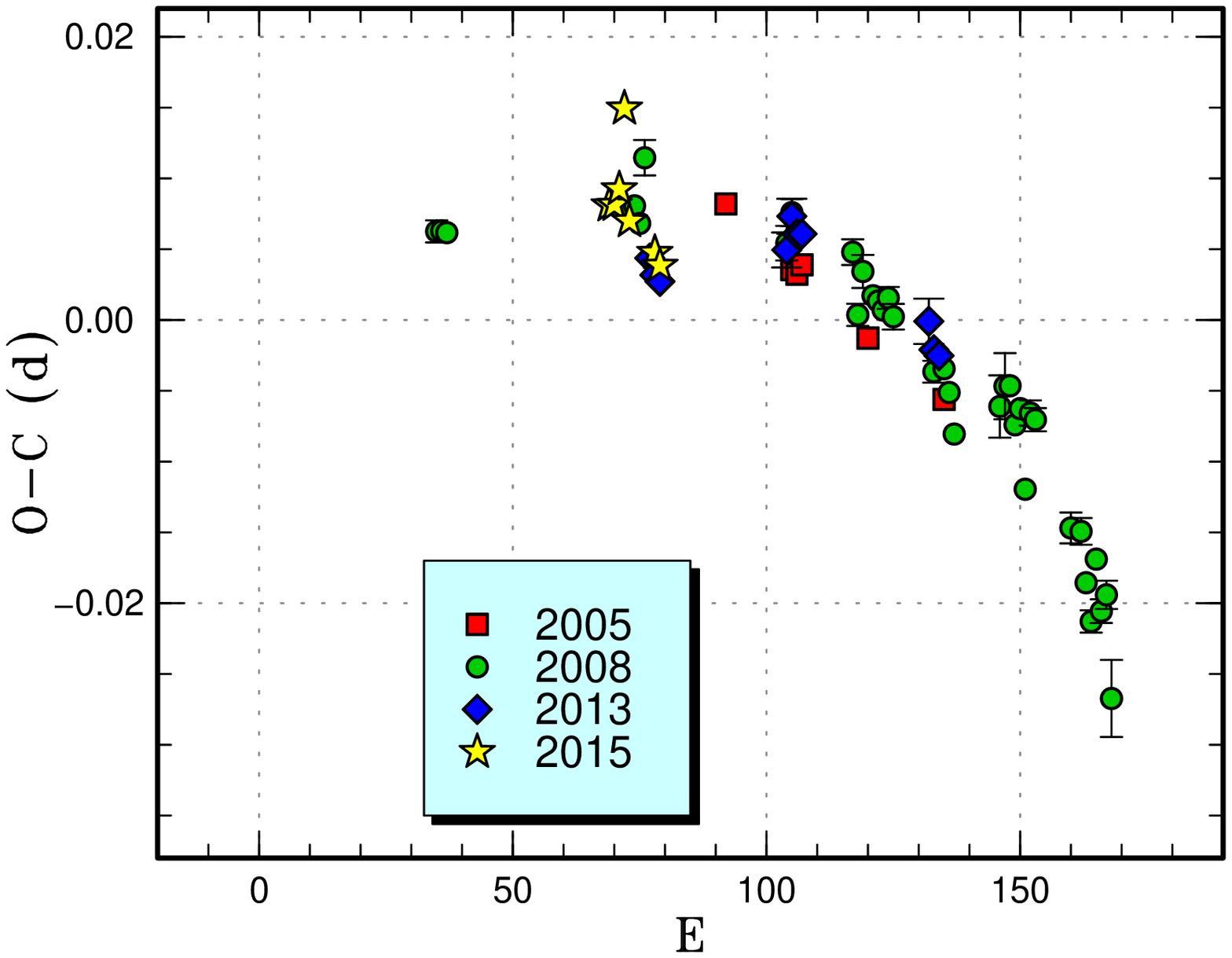}
  \end{center}
  \caption{Comparison of $O-C$ diagrams of TY Psc between different
  superoutbursts.  A period of 0.07066~d was used to draw this figure.
  Approximate cycle counts ($E$) after the start of the superoutburst
  were used.  Since the start of the 2013 superoutburst
  was not well constrained, we shifted the $O-C$ diagram
  to best fit the others.
  }
  \label{fig:typsccomp3}
\end{figure}

   For the history of this well-known SU UMa-type
object, see \citet{Pdot6}.
The 2015 superoutburst was visually detected by 
E. Muyllaert at a magnitude of 12.2 on October 29.
Our observations covered the later part of
the outburst and recorded superhumps in
table \ref{tab:typscoc2015}.
A comparison of $O-C$ diagrams of TY Psc between different
superoutbursts is given in figure \ref{fig:typsccomp3}.

% SI

\begin{table}
\caption{Superhump maxima of TY Psc (2015)}\label{tab:typscoc2015}
\begin{center}
\begin{tabular}{rp{55pt}p{40pt}r@{.}lr}
\hline
\multicolumn{1}{c}{$E$} & \multicolumn{1}{c}{max\commenta} & \multicolumn{1}{c}{error} & \multicolumn{2}{c}{$O-C$\commentb} & \multicolumn{1}{c}{$N$\commentc} \\
\hline
0 & 57330.2711 & 0.0004 & $-$0&0022 & 225 \\
1 & 57330.3418 & 0.0004 & $-$0&0016 & 225 \\
2 & 57330.4136 & 0.0004 & 0&0000 & 146 \\
3 & 57330.4899 & 0.0009 & 0&0063 & 39 \\
4 & 57330.5526 & 0.0006 & $-$0&0011 & 54 \\
9 & 57330.9037 & 0.0008 & $-$0&0005 & 88 \\
10 & 57330.9734 & 0.0005 & $-$0&0008 & 131 \\
\hline
  \multicolumn{6}{l}{\commenta BJD$-$2400000.} \\
  \multicolumn{6}{l}{\commentb Against max $= 2457330.2734 + 0.070093 E$.} \\
  \multicolumn{6}{l}{\commentc Number of points used to determine the maximum.} \\
\end{tabular}
\end{center}
\end{table}

\subsection{V493 Serpentis}\label{obj:v493ser}

\begin{figure}
  \begin{center}
%    \FigureFile(85mm,70mm){v493sercomp2.eps}
    \FigureFile(85mm,70mm){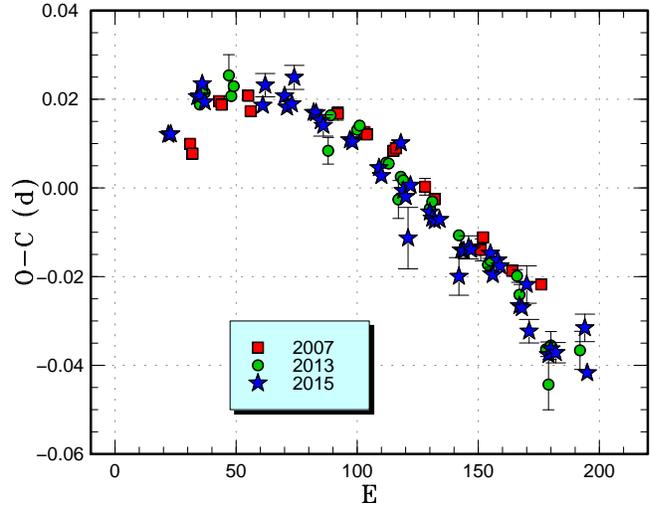}
  \end{center}
  \caption{Comparison of $O-C$ diagrams of V493 Ser between different
  superoutbursts.  A period of 0.08300~d was used to draw this figure.
  Approximate cycle counts ($E$) after the start of the superoutburst
  were used.  Since the start of the 2013 superoutburst
  was not well constrained, we shifted the $O-C$ diagram
  to best fit the better-recorded 2007 one.
  }
  \label{fig:v493sercomp2}
\end{figure}

   This object (=SDSS J155644.24$-$000950.2) was selected as
a dwarf nova by SDSS \citep{szk02SDSSCVs}.  The SU UMa-type
nature was confirmed during the 2006 and 2007
superoutbursts \citep{Pdot}.
See \citet{Pdot5} for more history.
The 2015 superoutburst was detected by the ASAS-SN team in its
early phase (vsnet-alert 18666).
Subsequent observations indeed recorded stage A superhumps
(vsnet-alert 18673, 18683) and later development
(vsnet-alert 18721).
The times of superhump maxima are listed in
table \ref{tab:v493seroc2015}.  Although transitions between
stages were rather smooth, we adopted stage classifications
listed in table \ref{tab:perlist} referring to the well-recorded
observation in \citet{Pdot} (figure \ref{fig:v493sercomp2}).
The resultant $\epsilon^*$=0.0449(13) for stage A superhumps
corresponds to $q$=0.129(5), which is in good agreement
with $q$=0.136(6) using the 2007 observation
\citep{kat13qfromstageA}.

% SI

\begin{table*}
\caption{Superhump maxima of V493 Ser (2015)}\label{tab:v493seroc2015}
\begin{center}
\begin{tabular}{rp{55pt}p{40pt}r@{.}lrrp{55pt}p{40pt}r@{.}lr}
\hline
\multicolumn{1}{c}{$E$} & \multicolumn{1}{c}{max\commenta} & \multicolumn{1}{c}{error} & \multicolumn{2}{c}{$O-C$\commentb} & \multicolumn{1}{c}{$N$\commentc} & \multicolumn{1}{c}{$E$} & \multicolumn{1}{c}{max\commenta} & \multicolumn{1}{c}{error} & \multicolumn{2}{c}{$O-C$\commentb} & \multicolumn{1}{c}{$N$\commentc} \\
\hline
0 & 57171.3874 & 0.0003 & $-$0&0203 & 85 & 100 & 57179.6859 & 0.0009 & 0&0060 & 25 \\
1 & 57171.4706 & 0.0004 & $-$0&0198 & 61 & 108 & 57180.3447 & 0.0010 & 0&0030 & 61 \\
12 & 57172.3931 & 0.0002 & $-$0&0072 & 85 & 109 & 57180.4264 & 0.0004 & 0&0020 & 84 \\
13 & 57172.4763 & 0.0002 & $-$0&0067 & 85 & 110 & 57180.5089 & 0.0004 & 0&0018 & 67 \\
14 & 57172.5622 & 0.0006 & $-$0&0036 & 38 & 112 & 57180.6754 & 0.0008 & 0&0028 & 18 \\
15 & 57172.6412 & 0.0010 & $-$0&0073 & 17 & 120 & 57181.3274 & 0.0042 & $-$0&0070 & 44 \\
39 & 57174.6348 & 0.0010 & 0&0010 & 16 & 121 & 57181.4164 & 0.0004 & $-$0&0006 & 84 \\
40 & 57174.7225 & 0.0026 & 0&0059 & 10 & 122 & 57181.4993 & 0.0005 & $-$0&0005 & 83 \\
48 & 57175.3849 & 0.0004 & 0&0065 & 62 & 124 & 57181.6663 & 0.0026 & 0&0011 & 20 \\
49 & 57175.4654 & 0.0004 & 0&0044 & 60 & 125 & 57181.7489 & 0.0013 & 0&0010 & 18 \\
51 & 57175.6324 & 0.0009 & 0&0059 & 20 & 133 & 57182.4129 & 0.0009 & 0&0031 & 69 \\
52 & 57175.7214 & 0.0027 & 0&0122 & 10 & 134 & 57182.4912 & 0.0005 & $-$0&0013 & 85 \\
60 & 57176.3783 & 0.0004 & 0&0073 & 80 & 136 & 57182.6606 & 0.0019 & 0&0028 & 20 \\
61 & 57176.4614 & 0.0003 & 0&0076 & 85 & 137 & 57182.7424 & 0.0016 & 0&0018 & 22 \\
63 & 57176.6258 & 0.0017 & 0&0066 & 20 & 145 & 57183.3983 & 0.0005 & $-$0&0041 & 84 \\
64 & 57176.7078 & 0.0024 & 0&0059 & 12 & 146 & 57183.4809 & 0.0006 & $-$0&0042 & 85 \\
75 & 57177.6186 & 0.0013 & 0&0068 & 17 & 148 & 57183.6523 & 0.0042 & 0&0018 & 19 \\
76 & 57177.7013 & 0.0012 & 0&0067 & 13 & 149 & 57183.7249 & 0.0027 & $-$0&0084 & 19 \\
87 & 57178.6095 & 0.0016 & 0&0051 & 13 & 157 & 57184.3844 & 0.0012 & $-$0&0106 & 45 \\
88 & 57178.6909 & 0.0007 & 0&0037 & 21 & 158 & 57184.4688 & 0.0009 & $-$0&0090 & 45 \\
96 & 57179.3631 & 0.0011 & 0&0141 & 43 & 160 & 57184.6342 & 0.0023 & $-$0&0091 & 19 \\
97 & 57179.4354 & 0.0003 & 0&0037 & 85 & 172 & 57185.6370 & 0.0031 & 0&0011 & 18 \\
98 & 57179.5171 & 0.0011 & 0&0027 & 39 & 173 & 57185.7099 & 0.0017 & $-$0&0087 & 20 \\
99 & 57179.5909 & 0.0069 & $-$0&0063 & 16 & \multicolumn{1}{c}{--} & \multicolumn{1}{c}{--} & \multicolumn{1}{c}{--} & \multicolumn{2}{c}{--} & \multicolumn{1}{c}{--}\\
\hline
  \multicolumn{12}{l}{\commenta BJD$-$2400000.} \\
  \multicolumn{12}{l}{\commentb Against max $= 2457171.4077 + 0.082722 E$.} \\
  \multicolumn{12}{l}{\commentc Number of points used to determine the maximum.} \\
\end{tabular}
\end{center}
\end{table*}

\subsection{V1212 Tauri}\label{obj:v1212tau}

   V1212 Tau was discovered as an eruptive object near M45
\citep{par83v1212tau}.  See \citet{Pdot3} and
\citet{Pdot5} for more history.
The 2016 superoutburst was detected on February 2
at $V$=15.91 by the ASAS-SN team.  The superoutburst was
also detected by M. Moriyama on February 7
at an unfiltered CCD magnitude of 15.6
(vsnet-alert 19462, 19465).  Only single-night observations
on February 8 were obtained and two superhumps
were recorded: BJD 2457427.0376(10) ($N$=103)
and 2457427.1028(28) ($N$=52).

\subsection{KK Telescopii}\label{obj:kktel}

   KK Tel was discovered as a dwarf nova by \citet{hof63VSS61}.
\citet{how91faintCV4} performed time-resolved photometry
in quiescence and recorded a period of 0.084~d.
The SU UMa-type nature was clarified during the 2002
superoutburst \citet{kat03v877arakktelpucma}.
\citet{kat03v877arakktelpucma} noticed an exceptionally
large period decrease of superhumps, and was considered
to be a prototypical SU UMa-type dwarf nova with
a large period decrease and this result was often
referred for comparison [cf. MN Dra \citep{nog03var73dra};
KS UMa \citep{ole03ksuma}; V419 Lyr \citep{rut07v419lyr}].
This detection of a large period decrease was before
the establishment of the common pattern of period variations
(stages A--C, \cite{Pdot}), and it has been clarified
that the large period decrease in KK Tel was likely
a result of stage A--B transition.

\begin{figure}
  \begin{center}
%    \FigureFile(85mm,110mm){kktel2015humpall.eps}
    \FigureFile(85mm,110mm){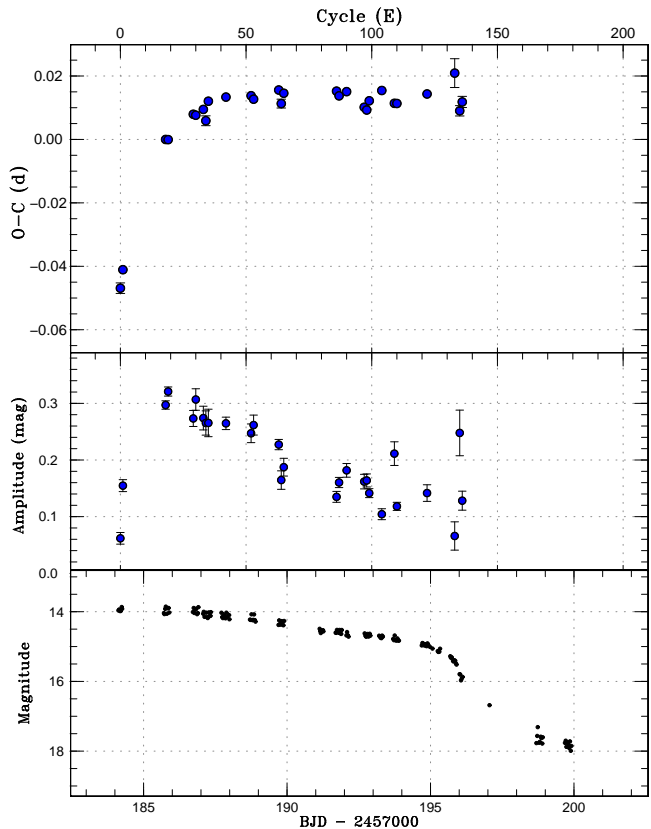}
  \end{center}
  \caption{$O-C$ diagram of superhumps in KK Tel (2015).
     (Upper:) $O-C$ diagram.
     We used a period of 0.08761~d for calculating the $O-C$ residuals.
     The superhump maxima up to $E=19$ are stage A superhumps.
     (Middle:) Amplitudes of superhumps.  The amplitudes were
     small around $E=0$.  It is worth noting that the epoch of
     the peak amplitude was earlier than the flattening of
     the $O-C$ diagram (pure stage B superhumps).
     (Lower:) Light curve.  The data were binned to 0.029~d.
     The initial outburst detection was on BJD 2457182.7,
     3~d before the start of our observation.  It took 4~d
     for this object to fully develop stage B superhumps.
  }
  \label{fig:kktel2015humpall}
\end{figure}

   The 2015 superoutburst was detected by the ASAS-SN team
on June 9 (cf. vsnet-alert 18713).  Subsequent observations
detected superhumps (vsnet-alert 18719, 18732, 18801).
The times of superhump maxima are listed in
table \ref{tab:kkteloc2015}.  The maxima up to $E$=19
were stage A superhumps (figure \ref{fig:kktel2015humpall}).
Since the 2015 observation
started 2~d later than the initial outburst detection,
it took $\sim$60 cycles for this object to develop
stage B superhumps.  It was not probably by chance that
stage A superhumps in this system were well recorded
both in 2002 and 2015, but is was likely a result of
relatively long-lasting stage A in this system.
In \citet{Pdot6} and \citet{kat16v1006cyg}, we proposed
that systems having $q$ close to the stability
border of the 3:1 resonance show slow evolution of
superhumps (see also subsection \ref{sec:longstagea}).
This is also probably the case for KK Tel and V419 Lyr.
It is also worth noting that the epoch of
the peak amplitude was earlier than the flattening of
the $O-C$ diagram (pure stage B superhumps;
figure \ref{fig:kktel2015humpall}).  This tendency was
also seen in the long-period system V1006 Cyg
with a long duration of stage A \citep{kat16v1006cyg}.

   Since the period of stage A superhumps is well determined
for KK Tel, precise determination of the orbital period
will lead to a measurement of $q$, providing a test
for this hypothesis.

\begin{figure}
  \begin{center}
%    \FigureFile(85mm,70mm){kktelcomp2.eps}
    \FigureFile(85mm,70mm){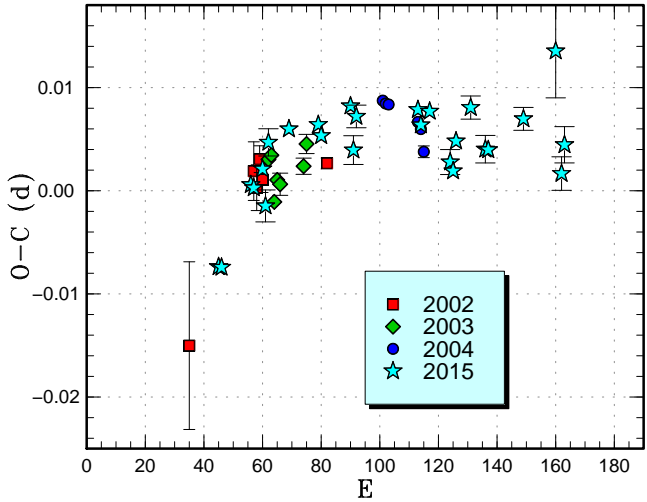}
  \end{center}
  \caption{Comparison of $O-C$ diagrams of KK Tel between different
  superoutbursts.  A period of 0.08761~d was used to draw this figure.
  Approximate cycle counts ($E$) after the start of the superoutburst
  were used.  In contrast to the diagram in \citet{Pdot},
  the epochs were shifted by 24, 30 and 20 cycles for
  the 2002, 2003 and 2004 superoutbursts, respectively,
  to best match the 2015 result.  These values suggests that
  either the true starts of the 2002--2004 superoutbursts were missed
  by 2--3~d, or the 2015 superoutburst started earlier
  than the others (such as a form of a precursor outburst).
  }
  \label{fig:kktelcomp2}
\end{figure}

% SI

\begin{table}
\caption{Superhump maxima of KK Tel (2015)}\label{tab:kkteloc2015}
\begin{center}
\begin{tabular}{rp{55pt}p{40pt}r@{.}lr}
\hline
\multicolumn{1}{c}{$E$} & \multicolumn{1}{c}{max\commenta} & \multicolumn{1}{c}{error} & \multicolumn{2}{c}{$O-C$\commentb} & \multicolumn{1}{c}{$N$\commentc} \\
\hline
0 & 57184.1440 & 0.0017 & $-$0&0392 & 53 \\
1 & 57184.2375 & 0.0009 & $-$0&0336 & 34 \\
18 & 57185.7679 & 0.0003 & 0&0038 & 20 \\
19 & 57185.8554 & 0.0003 & 0&0035 & 26 \\
29 & 57186.7396 & 0.0006 & 0&0093 & 31 \\
30 & 57186.8269 & 0.0006 & 0&0088 & 19 \\
33 & 57187.0915 & 0.0007 & 0&0100 & 41 \\
34 & 57187.1755 & 0.0015 & 0&0062 & 31 \\
35 & 57187.2693 & 0.0013 & 0&0121 & 42 \\
42 & 57187.8839 & 0.0004 & 0&0119 & 25 \\
52 & 57188.7604 & 0.0008 & 0&0102 & 37 \\
53 & 57188.8469 & 0.0008 & 0&0089 & 26 \\
63 & 57189.7259 & 0.0004 & 0&0096 & 50 \\
64 & 57189.8093 & 0.0014 & 0&0051 & 15 \\
65 & 57189.9001 & 0.0011 & 0&0082 & 19 \\
86 & 57191.7406 & 0.0008 & 0&0043 & 27 \\
87 & 57191.8267 & 0.0006 & 0&0026 & 38 \\
90 & 57192.0909 & 0.0006 & 0&0033 & 48 \\
97 & 57192.6992 & 0.0012 & $-$0&0032 & 18 \\
98 & 57192.7860 & 0.0007 & $-$0&0043 & 19 \\
99 & 57192.8765 & 0.0006 & $-$0&0016 & 49 \\
104 & 57193.3178 & 0.0011 & 0&0006 & 43 \\
109 & 57193.7518 & 0.0013 & $-$0&0045 & 17 \\
110 & 57193.8394 & 0.0006 & $-$0&0048 & 40 \\
122 & 57194.8937 & 0.0011 & $-$0&0044 & 38 \\
133 & 57195.8640 & 0.0045 & $-$0&0002 & 31 \\
135 & 57196.0273 & 0.0016 & $-$0&0125 & 44 \\
136 & 57196.1177 & 0.0018 & $-$0&0099 & 39 \\
\hline
  \multicolumn{6}{l}{\commenta BJD$-$2400000.} \\
  \multicolumn{6}{l}{\commentb Against max $= 2457184.1833 + 0.087826 E$.} \\
  \multicolumn{6}{l}{\commentc Number of points used to determine the maximum.} \\
\end{tabular}
\end{center}
\end{table}

\subsection{CI Ursae Majoris}\label{obj:ciuma}

   This object was discovered by \citet{gor72ciuma}
and was confirmed to be an SU UMa-type dwarf nova
by \citet{nog97ciuma}.  See \citet{Pdot5} for more
history.  The 2016 outburst was detected 
at an unfiltered CCD magnitude of 14.4 on March 14
by M. Hiraga (vsnet-alert 19626).  The outburst
did not receive special attention since it faded
rather quickly.  The outburst, however, turned out
to be a precursor based on the ASAS-SN observations
(vsnet-alert 19626).  Only single-night observations
were obtained yielding the superhump maxima
in table \ref{tab:ciumaoc2016}.

\begin{table}
\caption{Superhump maxima of CI UMa (2016)}\label{tab:ciumaoc2016}
\begin{center}
\begin{tabular}{rp{55pt}p{40pt}r@{.}lr}
\hline
\multicolumn{1}{c}{$E$} & \multicolumn{1}{c}{max\commenta} & \multicolumn{1}{c}{error} & \multicolumn{2}{c}{$O-C$\commentb} & \multicolumn{1}{c}{$N$\commentc} \\
\hline
0 & 57470.6381 & 0.0042 & $-$0&0017 & 44 \\
1 & 57470.7042 & 0.0017 & 0&0012 & 27 \\
2 & 57470.7665 & 0.0019 & 0&0001 & 17 \\
3 & 57470.8314 & 0.0021 & 0&0018 & 23 \\
4 & 57470.8925 & 0.0037 & $-$0&0004 & 26 \\
5 & 57470.9551 & 0.0105 & $-$0&0011 & 19 \\
\hline
  \multicolumn{6}{l}{\commenta BJD$-$2400000.} \\
  \multicolumn{6}{l}{\commentb Against max $= 2457470.6398 + 0.063283 E$.} \\
  \multicolumn{6}{l}{\commentc Number of points used to determine the maximum.} \\
\end{tabular}
\end{center}
\end{table}

\subsection{KS Ursae Majoris}\label{obj:ksuma}

   KS UMa was originally discovered as an emission-line
object (=SBS1017$+$533) \citep{bal97SBS2spec}.
The SU UMa-type nature was confirmed during
the 1998 outburst.  \citet{ole03ksuma} studied
the 2003 superoutburst in detail.
For more history, see \citet{Pdot}.

   The 2015 superoutburst was detected by K. Hirosawa
at $V$=13.0 on December 6.  Subsequent observations
detected superhumps (vsnet-alert 19330).
The times of superhump maxima are listed in
table \ref{tab:ksumaoc2015}.
A comparison of $O-C$ diagrams suggests that
we observed stage B (initial part) and stage C
with a long gap in the observation between them.

\begin{figure}
  \begin{center}
%    \FigureFile(85mm,70mm){ksumacomp3.eps}
    \FigureFile(85mm,70mm){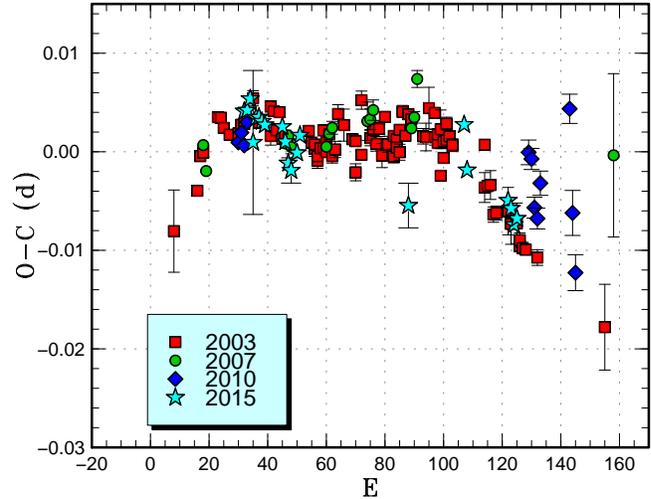}
  \end{center}
  \caption{Comparison of $O-C$ diagrams of KS UMa between different
  superoutbursts.
  A period of 0.07019~d was used to draw this figure.
  Approximate cycle counts ($E$) after the start of the superoutburst
  were used.  Since the 2003 superoutburst was very well
  recorded, we shifted the others to fit the 2003 one.
  We had to shift 18, 30 and 32 cycles for the 2007, 2010
  and 2015 data, respectively.  Note that this treatment
  is different from the corresponding figure in \citet{Pdot}.
  }
  \label{fig:ksumacomp3}
\end{figure}

% SI

\begin{table}
\caption{Superhump maxima of KS UMa (2015)}\label{tab:ksumaoc2015}
\begin{center}
\begin{tabular}{rp{55pt}p{40pt}r@{.}lr}
\hline
\multicolumn{1}{c}{$E$} & \multicolumn{1}{c}{max\commenta} & \multicolumn{1}{c}{error} & \multicolumn{2}{c}{$O-C$\commentb} & \multicolumn{1}{c}{$N$\commentc} \\
\hline
0 & 57364.2167 & 0.0002 & 0&0011 & 76 \\
1 & 57364.2870 & 0.0003 & 0&0013 & 75 \\
2 & 57364.3584 & 0.0007 & 0&0026 & 30 \\
3 & 57364.4241 & 0.0073 & $-$0&0018 & 34 \\
4 & 57364.4968 & 0.0005 & 0&0008 & 89 \\
5 & 57364.5671 & 0.0004 & 0&0010 & 94 \\
6 & 57364.6368 & 0.0004 & 0&0006 & 94 \\
7 & 57364.7067 & 0.0004 & 0&0004 & 93 \\
13 & 57365.1276 & 0.0005 & 0&0007 & 52 \\
14 & 57365.1963 & 0.0004 & $-$0&0007 & 175 \\
15 & 57365.2643 & 0.0005 & $-$0&0028 & 181 \\
16 & 57365.3337 & 0.0013 & $-$0&0034 & 60 \\
18 & 57365.4759 & 0.0003 & $-$0&0015 & 75 \\
19 & 57365.5479 & 0.0004 & 0&0004 & 69 \\
56 & 57368.1378 & 0.0023 & $-$0&0034 & 12 \\
75 & 57369.4796 & 0.0007 & 0&0065 & 131 \\
76 & 57369.5452 & 0.0007 & 0&0020 & 124 \\
90 & 57370.5248 & 0.0008 & 0&0002 & 77 \\
91 & 57370.5941 & 0.0005 & $-$0&0005 & 77 \\
92 & 57370.6627 & 0.0005 & $-$0&0021 & 77 \\
93 & 57370.7335 & 0.0007 & $-$0&0014 & 55 \\
\hline
  \multicolumn{6}{l}{\commenta BJD$-$2400000.} \\
  \multicolumn{6}{l}{\commentb Against max $= 2457364.2156 + 0.070100 E$.} \\
  \multicolumn{6}{l}{\commentc Number of points used to determine the maximum.} \\
\end{tabular}
\end{center}
\end{table}

\subsection{MR Ursae Majoris}\label{obj:mruma}

   This object is a well-known SU UMa-type dwarf nova.
See \citet{Pdot5} for the history.  The 2015 March
superoutburst had a precursor outburst on March 7
(visually detected by E. Muyllaert, vsnet-alert 18426).
We observed on one night during the fading part
of the precursor outburst and obtained
two superhump maxima: BJD 2457091.4918(3) ($N$=72)
and 2457091.5573(5) ($N$=72).  Although these superhumps
most likely correspond to stage A superhumps,
the period was not meaningfully determined.

   Recent outbursts of this object are listed in table
\ref{tab:mrumaout}.  Although observations up to 2005
inferred a supercycle of approximately a year, recent
observations suggest that the supercycles varied considerably
and they can be as short as $\sim$260~d.

\begin{table*}
\caption{List of recent outbursts of MR UMa}\label{tab:mrumaout}
\begin{center}
\begin{tabular}{cccccl}
\hline
Year & Month & max\commenta & magnitude & type & source \\
\hline
2002 & 3 & 52341 & 12.7 & super & VSNET, AAVSO, \citet{Pdot} \\
2003 & 3 & 52711 & 12.8 & super & VSNET, AAVSO, \citet{Pdot} \\
2004 & 3 & 53084 & 13.2 & super & VSNET, AAVSO \\
2005 & 3 & 53440 & 12.9 & super & VSNET, AAVSO \\
2005 & 4 & 53474 & 15.3 & normal & AAVSO \\
2006 & 5 & 53870 & 12.8 & super & VSNET, AAVSO \\
2006 & 11 & 54058 & 13.1\commentb & ? & VSNET \\
2007 & 4 & 54205 & 12.5 & super & VSNET, AAVSO, \citet{Pdot} \\
2008 & 2 & 54503 & 12.7 & super & VSNET, AAVSO \\
2008 & 6 & 54627 & 12.9 & normal & VSNET, AAVSO \\
2009 & 4 & 54948 & 13.2 & normal & VSNET, AAVSO \\
2010 & 4 & 55303 & 12.7 & super & VSNET, AAVSO, \citet{Pdot2} \\
2011 & 5 & 55684 & 12.9 & normal & VSNET, AAVSO \\
2012 & 6 & 56090 & 13.2 & super & VSNET, AAVSO, \citet{Pdot4} \\
2013 & 3 & 56354 & 13.2 & super & VSNET, AAVSO, \citet{Pdot5} \\
2013 & 11 & 56610 & 12.8 & super? & AAVSO \\
2014 & 2 & 56708 & 13.3\commentc & normal & AAVSO \\
2014 & 4 & 56770 & 13.5 & normal & VSNET, AAVSO \\
2015 & 3 & 57094 & 12.8\commentd & super & VSNET, AAVSO, this paper \\
\hline
  \multicolumn{6}{l}{\commenta JD$-$2400000.} \\
  \multicolumn{6}{l}{\commentb Single visual observation.} \\
  \multicolumn{6}{l}{\commentc Single CCD observation.} \\
  \multicolumn{6}{l}{\commentd Refers to the main superoutburst.} \\
\end{tabular}
\end{center}
\end{table*}

\subsection{NSV 2026}\label{obj:nsv2026}

\subsubsection{Introduction}

   This object was discovered as a variable star
(=HV 6907) by \citet{hof35newvar}.  The exact coordinates
were given by \citet{web02varpos}.
The object was identified to be a dwarf nova
by a detection of an outburst by CRTS Mount Lemmon survey
(=MLS101214:052959$+$184810) in 2010 (vsnet-alert 12503).
It has been monitored for outbursts since then.
Several outbursts were recorded by BAAVSS/AAVSO observers
(since 2012) and by the MISAO project (two outbursts
in 2011 and one in 2012).
There is an X-ray counterpart 1RXS J052954.9$+$184817.
There were two past relatively long outbursts
is 2012 February-March and 2014 February.
Although time-resolved photometry was undertaken,
no convincing superhumps were detected.\footnote{
  $<$http://www.britastro.org/vss/NSV2026.pdf$>$.
}

\subsubsection{2015 Superoutburst}

\begin{figure}
  \begin{center}
%    \FigureFile(85mm,110mm){nsv2026shpdm.eps}
    \FigureFile(85mm,110mm){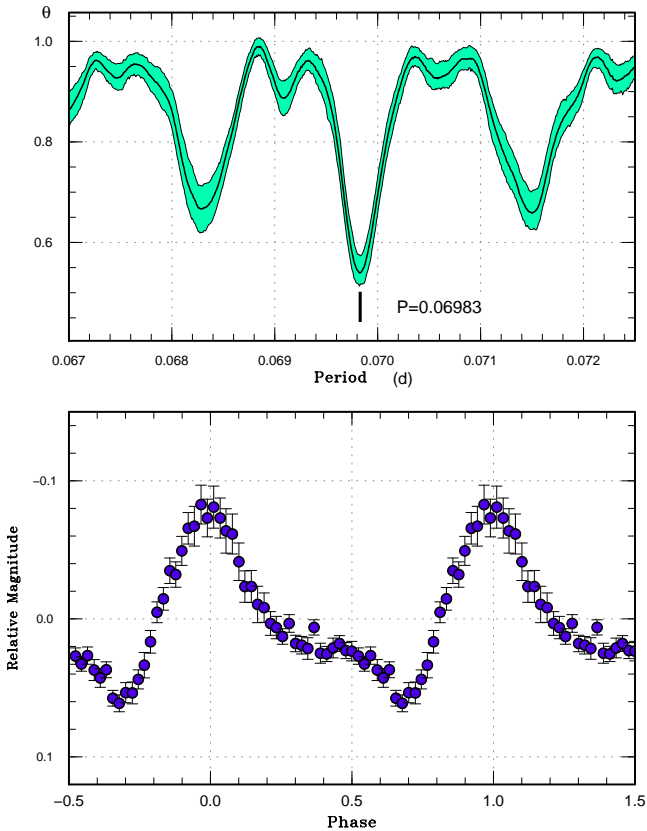}
  \end{center}
  \caption{Superhumps in NSV 2026 (2015).
     (Upper): PDM analysis.
     (Lower): Phase-averaged profile.}
  \label{fig:nsv2026shpdm}
\end{figure}

   The 2015 November bright outburst was detected
by J. Shears on November 8 at an unfiltered CCD
magnitude of 13.6 (vsnet-outburst 18873).\footnote{
   See also $<$http://www.britastro.org/vss/NSV2026.pdf$>$
   for the BAAVSS campaign.
}
This outburst turned out to be a superoutburst by
the detection of superhumps
(vsnet-alert 19258; figure \ref{fig:nsv2026shpdm}).
The times of superhump maxima are listed in
table \ref{tab:nsv2026oc2015}.
Although there was an apparent stage transition
(vsnet-alert 19282), we gave a globally averaged
period due to insufficient observations.

% SI

\begin{figure}
  \begin{center}
%    \FigureFile(85mm,70mm){nsv2026comp.eps}
    \FigureFile(85mm,70mm){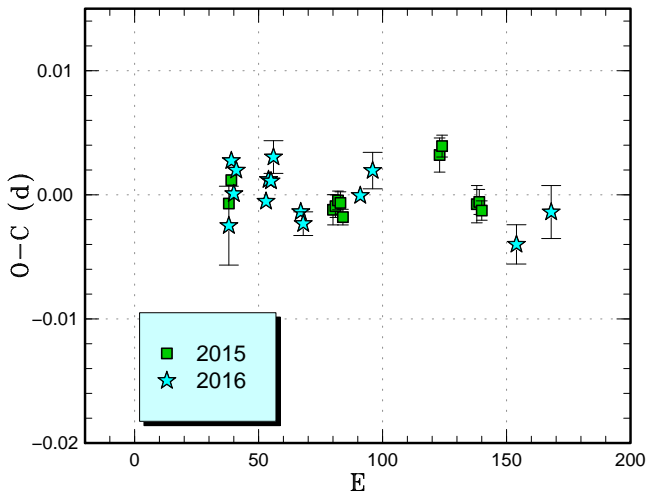}
  \end{center}
  \caption{Comparison of $O-C$ diagrams of NSV 2026 between different
  superoutbursts.  A period of 0.06982~d was used to draw this figure.
  Approximate cycle counts ($E$) after the starts of the outbursts
  were used.  The start of the 2016 outburst refers to
  the precursor outburst.  Since the start of the 2015 outburst
  was not well constrained, ther $O-C$ curve was shifted
  as in the 2016 one.
  }
  \label{fig:nsv2026comp}
\end{figure}

% SI

\begin{table}
\caption{Superhump maxima of NSV 2026 (2015)}\label{tab:nsv2026oc2015}
\begin{center}
\begin{tabular}{rp{55pt}p{40pt}r@{.}lr}
\hline
\multicolumn{1}{c}{$E$} & \multicolumn{1}{c}{max\commenta} & \multicolumn{1}{c}{error} & \multicolumn{2}{c}{$O-C$\commentb} & \multicolumn{1}{c}{$N$\commentc} \\
\hline
0 & 57335.4992 & 0.0004 & $-$0&0002 & 90 \\
1 & 57335.5709 & 0.0003 & 0&0017 & 91 \\
42 & 57338.4311 & 0.0012 & $-$0&0011 & 74 \\
43 & 57338.5012 & 0.0009 & $-$0&0008 & 93 \\
44 & 57338.5715 & 0.0008 & $-$0&0003 & 93 \\
45 & 57338.6411 & 0.0009 & $-$0&0005 & 77 \\
46 & 57338.7098 & 0.0006 & $-$0&0017 & 68 \\
85 & 57341.4378 & 0.0014 & 0&0030 & 55 \\
86 & 57341.5083 & 0.0009 & 0&0037 & 60 \\
100 & 57342.4811 & 0.0015 & $-$0&0011 & 34 \\
101 & 57342.5511 & 0.0010 & $-$0&0009 & 34 \\
102 & 57342.6203 & 0.0008 & $-$0&0016 & 34 \\
\hline
  \multicolumn{6}{l}{\commenta BJD$-$2400000.} \\
  \multicolumn{6}{l}{\commentb Against max $= 2457335.4994 + 0.069829 E$.} \\
  \multicolumn{6}{l}{\commentc Number of points used to determine the maximum.} \\
\end{tabular}
\end{center}
\end{table}

\subsubsection{2016 Superoutburst}

\begin{figure}
  \begin{center}
%    \FigureFile(85mm,110mm){nsv2026lc.eps}
    \FigureFile(85mm,110mm){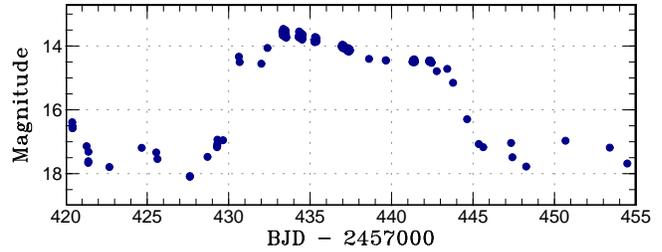}
  \end{center}
  \caption{Light curve of the superoutburst of NSV 2026
     (2016).  The data were binned to 0.01~d.
     The initial precursor part of the outburst and
     subsequent rise to the superoutburst maximum are
     clearly depicted.  The widths of the light curve
     mainly reflect the amplitudes of superhumps.}
  \label{fig:nsv2026lc}
\end{figure}

   The 2016 superoutburst was detected by J. Shears on
February 12 at an unfiltered CCD magnitude of 14.4
(BAAVSS alert 4308).  Subsequent observations detected
superhumps (vsnet-alert 19485, 19498, 19509).
The outburst started with a precursor
(figure \ref{fig:nsv2026lc}).
The times of superhump maxima are listed in
table \ref{tab:nsv2026oc2016}.
Although the initial part of the data ($E \le$58)
most likely refers to stage B, we gave a globally averaged
period since the observations were not sufficient
to determine $P_{\rm dot}$ for stage B.
A comparison of the $O-C$ diagrams (figure \ref{fig:nsv2026comp})
suggests that the later parts of 2015 and 2016
observations were likely stage C.

% SI

\begin{table}
\caption{Superhump maxima of NSV 2026 (2016)}\label{tab:nsv2026oc2016}
\begin{center}
\begin{tabular}{rp{55pt}p{40pt}r@{.}lr}
\hline
\multicolumn{1}{c}{$E$} & \multicolumn{1}{c}{max\commenta} & \multicolumn{1}{c}{error} & \multicolumn{2}{c}{$O-C$\commentb} & \multicolumn{1}{c}{$N$\commentc} \\
\hline
0 & 57433.2811 & 0.0032 & $-$0&0033 & 54 \\
1 & 57433.3561 & 0.0002 & 0&0019 & 187 \\
2 & 57433.4233 & 0.0002 & $-$0&0007 & 286 \\
3 & 57433.4950 & 0.0002 & 0&0012 & 159 \\
15 & 57434.3303 & 0.0005 & $-$0&0010 & 78 \\
16 & 57434.4019 & 0.0002 & 0&0008 & 197 \\
17 & 57434.4716 & 0.0004 & 0&0007 & 145 \\
18 & 57434.5433 & 0.0013 & 0&0026 & 33 \\
29 & 57435.3070 & 0.0005 & $-$0&0015 & 104 \\
30 & 57435.3758 & 0.0010 & $-$0&0024 & 57 \\
53 & 57436.9839 & 0.0007 & 0&0004 & 145 \\
58 & 57437.3351 & 0.0015 & 0&0025 & 66 \\
116 & 57441.3787 & 0.0016 & $-$0&0020 & 50 \\
130 & 57442.3588 & 0.0021 & 0&0009 & 196 \\
\hline
  \multicolumn{6}{l}{\commenta BJD$-$2400000.} \\
  \multicolumn{6}{l}{\commentb Against max $= 2457433.2844 + 0.069795 E$.} \\
  \multicolumn{6}{l}{\commentc Number of points used to determine the maximum.} \\
\end{tabular}
\end{center}
\end{table}

\subsubsection{Interpretation}

   Although the 2012 and 2014 outbursts were not
very well observed, the lack of prominent superhumps
would suggest that these outbursts may have been
long, normal outbursts as seen in TU Men
(\cite{war95suuma}; \cite{bat00tumen}), V1006 Cyg
\citep{kat16v1006cyg} and potentially NY Ser
\citep{pav14nyser}.  The big difference between
NSV 2026 and these objects is the orbital period --
TU Men, V1006 Cyg and NY Ser have long orbital periods
in or above the period gap, while NSV 2026 is not.
If the presence of long normal outbursts is
due to the difficulty in attaining the 3:1 resonance,
it would be easy to understand why most objects
showing this behavior have long-$P_{\rm orb}$
and the case of NSV 2026 might require another
explanation.  Since superhump amplitudes became
significantly smaller in the 2016 superoutburst
in the late phase, it may just have been that
the 2012 and 2014 observations did not record
the phases with large-amplitude superhumps.
Confirmation of superhumps in
every long outburst of NSV 2026 would be a task
to check these possibilities.

\begin{figure}
  \begin{center}
%    \FigureFile(85mm,110mm){nsv2026global.eps}
    \FigureFile(85mm,110mm){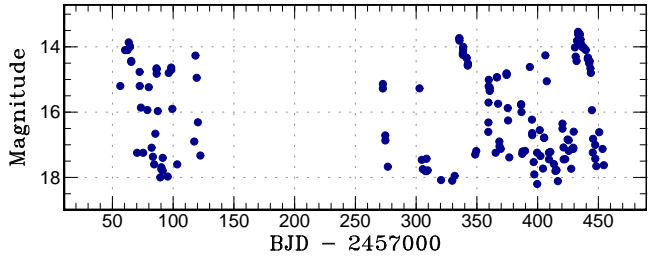}
  \end{center}
  \caption{Long-term curve of NSV 2026 from AAVSO
     observations.  The data were binned to 0.1~d.}
  \label{fig:nsv2026global}
\end{figure}

   AAVSO observations of this object show 
a relatively regular supercycle pattern: normal outbursts
with recurrence time of 6--14~d and the supercycle
length of $\sim$95~d (figure \ref{fig:nsv2026global}).
As judged from this light curve, NSV 2026 looks like
to be a fairly normal SU UMa-type dwarf nova
with frequent outbursts.

\subsection{ASASSN-13ah}\label{obj:asassn13ah}

   This object was detected as a transient
on 2013 April 23 by the ASAS-SN team \citep{sha13asassn13ahatel5052}.
The object was confirmed to be a dwarf nova
in outburst by spectroscopy \citep{sha13asassn13ahatel5052}.
Although there were observations during the 2015
outburst (cf. vsnet-alert 18619) by KU team,
the observations were insufficient to confirm
superhumps.

\begin{figure}
  \begin{center}
%    \FigureFile(85mm,110mm){asassn13ahshpdm.eps}
    \FigureFile(85mm,110mm){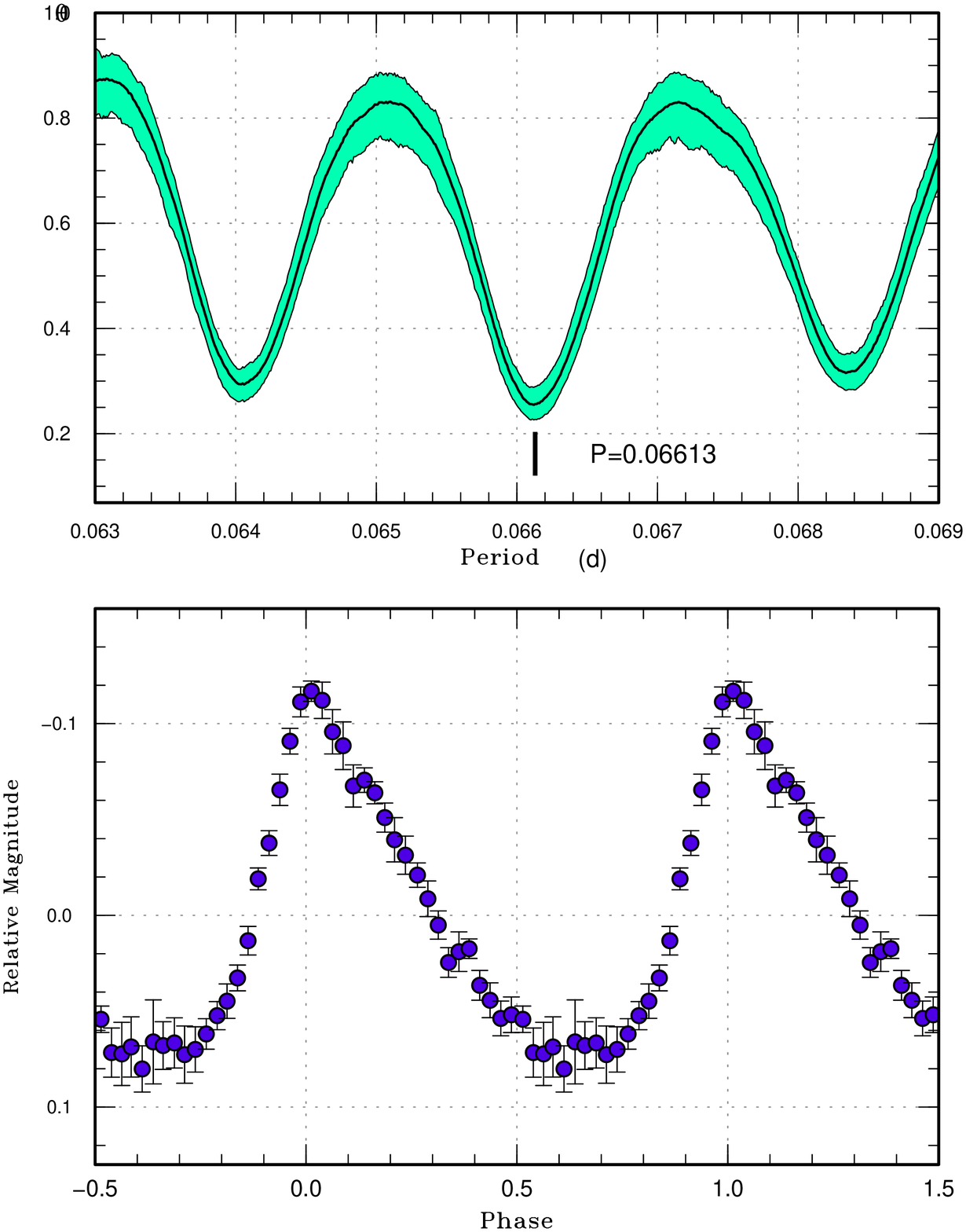}
  \end{center}
  \caption{Superhumps in ASASSN-13ah (2016).
     (Upper): PDM analysis.  The alias selection was
     based on $O-C$ analysis.
     (Lower): Phase-averaged profile.}
  \label{fig:asassn13ahshpdm}
\end{figure}

   The 2016 outburst was detected by the ASAS-SN team
on February 11 at $V$=16.4.  Subsequent observations
detected superhumps (vsnet-alert 19480, 19497;
figure \ref{fig:asassn13ahshpdm}).
The times of superhump maxima are listed in
table \ref{tab:asassn13ahoc2016}.

% SI

% SI

\begin{table}
\caption{Superhump maxima of ASASSN-13ah (2016)}\label{tab:asassn13ahoc2016}
\begin{center}
\begin{tabular}{rp{55pt}p{40pt}r@{.}lr}
\hline
\multicolumn{1}{c}{$E$} & \multicolumn{1}{c}{max\commenta} & \multicolumn{1}{c}{error} & \multicolumn{2}{c}{$O-C$\commentb} & \multicolumn{1}{c}{$N$\commentc} \\
\hline
0 & 57431.5184 & 0.0004 & 0&0006 & 67 \\
1 & 57431.5839 & 0.0003 & $-$0&0001 & 65 \\
2 & 57431.6495 & 0.0003 & $-$0&0006 & 67 \\
3 & 57431.7162 & 0.0004 & 0&0000 & 62 \\
31 & 57433.5679 & 0.0007 & $-$0&0003 & 62 \\
32 & 57433.6340 & 0.0010 & $-$0&0003 & 56 \\
33 & 57433.7011 & 0.0011 & 0&0006 & 64 \\
\hline
  \multicolumn{6}{l}{\commenta BJD$-$2400000.} \\
  \multicolumn{6}{l}{\commentb Against max $= 2457431.5178 + 0.066141 E$.} \\
  \multicolumn{6}{l}{\commentc Number of points used to determine the maximum.} \\
\end{tabular}
\end{center}
\end{table}

\subsection{ASASSN-13ak}\label{obj:asassn13ak}

   This object was detected as a transient at $V$=15.4
on 2013 May 23 by the ASAS-SN team \citep{sta13asassn13akatel5082}.
The MASTER network also detected this outburst
\citep{shu13asassn13akatel5083}.
There is an X-ray counterpart (1RXS J174827.1$+$505053).
There have been five outbursts (including the 2015 one)
in the CRTS data.  The SDSS colors suggested a short
orbital period (vsnet-alert 15742)

\begin{figure}
  \begin{center}
%    \FigureFile(85mm,110mm){asassn13akshpdm.eps}
    \FigureFile(85mm,110mm){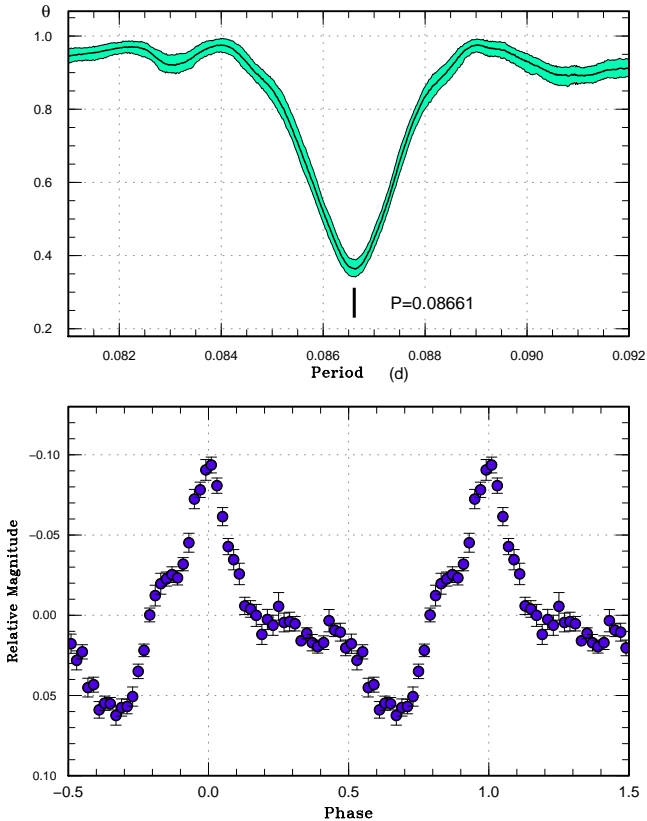}
  \end{center}
  \caption{Superhumps in ASASSN-13ak (2015).
     (Upper): PDM analysis.
     (Lower): Phase-averaged profile.}
  \label{fig:asassn13akshpdm}
\end{figure}

   The 2015 outburst was detected by E. Muyllaert
on May 8 at an unfiltered CCD magnitude of 15.1
(vsnet-alert 18607).  This object was confirmed to be
an SU UMa-type dwarf nova by the detection of
superhumps (vsnet-alert 18612, 18613, 18615, 18624;
figure \ref{fig:asassn13akshpdm}).
The times of superhump maxima are listed in
table \ref{tab:asassn13akoc2015}.
The superhump stage is unknown.

% SI

% SI

\begin{table}
\caption{Superhump maxima of ASASSN-13ak (2015)}\label{tab:asassn13akoc2015}
\begin{center}
\begin{tabular}{rp{55pt}p{40pt}r@{.}lr}
\hline
\multicolumn{1}{c}{$E$} & \multicolumn{1}{c}{max\commenta} & \multicolumn{1}{c}{error} & \multicolumn{2}{c}{$O-C$\commentb} & \multicolumn{1}{c}{$N$\commentc} \\
\hline
0 & 57154.6695 & 0.0005 & 0&0001 & 63 \\
9 & 57155.4500 & 0.0003 & 0&0007 & 187 \\
10 & 57155.5365 & 0.0003 & 0&0005 & 195 \\
11 & 57155.6226 & 0.0003 & $-$0&0001 & 114 \\
18 & 57156.2282 & 0.0003 & $-$0&0010 & 181 \\
20 & 57156.3995 & 0.0012 & $-$0&0030 & 56 \\
21 & 57156.4901 & 0.0009 & 0&0009 & 77 \\
22 & 57156.5767 & 0.0009 & 0&0009 & 73 \\
32 & 57157.4444 & 0.0008 & 0&0020 & 62 \\
33 & 57157.5289 & 0.0005 & $-$0&0002 & 84 \\
34 & 57157.6148 & 0.0006 & $-$0&0009 & 71 \\
\hline
  \multicolumn{6}{l}{\commenta BJD$-$2400000.} \\
  \multicolumn{6}{l}{\commentb Against max $= 2457154.6694 + 0.086655 E$.} \\
  \multicolumn{6}{l}{\commentc Number of points used to determine the maximum.} \\
\end{tabular}
\end{center}
\end{table}

\subsection{ASASSN-13az}\label{obj:asassn13az}

   This object was detected as a transient at $V$=14.4
on 2013 July 1 by the ASAS-SN team (vsnet-alert 15892).
Although the object was identified with a 20.8 mag
($B_j$) star in the USNO-B1.0 catalog,
this object was later found to be unrelated by
spectroscopy (cf. vsnet-alert 19555).
Although there was a 14.858 mag detection close to
the ASAS-SN position in URAT1 catalog \citep{URAT1},
the identification is unclear.  The object may have been
recorded in outburst.  If this identification is
confirmed, the coordinates are \timeform{18h 42m 58.21s},
\timeform{+73D 42' 28.4''}.

\begin{figure}
  \begin{center}
%    \FigureFile(85mm,70mm){asassn13azshlc.eps}
    \FigureFile(85mm,70mm){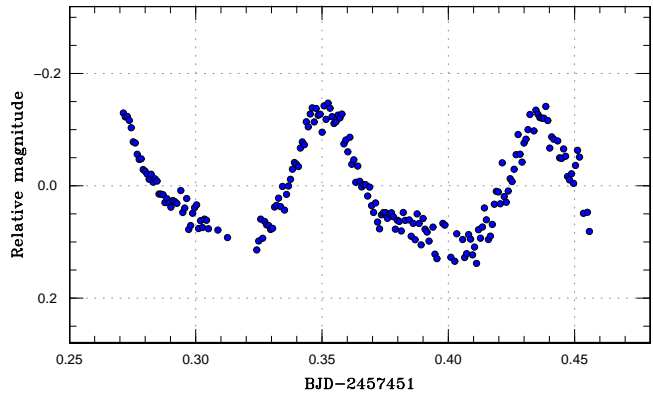}
  \end{center}
  \caption{Superhumps in ASASSN-13az (2016).
  }
  \label{fig:asassn13azshlc}
\end{figure}

   The 2016 outburst was detected by the ASAS-SN team
on March 1 at $V$=14.42.  Subsequent observations
detected superhumps (vsnet-alert 19554;
figure \ref{fig:asassn13azshlc}).
Two superhump maxima were recorded:
BJD 2457451.3514(3) ($N$=77) and 2457451.4374(5) ($N$=61).
The best superhump period with the PDM method is 
0.0843(3)~d.

\subsection{ASASSN-14ca}\label{obj:asassn14ca}

   This object was detected as a transient at $V$=15.5
on 2014 June 7 by the ASAS-SN team \citep{dav14asassn14caatel6211}.
The object was initially reported as an unusual long-lived
transient from a red source.  \citet{dav14asassn14caatel6211}
also reported an outburst in 2005 July in the CRTS data,
which lasted at least for 6~d.  Upon this report,
\citet{cao14asassn14caatel6221} examined
the PTF/iPTF archival data and found another brightening
in 2009 November, which lasted at least for 15~d.
\citet{cao14asassn14caatel6221} reported that the color
temperature is consistent with a Mira star.
\citet{pri14asassn14caatel6249} reported a spectrum taken
on June 9 and it had a strong blue continuum, Balmer lines
in absorption, H$\alpha$ line in double-peaked emission
and He\textsc{ii} 468.6nm in emission.  The object was
confirmed to be a dwarf nova in outburst.
The object was independently suggested to be a dwarf nova
(vsnet-alert 17369) and D. Denisenko detected an outburst
in 2009 November in the MASTER network observations,
the same one reported in \citet{cao14asassn14caatel6221}.
The $g$=20.6 SDSS counterpart
may be an unrelated unresolved red star.

\begin{figure}
  \begin{center}
%    \FigureFile(85mm,110mm){asassn14cashpdm.eps}
    \FigureFile(85mm,110mm){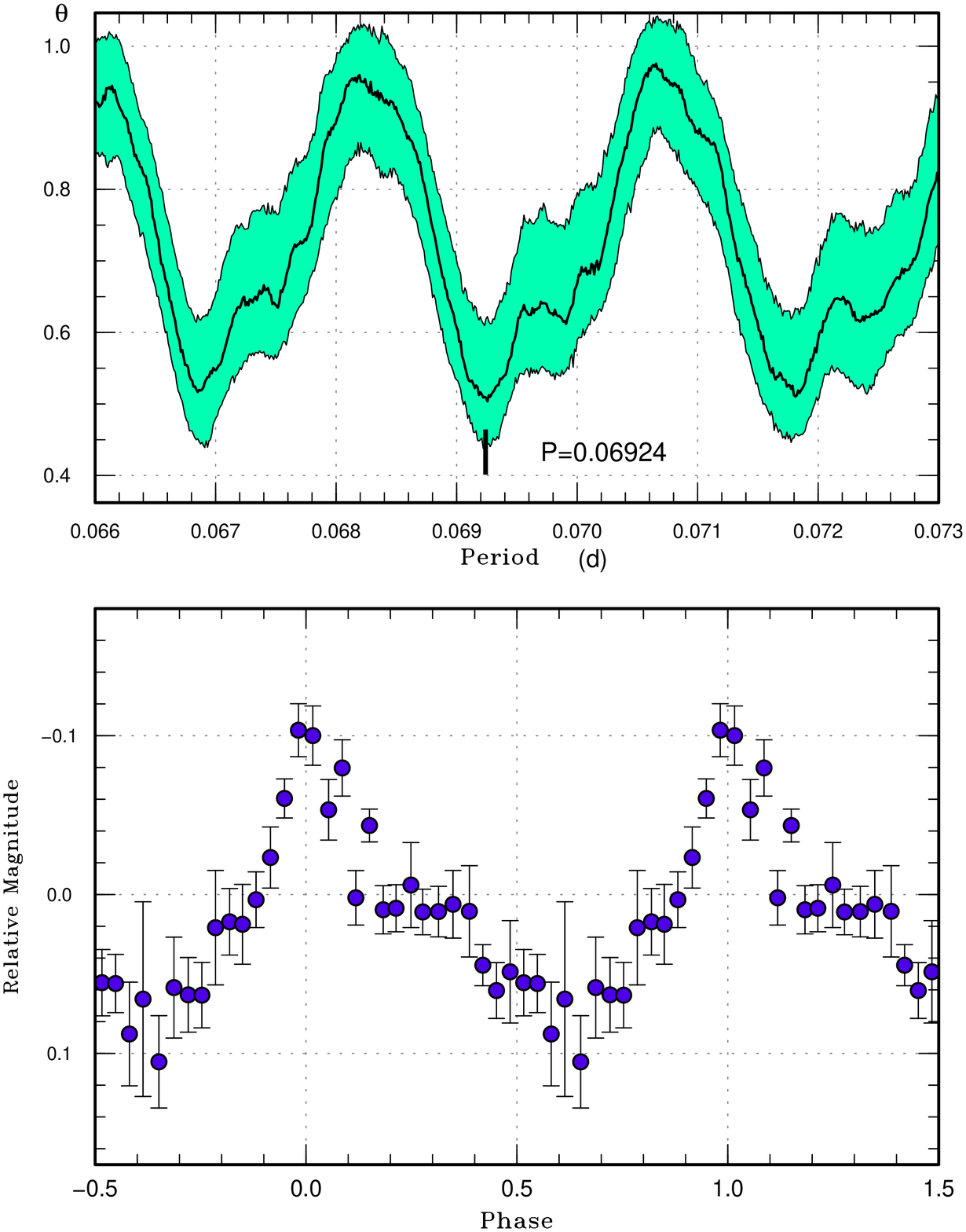}
  \end{center}
  \caption{Superhumps in ASASSN-14ca (2015).
     (Upper): PDM analysis.  The alias selection was based on
     the $O-C$ analysis.
     (Lower): Phase-averaged profile.}
  \label{fig:asassn14cashpdm}
\end{figure}

   The 2015 outburst was detected by the ASAS-SN team
on July 7 at $V$=15.53 (cf. vsnet-alert 18833).
Subsequent observations detected superhumps
(vsnet-alert 18852; figure \ref{fig:asassn14cashpdm}).
The times of superhump maxima are listed in
table \ref{tab:asassn14caoc2015}.
The alias selection was based on the $O-C$ analysis
of the second night.

% SI

% SI

\begin{table}
\caption{Superhump maxima of ASASSN-14ca (2015)}\label{tab:asassn14caoc2015}
\begin{center}
\begin{tabular}{rp{55pt}p{40pt}r@{.}lr}
\hline
\multicolumn{1}{c}{$E$} & \multicolumn{1}{c}{max\commenta} & \multicolumn{1}{c}{error} & \multicolumn{2}{c}{$O-C$\commentb} & \multicolumn{1}{c}{$N$\commentc} \\
\hline
0 & 57219.5689 & 0.0009 & $-$0&0000 & 50 \\
28 & 57221.5070 & 0.0017 & 0&0002 & 64 \\
29 & 57221.5758 & 0.0011 & $-$0&0002 & 59 \\
\hline
  \multicolumn{6}{l}{\commenta BJD$-$2400000.} \\
  \multicolumn{6}{l}{\commentb Against max $= 2457219.5689 + 0.069210 E$.} \\
  \multicolumn{6}{l}{\commentc Number of points used to determine the maximum.} \\
\end{tabular}
\end{center}
\end{table}

\subsection{ASASSN-14dh}\label{obj:asassn14dh}

    This object was detected as a transient at $V$=15.93
on 2014 June 27 by the ASAS-SN team.
There were past known outbursts in the CRTS data
and ASAS-3 data (see also vsnet-alert 17424, 18823).

\begin{figure}
  \begin{center}
%    \FigureFile(85mm,110mm){asassn14dhshpdm.eps}
    \FigureFile(85mm,110mm){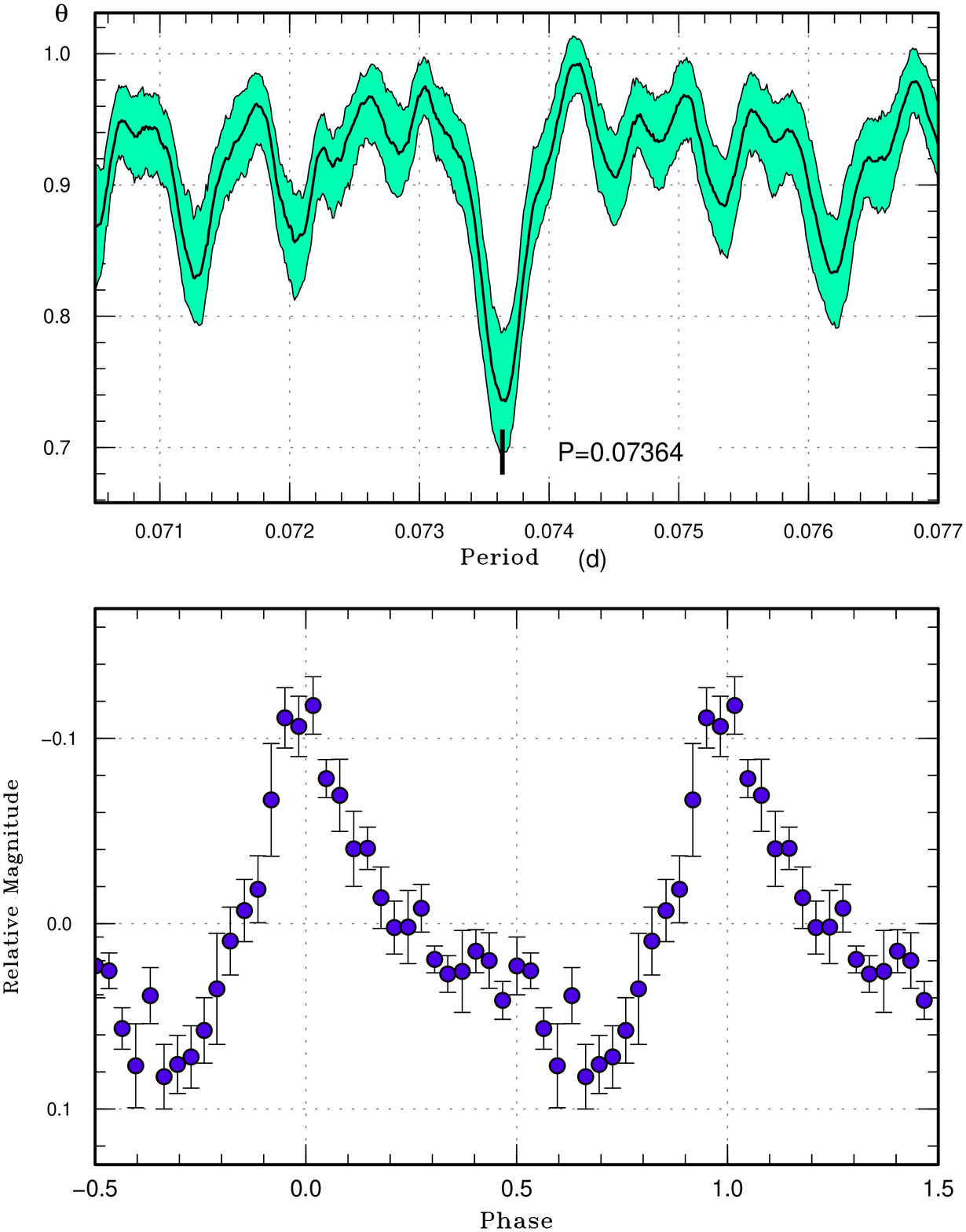}
  \end{center}
  \caption{Superhumps in ASASSN-14dh (2015).
     (Upper): PDM analysis.
     (Lower): Phase-averaged profile.}
  \label{fig:asassn14dhshpdm}
\end{figure}

   The 2015 outburst was detected by ASSN-SN on July 2
at $V$=13.3.  Subsequent observations detected superhumps
(vsnet-alert 18840, 18842; figure \ref{fig:asassn14dhshpdm}).
The times of superhump maxima are listed in
table \ref{tab:asassn14dhoc2015}.  The maxima for
$E \le$ 89 corresponds to superhumps after
the rapid fading.  This outburst was
observed only during the late phase and we probably
observed only stage C superhumps.

% SI

% SI

\begin{table}
\caption{Superhump maxima of ASASSN-14dh (2015)}\label{tab:asassn14dhoc2015}
\begin{center}
\begin{tabular}{rp{55pt}p{40pt}r@{.}lr}
\hline
\multicolumn{1}{c}{$E$} & \multicolumn{1}{c}{max\commenta} & \multicolumn{1}{c}{error} & \multicolumn{2}{c}{$O-C$\commentb} & \multicolumn{1}{c}{$N$\commentc} \\
\hline
0 & 57212.2003 & 0.0003 & 0&0016 & 153 \\
1 & 57212.2739 & 0.0003 & 0&0015 & 151 \\
28 & 57214.2575 & 0.0024 & $-$0&0029 & 88 \\
35 & 57214.7765 & 0.0009 & 0&0007 & 28 \\
36 & 57214.8489 & 0.0008 & $-$0&0005 & 30 \\
49 & 57215.8065 & 0.0006 & $-$0&0002 & 44 \\
50 & 57215.8824 & 0.0010 & 0&0021 & 24 \\
62 & 57216.7595 & 0.0017 & $-$0&0043 & 29 \\
63 & 57216.8343 & 0.0012 & $-$0&0031 & 40 \\
64 & 57216.9116 & 0.0017 & 0&0005 & 12 \\
89 & 57218.7578 & 0.0072 & 0&0060 & 31 \\
90 & 57218.8289 & 0.0026 & 0&0035 & 40 \\
91 & 57218.8942 & 0.0064 & $-$0&0048 & 20 \\
\hline
  \multicolumn{6}{l}{\commenta BJD$-$2400000.} \\
  \multicolumn{6}{l}{\commentb Against max $= 2457212.1988 + 0.073629 E$.} \\
  \multicolumn{6}{l}{\commentc Number of points used to determine the maximum.} \\
\end{tabular}
\end{center}
\end{table}

\subsection{ASASSN-14fz}\label{obj:asassn14fz}

    This object was detected as a transient at $V$=15.18
on 2014 August 20 by the ASAS-SN team.
The light curve of the 2014
outburst was suggestive of an SU UMa-type precursor
outburst (vsnet-alert 17647).

\begin{figure}
  \begin{center}
%    \FigureFile(85mm,110mm){asassn14fzshpdm.eps}
    \FigureFile(85mm,110mm){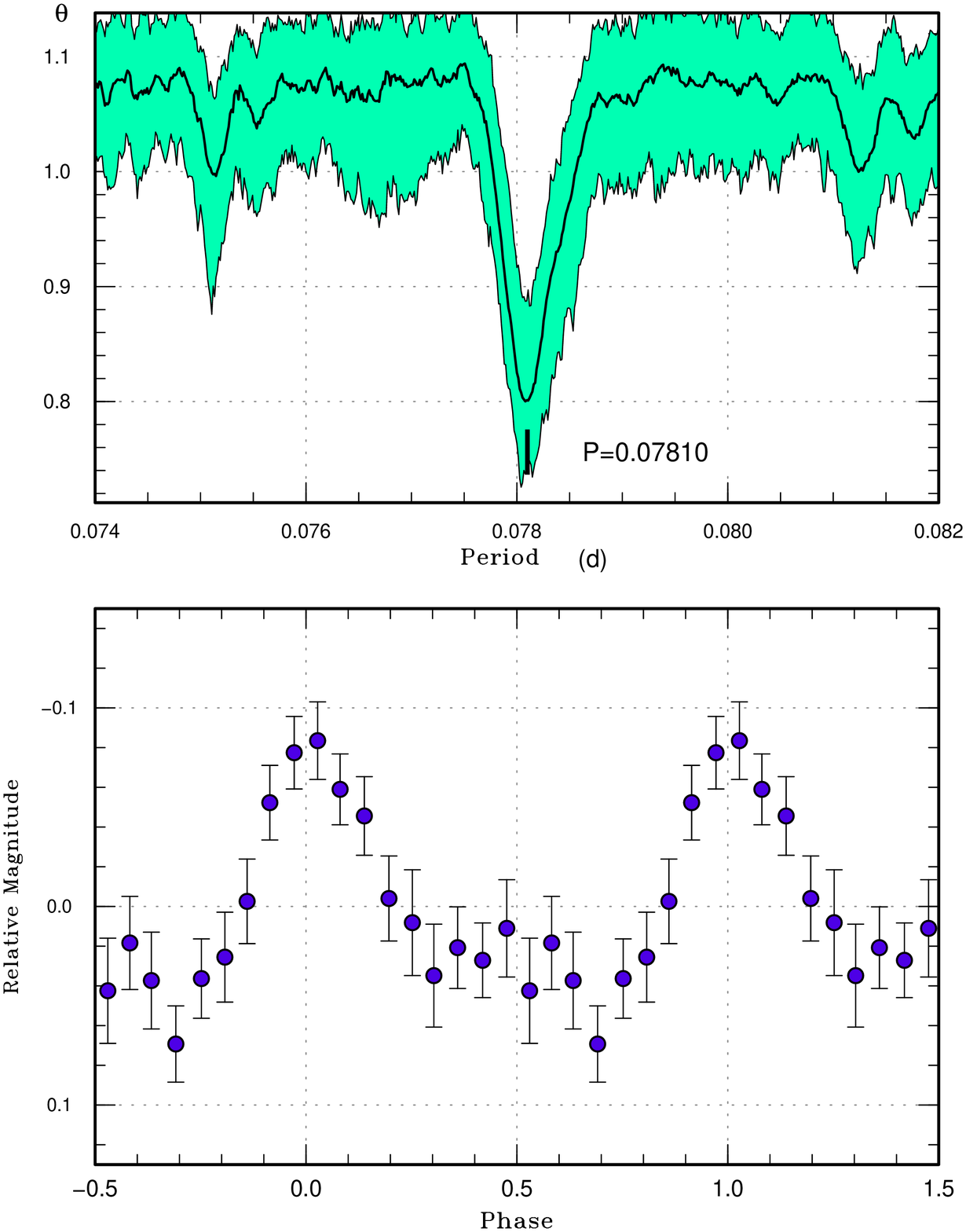}
  \end{center}
  \caption{Superhumps in ASASSN-14fz during the plateau phase
     (2015).
     (Upper): PDM analysis.
     (Lower): Phase-averaged profile.}
  \label{fig:asassn14fzshpdm}
\end{figure}

   The 2015 outburst was detected by the ASAS-SN team on May 27
at $V$=14.21.  Subsequent observations detected superhumps
(vsnet-alert 18672, 18685; figure \ref{fig:asassn14fzshpdm}).
The times of superhump maxima are listed in table
\ref{tab:asassn14fzoc2015}.  The epochs $E \le$ 1 correspond
to stage A superhumps.  It was not clear whether there
was a stage transition in the later part of the outburst,
and we listed a global $P_{\rm dot}$ in table \ref{tab:perlist}.

% SI

% SI

\begin{table}
\caption{Superhump maxima of ASASSN-14fz (2015)}\label{tab:asassn14fzoc2015}
\begin{center}
\begin{tabular}{rp{55pt}p{40pt}r@{.}lr}
\hline
\multicolumn{1}{c}{$E$} & \multicolumn{1}{c}{max\commenta} & \multicolumn{1}{c}{error} & \multicolumn{2}{c}{$O-C$\commentb} & \multicolumn{1}{c}{$N$\commentc} \\
\hline
0 & 57171.7282 & 0.0043 & $-$0&0063 & 9 \\
1 & 57171.8107 & 0.0005 & $-$0&0020 & 15 \\
13 & 57172.7518 & 0.0013 & 0&0019 & 14 \\
26 & 57173.7685 & 0.0007 & 0&0034 & 17 \\
39 & 57174.7837 & 0.0008 & 0&0033 & 17 \\
52 & 57175.7996 & 0.0023 & 0&0040 & 11 \\
64 & 57176.7362 & 0.0021 & 0&0034 & 13 \\
77 & 57177.7435 & 0.0024 & $-$0&0045 & 17 \\
90 & 57178.7621 & 0.0017 & $-$0&0012 & 18 \\
103 & 57179.7765 & 0.0016 & $-$0&0020 & 17 \\
\hline
  \multicolumn{6}{l}{\commenta BJD$-$2400000.} \\
  \multicolumn{6}{l}{\commentb Against max $= 2457171.7345 + 0.078097 E$.} \\
  \multicolumn{6}{l}{\commentc Number of points used to determine the maximum.} \\
\end{tabular}
\end{center}
\end{table}

\subsection{ASASSN-14le}\label{obj:asassn14le}

\begin{figure}
  \begin{center}
%    \FigureFile(85mm,70mm){asassn14leshlc.eps}
    \FigureFile(85mm,70mm){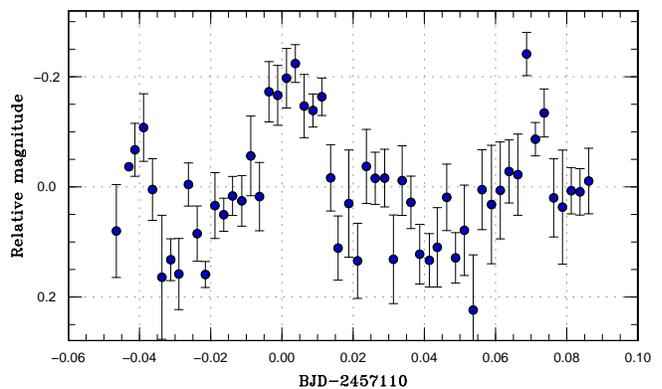}
  \end{center}
  \caption{Superhumps in ASASSN-14le (2014).
  The data were binned to 0.0025~d.
  }
  \label{fig:asassn14leshlc}
\end{figure}

   This object was detected as a transient at $V$=16.4
on 2014 November 29 by the ASAS-SN team.
Observations on December 5 detected superhumps
(figure \ref{fig:asassn14leshlc}).
The times of maxima are BJD 2456997.0031(12) ($N$=135)
and 2456997.0719(20) ($N$=113).  The superhump period
was estimated to be 0.068(1)~d using these times of
maxima and with the PDM method.

\subsection{ASASSN-15cl}\label{obj:asassn15cl}

   This object was detected as a transient at $V$=13.9
on 2015 February 1 by the ASAS-SN team.
The object was found to be already in outburst
at $V$=14.8 on January 31.
There were three past outbursts
reaching $V$=13.3 in the ASAS-3 data (cf. vsnet-alert 18256).
Although the 2015 outburst was observed on two nights
by S. Kiyota, no definite superhump signal was
detected.

\begin{figure}
  \begin{center}
%    \FigureFile(85mm,70mm){asassn15clshlc.eps}
    \FigureFile(85mm,70mm){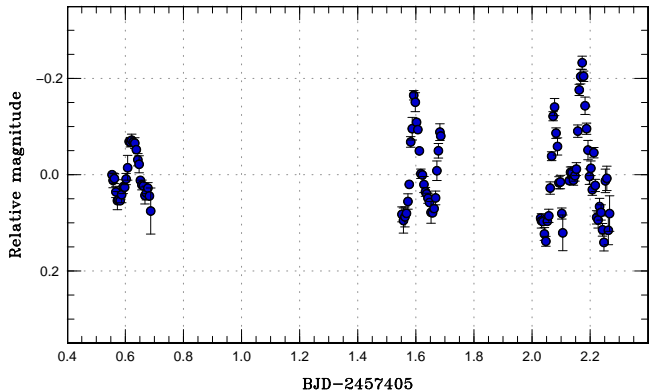}
  \end{center}
  \caption{Growing superhumps in ASASSN-15cl (2016).
  The data were binned to 0.005~d.
  }
  \label{fig:asassn15clshlc}
\end{figure}

\begin{figure}
  \begin{center}
%    \FigureFile(85mm,110mm){asassn15clshpdm.eps}
    \FigureFile(85mm,110mm){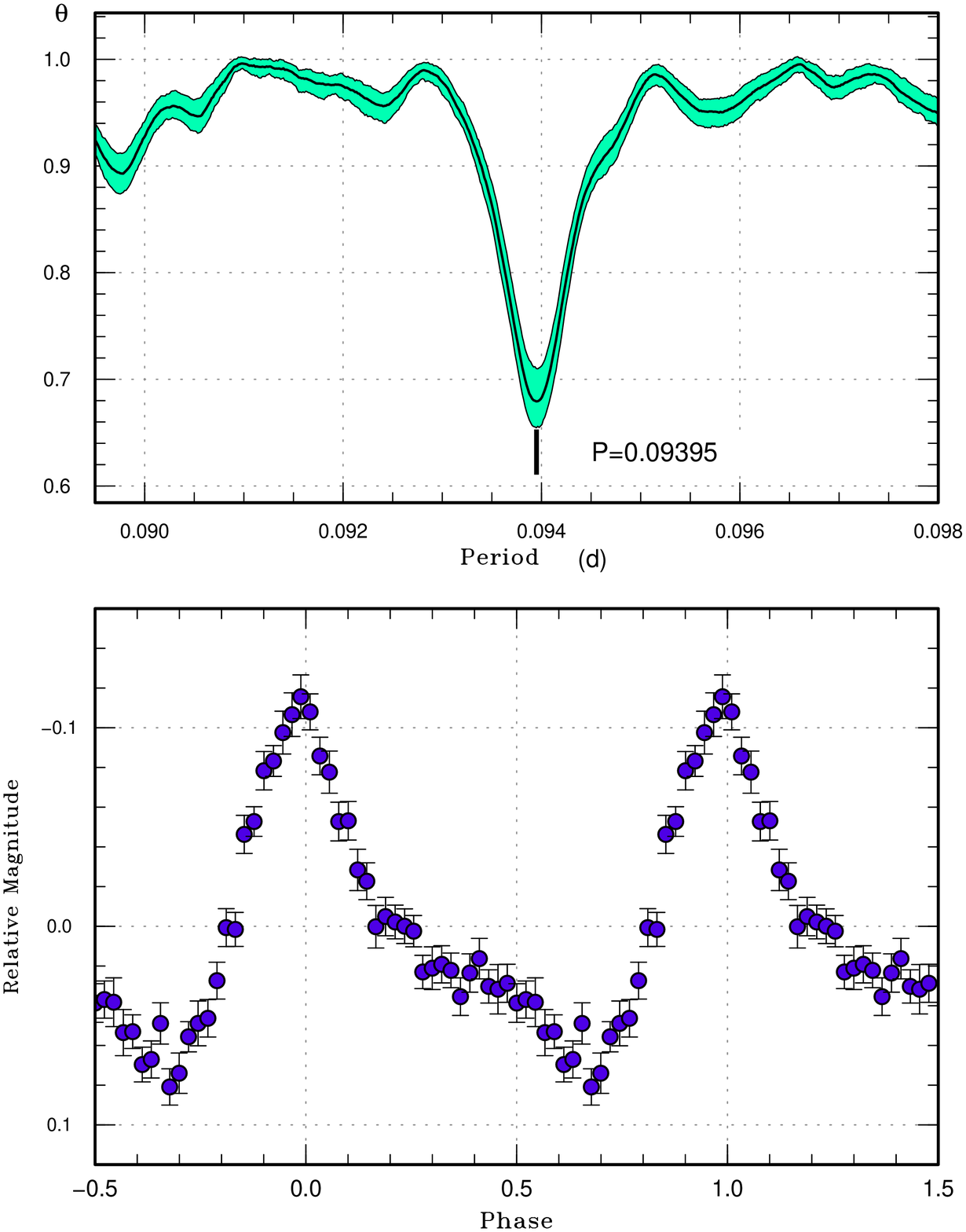}
  \end{center}
  \caption{Superhumps in ASASSN-15cl (2016).
     The data after BJD 2457407.5 (excluding stage A)
     were used.
     (Upper): PDM analysis.
     (Lower): Phase-averaged profile.}
  \label{fig:asassn15clshpdm}
\end{figure}

\begin{figure}
  \begin{center}
%    \FigureFile(70mm,88mm){asassn15clhumpall.eps}
    \FigureFile(70mm,88mm){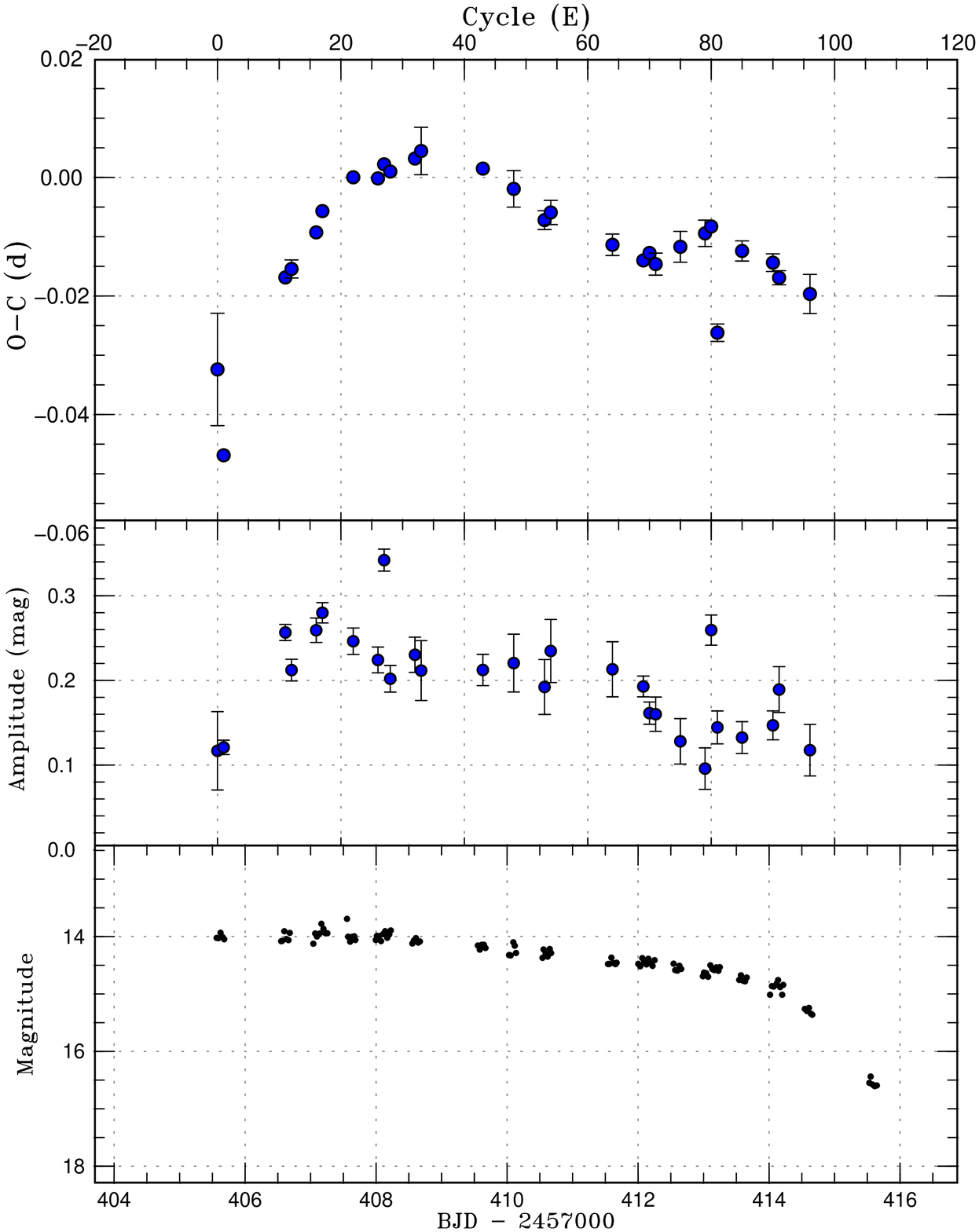}
  \end{center}
  \caption{$O-C$ diagram of superhumps in ASASSN-15cl (2016).
     (Upper:) $O-C$ diagram.
     We used a period of 0.09423~d for calculating the $O-C$ residuals.
     The superhump maxima up to $E$=22 are stage A superhumps, maxima
     between $E$=22 and $E$=33 are identified as stage B superhumps.
     After this, the period decreased to a constant one
     (stage C superhumps).
     (Middle:) Amplitudes of superhumps.  The amplitudes were
     small around $E=0$.
     (Lower:) Light curve.  The data were binned to 0.031~d.
  }
  \label{fig:asassn15clhumpall}
\end{figure}

   The 2016 outburst was detected by the ASAS-SN team
at $V$=13.8 on January 17 (vsnet-alert 19421).
Subsequent observations detected long-period
superhumps (vsnet-alert 19427, 19431).
The superhumps were still growing
(see also figure \ref{fig:asassn15clshlc})
and the period then dramatically shortened
(vsnet-alert 19431, 19436, 19442).
The times of superhump maxima are listed in
table \ref{tab:asassn15cloc2016}.  The behavior
was very similar to another long-$P_{\rm orb}$ system
V1006 Cyg \citep{kat16v1006cyg} and we identified
$E \le$22 to be stage A superhumps
(figure \ref{fig:asassn15clhumpall}).
The sharp decrease in the period was likely
due to stage B to C transition around $E$=33.
The periods given in table \ref{tab:perlist}
refers to these period identifications.
The mean profile excluding stage A superhumps
is shown in figure \ref{fig:asassn15clshpdm}.

   \citet{kat16v1006cyg} discussed that the long
duration of the growing stage of superhumps
in a long-$P_{\rm orb}$ system reflects the slow
growth rate of the 3:1 resonance when the
mass ratio is close to the stability limit
of the 3:1 resonance (see also subsection
\ref{sec:longstagea}).
Yet another example ASASSN-15cl also supports that
this mechanism is likely working
in many long-$P_{\rm orb}$ systems.
Future determination of the orbital period in
this system will allow the measurement of $q$
using the period of stage A superhumps in this study.
Such a measurement will test the hypothesis whether
the system has a borderline $q$.

% SI

% SI

\begin{table}
\caption{Superhump maxima of ASASSN-15cl (2016)}\label{tab:asassn15cloc2016}
\begin{center}
\begin{tabular}{rp{55pt}p{40pt}r@{.}lr}
\hline
\multicolumn{1}{c}{$E$} & \multicolumn{1}{c}{max\commenta} & \multicolumn{1}{c}{error} & \multicolumn{2}{c}{$O-C$\commentb} & \multicolumn{1}{c}{$N$\commentc} \\
\hline
0 & 57405.5459 & 0.0095 & $-$0&0218 & 21 \\
1 & 57405.6256 & 0.0008 & $-$0&0363 & 44 \\
11 & 57406.5979 & 0.0004 & $-$0&0063 & 47 \\
12 & 57406.6936 & 0.0015 & $-$0&0048 & 19 \\
16 & 57407.0767 & 0.0006 & 0&0014 & 150 \\
17 & 57407.1745 & 0.0005 & 0&0050 & 176 \\
22 & 57407.6513 & 0.0007 & 0&0107 & 30 \\
26 & 57408.0281 & 0.0008 & 0&0105 & 167 \\
27 & 57408.1247 & 0.0005 & 0&0129 & 164 \\
28 & 57408.2177 & 0.0010 & 0&0117 & 117 \\
32 & 57408.5968 & 0.0010 & 0&0139 & 28 \\
33 & 57408.6923 & 0.0040 & 0&0152 & 11 \\
43 & 57409.6316 & 0.0010 & 0&0122 & 32 \\
48 & 57410.0993 & 0.0031 & 0&0088 & 106 \\
53 & 57410.5652 & 0.0016 & 0&0036 & 27 \\
54 & 57410.6608 & 0.0020 & 0&0049 & 23 \\
64 & 57411.5976 & 0.0018 & $-$0&0006 & 32 \\
69 & 57412.0661 & 0.0007 & $-$0&0032 & 220 \\
70 & 57412.1616 & 0.0010 & $-$0&0019 & 161 \\
71 & 57412.2539 & 0.0019 & $-$0&0038 & 73 \\
75 & 57412.6338 & 0.0026 & $-$0&0009 & 33 \\
79 & 57413.0130 & 0.0022 & 0&0014 & 130 \\
80 & 57413.1084 & 0.0009 & 0&0026 & 156 \\
81 & 57413.1847 & 0.0015 & $-$0&0153 & 173 \\
85 & 57413.5754 & 0.0017 & $-$0&0015 & 30 \\
90 & 57414.0446 & 0.0015 & $-$0&0035 & 175 \\
91 & 57414.1363 & 0.0012 & $-$0&0060 & 132 \\
96 & 57414.6047 & 0.0033 & $-$0&0087 & 31 \\
\hline
  \multicolumn{6}{l}{\commenta BJD$-$2400000.} \\
  \multicolumn{6}{l}{\commentb Against max $= 2457405.5677 + 0.094226 E$.} \\
  \multicolumn{6}{l}{\commentc Number of points used to determine the maximum.} \\
\end{tabular}
\end{center}
\end{table}

\subsection{ASASSN-15cy}\label{obj:asassn15cy}

   This object was detected as a transient at $V$=14.6
on 2015 February 16 by the ASAS-SN team.
The coordinates are \timeform{08h 11m 50.53s},
\timeform{-12D 27' 51.5''} (based on S. Kiyota's image
on February 20, UCAC4 reference stars).
There is no quiescent counterpart in available
catalogs.

\begin{figure}
  \begin{center}
%    \FigureFile(85mm,110mm){asassn15cyshpdm.eps}
    \FigureFile(85mm,110mm){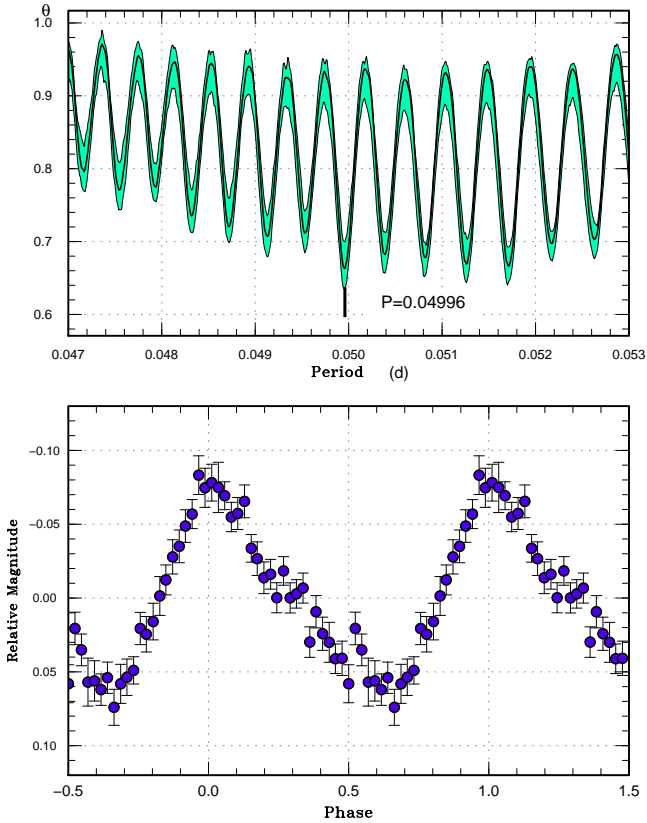}
  \end{center}
  \caption{Superhumps in ASASSN-15cy between
     BJD 2457076 and 2457083 (2015).
     (Upper): PDM analysis.  The alias selection was based
     on the continuous data on BJD 2457076 and the selection
     of the cycle counts to minimize the scatter in
     the $O-C$ diagram.
     (Lower): Phase-averaged profile.}
  \label{fig:asassn15cyshpdm}
\end{figure}

   Up to February 20, the object showed only little
variations.  On February 22 (6~d after the initial
detection), it showed prominent superhumps with
a very short ($\sim$0.050~d) period (vsnet-alert
18326, 18327).  There were long gaps in the observation
and the next continuous run was on February 28.
There was only one superhump detection on February 26.
Using the period determined from the continuous run
on February 22 and the single epoch on February 26,
we have tried several possible periods and obtained
the smallest $O-C$ scatter using a period of
0.04996~d up to February 28.  The cycle counts and
$O-C$ values assuming this period is shown in
table \ref{tab:asassn15cyoc2015}.
The mean superhump profile is shown in figure
\ref{fig:asassn15cyshpdm}.  We should note,
however, this choice may not be correct, particularly
if there was strong period variation when there were
gaps in the observation.
The times for $E \ge 161$ were uncertain since the amplitudes
of superhumps were already small ($\sim$0.07 mag)
and the object already faded close to 16 mag.
The phase jump between $E=122$ and $E=161$ was too
large to be identified as stage B-C transition.
Although there may have been a true phase jump,
the conclusion is unclear due to the poor quality
of the data around these epochs.

   The resultant superhump period suggests that
this object belong to EI Psc-type objects
with compact secondaries having an evolved core
(cf. \Ohtprep).  Since the quiescent brightness
is below the limit of photographic surveys,
the outburst amplitude is likely larger than 6 mag.
This object shares properties with 
CSS J174033.5$+$414756
($P_{\rm orb}$=0.04505~d, outburst amplitude
$\sim$6.7 mag; \cite{Pdot5}; \cite{Pdot7}; \Ohtprep;
\cite{pri13j1740asassn13adatel4999};
\cite{nes13j1740ibvs6059}), which is classified as
a WZ Sge-type object in \citet{kat15wzsge}.
Although a period of 0.0494(2)~d was inferred
from the observations on the first two nights
(cf. vsnet-alert 18327), this period is uncertain
due to the low amplitude (less than 0.02 mag)
and limited coverage.

% SI

% SI

\begin{table}
\caption{Superhump maxima of ASASSN-15cy (2015)}\label{tab:asassn15cyoc2015}
\begin{center}
\begin{tabular}{rp{55pt}p{40pt}r@{.}lr}
\hline
\multicolumn{1}{c}{$E$} & \multicolumn{1}{c}{max\commenta} & \multicolumn{1}{c}{error} & \multicolumn{2}{c}{$O-C$\commentb} & \multicolumn{1}{c}{$N$\commentc} \\
\hline
0 & 57076.3205 & 0.0019 & $-$0&0024 & 61 \\
1 & 57076.3693 & 0.0004 & $-$0&0034 & 115 \\
2 & 57076.4190 & 0.0004 & $-$0&0035 & 114 \\
3 & 57076.4685 & 0.0004 & $-$0&0038 & 115 \\
4 & 57076.5188 & 0.0004 & $-$0&0033 & 115 \\
5 & 57076.5670 & 0.0018 & $-$0&0048 & 55 \\
66 & 57079.6148 & 0.0007 & 0&0063 & 13 \\
86 & 57080.6117 & 0.0016 & 0&0076 & 12 \\
120 & 57082.3159 & 0.0018 & 0&0193 & 115 \\
121 & 57082.3637 & 0.0009 & 0&0172 & 115 \\
122 & 57082.4132 & 0.0013 & 0&0170 & 114 \\
161 & 57084.3262 & 0.0019 & $-$0&0115 & 115 \\
162 & 57084.3724 & 0.0024 & $-$0&0151 & 115 \\
163 & 57084.4177 & 0.0025 & $-$0&0195 & 113 \\
\hline
  \multicolumn{6}{l}{\commenta BJD$-$2400000.} \\
  \multicolumn{6}{l}{\commentb Against max $= 2457076.3230 + 0.049781 E$.} \\
  \multicolumn{6}{l}{\commentc Number of points used to determine the maximum.} \\
\end{tabular}
\end{center}
\end{table}

\subsection{ASASSN-15dh}\label{obj:asassn15dh}

\begin{figure}
  \begin{center}
%    \FigureFile(85mm,110mm){asassn15dhshpdm.eps}
    \FigureFile(85mm,110mm){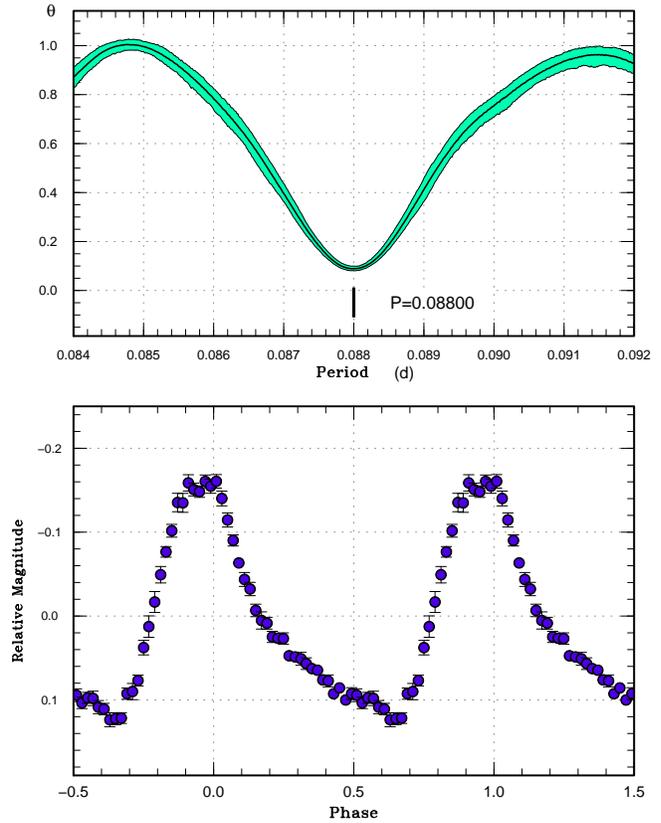}
  \end{center}
  \caption{Superhumps in ASASSN-15dh (2015).
     (Upper): PDM analysis.
     (Lower): Phase-averaged profile.}
  \label{fig:asassn15dhshpdm}
\end{figure}

   This object was detected as a transient at $V$=15.2
on 2015 February 12 by the ASAS-SN team.
There was another outburst on October 29
at $V$=15.03 detected by the ASAS-SN team
(see also vsnet-alert 19228).
Subsequent observations detected superhumps
(vsnet-alert 19218, 19228;
figure \ref{fig:asassn15dhshpdm}).
The times of superhump maxima are listed in
table \ref{tab:asassn15dhoc2015}.
There was probably a stage transition at around
$E$=11.  We were not able to determine the 
type of transition.  In table \ref{tab:perlist},
we listed a global value.

% SI

% SI

\begin{table}
\caption{Superhump maxima of ASASSN-15dh (2015)}\label{tab:asassn15dhoc2015}
\begin{center}
\begin{tabular}{rp{55pt}p{40pt}r@{.}lr}
\hline
\multicolumn{1}{c}{$E$} & \multicolumn{1}{c}{max\commenta} & \multicolumn{1}{c}{error} & \multicolumn{2}{c}{$O-C$\commentb} & \multicolumn{1}{c}{$N$\commentc} \\
\hline
0 & 57328.2626 & 0.0003 & $-$0&0014 & 78 \\
1 & 57328.3508 & 0.0003 & $-$0&0012 & 77 \\
2 & 57328.4407 & 0.0003 & 0&0007 & 79 \\
3 & 57328.5281 & 0.0004 & 0&0001 & 81 \\
11 & 57329.2343 & 0.0007 & 0&0022 & 52 \\
12 & 57329.3226 & 0.0003 & 0&0024 & 81 \\
18 & 57329.8467 & 0.0003 & $-$0&0015 & 74 \\
19 & 57329.9348 & 0.0004 & $-$0&0014 & 88 \\
\hline
  \multicolumn{6}{l}{\commenta BJD$-$2400000.} \\
  \multicolumn{6}{l}{\commentb Against max $= 2457328.2639 + 0.088014 E$.} \\
  \multicolumn{6}{l}{\commentc Number of points used to determine the maximum.} \\
\end{tabular}
\end{center}
\end{table}

\subsection{ASASSN-15dp}\label{obj:asassn15dp}

   This object was detected as a transient at $V$=14.1
on 2015 February 22 by the ASAS-SN team.
The object was recorded in outburst in 1989
at a red magnitude of 15.2 in GSC 2.3 (GSC2.3NCCX024953).
The quiescent magnitude was 19.4(4) (green magnitude,
Initial Gaia Source List).

\begin{figure}
  \begin{center}
%    \FigureFile(85mm,110mm){asassn15dpshpdm.eps}
    \FigureFile(85mm,110mm){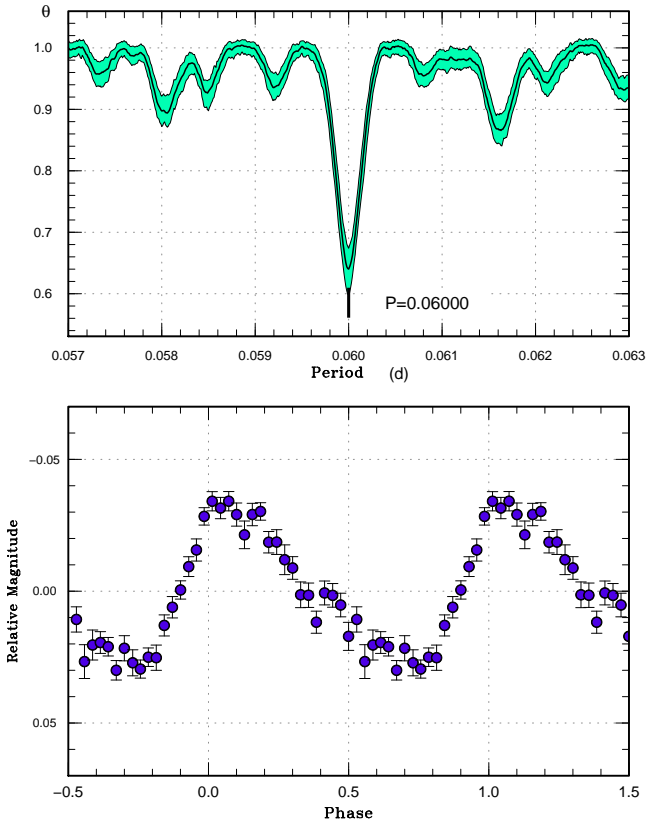}
  \end{center}
  \caption{Superhumps in ASASSN-15dp after BJD 2457084 (2015).
     (Upper): PDM analysis.
     (Lower): Phase-averaged profile.}
  \label{fig:asassn15dpshpdm}
\end{figure}

   Early observations soon detected superhumps
(vsnet-alert 18350).  It soon became apparent that early
observations recorded the final part of the precursor
outburst with a relatively rapid fading, and 
stage A superhumps were observed on
the first two nights (vsnet-alert 18363, 18417).
Figure \ref{fig:asassn15dpshpdm}
shows the profile of stage B superhumps.
The times of superhump maxima are listed in table
\ref{tab:asassn15dpoc2015}.  The maxima for $E \le 24$
correspond to stage A superhumps.  There was no
observed transition to stage C superhumps.

   The object entered the rapid fading stage on March 11,
17~d after the initial outburst detection.
It took $\sim$8~d to develop stage B superhumps,
which is relatively long.  It was also somewhat
unusual that the fading part of the precursor outburst
was observed even 5~d after the initial detection.
The lack of stage C superhumps and small $P_{\rm dot}$
are usual characteristics of WZ Sge-type dwarf novae
\citep{kat15wzsge}.  Since the system has a relatively
long superhump period, there could even be a chance of
a candidate period bouncer, if our measurement of $P_{\rm dot}$
is correct.  Since the present observations lacks
the earliest data and post-outburst data,
this object should require further observation
on the next outburst occasion.

% SI

% SI

\begin{table}
\caption{Superhump maxima of ASASSN-15dp (2015)}\label{tab:asassn15dpoc2015}
\begin{center}
\begin{tabular}{rp{55pt}p{40pt}r@{.}lr}
\hline
\multicolumn{1}{c}{$E$} & \multicolumn{1}{c}{max\commenta} & \multicolumn{1}{c}{error} & \multicolumn{2}{c}{$O-C$\commentb} & \multicolumn{1}{c}{$N$\commentc} \\
\hline
0 & 57081.3059 & 0.0013 & $-$0&0242 & 61 \\
1 & 57081.3670 & 0.0019 & $-$0&0233 & 61 \\
2 & 57081.4320 & 0.0013 & $-$0&0185 & 61 \\
23 & 57082.7238 & 0.0014 & 0&0093 & 61 \\
24 & 57082.7780 & 0.0029 & 0&0034 & 30 \\
49 & 57084.2959 & 0.0003 & 0&0165 & 54 \\
50 & 57084.3580 & 0.0004 & 0&0185 & 44 \\
66 & 57085.3171 & 0.0009 & 0&0145 & 64 \\
67 & 57085.3733 & 0.0011 & 0&0106 & 67 \\
84 & 57086.3971 & 0.0012 & 0&0112 & 51 \\
115 & 57088.2596 & 0.0008 & 0&0078 & 91 \\
116 & 57088.3171 & 0.0006 & 0&0052 & 186 \\
117 & 57088.3781 & 0.0007 & 0&0059 & 170 \\
132 & 57089.2793 & 0.0008 & 0&0043 & 62 \\
133 & 57089.3371 & 0.0016 & 0&0020 & 59 \\
134 & 57089.3971 & 0.0017 & 0&0018 & 61 \\
150 & 57090.3488 & 0.0009 & $-$0&0096 & 42 \\
151 & 57090.4124 & 0.0009 & $-$0&0061 & 58 \\
165 & 57091.2587 & 0.0012 & $-$0&0025 & 61 \\
166 & 57091.3161 & 0.0010 & $-$0&0053 & 62 \\
167 & 57091.3789 & 0.0013 & $-$0&0027 & 62 \\
199 & 57093.2971 & 0.0018 & $-$0&0105 & 61 \\
200 & 57093.3595 & 0.0031 & $-$0&0083 & 61 \\
\hline
  \multicolumn{6}{l}{\commenta BJD$-$2400000.} \\
  \multicolumn{6}{l}{\commentb Against max $= 2457081.3276 + 0.060201 E$.} \\
  \multicolumn{6}{l}{\commentc Number of points used to determine the maximum.} \\
\end{tabular}
\end{center}
\end{table}

\subsection{ASASSN-15dr}\label{obj:asassn15dr}

\begin{figure}
  \begin{center}
%    \FigureFile(85mm,110mm){asassn15drshpdm.eps}
    \FigureFile(85mm,110mm){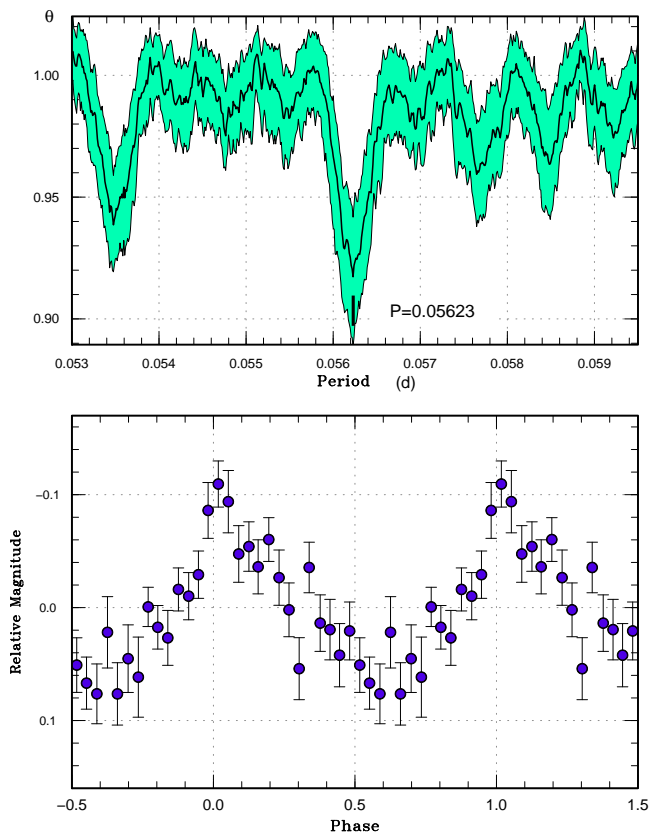}
  \end{center}
  \caption{Superhumps in ASASSN-15dr between
     BJD 2457082 and 2457087 (2015).
     (Upper): PDM analysis.
     (Lower): Phase-averaged profile.}
  \label{fig:asassn15drshpdm}
\end{figure}

   This object was detected as a transient at $V$=14.9
on 2015 February 22 by the ASAS-SN team.
On February 28, the object started to show growing
superhumps (vsnet-alert 18366).  Stable superhumps
were observed later (vsnet-alert 18384;
figure \ref{fig:asassn15drshpdm}).
The times of superhump maxima are listed in table
\ref{tab:asassn15droc2015}.  Although the times
for $E \le 2$ apparently correspond to stage A
superhumps, the period of stage A superhump could not
be determined due to the lack of the data.
The photometric quality for this object was not
good enough due to its faintness (16.3 mag on
February 26, which was much fainter than the ASAS-SN
report).

% SI

% SI

\begin{table}
\caption{Superhump maxima of ASASSN-15dr (2015)}\label{tab:asassn15droc2015}
\begin{center}
\begin{tabular}{rp{55pt}p{40pt}r@{.}lr}
\hline
\multicolumn{1}{c}{$E$} & \multicolumn{1}{c}{max\commenta} & \multicolumn{1}{c}{error} & \multicolumn{2}{c}{$O-C$\commentb} & \multicolumn{1}{c}{$N$\commentc} \\
\hline
0 & 57082.2958 & 0.0021 & $-$0&0146 & 129 \\
1 & 57082.3476 & 0.0030 & $-$0&0196 & 130 \\
2 & 57082.4085 & 0.0038 & $-$0&0155 & 130 \\
24 & 57083.6840 & 0.0020 & 0&0107 & 15 \\
25 & 57083.7430 & 0.0015 & 0&0129 & 17 \\
26 & 57083.8002 & 0.0012 & 0&0133 & 22 \\
27 & 57083.8568 & 0.0011 & 0&0132 & 22 \\
40 & 57084.5919 & 0.0053 & 0&0100 & 12 \\
43 & 57084.7647 & 0.0022 & 0&0125 & 20 \\
44 & 57084.8134 & 0.0023 & 0&0044 & 21 \\
45 & 57084.8727 & 0.0016 & 0&0069 & 16 \\
59 & 57085.6680 & 0.0065 & 0&0072 & 13 \\
60 & 57085.7189 & 0.0012 & 0&0014 & 15 \\
61 & 57085.7724 & 0.0017 & $-$0&0020 & 17 \\
62 & 57085.8234 & 0.0104 & $-$0&0078 & 15 \\
76 & 57086.6181 & 0.0042 & $-$0&0080 & 14 \\
77 & 57086.6746 & 0.0033 & $-$0&0084 & 19 \\
78 & 57086.7369 & 0.0037 & $-$0&0028 & 25 \\
79 & 57086.7883 & 0.0086 & $-$0&0083 & 29 \\
80 & 57086.8477 & 0.0063 & $-$0&0057 & 28 \\
\hline
  \multicolumn{6}{l}{\commenta BJD$-$2400000.} \\
  \multicolumn{6}{l}{\commentb Against max $= 2457082.3104 + 0.056786 E$.} \\
  \multicolumn{6}{l}{\commentc Number of points used to determine the maximum.} \\
\end{tabular}
\end{center}
\end{table}

\subsection{ASASSN-15ea}\label{obj:asassn15ea}

   This object was detected as a transient at $V$=16.1
on 2015 February 25 by the ASAS-SN team.
There was one previous outburst reaching 14.15 mag
on 2006 October 4 in the CRTS data.
Although T. Vanmunster reported the detection of superhumps
(vsnet-alert 18357), this period was later rejected
(vsnet-alert 18373).

\begin{figure}
  \begin{center}
%    \FigureFile(85mm,110mm){asassn15eashpdm.eps}
    \FigureFile(85mm,110mm){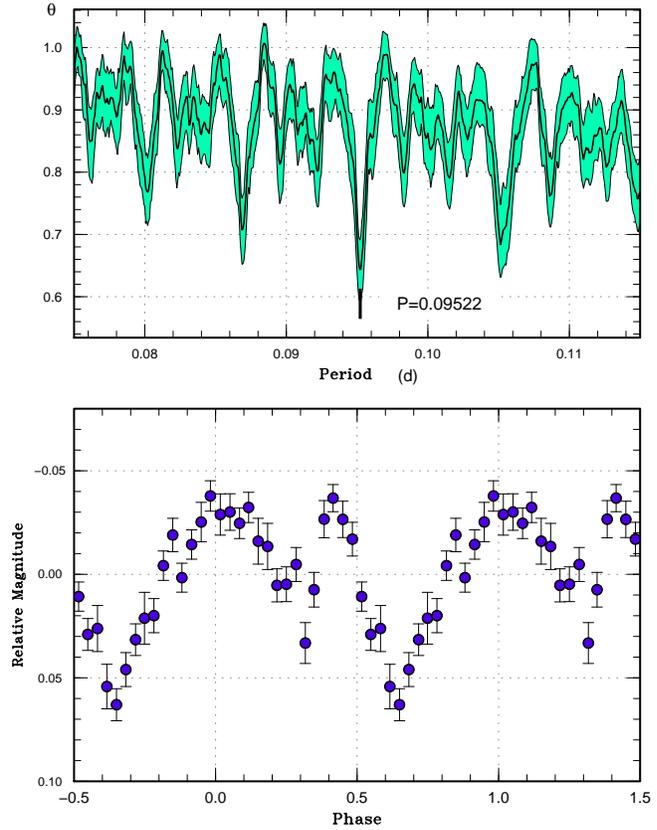}
  \end{center}
  \caption{Superhumps in ASASSN-15ea (2015).
     (Upper): PDM analysis.
     (Lower): Phase-averaged profile.}
  \label{fig:asassn15eashpdm}
\end{figure}

   Although individual superhumps maxima were
difficult to determine due to the poor coverage
(we do not give a table of superhump maxima),
a PDM analysis yielded a strong signal of
superhumps (figure \ref{fig:asassn15eashpdm}).
Although the best period is 0.09522(8)~d, one-day
aliases are still possible.  If this period is
confirmed, the object is located on the lower edge
of the period gap.

% SI

\subsection{ASASSN-15ee}\label{obj:asassn15ee}

\begin{figure}
  \begin{center}
%    \FigureFile(85mm,110mm){asassn15eeshpdm.eps}
    \FigureFile(85mm,110mm){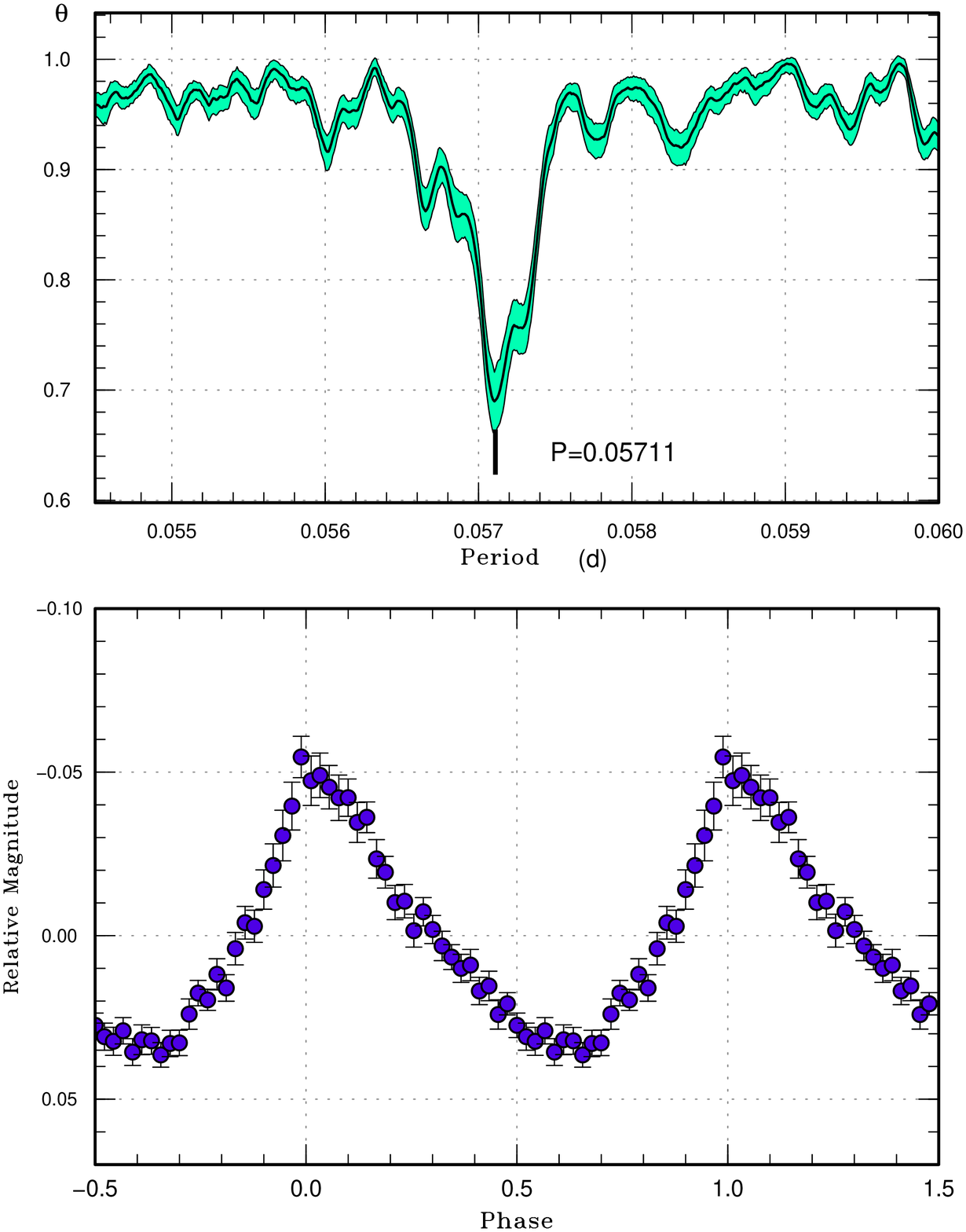}
  \end{center}
  \caption{Superhumps in ASASSN-15ee (2015).
     (Upper): PDM analysis.
     (Lower): Phase-averaged profile.}
  \label{fig:asassn15eeshpdm}
\end{figure}

\begin{figure}
  \begin{center}
%    \FigureFile(85mm,100mm){asassn15eehumpall.eps}
    \FigureFile(85mm,100mm){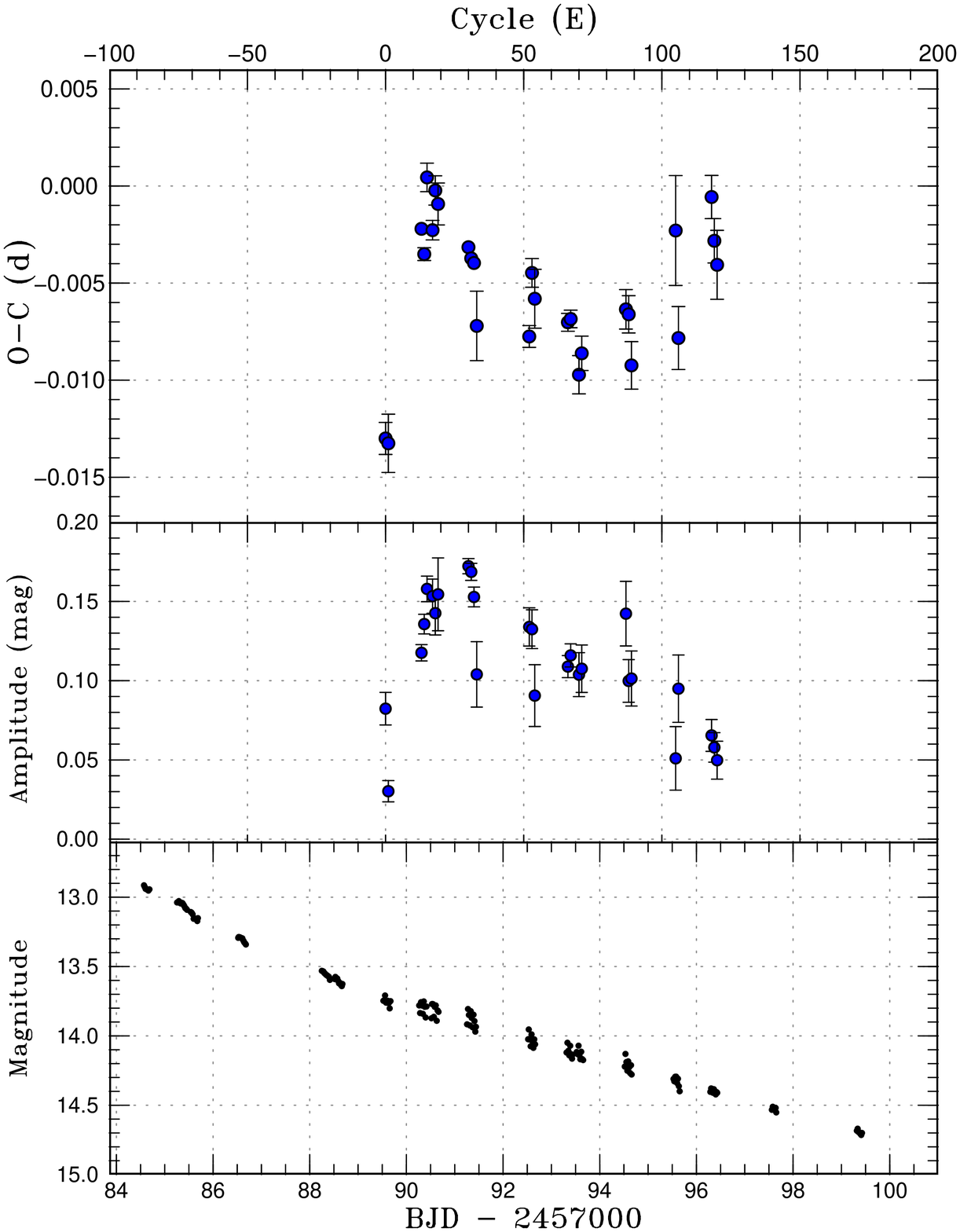}
  \end{center}
  \caption{$O-C$ diagram of superhumps in ASASSN-15ee (2015).
     (Upper:) $O-C$ diagram.
     We used a period of 0.05716~d for calculating the $O-C$ residuals.
     (Middle:) Amplitudes of superhumps.
     (Lower:) Light curve.  The data were binned to 0.019~d.
  }
  \label{fig:asassn15eehumpall}
\end{figure}

   This object was detected as a transient at $V$=12.6
on 2015 March 2 by the ASAS-SN team.
The outburst amplitude exceeded 7 mag, and was
considered to be a good candidate for a WZ Sge-type
dwarf nova (vsnet-alert 18364).
The object initially faded rather rapidly without strong
modulations (vsnet-alert 18381).
Ordinary superhump started to appear on March 8,
6~d after the initial detection and reached
a peak amplitude of 0.17 mag within 1~d.
The superhumps started to decay slowly
(vsnet-alert 18392, 18394, 18415, 18423, 18428, 18437;
figure \ref{fig:asassn15eeshpdm})
The times of superhump maxima are listed in table
\ref{tab:asassn15eeoc2015}.  The epochs $E \le 14$
likely correspond to stage A superhumps since
the amplitudes grew up to $E=14$
(figure \ref{fig:asassn15eehumpall}).
The epoch for $E \ge$200 probably correspond to stage C
superhumps.
There was no strong indication of early superhumps
before the development of ordinary superhumps.

   Although the outburst amplitude is large,
the object is probably not an extreme WZ Sge-type
dwarf nova since the growth of superhumps is quick
and the $P_{\rm dot}$ for stage B superhumps is
large [$+8.1(1.2) \times 10^{-5}$].
The inclination of this object is probably low,
as suggested from the lack of early superhumps,
which would have made the outburst amplitude larger.

% SI

% SI

\begin{table}
\caption{Superhump maxima of ASASSN-15ee (2015)}\label{tab:asassn15eeoc2015}
\begin{center}
\begin{tabular}{rp{55pt}p{40pt}r@{.}lr}
\hline
\multicolumn{1}{c}{$E$} & \multicolumn{1}{c}{max\commenta} & \multicolumn{1}{c}{error} & \multicolumn{2}{c}{$O-C$\commentb} & \multicolumn{1}{c}{$N$\commentc} \\
\hline
0 & 57089.5560 & 0.0008 & $-$0&0115 & 12 \\
1 & 57089.6130 & 0.0015 & $-$0&0116 & 13 \\
13 & 57090.3099 & 0.0003 & 0&0002 & 132 \\
14 & 57090.3657 & 0.0003 & $-$0&0010 & 131 \\
15 & 57090.4268 & 0.0007 & 0&0030 & 62 \\
17 & 57090.5384 & 0.0005 & 0&0004 & 12 \\
18 & 57090.5976 & 0.0008 & 0&0025 & 12 \\
19 & 57090.6541 & 0.0011 & 0&0019 & 13 \\
30 & 57091.2809 & 0.0002 & 0&0007 & 132 \\
31 & 57091.3374 & 0.0002 & 0&0002 & 132 \\
32 & 57091.3940 & 0.0003 & $-$0&0004 & 132 \\
33 & 57091.4482 & 0.0019 & $-$0&0033 & 30 \\
52 & 57092.5336 & 0.0006 & $-$0&0026 & 17 \\
53 & 57092.5940 & 0.0007 & 0&0008 & 18 \\
54 & 57092.6498 & 0.0015 & $-$0&0005 & 18 \\
66 & 57093.3340 & 0.0005 & $-$0&0013 & 132 \\
67 & 57093.3914 & 0.0005 & $-$0&0010 & 131 \\
70 & 57093.5607 & 0.0009 & $-$0&0030 & 15 \\
71 & 57093.6188 & 0.0009 & $-$0&0020 & 17 \\
87 & 57094.5356 & 0.0010 & 0&0013 & 19 \\
88 & 57094.5925 & 0.0010 & 0&0011 & 17 \\
89 & 57094.6470 & 0.0012 & $-$0&0014 & 17 \\
105 & 57095.5685 & 0.0028 & 0&0067 & 15 \\
106 & 57095.6201 & 0.0016 & 0&0012 & 16 \\
118 & 57096.3134 & 0.0011 & 0&0094 & 132 \\
119 & 57096.3683 & 0.0011 & 0&0072 & 131 \\
120 & 57096.4242 & 0.0018 & 0&0060 & 88 \\
130 & 57097.0003 & 0.0014 & 0&0112 & 48 \\
131 & 57097.0554 & 0.0009 & 0&0093 & 65 \\
200 & 57100.9730 & 0.0021 & $-$0&0122 & 30 \\
287 & 57105.9418 & 0.0018 & $-$0&0102 & 65 \\
288 & 57106.0081 & 0.0017 & $-$0&0010 & 31 \\
\hline
  \multicolumn{6}{l}{\commenta BJD$-$2400000.} \\
  \multicolumn{6}{l}{\commentb Against max $= 2457089.5675 + 0.057089 E$.} \\
  \multicolumn{6}{l}{\commentc Number of points used to determine the maximum.} \\
\end{tabular}
\end{center}
\end{table}

\subsection{ASASSN-15eh}\label{obj:asassn15eh}

\begin{figure}
  \begin{center}
%    \FigureFile(85mm,110mm){asassn15ehshpdm.eps}
    \FigureFile(85mm,110mm){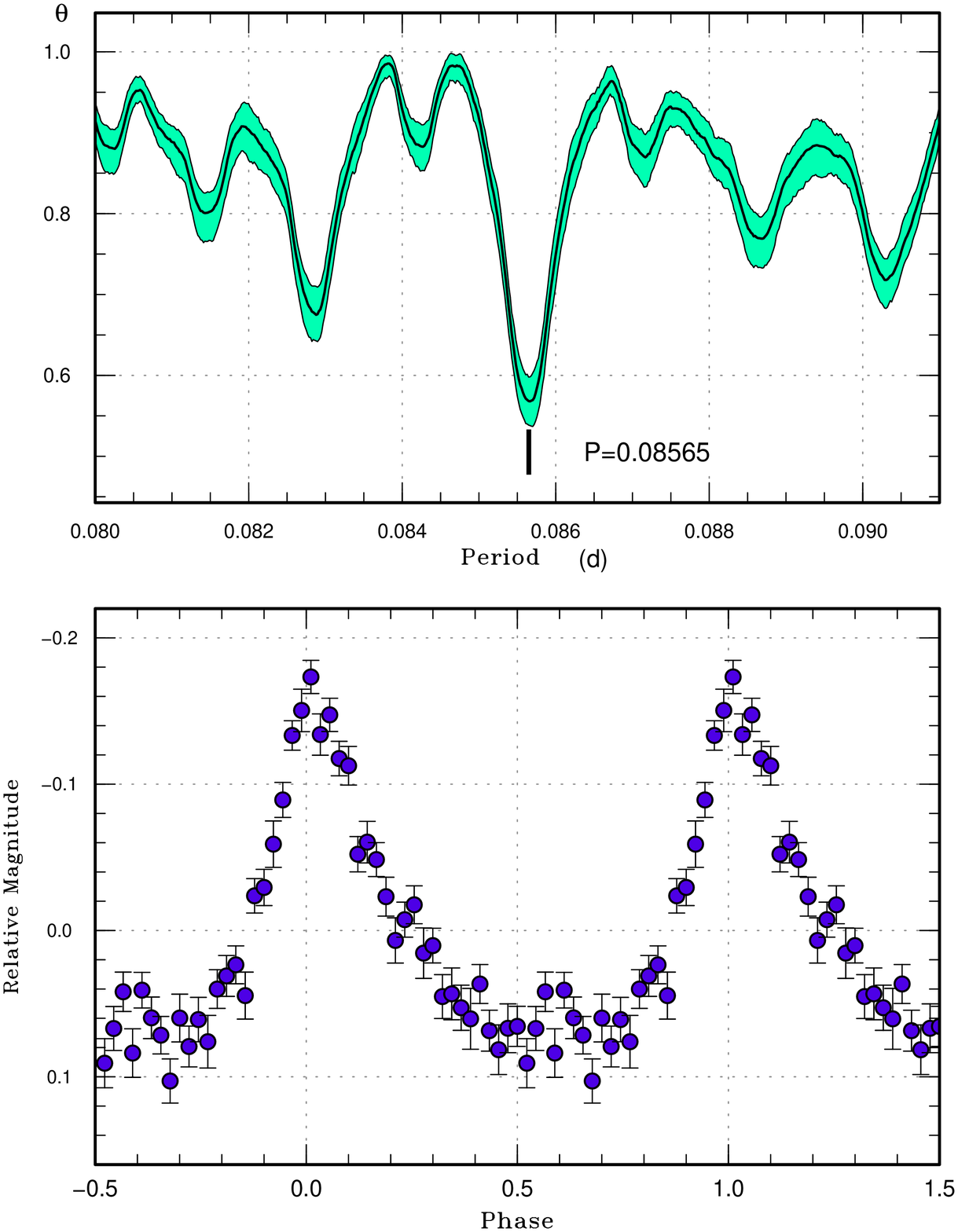}
  \end{center}
  \caption{Superhumps in ASASSN-15eh (2015).
     (Upper): PDM analysis.
     (Lower): Phase-averaged profile.}
  \label{fig:asassn15ehshpdm}
\end{figure}

   This object was detected as a transient at $V$=15.6
on 2015 March 3 by the ASAS-SN team.
The object was apparent detected during a precursor
outburst, and the SU UMa-type nature was suspected
(vsnet-alert 18380).  Superhump were subsequently
detected (vsnet-alert 18391, 18393;
figure \ref{fig:asassn15ehshpdm}).
The times of superhump maxima are listed in
table \ref{tab:asassn15ehoc2015}.  Although there
may have been a stage transition between the second
and the final night, we listed the mean value in
table \ref{tab:perlist} due to the lack of the data.

% SI

% SI

\begin{table}
\caption{Superhump maxima of ASASSN-15eh (2015)}\label{tab:asassn15ehoc2015}
\begin{center}
\begin{tabular}{rp{55pt}p{40pt}r@{.}lr}
\hline
\multicolumn{1}{c}{$E$} & \multicolumn{1}{c}{max\commenta} & \multicolumn{1}{c}{error} & \multicolumn{2}{c}{$O-C$\commentb} & \multicolumn{1}{c}{$N$\commentc} \\
\hline
0 & 57088.4648 & 0.0008 & $-$0&0014 & 178 \\
1 & 57088.5504 & 0.0011 & $-$0&0015 & 197 \\
2 & 57088.6375 & 0.0012 & $-$0&0001 & 132 \\
24 & 57090.5234 & 0.0008 & 0&0012 & 198 \\
25 & 57090.6117 & 0.0008 & 0&0039 & 196 \\
59 & 57093.5196 & 0.0009 & $-$0&0008 & 197 \\
60 & 57093.6049 & 0.0009 & $-$0&0013 & 165 \\
\hline
  \multicolumn{6}{l}{\commenta BJD$-$2400000.} \\
  \multicolumn{6}{l}{\commentb Against max $= 2457088.4662 + 0.085665 E$.} \\
  \multicolumn{6}{l}{\commentc Number of points used to determine the maximum.} \\
\end{tabular}
\end{center}
\end{table}

\subsection{ASASSN-15ev}\label{obj:asassn15ev}

\begin{figure}
  \begin{center}
%    \FigureFile(85mm,110mm){asassn15evshpdm.eps}
    \FigureFile(85mm,110mm){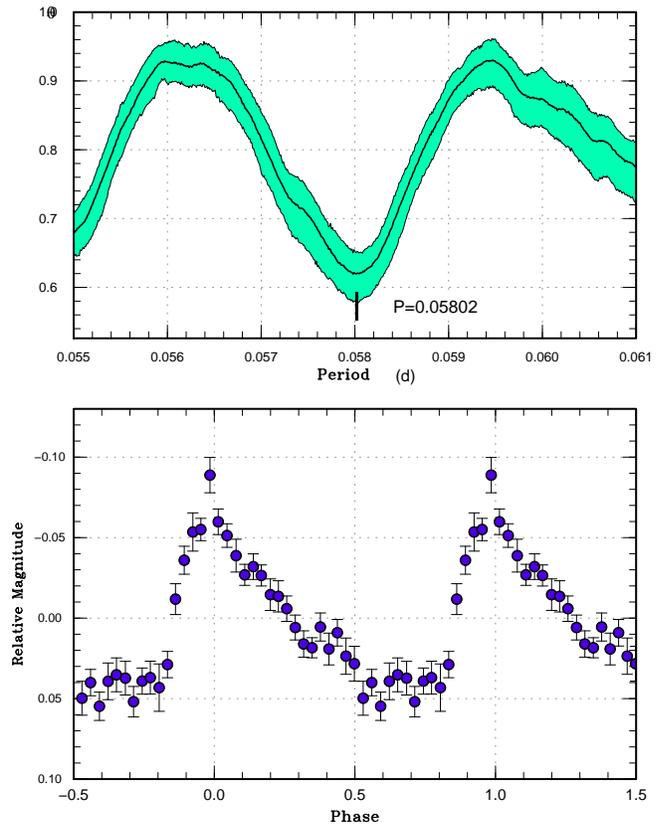}
  \end{center}
  \caption{Superhumps in ASASSN-15ev during the plateau phase (2015).
     (Upper): PDM analysis.
     (Lower): Phase-averaged profile.}
  \label{fig:asassn15evshpdm}
\end{figure}

   This object was detected as a transient at $V$=15.1
on 2015 March 16 by the ASAS-SN team.
The object has an X-ray counterpart
1SXPS J073819.6$-$825039.
Subsequent observations detected
superhumps (vsnet-alert 18465, 18482;
figure \ref{fig:asassn15evshpdm}).
The times of superhump maxima are listed in
table \ref{tab:asassn15evoc2015}.
The superhump period in table \ref{tab:perlist}
was determined with the PDM method.
The object started fading rapidly on March 26--27,
$\sim$10~d after the outburst detection.
The upper limit of the duration of the plateau phase
was 13~d, which is relatively short for an SU UMa-type
dwarf nova with this superhump period.

% SI

% SI

\begin{table}
\caption{Superhump maxima of ASASSN-15ev (2015)}\label{tab:asassn15evoc2015}
\begin{center}
\begin{tabular}{rp{55pt}p{40pt}r@{.}lr}
\hline
\multicolumn{1}{c}{$E$} & \multicolumn{1}{c}{max\commenta} & \multicolumn{1}{c}{error} & \multicolumn{2}{c}{$O-C$\commentb} & \multicolumn{1}{c}{$N$\commentc} \\
\hline
0 & 57100.2449 & 0.0022 & $-$0&0016 & 79 \\
1 & 57100.3036 & 0.0006 & $-$0&0008 & 134 \\
2 & 57100.3661 & 0.0024 & 0&0037 & 77 \\
3 & 57100.4191 & 0.0006 & $-$0&0013 & 133 \\
18 & 57101.2933 & 0.0014 & 0&0035 & 109 \\
19 & 57101.3448 & 0.0012 & $-$0&0029 & 99 \\
20 & 57101.4052 & 0.0009 & $-$0&0005 & 61 \\
\hline
  \multicolumn{6}{l}{\commenta BJD$-$2400000.} \\
  \multicolumn{6}{l}{\commentb Against max $= 2457100.2465 + 0.057961 E$.} \\
  \multicolumn{6}{l}{\commentc Number of points used to determine the maximum.} \\
\end{tabular}
\end{center}
\end{table}

\subsection{ASASSN-15fo}\label{obj:asassn15fo}

\begin{figure}
  \begin{center}
%    \FigureFile(85mm,70mm){asassn15foshlc.eps}
    \FigureFile(85mm,70mm){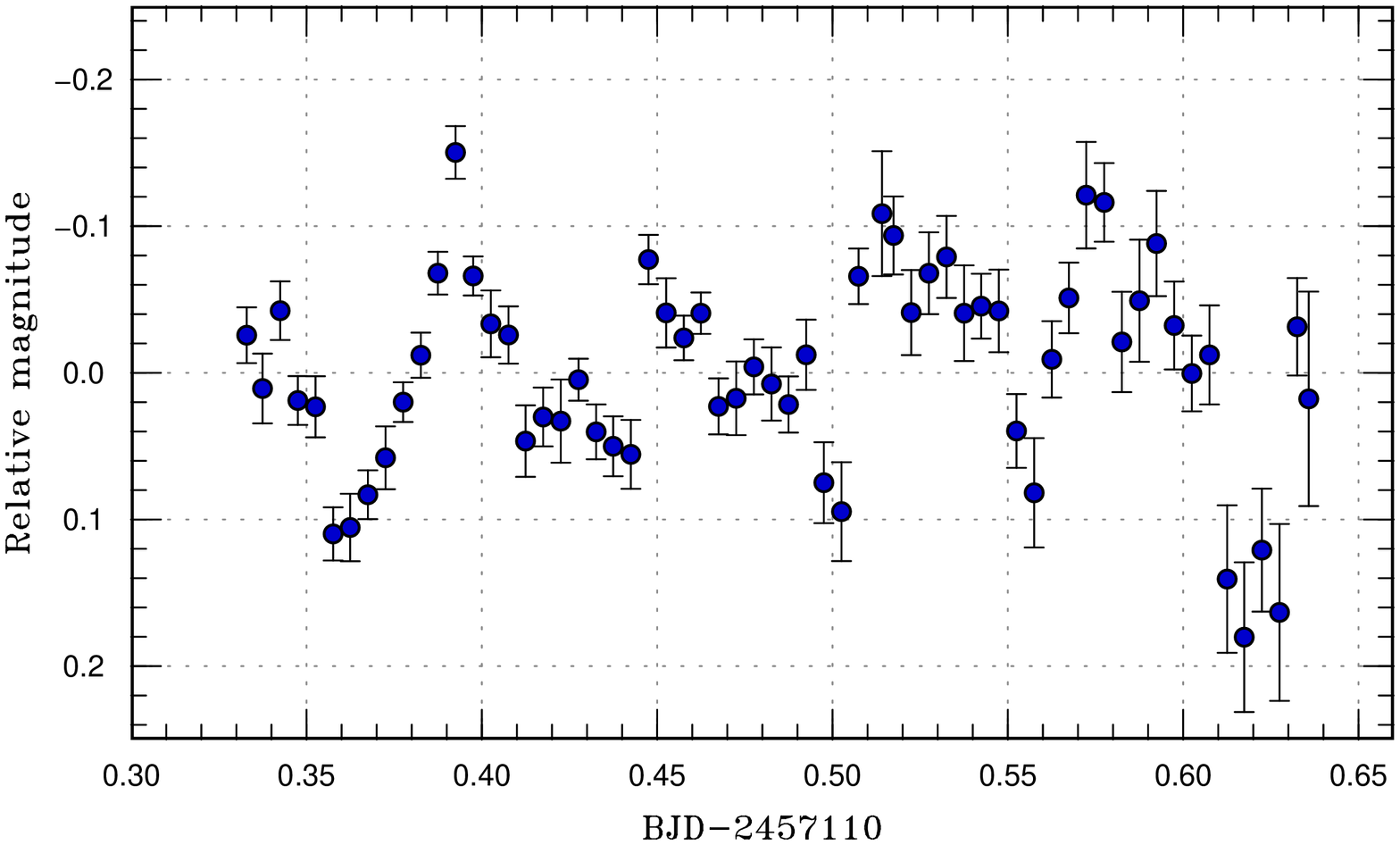}
  \end{center}
  \caption{Superhumps in ASASSN-15fo (2015).
  The data were binned to 0.005~d.
  }
  \label{fig:asassn15foshlc}
\end{figure}

   This object was detected as a transient at $V$=15.4
on 2015 March 19 by the ASAS-SN team.  It further
brightened to $V$=14.7 on March 20.
Subsequent observations detected
superhumps (vsnet-alert 18498).
The superhumps were clearly seen only on the first
night of our observations
(figure \ref{fig:asassn15foshlc}).
Although there were observations on later nights,
they did not yield a meaningful superhump signal
due to the faintness of the object (fainter than
16 mag after March 31).
We restricted our analysis to the first-night observation.
The times of superhump maxima are listed in
table \ref{tab:asassn15fooc2015}.
The period given in table \ref{tab:perlist} is
by the PDM analysis.
The object started fading rapidly on April 4,
giving 16~d for the duration of the superoutburst.

% SI

\begin{table}
\caption{Superhump maxima of ASASSN-15fo (2015)}\label{tab:asassn15fooc2015}
\begin{center}
\begin{tabular}{rp{55pt}p{40pt}r@{.}lr}
\hline
\multicolumn{1}{c}{$E$} & \multicolumn{1}{c}{max\commenta} & \multicolumn{1}{c}{error} & \multicolumn{2}{c}{$O-C$\commentb} & \multicolumn{1}{c}{$N$\commentc} \\
\hline
0 & 57110.3405 & 0.0012 & 0&0037 & 97 \\
1 & 57110.3927 & 0.0008 & $-$0&0045 & 145 \\
2 & 57110.4549 & 0.0017 & $-$0&0025 & 145 \\
3 & 57110.5215 & 0.0018 & 0&0038 & 135 \\
4 & 57110.5776 & 0.0018 & $-$0&0005 & 144 \\
\hline
  \multicolumn{6}{l}{\commenta BJD$-$2400000.} \\
  \multicolumn{6}{l}{\commentb Against max $= 2457110.3368 + 0.060301 E$.} \\
  \multicolumn{6}{l}{\commentc Number of points used to determine the maximum.} \\
\end{tabular}
\end{center}
\end{table}

\subsection{ASASSN-15fu}\label{obj:asassn15fu}

\begin{figure}
  \begin{center}
%    \FigureFile(85mm,110mm){asassn15fushpdm.eps}
    \FigureFile(85mm,110mm){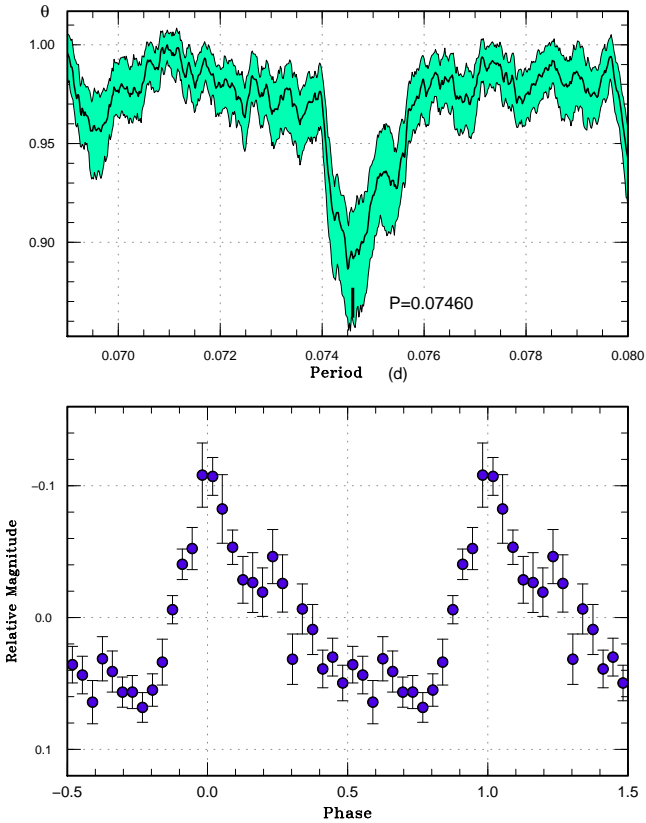}
  \end{center}
  \caption{Superhumps in ASASSN-15fu (2015).
     (Upper): PDM analysis.
     (Lower): Phase-averaged profile.}
  \label{fig:asassn15fushpdm}
\end{figure}

   This object was detected as a transient at $V$=15.6
on 2015 March 27 by the ASAS-SN team.
Superhumps were immediately detected
(vsnet-alert 18503, 18506, 18522;
figure \ref{fig:asassn15fushpdm}).
The times of superhump maxima are listed in
table \ref{tab:asassn15fuoc2015}.
Although there was a stage B-C transition in
the later part of the observation, the period of
stage C superhumps (table \ref{tab:perlist})
was uncertain due to the limited 
quality of the observation.

% SI

% SI

\begin{table}
\caption{Superhump maxima of ASASSN-15fu (2015)}\label{tab:asassn15fuoc2015}
\begin{center}
\begin{tabular}{rp{55pt}p{40pt}r@{.}lr}
\hline
\multicolumn{1}{c}{$E$} & \multicolumn{1}{c}{max\commenta} & \multicolumn{1}{c}{error} & \multicolumn{2}{c}{$O-C$\commentb} & \multicolumn{1}{c}{$N$\commentc} \\
\hline
0 & 57111.2945 & 0.0013 & $-$0&0027 & 121 \\
1 & 57111.3692 & 0.0007 & $-$0&0027 & 172 \\
2 & 57111.4465 & 0.0008 & 0&0001 & 173 \\
3 & 57111.5189 & 0.0007 & $-$0&0021 & 173 \\
4 & 57111.5984 & 0.0030 & 0&0028 & 71 \\
13 & 57112.2661 & 0.0010 & $-$0&0006 & 152 \\
14 & 57112.3396 & 0.0010 & $-$0&0016 & 172 \\
15 & 57112.4145 & 0.0012 & $-$0&0013 & 162 \\
43 & 57114.5155 & 0.0011 & 0&0118 & 38 \\
44 & 57114.5813 & 0.0034 & 0&0030 & 35 \\
57 & 57115.5566 & 0.0029 & 0&0089 & 44 \\
70 & 57116.5070 & 0.0051 & $-$0&0100 & 16 \\
71 & 57116.5860 & 0.0031 & $-$0&0056 & 19 \\
\hline
  \multicolumn{6}{l}{\commenta BJD$-$2400000.} \\
  \multicolumn{6}{l}{\commentb Against max $= 2457111.2973 + 0.074568 E$.} \\
  \multicolumn{6}{l}{\commentc Number of points used to determine the maximum.} \\
\end{tabular}
\end{center}
\end{table}

\subsection{ASASSN-15gf}\label{obj:asassn15gf}

\begin{figure}
  \begin{center}
%    \FigureFile(85mm,110mm){asassn15gfshpdm.eps}
    \FigureFile(85mm,110mm){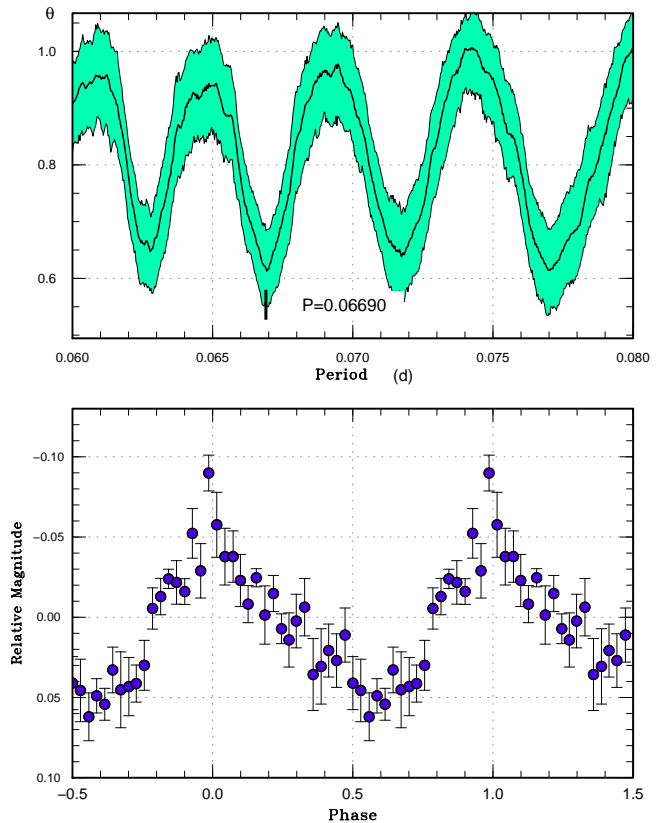}
  \end{center}
  \caption{Superhumps in ASASSN-15gf (2015).
     (Upper): PDM analysis of the first two nights.
     The alias selection was based on the $O-C$ analysis.
     The other one-day aliases are still possible.
     (Lower): Phase-averaged profile.}
  \label{fig:asassn15gfshpdm}
\end{figure}

   This object was detected as a transient at $V$=15.2
on 2015 April 2 by the ASAS-SN team.
There is an $r$=21.5 mag counterpart in IPHAS DR2.
Subsequent observations detected superhumps
(vsnet-alert 18520, 18525).
Due to the shortness of the runs, there were many
equally acceptable one-day aliases by the PDM analysis
(figure \ref{fig:asassn15gfshpdm}).
We have chosen the one which give the smallest 
$O-C$ scatter (table \ref{tab:asassn15gfoc2015}).
We should note that other one-day aliases are
still possible.  In table \ref{tab:perlist}, we gave
the period selected by the $O-C$ method and refined
by the PDM analysis.

% SI

% SI

\begin{table}
\caption{Superhump maxima of ASASSN-15gf (2015)}\label{tab:asassn15gfoc2015}
\begin{center}
\begin{tabular}{rp{55pt}p{40pt}r@{.}lr}
\hline
\multicolumn{1}{c}{$E$} & \multicolumn{1}{c}{max\commenta} & \multicolumn{1}{c}{error} & \multicolumn{2}{c}{$O-C$\commentb} & \multicolumn{1}{c}{$N$\commentc} \\
\hline
0 & 57118.3250 & 0.0010 & 0&0007 & 96 \\
1 & 57118.3904 & 0.0017 & $-$0&0008 & 67 \\
15 & 57119.3282 & 0.0009 & 0&0001 & 58 \\
\hline
  \multicolumn{6}{l}{\commenta BJD$-$2400000.} \\
  \multicolumn{6}{l}{\commentb Against max $= 2457118.3243 + 0.066928 E$.} \\
  \multicolumn{6}{l}{\commentc Number of points used to determine the maximum.} \\
\end{tabular}
\end{center}
\end{table}

\subsection{ASASSN-15gh}\label{obj:asassn15gh}

\begin{figure}
  \begin{center}
%    \FigureFile(85mm,110mm){asassn15ghshpdm.eps}
    \FigureFile(85mm,110mm){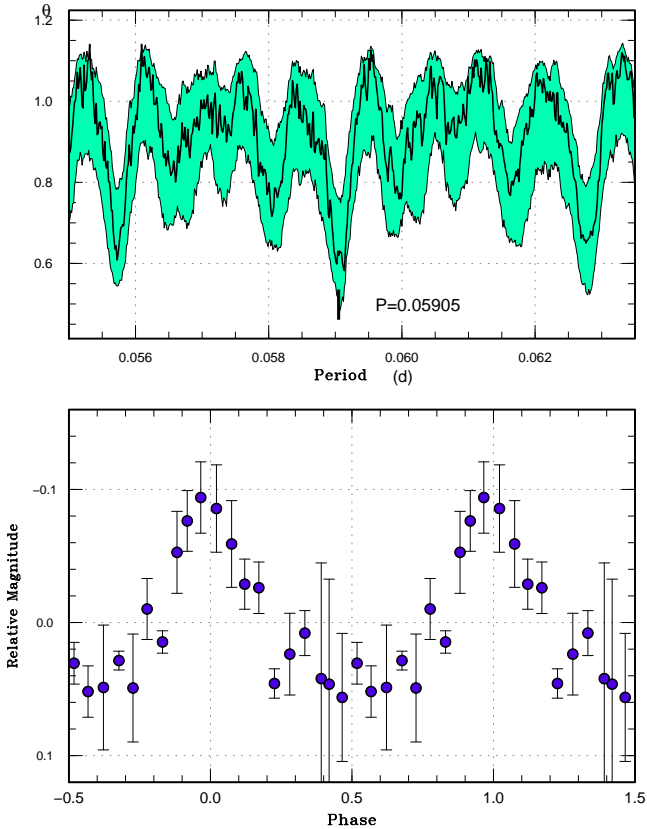}
  \end{center}
  \caption{Superhumps in ASASSN-15gh (2015).
     (Upper): PDM analysis for the interval
     BJD 2457122--2457127 when superhumps were most
     clearly visible.
     The alias selection was based on the $O-C$ analysis.
     The other one-day aliases are still possible.
     (Lower): Phase-averaged profile.}
  \label{fig:asassn15ghshpdm}
\end{figure}

   This object was detected as a transient at $V$=14.6
on 2015 April 1 by the ASAS-SN team.
No quiescent counterpart was recorded.
Before April 9, there was little indication of
hump-like variations.  On April 10, superhumps became
apparent.
Since the object had very low signal-to-noise due to
the faintness (the object was already at 16.2 mag on
April 10; it was only 15.5 mag on April 4 and
the ASAS-SN detection magnitude may have been
too bright), only the data between April 10 and 14
(BJD 2457122--2457127) were used to determine superhumps.
Among the potential aliases, we selected the one
which minimizes the scatter in the $O-C$ diagram
(figure \ref{fig:asassn15ghshpdm}).
Other possibilities still remain.
The times of superhump maxima are listed in
table \ref{tab:asassn15ghoc2015}.
In table \ref{tab:perlist}, we gave
the period selected by the $O-C$ method and refined
by the PDM analysis.

% SI

% SI

\begin{table}
\caption{Superhump maxima of ASASSN-15gh (2015)}\label{tab:asassn15ghoc2015}
\begin{center}
\begin{tabular}{rp{55pt}p{40pt}r@{.}lr}
\hline
\multicolumn{1}{c}{$E$} & \multicolumn{1}{c}{max\commenta} & \multicolumn{1}{c}{error} & \multicolumn{2}{c}{$O-C$\commentb} & \multicolumn{1}{c}{$N$\commentc} \\
\hline
0 & 57122.8319 & 0.0015 & 0&0006 & 15 \\
1 & 57122.8894 & 0.0021 & $-$0&0009 & 14 \\
51 & 57125.8444 & 0.0013 & 0&0015 & 13 \\
68 & 57126.8457 & 0.0022 & $-$0&0011 & 13 \\
\hline
  \multicolumn{6}{l}{\commenta BJD$-$2400000.} \\
  \multicolumn{6}{l}{\commentb Against max $= 2457122.8313 + 0.059050 E$.} \\
  \multicolumn{6}{l}{\commentc Number of points used to determine the maximum.} \\
\end{tabular}
\end{center}
\end{table}

\subsection{ASASSN-15gi}\label{obj:asassn15gi}

\begin{figure}
  \begin{center}
%    \FigureFile(85mm,110mm){asassn15gishpdm.eps}
    \FigureFile(85mm,110mm){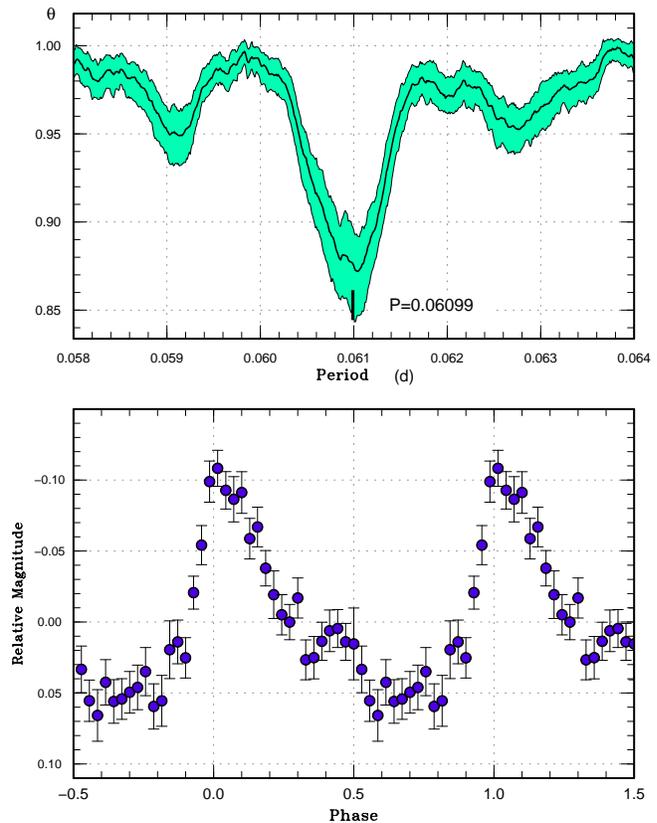}
  \end{center}
  \caption{Superhumps in ASASSN-15gi (2015).
     (Upper): PDM analysis.
     (Lower): Phase-averaged profile.}
  \label{fig:asassn15gishpdm}
\end{figure}

   This object was detected as a transient at $V$=15.4
on 2015 April 1 by the ASAS-SN team.
Subsequent observations detected superhumps
(vsnet-alert 18521, 18529; figure \ref{fig:asassn15gishpdm}).
The times of superhump maxima are listed in 
table \ref{tab:asassn15gioc2015}.
The $O-C$ data indicate the presence of stage B-C
transition around $E$=65.
At the time of observation on April 4, the object
was at 15.8 mag.  The initial report on the ASAS-SN Transients
page ($V$=14.6) was probably erroneous.

% SI

% SI

\begin{table}
\caption{Superhump maxima of ASASSN-15gi (2015)}\label{tab:asassn15gioc2015}
\begin{center}
\begin{tabular}{rp{55pt}p{40pt}r@{.}lr}
\hline
\multicolumn{1}{c}{$E$} & \multicolumn{1}{c}{max\commenta} & \multicolumn{1}{c}{error} & \multicolumn{2}{c}{$O-C$\commentb} & \multicolumn{1}{c}{$N$\commentc} \\
\hline
0 & 57116.5035 & 0.0016 & 0&0018 & 15 \\
1 & 57116.5586 & 0.0023 & $-$0&0042 & 17 \\
2 & 57116.6204 & 0.0010 & $-$0&0035 & 15 \\
17 & 57117.5374 & 0.0016 & $-$0&0037 & 22 \\
18 & 57117.5978 & 0.0014 & $-$0&0044 & 23 \\
33 & 57118.5149 & 0.0018 & $-$0&0044 & 20 \\
34 & 57118.5794 & 0.0021 & $-$0&0011 & 23 \\
64 & 57120.4192 & 0.0021 & 0&0045 & 141 \\
65 & 57120.4873 & 0.0040 & 0&0115 & 146 \\
66 & 57120.5380 & 0.0053 & 0&0010 & 56 \\
78 & 57121.2742 & 0.0013 & 0&0035 & 80 \\
79 & 57121.3386 & 0.0006 & 0&0067 & 141 \\
80 & 57121.3990 & 0.0008 & 0&0060 & 140 \\
88 & 57121.8799 & 0.0035 & $-$0&0022 & 15 \\
94 & 57122.2509 & 0.0007 & 0&0019 & 141 \\
95 & 57122.3140 & 0.0007 & 0&0039 & 140 \\
96 & 57122.3739 & 0.0009 & 0&0026 & 141 \\
97 & 57122.4354 & 0.0011 & 0&0031 & 141 \\
99 & 57122.5539 & 0.0034 & $-$0&0008 & 23 \\
100 & 57122.6164 & 0.0016 & 0&0006 & 15 \\
128 & 57124.3175 & 0.0023 & $-$0&0102 & 139 \\
129 & 57124.3846 & 0.0016 & $-$0&0042 & 141 \\
130 & 57124.4416 & 0.0031 & $-$0&0084 & 141 \\
\hline
  \multicolumn{6}{l}{\commenta BJD$-$2400000.} \\
  \multicolumn{6}{l}{\commentb Against max $= 2457116.5017 + 0.061141 E$.} \\
  \multicolumn{6}{l}{\commentc Number of points used to determine the maximum.} \\
\end{tabular}
\end{center}
\end{table}

\subsection{ASASSN-15gn}\label{obj:asassn15gn}

\begin{figure}
  \begin{center}
%    \FigureFile(85mm,110mm){asassn15gnshpdm.eps}
    \FigureFile(85mm,110mm){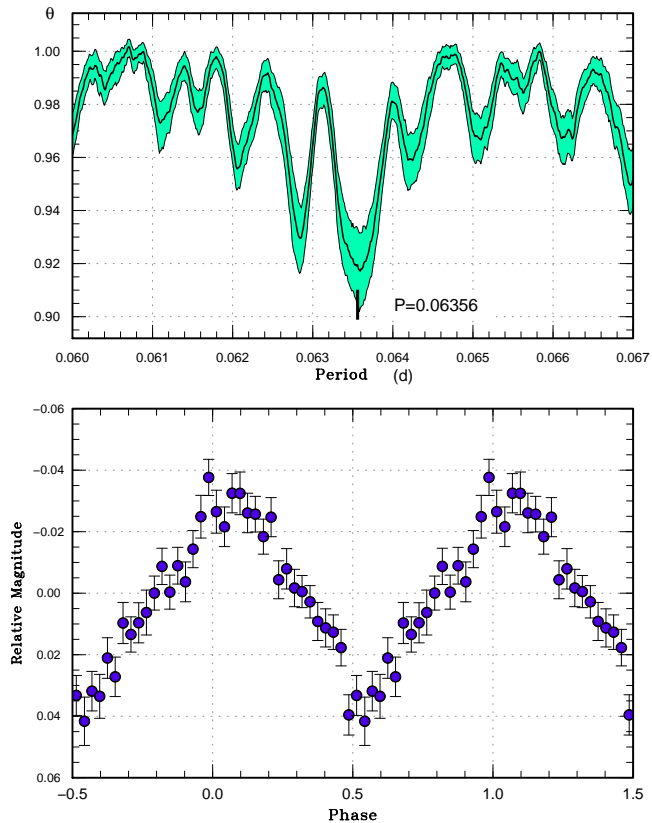}
  \end{center}
  \caption{Ordinary superhumps in ASASSN-15gn (2015).
     (Upper): PDM analysis.
     (Lower): Phase-averaged profile.}
  \label{fig:asassn15gnshpdm}
\end{figure}

\begin{figure}
  \begin{center}
%    \FigureFile(85mm,100mm){asassn15gnhumpall.eps}
    \FigureFile(85mm,100mm){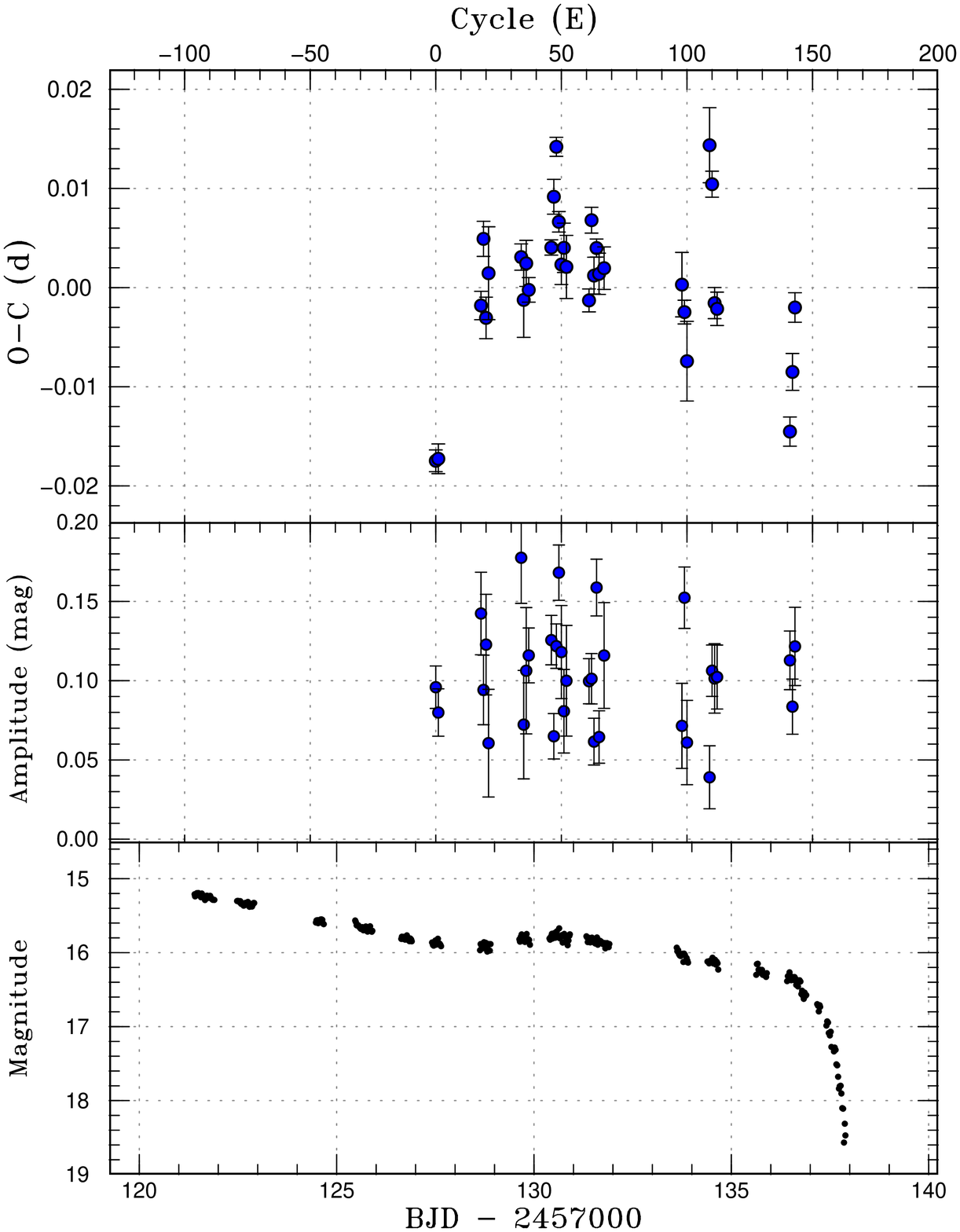}
  \end{center}
  \caption{$O-C$ diagram of superhumps in ASASSN-15gn (2015).
     (Upper:) $O-C$ diagram.
     We used a period of 0.06364~d for calculating the $O-C$ residuals.
     (Middle:) Amplitudes of superhumps.
     (Lower:) Light curve.  The data were binned to 0.021~d.
  }
  \label{fig:asassn15gnhumpall}
\end{figure}

   This object was detected as a transient at $V$=14.2
on 2015 April 3 by the ASAS-SN team (vsnet-alert 18518).
The likely quiescent counterpart was recorded very faint
(22.4 mag on J plate).
The object initially did not show strong short-term
variations.  On April 14 (11~d after the outburst
detection), it started to show superhumps
(vsnet-alert 18546, 18548, 18550, 18556, 18559;
figure \ref{fig:asassn15gnshpdm}).
On April 25, it started to fade rapidly
(figure \ref{fig:asassn15gnhumpall}).
The times of superhump maxima are listed in
table \ref{tab:asassn15gnoc2015}.
Although it was initially suggested that superhumps
in this systems grew slowly (vsnet-alert 18550),
the $O-C$ data would indicate that stage A superhumps
were likely only recorded for the initial night when
superhumps started to appear.
In table \ref{tab:perlist}, we listed a period of
stage A superhumps with an assumption that $E$=18
corresponds to the end of stage A.
There was some indication of stage C after $E$=112.

   This object has a large outburst amplitude,
a long superhump period, a long waiting time before
the appearance of superhumps
and a small $P_{\rm dot}$.  These features are suggestive
of a period bouncer (cf. \cite{kat15wzsge}).
The small amplitude of superhumps is also suggestive
of a small tidal effect and the superhump profile
is more symmetric than in other SU UMa-type
dwarf novae (figure \ref{fig:asassn15gnshpdm}).
The growth time of superhumps, however, is short,
which is not compatible with a small $q$ if this object
is a period bouncer.  It may be possible either that
the growing stage of superhumps was not well recorded
due to the faintness of the object (15.5 mag at
the time of emergence of superhumps) and low sampling
rates on some nights or that the stage identification is
not correct and the entire superhumps are stage A
superhumps.  The either possibility is not
ruled out because superhumps were close to the detection
limit and the data were not so densely obtained.
The small amplitude and relatively symmetric profile
of the superhumps may suggest that the outburst terminated 
before superhump developed fully.  Since the behavior
of superhumps is known to be complex in period bouncers
(cf. \Nakataprep, \cite{kat15wzsge}), this possibility
may deserve consideration.

% SI

% SI

\begin{table}
\caption{Superhump maxima of ASASSN-15gn (2015)}\label{tab:asassn15gnoc2015}
\begin{center}
\begin{tabular}{rp{55pt}p{40pt}r@{.}lr}
\hline
\multicolumn{1}{c}{$E$} & \multicolumn{1}{c}{max\commenta} & \multicolumn{1}{c}{error} & \multicolumn{2}{c}{$O-C$\commentb} & \multicolumn{1}{c}{$N$\commentc} \\
\hline
0 & 57127.4945 & 0.0011 & $-$0&0179 & 146 \\
1 & 57127.5584 & 0.0015 & $-$0&0177 & 147 \\
18 & 57128.6557 & 0.0014 & $-$0&0022 & 17 \\
19 & 57128.7261 & 0.0018 & 0&0045 & 17 \\
20 & 57128.7817 & 0.0021 & $-$0&0035 & 15 \\
21 & 57128.8499 & 0.0047 & 0&0010 & 16 \\
34 & 57129.6788 & 0.0013 & 0&0027 & 17 \\
35 & 57129.7382 & 0.0038 & $-$0&0017 & 17 \\
36 & 57129.8055 & 0.0023 & 0&0020 & 13 \\
37 & 57129.8665 & 0.0012 & $-$0&0006 & 17 \\
46 & 57130.4435 & 0.0008 & 0&0036 & 114 \\
47 & 57130.5122 & 0.0018 & 0&0088 & 147 \\
48 & 57130.5809 & 0.0010 & 0&0138 & 127 \\
49 & 57130.6370 & 0.0010 & 0&0062 & 75 \\
50 & 57130.6963 & 0.0020 & 0&0019 & 17 \\
51 & 57130.7617 & 0.0025 & 0&0036 & 15 \\
52 & 57130.8234 & 0.0032 & 0&0017 & 14 \\
61 & 57131.3927 & 0.0012 & $-$0&0017 & 147 \\
62 & 57131.4645 & 0.0013 & 0&0064 & 145 \\
63 & 57131.5225 & 0.0019 & 0&0008 & 146 \\
64 & 57131.5890 & 0.0009 & 0&0036 & 147 \\
65 & 57131.6500 & 0.0021 & 0&0010 & 162 \\
67 & 57131.7778 & 0.0021 & 0&0015 & 15 \\
98 & 57133.7490 & 0.0033 & $-$0&0001 & 18 \\
99 & 57133.8099 & 0.0012 & $-$0&0029 & 15 \\
100 & 57133.8686 & 0.0040 & $-$0&0079 & 17 \\
109 & 57134.4631 & 0.0038 & 0&0139 & 126 \\
110 & 57134.5228 & 0.0013 & 0&0100 & 147 \\
111 & 57134.5745 & 0.0016 & $-$0&0020 & 146 \\
112 & 57134.6375 & 0.0017 & $-$0&0026 & 147 \\
141 & 57136.4707 & 0.0015 & $-$0&0150 & 146 \\
142 & 57136.5404 & 0.0019 & $-$0&0090 & 147 \\
143 & 57136.6105 & 0.0015 & $-$0&0025 & 151 \\
\hline
  \multicolumn{6}{l}{\commenta BJD$-$2400000.} \\
  \multicolumn{6}{l}{\commentb Against max $= 2457127.5124 + 0.063640 E$.} \\
  \multicolumn{6}{l}{\commentc Number of points used to determine the maximum.} \\
\end{tabular}
\end{center}
\end{table}

\subsection{ASASSN-15gq}\label{obj:asassn15gq}

   This object was detected as a transient at $V$=15.4
on 2015 April 10 by the ASAS-SN team.
There is a $g$=21.6 mag SDSS object (SDSS J101510.86$+$812418.7)  
\timeform{1.5''} distant from this position.
This SDSS object is less likely the quiescent
counterpart since it has colors (e.g. $u-g$=0.76)
unlike a CV.

\begin{figure}
  \begin{center}
%    \FigureFile(85mm,110mm){asassn15gqeshpdm.eps}
    \FigureFile(85mm,110mm){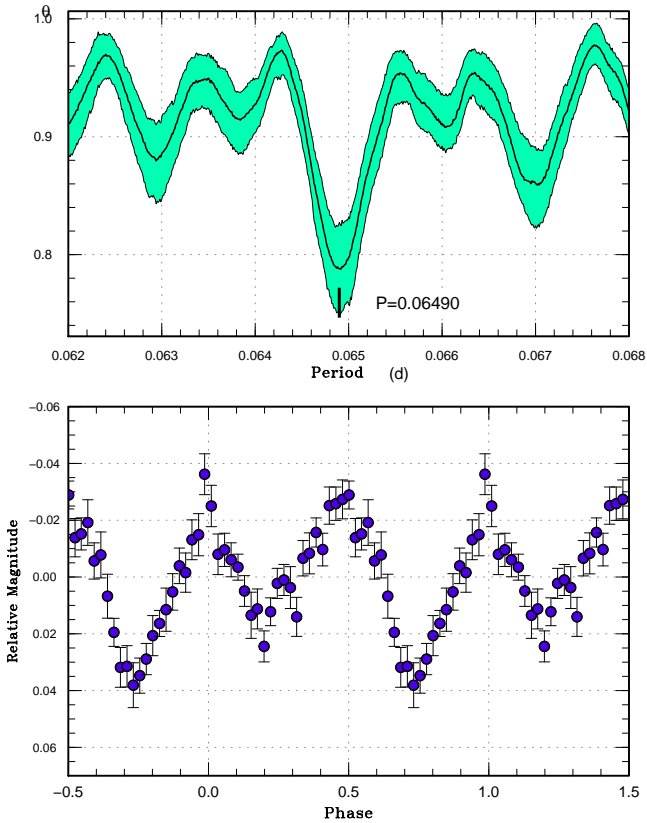}
  \end{center}
  \caption{Early superhumps in ASASSN-15gq (2015).
     (Upper): PDM analysis.
     (Lower): Phase-averaged profile.}
  \label{fig:asassn15gqeshpdm}
\end{figure}

\begin{figure}
  \begin{center}
%    \FigureFile(85mm,110mm){asassn15gqshpdm.eps}
    \FigureFile(85mm,110mm){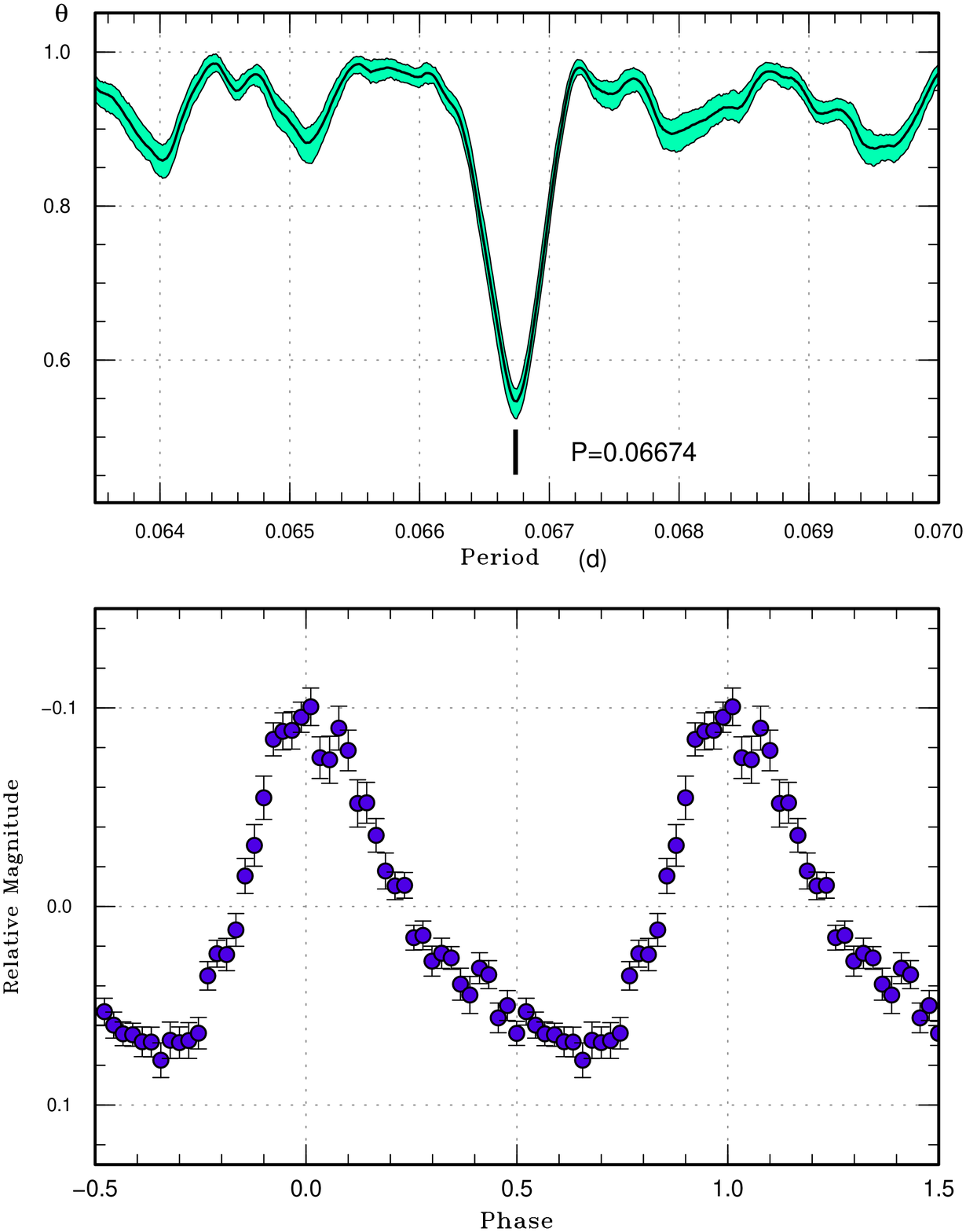}
  \end{center}
  \caption{Ordinary superhumps in ASASSN-15gq (2015).
     (Upper): PDM analysis.
     (Lower): Phase-averaged profile.}
  \label{fig:asassn15gqshpdm}
\end{figure}

   The object initially showed double-wave modulations
(vsnet-alert 18526, 18528, 18531, 18539),
which we identify to be early superhumps
(figure \ref{fig:asassn15gqeshpdm}).  The object
is thus identified as a WZ Sge-type dwarf nova.
Later the object showed fully developed ordinary
superhumps (vsnet-alert 18549, 18558, 18593;
figure \ref{fig:asassn15gqshpdm}).
The times of maxima of ordinary superhumps are
listed in table \ref{tab:asassn15gqoc2015}.
The times for $E \le 1$ are clearly of stage A
superhumps.  There is a suggestion for stage B-C
transition after $E$=120.
In table \ref{tab:perlist}, we listed a period of
stage A superhumps with an assumption that $E$=15
corresponds to the end of stage A.
This fractional superhump excess [$\epsilon^*$=0.038(2)]
corresponds to $q$=0.107(8).  Although we could not
observe the termination of stage A superhumps,
the above value is likely close to the actual value
since if stage A terminated much earlier than $E$=15,
the value of $\epsilon^*$ should be larger, which
will give an unacceptably large $q$ for a WZ Sge-type
object.

   Considering the large positive $P_{\rm dot}$ for
stage B superhumps, the presence of stage C superhumps,
the relatively long orbital period
and the short duration of stage A, this object is
probably a borderline WZ Sge/SU UMa-type object
rather than an extreme WZ Sge-type object
(cf. \cite{kat15wzsge}).

% SI

% SI

% SI

\begin{table}
\caption{Superhump maxima of ASASSN-15gq (2015)}\label{tab:asassn15gqoc2015}
\begin{center}
\begin{tabular}{rp{55pt}p{40pt}r@{.}lr}
\hline
\multicolumn{1}{c}{$E$} & \multicolumn{1}{c}{max\commenta} & \multicolumn{1}{c}{error} & \multicolumn{2}{c}{$O-C$\commentb} & \multicolumn{1}{c}{$N$\commentc} \\
\hline
0 & 57127.3844 & 0.0012 & $-$0&0022 & 67 \\
1 & 57127.4489 & 0.0008 & $-$0&0046 & 63 \\
15 & 57128.3953 & 0.0005 & 0&0069 & 95 \\
16 & 57128.4610 & 0.0004 & 0&0058 & 114 \\
17 & 57128.5292 & 0.0008 & 0&0072 & 63 \\
18 & 57128.5953 & 0.0005 & 0&0065 & 57 \\
45 & 57130.3917 & 0.0004 & $-$0&0003 & 62 \\
46 & 57130.4578 & 0.0004 & $-$0&0009 & 51 \\
60 & 57131.3893 & 0.0004 & $-$0&0044 & 98 \\
61 & 57131.4569 & 0.0004 & $-$0&0035 & 122 \\
62 & 57131.5233 & 0.0006 & $-$0&0040 & 56 \\
65 & 57131.7224 & 0.0019 & $-$0&0052 & 86 \\
66 & 57131.7922 & 0.0007 & $-$0&0022 & 104 \\
75 & 57132.3904 & 0.0004 & $-$0&0050 & 61 \\
76 & 57132.4575 & 0.0006 & $-$0&0048 & 51 \\
90 & 57133.3939 & 0.0005 & $-$0&0033 & 128 \\
91 & 57133.4619 & 0.0007 & $-$0&0021 & 116 \\
105 & 57134.3994 & 0.0013 & 0&0004 & 86 \\
106 & 57134.4647 & 0.0008 & $-$0&0010 & 92 \\
120 & 57135.4089 & 0.0008 & 0&0081 & 45 \\
134 & 57136.3392 & 0.0012 & 0&0035 & 29 \\
135 & 57136.4076 & 0.0008 & 0&0051 & 45 \\
\hline
  \multicolumn{6}{l}{\commenta BJD$-$2400000.} \\
  \multicolumn{6}{l}{\commentb Against max $= 2457127.3867 + 0.066784 E$.} \\
  \multicolumn{6}{l}{\commentc Number of points used to determine the maximum.} \\
\end{tabular}
\end{center}
\end{table}

\subsection{ASASSN-15gs}\label{obj:asassn15gs}

\begin{figure}
  \begin{center}
%    \FigureFile(85mm,70mm){asassn15gsshlc.eps}
    \FigureFile(85mm,70mm){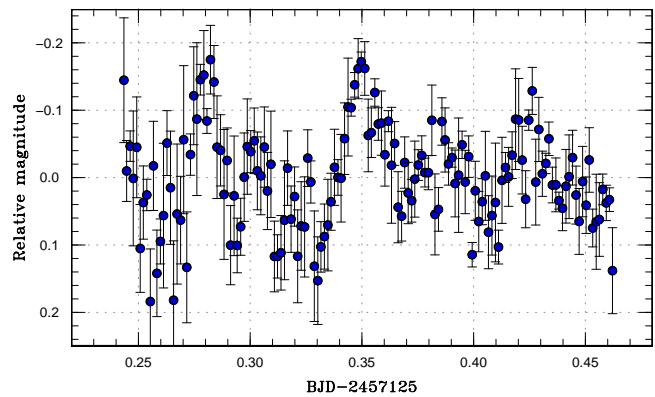}
  \end{center}
  \caption{Superhumps in ASASSN-15gs (2015).
  The data were binned to 0.0015~d.
  }
  \label{fig:asassn15gsshlc}
\end{figure}

   This object was detected as a transient at $V$=15.7
on 2015 April 8 by the ASAS-SN team.
There is an X-ray counterpart (1SXPS J135917.3$-$375242).
The object was observed on one night and superhumps were
detected (vsnet-alert 18541, figure
\ref{fig:asassn15gsshlc}).
The times of superhump maxima are listed in 
table \ref{tab:asassn15gsoc2015}.
The best period with the PDM method was 0.0719(8)~d.

% SI

\begin{table}
\caption{Superhump maxima of ASASSN-15gs (2015)}\label{tab:asassn15gsoc2015}
\begin{center}
\begin{tabular}{rp{55pt}p{40pt}r@{.}lr}
\hline
\multicolumn{1}{c}{$E$} & \multicolumn{1}{c}{max\commenta} & \multicolumn{1}{c}{error} & \multicolumn{2}{c}{$O-C$\commentb} & \multicolumn{1}{c}{$N$\commentc} \\
\hline
1 & 57125.2821 & 0.0018 & $-$0&0009 & 165 \\
2 & 57125.3529 & 0.0009 & $-$0&0020 & 165 \\
3 & 57125.4264 & 0.0018 & $-$0&0004 & 166 \\
\hline
  \multicolumn{6}{l}{\commenta BJD$-$2400000.} \\
  \multicolumn{6}{l}{\commentb Against max $= 2457125.2818 + 0.072165 E$.} \\
  \multicolumn{6}{l}{\commentc Number of points used to determine the maximum.} \\
\end{tabular}
\end{center}
\end{table}

\subsection{ASASSN-15hd}\label{obj:asassn15hd}

   This object was detected as a transient at $V$=14.0
on 2015 April 15 by the ASAS-SN team (vsnet-alert 18547).
The large outburst amplitude inferred from the faint
($g$=21.4--21.9 mag) SDSS counterpart already suggested
a WZ Sge-type object.

\begin{figure}
  \begin{center}
%    \FigureFile(85mm,110mm){asassn15hdeshpdm.eps}
    \FigureFile(85mm,110mm){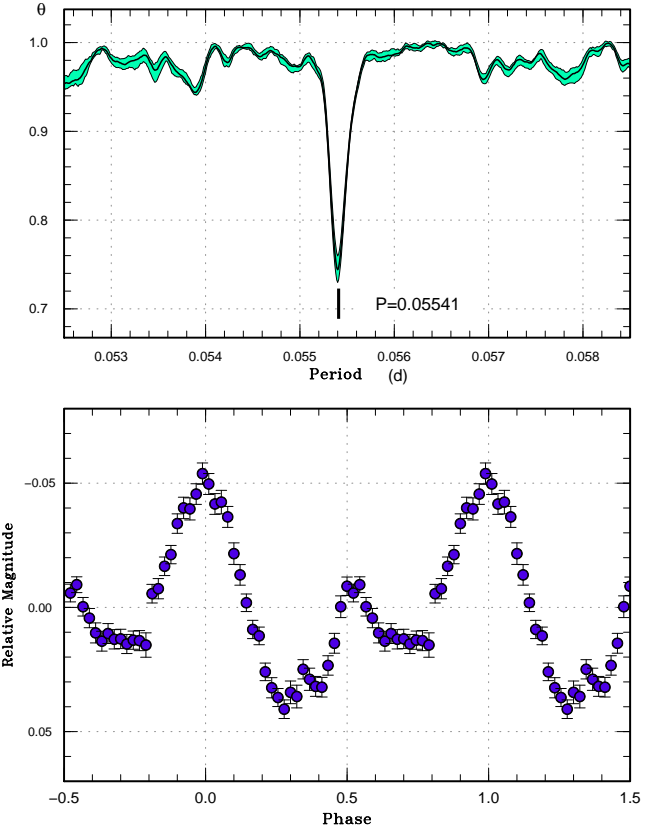}
  \end{center}
  \caption{Early superhumps in ASASSN-15hd (2015).
     (Upper): PDM analysis.
     (Lower): Phase-averaged profile.}
  \label{fig:asassn15hdeshpdm}
\end{figure}

\begin{figure}
  \begin{center}
%    \FigureFile(85mm,110mm){asassn15hdshpdm.eps}
    \FigureFile(85mm,110mm){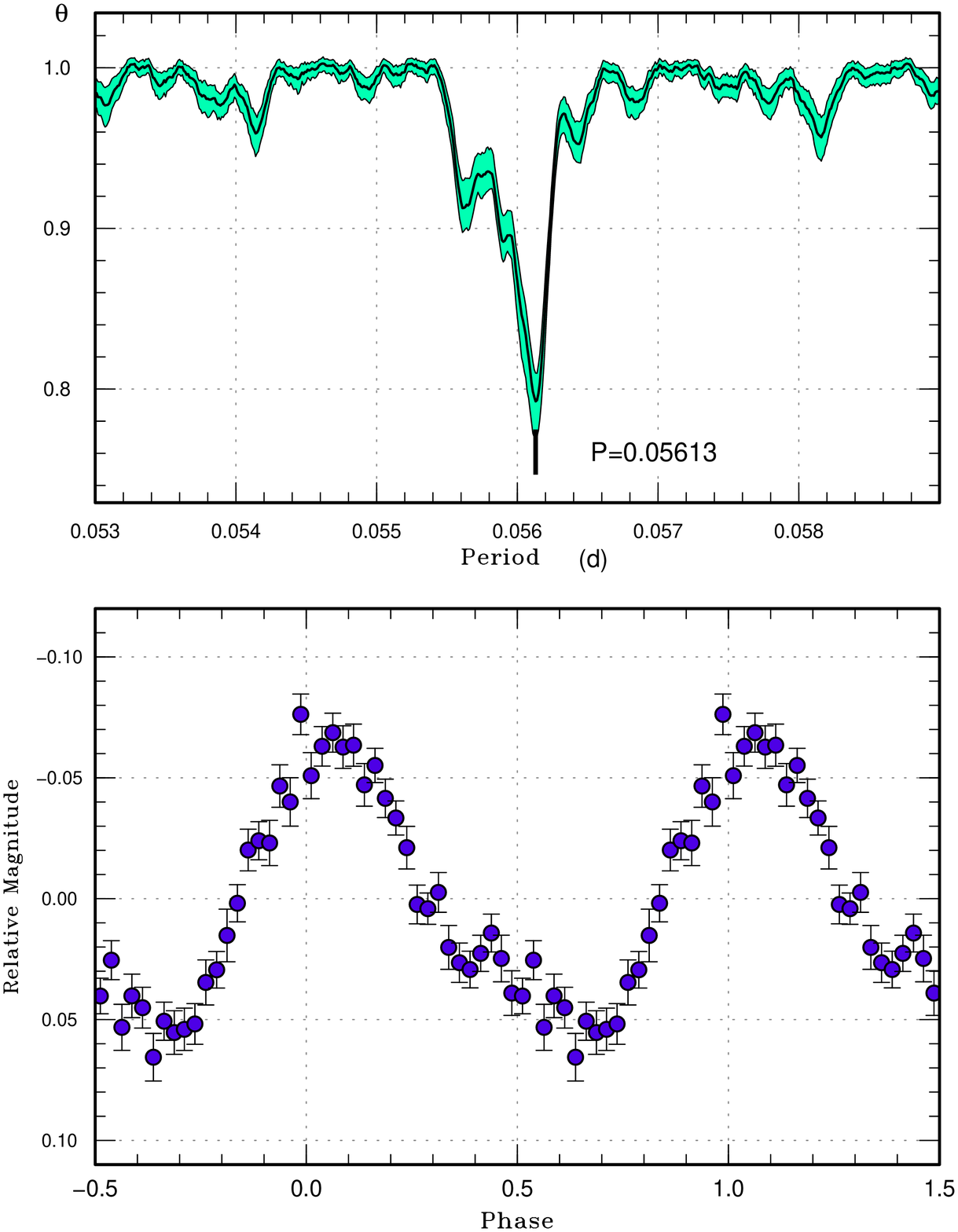}
  \end{center}
  \caption{Ordinary superhumps in ASASSN-15hd (2015).
     (Upper): PDM analysis.
     (Lower): Phase-averaged profile.}
  \label{fig:asassn15hdshpdm}
\end{figure}

\begin{figure}
  \begin{center}
%    \FigureFile(85mm,100mm){asassn15hdhumpall.eps}
    \FigureFile(85mm,100mm){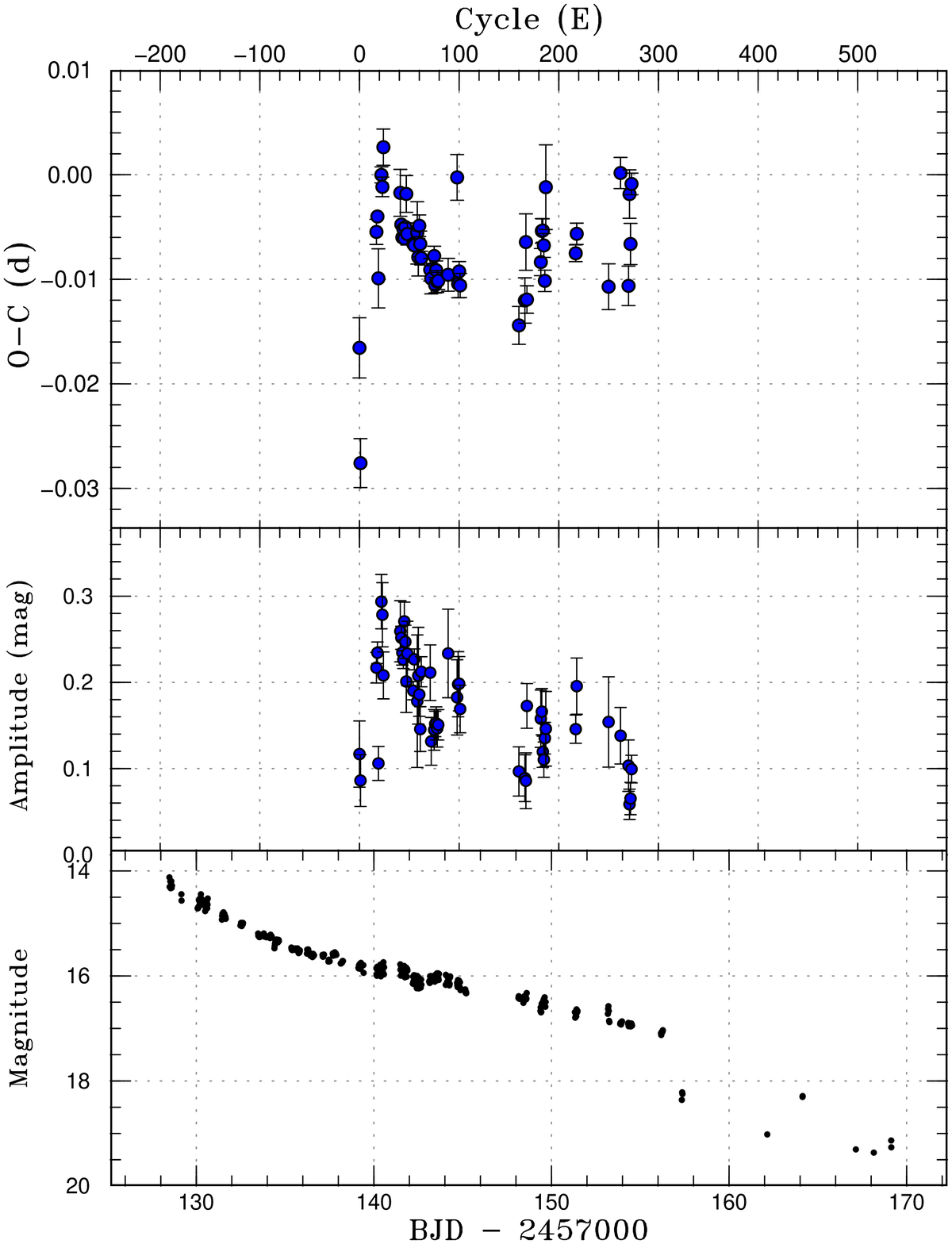}
  \end{center}
  \caption{$O-C$ diagram of superhumps in ASASSN-15hd (2015).
     (Upper:) $O-C$ diagram.
     We used a period of 0.05611~d for calculating the $O-C$ residuals.
     (Middle:) Amplitudes of superhumps.
     (Lower:) Light curve.  The data were binned to 0.019~d.
  }
  \label{fig:asassn15hdhumpall}
\end{figure}

   The object initially showed large-amplitude (0.3 mag)
variations which resembled ordinary superhumps, but
they gradually became double-wave modulations characteristic
to early superhumps (vsnet-alert 18552, 18555, 18557,
18573; figure \ref{fig:asassn15hdeshpdm}).
The object then showed ordinary superhumps
(vsnet-alert 18582, 18585, 18594, 18604, 18609;
figure \ref{fig:asassn15hdshpdm}).
The times of superhump maxima are listed in
table \ref{tab:asassn15hdoc2015}.
There was a clear transition from stage A to B around $E$=22.
There was no strong indication of transition to stage C
(figure \ref{fig:asassn15hdhumpall}).

   The $\epsilon^*$ of 0.028(4) for stage A superhumps
corresponds to $q$=0.076(12).  This relatively low $q$
value appears to be consistent with the small $P_{\rm dot}$
of stage B superhumps (cf. \cite{kat15wzsge}).
There was one post-superoutburst rebrightening
(figure \ref{fig:asassn15hdhumpall}).  We cannot, however,
exclude the possibility of more rebrightenings since
the object was not well observed after the superoutburst.

   The object is notable for its initially large amplitude of
early superhumps and the singly-peaked ``saw-tooth''
profile at the beginning (figure \ref{fig:asassn15hdhumpall}).
The mean amplitude of early superhumps of 0.09 mag
implies that ASASSN-15hd belongs to a group of
WZ Sge-type objects with largest amplitudes of
early superhumps \citep{kat15wzsge}.  The initial
``saw-tooth''-like profile was also recorded in
V455 And (cf. \cite{kat15wzsge}).  Since the large
amplitude of early superhumps strongly suggests
a high orbital inclination, the common presence
of the ``saw-tooth''-like profile in this system
and an eclipsing system V455 And suggests that
this feature is seen only in high-inclination systems.

% SI

% SI

\begin{figure}
  \begin{center}
%    \FigureFile(85mm,180mm){asassn15hdprof.eps}
    \FigureFile(85mm,180mm){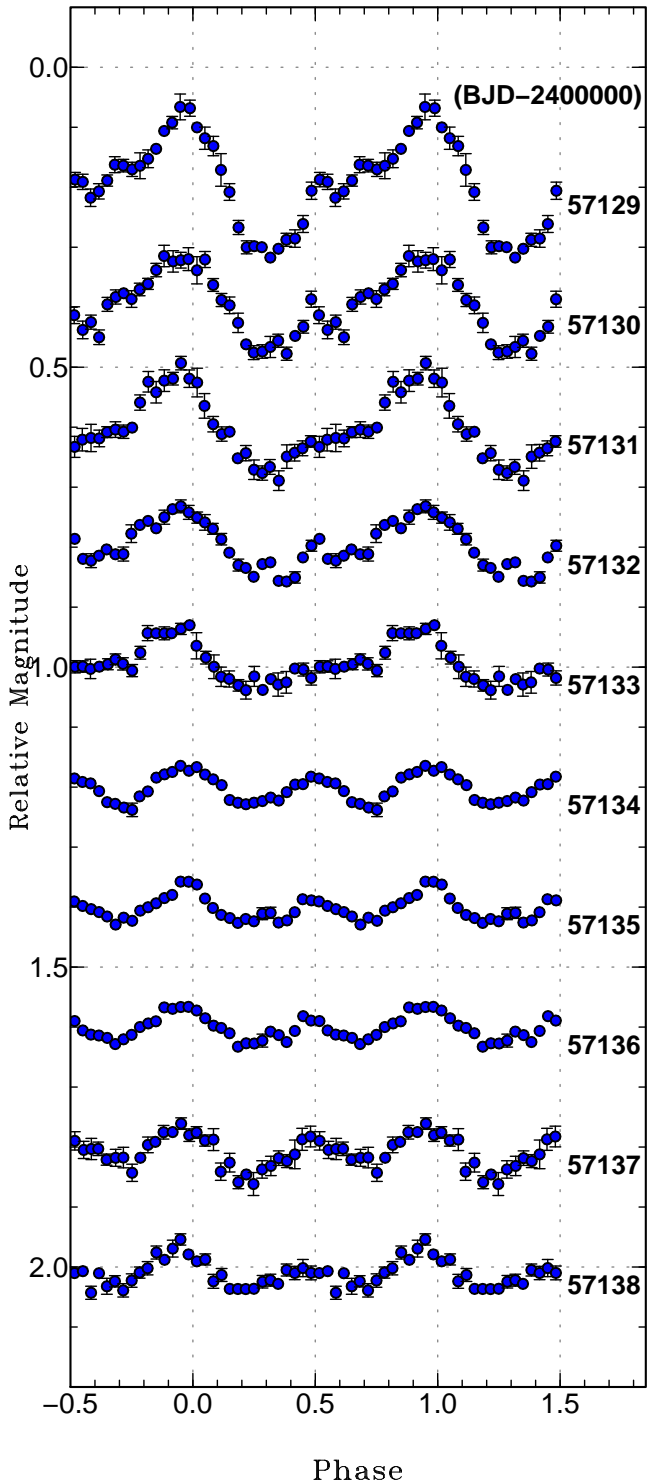}
  \end{center}
  \caption{Evolution of profile of early superhumps in 
     ASASSN-15hd (2015).  A period of 0.055410~d was
     used to draw this figure.}
  \label{fig:asassn15hdprof}
\end{figure}

% SI

\begin{table*}
\caption{Superhump maxima of ASASSN-15hd (2015)}\label{tab:asassn15hdoc2015}
\begin{center}
\begin{tabular}{rp{55pt}p{40pt}r@{.}lrrp{55pt}p{40pt}r@{.}lr}
\hline
\multicolumn{1}{c}{$E$} & \multicolumn{1}{c}{max\commenta} & \multicolumn{1}{c}{error} & \multicolumn{2}{c}{$O-C$\commentb} & \multicolumn{1}{c}{$N$\commentc} & \multicolumn{1}{c}{$E$} & \multicolumn{1}{c}{max\commenta} & \multicolumn{1}{c}{error} & \multicolumn{2}{c}{$O-C$\commentb} & \multicolumn{1}{c}{$N$\commentc} \\
\hline
0 & 57139.1728 & 0.0029 & $-$0&0089 & 64 & 77 & 57143.5007 & 0.0009 & $-$0&0019 & 51 \\
1 & 57139.2179 & 0.0023 & $-$0&0199 & 88 & 78 & 57143.5557 & 0.0011 & $-$0&0029 & 51 \\
17 & 57140.1378 & 0.0012 & 0&0021 & 24 & 79 & 57143.6119 & 0.0008 & $-$0&0029 & 50 \\
18 & 57140.1954 & 0.0004 & 0&0036 & 42 & 89 & 57144.1736 & 0.0016 & $-$0&0024 & 55 \\
19 & 57140.2456 & 0.0028 & $-$0&0023 & 17 & 98 & 57144.6879 & 0.0022 & 0&0069 & 37 \\
22 & 57140.4238 & 0.0008 & 0&0076 & 32 & 99 & 57144.7338 & 0.0013 & $-$0&0033 & 56 \\
23 & 57140.4787 & 0.0009 & 0&0064 & 31 & 100 & 57144.7911 & 0.0009 & $-$0&0021 & 51 \\
24 & 57140.5387 & 0.0017 & 0&0102 & 16 & 101 & 57144.8459 & 0.0012 & $-$0&0034 & 57 \\
41 & 57141.4882 & 0.0022 & 0&0057 & 25 & 160 & 57148.1526 & 0.0018 & $-$0&0076 & 39 \\
42 & 57141.5412 & 0.0004 & 0&0027 & 52 & 166 & 57148.4916 & 0.0022 & $-$0&0052 & 38 \\
43 & 57141.5961 & 0.0005 & 0&0015 & 43 & 167 & 57148.5533 & 0.0027 & 0&0004 & 37 \\
44 & 57141.6531 & 0.0003 & 0&0023 & 50 & 168 & 57148.6039 & 0.0013 & $-$0&0051 & 22 \\
45 & 57141.7082 & 0.0005 & 0&0014 & 55 & 182 & 57149.3930 & 0.0013 & $-$0&0016 & 35 \\
46 & 57141.7654 & 0.0006 & 0&0024 & 56 & 183 & 57149.4521 & 0.0012 & 0&0013 & 34 \\
47 & 57141.8247 & 0.0018 & 0&0056 & 13 & 184 & 57149.5083 & 0.0011 & 0&0014 & 78 \\
48 & 57141.8770 & 0.0011 & 0&0018 & 30 & 185 & 57149.5630 & 0.0011 & $-$0&0001 & 72 \\
54 & 57142.2127 & 0.0004 & 0&0008 & 25 & 186 & 57149.6157 & 0.0010 & $-$0&0034 & 60 \\
55 & 57142.2687 & 0.0004 & 0&0006 & 25 & 187 & 57149.6808 & 0.0041 & 0&0055 & 19 \\
58 & 57142.4382 & 0.0030 & 0&0018 & 30 & 217 & 57151.3577 & 0.0008 & $-$0&0010 & 30 \\
59 & 57142.4920 & 0.0018 & $-$0&0005 & 27 & 218 & 57151.4157 & 0.0010 & 0&0009 & 23 \\
60 & 57142.5511 & 0.0010 & 0&0025 & 32 & 250 & 57153.2062 & 0.0022 & $-$0&0043 & 42 \\
61 & 57142.6055 & 0.0012 & 0&0007 & 40 & 262 & 57153.8904 & 0.0015 & 0&0065 & 92 \\
62 & 57142.6602 & 0.0007 & $-$0&0006 & 30 & 270 & 57154.3285 & 0.0019 & $-$0&0044 & 30 \\
71 & 57143.1641 & 0.0011 & $-$0&0018 & 87 & 271 & 57154.3933 & 0.0023 & 0&0044 & 31 \\
72 & 57143.2194 & 0.0015 & $-$0&0026 & 87 & 272 & 57154.4447 & 0.0020 & $-$0&0004 & 30 \\
75 & 57143.3899 & 0.0009 & $-$0&0005 & 53 & 273 & 57154.5065 & 0.0010 & 0&0054 & 30 \\
76 & 57143.4432 & 0.0007 & $-$0&0033 & 45 & \multicolumn{1}{c}{--} & \multicolumn{1}{c}{--} & \multicolumn{1}{c}{--} & \multicolumn{2}{c}{--} & \multicolumn{1}{c}{--}\\
\hline
  \multicolumn{12}{l}{\commenta BJD$-$2400000.} \\
  \multicolumn{12}{l}{\commentb Against max $= 2457139.1817 + 0.056115 E$.} \\
  \multicolumn{12}{l}{\commentc Number of points used to determine the maximum.} \\
\end{tabular}
\end{center}
\end{table*}

\subsection{ASASSN-15hl}\label{obj:asassn15hl}

\begin{figure}
  \begin{center}
%    \FigureFile(85mm,110mm){asassn15hlshpdm.eps}
    \FigureFile(85mm,110mm){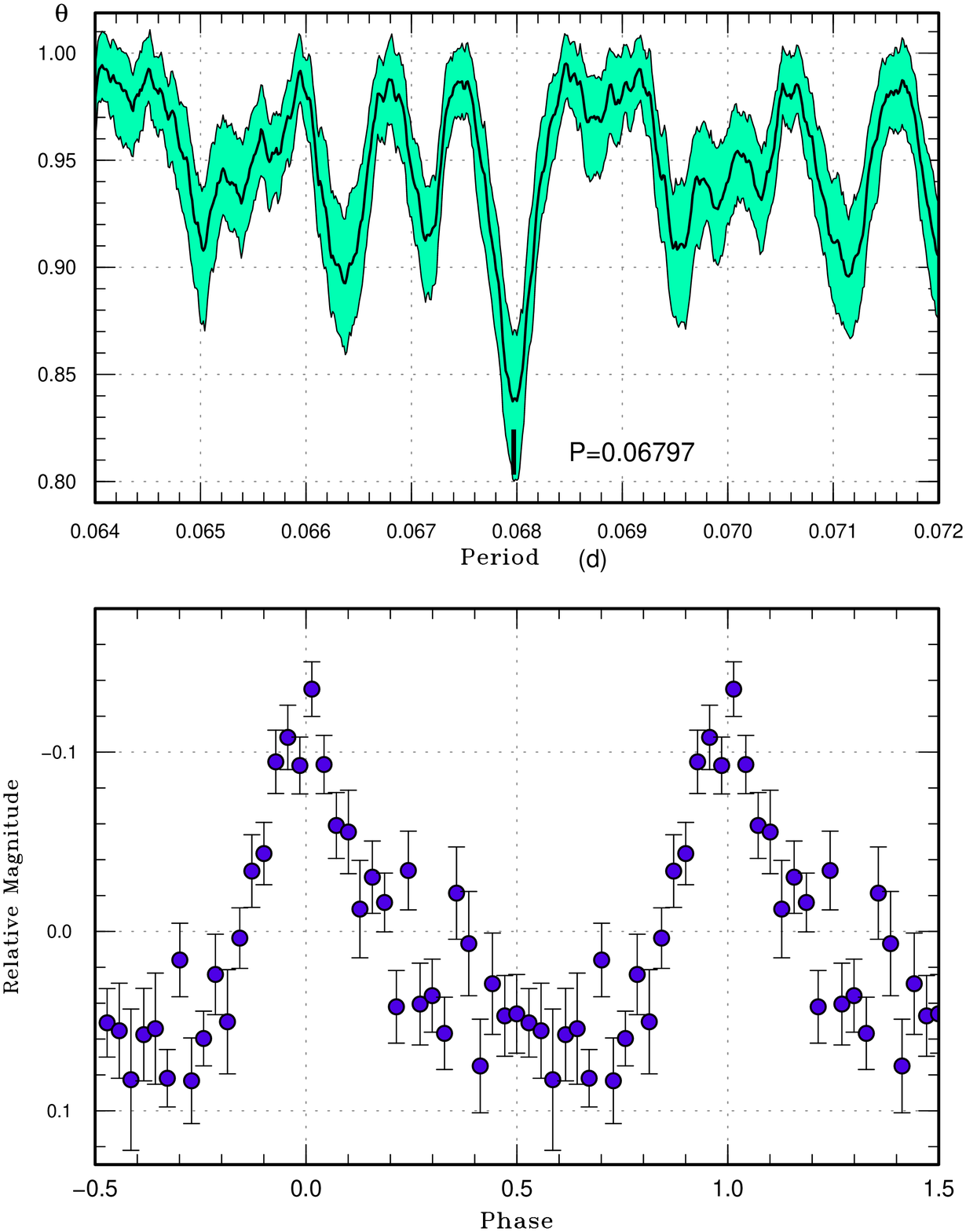}
  \end{center}
  \caption{Superhumps in ASASSN-15hl (2015).
     (Upper): PDM analysis.
     (Lower): Phase-averaged profile.}
  \label{fig:asassn15hlshpdm}
\end{figure}

   This object was detected as a transient at $V$=16.2
on 2015 April 19 by the ASAS-SN team.
Superhumps (figure \ref{fig:asassn15hlshpdm}) were
soon detected (vsnet-alert 18568).
The times of superhump maxima are listed in
table \ref{tab:asassn15hloc2015}.  The superhump
stage is unknown.

% SI

% SI

\begin{table}
\caption{Superhump maxima of ASASSN-15hl (2015)}\label{tab:asassn15hloc2015}
\begin{center}
\begin{tabular}{rp{55pt}p{40pt}r@{.}lr}
\hline
\multicolumn{1}{c}{$E$} & \multicolumn{1}{c}{max\commenta} & \multicolumn{1}{c}{error} & \multicolumn{2}{c}{$O-C$\commentb} & \multicolumn{1}{c}{$N$\commentc} \\
\hline
0 & 57134.2526 & 0.0010 & $-$0&0007 & 116 \\
1 & 57134.3197 & 0.0007 & $-$0&0016 & 147 \\
44 & 57137.2457 & 0.0012 & 0&0028 & 157 \\
45 & 57137.3114 & 0.0015 & 0&0005 & 126 \\
74 & 57139.2854 & 0.0017 & 0&0040 & 156 \\
88 & 57140.2276 & 0.0013 & $-$0&0050 & 123 \\
89 & 57140.3006 & 0.0030 & 0&0000 & 124 \\
\hline
  \multicolumn{6}{l}{\commenta BJD$-$2400000.} \\
  \multicolumn{6}{l}{\commentb Against max $= 2457134.2533 + 0.067947 E$.} \\
  \multicolumn{6}{l}{\commentc Number of points used to determine the maximum.} \\
\end{tabular}
\end{center}
\end{table}

\subsection{ASASSN-15hm}\label{obj:asassn15hm}

\begin{figure}
  \begin{center}
%    \FigureFile(85mm,110mm){asassn15hmshpdm.eps}
    \FigureFile(85mm,110mm){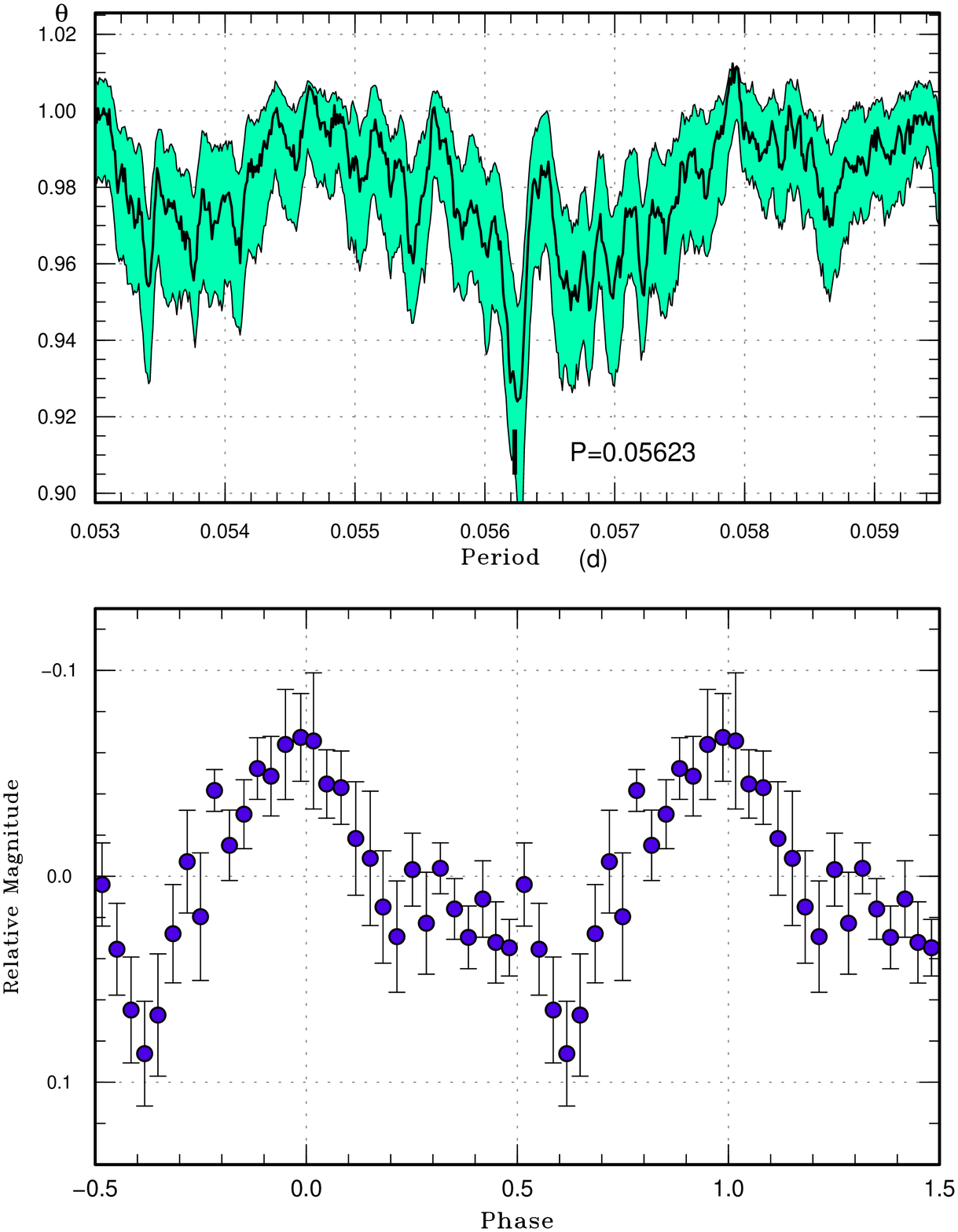}
  \end{center}
  \caption{Superhumps in ASASSN-15hm (2015).
     (Upper): PDM analysis.
     (Lower): Phase-averaged profile.}
  \label{fig:asassn15hmshpdm}
\end{figure}

   This object was detected as a transient at $V$=13.6
on 2015 April 18 by the ASAS-SN team (vsnet-alert 18563).
On April 26 (8~d after the outburst detection),
this object started to show superhumps
(vsnet-alert 18578, 18587; figure \ref{fig:asassn15hmshpdm}).
The times of superhump maxima are listed in
table \ref{tab:asassn15hmoc2015}.
The $O-C$ values clearly indicate that stage B started
relatively late (around $E$=34).
The slow evolution of superhumps suggests
a relatively low $q$, although it was impossible to
determine the $q$ value from stage A superhump
due to the lack of information about the orbital period.
Although the large outburst amplitude suggests
a WZ Sge-type object, we do not have evidence for it.

% SI

% SI

\begin{table}
\caption{Superhump maxima of ASASSN-15hm (2015)}\label{tab:asassn15hmoc2015}
\begin{center}
\begin{tabular}{rp{55pt}p{40pt}r@{.}lr}
\hline
\multicolumn{1}{c}{$E$} & \multicolumn{1}{c}{max\commenta} & \multicolumn{1}{c}{error} & \multicolumn{2}{c}{$O-C$\commentb} & \multicolumn{1}{c}{$N$\commentc} \\
\hline
0 & 57138.5846 & 0.0010 & $-$0&0089 & 15 \\
9 & 57139.0952 & 0.0028 & $-$0&0042 & 85 \\
16 & 57139.4893 & 0.0054 & $-$0&0037 & 8 \\
17 & 57139.5550 & 0.0004 & 0&0057 & 19 \\
25 & 57139.9984 & 0.0059 & $-$0&0006 & 53 \\
26 & 57140.0586 & 0.0013 & 0&0034 & 101 \\
34 & 57140.5112 & 0.0013 & 0&0062 & 17 \\
35 & 57140.5667 & 0.0009 & 0&0056 & 19 \\
52 & 57141.5179 & 0.0021 & 0&0010 & 19 \\
53 & 57141.5761 & 0.0020 & 0&0030 & 16 \\
70 & 57142.5275 & 0.0029 & $-$0&0013 & 18 \\
71 & 57142.5879 & 0.0061 & 0&0029 & 12 \\
88 & 57143.5334 & 0.0030 & $-$0&0074 & 15 \\
105 & 57144.4941 & 0.0015 & $-$0&0024 & 13 \\
106 & 57144.5528 & 0.0036 & 0&0002 & 23 \\
159 & 57147.5328 & 0.0023 & 0&0005 & 19 \\
\hline
  \multicolumn{6}{l}{\commenta BJD$-$2400000.} \\
  \multicolumn{6}{l}{\commentb Against max $= 2457138.5935 + 0.056219 E$.} \\
  \multicolumn{6}{l}{\commentc Number of points used to determine the maximum.} \\
\end{tabular}
\end{center}
\end{table}

\subsection{ASASSN-15hn}\label{obj:asassn15hn}

\begin{figure}
  \begin{center}
%    \FigureFile(85mm,100mm){asassn15hnhumpall.eps}
    \FigureFile(85mm,100mm){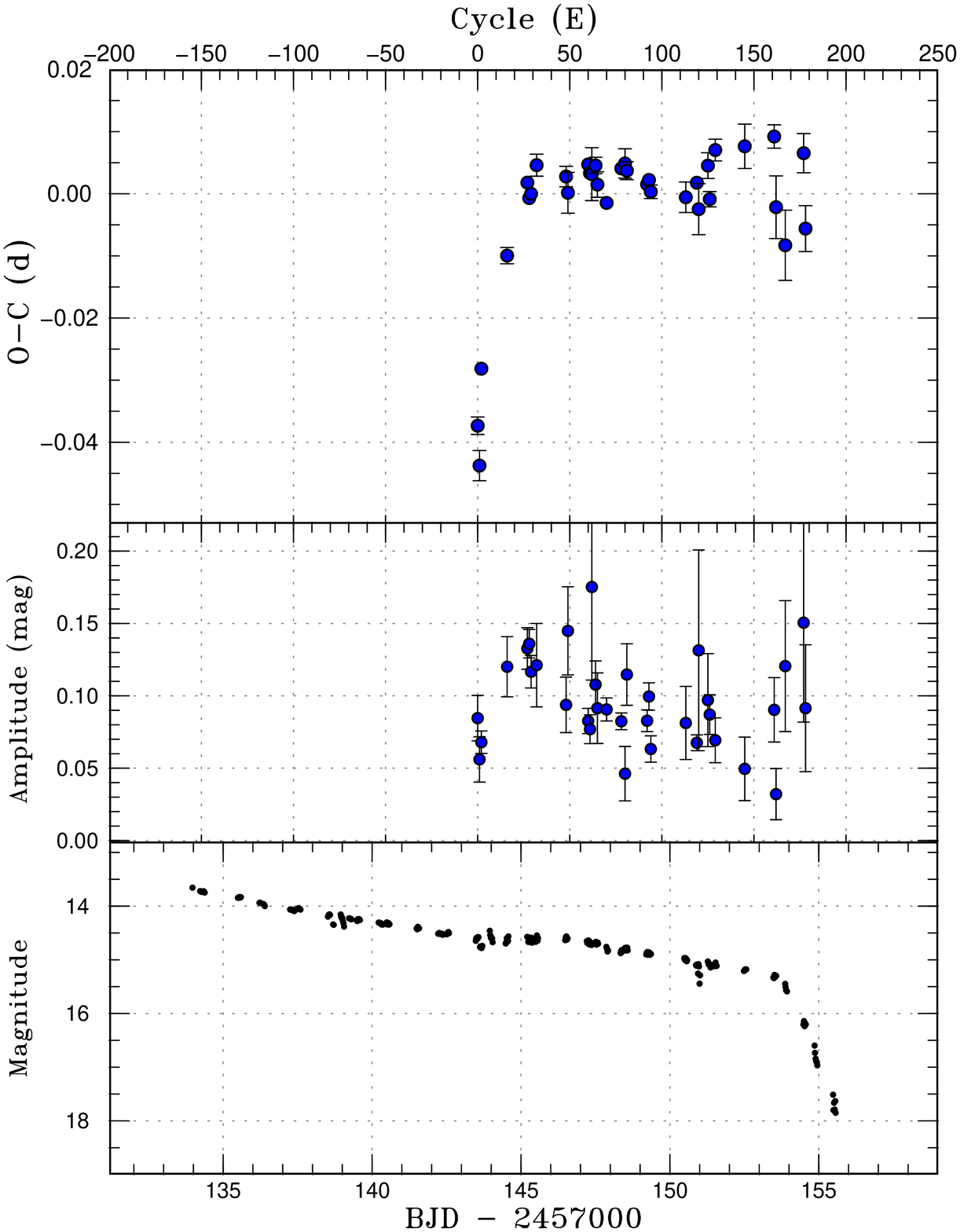}
  \end{center}
  \caption{$O-C$ diagram of superhumps in ASASSN-15hn (2015).
     (Upper:) $O-C$ diagram.
     We used a period of 0.06185~d for calculating the $O-C$ residuals.
     (Middle:) Amplitudes of superhumps.
     (Lower:) Light curve.  The data were binned to 0.021~d.
  }
  \label{fig:asassn15hnhumpall}
\end{figure}

\begin{figure}
  \begin{center}
%    \FigureFile(85mm,110mm){asassn15hnshpdm.eps}
    \FigureFile(85mm,110mm){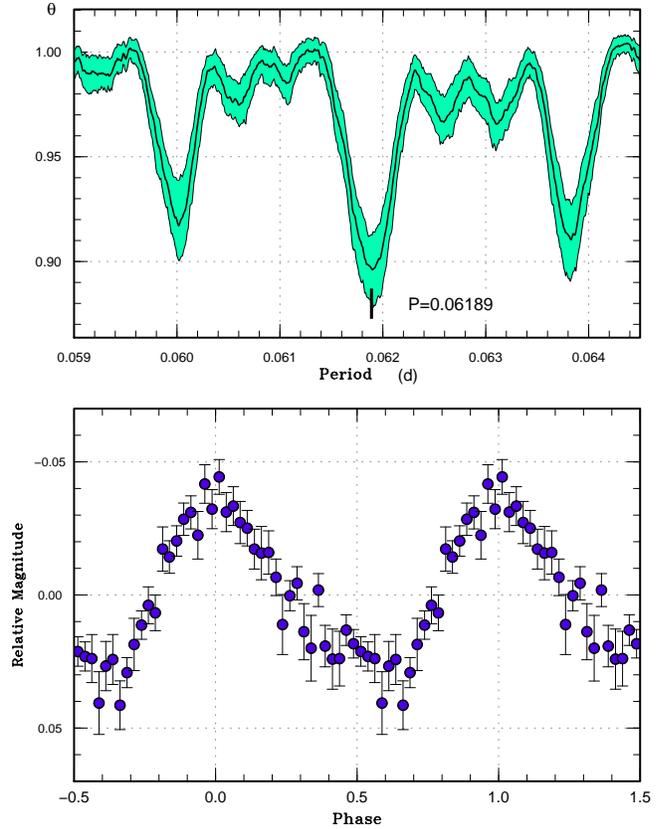}
  \end{center}
  \caption{Ordinary superhumps in ASASSN-15hn during the plateau phase (2015).
     (Upper): PDM analysis.
     (Lower): Phase-averaged profile.}
  \label{fig:asassn15hnshpdm}
\end{figure}

   This object was detected as a transient at $V$=12.9
on 2015 April 17 by the ASAS-SN team (vsnet-alert 18564).
The quiescent counterpart was very faint (21.9 mag
in GSC 2.3.2 on J plate) and the large outburst
amplitude suggested a WZ Sge-type dwarf nova.
Ordinary superhumps started to appear on
April 30--May 1 (13--14~d after the outburst detection,
vsnet-alert 18592, 18599, 18605, 18610;
figure \ref{fig:asassn15hnshpdm}).
The object started to fade rapidly on May 11,
24~d after the outburst detection.
The times of superhump maxima are listed in
table \ref{tab:asassn15hnoc2015}.
Stage A lasted nearly 30 cycles
(figure \ref{fig:asassn15hnhumpall}), which implies
a small $q$ \citep{kat15wzsge}.

   There was no indication of early superhumps before
the appearance of ordinary superhumps.  The upper limit
for the amplitude of early superhumps was 0.005 mag,
probably suggesting a low orbital inclination.
Although this object did not show early superhumps,
all the observed features suggest the WZ Sge-type
classification: long duration (13--14~d) before 
the appearance of ordinary superhumps, long duration
of stage A ($\sim$30 cycles), small $P_{\rm dot}$
for stage B superhumps, low amplitude of superhumps
and the lack of stage C superhumps.
The empirical relation between $P_{\rm dot}$
for stage B superhumps and $q$ (equation 6 in
\cite{kat15wzsge}) gives a $q$ of 0.058(9).
Since the object has a relatively long superhump
period, this estimated $q$ would place it
in a region of period bouncers.  The apparent
large outburst amplitude would favor this
interpretation.  Ordinary superhumps started to appear
at 14.7 mag.  Although quiescent magnitude is highly
uncertain, the ``amplitude'' when ordinary superhumps
appear (7.2 mag) is also in the region of period bouncers
(see figure 23 of \cite{kat15wzsge}).

% SI

% SI

\begin{table}
\caption{Superhump maxima of ASASSN-15hn (2015)}\label{tab:asassn15hnoc2015}
\begin{center}
\begin{tabular}{rp{55pt}p{40pt}r@{.}lr}
\hline
\multicolumn{1}{c}{$E$} & \multicolumn{1}{c}{max\commenta} & \multicolumn{1}{c}{error} & \multicolumn{2}{c}{$O-C$\commentb} & \multicolumn{1}{c}{$N$\commentc} \\
\hline
0 & 57143.5085 & 0.0014 & $-$0&0266 & 15 \\
1 & 57143.5639 & 0.0024 & $-$0&0331 & 17 \\
2 & 57143.6414 & 0.0009 & $-$0&0176 & 90 \\
16 & 57144.5255 & 0.0013 & $-$0&0009 & 23 \\
27 & 57145.2176 & 0.0007 & 0&0097 & 111 \\
28 & 57145.2770 & 0.0006 & 0&0071 & 143 \\
29 & 57145.3395 & 0.0008 & 0&0077 & 143 \\
32 & 57145.5297 & 0.0018 & 0&0119 & 24 \\
48 & 57146.5174 & 0.0016 & 0&0084 & 23 \\
49 & 57146.5767 & 0.0033 & 0&0057 & 13 \\
60 & 57147.2616 & 0.0008 & 0&0091 & 142 \\
61 & 57147.3220 & 0.0010 & 0&0076 & 142 \\
62 & 57147.3837 & 0.0043 & 0&0073 & 36 \\
64 & 57147.5088 & 0.0013 & 0&0085 & 18 \\
65 & 57147.5676 & 0.0020 & 0&0053 & 20 \\
70 & 57147.8739 & 0.0007 & 0&0018 & 20 \\
78 & 57148.3742 & 0.0005 & 0&0065 & 83 \\
80 & 57148.4987 & 0.0024 & 0&0071 & 16 \\
81 & 57148.5594 & 0.0014 & 0&0058 & 21 \\
92 & 57149.2376 & 0.0007 & 0&0025 & 143 \\
93 & 57149.3001 & 0.0007 & 0&0031 & 143 \\
94 & 57149.3601 & 0.0011 & 0&0011 & 154 \\
113 & 57150.5343 & 0.0025 & $-$0&0019 & 21 \\
119 & 57150.9078 & 0.0007 & $-$0&0002 & 22 \\
120 & 57150.9654 & 0.0041 & $-$0&0045 & 56 \\
125 & 57151.2816 & 0.0021 & 0&0020 & 54 \\
126 & 57151.3381 & 0.0012 & $-$0&0036 & 143 \\
129 & 57151.5315 & 0.0018 & 0&0040 & 20 \\
145 & 57152.5217 & 0.0036 & 0&0029 & 20 \\
161 & 57153.5129 & 0.0019 & 0&0028 & 20 \\
162 & 57153.5634 & 0.0050 & $-$0&0087 & 16 \\
167 & 57153.8665 & 0.0056 & $-$0&0153 & 12 \\
177 & 57154.4998 & 0.0032 & $-$0&0016 & 23 \\
178 & 57154.5495 & 0.0037 & $-$0&0138 & 25 \\
\hline
  \multicolumn{6}{l}{\commenta BJD$-$2400000.} \\
  \multicolumn{6}{l}{\commentb Against max $= 2457143.5351 + 0.061957 E$.} \\
  \multicolumn{6}{l}{\commentc Number of points used to determine the maximum.} \\
\end{tabular}
\end{center}
\end{table}

\subsection{ASASSN-15ia}\label{obj:asassn15ia}

\begin{figure}
  \begin{center}
%    \FigureFile(85mm,110mm){asassn15iashpdm.eps}
    \FigureFile(85mm,110mm){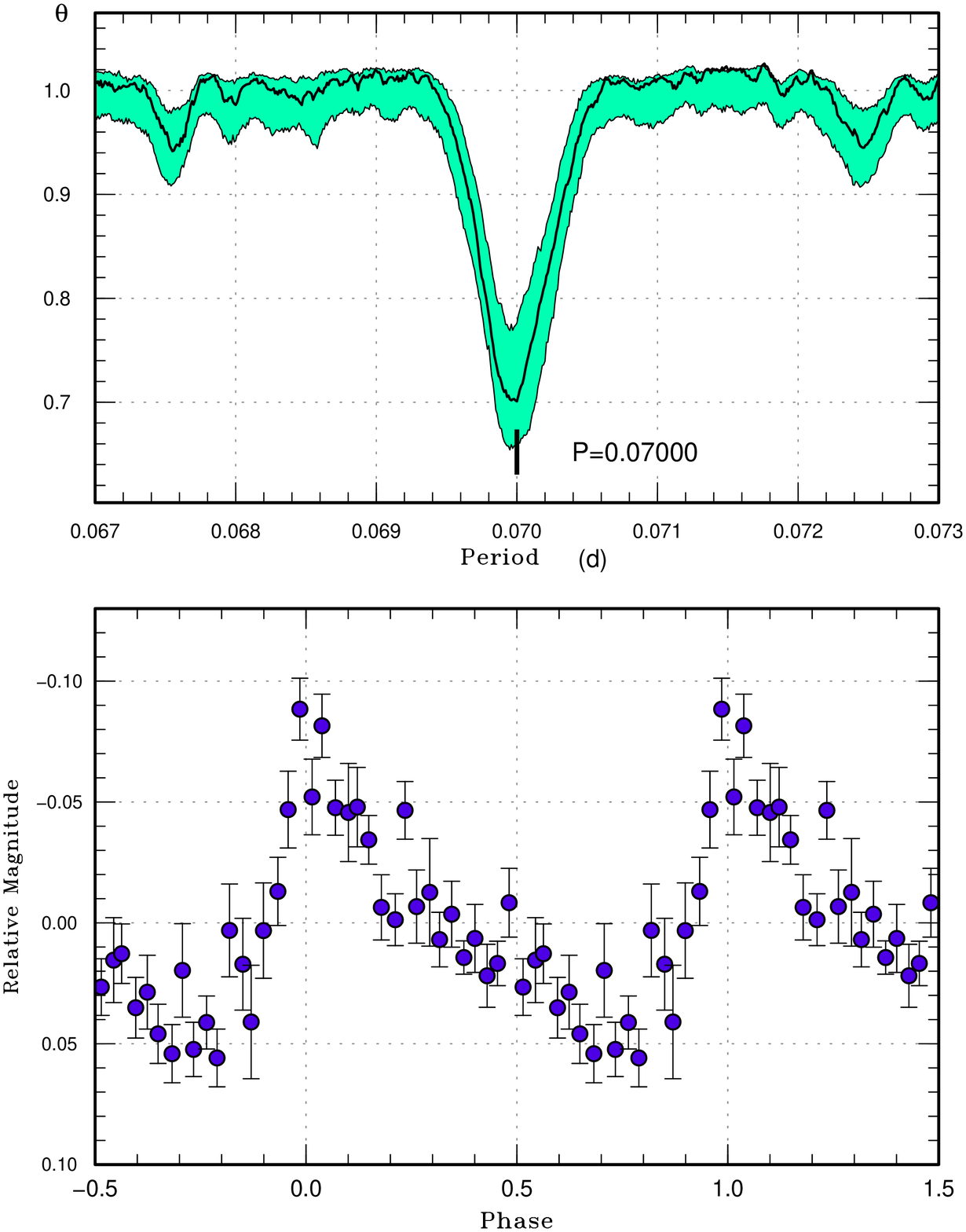}
  \end{center}
  \caption{Superhumps in ASASSN-15ia during the plateau phase (2015).
     (Upper): PDM analysis.
     (Lower): Phase-averaged profile.}
  \label{fig:asassn15iashpdm}
\end{figure}

   This object was detected as a transient at $V$=15.3
on 2015 April 25 by the ASAS-SN team.
The object faded to $V$=15.6 on April 26 and brightened
to $V$=15.3 on April 27, suggesting a precursor outburst.
Superhumps were detected in observations starting 4~d
after the outburst detection (figure \ref{fig:asassn15iashpdm}).
The times of superhump maxima are listed in
table \ref{tab:asassn15iaoc2015}.
Both stages B and C were recorded.  The early part of
stage B was not observed.
The rapid fading from the superoutburst plateau apparently
took place on May 7.  The total duration of the superoutburst
was 11~d.

% SI

% SI

\begin{table}
\caption{Superhump maxima of ASASSN-15ia (2015)}\label{tab:asassn15iaoc2015}
\begin{center}
\begin{tabular}{rp{55pt}p{40pt}r@{.}lr}
\hline
\multicolumn{1}{c}{$E$} & \multicolumn{1}{c}{max\commenta} & \multicolumn{1}{c}{error} & \multicolumn{2}{c}{$O-C$\commentb} & \multicolumn{1}{c}{$N$\commentc} \\
\hline
0 & 57141.8193 & 0.0012 & $-$0&0008 & 22 \\
1 & 57141.8864 & 0.0013 & $-$0&0037 & 19 \\
14 & 57142.7982 & 0.0014 & $-$0&0025 & 21 \\
15 & 57142.8698 & 0.0010 & $-$0&0009 & 23 \\
28 & 57143.7852 & 0.0017 & 0&0039 & 18 \\
29 & 57143.8562 & 0.0016 & 0&0048 & 18 \\
42 & 57144.7618 & 0.0018 & $-$0&0001 & 18 \\
43 & 57144.8377 & 0.0011 & 0&0058 & 23 \\
44 & 57144.9031 & 0.0016 & 0&0011 & 12 \\
57 & 57145.8078 & 0.0017 & $-$0&0047 & 19 \\
58 & 57145.8838 & 0.0030 & 0&0012 & 18 \\
71 & 57146.7892 & 0.0018 & $-$0&0040 & 27 \\
72 & 57146.8631 & 0.0064 & $-$0&0001 & 29 \\
\hline
  \multicolumn{6}{l}{\commenta BJD$-$2400000.} \\
  \multicolumn{6}{l}{\commentb Against max $= 2457141.8201 + 0.070043 E$.} \\
  \multicolumn{6}{l}{\commentc Number of points used to determine the maximum.} \\
\end{tabular}
\end{center}
\end{table}

\subsection{ASASSN-15ie}\label{obj:asassn15ie}

\begin{figure}
  \begin{center}
%    \FigureFile(85mm,110mm){asassn15ieshpdm.eps}
    \FigureFile(85mm,110mm){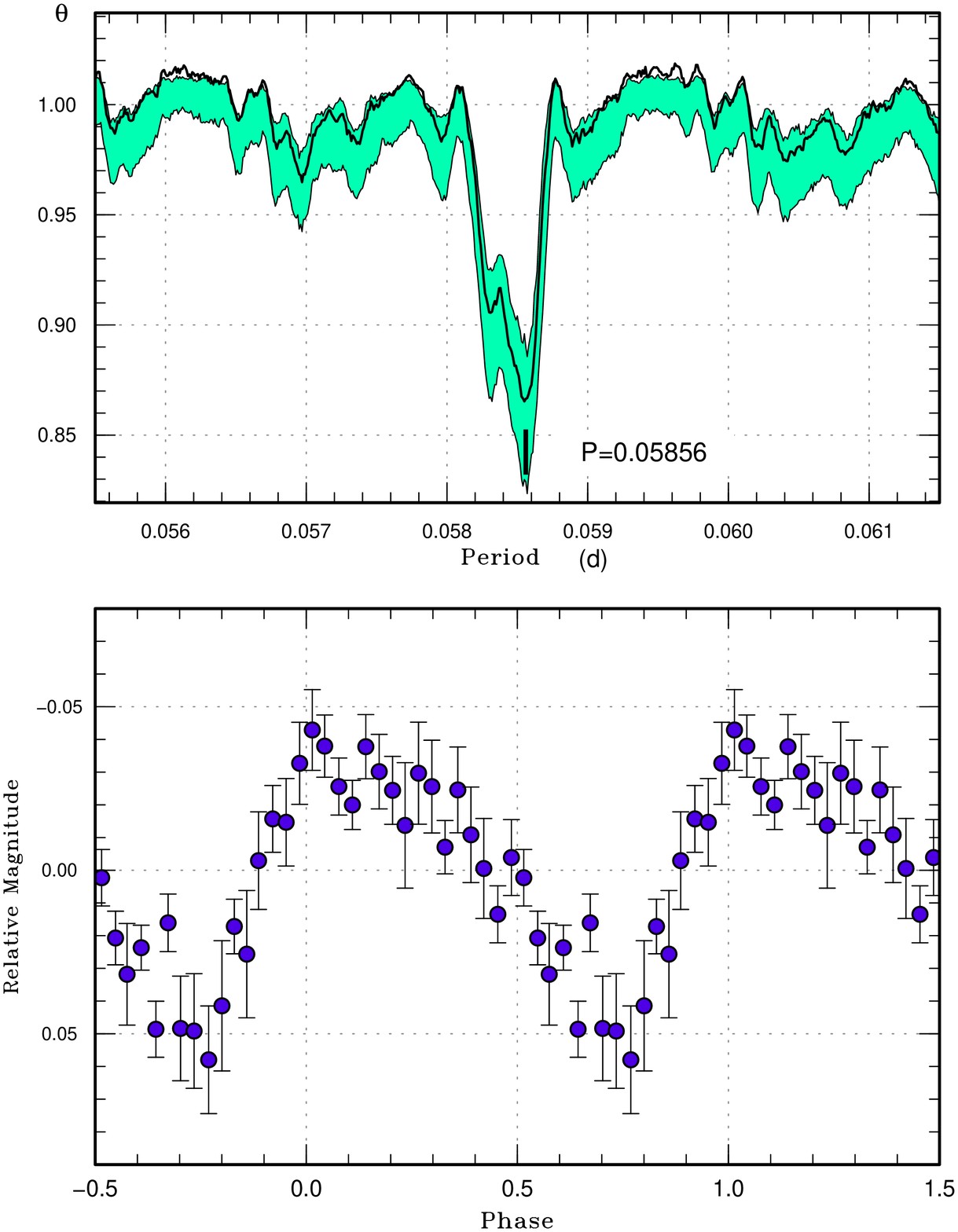}
  \end{center}
  \caption{Superhumps in ASASSN-15ie (2015).
     (Upper): PDM analysis.
     (Lower): Phase-averaged profile.}
  \label{fig:asassn15ieshpdm}
\end{figure}

   This object was detected as a transient at $V$=14.0
on 2015 May 4 by the ASAS-SN team.
On May 9 (5~d after the outburst detection), superhumps
started to appear (vsnet-alert 18606, 18611, 18616, 18634;
figure \ref{fig:asassn15ieshpdm}).
The times of superhump maxima are listed in
table \ref{tab:asassn15ieoc2015}.
The maxima for $E \le 2$ recorded stage A superhumps.
Although there were observations, the times of
maxima between $E$=138 and $E$=222 could not be determined
due to the low amplitudes of superhumps and the faintness
(16.2 mag) of the object.
Tha maxima around $E$=223 correspond to stage B-C transition,
when superhump again became apparent and the object
slightly brightened.
The object started fading rapidly on May 25 and the total
duration of the superoutburst was 20~d.

   Although the apparent large outburst amplitude,
the short delay before the appearance of superhumps
and the presence of stage C make this object less likely
an extreme WZ Sge-type object.

% SI

% SI

\begin{table}
\caption{Superhump maxima of ASASSN-15ie (2015)}\label{tab:asassn15ieoc2015}
\begin{center}
\begin{tabular}{rp{55pt}p{40pt}r@{.}lr}
\hline
\multicolumn{1}{c}{$E$} & \multicolumn{1}{c}{max\commenta} & \multicolumn{1}{c}{error} & \multicolumn{2}{c}{$O-C$\commentb} & \multicolumn{1}{c}{$N$\commentc} \\
\hline
0 & 57151.7387 & 0.0010 & 0&0070 & 15 \\
1 & 57151.7990 & 0.0007 & 0&0087 & 15 \\
2 & 57151.8578 & 0.0012 & 0&0089 & 13 \\
35 & 57153.7938 & 0.0011 & 0&0121 & 19 \\
52 & 57154.7765 & 0.0007 & $-$0&0009 & 16 \\
53 & 57154.8370 & 0.0010 & 0&0009 & 16 \\
69 & 57155.7686 & 0.0007 & $-$0&0046 & 16 \\
70 & 57155.8261 & 0.0008 & $-$0&0056 & 15 \\
71 & 57155.8853 & 0.0010 & $-$0&0050 & 16 \\
86 & 57156.7624 & 0.0008 & $-$0&0065 & 23 \\
87 & 57156.8225 & 0.0007 & $-$0&0049 & 20 \\
88 & 57156.8801 & 0.0012 & $-$0&0059 & 17 \\
103 & 57157.7586 & 0.0011 & $-$0&0060 & 28 \\
104 & 57157.8169 & 0.0013 & $-$0&0063 & 27 \\
105 & 57157.8766 & 0.0015 & $-$0&0051 & 18 \\
120 & 57158.7574 & 0.0008 & $-$0&0029 & 28 \\
121 & 57158.8150 & 0.0013 & $-$0&0039 & 27 \\
122 & 57158.8727 & 0.0020 & $-$0&0048 & 19 \\
137 & 57159.7568 & 0.0015 & 0&0007 & 17 \\
138 & 57159.8168 & 0.0027 & 0&0022 & 14 \\
222 & 57164.7413 & 0.0017 & 0&0066 & 15 \\
223 & 57164.8031 & 0.0022 & 0&0099 & 15 \\
224 & 57164.8565 & 0.0016 & 0&0047 & 8 \\
240 & 57165.7930 & 0.0024 & 0&0041 & 14 \\
241 & 57165.8442 & 0.0022 & $-$0&0033 & 11 \\
\hline
  \multicolumn{6}{l}{\commenta BJD$-$2400000.} \\
  \multicolumn{6}{l}{\commentb Against max $= 2457151.7317 + 0.058572 E$.} \\
  \multicolumn{6}{l}{\commentc Number of points used to determine the maximum.} \\
\end{tabular}
\end{center}
\end{table}

\subsection{ASASSN-15iv}\label{obj:asassn15iv}

\begin{figure}
  \begin{center}
%    \FigureFile(85mm,110mm){asassn15ivshpdm.eps}
    \FigureFile(85mm,110mm){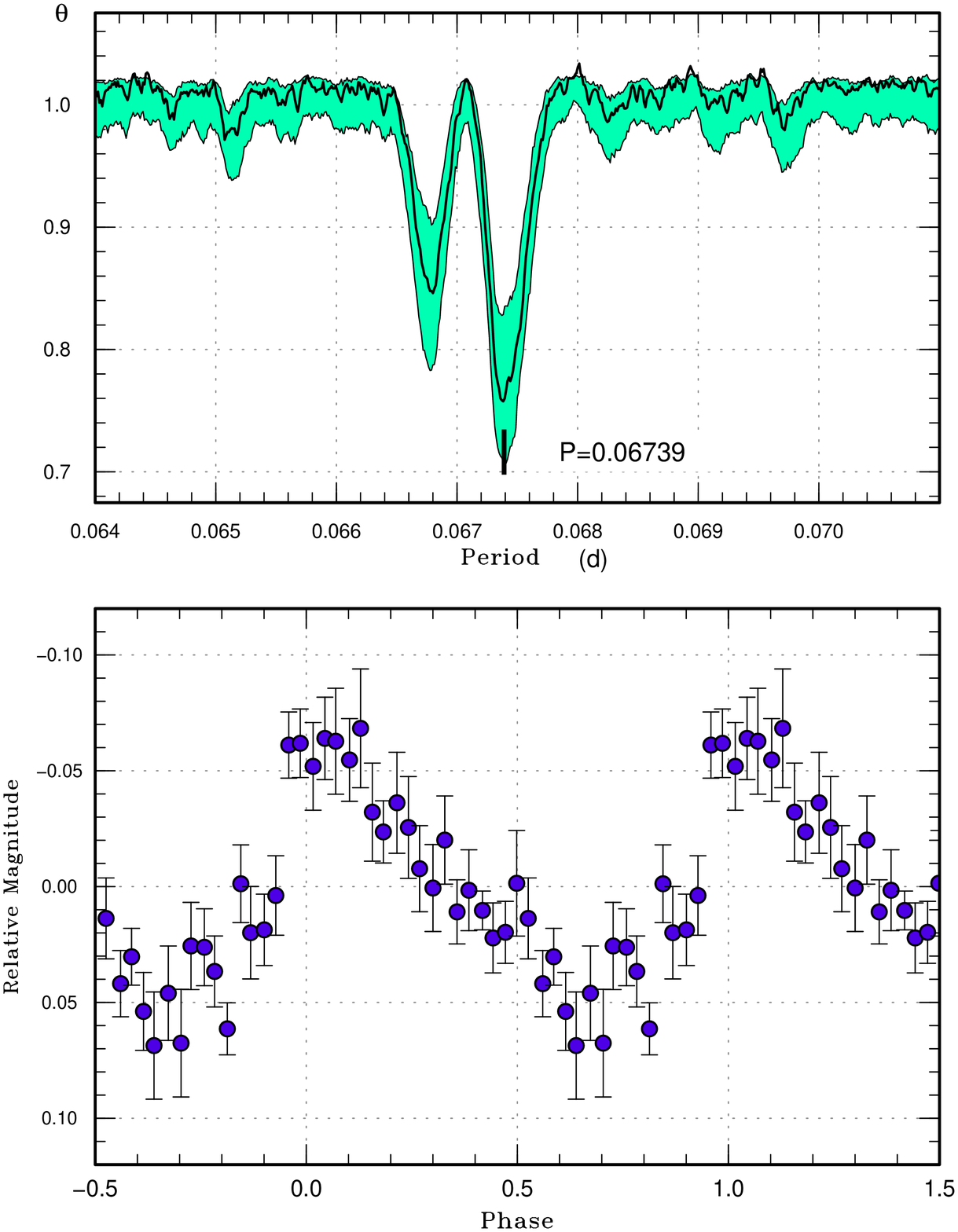}
  \end{center}
  \caption{Superhumps in ASASSN-15iv (2015).
     (Upper): PDM analysis.  The apparent signal around
     0.0668~d was most likely an artefact produced by
     a combination of stage B and C superhumps and
     observational intervals.
     This signal did not appear when we analyzed
     the segments of stage B and C individually. 
     (Lower): Phase-averaged profile.}
  \label{fig:asassn15ivshpdm}
\end{figure}

   This object was detected as a transient at $V$=15.8
on 2015 May 11 by the ASAS-SN team.
Superhumps were present already 2~d after the outburst
detection (vsnet-alert 18614, 18643;
figure \ref{fig:asassn15ivshpdm}).
The times of superhump maxima are listed in
table \ref{tab:asassn15ivoc2015}.
There was clear stage B-C transition around $E$=88.
The object started fading rapidly between May 23 and 25.
The total duration of the superoutburst was about 13~d.

% SI

% SI

\begin{table}
\caption{Superhump maxima of ASASSN-15iv (2015)}\label{tab:asassn15ivoc2015}
\begin{center}
\begin{tabular}{rp{55pt}p{40pt}r@{.}lr}
\hline
\multicolumn{1}{c}{$E$} & \multicolumn{1}{c}{max\commenta} & \multicolumn{1}{c}{error} & \multicolumn{2}{c}{$O-C$\commentb} & \multicolumn{1}{c}{$N$\commentc} \\
\hline
0 & 57155.6363 & 0.0014 & 0&0004 & 19 \\
1 & 57155.7062 & 0.0013 & 0&0029 & 11 \\
15 & 57156.6463 & 0.0011 & $-$0&0003 & 21 \\
30 & 57157.6545 & 0.0015 & $-$0&0028 & 21 \\
44 & 57158.5963 & 0.0022 & $-$0&0044 & 13 \\
45 & 57158.6608 & 0.0019 & $-$0&0073 & 18 \\
59 & 57159.6047 & 0.0040 & $-$0&0067 & 12 \\
60 & 57159.6701 & 0.0032 & $-$0&0087 & 21 \\
74 & 57160.6241 & 0.0044 & 0&0020 & 18 \\
75 & 57160.6871 & 0.0031 & $-$0&0024 & 13 \\
87 & 57161.5077 & 0.0040 & 0&0096 & 16 \\
88 & 57161.5715 & 0.0076 & 0&0060 & 11 \\
89 & 57161.6393 & 0.0026 & 0&0065 & 19 \\
102 & 57162.5164 & 0.0021 & 0&0075 & 16 \\
103 & 57162.5818 & 0.0013 & 0&0056 & 11 \\
104 & 57162.6512 & 0.0038 & 0&0076 & 17 \\
117 & 57163.5209 & 0.0023 & 0&0014 & 15 \\
118 & 57163.6024 & 0.0023 & 0&0154 & 12 \\
119 & 57163.6494 & 0.0025 & $-$0&0049 & 17 \\
132 & 57164.5263 & 0.0074 & $-$0&0040 & 14 \\
134 & 57164.6588 & 0.0039 & $-$0&0063 & 15 \\
148 & 57165.6000 & 0.0033 & $-$0&0084 & 14 \\
149 & 57165.6671 & 0.0050 & $-$0&0087 & 13 \\
\hline
  \multicolumn{6}{l}{\commenta BJD$-$2400000.} \\
  \multicolumn{6}{l}{\commentb Against max $= 2457155.6359 + 0.067382 E$.} \\
  \multicolumn{6}{l}{\commentc Number of points used to determine the maximum.} \\
\end{tabular}
\end{center}
\end{table}

\subsection{ASASSN-15iz}\label{obj:asassn15iz}

\begin{figure}
  \begin{center}
%    \FigureFile(85mm,110mm){asassn15izshpdm.eps}
    \FigureFile(85mm,110mm){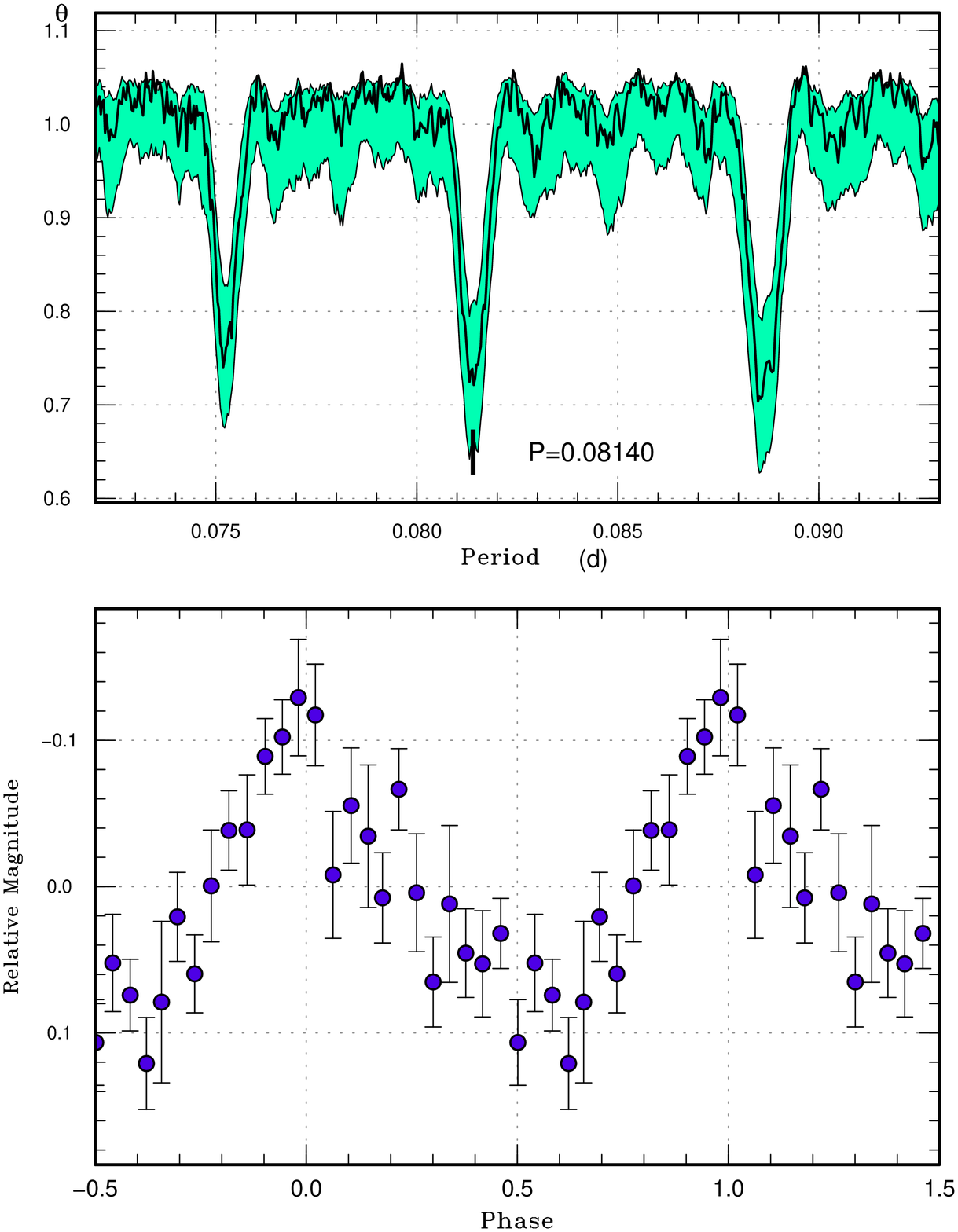}
  \end{center}
  \caption{Superhumps in ASASSN-15iz (2015).
     (Upper): PDM analysis.
     The alias selection was based on the $O-C$ analysis.
     The other one-day aliases [particularly 0.08863(7)~d]
     are still possible.
     (Lower): Phase-averaged profile.}
  \label{fig:asassn15izshpdm}
\end{figure}

   This object was detected as a transient at $V$=16.7
on 2015 May 11 by the ASAS-SN team.
It was initially suspected that this object could be
identified with a cataloged high proper motion object
(cf. vsnet-alert 18620).  This high proper motion,
however, was found to be spurious by examination
of archival plates (B. Skiff, vsnet-alert 18623).
The object showed large-amplitude superhumps
(vsnet-alert 18626, 18644; figure \ref{fig:asassn15izshpdm}).
The times of superhump maxima are listed in
table \ref{tab:asassn15izoc2015}.  In this table,
we adopted a period [0.08140(6)~d] which appears to give
the smallest scatter in the $O-C$ diagram.
Other longer one-day aliases [particularly 0.08863(7)~d]
are still possible, and the cycle numbers
in this table may be subject to correction.

% SI

% SI

\begin{table}
\caption{Superhump maxima of ASASSN-15iz (2015)}\label{tab:asassn15izoc2015}
\begin{center}
\begin{tabular}{rp{55pt}p{40pt}r@{.}lr}
\hline
\multicolumn{1}{c}{$E$} & \multicolumn{1}{c}{max\commenta} & \multicolumn{1}{c}{error} & \multicolumn{2}{c}{$O-C$\commentb} & \multicolumn{1}{c}{$N$\commentc} \\
\hline
0 & 57159.6872 & 0.0016 & 0&0013 & 16 \\
12 & 57160.6601 & 0.0020 & $-$0&0030 & 22 \\
24 & 57161.6415 & 0.0014 & 0&0012 & 23 \\
48 & 57163.5967 & 0.0037 & 0&0019 & 13 \\
61 & 57164.6520 & 0.0031 & $-$0&0014 & 17 \\
\hline
  \multicolumn{6}{l}{\commenta BJD$-$2400000.} \\
  \multicolumn{6}{l}{\commentb Against max $= 2457159.6859 + 0.081434 E$.} \\
  \multicolumn{6}{l}{\commentc Number of points used to determine the maximum.} \\
\end{tabular}
\end{center}
\end{table}

\subsection{ASASSN-15jj}\label{obj:asassn15jj}

\begin{figure}
  \begin{center}
%    \FigureFile(85mm,110mm){asassn15jjshpdm.eps}
    \FigureFile(85mm,110mm){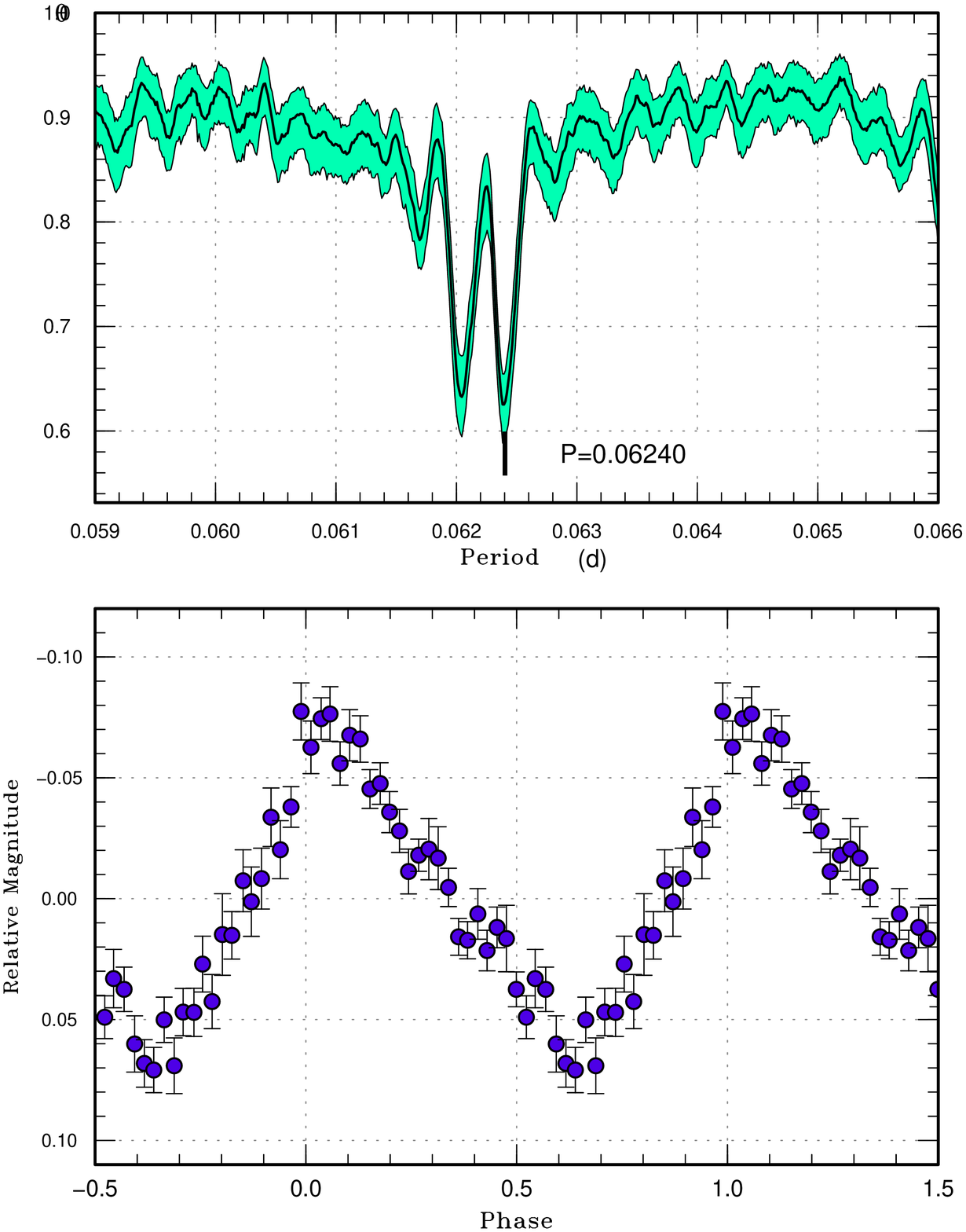}
  \end{center}
  \caption{Superhumps in ASASSN-15jj (2015).
     (Upper): PDM analysis.  The apparent signal around
     0.0620~d was most likely an artefact produced by
     a combination of the effect of strongly varying
     superhump periods and observational intervals
     rather than the signal of stage C superhumps.
     (Lower): Phase-averaged profile.}
  \label{fig:asassn15jjshpdm}
\end{figure}

   This object was detected as a transient at $V$=14.5
on 2015 May 16 by the ASAS-SN team.
Superhumps were already present on May 19 and were
continuously observed (vsnet-alert 18637, 18642,
18655; figure \ref{fig:asassn15jjshpdm}).
The times of superhump maxima are listed in
table \ref{tab:asassn15jjoc2015}.
The $O-C$ diagram shows a clear pattern of
stages B and C.  The large positive $P_{\rm dot}$
[$+8.1(0.6) \times 10^{-5}$] is typical for this
superhump period.

% SI

% SI

\begin{table}
\caption{Superhump maxima of ASASSN-15jj (2015)}\label{tab:asassn15jjoc2015}
\begin{center}
\begin{tabular}{rp{55pt}p{40pt}r@{.}lr}
\hline
\multicolumn{1}{c}{$E$} & \multicolumn{1}{c}{max\commenta} & \multicolumn{1}{c}{error} & \multicolumn{2}{c}{$O-C$\commentb} & \multicolumn{1}{c}{$N$\commentc} \\
\hline
0 & 57161.7489 & 0.0006 & 0&0055 & 16 \\
1 & 57161.8120 & 0.0007 & 0&0063 & 13 \\
2 & 57161.8757 & 0.0007 & 0&0076 & 14 \\
12 & 57162.4961 & 0.0021 & 0&0039 & 69 \\
13 & 57162.5581 & 0.0004 & 0&0035 & 144 \\
14 & 57162.6198 & 0.0005 & 0&0028 & 120 \\
16 & 57162.7441 & 0.0008 & 0&0023 & 17 \\
17 & 57162.8056 & 0.0011 & 0&0013 & 14 \\
18 & 57162.8688 & 0.0008 & 0&0022 & 16 \\
32 & 57163.7402 & 0.0008 & $-$0&0001 & 16 \\
33 & 57163.8002 & 0.0007 & $-$0&0025 & 15 \\
34 & 57163.8631 & 0.0010 & $-$0&0019 & 16 \\
48 & 57164.7341 & 0.0008 & $-$0&0046 & 16 \\
49 & 57164.7990 & 0.0021 & $-$0&0021 & 14 \\
50 & 57164.8590 & 0.0010 & $-$0&0045 & 15 \\
64 & 57165.7303 & 0.0008 & $-$0&0069 & 15 \\
65 & 57165.7932 & 0.0014 & $-$0&0064 & 15 \\
66 & 57165.8536 & 0.0011 & $-$0&0084 & 16 \\
80 & 57166.7314 & 0.0015 & $-$0&0042 & 13 \\
81 & 57166.7860 & 0.0043 & $-$0&0120 & 15 \\
82 & 57166.8542 & 0.0076 & $-$0&0062 & 17 \\
96 & 57167.7288 & 0.0022 & $-$0&0053 & 21 \\
97 & 57167.7961 & 0.0069 & $-$0&0003 & 21 \\
144 & 57170.7385 & 0.0021 & 0&0091 & 20 \\
145 & 57170.7985 & 0.0015 & 0&0066 & 20 \\
146 & 57170.8587 & 0.0018 & 0&0044 & 17 \\
161 & 57171.7954 & 0.0018 & 0&0052 & 14 \\
162 & 57171.8605 & 0.0009 & 0&0078 & 24 \\
178 & 57172.8551 & 0.0059 & 0&0040 & 17 \\
194 & 57173.8510 & 0.0030 & 0&0014 & 16 \\
209 & 57174.7771 & 0.0016 & $-$0&0085 & 14 \\
\hline
  \multicolumn{6}{l}{\commenta BJD$-$2400000.} \\
  \multicolumn{6}{l}{\commentb Against max $= 2457161.7433 + 0.062403 E$.} \\
  \multicolumn{6}{l}{\commentc Number of points used to determine the maximum.} \\
\end{tabular}
\end{center}
\end{table}

\subsection{ASASSN-15kf}\label{obj:asassn15kf}

\begin{figure}
  \begin{center}
%    \FigureFile(85mm,70mm){asassn15kfshlc.eps}
    \FigureFile(85mm,70mm){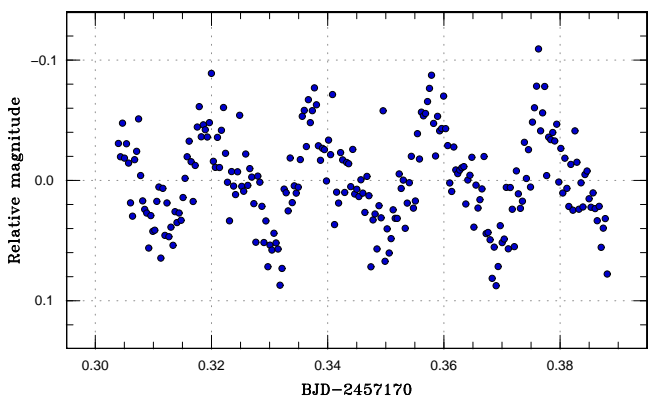}
  \end{center}
  \caption{Superhumps in ASASSN-15kf (2015).
  This object is likely an AM CVn-type system.
  }
  \label{fig:asassn15kfshlc}
\end{figure}

   This object was detected as a transient at $V$=15.0
on 2015 May 27 by the ASAS-SN team.
Subsequent observations detected very short-period superhumps
[period 0.0192(1)~d, figure \ref{fig:asassn15kfshlc}],
making this object a likely AM CVn-type object
(vsnet-alert 18669).  The superhumps were clearly detected
only on this night, and the object faded after a 6~d gap
in observation.  The times of superhumps maxima are listed
in table \ref{tab:asassn15kfoc2015}.
The period given in table \ref{tab:perlist} refers to
the one determined with the PDM method.

\begin{figure}
  \begin{center}
%    \FigureFile(85mm,70mm){asassn15kflc.eps}
    \FigureFile(85mm,70mm){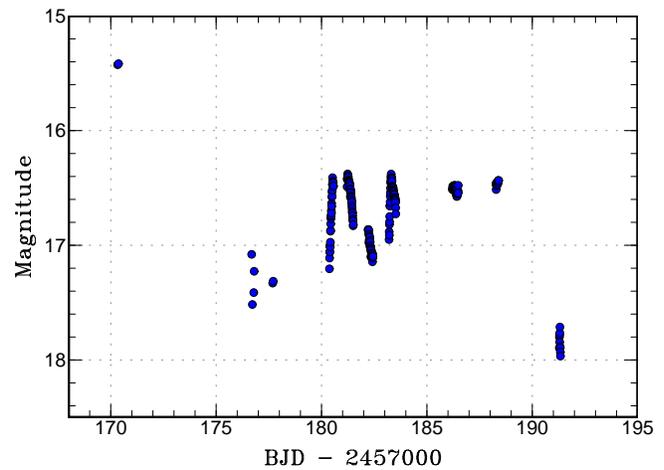}
  \end{center}
  \caption{Outburst light curve of ASASSN-15kf (2015).
  The data for BJD 2457176--2457178 were binned to 0.05~d.
  The other data were binned to 0.0064~d.
  }
  \label{fig:asassn15kflc}
\end{figure}

\begin{figure}
  \begin{center}
%    \FigureFile(85mm,110mm){asassn15kfshpdm.eps}
    \FigureFile(85mm,110mm){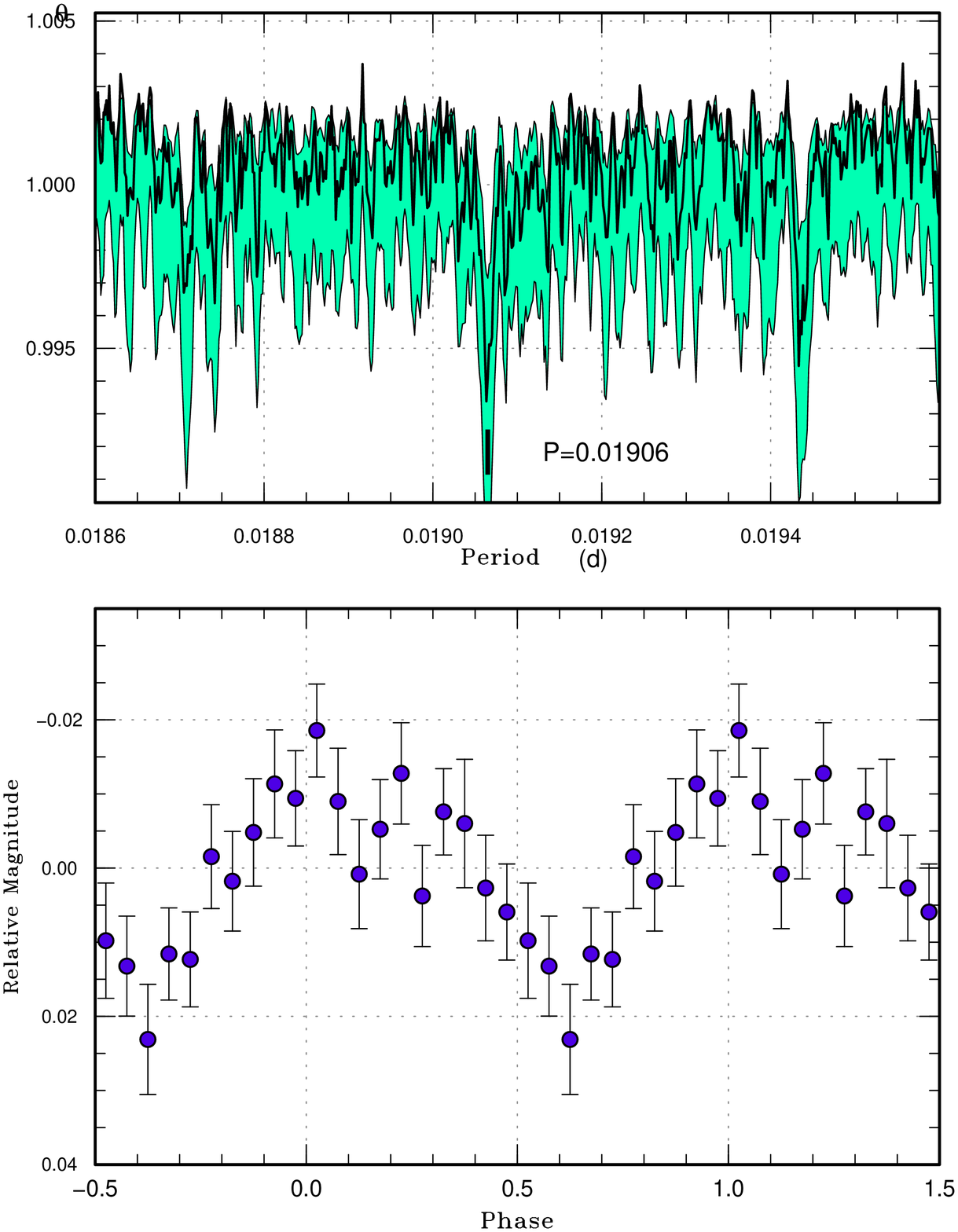}
  \end{center}
  \caption{Late-stage variations in ASASSN-15kf (2015).
     (Upper): PDM analysis.
     (Lower): Phase-averaged profile.}
  \label{fig:asassn15kfshpdm}
\end{figure}

   The object then showed (probably damping) oscillations
(vsnet-alert 18712, 18724; figure \ref{fig:asassn15kflc}),
which are characteristic to AM CVn-type superoutbursts
(cf. \cite{lev15amcvn}, \cite{kat04v803cen}, \cite{nog04v406hya},
\cite{Pdot4}, \cite{Pdot5}).
During this phase, there were weak variations with a period
of 0.01906(1)~d, which may be late stage superhumps
or the orbital variation (figure \ref{fig:asassn15kfshpdm}).

% SI

% SI

\begin{table}
\caption{Superhump maxima of ASASSN-15kf (2015)}\label{tab:asassn15kfoc2015}
\begin{center}
\begin{tabular}{rp{55pt}p{40pt}r@{.}lr}
\hline
\multicolumn{1}{c}{$E$} & \multicolumn{1}{c}{max\commenta} & \multicolumn{1}{c}{error} & \multicolumn{2}{c}{$O-C$\commentb} & \multicolumn{1}{c}{$N$\commentc} \\
\hline
0 & 57170.3198 & 0.0003 & 0&0004 & 44 \\
1 & 57170.3380 & 0.0003 & $-$0&0007 & 45 \\
2 & 57170.3584 & 0.0003 & 0&0003 & 44 \\
3 & 57170.3773 & 0.0003 & 0&0000 & 44 \\
\hline
  \multicolumn{6}{l}{\commenta BJD$-$2400000.} \\
  \multicolumn{6}{l}{\commentb Against max $= 2457170.3195 + 0.019272 E$.} \\
  \multicolumn{6}{l}{\commentc Number of points used to determine the maximum.} \\
\end{tabular}
\end{center}
\end{table}

\subsection{ASASSN-15kh}\label{obj:asassn15kh}

\begin{figure}
  \begin{center}
%    \FigureFile(85mm,110mm){asassn15khshpdm.eps}
    \FigureFile(85mm,110mm){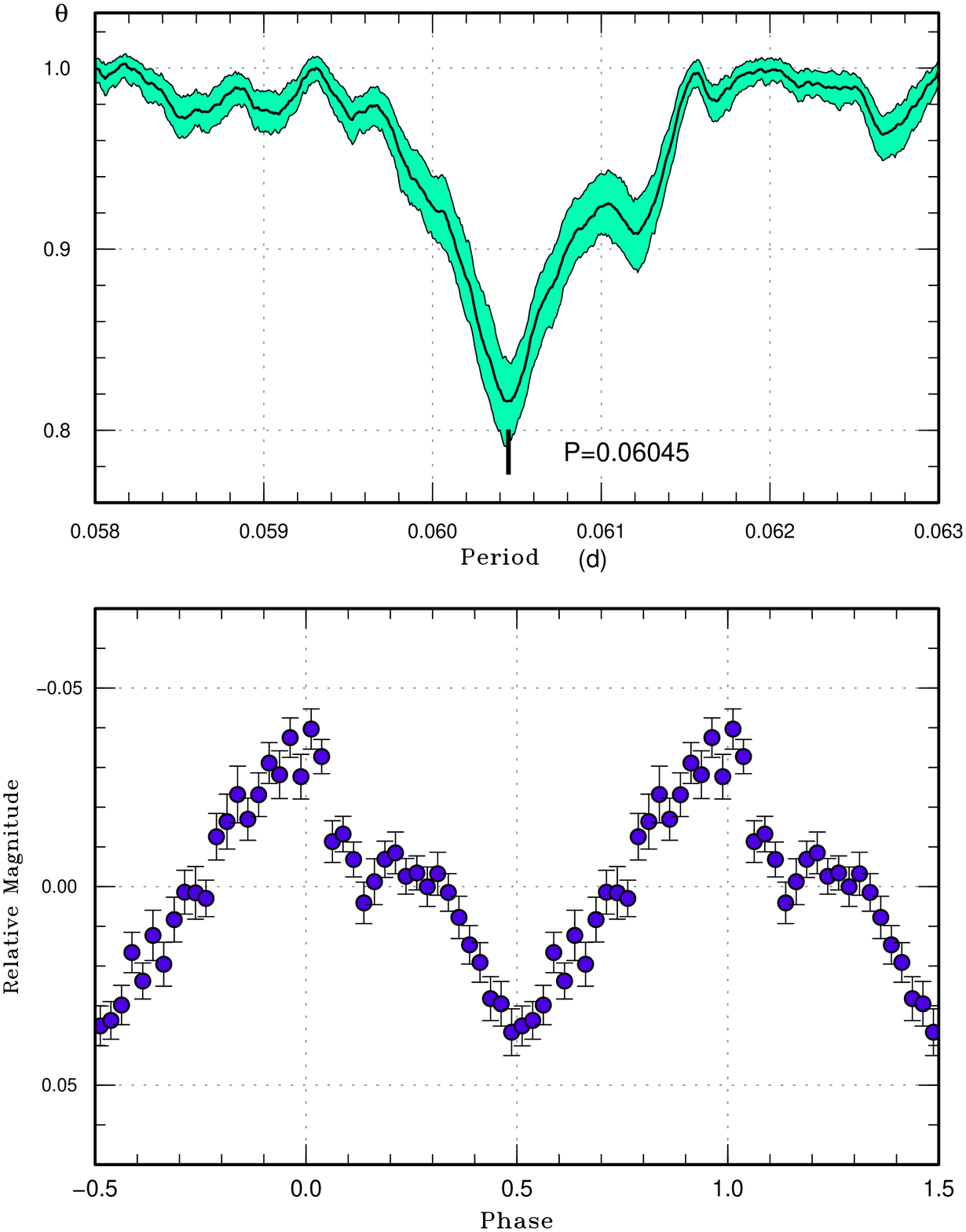}
  \end{center}
  \caption{Superhumps in ASASSN-15kh (2015).
     (Upper): PDM analysis.
     (Lower): Phase-averaged profile.}
  \label{fig:asassn15khshpdm}
\end{figure}

\begin{figure}
  \begin{center}
%    \FigureFile(85mm,100mm){asassn15khhumpall.eps}
    \FigureFile(85mm,100mm){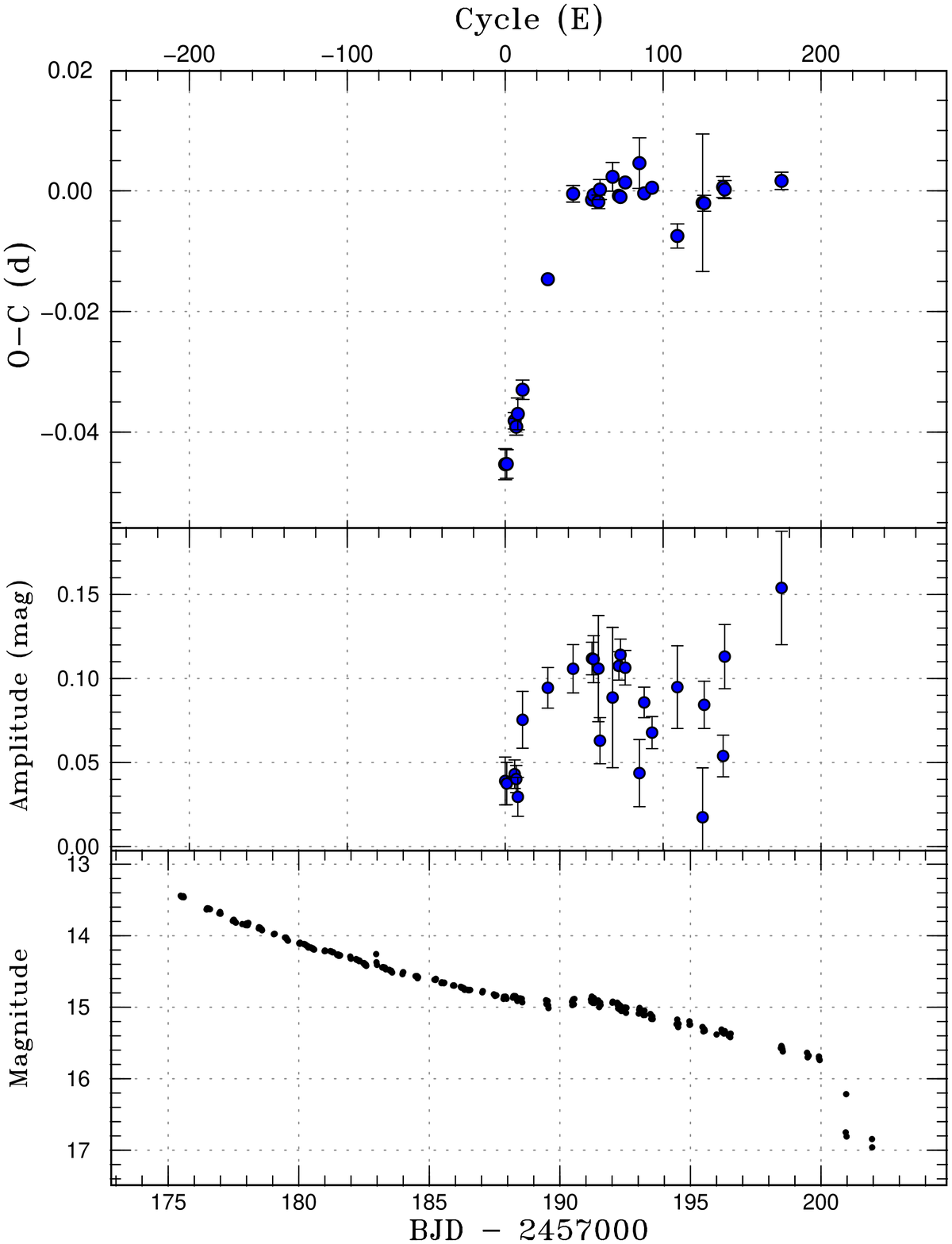}
  \end{center}
  \caption{$O-C$ diagram of superhumps in ASASSN-15kh (2015).
     (Upper:) $O-C$ diagram.
     We used a period of 0.06048~d for calculating the $O-C$ residuals.
     (Middle:) Amplitudes of superhumps.
     (Lower:) Light curve.  The data were binned to 0.020~d.
  }
  \label{fig:asassn15khhumpall}
\end{figure}

   This object was detected as a transient at $V$=13.2
on 2015 June 1 by the ASAS-SN team.
No quiescent counterpart is known.
Emerging (ordinary) superhumps were observed on June 14
(13~d after the outburst detection,
vsnet-alert 18734, 18758, 18799;
figure \ref{fig:asassn15khshpdm}).
The times of superhump maxima are listed in
table \ref{tab:asassn15khoc2015}.
The $O-C$ diagram shows clear stages A and B.
Stage A lasted for 43 cycles and there was no indication
of stage C (figure \ref{fig:asassn15khhumpall}).
The object started fading rapidly on June 27,
26~d after the outburst detection.

   The amplitude of early superhumps was below
the limit (0.01 mag) of our detection although the long
duration before the appearance of stage A superhumps
(figure \ref{fig:asassn15khhumpall})
strongly suggests that the 2:1 resonance was working.

   The small $P_{\rm dot}$ [$+1.2(1.6) \times 10^{-5}$] 
suggests that the object has a small $q$.  
The empirical relation between $P_{\rm dot}$
for stage B superhumps and $q$ (equation 6 in
\cite{kat15wzsge}) gives a $q$ of 0.065(9).
Combined with the long duration of stage A superhumps,
low amplitude of superhumps (probably reflecting
the small tidal torque) and the long superhump period,
this object is a good candidate for the period bouncer.
The brightness when superhump appeared was 14.9 mag.
\citet{kat15wzsge} has shown that quiescent brightness
is 6.4 and 7.2 mag (in average) fainter in WZ Sge-type
objects and period bouncers, respectively.
The expected quiescent brightness is 21.3 and 22.1 mag,
respectively, and it is not a surprise that there was
no previous detection of the quiescent counterpart.

% SI

% SI

\begin{table}
\caption{Superhump maxima of ASASSN-15kh (2015)}\label{tab:asassn15khoc2015}
\begin{center}
\begin{tabular}{rp{55pt}p{40pt}r@{.}lr}
\hline
\multicolumn{1}{c}{$E$} & \multicolumn{1}{c}{max\commenta} & \multicolumn{1}{c}{error} & \multicolumn{2}{c}{$O-C$\commentb} & \multicolumn{1}{c}{$N$\commentc} \\
\hline
0 & 57187.8600 & 0.0026 & $-$0&0167 & 36 \\
1 & 57187.9206 & 0.0024 & $-$0&0169 & 38 \\
6 & 57188.2301 & 0.0014 & $-$0&0111 & 137 \\
7 & 57188.2896 & 0.0014 & $-$0&0124 & 138 \\
8 & 57188.3522 & 0.0026 & $-$0&0105 & 118 \\
11 & 57188.5377 & 0.0016 & $-$0&0073 & 16 \\
27 & 57189.5237 & 0.0009 & 0&0067 & 15 \\
43 & 57190.5055 & 0.0014 & 0&0166 & 13 \\
55 & 57191.2303 & 0.0007 & 0&0124 & 140 \\
56 & 57191.2916 & 0.0010 & 0&0129 & 139 \\
59 & 57191.4718 & 0.0011 & 0&0109 & 11 \\
60 & 57191.5344 & 0.0017 & 0&0127 & 16 \\
68 & 57192.0203 & 0.0024 & 0&0127 & 21 \\
72 & 57192.2591 & 0.0006 & 0&0085 & 139 \\
73 & 57192.3194 & 0.0006 & 0&0080 & 139 \\
76 & 57192.5032 & 0.0011 & 0&0096 & 13 \\
85 & 57193.0508 & 0.0042 & 0&0104 & 26 \\
88 & 57193.2272 & 0.0008 & 0&0046 & 138 \\
93 & 57193.5305 & 0.0010 & 0&0042 & 16 \\
109 & 57194.4902 & 0.0020 & $-$0&0081 & 14 \\
125 & 57195.4634 & 0.0114 & $-$0&0069 & 10 \\
126 & 57195.5238 & 0.0013 & $-$0&0073 & 17 \\
138 & 57196.2522 & 0.0017 & $-$0&0078 & 138 \\
139 & 57196.3123 & 0.0015 & $-$0&0085 & 94 \\
175 & 57198.4910 & 0.0014 & $-$0&0167 & 13 \\
\hline
  \multicolumn{6}{l}{\commenta BJD$-$2400000.} \\
  \multicolumn{6}{l}{\commentb Against max $= 2457187.8767 + 0.060748 E$.} \\
  \multicolumn{6}{l}{\commentc Number of points used to determine the maximum.} \\
\end{tabular}
\end{center}
\end{table}

\subsection{ASASSN-15le}\label{obj:asassn15le}

\begin{figure}
  \begin{center}
%    \FigureFile(85mm,110mm){asassn15leshpdm.eps}
    \FigureFile(85mm,110mm){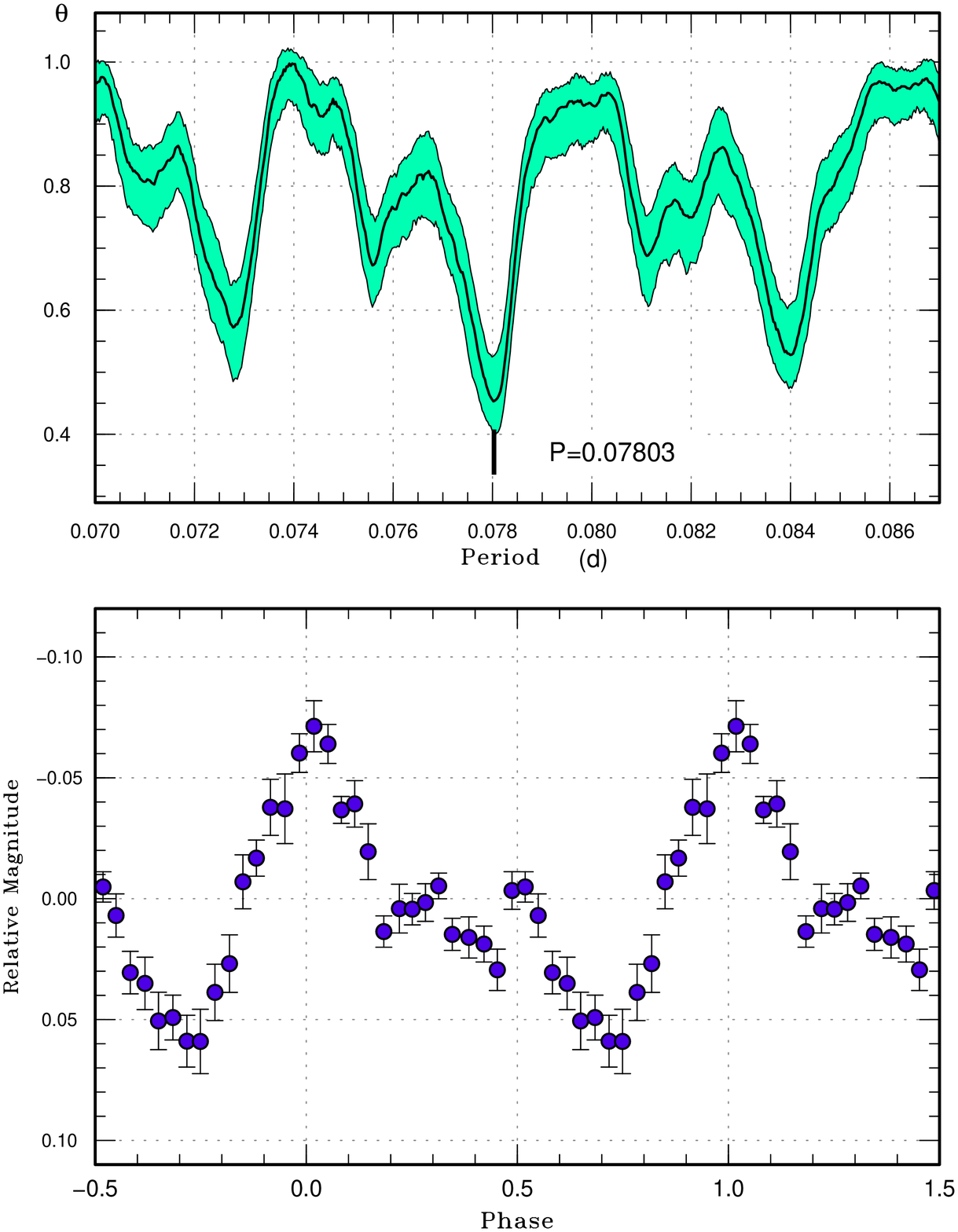}
  \end{center}
  \caption{Superhumps in ASASSN-15le for the first three nights
     (BJD 2457189--2457193) (2015).
     (Upper): PDM analysis.  The alias selection was the most
     likely one based on the $O-C$ analysis.
     (Lower): Phase-averaged profile.}
  \label{fig:asassn15leshpdm}
\end{figure}

   This object was detected as a transient at $V$=14.9
on 2015 June 12 by the ASAS-SN team.
There is a $V$=14.9 mag star \timeform{5''} from
this position.  There is a GALEX counterpart with
an NUV magnitude of 19.5(1).
There was at least one long outburst reaching $V$=13.6 in
2007 October in the ASAS-3 data.
Superhumps were soon detected (vsnet-alert 18736, 18746;
figure \ref{fig:asassn15leshpdm}).
Although there were observations after BJD 2457193,
we could not detect a confident superhump signal on later
nights, probably due to the contamination
by the nearby star.  We only listed superhump maxima for
the initial three nights in table \ref{tab:asassn15leoc2015}.

% SI

% SI

\begin{table}
\caption{Superhump maxima of ASASSN-15le (2015)}\label{tab:asassn15leoc2015}
\begin{center}
\begin{tabular}{rp{55pt}p{40pt}r@{.}lr}
\hline
\multicolumn{1}{c}{$E$} & \multicolumn{1}{c}{max\commenta} & \multicolumn{1}{c}{error} & \multicolumn{2}{c}{$O-C$\commentb} & \multicolumn{1}{c}{$N$\commentc} \\
\hline
0 & 57189.4670 & 0.0008 & 0&0020 & 57 \\
13 & 57190.4796 & 0.0006 & 0&0006 & 73 \\
14 & 57190.5532 & 0.0007 & $-$0&0037 & 59 \\
41 & 57192.6640 & 0.0044 & 0&0011 & 18 \\
42 & 57192.7411 & 0.0024 & 0&0001 & 21 \\
43 & 57192.8189 & 0.0035 & $-$0&0001 & 27 \\
\hline
  \multicolumn{6}{l}{\commenta BJD$-$2400000.} \\
  \multicolumn{6}{l}{\commentb Against max $= 2457189.4650 + 0.078000 E$.} \\
  \multicolumn{6}{l}{\commentc Number of points used to determine the maximum.} \\
\end{tabular}
\end{center}
\end{table}

\subsection{ASASSN-15lt}\label{obj:asassn15lt}

\begin{figure}
  \begin{center}
%    \FigureFile(85mm,110mm){asassn15ltshpdm.eps}
    \FigureFile(85mm,110mm){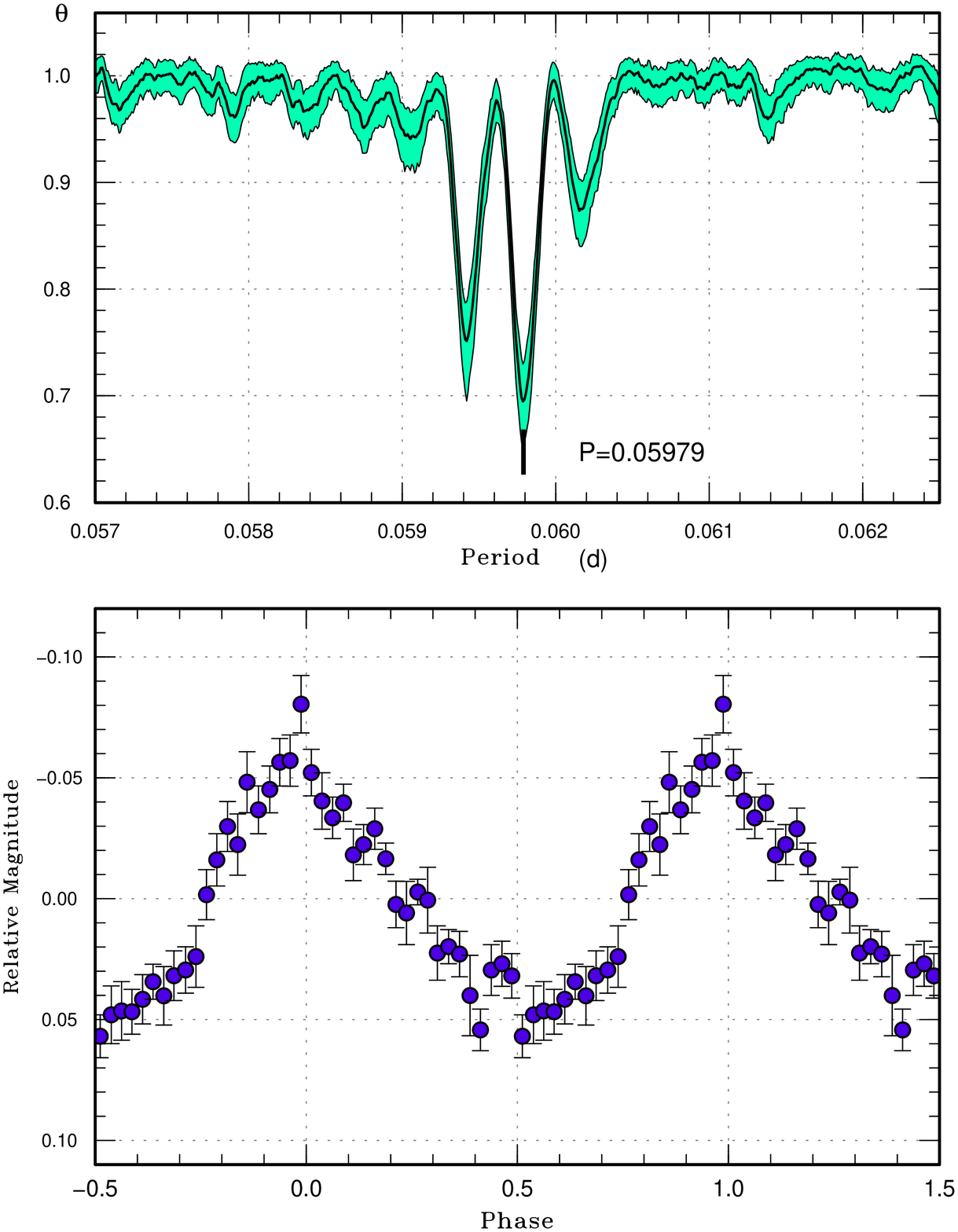}
  \end{center}
  \caption{Superhumps in ASASSN-15lt (2015).
     (Upper): PDM analysis.
     (Lower): Phase-averaged profile.}
  \label{fig:asassn15ltshpdm}
\end{figure}

   This object was detected as a transient at $V$=13.8
on 2015 June 21 by the ASAS-SN team.
The object soon developed superhumps (vsnet-alert 18797;
figure \ref{fig:asassn15ltshpdm}).
The early evolution (3~d after the outburst detection,
less than 7~d after the outburst maximum even considering
the observational gap in the ASAS-SN data) of superhumps apparently
excludes the possibility of a WZ Sge-type object.
The times of superhump maxima are listed in
table \ref{tab:asassn15ltoc2015}.
There was an apparent stage A-B transition around
$E$=50.  The maxima after $E$=188 were likely stage C
superhumps, which were associated with the small
brightening of the object.
Due to the observational gap, the $P_{\rm dot}$ for
stage B superhumps could not be determined.
We adopted the period of stage A superhumps
determined from the maxima $E \le 17$.

% SI

% SI

\begin{table}
\caption{Superhump maxima of ASASSN-15lt (2015)}\label{tab:asassn15ltoc2015}
\begin{center}
\begin{tabular}{rp{55pt}p{40pt}r@{.}lr}
\hline
\multicolumn{1}{c}{$E$} & \multicolumn{1}{c}{max\commenta} & \multicolumn{1}{c}{error} & \multicolumn{2}{c}{$O-C$\commentb} & \multicolumn{1}{c}{$N$\commentc} \\
\hline
0 & 57197.8197 & 0.0026 & $-$0&0328 & 18 \\
1 & 57197.8748 & 0.0015 & $-$0&0375 & 23 \\
16 & 57198.8121 & 0.0009 & 0&0022 & 17 \\
17 & 57198.8711 & 0.0009 & 0&0015 & 23 \\
33 & 57199.8364 & 0.0012 & 0&0093 & 20 \\
34 & 57199.8983 & 0.0007 & 0&0113 & 20 \\
49 & 57200.7980 & 0.0011 & 0&0135 & 23 \\
50 & 57200.8585 & 0.0006 & 0&0142 & 36 \\
58 & 57201.3360 & 0.0008 & 0&0129 & 28 \\
66 & 57201.8077 & 0.0012 & 0&0059 & 24 \\
67 & 57201.8675 & 0.0017 & 0&0059 & 29 \\
71 & 57202.1063 & 0.0007 & 0&0053 & 37 \\
188 & 57209.1090 & 0.0008 & 0&0070 & 37 \\
189 & 57209.1701 & 0.0010 & 0&0082 & 34 \\
204 & 57210.0625 & 0.0015 & 0&0031 & 29 \\
205 & 57210.1235 & 0.0010 & 0&0043 & 32 \\
222 & 57211.1390 & 0.0008 & 0&0025 & 32 \\
223 & 57211.1929 & 0.0006 & $-$0&0034 & 35 \\
233 & 57211.7859 & 0.0016 & $-$0&0088 & 26 \\
234 & 57211.8503 & 0.0023 & $-$0&0043 & 30 \\
250 & 57212.8038 & 0.0021 & $-$0&0082 & 31 \\
251 & 57212.8602 & 0.0015 & $-$0&0116 & 25 \\
266 & 57213.7687 & 0.0046 & $-$0&0007 & 23 \\
\hline
  \multicolumn{6}{l}{\commenta BJD$-$2400000.} \\
  \multicolumn{6}{l}{\commentb Against max $= 2457197.8524 + 0.059838 E$.} \\
  \multicolumn{6}{l}{\commentc Number of points used to determine the maximum.} \\
\end{tabular}
\end{center}
\end{table}

\subsection{ASASSN-15mb}\label{obj:asassn15mb}

   This object was detected as a transient at $V$=13.0
on 2015 June 30 by the ASAS-SN team.
There is an likely X-ray counterpart 1RXS J025246.3$-$395853.
There was a past outburst in 2009, which was
initially detected at $V$=12.74 by ASAS-3 on October 31,
and was observed at 14.2 mag (unfiltered CCD) by CRTS
on November 18 (vsnet-alert 18818, 18821).

\begin{figure}
  \begin{center}
%    \FigureFile(85mm,110mm){asassn15mbshpdm.eps}
    \FigureFile(85mm,110mm){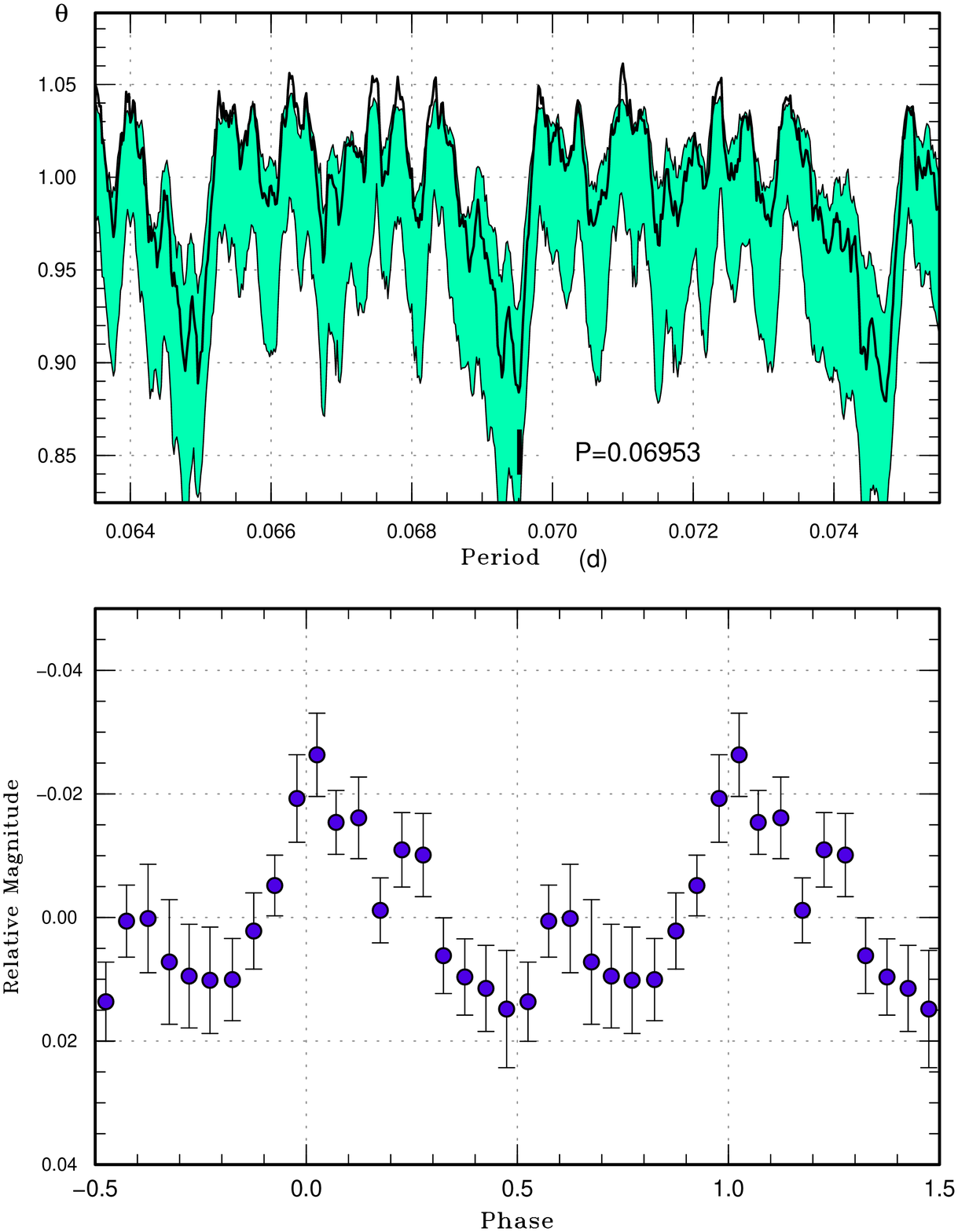}
  \end{center}
  \caption{Superhumps in ASASSN-15mb during the plateau phase (2015).
     (Upper): PDM analysis.
     (Lower): Phase-averaged profile.}
  \label{fig:asassn15mbshpdm}
\end{figure}

\begin{figure}
  \begin{center}
%    \FigureFile(85mm,110mm){asassn15mbshpostpdm.eps}
    \FigureFile(85mm,110mm){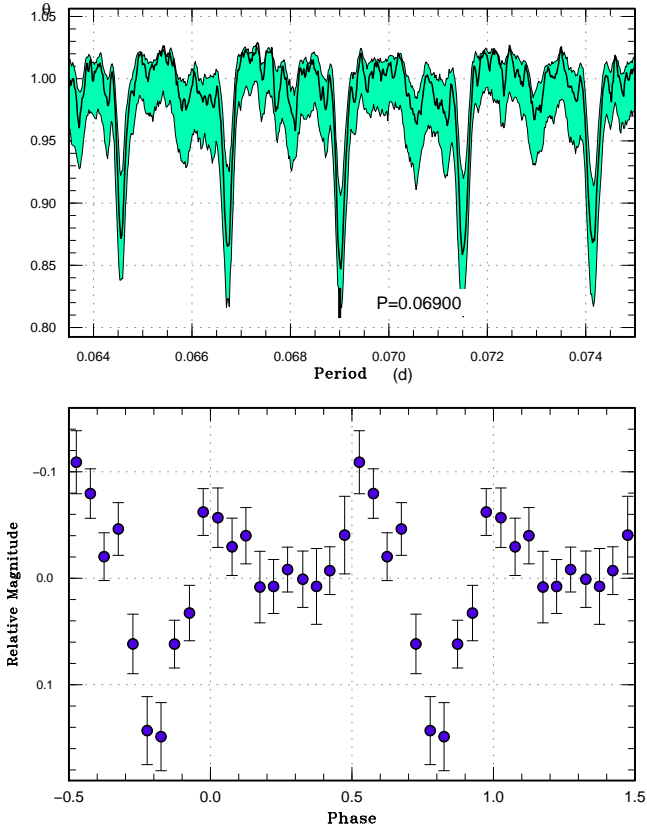}
  \end{center}
  \caption{Superhumps in ASASSN-15mb during the post-superoutburst
     phase (BJD 2457222--2457232) (2015).
     (Upper): PDM analysis.
     (Lower): Phase-averaged profile.}
  \label{fig:asassn15mbshpostpdm}
\end{figure}

   During the 2015 outburst, superhumps were detected
(vsnet-alert 18838, 18843).
Due to the short nightly observations, we could not
detect many superhumps.  The data, however, showed
periodic signals in different segments by PDM analysis
(figures \ref{fig:asassn15mbshpdm},
\ref{fig:asassn15mbshpostpdm}).
Since the $O-C$ analysis
is not helpful in identifying the alias, we selected
the strongest signal in the PDM analysis and assigned
the cycle counts (table \ref{tab:asassn15mboc2015}).
The maxima for $E \le$173 were recorded in
the post-superoutburst phase.  Since there was
significant brightening during the superoutburst
plateau around BJD 2457217, the maxima after this
can be attributed to stage C superhumps.
We included $E$=43 to stage C in table \ref{tab:perlist}
based on the continuity of the $O-C$ curve.
There remain possibilities of other one-day aliases.

   The superoutburst lasted at least 19~d (but less than
23~d).  The object showed a post-superoutburst rebrightening
on July 29, 11~d after the rapid fading from
the superoutburst plateau.  The second observation
during the 2009 outburst likely detected the rapidly
fading part or a rebrightening.

% SI

% SI

% SI

\begin{table}
\caption{Superhump maxima of ASASSN-15mb (2015)}\label{tab:asassn15mboc2015}
\begin{center}
\begin{tabular}{rp{55pt}p{40pt}r@{.}lr}
\hline
\multicolumn{1}{c}{$E$} & \multicolumn{1}{c}{max\commenta} & \multicolumn{1}{c}{error} & \multicolumn{2}{c}{$O-C$\commentb} & \multicolumn{1}{c}{$N$\commentc} \\
\hline
0 & 57212.8860 & 0.0011 & $-$0&0217 & 18 \\
43 & 57215.8791 & 0.0015 & 0&0059 & 23 \\
86 & 57218.8477 & 0.0021 & 0&0091 & 17 \\
101 & 57219.8807 & 0.0021 & 0&0076 & 29 \\
115 & 57220.8560 & 0.0025 & 0&0174 & 20 \\
173 & 57224.8420 & 0.0017 & 0&0036 & 30 \\
202 & 57226.8329 & 0.0013 & $-$0&0055 & 35 \\
203 & 57226.8910 & 0.0022 & $-$0&0163 & 38 \\
\hline
  \multicolumn{6}{l}{\commenta BJD$-$2400000.} \\
  \multicolumn{6}{l}{\commentb Against max $= 2457212.9078 + 0.068964 E$.} \\
  \multicolumn{6}{l}{\commentc Number of points used to determine the maximum.} \\
\end{tabular}
\end{center}
\end{table}

\subsection{ASASSN-15mt}\label{obj:asassn15mt}

   This object was detected as a transient at $V$=13.7
on 2015 July 18 by the ASAS-SN team \citep{sim15asassn15mtatel7809}.
The object has a strong UV excess in quiescence.
($U-g=-$0.74, \cite{gre12KeplerINT}).

\begin{figure}
  \begin{center}
%    \FigureFile(85mm,110mm){asassn15mtshpdm.eps}
    \FigureFile(85mm,110mm){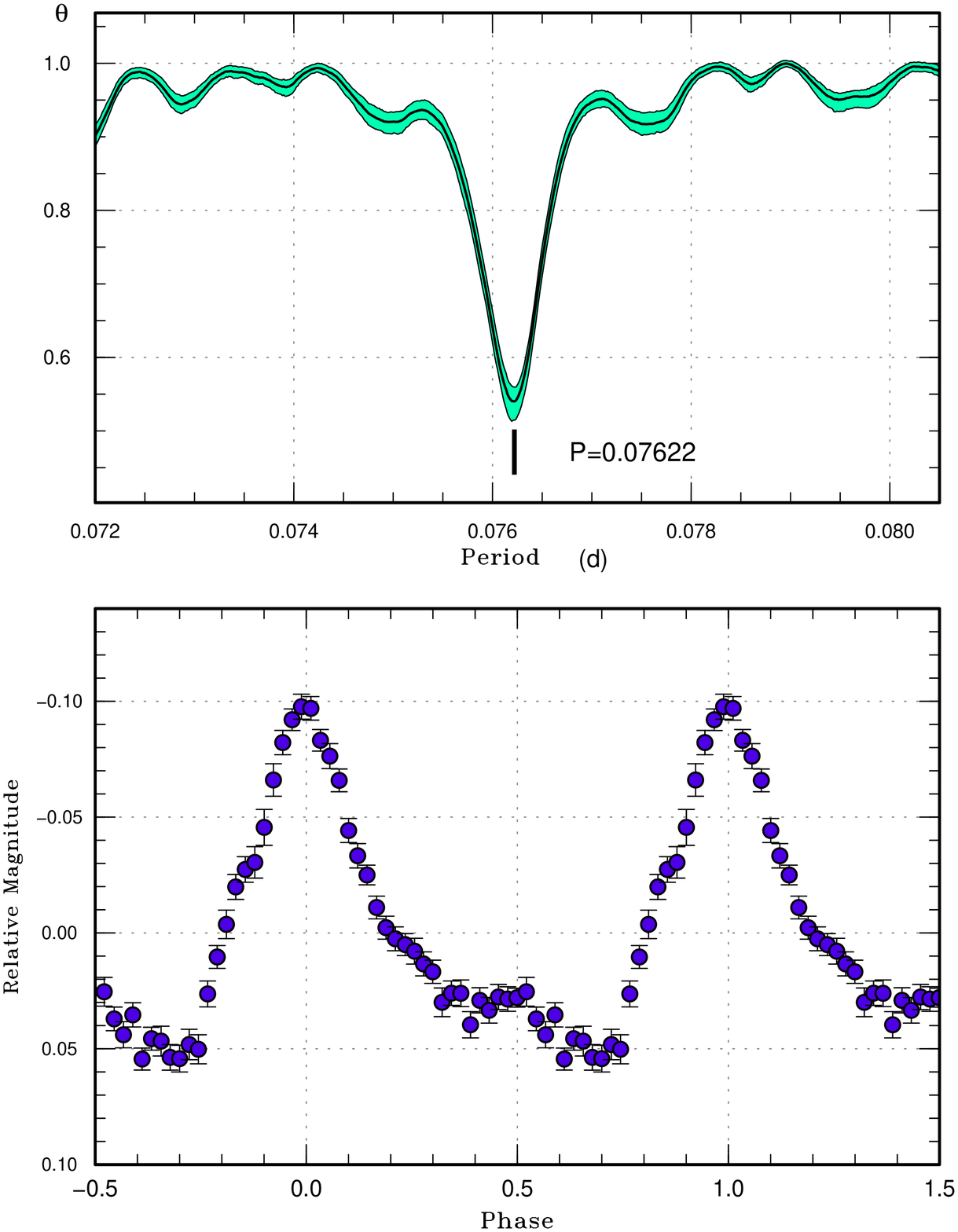}
  \end{center}
  \caption{Superhumps in ASASSN-15mt during the superoutburst
     plateau (2015).
     (Upper): PDM analysis.
     (Lower): Phase-averaged profile.}
  \label{fig:asassn15mtshpdm}
\end{figure}

   Subsequent observations immediately detected
superhumps (vsnet-alert 18878, 18882, 18890;
figure \ref{fig:asassn15mtshpdm}).
The times of superhump maxima are listed in
table \ref{tab:asassn15mtoc2015}.
After $E$=42, there was significant reduction of
the superhump amplitudes and we identified this epoch
to be stage B-C transition.  As is usual for a system
with a long superhump period, the transition is not
sharp as in short-period systems.
The object started fading rapidly on July 28--29,
and the duration of the superoutburst was at least
11~d (but shorter than 16~d).

% SI

% SI

\begin{table}
\caption{Superhump maxima of ASASSN-15mt (2015)}\label{tab:asassn15mtoc2015}
\begin{center}
\begin{tabular}{rp{55pt}p{40pt}r@{.}lr}
\hline
\multicolumn{1}{c}{$E$} & \multicolumn{1}{c}{max\commenta} & \multicolumn{1}{c}{error} & \multicolumn{2}{c}{$O-C$\commentb} & \multicolumn{1}{c}{$N$\commentc} \\
\hline
0 & 57225.0985 & 0.0004 & $-$0&0036 & 116 \\
1 & 57225.1739 & 0.0004 & $-$0&0043 & 140 \\
2 & 57225.2493 & 0.0005 & $-$0&0051 & 140 \\
19 & 57226.5482 & 0.0004 & $-$0&0003 & 131 \\
29 & 57227.3108 & 0.0005 & 0&0010 & 43 \\
30 & 57227.3878 & 0.0006 & 0&0018 & 49 \\
31 & 57227.4647 & 0.0006 & 0&0026 & 29 \\
42 & 57228.3007 & 0.0015 & 0&0011 & 22 \\
43 & 57228.3782 & 0.0004 & 0&0026 & 45 \\
44 & 57228.4553 & 0.0003 & 0&0035 & 45 \\
45 & 57228.5315 & 0.0003 & 0&0036 & 35 \\
48 & 57228.7580 & 0.0003 & 0&0016 & 215 \\
52 & 57229.0632 & 0.0006 & 0&0023 & 111 \\
53 & 57229.1350 & 0.0011 & $-$0&0020 & 180 \\
54 & 57229.2118 & 0.0010 & $-$0&0013 & 139 \\
55 & 57229.2925 & 0.0014 & 0&0032 & 66 \\
56 & 57229.3684 & 0.0004 & 0&0030 & 44 \\
57 & 57229.4432 & 0.0006 & 0&0017 & 45 \\
58 & 57229.5179 & 0.0004 & 0&0003 & 43 \\
64 & 57229.9772 & 0.0014 & 0&0028 & 82 \\
65 & 57230.0540 & 0.0006 & 0&0034 & 231 \\
66 & 57230.1272 & 0.0007 & 0&0006 & 191 \\
67 & 57230.2009 & 0.0011 & $-$0&0019 & 139 \\
68 & 57230.2674 & 0.0063 & $-$0&0115 & 81 \\
69 & 57230.3554 & 0.0011 & 0&0003 & 27 \\
70 & 57230.4316 & 0.0007 & 0&0004 & 44 \\
71 & 57230.5084 & 0.0004 & 0&0011 & 46 \\
83 & 57231.4217 & 0.0014 & 0&0008 & 29 \\
84 & 57231.5024 & 0.0007 & 0&0053 & 43 \\
95 & 57232.3306 & 0.0009 & $-$0&0039 & 42 \\
96 & 57232.4061 & 0.0017 & $-$0&0045 & 38 \\
97 & 57232.4818 & 0.0015 & $-$0&0049 & 45 \\
\hline
  \multicolumn{6}{l}{\commenta BJD$-$2400000.} \\
  \multicolumn{6}{l}{\commentb Against max $= 2457225.1021 + 0.076131 E$.} \\
  \multicolumn{6}{l}{\commentc Number of points used to determine the maximum.} \\
\end{tabular}
\end{center}
\end{table}

\subsection{ASASSN-15na}\label{obj:asassn15na}

\begin{figure}
  \begin{center}
%    \FigureFile(85mm,110mm){asassn15nashpdm.eps}
    \FigureFile(85mm,110mm){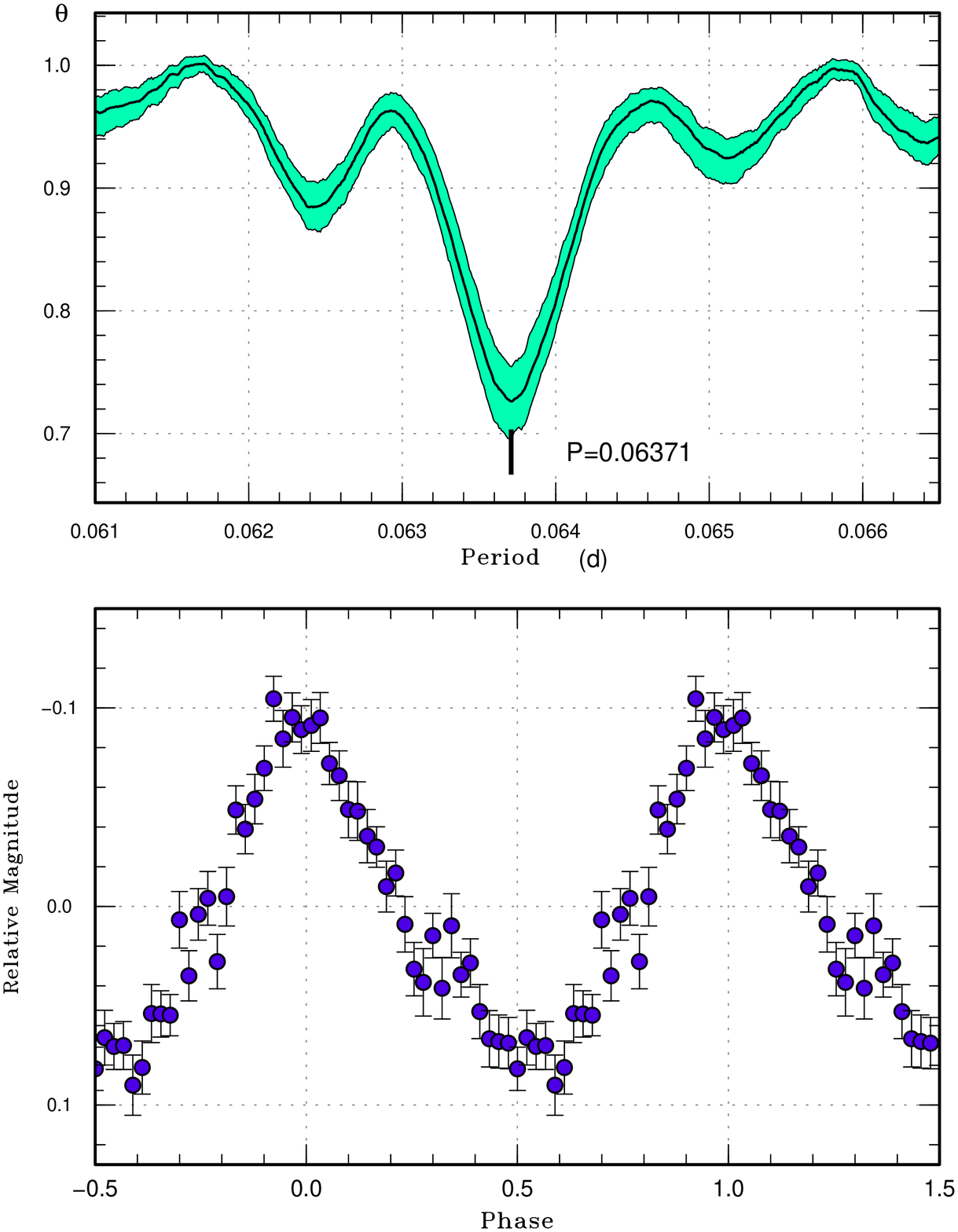}
  \end{center}
  \caption{Superhumps in ASASSN-15na (2015).
     (Upper): PDM analysis of the segment BJD 2457234--2457242.
     (Lower): Phase-averaged profile.}
  \label{fig:asassn15nashpdm}
\end{figure}

   This object was detected as a transient at $V$=14.8
on 2015 July 20 by the ASAS-SN team.
Double-wave modulations were immediately detected
(vsnet-alert 18884).  The object was initially suspected
as an AM CVn-type dwarf nova (cf. vsnet-alert 18910),
which was incorrect due to the mistaken identity of
an non-existent rapid fading (cf. vsnet-alert 18923).
The object showed growth of superhumps associated with
the brightening of the system (vsnet-alert 18923, 18933;
figure \ref{fig:asassn15nashpdm}).

\begin{figure}
  \begin{center}
%    \FigureFile(85mm,110mm){asassn15naeshpdm.eps}
    \FigureFile(85mm,110mm){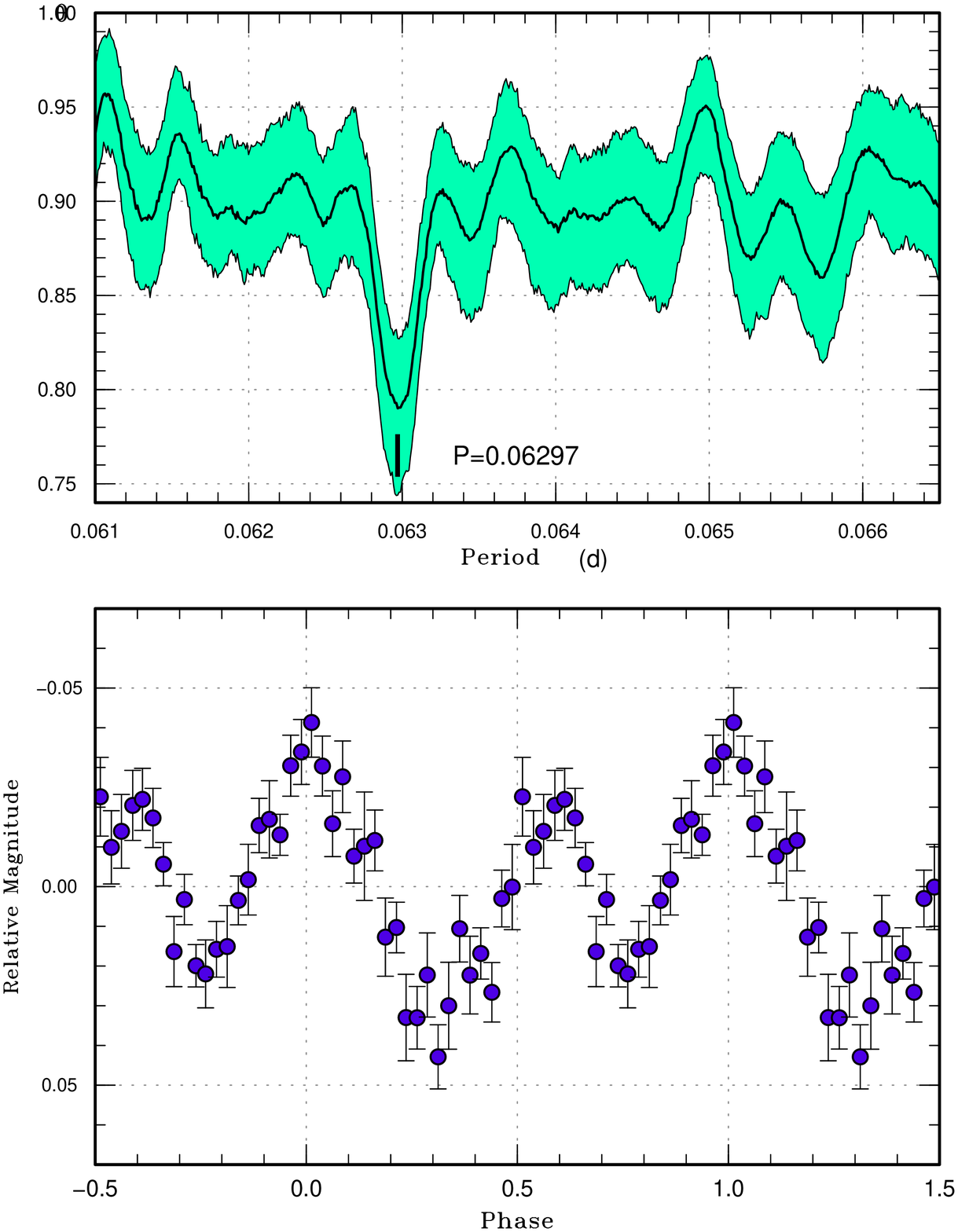}
  \end{center}
  \caption{Early superhumps in ASASSN-15na (2015).
     (Upper): PDM analysis.
     (Lower): Phase-averaged profile.}
  \label{fig:asassn15naeshpdm}
\end{figure}

   The times of superhump maxima after the development
of superhumps are listed in table \ref{tab:asassn15naoc2015}.
Although there were modulations before these epochs,
we have not been able to determine the individual maxima
due to the poor statistics (due to the faintness of the object).
The epoch $E$=174 probably corresponds to a stage C
superhump.  The cycle count between $E$=111 and $E$=174
is somewhat ambiguous.
Using the data between BJD 2457231.6 and 2457233.8,
a period of 0.06491(12)~d was obtained, which we identified
to be the period of stage A superhumps.
The data before BJD 2457231.5 were well expressed by
a shorter period of 0.06297(2)~d, which we identified
to be the period of early superhumps
(figure \ref{fig:asassn15naeshpdm}).
By using these periods, we have obtained $q$=0.081(5).

   Although the orbital period (approximated by the period
of early superhumps) is much longer than the period minimum,
the object does not have a very low $q$, which is expected
for a period bouncer.  The short duration of stage A and
large amplitude of superhumps (figure \ref{fig:asassn15nashpdm})
are also consistent with a high $q$ \citep{kat15wzsge}.
The system may be similar to WZ Sge-type dwarf novae
with multiple rebrightenings (MASTER OT J211258.65$+$242145.4,
MASTER OT J203749.39$+$552210.3: \cite{nak13j2112j2037}).

% SI

% SI

% SI

\begin{table}
\caption{Superhump maxima of ASASSN-15na (2015)}\label{tab:asassn15naoc2015}
\begin{center}
\begin{tabular}{rp{55pt}p{40pt}r@{.}lr}
\hline
\multicolumn{1}{c}{$E$} & \multicolumn{1}{c}{max\commenta} & \multicolumn{1}{c}{error} & \multicolumn{2}{c}{$O-C$\commentb} & \multicolumn{1}{c}{$N$\commentc} \\
\hline
0 & 57234.6800 & 0.0017 & $-$0&0041 & 16 \\
1 & 57234.7438 & 0.0013 & $-$0&0039 & 24 \\
9 & 57235.2544 & 0.0009 & $-$0&0022 & 146 \\
10 & 57235.3208 & 0.0009 & 0&0006 & 143 \\
11 & 57235.3826 & 0.0007 & $-$0&0012 & 146 \\
12 & 57235.4460 & 0.0006 & $-$0&0014 & 146 \\
13 & 57235.5114 & 0.0011 & 0&0004 & 97 \\
16 & 57235.6973 & 0.0008 & $-$0&0045 & 24 \\
17 & 57235.7603 & 0.0014 & $-$0&0052 & 24 \\
25 & 57236.2735 & 0.0010 & $-$0&0009 & 147 \\
26 & 57236.3380 & 0.0006 & 0&0000 & 147 \\
27 & 57236.4012 & 0.0012 & $-$0&0004 & 133 \\
32 & 57236.7144 & 0.0024 & $-$0&0053 & 14 \\
47 & 57237.6768 & 0.0022 & 0&0030 & 11 \\
48 & 57237.7390 & 0.0021 & 0&0015 & 14 \\
56 & 57238.2528 & 0.0023 & 0&0064 & 128 \\
57 & 57238.3102 & 0.0012 & 0&0002 & 147 \\
58 & 57238.3777 & 0.0055 & 0&0040 & 133 \\
59 & 57238.4457 & 0.0033 & 0&0085 & 65 \\
63 & 57238.6917 & 0.0027 & 0&0000 & 14 \\
64 & 57238.7601 & 0.0019 & 0&0048 & 14 \\
79 & 57239.6989 & 0.0036 & $-$0&0106 & 18 \\
80 & 57239.7748 & 0.0053 & 0&0016 & 11 \\
95 & 57240.7301 & 0.0042 & 0&0028 & 17 \\
110 & 57241.6970 & 0.0016 & 0&0155 & 18 \\
111 & 57241.7580 & 0.0030 & 0&0128 & 15 \\
174 & 57245.7304 & 0.0019 & $-$0&0224 & 26 \\
\hline
  \multicolumn{6}{l}{\commenta BJD$-$2400000.} \\
  \multicolumn{6}{l}{\commentb Against max $= 2457234.6841 + 0.063613 E$.} \\
  \multicolumn{6}{l}{\commentc Number of points used to determine the maximum.} \\
\end{tabular}
\end{center}
\end{table}

\subsection{ASASSN-15ni}\label{obj:asassn15ni}

   This object was detected as a transient at $V$=12.9
on 2015 July 28 by the ASAS-SN team \citep{don15asassn15niatel7850}.
The dwarf nova-type nature was spectroscopically confirmed
\citep{ber15asassn15niatel7854}.

\begin{figure}
  \begin{center}
%    \FigureFile(85mm,110mm){asassn15nishpdm.eps}
    \FigureFile(85mm,110mm){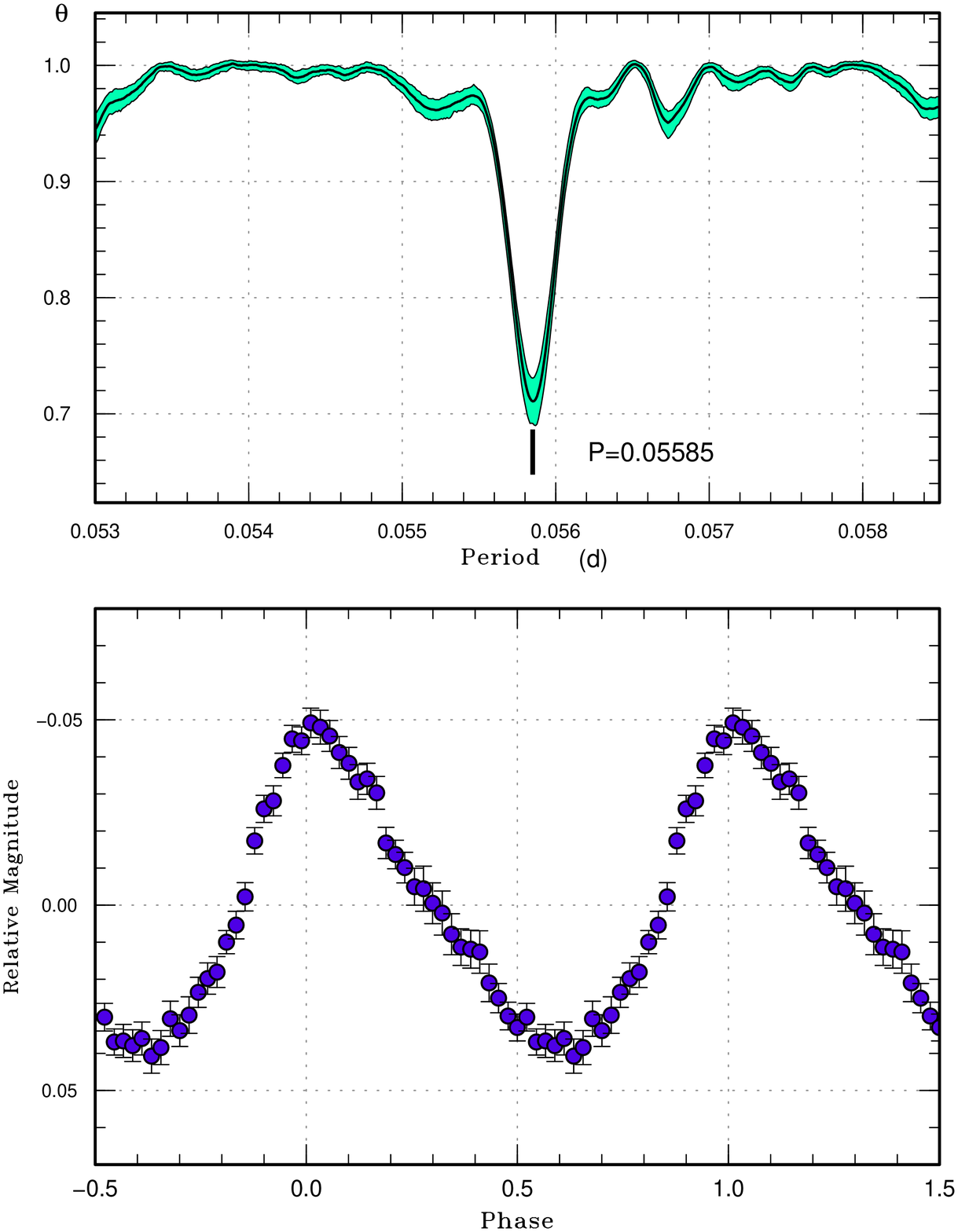}
  \end{center}
  \caption{Superhumps in ASASSN-15ni (2015).
     (Upper): PDM analysis of the segment BJD 2457241--2457254.
     (Lower): Phase-averaged profile.}
  \label{fig:asassn15nishpdm}
\end{figure}

\begin{figure}
  \begin{center}
%    \FigureFile(85mm,110mm){asassn15nieshpdm.eps}
    \FigureFile(85mm,110mm){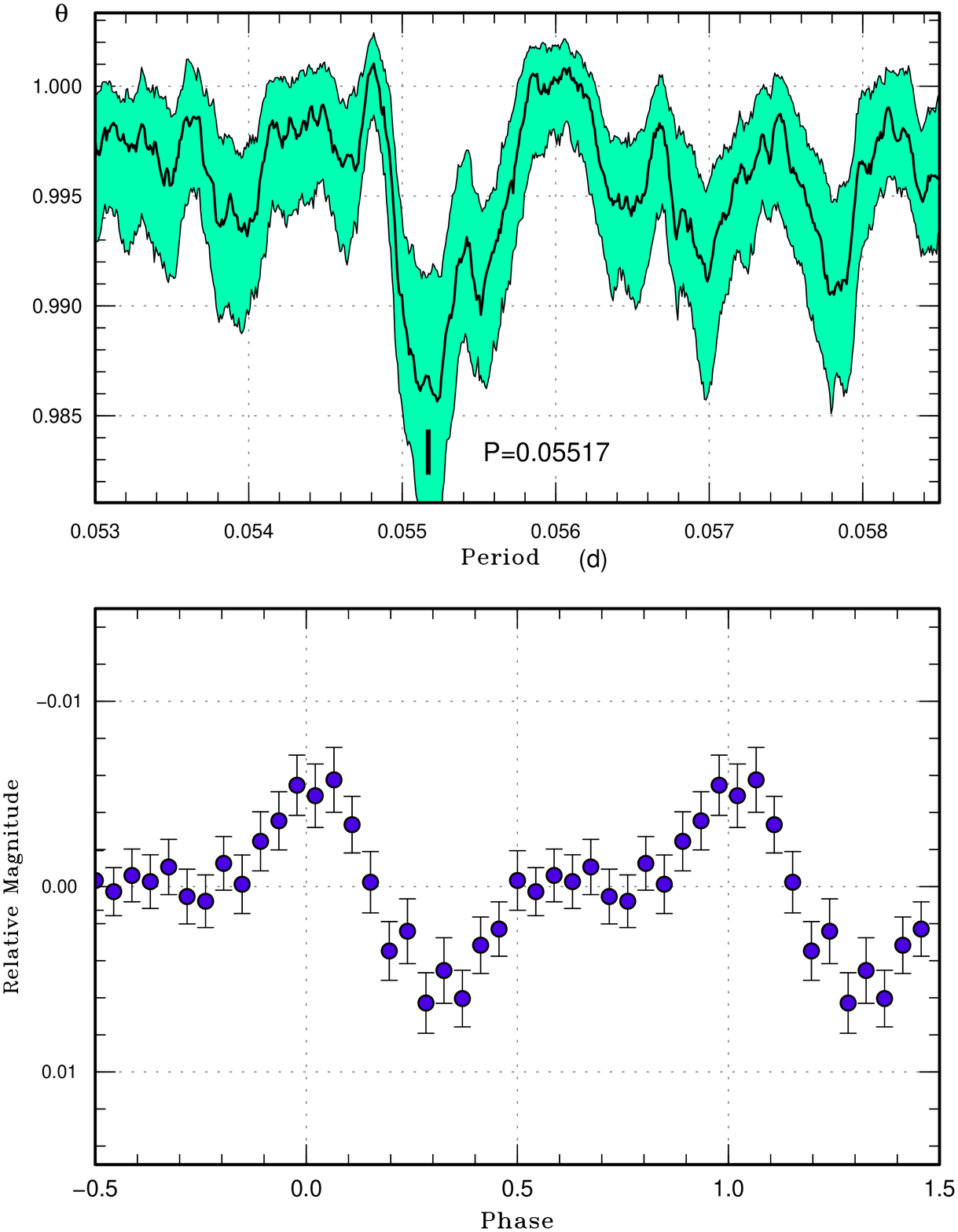}
  \end{center}
  \caption{Early superhumps in ASASSN-15ni (2015).
     (Upper): PDM analysis.
     (Lower): Phase-averaged profile.}
  \label{fig:asassn15nieshpdm}
\end{figure}

   Although no clear superhumps were visible before August 7
(10~d after the outburst detection), ordinary superhumps
appeared (vsnet-alert 18945, 18950, 18959, 18968, 18976;
figure \ref{fig:asassn15nishpdm}).
The times of superhump maxima are listed in
table \ref{tab:asassn15nioc2015}.  Clear stages A and B
can be recognized.  Stage A lasted at least 26 cycles.
A retrospective analysis of the early data detected
low-amplitude double-wave early superhumps
(figure \ref{fig:asassn15nieshpdm}) confirming
the WZ Sge-type classification.
The period of early superhump with the PDM method
is 0.05517(4)~d.  The fractional superhump excess 
$\epsilon^*$=0.027(2) of stage A corresponds
to $q$=0.074(2).  The low value of $q$ is consistent
with the long duration of stage A, relatively small
$P_{\rm dot}$ and low amplitude of ordinary superhumps.

% SI

% SI

% SI

\begin{table*}
\caption{Superhump maxima of ASASSN-15ni (2015)}\label{tab:asassn15nioc2015}
\begin{center}
\begin{tabular}{rp{55pt}p{40pt}r@{.}lrrp{55pt}p{40pt}r@{.}lr}
\hline
\multicolumn{1}{c}{$E$} & \multicolumn{1}{c}{max\commenta} & \multicolumn{1}{c}{error} & \multicolumn{2}{c}{$O-C$\commentb} & \multicolumn{1}{c}{$N$\commentc} & \multicolumn{1}{c}{$E$} & \multicolumn{1}{c}{max\commenta} & \multicolumn{1}{c}{error} & \multicolumn{2}{c}{$O-C$\commentb} & \multicolumn{1}{c}{$N$\commentc} \\
\hline
0 & 57241.5578 & 0.0013 & $-$0&0152 & 54 & 86 & 57246.3778 & 0.0004 & $-$0&0002 & 29 \\
9 & 57242.0711 & 0.0004 & $-$0&0047 & 120 & 87 & 57246.4340 & 0.0008 & 0&0002 & 134 \\
10 & 57242.1227 & 0.0004 & $-$0&0090 & 119 & 88 & 57246.4893 & 0.0007 & $-$0&0005 & 135 \\
13 & 57242.3053 & 0.0035 & 0&0060 & 17 & 89 & 57246.5451 & 0.0004 & $-$0&0005 & 90 \\
14 & 57242.3528 & 0.0004 & $-$0&0024 & 55 & 90 & 57246.5988 & 0.0021 & $-$0&0027 & 26 \\
15 & 57242.4116 & 0.0005 & 0&0005 & 76 & 104 & 57247.3815 & 0.0011 & $-$0&0022 & 27 \\
16 & 57242.4655 & 0.0004 & $-$0&0015 & 96 & 105 & 57247.4353 & 0.0006 & $-$0&0043 & 50 \\
17 & 57242.5225 & 0.0004 & $-$0&0003 & 56 & 106 & 57247.4934 & 0.0007 & $-$0&0021 & 58 \\
18 & 57242.5811 & 0.0005 & 0&0025 & 55 & 107 & 57247.5451 & 0.0016 & $-$0&0062 & 51 \\
26 & 57243.0327 & 0.0003 & 0&0070 & 64 & 122 & 57248.3867 & 0.0010 & $-$0&0027 & 28 \\
32 & 57243.3684 & 0.0003 & 0&0075 & 56 & 123 & 57248.4434 & 0.0011 & $-$0&0019 & 27 \\
33 & 57243.4229 & 0.0003 & 0&0062 & 54 & 124 & 57248.4983 & 0.0011 & $-$0&0029 & 29 \\
44 & 57244.0334 & 0.0011 & 0&0020 & 38 & 134 & 57249.0532 & 0.0014 & $-$0&0067 & 31 \\
45 & 57244.0903 & 0.0007 & 0&0031 & 45 & 135 & 57249.1132 & 0.0004 & $-$0&0025 & 100 \\
49 & 57244.3122 & 0.0003 & 0&0015 & 42 & 139 & 57249.3467 & 0.0031 & 0&0075 & 22 \\
50 & 57244.3689 & 0.0003 & 0&0023 & 55 & 140 & 57249.3927 & 0.0016 & $-$0&0024 & 25 \\
51 & 57244.4252 & 0.0003 & 0&0027 & 58 & 141 & 57249.4490 & 0.0006 & $-$0&0019 & 28 \\
52 & 57244.4806 & 0.0007 & 0&0022 & 82 & 142 & 57249.5104 & 0.0032 & 0&0036 & 17 \\
53 & 57244.5363 & 0.0008 & 0&0021 & 75 & 152 & 57250.0641 & 0.0012 & $-$0&0014 & 144 \\
54 & 57244.5937 & 0.0022 & 0&0036 & 9 & 153 & 57250.1208 & 0.0005 & $-$0&0006 & 113 \\
64 & 57245.1502 & 0.0002 & 0&0014 & 105 & 154 & 57250.1784 & 0.0007 & 0&0011 & 85 \\
65 & 57245.2057 & 0.0005 & 0&0010 & 119 & 157 & 57250.3434 & 0.0013 & $-$0&0015 & 28 \\
67 & 57245.3167 & 0.0007 & 0&0003 & 20 & 158 & 57250.3997 & 0.0009 & $-$0&0011 & 30 \\
68 & 57245.3717 & 0.0006 & $-$0&0006 & 67 & 159 & 57250.4554 & 0.0021 & $-$0&0013 & 25 \\
69 & 57245.4302 & 0.0005 & 0&0021 & 85 & 160 & 57250.5106 & 0.0013 & $-$0&0020 & 26 \\
70 & 57245.4857 & 0.0005 & 0&0017 & 76 & 175 & 57251.3568 & 0.0024 & 0&0062 & 22 \\
71 & 57245.5408 & 0.0018 & 0&0009 & 14 & 176 & 57251.4103 & 0.0031 & 0&0039 & 29 \\
80 & 57246.0407 & 0.0022 & $-$0&0020 & 49 & 177 & 57251.4603 & 0.0015 & $-$0&0021 & 27 \\
81 & 57246.0979 & 0.0008 & $-$0&0007 & 76 & 190 & 57252.1870 & 0.0020 & $-$0&0016 & 51 \\
82 & 57246.1594 & 0.0021 & 0&0049 & 28 & 205 & 57253.0315 & 0.0050 & 0&0047 & 14 \\
85 & 57246.3210 & 0.0005 & $-$0&0012 & 24 & \multicolumn{1}{c}{--} & \multicolumn{1}{c}{--} & \multicolumn{1}{c}{--} & \multicolumn{2}{c}{--} & \multicolumn{1}{c}{--}\\
\hline
  \multicolumn{12}{l}{\commenta BJD$-$2400000.} \\
  \multicolumn{12}{l}{\commentb Against max $= 2457241.5730 + 0.055872 E$.} \\
  \multicolumn{12}{l}{\commentc Number of points used to determine the maximum.} \\
\end{tabular}
\end{center}
\end{table*}

\subsection{ASASSN-15nl}\label{obj:asassn15nl}

\begin{figure}
  \begin{center}
%    \FigureFile(85mm,110mm){asassn15nlshpdm.eps}
    \FigureFile(85mm,110mm){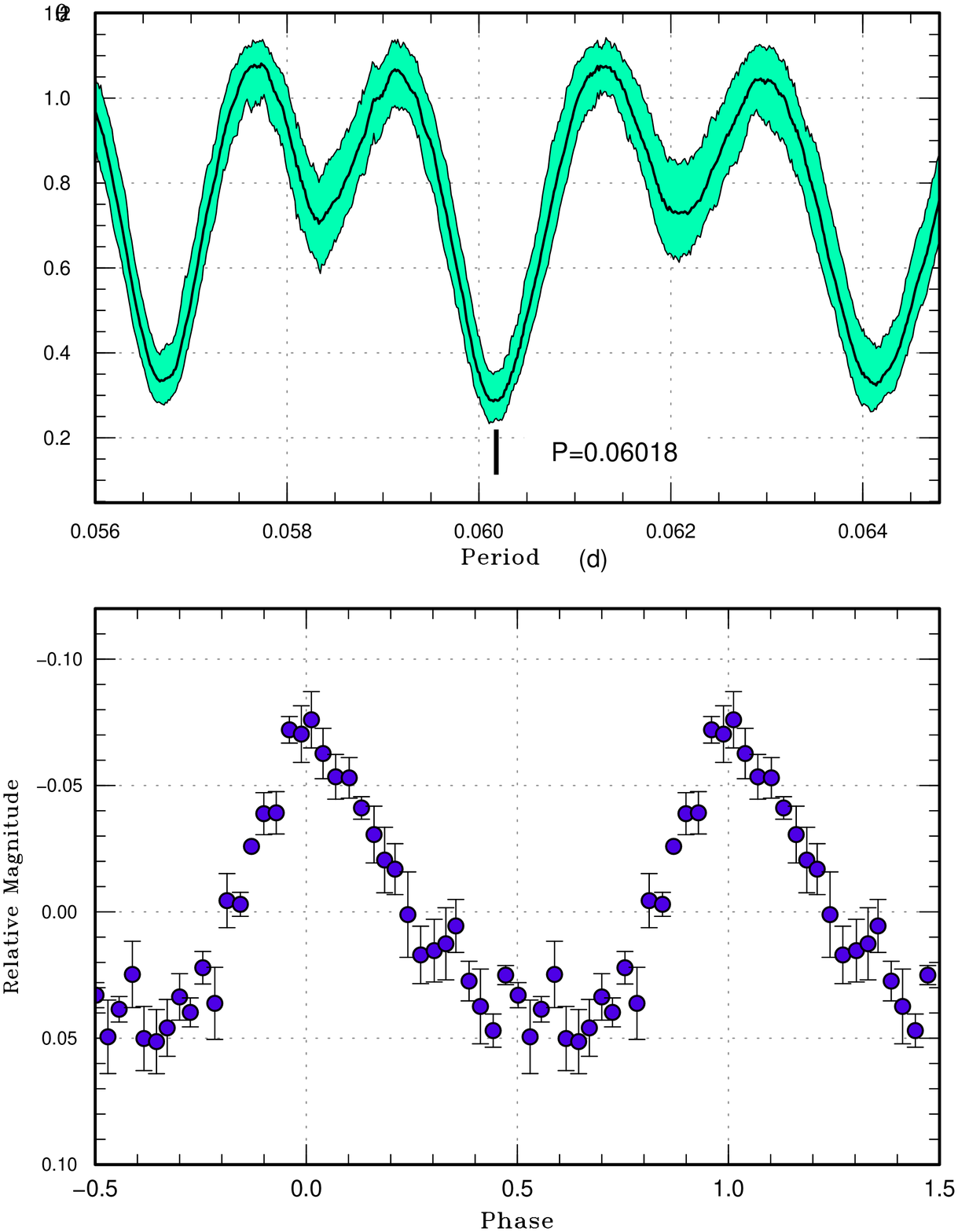}
  \end{center}
  \caption{Superhumps in ASASSN-15nl (2015).
     (Upper): PDM analysis of the segment after BJD 2457244.
     The alias selection was based on $O-C$ analysis of two
     nights.
     (Lower): Phase-averaged profile.}
  \label{fig:asassn15nlshpdm}
\end{figure}

   This object was detected as a transient at $V$=14.1
on 2015 August 1 by the ASAS-SN team.  The object was
found to be already bright ($V$=13.3) on July 29.
Although no superhump-like modulations were recorded
in our early observations, the object showed superhumps
on August 9 (vsnet-alert 18958, 18966;
figure \ref{fig:asassn15nlshpdm}).
Since there was a gap in the observation
between August 5 and 9, we could not determine when
superhumps started to appear but it took more than
7~d to develop superhumps.
The times of superhump maxima are listed in
table \ref{tab:asassn15nloc2015}.

% SI

% SI

\begin{table}
\caption{Superhump maxima of ASASSN-15nl (2015)}\label{tab:asassn15nloc2015}
\begin{center}
\begin{tabular}{rp{55pt}p{40pt}r@{.}lr}
\hline
\multicolumn{1}{c}{$E$} & \multicolumn{1}{c}{max\commenta} & \multicolumn{1}{c}{error} & \multicolumn{2}{c}{$O-C$\commentb} & \multicolumn{1}{c}{$N$\commentc} \\
\hline
0 & 57244.3661 & 0.0006 & 0&0006 & 33 \\
1 & 57244.4252 & 0.0008 & $-$0&0004 & 33 \\
16 & 57245.3269 & 0.0005 & $-$0&0002 & 48 \\
32 & 57246.2848 & 0.0043 & $-$0&0038 & 13 \\
33 & 57246.3524 & 0.0006 & 0&0038 & 44 \\
\hline
  \multicolumn{6}{l}{\commenta BJD$-$2400000.} \\
  \multicolumn{6}{l}{\commentb Against max $= 2457244.3655 + 0.060095 E$.} \\
  \multicolumn{6}{l}{\commentc Number of points used to determine the maximum.} \\
\end{tabular}
\end{center}
\end{table}

\subsection{ASASSN-15ob}\label{obj:asassn15ob}

\begin{figure}
  \begin{center}
%    \FigureFile(85mm,110mm){asassn15obshpdm.eps}
    \FigureFile(85mm,110mm){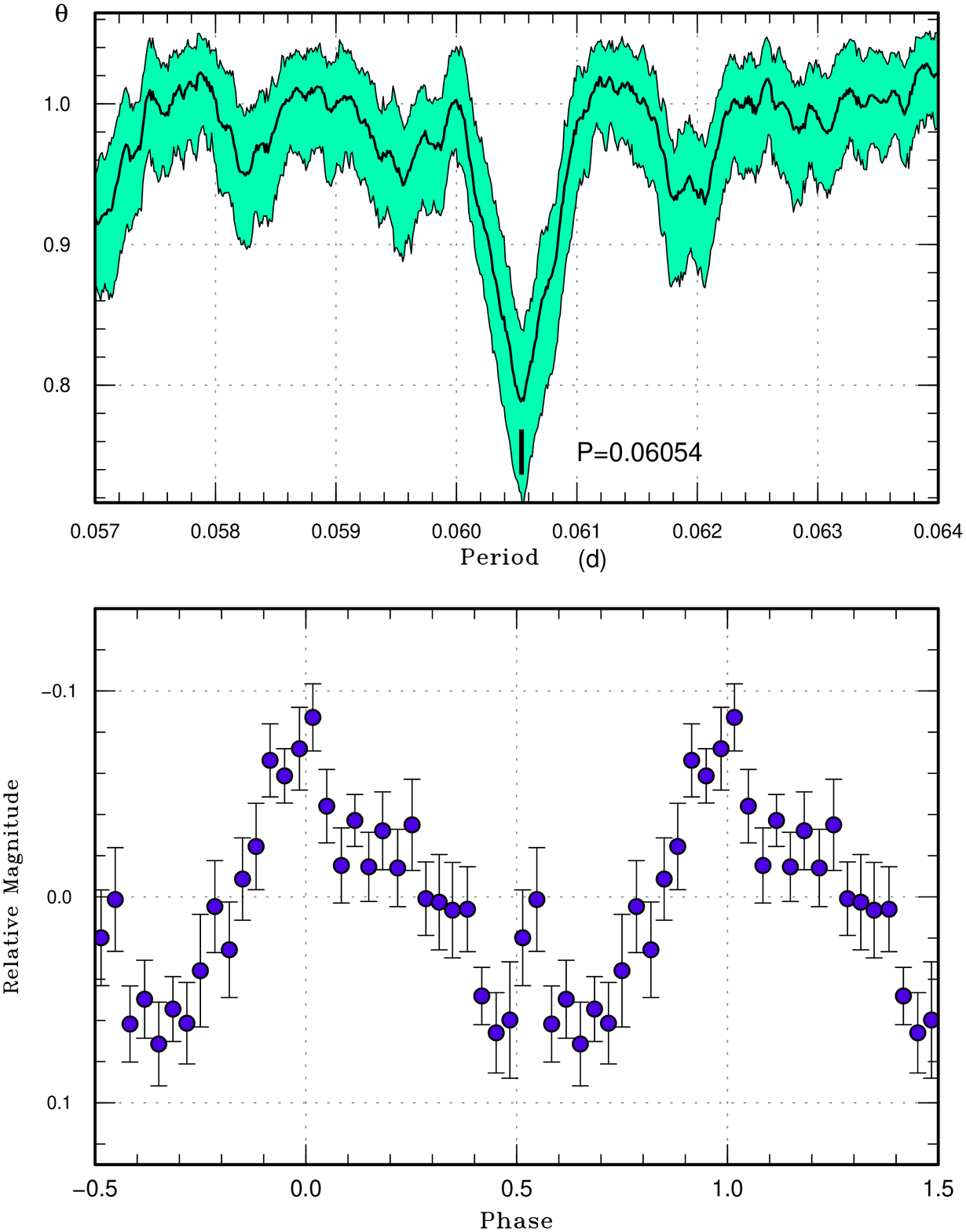}
  \end{center}
  \caption{Superhumps in ASASSN-15ob (2015).
     (Upper): PDM analysis.
     (Lower): Phase-averaged profile.}
  \label{fig:asassn15obshpdm}
\end{figure}

   This object was detected as a transient at $V$=15.2
on 2015 August 9 by the ASAS-SN team.
There were previous outbursts in the CRTS data.
Subsequent observations detected superhumps
(vsnet-alert 18975, 18982, 18991;
figure \ref{fig:asassn15obshpdm}).
The times of superhump maxima are listed in
table \ref{tab:asassn15oboc2015}.

% SI

% SI

\begin{table}
\caption{Superhump maxima of ASASSN-15ob (2015)}\label{tab:asassn15oboc2015}
\begin{center}
\begin{tabular}{rp{55pt}p{40pt}r@{.}lr}
\hline
\multicolumn{1}{c}{$E$} & \multicolumn{1}{c}{max\commenta} & \multicolumn{1}{c}{error} & \multicolumn{2}{c}{$O-C$\commentb} & \multicolumn{1}{c}{$N$\commentc} \\
\hline
0 & 57247.5669 & 0.0008 & 0&0047 & 19 \\
16 & 57248.5307 & 0.0040 & 0&0002 & 11 \\
17 & 57248.5921 & 0.0028 & 0&0010 & 15 \\
29 & 57249.3127 & 0.0012 & $-$0&0047 & 83 \\
33 & 57249.5580 & 0.0008 & $-$0&0015 & 19 \\
45 & 57250.2821 & 0.0009 & $-$0&0037 & 32 \\
83 & 57252.5897 & 0.0021 & 0&0040 & 12 \\
\hline
  \multicolumn{6}{l}{\commenta BJD$-$2400000.} \\
  \multicolumn{6}{l}{\commentb Against max $= 2457247.5622 + 0.060525 E$.} \\
  \multicolumn{6}{l}{\commentc Number of points used to determine the maximum.} \\
\end{tabular}
\end{center}
\end{table}

\subsection{ASASSN-15oj}\label{obj:asassn15oj}

\begin{figure}
  \begin{center}
%    \FigureFile(85mm,110mm){asassn15ojshpdm.eps}
    \FigureFile(85mm,110mm){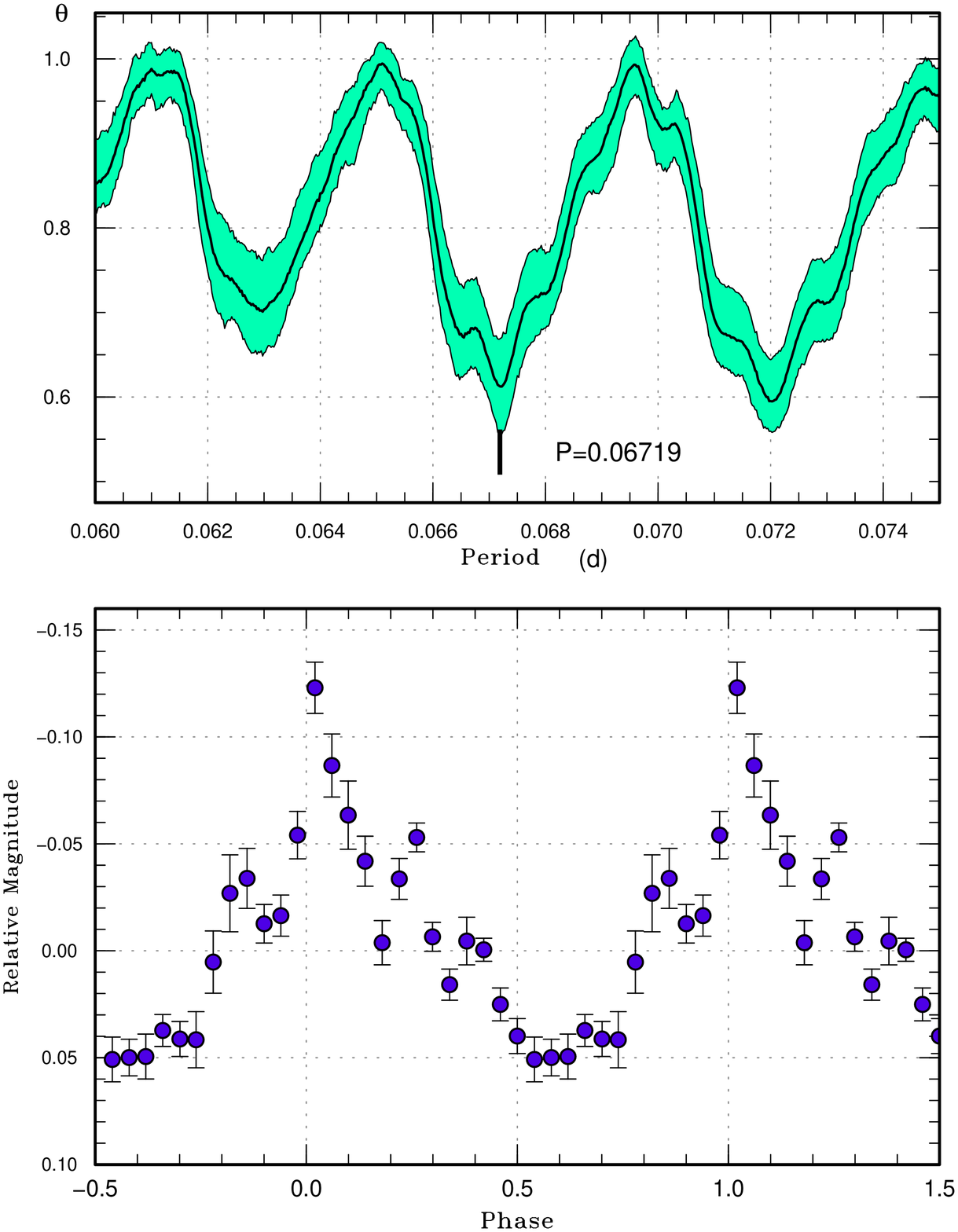}
  \end{center}
  \caption{Superhumps in ASASSN-15oj (2015).
     (Upper): PDM analysis.  The alias selection was to minimize
     the $O-C$ scatter.  Other aliases are also viable.
     (Lower): Phase-averaged profile.}
  \label{fig:asassn15ojshpdm}
\end{figure}

   This object was detected as a transient at $V$=13.8
on 2015 August 15 by the ASAS-SN team.
There were four previous outbursts in the CRTS data.
Subsequent observations detected superhumps
(vsnet-alert 19007; figure \ref{fig:asassn15ojshpdm}).
The times of superhump maxima are listed in
table \ref{tab:asassn15ojoc2015}.
This table is based on a candidate period.
Other aliases [0.07203(1)~d, 0.07764(1)~d] are also
viable.

% SI

% SI

\begin{table}
\caption{Superhump maxima of ASASSN-15oj (2015)}\label{tab:asassn15ojoc2015}
\begin{center}
\begin{tabular}{rp{55pt}p{40pt}r@{.}lr}
\hline
\multicolumn{1}{c}{$E$} & \multicolumn{1}{c}{max\commenta} & \multicolumn{1}{c}{error} & \multicolumn{2}{c}{$O-C$\commentb} & \multicolumn{1}{c}{$N$\commentc} \\
\hline
0 & 57259.2591 & 0.0010 & $-$0&0006 & 154 \\
14 & 57260.2143 & 0.0016 & 0&0083 & 99 \\
15 & 57260.2658 & 0.0011 & $-$0&0077 & 155 \\
\hline
  \multicolumn{6}{l}{\commenta BJD$-$2400000.} \\
  \multicolumn{6}{l}{\commentb Against max $= 2457259.2596 + 0.067596 E$.} \\
  \multicolumn{6}{l}{\commentc Number of points used to determine the maximum.} \\
\end{tabular}
\end{center}
\end{table}

\subsection{ASASSN-15ok}\label{obj:asassn15ok}

\begin{figure}
  \begin{center}
%    \FigureFile(85mm,110mm){asassn15okshpdm.eps}
    \FigureFile(85mm,110mm){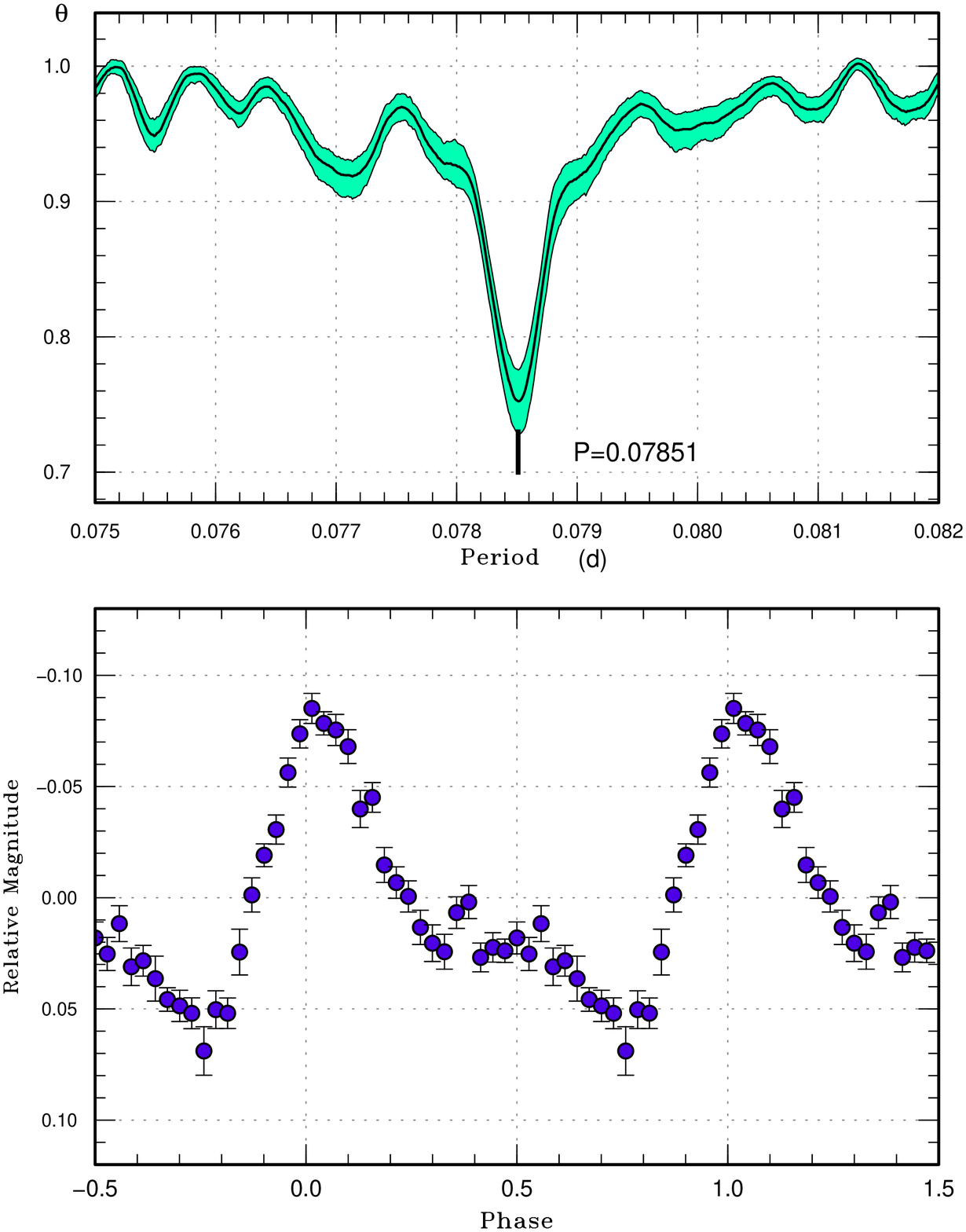}
  \end{center}
  \caption{Superhumps in ASASSN-15ok (2015).
     (Upper): PDM analysis.
     (Lower): Phase-averaged profile.}
  \label{fig:asassn15okshpdm}
\end{figure}

   This object was detected as a transient at $V$=13.7
on 2015 August 15 by the ASAS-SN team.
There was at least one well-recorded outburst
(2008 February) in ASAS-3 data (vsnet-alert 18980).
Although the object was initially suspected to be
an SS Cyg-type due to the relatively bright counterpart
in 2MASS (vsnet-alert 18980), the 2MASS colors are
consistent (such as $J-K \sim$0) with those of
an outbursting dwarf nova.  The object was accidentally
in outburst in 2MASS scans.
Subsequent observations detected superhumps
(vsnet-alert 18987, 18994, 19008; 
figure \ref{fig:asassn15okshpdm}).
The times of superhump maxima are listed in
table \ref{tab:asassn15okoc2015}.
The observations apparently covered the later phase of
the superoutburst (the object faded rapidly 9~d
after the initial observation) and the maxima
for $E \ge$25 likely represent stage C superhumps.
The maxima $E <$25 were likely the final phase of
stage B superhumps.

% SI

% SI

\begin{table}
\caption{Superhump maxima of ASASSN-15ok (2015)}\label{tab:asassn15okoc2015}
\begin{center}
\begin{tabular}{rp{55pt}p{40pt}r@{.}lr}
\hline
\multicolumn{1}{c}{$E$} & \multicolumn{1}{c}{max\commenta} & \multicolumn{1}{c}{error} & \multicolumn{2}{c}{$O-C$\commentb} & \multicolumn{1}{c}{$N$\commentc} \\
\hline
0 & 57251.8177 & 0.0006 & $-$0&0068 & 29 \\
1 & 57251.8954 & 0.0008 & $-$0&0076 & 23 \\
12 & 57252.7606 & 0.0013 & $-$0&0065 & 20 \\
25 & 57253.7910 & 0.0007 & 0&0027 & 30 \\
26 & 57253.8680 & 0.0008 & 0&0012 & 30 \\
38 & 57254.8116 & 0.0008 & 0&0021 & 32 \\
39 & 57254.8906 & 0.0008 & 0&0026 & 26 \\
46 & 57255.4404 & 0.0003 & 0&0025 & 175 \\
47 & 57255.5206 & 0.0003 & 0&0042 & 182 \\
48 & 57255.5980 & 0.0003 & 0&0030 & 181 \\
51 & 57255.8345 & 0.0008 & 0&0039 & 23 \\
58 & 57256.3828 & 0.0004 & 0&0023 & 128 \\
59 & 57256.4624 & 0.0003 & 0&0034 & 181 \\
60 & 57256.5401 & 0.0003 & 0&0025 & 181 \\
61 & 57256.6189 & 0.0003 & 0&0027 & 181 \\
76 & 57257.7925 & 0.0015 & $-$0&0019 & 25 \\
77 & 57257.8716 & 0.0015 & $-$0&0013 & 25 \\
89 & 57258.8191 & 0.0027 & 0&0036 & 25 \\
90 & 57258.8946 & 0.0032 & 0&0005 & 16 \\
96 & 57259.3696 & 0.0009 & 0&0042 & 181 \\
97 & 57259.4450 & 0.0013 & 0&0011 & 182 \\
98 & 57259.5276 & 0.0008 & 0&0051 & 135 \\
102 & 57259.8361 & 0.0024 & $-$0&0006 & 24 \\
109 & 57260.3840 & 0.0028 & $-$0&0026 & 181 \\
110 & 57260.4631 & 0.0021 & $-$0&0020 & 181 \\
111 & 57260.5349 & 0.0035 & $-$0&0088 & 181 \\
115 & 57260.8485 & 0.0050 & $-$0&0093 & 22 \\
\hline
  \multicolumn{6}{l}{\commenta BJD$-$2400000.} \\
  \multicolumn{6}{l}{\commentb Against max $= 2457251.8245 + 0.078551 E$.} \\
  \multicolumn{6}{l}{\commentc Number of points used to determine the maximum.} \\
\end{tabular}
\end{center}
\end{table}

\subsection{ASASSN-15pi}\label{obj:asassn15pi}

\begin{figure}
  \begin{center}
%    \FigureFile(85mm,110mm){asassn15pishpdm.eps}
    \FigureFile(85mm,110mm){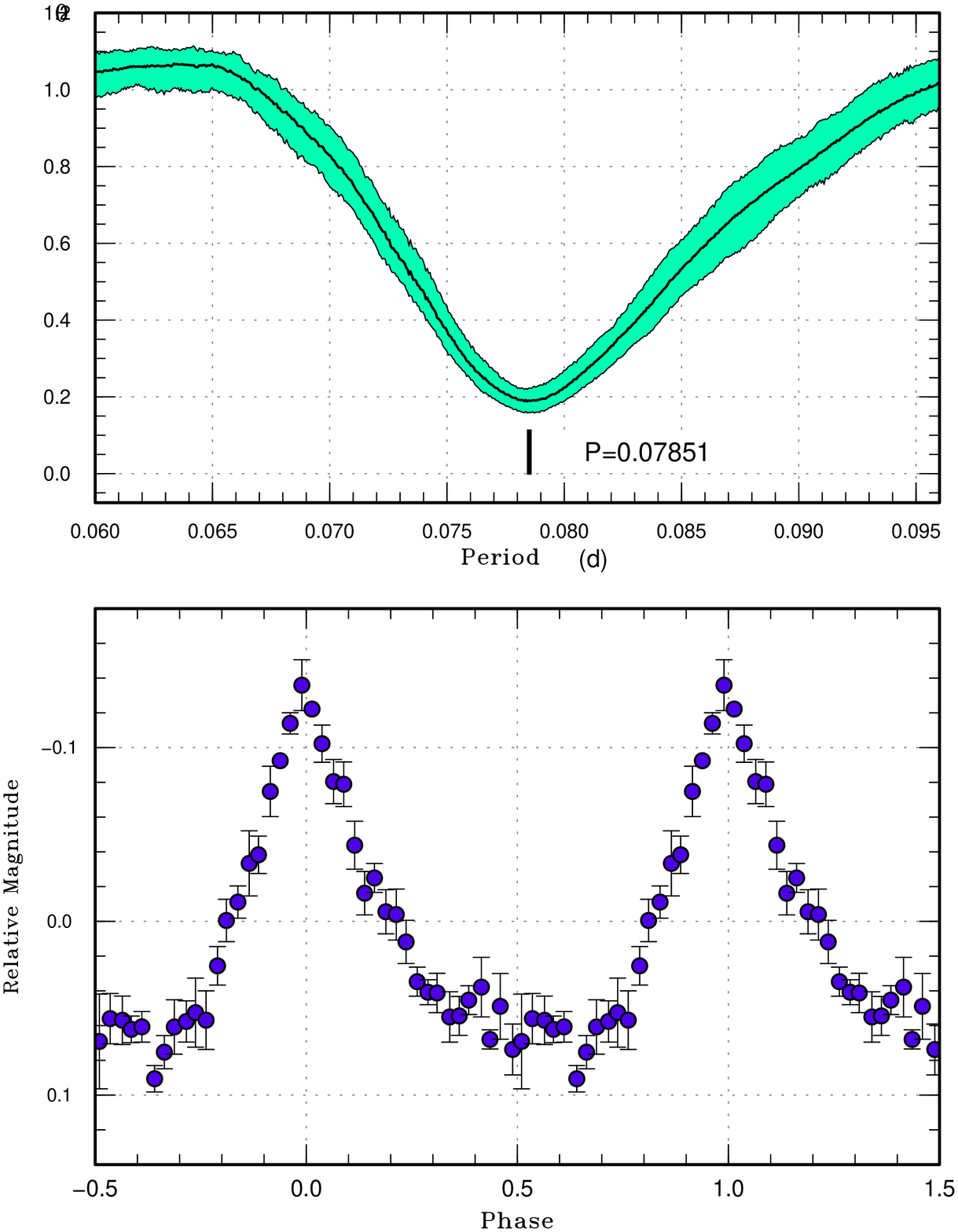}
  \end{center}
  \caption{Superhumps in ASASSN-15pi (2015).
     (Upper): PDM analysis.
     (Lower): Phase-averaged profile.}
  \label{fig:asassn15pishpdm}
\end{figure}

   This object was detected as a transient at $V$=16.4
on 2015 September 8 by the ASAS-SN team.  The object was
already $V$=16.1 on September 5 (ASAS-SN data).
There is no known quiescent counterpart.
Superhumps were detected on single-night observation
on September 10 (vsnet-alert 19046;
figure \ref{fig:asassn15pishpdm}).
The best period determined by the PDM method
is 0.0785(2)~d.
The times of superhump maxima are listed in
table \ref{tab:asassn15pioc2015}.

% SI

% SI

\begin{table}
\caption{Superhump maxima of ASASSN-15pi (2015)}\label{tab:asassn15pioc2015}
\begin{center}
\begin{tabular}{rp{55pt}p{40pt}r@{.}lr}
\hline
\multicolumn{1}{c}{$E$} & \multicolumn{1}{c}{max\commenta} & \multicolumn{1}{c}{error} & \multicolumn{2}{c}{$O-C$\commentb} & \multicolumn{1}{c}{$N$\commentc} \\
\hline
0 & 57276.3205 & 0.0008 & $-$0&0006 & 36 \\
1 & 57276.4000 & 0.0007 & 0&0006 & 51 \\
2 & 57276.4784 & 0.0006 & 0&0007 & 50 \\
3 & 57276.5548 & 0.0009 & $-$0&0012 & 51 \\
4 & 57276.6346 & 0.0007 & 0&0003 & 45 \\
\hline
  \multicolumn{6}{l}{\commenta BJD$-$2400000.} \\
  \multicolumn{6}{l}{\commentb Against max $= 2457276.3211 + 0.078307 E$.} \\
  \multicolumn{6}{l}{\commentc Number of points used to determine the maximum.} \\
\end{tabular}
\end{center}
\end{table}

\subsection{ASASSN-15pu}\label{obj:asassn15pu}

\begin{figure}
  \begin{center}
%    \FigureFile(85mm,110mm){asassn15pueshpdm.eps}
    \FigureFile(85mm,110mm){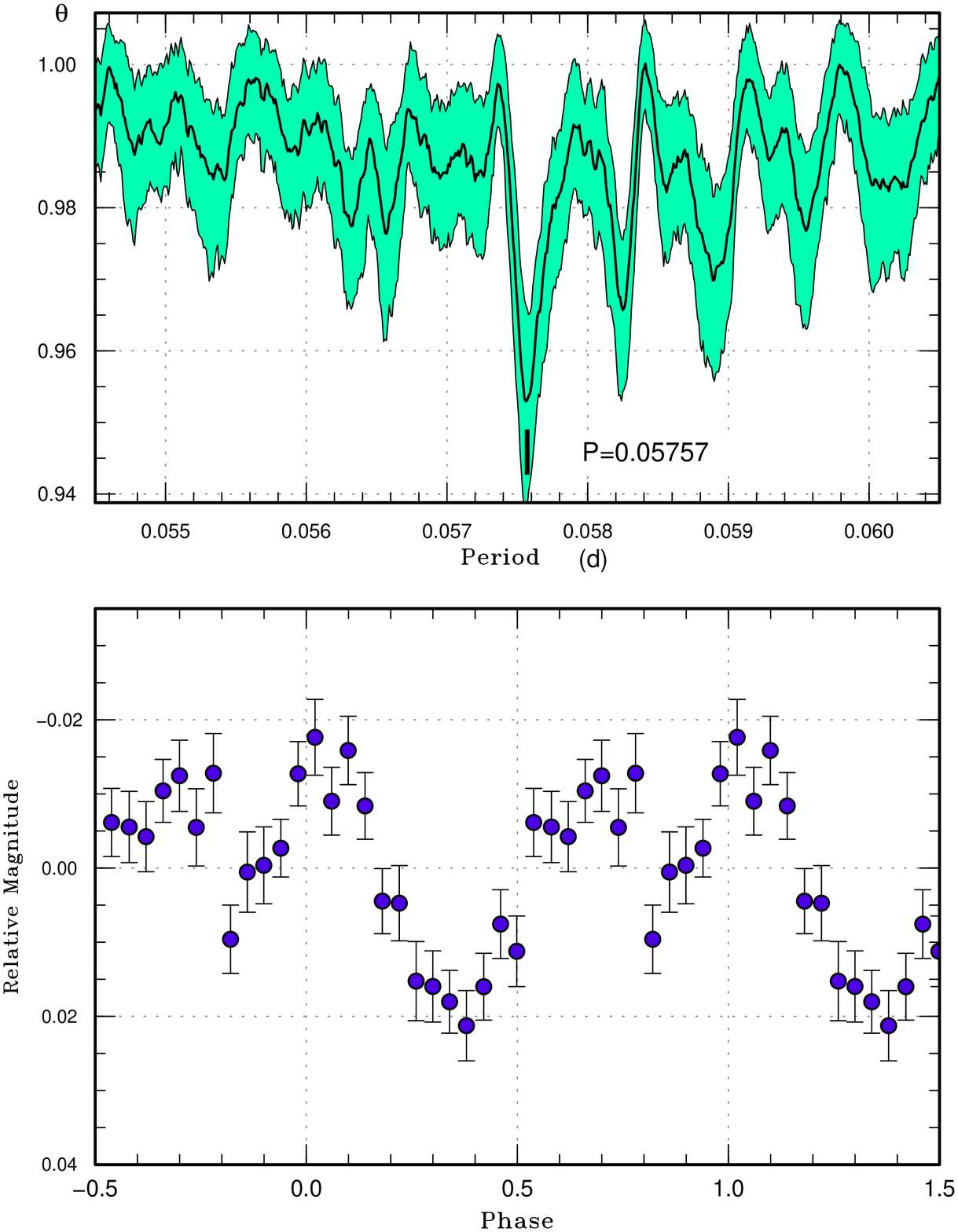}
  \end{center}
  \caption{Early superhumps in ASASSN-15pu (2015).
     (Upper): PDM analysis.
     (Lower): Phase-averaged profile.}
  \label{fig:asassn15pueshpdm}
\end{figure}

   This object was detected as a transient at $V$=13.7
on 2015 September 18 by the ASAS-SN team
\citep{sta15asassn15puatel8068}.
There is a $B_j$=22.1 mag counterpart in GSC 2.3.2.
The large outburst amplitude already suggested
a WZ Sge-type dwarf nova \citep{sta15asassn15puatel8068}.
Early superhumps were immediately recorded
(vsnet-alert 19074; the period was corrected
in vsnet-alert 19095; figure \ref{fig:asassn15pueshpdm}),
Ordinary superhumps started to grow
on September 28 (vsnet-alert 19095, 19125).
The times of superhump maxima are listed in
table \ref{tab:asassn15puoc2015}.  The maxima for
$E \le$18 correspond to stage A superhumps.
The $\epsilon^*$ for stage A superhumps [0.028(5)]
corresponds to $q$=0.074(16).

% SI

% SI

\begin{figure}
  \begin{center}
%    \FigureFile(85mm,110mm){asassn15pushpdm.eps}
    \FigureFile(85mm,110mm){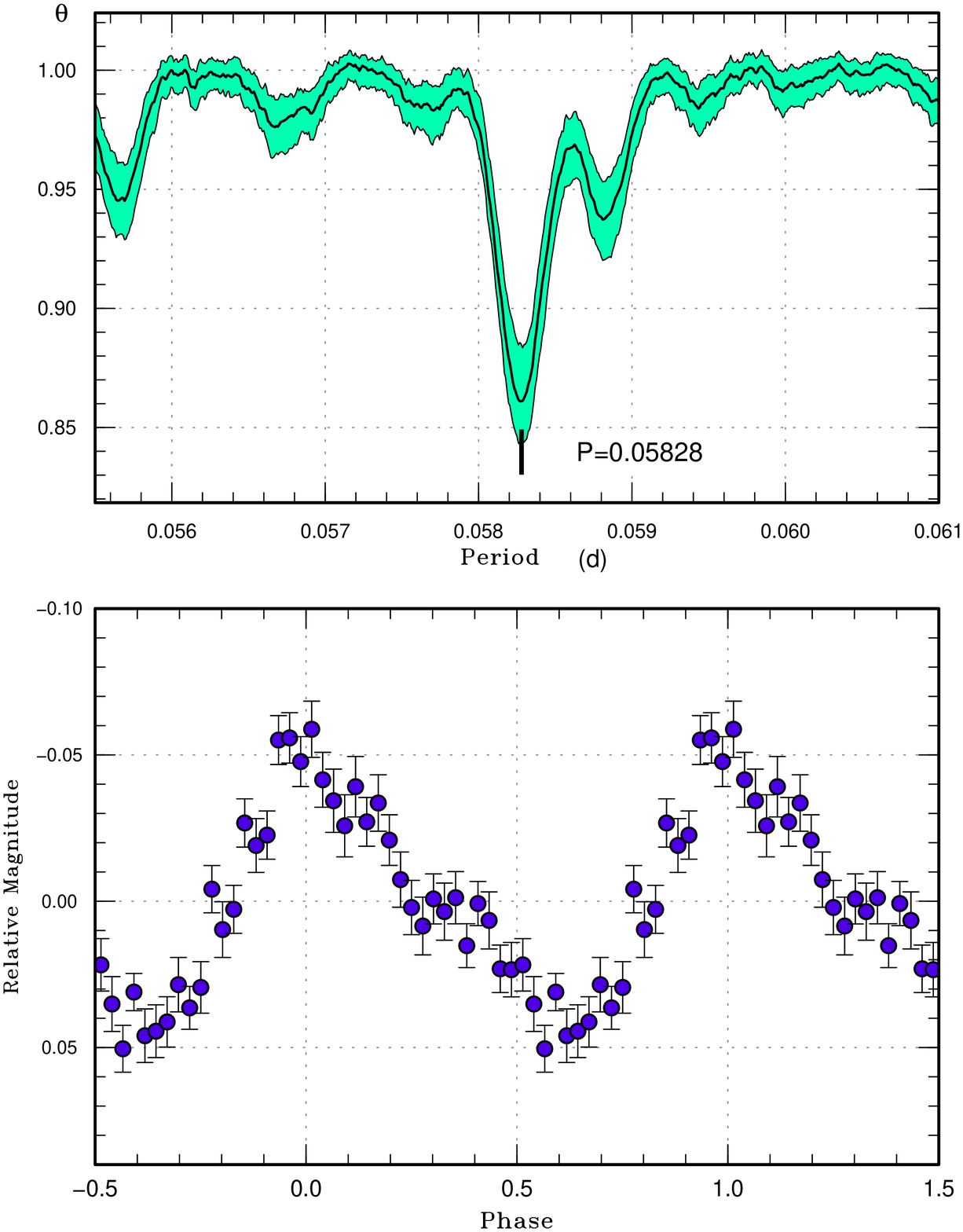}
  \end{center}
  \caption{Ordinary superhumps in ASASSN-15pu (2015).
     (Upper): PDM analysis.
     (Lower): Phase-averaged profile.}
  \label{fig:asassn15pushpdm}
\end{figure}

% SI

\begin{table}
\caption{Superhump maxima of ASASSN-15pu (2015)}\label{tab:asassn15puoc2015}
\begin{center}
\begin{tabular}{rp{55pt}p{40pt}r@{.}lr}
\hline
\multicolumn{1}{c}{$E$} & \multicolumn{1}{c}{max\commenta} & \multicolumn{1}{c}{error} & \multicolumn{2}{c}{$O-C$\commentb} & \multicolumn{1}{c}{$N$\commentc} \\
\hline
0 & 57293.0992 & 0.0029 & $-$0&0064 & 21 \\
1 & 57293.1490 & 0.0023 & $-$0&0149 & 36 \\
16 & 57294.0433 & 0.0013 & 0&0048 & 25 \\
17 & 57294.1001 & 0.0014 & 0&0033 & 37 \\
18 & 57294.1605 & 0.0015 & 0&0054 & 36 \\
34 & 57295.0918 & 0.0010 & 0&0038 & 36 \\
35 & 57295.1552 & 0.0016 & 0&0089 & 36 \\
69 & 57297.1285 & 0.0016 & $-$0&0003 & 36 \\
70 & 57297.1919 & 0.0035 & 0&0048 & 17 \\
84 & 57298.0006 & 0.0007 & $-$0&0028 & 23 \\
86 & 57298.1174 & 0.0015 & $-$0&0026 & 37 \\
87 & 57298.1786 & 0.0014 & 0&0004 & 30 \\
103 & 57299.1129 & 0.0013 & 0&0017 & 37 \\
104 & 57299.1685 & 0.0018 & $-$0&0010 & 30 \\
106 & 57299.2841 & 0.0012 & $-$0&0020 & 134 \\
107 & 57299.3435 & 0.0012 & $-$0&0009 & 134 \\
108 & 57299.4008 & 0.0025 & $-$0&0019 & 135 \\
109 & 57299.4589 & 0.0053 & $-$0&0021 & 54 \\
123 & 57300.2816 & 0.0017 & 0&0043 & 135 \\
124 & 57300.3375 & 0.0017 & 0&0019 & 134 \\
125 & 57300.3929 & 0.0021 & $-$0&0011 & 134 \\
141 & 57301.3274 & 0.0018 & 0&0006 & 134 \\
142 & 57301.3884 & 0.0051 & 0&0033 & 60 \\
146 & 57301.6112 & 0.0042 & $-$0&0072 & 27 \\
\hline
  \multicolumn{6}{l}{\commenta BJD$-$2400000.} \\
  \multicolumn{6}{l}{\commentb Against max $= 2457293.1056 + 0.058307 E$.} \\
  \multicolumn{6}{l}{\commentc Number of points used to determine the maximum.} \\
\end{tabular}
\end{center}
\end{table}

\subsection{ASASSN-15qe}\label{obj:asassn15qe}

\begin{figure}
  \begin{center}
%    \FigureFile(85mm,110mm){asassn15qeshpdm.eps}
    \FigureFile(85mm,110mm){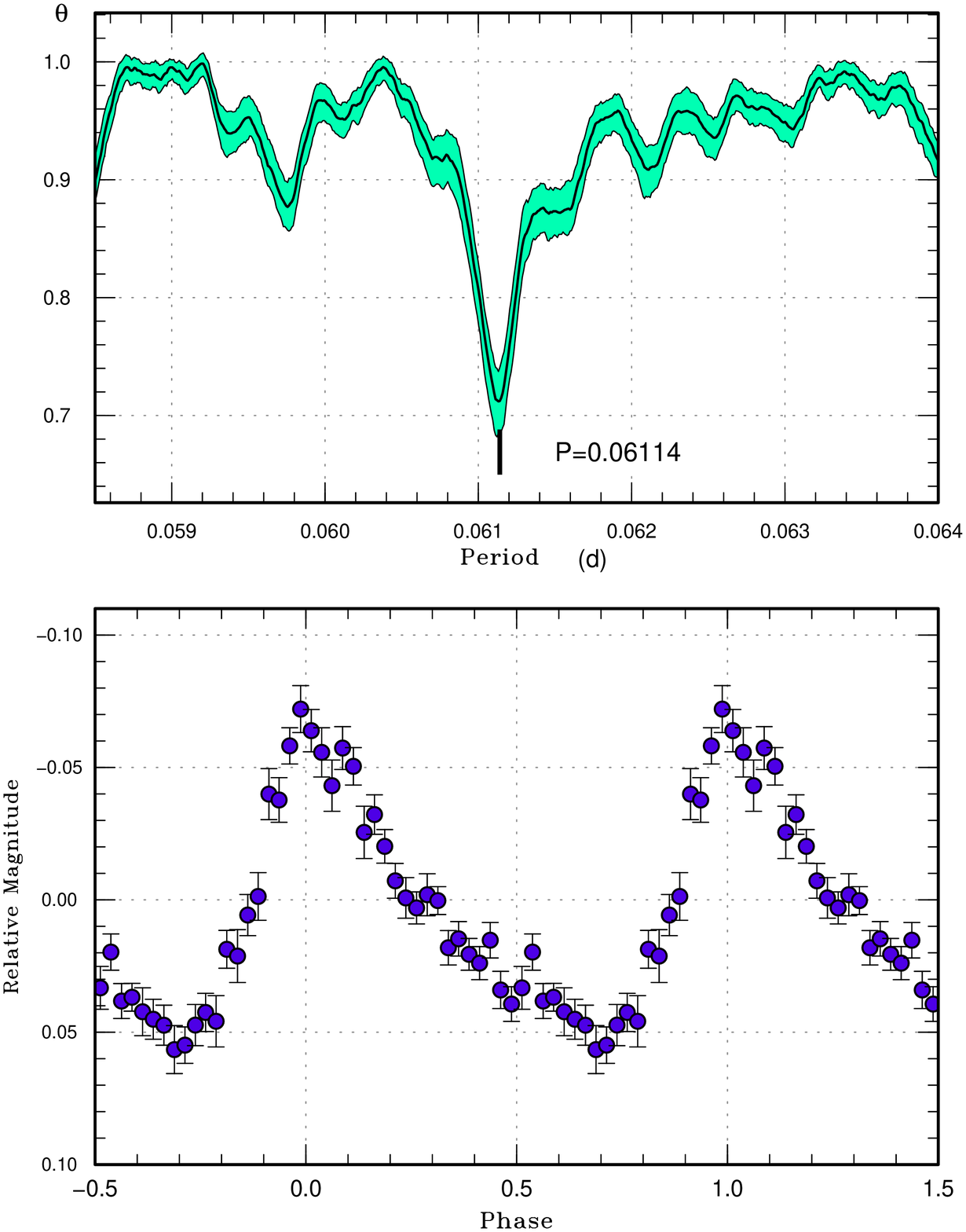}
  \end{center}
  \caption{Ordinary superhumps in ASASSN-15qe (2015).
     (Upper): PDM analysis.
     (Lower): Phase-averaged profile.}
  \label{fig:asassn15qeshpdm}
\end{figure}

   This object was detected as a transient at $V$=15.0
on 2015 October 2 by the ASAS-SN team.  The object was
found to be rising at $V$=16.3 on October 1.
There is a $B_j$=21.2 mag counterpart in GSC 2.3.2.
Subsequent observations detected
superhumps (vsnet-alert 19112, 19131;
figure \ref{fig:asassn15qeshpdm}).
The times of superhump maxima are listed in
table \ref{tab:asassn15qeoc2015}.  Due to the gap
in the observation between $E$=62 and $E$=175,
the cycle count between them is uncertain.
We used maxima for $E \le$62 for determining
the period in table \ref{tab:perlist}.

% SI

% SI

\begin{table}
\caption{Superhump maxima of ASASSN-15qe (2015)}\label{tab:asassn15qeoc2015}
\begin{center}
\begin{tabular}{rp{55pt}p{40pt}r@{.}lr}
\hline
\multicolumn{1}{c}{$E$} & \multicolumn{1}{c}{max\commenta} & \multicolumn{1}{c}{error} & \multicolumn{2}{c}{$O-C$\commentb} & \multicolumn{1}{c}{$N$\commentc} \\
\hline
0 & 57299.3882 & 0.0002 & 0&0015 & 54 \\
1 & 57299.4498 & 0.0003 & 0&0019 & 57 \\
42 & 57301.9524 & 0.0005 & $-$0&0017 & 88 \\
43 & 57302.0165 & 0.0003 & 0&0013 & 171 \\
44 & 57302.0747 & 0.0002 & $-$0&0017 & 229 \\
45 & 57302.1361 & 0.0004 & $-$0&0014 & 129 \\
59 & 57302.9929 & 0.0007 & $-$0&0004 & 101 \\
60 & 57303.0546 & 0.0005 & 0&0002 & 131 \\
61 & 57303.1160 & 0.0004 & 0&0004 & 131 \\
62 & 57303.1751 & 0.0014 & $-$0&0015 & 63 \\
175 & 57310.0869 & 0.0023 & 0&0027 & 99 \\
176 & 57310.1439 & 0.0016 & $-$0&0014 & 129 \\
\hline
  \multicolumn{6}{l}{\commenta BJD$-$2400000.} \\
  \multicolumn{6}{l}{\commentb Against max $= 2457299.3867 + 0.061128 E$.} \\
  \multicolumn{6}{l}{\commentc Number of points used to determine the maximum.} \\
\end{tabular}
\end{center}
\end{table}

\subsection{ASASSN-15ql}\label{obj:asassn15ql}

\begin{figure}
  \begin{center}
%    \FigureFile(85mm,70mm){asassn15qlshlc.eps}
    \FigureFile(85mm,70mm){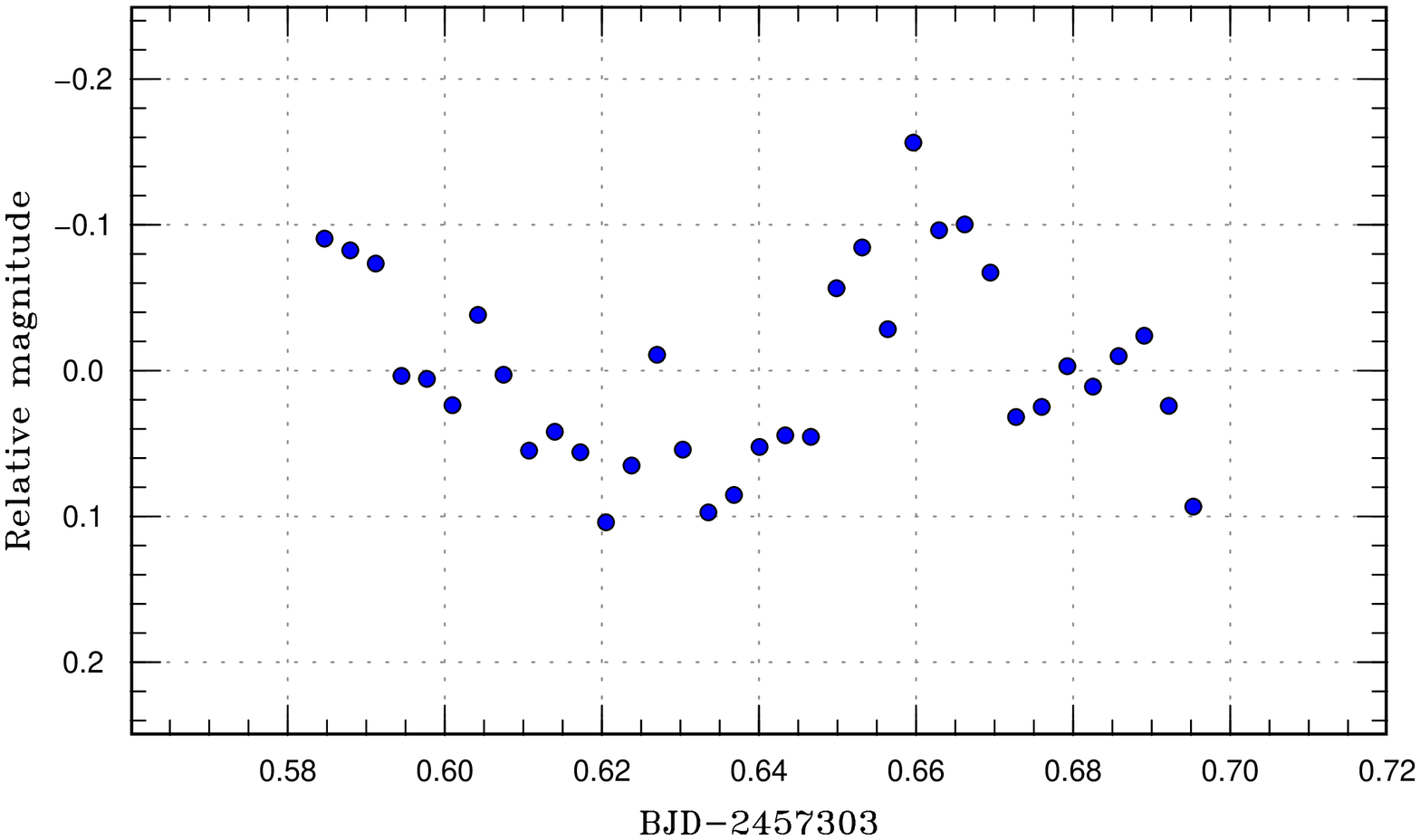}
  \end{center}
  \caption{Likely superhumps in ASASSN-15ql (2015).
  }
  \label{fig:asassn15qlshlc}
\end{figure}

   This object was detected as a transient at $V$=15.4
on 2015 October 4 by the ASAS-SN team.  The object was
found to be already bright at $V$=14.8 on September 30.
Although the observation on October 8 suggested
the presence of superhumps (vsnet-alert 19141;
figure \ref{fig:asassn15qlshlc}),
the object was already rapidly fading 4~d later,
and no confirmatory observations were obtained.

\subsection{ASASSN-15qo}\label{obj:asassn15qo}

\begin{figure}
  \begin{center}
%    \FigureFile(85mm,70mm){asassn15qoshlc.eps}
    \FigureFile(85mm,70mm){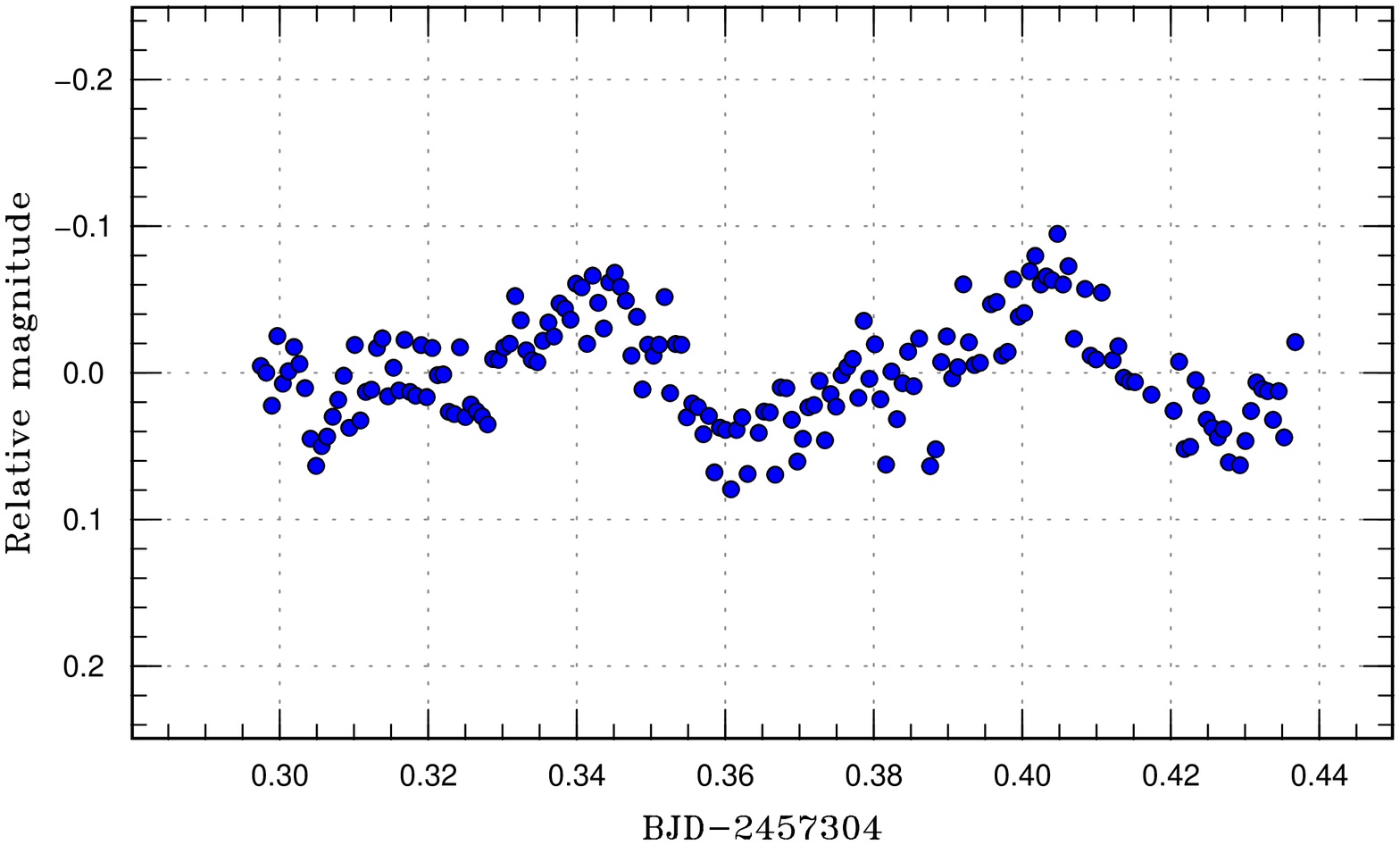}
  \end{center}
  \caption{Likely early superhumps in ASASSN-15qo (2015).
  }
  \label{fig:asassn15qoshlc}
\end{figure}

   This object was detected as a transient at $V$=15.2
on 2015 October 7 by the ASAS-SN team.
Observations on October 8 detected modulations
(vsnet-alert 19142; figure \ref{fig:asassn15qoshlc}).
The data, however, on later nights were fragmentary
and not conclusive.  As judged from the large outburst
amplitude and the profile of the humps (slower rise),
we consider that these modulations were early superhumps
and the object is likely a WZ Sge-type dwarf nova.
The period determined from the single-night observations
is 0.062(1)~d, which is long for a WZ Sge-type dwarf nova.

\subsection{ASASSN-15qq}\label{obj:asassn15qq}

\begin{figure}
  \begin{center}
%    \FigureFile(85mm,110mm){asassn15qqshpdm.eps}
    \FigureFile(85mm,110mm){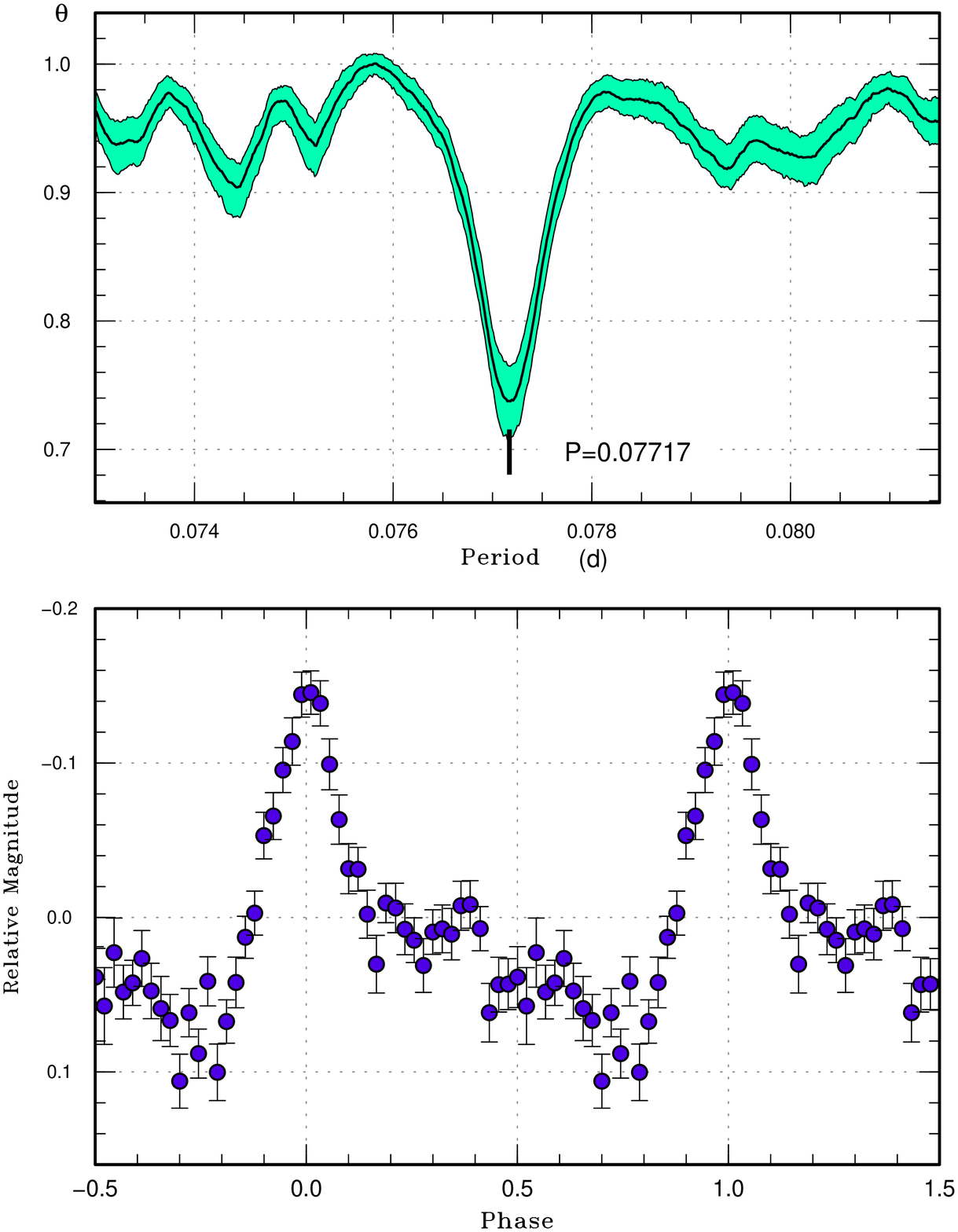}
  \end{center}
  \caption{Superhumps in ASASSN-15qq (2015).
     (Upper): PDM analysis.
     (Lower): Phase-averaged profile.}
  \label{fig:asassn15qqshpdm}
\end{figure}

   This object was detected as a transient at $V$=16.2
on 2015 October 7 by the ASAS-SN team.  The object was found
to be already bright at $V$=15.7 on October 4.
Superhumps were immediately detected
(vsnet-alert 19152, 19163, 19171;
figure \ref{fig:asassn15qqshpdm}).
The times of superhump maxima are listed in
table \ref{tab:asassn15qqoc2015}.
The stage of superhumps was not clear.  It is possible
that the data are a combination of stages B and C,
since the outburst detection was not made early
enough.  In table \ref{tab:perlist}, we gave a globally
averaged value.

% SI

% SI

\begin{table}
\caption{Superhump maxima of ASASSN-15qq (2015)}\label{tab:asassn15qqoc2015}
\begin{center}
\begin{tabular}{rp{55pt}p{40pt}r@{.}lr}
\hline
\multicolumn{1}{c}{$E$} & \multicolumn{1}{c}{max\commenta} & \multicolumn{1}{c}{error} & \multicolumn{2}{c}{$O-C$\commentb} & \multicolumn{1}{c}{$N$\commentc} \\
\hline
0 & 57305.4081 & 0.0010 & $-$0&0012 & 178 \\
1 & 57305.4876 & 0.0016 & 0&0011 & 112 \\
11 & 57306.2561 & 0.0007 & $-$0&0017 & 138 \\
12 & 57306.3352 & 0.0008 & 0&0003 & 112 \\
25 & 57307.3363 & 0.0010 & $-$0&0013 & 178 \\
26 & 57307.4128 & 0.0013 & $-$0&0019 & 142 \\
29 & 57307.6452 & 0.0019 & $-$0&0009 & 19 \\
30 & 57307.7278 & 0.0038 & 0&0045 & 9 \\
37 & 57308.2654 & 0.0009 & 0&0023 & 149 \\
39 & 57308.4183 & 0.0009 & 0&0009 & 159 \\
42 & 57308.6502 & 0.0025 & 0&0013 & 19 \\
55 & 57309.6499 & 0.0034 & $-$0&0016 & 26 \\
68 & 57310.6507 & 0.0045 & $-$0&0035 & 26 \\
81 & 57311.6587 & 0.0028 & 0&0018 & 22 \\
\hline
  \multicolumn{6}{l}{\commenta BJD$-$2400000.} \\
  \multicolumn{6}{l}{\commentb Against max $= 2457305.4093 + 0.077131 E$.} \\
  \multicolumn{6}{l}{\commentc Number of points used to determine the maximum.} \\
\end{tabular}
\end{center}
\end{table}

\subsection{ASASSN-15rf}\label{obj:asassn15rf}

\begin{figure}
  \begin{center}
%    \FigureFile(85mm,70mm){asassn15rfshlc.eps}
    \FigureFile(85mm,70mm){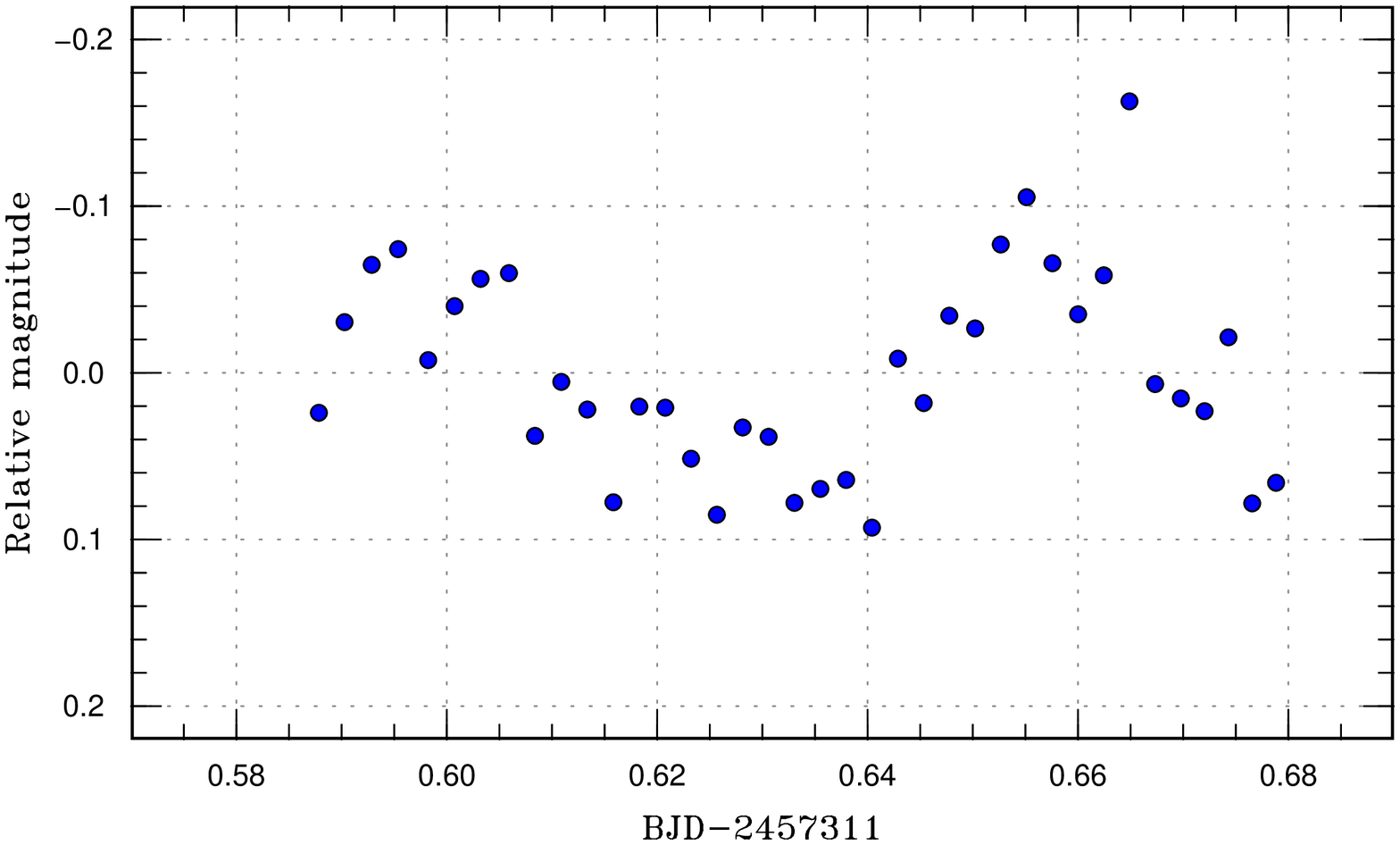}
  \end{center}
  \caption{Likely superhumps in ASASSN-15rf (2015).
  }
  \label{fig:asassn15rfshlc}
\end{figure}

   This object was detected as a transient at $V$=15.7
on 2015 October 13 by the ASAS-SN team.
Although observations on the initial night (October 16,
figure \ref{fig:asassn15rfshlc})
detected likely superhumps, we could not detect
a significant signal on four nights since October 22,
probably due to the low signal-to-noise ratio.

\subsection{ASASSN-15rj}\label{obj:asassn15rj}

\begin{figure}
  \begin{center}
%    \FigureFile(85mm,110mm){asassn15rjshpdm.eps}
    \FigureFile(85mm,110mm){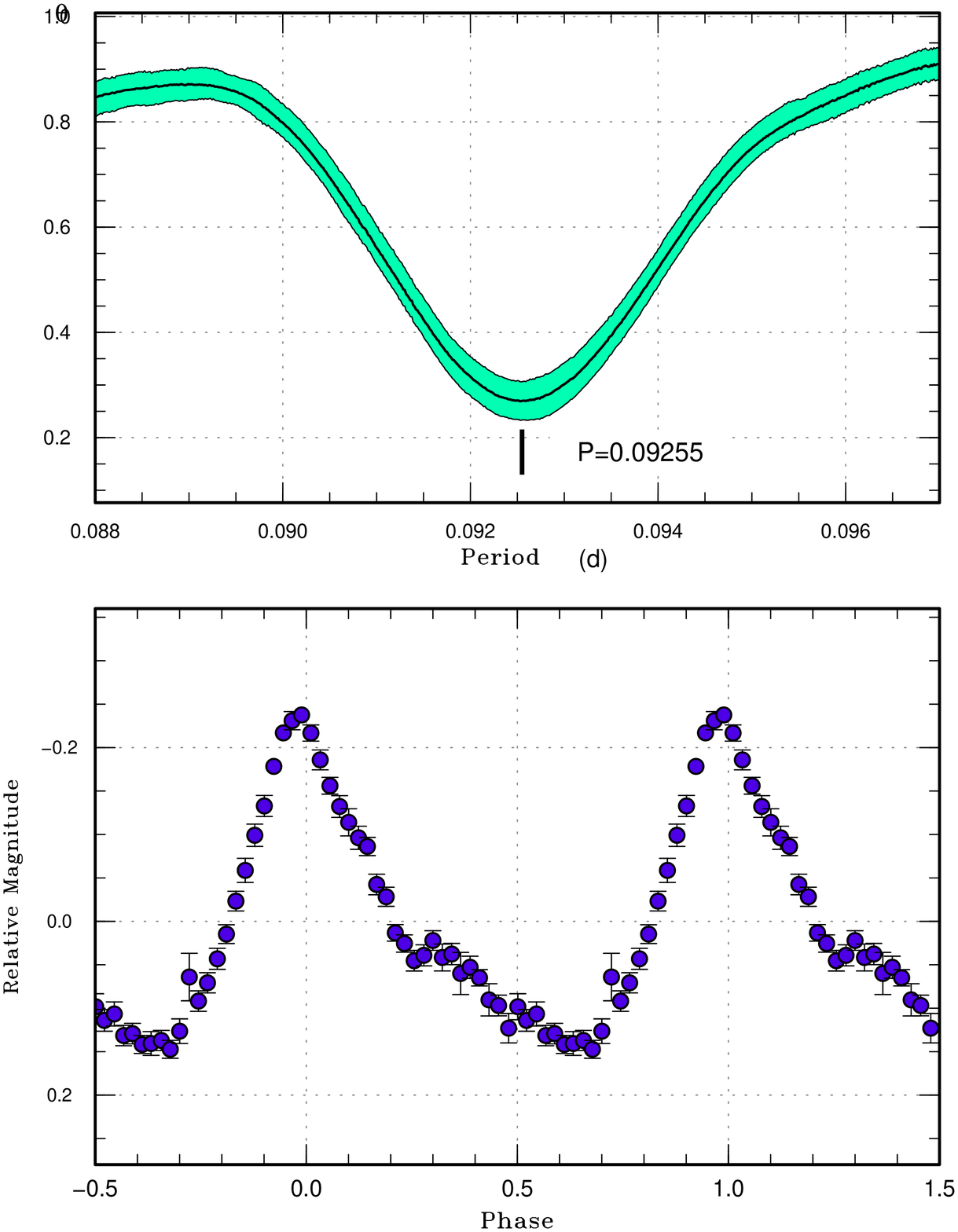}
  \end{center}
  \caption{Superhumps in ASASSN-15rj (2015).
     (Upper): PDM analysis.
     (Lower): Phase-averaged profile.}
  \label{fig:asassn15rjshpdm}
\end{figure}

   This object was detected as a transient at $V$=16.0
on 2015 October 14 by the ASAS-SN team.
Subsequent observations detected superhumps
(vsnet-alert 19168, 19174; figure \ref{fig:asassn15rjshpdm}).
The times of superhump maxima are listed in
table \ref{tab:asassn15rjoc2015}.
The mean superhump period with the PDM method is
0.09255(4)~d.

% SI

% SI

\begin{table}
\caption{Superhump maxima of ASASSN-15rj (2015)}\label{tab:asassn15rjoc2015}
\begin{center}
\begin{tabular}{rp{55pt}p{40pt}r@{.}lr}
\hline
\multicolumn{1}{c}{$E$} & \multicolumn{1}{c}{max\commenta} & \multicolumn{1}{c}{error} & \multicolumn{2}{c}{$O-C$\commentb} & \multicolumn{1}{c}{$N$\commentc} \\
\hline
0 & 57311.1576 & 0.0003 & $-$0&0001 & 186 \\
1 & 57311.2493 & 0.0003 & $-$0&0009 & 179 \\
11 & 57312.1768 & 0.0007 & 0&0019 & 255 \\
12 & 57312.2679 & 0.0005 & 0&0006 & 305 \\
13 & 57312.3609 & 0.0009 & 0&0012 & 86 \\
14 & 57312.4541 & 0.0011 & 0&0019 & 34 \\
15 & 57312.5400 & 0.0017 & $-$0&0046 & 34 \\
\hline
  \multicolumn{6}{l}{\commenta BJD$-$2400000.} \\
  \multicolumn{6}{l}{\commentb Against max $= 2457311.1577 + 0.092463 E$.} \\
  \multicolumn{6}{l}{\commentc Number of points used to determine the maximum.} \\
\end{tabular}
\end{center}
\end{table}

\subsection{ASASSN-15ro}\label{obj:asassn15ro}

   This object was detected as a transient at $V$=15.5
on 2015 October 18 by the ASAS-SN team.
Subsequent observations detected superhumps
(vsnet-alert 19181, 19185).
The times of superhump maxima are listed in
table \ref{tab:asassn15rooc2015}.
The large superhump amplitudes suggest that
we observed stage B superhumps.

% SI

\begin{figure}
  \begin{center}
%    \FigureFile(85mm,110mm){asassn15roshpdm.eps}
    \FigureFile(85mm,110mm){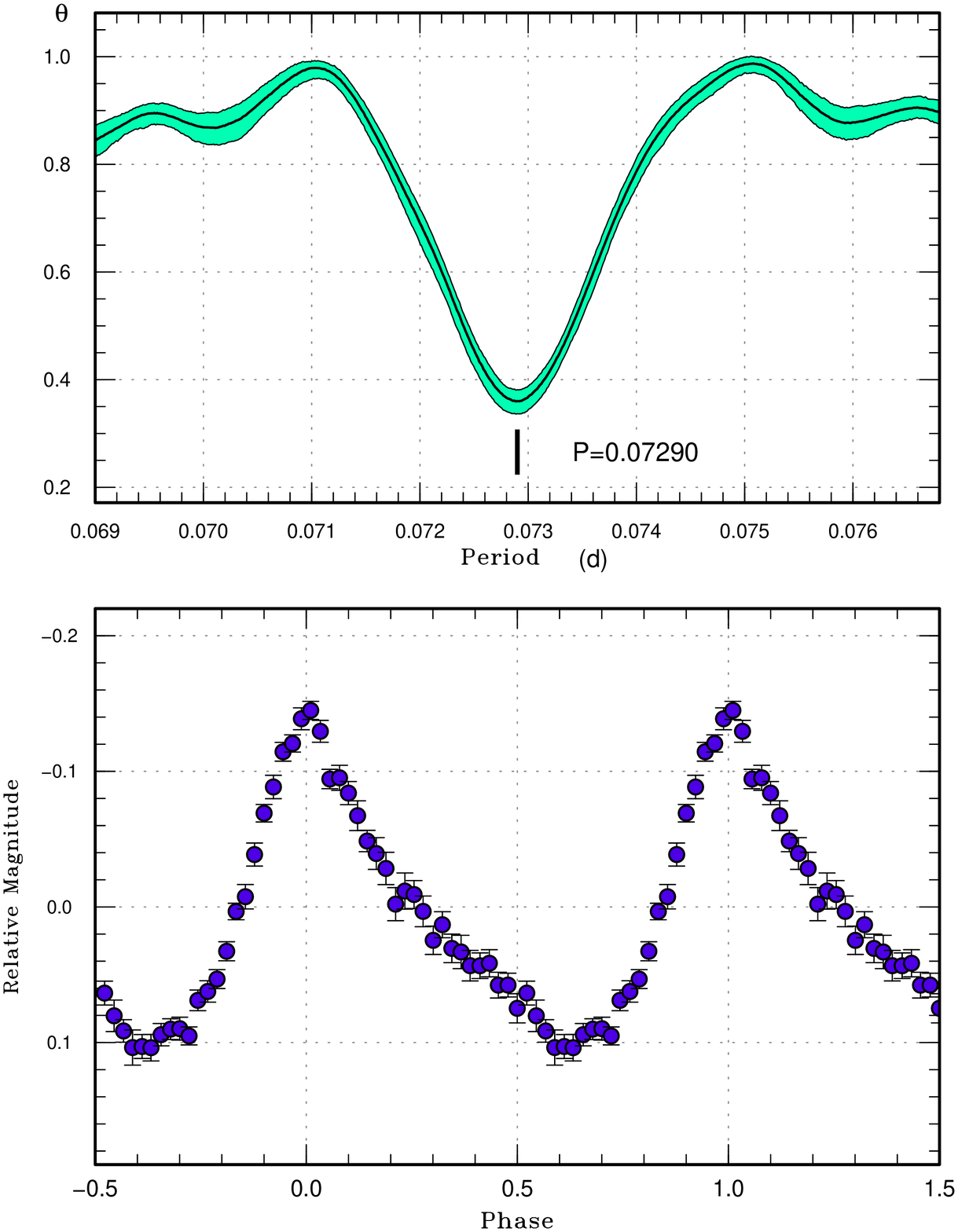}
  \end{center}
  \caption{Superhumps in ASASSN-15ro (2015).
     (Upper): PDM analysis.
     (Lower): Phase-averaged profile.}
  \label{fig:asassn15roshpdm}
\end{figure}

% SI

\begin{table}
\caption{Superhump maxima of ASASSN-15ro (2015)}\label{tab:asassn15rooc2015}
\begin{center}
\begin{tabular}{rp{55pt}p{40pt}r@{.}lr}
\hline
\multicolumn{1}{c}{$E$} & \multicolumn{1}{c}{max\commenta} & \multicolumn{1}{c}{error} & \multicolumn{2}{c}{$O-C$\commentb} & \multicolumn{1}{c}{$N$\commentc} \\
\hline
0 & 57316.0134 & 0.0033 & $-$0&0005 & 37 \\
1 & 57316.0885 & 0.0002 & 0&0016 & 148 \\
2 & 57316.1596 & 0.0004 & $-$0&0002 & 141 \\
3 & 57316.2336 & 0.0005 & 0&0009 & 130 \\
4 & 57316.3065 & 0.0006 & 0&0009 & 124 \\
15 & 57317.1006 & 0.0031 & $-$0&0070 & 72 \\
16 & 57317.1802 & 0.0005 & $-$0&0003 & 127 \\
17 & 57317.2508 & 0.0005 & $-$0&0026 & 140 \\
18 & 57317.3322 & 0.0008 & 0&0058 & 105 \\
27 & 57317.9833 & 0.0006 & 0&0008 & 127 \\
28 & 57318.0539 & 0.0007 & $-$0&0015 & 137 \\
29 & 57318.1306 & 0.0017 & 0&0022 & 41 \\
\hline
  \multicolumn{6}{l}{\commenta BJD$-$2400000.} \\
  \multicolumn{6}{l}{\commentb Against max $= 2457316.0140 + 0.072909 E$.} \\
  \multicolumn{6}{l}{\commentc Number of points used to determine the maximum.} \\
\end{tabular}
\end{center}
\end{table}

\subsection{ASASSN-15rr}\label{obj:asassn15rr}

   This object was detected as a transient at $V$=14.6
on 2015 October 17 by the ASAS-SN team (the detection
announcement was after the observation on October 21
at $V$=15.3).
The quiescent counterpart is very faint ($B_j$=22.0)
and the large outburst amplitude received attention.
Subsequent observations detected superhumps
(vsnet-alert 19196).
Due to the shortness of the observing runs and
the gap in the observation, there remained
possibilities of aliases.
The times of superhump maxima are listed in
table \ref{tab:asassn15rroc2015} with an assumption
of one of the alias which gives the smallest
absolute $O-C$ values.

% SI

\begin{figure}
  \begin{center}
%    \FigureFile(85mm,110mm){asassn15rrshpdm.eps}
    \FigureFile(85mm,110mm){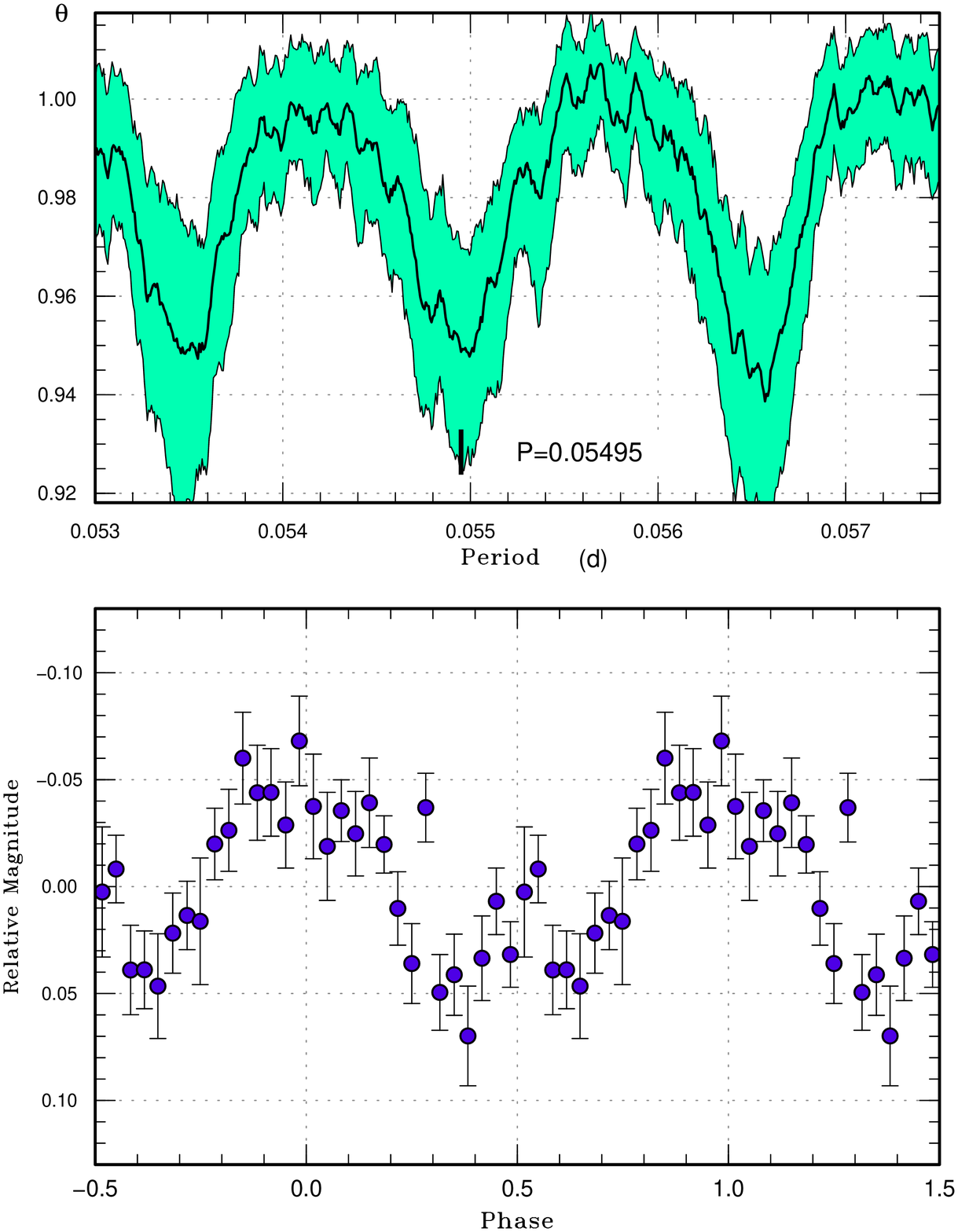}
  \end{center}
  \caption{Superhumps in ASASSN-15rr (2015).
     (Upper): PDM analysis.  The alias selection was based
     on $O-C$ analysis.  The other aliases, particularly
     0.05653(4)~d, are also viable.
     (Lower): Phase-averaged profile.}
  \label{fig:asassn15rrshpdm}
\end{figure}

% SI

\begin{table}
\caption{Superhump maxima of ASASSN-15rr (2015)}\label{tab:asassn15rroc2015}
\begin{center}
\begin{tabular}{rp{55pt}p{40pt}r@{.}lr}
\hline
\multicolumn{1}{c}{$E$} & \multicolumn{1}{c}{max\commenta} & \multicolumn{1}{c}{error} & \multicolumn{2}{c}{$O-C$\commentb} & \multicolumn{1}{c}{$N$\commentc} \\
\hline
0 & 57318.2426 & 0.0031 & 0&0043 & 69 \\
1 & 57318.2911 & 0.0010 & $-$0&0022 & 66 \\
37 & 57320.2687 & 0.0010 & $-$0&0024 & 115 \\
38 & 57320.3242 & 0.0020 & $-$0&0018 & 126 \\
73 & 57322.2527 & 0.0022 & 0&0038 & 73 \\
74 & 57322.3025 & 0.0023 & $-$0&0013 & 127 \\
75 & 57322.3584 & 0.0027 & $-$0&0003 & 106 \\
\hline
  \multicolumn{6}{l}{\commenta BJD$-$2400000.} \\
  \multicolumn{6}{l}{\commentb Against max $= 2457318.2384 + 0.054938 E$.} \\
  \multicolumn{6}{l}{\commentc Number of points used to determine the maximum.} \\
\end{tabular}
\end{center}
\end{table}

\subsection{ASASSN-15rs}\label{obj:asassn15rs}

\begin{figure}
  \begin{center}
%    \FigureFile(85mm,110mm){asassn15rsshpdm.eps}
    \FigureFile(85mm,110mm){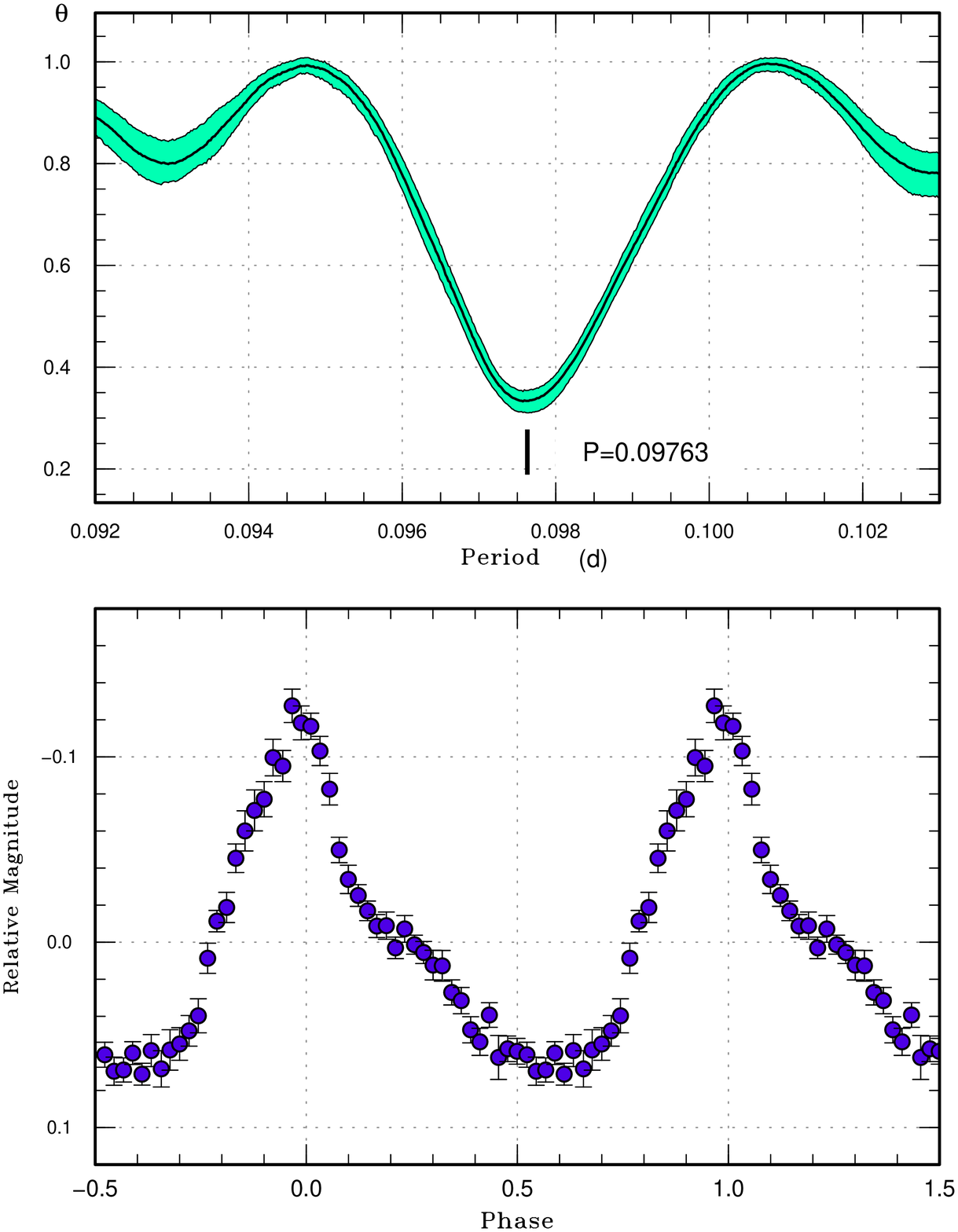}
  \end{center}
  \caption{Superhumps in ASASSN-15rs (2015).
     (Upper): PDM analysis before BJD 2457321.
     (Lower): Phase-averaged profile.}
  \label{fig:asassn15rsshpdm}
\end{figure}

   This object was detected as a transient at $V$=14.5
on 2015 October 21 by the ASAS-SN team.
The object was also cataloged as an H$\alpha$
emission-line object IPHAS J044633.68$+$485755.6 \citep{IPHAS}.
Subsequent observations detected long-period superhumps
(vsnet-alert 19195; figure \ref{fig:asassn15rsshpdm}).
The times of superhump maxima are listed in
table \ref{tab:asassn15rsoc2015}.  Due to the gap
in the observation, the cycle count between $E$=22
and $E$=92 is somewhat uncertain.  Since the superhump
period of this system is long, the superhump period
may have strongly decreased during the observation
(\cite{Pdot}; \cite{kat16v1006cyg}; see also
subsection \ref{sec:longstagea}).
We restricted the period analysis for $E \le$22
in table \ref{tab:perlist} and figure \ref{fig:asassn15rsshpdm}).

% SI

% SI

\begin{table}
\caption{Superhump maxima of ASASSN-15rs (2015)}\label{tab:asassn15rsoc2015}
\begin{center}
\begin{tabular}{rp{55pt}p{40pt}r@{.}lr}
\hline
\multicolumn{1}{c}{$E$} & \multicolumn{1}{c}{max\commenta} & \multicolumn{1}{c}{error} & \multicolumn{2}{c}{$O-C$\commentb} & \multicolumn{1}{c}{$N$\commentc} \\
\hline
0 & 57318.2008 & 0.0003 & $-$0&0018 & 197 \\
1 & 57318.2951 & 0.0004 & $-$0&0049 & 195 \\
12 & 57319.3681 & 0.0009 & $-$0&0035 & 104 \\
13 & 57319.4651 & 0.0008 & $-$0&0040 & 107 \\
14 & 57319.5678 & 0.0013 & 0&0012 & 69 \\
20 & 57320.1560 & 0.0010 & 0&0049 & 182 \\
21 & 57320.2473 & 0.0007 & $-$0&0013 & 109 \\
22 & 57320.3579 & 0.0036 & 0&0119 & 36 \\
92 & 57327.1743 & 0.0017 & 0&0084 & 80 \\
93 & 57327.2523 & 0.0014 & $-$0&0110 & 93 \\
\hline
  \multicolumn{6}{l}{\commenta BJD$-$2400000.} \\
  \multicolumn{6}{l}{\commentb Against max $= 2457318.2026 + 0.097427 E$.} \\
  \multicolumn{6}{l}{\commentc Number of points used to determine the maximum.} \\
\end{tabular}
\end{center}
\end{table}

\subsection{ASASSN-15ry}\label{obj:asassn15ry}

\begin{figure}
  \begin{center}
%    \FigureFile(85mm,110mm){asassn15ryshpdm.eps}
    \FigureFile(85mm,110mm){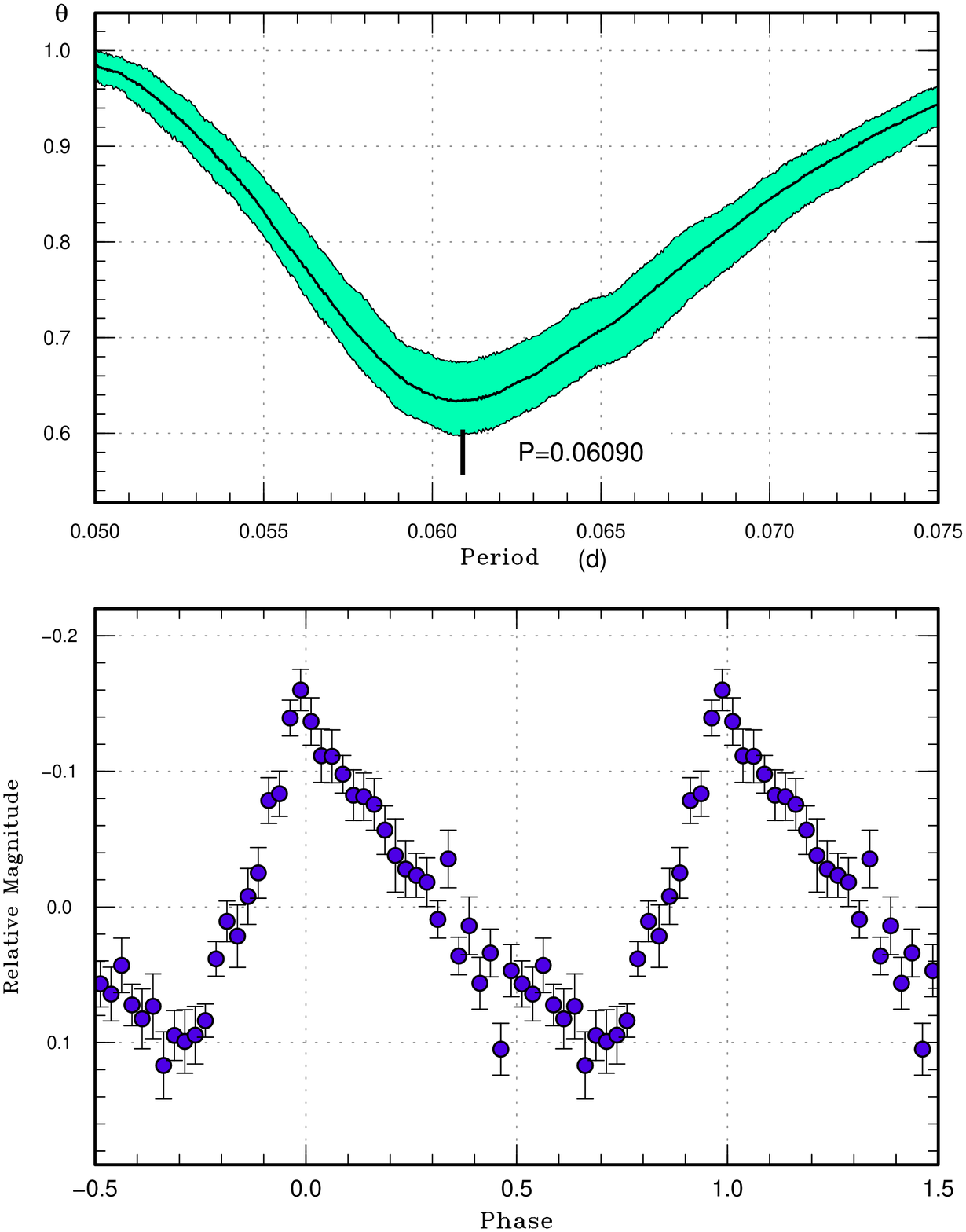}
  \end{center}
  \caption{Superhumps in ASASSN-15ry (2015).
     (Upper): PDM analysis.
     (Lower): Phase-averaged profile.}
  \label{fig:asassn15ryshpdm}
\end{figure}

   This object was detected as a transient at $V$=15.9
on 2015 October 22 by the ASAS-SN team (the detection
announcement was after the observation on October 24
at $V$=15.3).
The object was also independently detected by
S. Ueda (TCP J05285567$+$3618388) on October 24
at an unfiltered CCD magnitude of 14.2.\footnote{
  $<$http://www.cbat.eps.harvard.edu/unconf/\\
followups/J05285567$+$3618388.html$>$.
} (Our observations suggest that this estimate was
too bright by $\sim$1 mag).
The object was also cataloged as an H$\alpha$
emission-line object
IPHAS2 J052855.67$+$361839.0 \citep{IPHAS2}.
Subsequent observations detected superhumps
(vsnet-alert 19193; figure \ref{fig:asassn15ryshpdm}).
Only single-night observations are available and
the times of superhump maxima are listed in
table \ref{tab:asassn15ryoc2015}.
The superhump period given in table \ref{tab:perlist}
is based on the PDM analysis.

% SI

% SI

\begin{table}
\caption{Superhump maxima of ASASSN-15ry (2015)}\label{tab:asassn15ryoc2015}
\begin{center}
\begin{tabular}{rp{55pt}p{40pt}r@{.}lr}
\hline
\multicolumn{1}{c}{$E$} & \multicolumn{1}{c}{max\commenta} & \multicolumn{1}{c}{error} & \multicolumn{2}{c}{$O-C$\commentb} & \multicolumn{1}{c}{$N$\commentc} \\
\hline
0 & 57321.1181 & 0.0008 & $-$0&0004 & 170 \\
1 & 57321.1811 & 0.0005 & 0&0011 & 282 \\
2 & 57321.2405 & 0.0006 & $-$0&0009 & 228 \\
3 & 57321.3032 & 0.0007 & 0&0003 & 198 \\
\hline
  \multicolumn{6}{l}{\commenta BJD$-$2400000.} \\
  \multicolumn{6}{l}{\commentb Against max $= 2457321.1185 + 0.061469 E$.} \\
  \multicolumn{6}{l}{\commentc Number of points used to determine the maximum.} \\
\end{tabular}
\end{center}
\end{table}

\subsection{ASASSN-15sc}\label{obj:asassn15sc}

\begin{figure}
  \begin{center}
%    \FigureFile(85mm,110mm){asassn15scshpdm.eps}
    \FigureFile(85mm,110mm){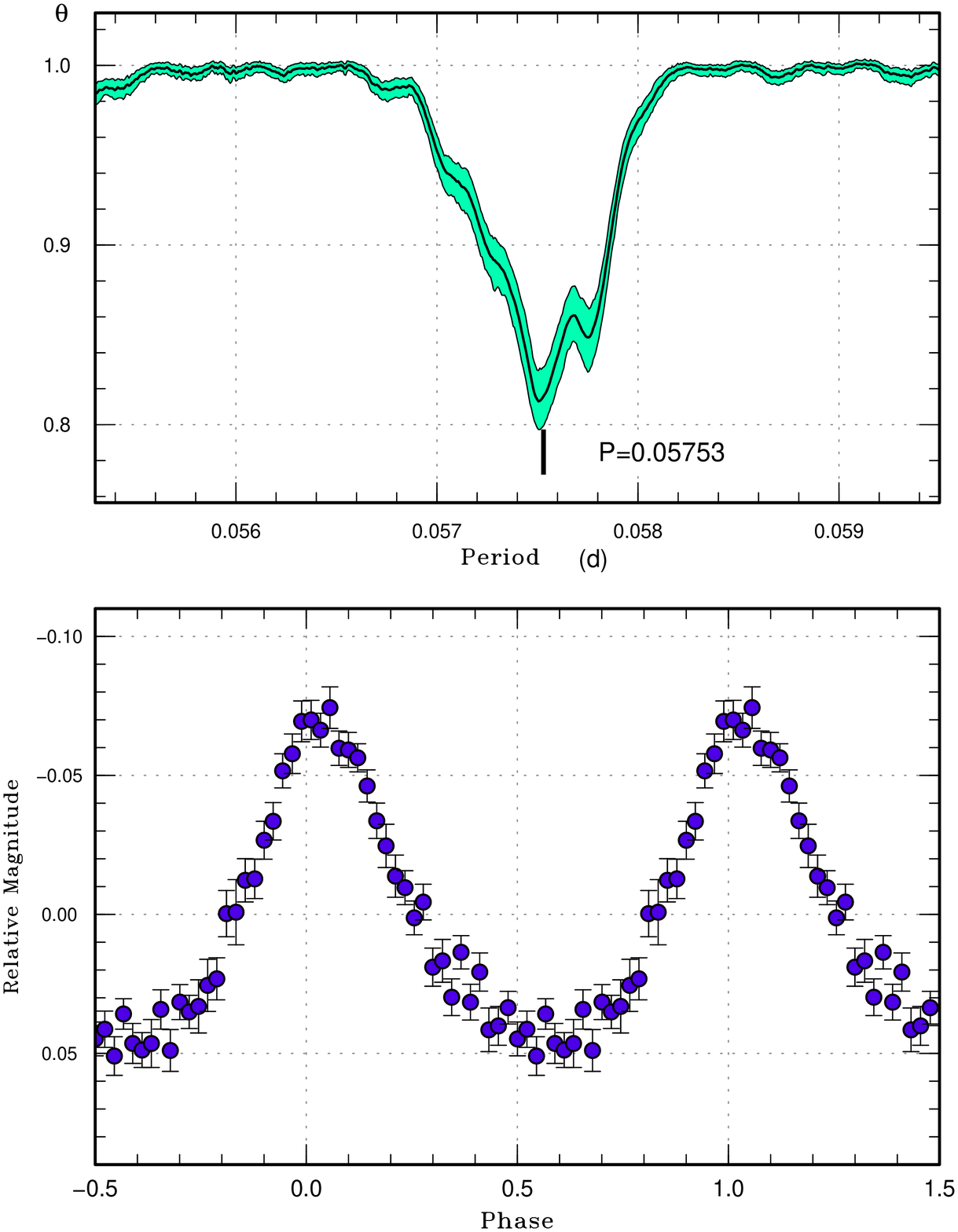}
  \end{center}
  \caption{Superhumps in ASASSN-15sc (2015).
     (Upper): PDM analysis.
     (Lower): Phase-averaged profile.}
  \label{fig:asassn15scshpdm}
\end{figure}

   This object was detected as a transient at $V$=14.6
on 2015 October 24 by the ASAS-SN team.  The object was
still rising at $V$=16.2 on October 22.
Although there is a 19.9-mag star \timeform{5''}
distant in the Initial Gaia Source List,
this object appears too bright and too red for
the quiescent counterpart.
Time-resolved photometry started on October 28.
There were no periodic modulations on the first
two night.  Growing likely superhumps were detected
on October 30 and further observations confirmed
superhumps (vsnet-alert 19215, 19217, 19219,
19226, 19236, 19238, 19249, 19277;
figure \ref{fig:asassn15scshpdm}).
The times of superhump maxima are listed in
table \ref{tab:asassn15scoc2015}.
The $O-C$ data clearly indicate the stages A-B-C
with a strongly positive $P_{\rm dot}$ typical
for this short superhump period.
Although stage C was present, the short duration
of stage C (typical for a short-period system)
made it difficult to determine the period
of stage C superhumps.

% SI

% SI

\begin{table*}
\caption{Superhump maxima of ASASSN-15sc (2015)}\label{tab:asassn15scoc2015}
\begin{center}
\begin{tabular}{rp{55pt}p{40pt}r@{.}lrrp{55pt}p{40pt}r@{.}lr}
\hline
\multicolumn{1}{c}{$E$} & \multicolumn{1}{c}{max\commenta} & \multicolumn{1}{c}{error} & \multicolumn{2}{c}{$O-C$\commentb} & \multicolumn{1}{c}{$N$\commentc} & \multicolumn{1}{c}{$E$} & \multicolumn{1}{c}{max\commenta} & \multicolumn{1}{c}{error} & \multicolumn{2}{c}{$O-C$\commentb} & \multicolumn{1}{c}{$N$\commentc} \\
\hline
0 & 57326.3985 & 0.0014 & $-$0&0058 & 66 & 79 & 57330.9645 & 0.0029 & $-$0&0025 & 50 \\
1 & 57326.4508 & 0.0027 & $-$0&0112 & 74 & 80 & 57331.0211 & 0.0017 & $-$0&0037 & 57 \\
15 & 57327.2772 & 0.0007 & 0&0066 & 45 & 81 & 57331.0776 & 0.0018 & $-$0&0049 & 56 \\
16 & 57327.3350 & 0.0004 & 0&0066 & 56 & 82 & 57331.1348 & 0.0011 & $-$0&0055 & 57 \\
17 & 57327.3937 & 0.0004 & 0&0076 & 56 & 83 & 57331.1934 & 0.0012 & $-$0&0047 & 81 \\
18 & 57327.4508 & 0.0005 & 0&0069 & 74 & 84 & 57331.2499 & 0.0011 & $-$0&0059 & 87 \\
19 & 57327.5100 & 0.0004 & 0&0084 & 115 & 85 & 57331.3072 & 0.0009 & $-$0&0063 & 36 \\
20 & 57327.5681 & 0.0002 & 0&0088 & 106 & 86 & 57331.3687 & 0.0007 & $-$0&0026 & 30 \\
21 & 57327.6274 & 0.0004 & 0&0103 & 41 & 87 & 57331.4236 & 0.0007 & $-$0&0055 & 30 \\
33 & 57328.3181 & 0.0004 & 0&0079 & 86 & 88 & 57331.4825 & 0.0007 & $-$0&0043 & 31 \\
34 & 57328.3737 & 0.0003 & 0&0057 & 86 & 89 & 57331.5392 & 0.0006 & $-$0&0054 & 31 \\
41 & 57328.7773 & 0.0002 & 0&0050 & 99 & 90 & 57331.5954 & 0.0009 & $-$0&0070 & 31 \\
42 & 57328.8341 & 0.0002 & 0&0041 & 116 & 101 & 57332.2317 & 0.0013 & $-$0&0060 & 30 \\
43 & 57328.8916 & 0.0003 & 0&0038 & 115 & 102 & 57332.2918 & 0.0008 & $-$0&0036 & 31 \\
44 & 57328.9488 & 0.0003 & 0&0032 & 110 & 103 & 57332.3485 & 0.0009 & $-$0&0046 & 19 \\
47 & 57329.1212 & 0.0011 & 0&0024 & 53 & 104 & 57332.4047 & 0.0011 & $-$0&0062 & 30 \\
48 & 57329.1774 & 0.0011 & 0&0008 & 114 & 105 & 57332.4633 & 0.0007 & $-$0&0054 & 31 \\
49 & 57329.2375 & 0.0010 & 0&0031 & 126 & 124 & 57333.5615 & 0.0009 & $-$0&0046 & 54 \\
50 & 57329.2929 & 0.0010 & 0&0008 & 90 & 125 & 57333.6189 & 0.0011 & $-$0&0049 & 54 \\
51 & 57329.3502 & 0.0008 & 0&0004 & 88 & 126 & 57333.6779 & 0.0010 & $-$0&0037 & 54 \\
52 & 57329.4094 & 0.0007 & 0&0018 & 31 & 139 & 57334.4304 & 0.0010 & $-$0&0020 & 54 \\
56 & 57329.6383 & 0.0004 & $-$0&0003 & 60 & 140 & 57334.4860 & 0.0013 & $-$0&0042 & 31 \\
57 & 57329.6966 & 0.0004 & 0&0002 & 61 & 158 & 57335.5288 & 0.0015 & $-$0&0010 & 54 \\
58 & 57329.7536 & 0.0003 & $-$0&0005 & 61 & 159 & 57335.5895 & 0.0021 & 0&0020 & 31 \\
59 & 57329.8123 & 0.0004 & 0&0004 & 41 & 207 & 57338.3652 & 0.0016 & 0&0053 & 189 \\
62 & 57329.9841 & 0.0016 & $-$0&0011 & 65 & 208 & 57338.4269 & 0.0013 & 0&0092 & 106 \\
63 & 57330.0436 & 0.0010 & 0&0007 & 66 & 262 & 57341.5460 & 0.0015 & 0&0095 & 54 \\
64 & 57330.1011 & 0.0013 & 0&0004 & 66 & 263 & 57341.6033 & 0.0027 & 0&0090 & 44 \\
65 & 57330.1548 & 0.0015 & $-$0&0037 & 91 & 269 & 57341.9350 & 0.0041 & $-$0&0058 & 54 \\
66 & 57330.2129 & 0.0006 & $-$0&0032 & 261 & 270 & 57341.9970 & 0.0059 & $-$0&0015 & 57 \\
67 & 57330.2732 & 0.0011 & $-$0&0007 & 261 & 278 & 57342.4586 & 0.0108 & $-$0&0020 & 19 \\
69 & 57330.3890 & 0.0013 & $-$0&0004 & 26 & 279 & 57342.5236 & 0.0016 & 0&0052 & 53 \\
70 & 57330.4480 & 0.0009 & 0&0008 & 31 & 280 & 57342.5801 & 0.0014 & 0&0039 & 54 \\
\hline
  \multicolumn{12}{l}{\commenta BJD$-$2400000.} \\
  \multicolumn{12}{l}{\commentb Against max $= 2457326.4042 + 0.057757 E$.} \\
  \multicolumn{12}{l}{\commentc Number of points used to determine the maximum.} \\
\end{tabular}
\end{center}
\end{table*}

\subsection{ASASSN-15sd}\label{obj:asassn15sd}

\begin{figure}
  \begin{center}
%    \FigureFile(85mm,110mm){asassn15sdshpdm.eps}
    \FigureFile(85mm,110mm){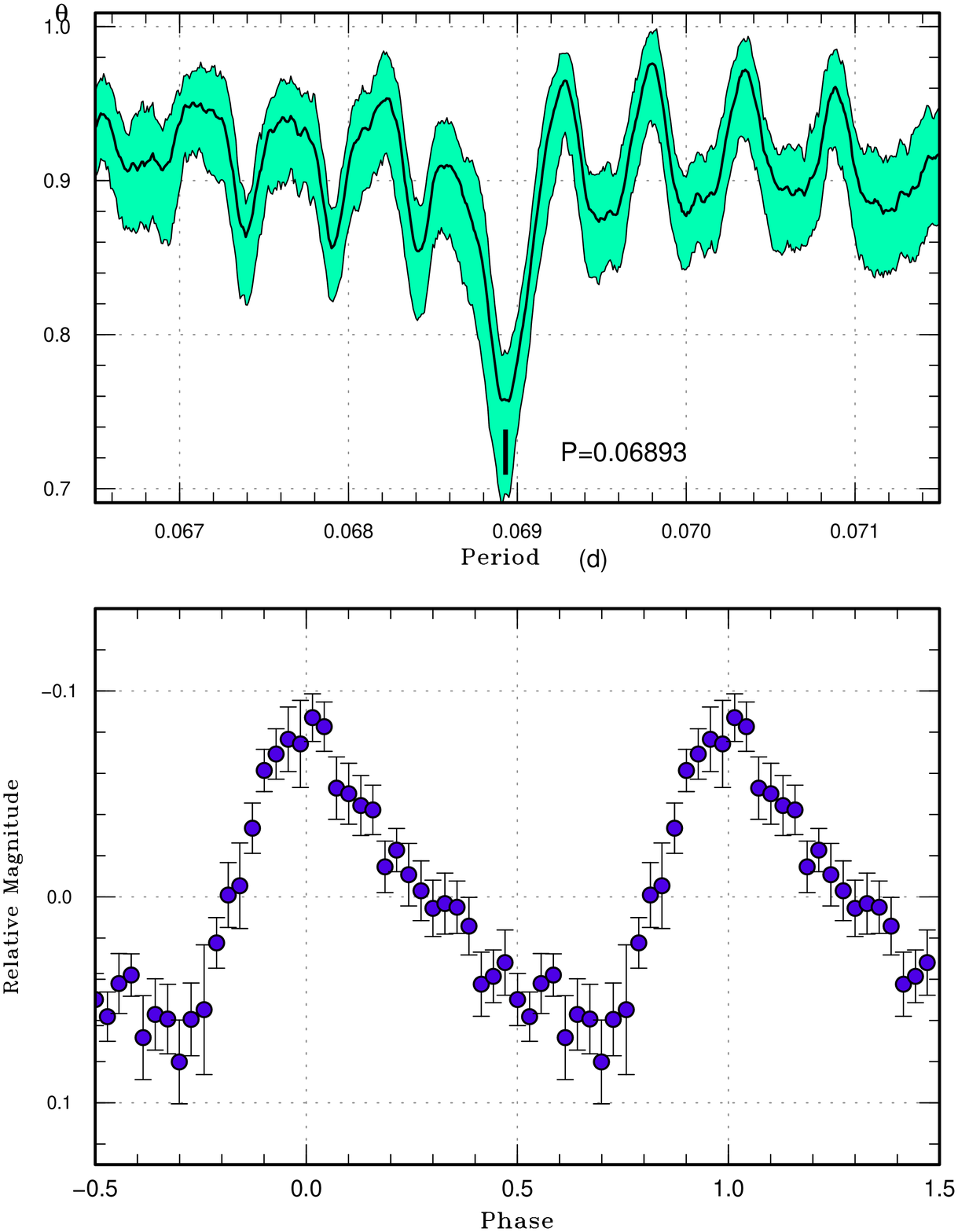}
  \end{center}
  \caption{Superhumps in ASASSN-15sd (2015).
     (Upper): PDM analysis.
     (Lower): Phase-averaged profile.}
  \label{fig:asassn15sdshpdm}
\end{figure}

   This object was detected as a transient at $V$=14.8
on 2015 October 27 by the ASAS-SN team.
Subsequent observations detected superhumps
(vsnet-alert 19210; figure \ref{fig:asassn15sdshpdm}).
The object rapidly faded on November 5.
The times of superhump maxima are listed in
table \ref{tab:asassn15sdoc2015}.
The lack of a significantly positive $P_{\rm dot}$
expected for this superhump period suggests that
we only observed stage C superhumps.
The duration of the superoutburst plateau was shorter
than 11~d, considering the negative observation in
the ASAS-SN data on October 24.  The development of superhumps
and the duration of the superoutburst may be somewhat
atypical for this superhump period and further observations
are needed to understand the object better.

% SI

% SI

\begin{table}
\caption{Superhump maxima of ASASSN-15sd (2015)}\label{tab:asassn15sdoc2015}
\begin{center}
\begin{tabular}{rp{55pt}p{40pt}r@{.}lr}
\hline
\multicolumn{1}{c}{$E$} & \multicolumn{1}{c}{max\commenta} & \multicolumn{1}{c}{error} & \multicolumn{2}{c}{$O-C$\commentb} & \multicolumn{1}{c}{$N$\commentc} \\
\hline
0 & 57324.2650 & 0.0003 & $-$0&0000 & 129 \\
1 & 57324.3366 & 0.0010 & 0&0027 & 94 \\
2 & 57324.4052 & 0.0004 & 0&0024 & 154 \\
19 & 57325.5669 & 0.0059 & $-$0&0071 & 18 \\
20 & 57325.6434 & 0.0020 & 0&0005 & 19 \\
34 & 57326.6089 & 0.0008 & 0&0015 & 40 \\
48 & 57327.5692 & 0.0017 & $-$0&0027 & 29 \\
49 & 57327.6426 & 0.0025 & 0&0018 & 14 \\
63 & 57328.6043 & 0.0011 & $-$0&0010 & 41 \\
92 & 57330.6048 & 0.0038 & 0&0016 & 42 \\
93 & 57330.6724 & 0.0163 & 0&0002 & 14 \\
\hline
  \multicolumn{6}{l}{\commenta BJD$-$2400000.} \\
  \multicolumn{6}{l}{\commentb Against max $= 2457324.2650 + 0.068894 E$.} \\
  \multicolumn{6}{l}{\commentc Number of points used to determine the maximum.} \\
\end{tabular}
\end{center}
\end{table}

\subsection{ASASSN-15se}\label{obj:asassn15se}

\begin{figure}
  \begin{center}
%    \FigureFile(85mm,110mm){asassn15seshpdm.eps}
    \FigureFile(85mm,110mm){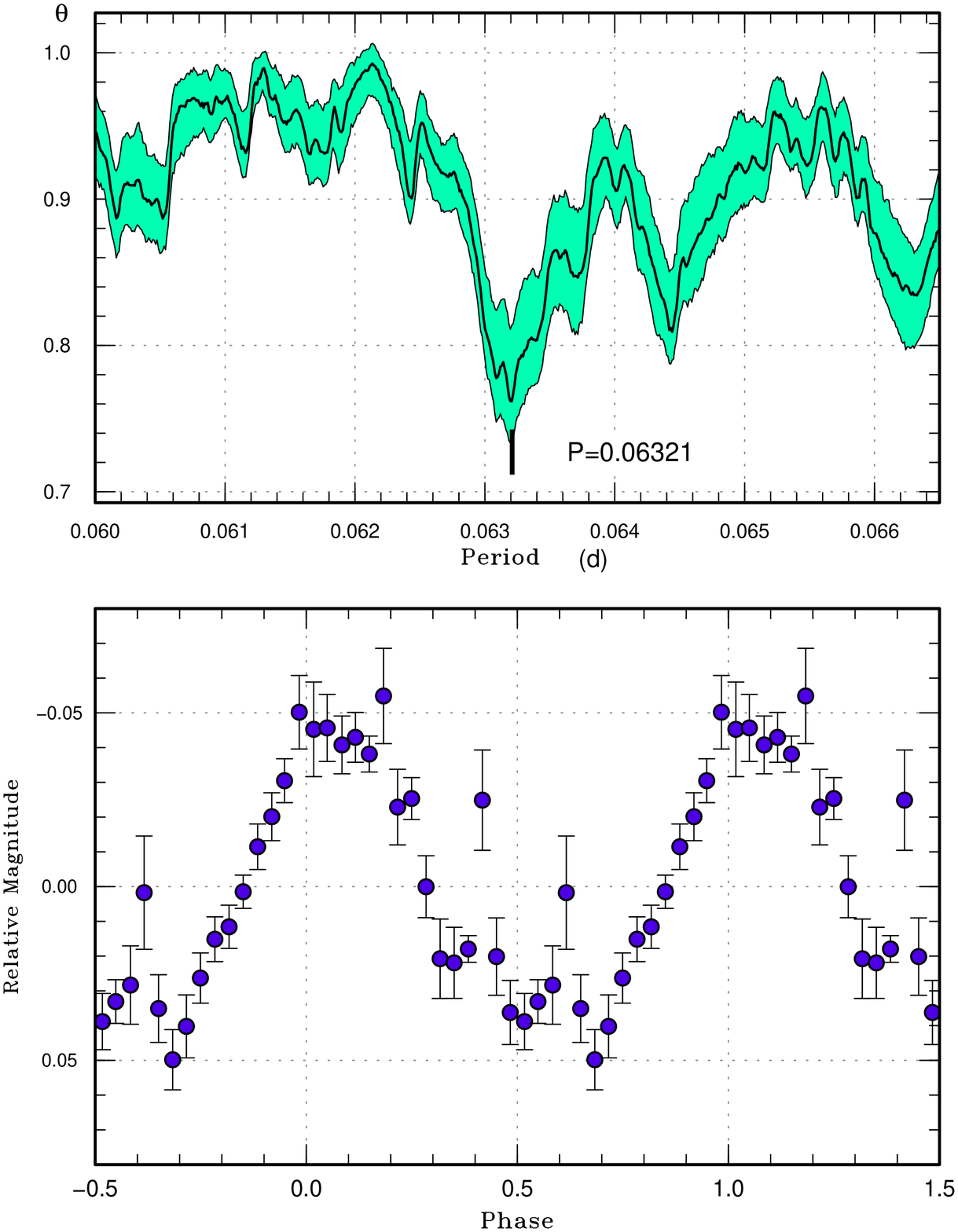}
  \end{center}
  \caption{Superhumps in ASASSN-15se (2015).
     (Upper): PDM analysis.
     (Lower): Phase-averaged profile.}
  \label{fig:asassn15seshpdm}
\end{figure}

\begin{figure}
  \begin{center}
%    \FigureFile(85mm,100mm){asassn15sehumpall.eps}
    \FigureFile(85mm,100mm){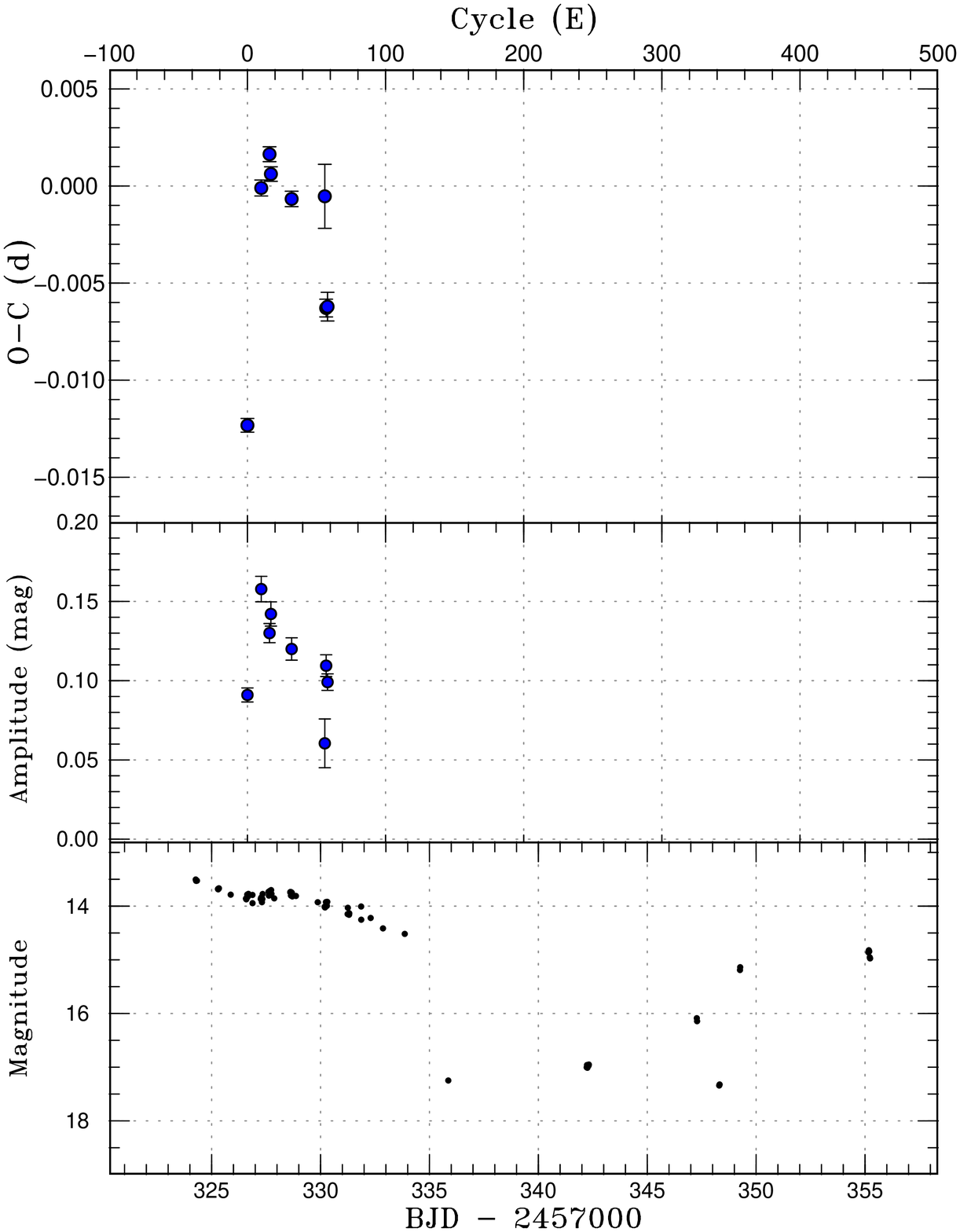}
  \end{center}
  \caption{$O-C$ diagram of superhumps in ASASSN-15se (2015).
     (Upper:) $O-C$ diagram.
     We used a period of 0.06343~d for calculating the $O-C$ residuals.
     (Middle:) Amplitudes of superhumps.
     (Lower:) Light curve.  The data were binned to 0.021~d.
  }
  \label{fig:asassn15sehumpall}
\end{figure}

   This object was detected as a transient at $V$=13.04
on 2015 October 25 by the ASAS-SN team.
The outburst announcement was after $V$=13.46 observation 
on October 28 (vsnet-alert 19208).
No superhumps were detected on October 28 and 29.
Growing superhumps were detected on October 31
(vsnet-alert 19216).  Further data confirmed
the superhumps (vsnet-alert 19220, 19227; the period
given in vsnet-alert 19227 was in error due to
the confusion with ASASSN-15sc;
figure \ref{fig:asassn15seshpdm}).
The times of superhump maxima are listed in
table \ref{tab:asassn15seoc2015}.
Although $E$=0 corresponded a stage A superhump,
we could not determine the period of stage A
superhumps.
The object showed at least two rebrightenings
on November 15 (vsnet-alert 19286) and
November 20--22 (vsnet-alert 19294)
(figure \ref{fig:asassn15sehumpall}).
It is not clear whether the type of
the rebrightenings is type A/B (multiple
rebrightenings with small amplitudes) or
type B (discrete multiple rebrightenings)
in \citet{kat15wzsge}.  Since the superhump period
is relatively long, this object may belong to
WZ Sge-type dwarf novae with multiple rebrightenings
as described by \citet{nak14j0754j2304}.

% SI

% SI

\begin{table}
\caption{Superhump maxima of ASASSN-15se (2015)}\label{tab:asassn15seoc2015}
\begin{center}
\begin{tabular}{rp{55pt}p{40pt}r@{.}lr}
\hline
\multicolumn{1}{c}{$E$} & \multicolumn{1}{c}{max\commenta} & \multicolumn{1}{c}{error} & \multicolumn{2}{c}{$O-C$\commentb} & \multicolumn{1}{c}{$N$\commentc} \\
\hline
0 & 57326.6294 & 0.0004 & $-$0&0093 & 59 \\
10 & 57327.2759 & 0.0004 & 0&0029 & 114 \\
16 & 57327.6582 & 0.0004 & 0&0046 & 97 \\
17 & 57327.7206 & 0.0004 & 0&0036 & 84 \\
32 & 57328.6708 & 0.0004 & 0&0023 & 57 \\
56 & 57330.1932 & 0.0016 & 0&0025 & 35 \\
57 & 57330.2509 & 0.0005 & $-$0&0033 & 46 \\
58 & 57330.3144 & 0.0007 & $-$0&0032 & 27 \\
\hline
  \multicolumn{6}{l}{\commenta BJD$-$2400000.} \\
  \multicolumn{6}{l}{\commentb Against max $= 2457326.6387 + 0.063430 E$.} \\
  \multicolumn{6}{l}{\commentc Number of points used to determine the maximum.} \\
\end{tabular}
\end{center}
\end{table}

\subsection{ASASSN-15sl}\label{obj:asassn15sl}

\begin{figure}
  \begin{center}
%    \FigureFile(85mm,120mm){asassn15sllc.eps}
    \FigureFile(85mm,120mm){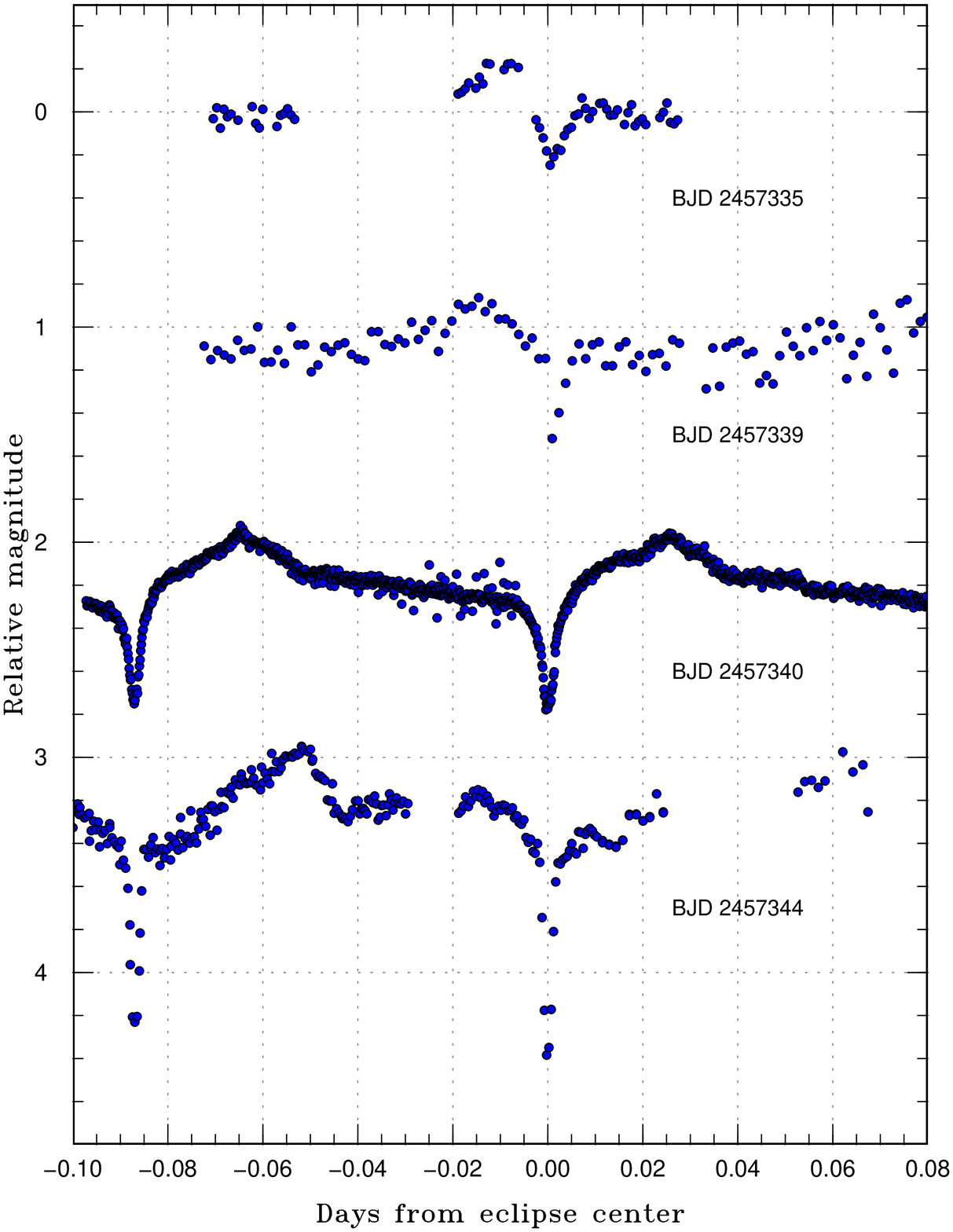}
  \end{center}
  \caption{Light curve of ASASSN-15sl.
  Superposition of superhumps and eclipses is well visible.}
  \label{fig:asassn15sllc}
\end{figure}

   This object was detected as a transient at $V$=15.0
on 2015 November 3 by the ASAS-SN team.
The object soon turned out to be a deeply eclipsing
SU UMa-type dwarf nova (vsnet-alert 19254, 19259,
19264, 19278, 19279; figure \ref{fig:asassn15sllc}).
The eclipse ephemeris was determined
by using MCMC analysis \citep{Pdot4}
of the observations:
\begin{equation}
{\rm Min(BJD)} = 2457341.23671(7) + 0.0870484(7) E .
\label{equ:asassn15slecl}
\end{equation}
This ephemeris is not intended for long-term prediction
of eclipses.  The epoch refers to the center of
the observation.
The times of superhump maxima outside the eclipses
are listed in table \ref{tab:asassn15sloc2015}.
Although the superhump stage is unknown, we should
note that most of our observations were in
the later phase of the superoutburst.
These superhumps may be mostly stage C superhumps.

% SI

\begin{table}
\caption{Superhump maxima of ASASSN-15sl (2015)}\label{tab:asassn15sloc2015}
\begin{center}
\begin{tabular}{rp{50pt}p{30pt}r@{.}lcr}
\hline
$E$ & max\commenta & error & \multicolumn{2}{c}{$O-C$\commentb} & phase\commentc & $N$\commentd \\
\hline
0 & 57335.5707 & 0.0014 & 0&0058 & 0.91 & 33 \\
43 & 57339.4808 & 0.0015 & 0&0002 & 0.83 & 41 \\
44 & 57339.5707 & 0.0073 & $-$0&0010 & 0.86 & 29 \\
54 & 57340.4768 & 0.0002 & $-$0&0055 & 0.27 & 443 \\
55 & 57340.5659 & 0.0002 & $-$0&0076 & 0.29 & 430 \\
72 & 57342.1247 & 0.0041 & 0&0032 & 0.20 & 62 \\
73 & 57342.2111 & 0.0067 & $-$0&0015 & 0.19 & 70 \\
76 & 57342.4854 & 0.0004 & $-$0&0004 & 0.34 & 253 \\
97 & 57344.4050 & 0.0006 & 0&0069 & 0.40 & 131 \\
\hline
  \multicolumn{7}{l}{\commenta BJD$-$2400000.} \\
  \multicolumn{7}{l}{\commentb Against max $= 2457335.5649 + 0.091065 E$.} \\
  \multicolumn{7}{l}{\commentc Orbital phase.} \\
  \multicolumn{7}{l}{\commentd Number of points used to determine the maximum.} \\
\end{tabular}
\end{center}
\end{table}

\subsection{ASASSN-15sn}\label{obj:asassn15sn}

\begin{figure}
  \begin{center}
%    \FigureFile(85mm,110mm){asassn15snshpdm.eps}
    \FigureFile(85mm,110mm){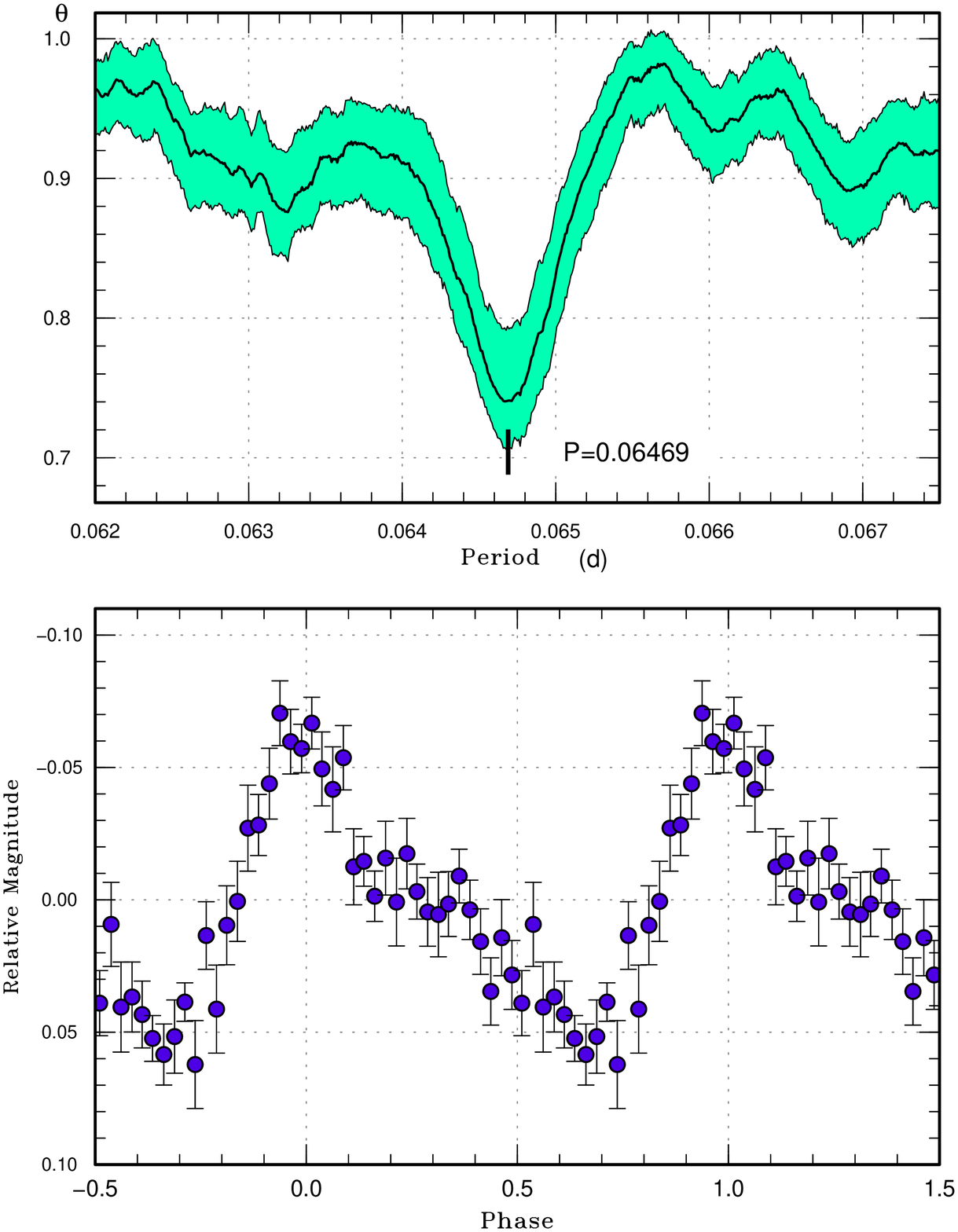}
  \end{center}
  \caption{Superhumps in ASASSN-15sn (2015).
     (Upper): PDM analysis.
     (Lower): Phase-averaged profile.}
  \label{fig:asassn15snshpdm}
\end{figure}

   This object was detected as a transient at $V$=15.4
on 2015 November 4 by the ASAS-SN team.
The object was in outburst at $g$=15.57
in the Kepler Input Catalog.
The object was recorded at $g$=20.85 and $U$=19.94
in quiescence in the Extended Kepler-INT Survey
\citep{gre12KeplerINT}.
Subsequent observations detected superhumps
(vsnet-alert 19247, 19257;
figure \ref{fig:asassn15snshpdm}).
The times of superhump maxima are listed in
table \ref{tab:asassn15snoc2015}.
Since we observed the relatively late stage of
the superoutburst, the break in the $O-C$ diagram
around $E$=27 probably reflects stage B-C transition.
We gave a global value in table \ref{tab:asassn15snoc2015}.

% SI

% SI

\begin{table}
\caption{Superhump maxima of ASASSN-15sn (2015)}\label{tab:asassn15snoc2015}
\begin{center}
\begin{tabular}{rp{55pt}p{40pt}r@{.}lr}
\hline
\multicolumn{1}{c}{$E$} & \multicolumn{1}{c}{max\commenta} & \multicolumn{1}{c}{error} & \multicolumn{2}{c}{$O-C$\commentb} & \multicolumn{1}{c}{$N$\commentc} \\
\hline
0 & 57335.2699 & 0.0018 & $-$0&0030 & 65 \\
1 & 57335.3348 & 0.0015 & $-$0&0028 & 65 \\
2 & 57335.3997 & 0.0019 & $-$0&0026 & 59 \\
27 & 57337.0303 & 0.0012 & 0&0110 & 47 \\
28 & 57337.0891 & 0.0049 & 0&0051 & 25 \\
31 & 57337.2802 & 0.0008 & 0&0022 & 69 \\
32 & 57337.3434 & 0.0014 & 0&0007 & 70 \\
33 & 57337.4089 & 0.0012 & 0&0014 & 66 \\
46 & 57338.2433 & 0.0015 & $-$0&0050 & 70 \\
47 & 57338.3121 & 0.0013 & $-$0&0009 & 68 \\
48 & 57338.3716 & 0.0011 & $-$0&0061 & 68 \\
\hline
  \multicolumn{6}{l}{\commenta BJD$-$2400000.} \\
  \multicolumn{6}{l}{\commentb Against max $= 2457335.2729 + 0.064684 E$.} \\
  \multicolumn{6}{l}{\commentc Number of points used to determine the maximum.} \\
\end{tabular}
\end{center}
\end{table}

\subsection{ASASSN-15sp}\label{obj:asassn15sp}

\begin{figure}
  \begin{center}
%    \FigureFile(85mm,110mm){asassn15spshpdm.eps}
    \FigureFile(85mm,110mm){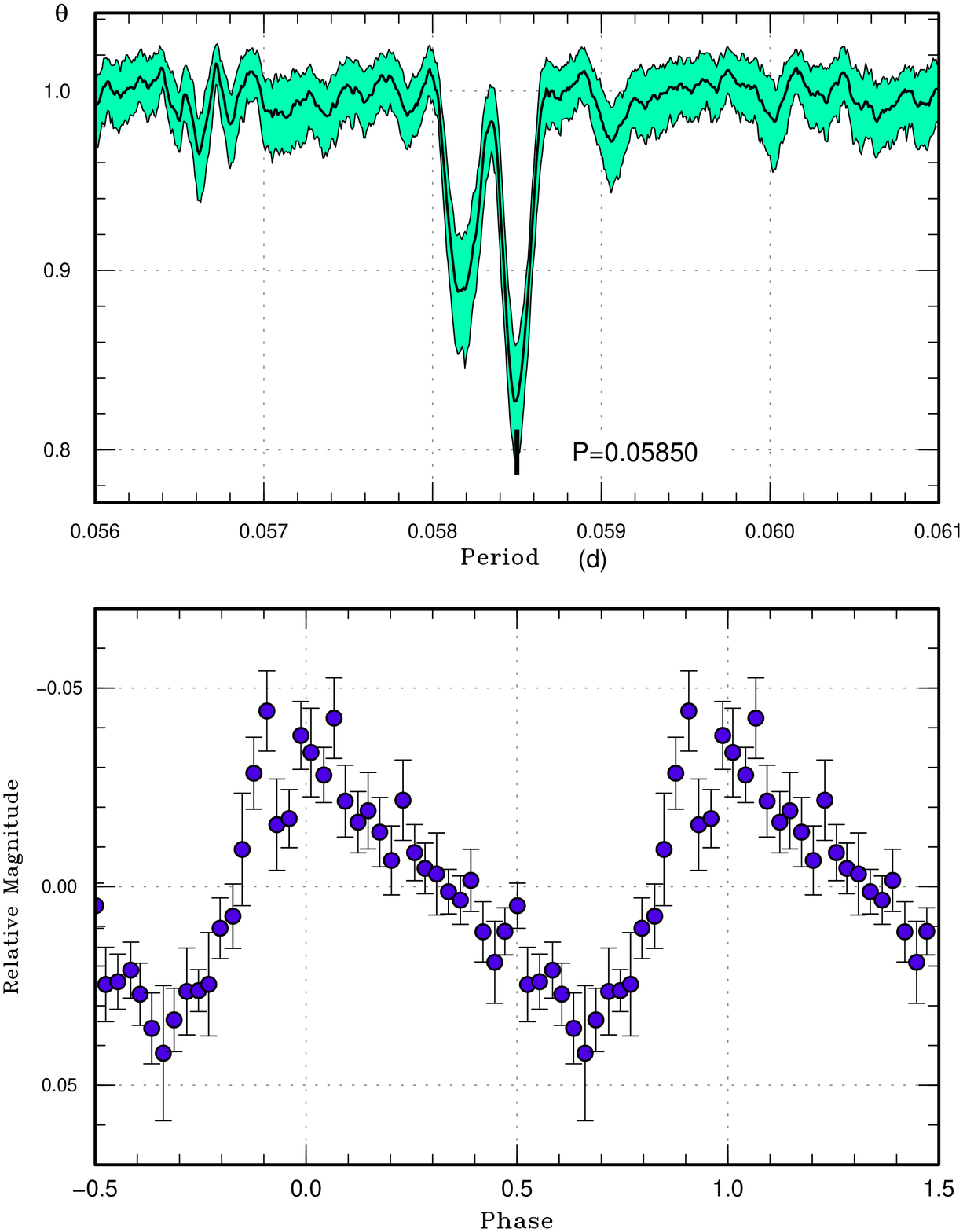}
  \end{center}
  \caption{Superhumps in ASASSN-15sp during
     the plateau phase (2015).
     (Upper): PDM analysis.
     (Lower): Phase-averaged profile.}
  \label{fig:asassn15spshpdm}
\end{figure}

   This object was detected as a transient at $V$=13.9
on 2015 November 8 by the ASAS-SN team.
There is an X-ray counterpart 1RXS J075806.5$-$572239
There was an outburst on 2008 January 26--28 reaching
$V$=14.02 detected by ASAS-3 (vsnet-alert 19246).
There were double-humped variations on November 9
(vsnet-alert 19252), which developed into full
superhumps on November 12 (vsnet-alert 19256,
19260, 19280; figure \ref{fig:asassn15spshpdm}).
The initial variation probably reflected growing
phase superhumps (part of stage A).
The times of superhump maxima are listed in
table \ref{tab:asassn15spoc2015}.
The maxima for $E \le$1 correspond to stage A
superhumps.  Due to the gap in the observation
before $E$=33, we could not determine the period
of stage A superhumps.
The $P_{\rm dot}$ was positive, as is expected
for this superhump period.
Although there were observations after $E$=138,
we could not determine individual times of
maxima.  A PDM analysis of the data between
BJD 2457345 and 2457352 (late part of the plateau
phase with slight brightening) yielded a strong
signal of 0.05829(4)~d.  This period is adopted
as that of stage C superhumps in table \ref{tab:perlist}.

% SI

% SI

\begin{table}
\caption{Superhump maxima of ASASSN-15sp (2015)}\label{tab:asassn15spoc2015}
\begin{center}
\begin{tabular}{rp{55pt}p{40pt}r@{.}lr}
\hline
\multicolumn{1}{c}{$E$} & \multicolumn{1}{c}{max\commenta} & \multicolumn{1}{c}{error} & \multicolumn{2}{c}{$O-C$\commentb} & \multicolumn{1}{c}{$N$\commentc} \\
\hline
0 & 57336.7419 & 0.0025 & $-$0&0168 & 18 \\
1 & 57336.8110 & 0.0015 & $-$0&0062 & 19 \\
33 & 57338.6977 & 0.0005 & 0&0099 & 13 \\
34 & 57338.7533 & 0.0006 & 0&0071 & 17 \\
35 & 57338.8129 & 0.0004 & 0&0082 & 19 \\
50 & 57339.6874 & 0.0019 & 0&0059 & 12 \\
51 & 57339.7435 & 0.0005 & 0&0035 & 20 \\
52 & 57339.8018 & 0.0006 & 0&0033 & 19 \\
85 & 57341.7263 & 0.0006 & $-$0&0013 & 20 \\
86 & 57341.7847 & 0.0009 & $-$0&0013 & 20 \\
87 & 57341.8415 & 0.0016 & $-$0&0029 & 11 \\
102 & 57342.7182 & 0.0012 & $-$0&0031 & 21 \\
103 & 57342.7777 & 0.0009 & $-$0&0020 & 19 \\
104 & 57342.8366 & 0.0016 & $-$0&0017 & 13 \\
119 & 57343.7153 & 0.0010 & 0&0002 & 21 \\
120 & 57343.7728 & 0.0023 & $-$0&0007 & 19 \\
121 & 57343.8295 & 0.0016 & $-$0&0025 & 16 \\
137 & 57344.7673 & 0.0020 & 0&0000 & 20 \\
138 & 57344.8261 & 0.0018 & 0&0003 & 16 \\
\hline
  \multicolumn{6}{l}{\commenta BJD$-$2400000.} \\
  \multicolumn{6}{l}{\commentb Against max $= 2457336.7587 + 0.058457 E$.} \\
  \multicolumn{6}{l}{\commentc Number of points used to determine the maximum.} \\
\end{tabular}
\end{center}
\end{table}

\subsection{ASASSN-15su}\label{obj:asassn15su}

\begin{figure}
  \begin{center}
%    \FigureFile(85mm,70mm){asassn15sushlc.eps}
    \FigureFile(85mm,70mm){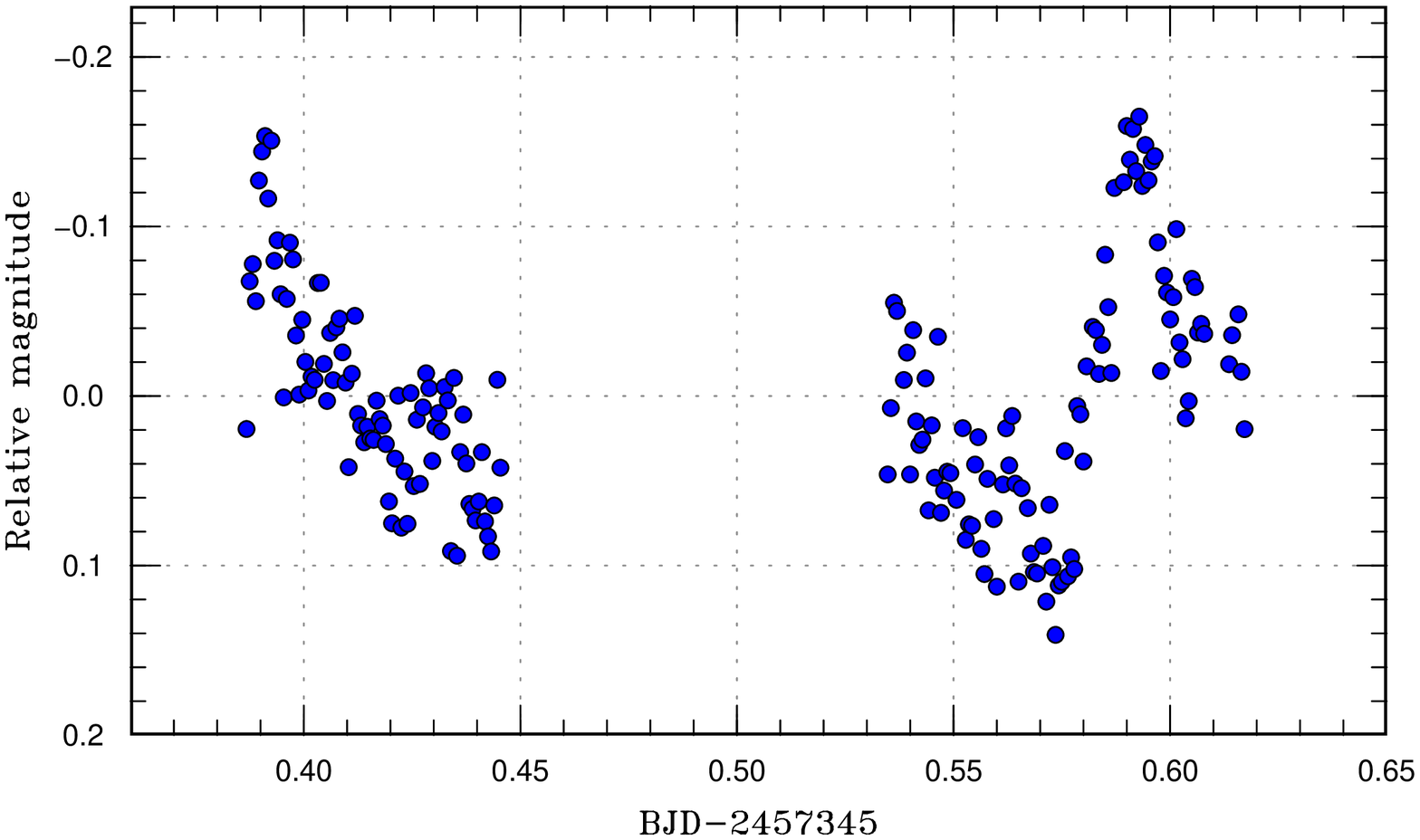}
  \end{center}
  \caption{Superhumps in ASASSN-15su (2015).
  }
  \label{fig:asassn15sushlc}
\end{figure}

\begin{figure}
  \begin{center}
%    \FigureFile(85mm,110mm){asassn15sushpdm.eps}
    \FigureFile(85mm,110mm){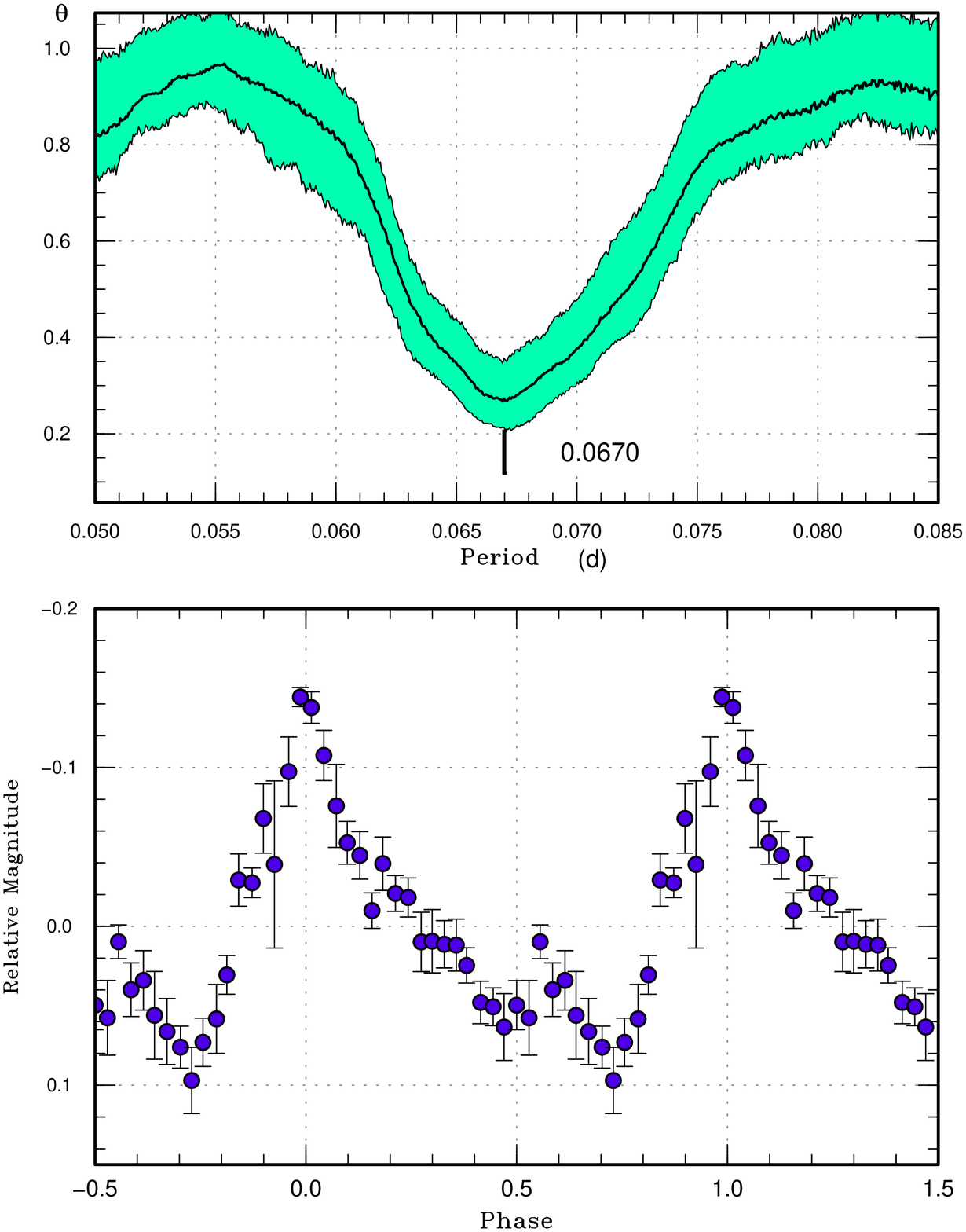}
  \end{center}
  \caption{Superhumps in ASASSN-15su (2015).
     (Upper): PDM analysis.
     (Lower): Phase-averaged profile.}
  \label{fig:asassn15sushpdm}
\end{figure}

   This object was detected as a transient at $V$=15.0
on 2015 November 15 by the ASAS-SN team.
Previous outbursts were recorded in the CRTS data.
Superhumps were immediately detected (vsnet-alert 19285).
Although only one superhump maximum was measured
at BJD 2457345.5947(6) ($N$=57), the period has been
reasonably determined as 0.0670(3)~d
(figures \ref{fig:asassn15sushlc}, \ref{fig:asassn15sushpdm}).

% SI

\subsection{ASASSN-15sv}\label{obj:asassn15sv}

\begin{figure}
  \begin{center}
%    \FigureFile(85mm,70mm){asassn15svshlc.eps}
    \FigureFile(85mm,70mm){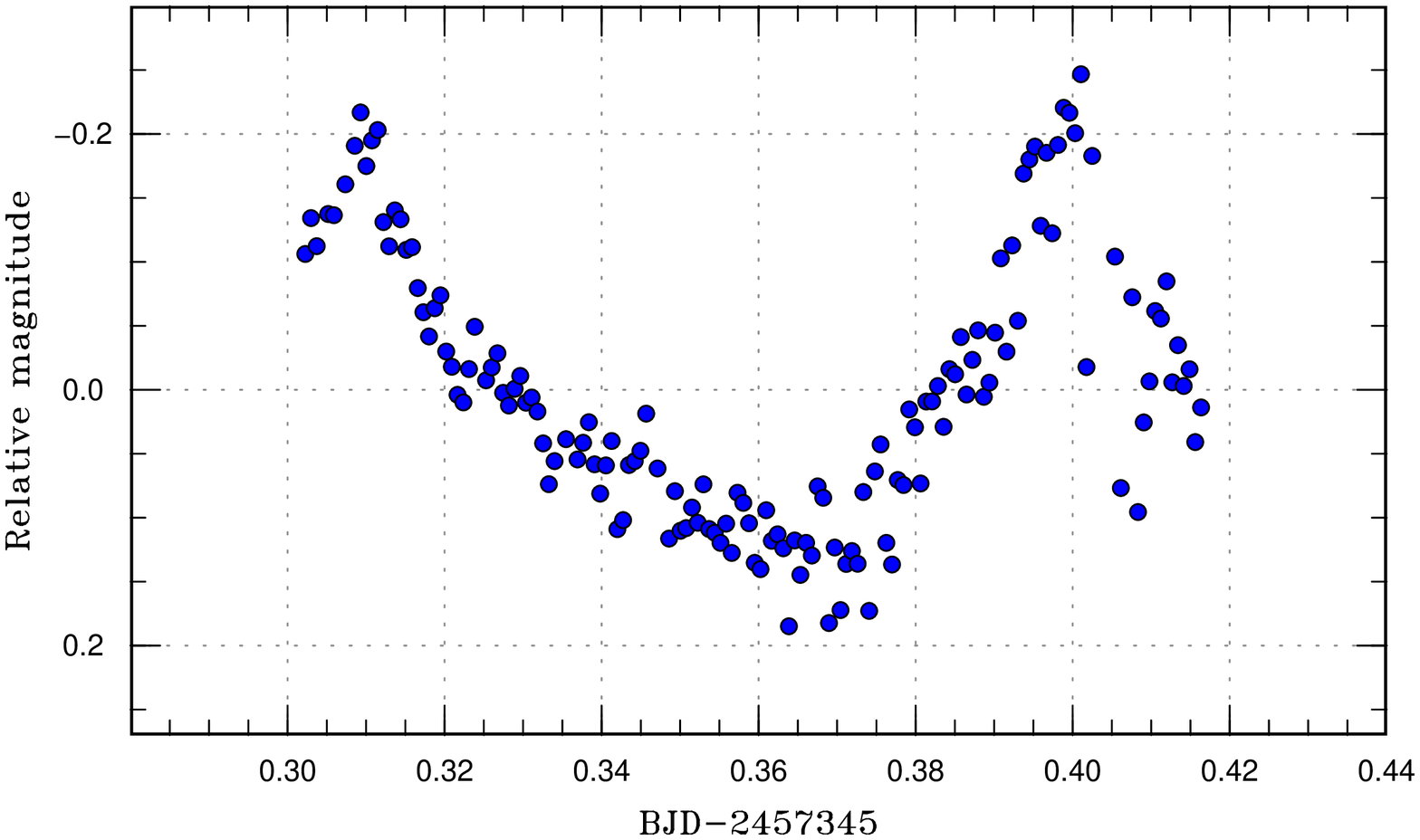}
  \end{center}
  \caption{Superhumps in ASASSN-15sv (2015).
  }
  \label{fig:asassn15svshlc}
\end{figure}

   This object was detected as a transient at $V$=15.8
on 2015 November 16 by the ASAS-SN team.
Superhumps were immediately detected (vsnet-alert 19284;
figure \ref{fig:asassn15svshlc}).
The long superhump period placed the object
in the period gap.
Two superhump maxima were recorded:
BJD 2457345.3086(9) ($N$=52) and 2457345.3998(9) ($N$=72).
The superhump period is 0.091(1)~d.

\subsection{ASASSN-15ud}\label{obj:asassn15ud}

\begin{figure}
  \begin{center}
%    \FigureFile(85mm,110mm){asassn15udshpdm.eps}
    \FigureFile(85mm,110mm){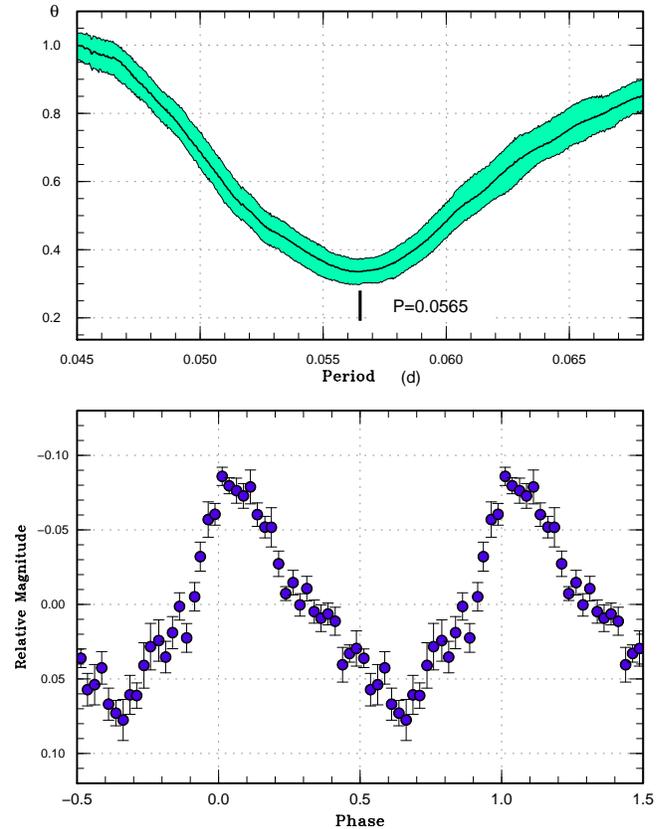}
  \end{center}
  \caption{Superhumps in ASASSN-15ud (2015).
     (Upper): PDM analysis.
     (Lower): Phase-averaged profile.}
  \label{fig:asassn15udshpdm}
\end{figure}

   This object was detected as a large-amplitude transient
at $V$=15.3 on 2015 December 14 by the ASAS-SN team.
There is a $B_j$=22.1 mag counterpart in GSC 2.3.2.
Single-night observations detected superhumps
(vsnet-alert 19353; figure \ref{fig:asassn15udshpdm}).
The times of superhump maxima are listed in
table \ref{tab:asassnoc2015}.
Although the outburst amplitude is large, the object
is likely an ordinary SU UMa-type dwarf nova
as judged from the early appearance of superhumps.
The period given in table \ref{tab:perlist}
is based on the PDM analysis.

% SI

% SI

\begin{table}
\caption{Superhump maxima of ASASSN-15ud (2015)}\label{tab:asassnoc2015}
\begin{center}
\begin{tabular}{rp{55pt}p{40pt}r@{.}lr}
\hline
\multicolumn{1}{c}{$E$} & \multicolumn{1}{c}{max\commenta} & \multicolumn{1}{c}{error} & \multicolumn{2}{c}{$O-C$\commentb} & \multicolumn{1}{c}{$N$\commentc} \\
\hline
0 & 57374.1940 & 0.0005 & 0&0005 & 89 \\
1 & 57374.2480 & 0.0006 & $-$0&0015 & 95 \\
2 & 57374.3070 & 0.0004 & 0&0015 & 108 \\
3 & 57374.3611 & 0.0006 & $-$0&0005 & 108 \\
\hline
  \multicolumn{6}{l}{\commenta BJD$-$2400000.} \\
  \multicolumn{6}{l}{\commentb Against max $= 2457374.1935 + 0.056030 E$.} \\
  \multicolumn{6}{l}{\commentc Number of points used to determine the maximum.} \\
\end{tabular}
\end{center}
\end{table}

\subsection{ASASSN-15uj}\label{obj:asassn15uj}

\begin{figure}
  \begin{center}
%    \FigureFile(85mm,110mm){asassn15ujeshpdm.eps}
    \FigureFile(85mm,110mm){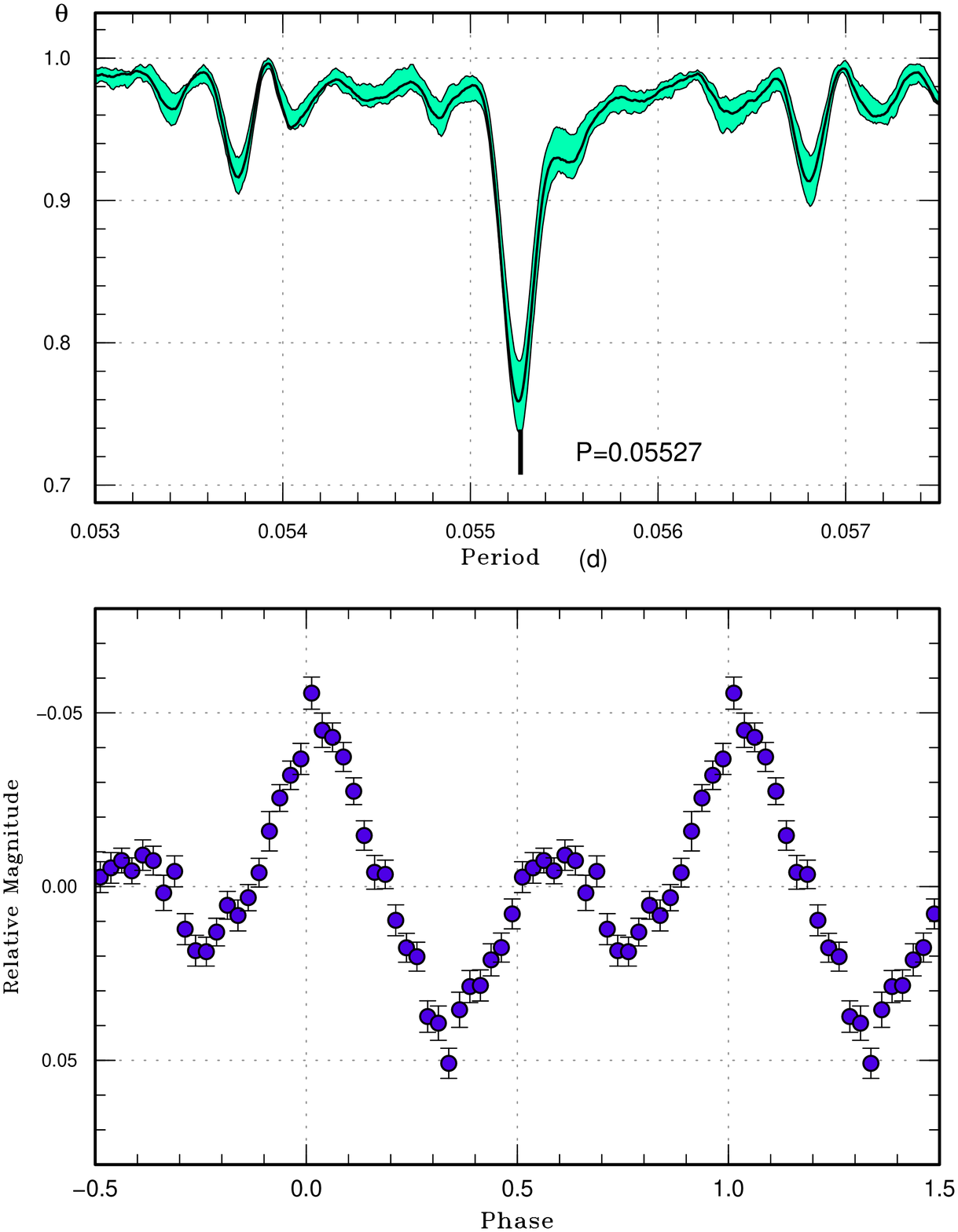}
  \end{center}
  \caption{Early superhumps in ASASSN-15uj (2015).
     (Upper): PDM analysis.
     (Lower): Phase-averaged profile.}
  \label{fig:asassn15ujeshpdm}
\end{figure}

\begin{figure}
  \begin{center}
%    \FigureFile(85mm,110mm){asassn15ujshpdm.eps}
    \FigureFile(85mm,110mm){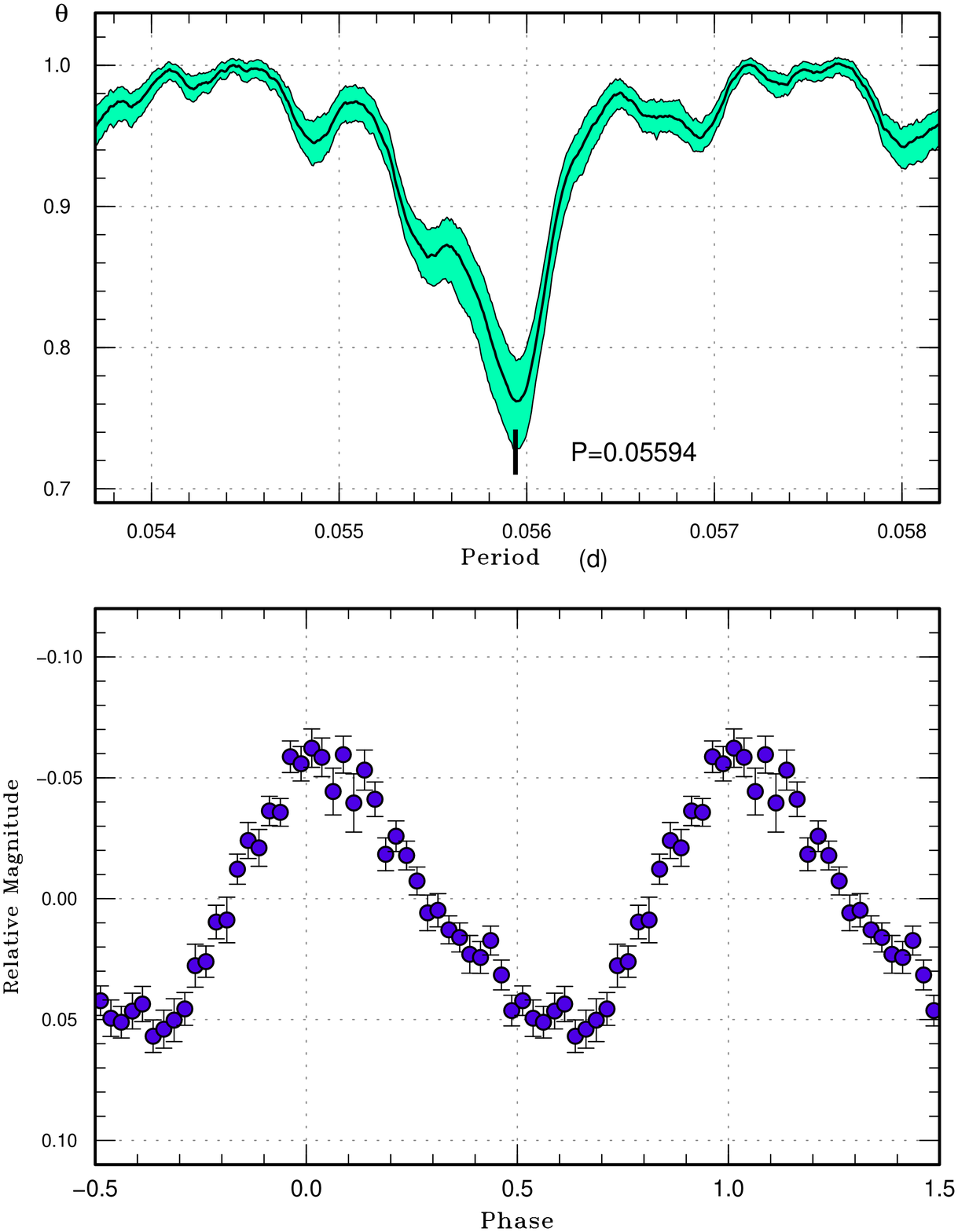}
  \end{center}
  \caption{Ordinary superhumps in ASASSN-15uj (2015).
     (Upper): PDM analysis.
     (Lower): Phase-averaged profile.}
  \label{fig:asassn15ujshpdm}
\end{figure}

\begin{figure}
  \begin{center}
%    \FigureFile(85mm,100mm){asassn15ujhumpall.eps}
    \FigureFile(85mm,100mm){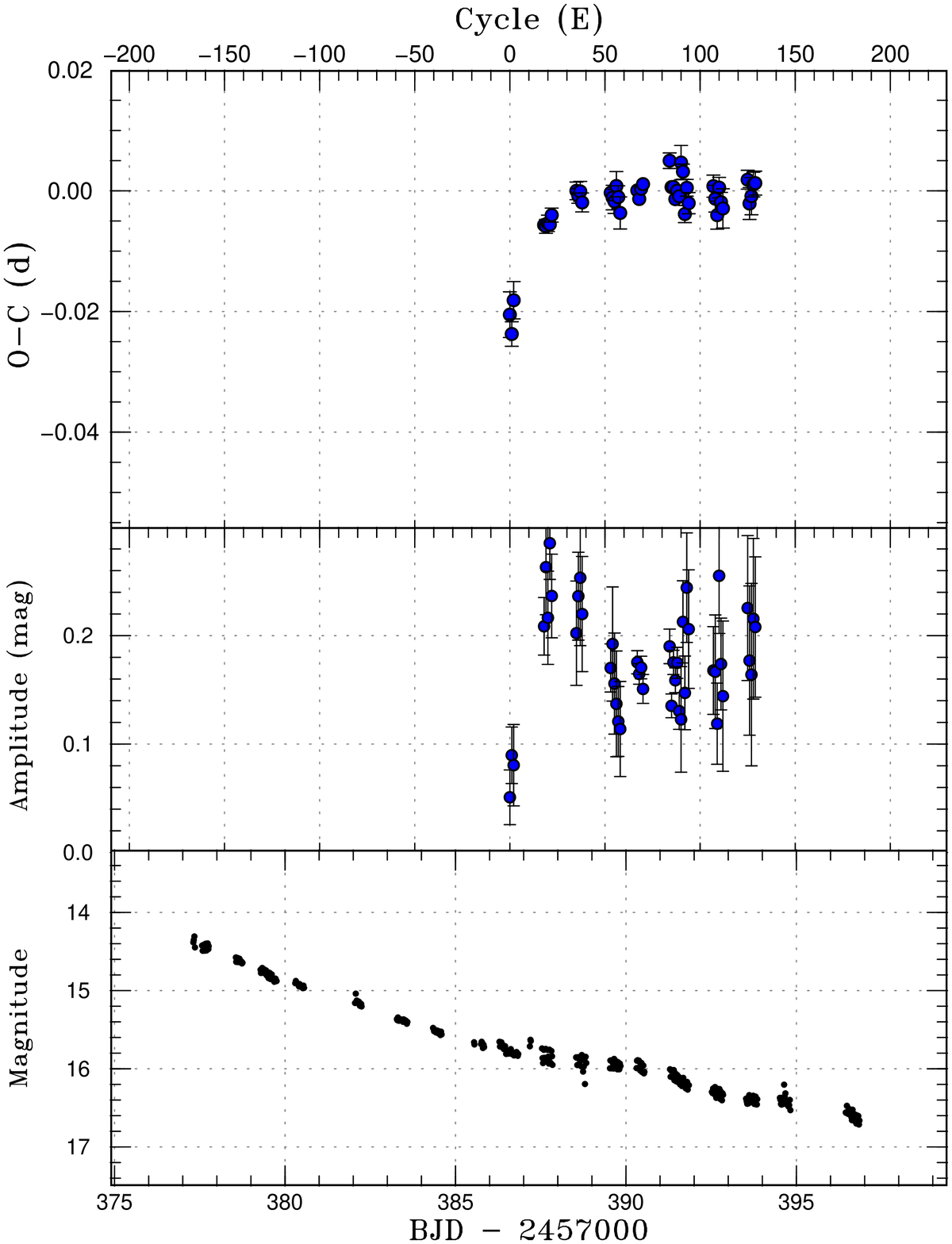}
  \end{center}
  \caption{$O-C$ diagram of superhumps in ASASSN-15uj (2015).
     (Upper:) $O-C$ diagram.
     We used a period of 0.05580~d for calculating the $O-C$ residuals.
     (Middle:) Amplitudes of superhumps.
     (Lower:) Light curve.  The data were binned to 0.018~d.
  }
  \label{fig:asassn15ujhumpall}
\end{figure}

   This object was detected as a large-amplitude transient
at $V$=14.3 on 2015 December 20 by the ASAS-SN team
(cf. vsnet-alert 19352).
Subsequent observations detected double-wave
early superhumps (vsnet-alert 19359, 19361, 19368;
figure \ref{fig:asassn15ujeshpdm}) confirming
the WZ Sge-type classification.
The best period of early superhumps with the PDM method
is 0.055266(7)~d.
The object then showed growing ordinary superhumps
(vsnet-alert 19392; figure \ref{fig:asassn15ujshpdm}).
The outburst lasted at least up to 2016 January 17
($V$=16.43, part of the rebrightening phase?,
ASAS-SN data).
The times of superhump maxima are listed in
table \ref{tab:asassn15ujoc2015}.
The maxima for $E \le$22 correspond to stage A
superhumps (figure \ref{fig:asassn15ujhumpall}).
The measured $\epsilon^*$=0.0243(13)
for stage A superhumps corresponds to $q$=0.064(4).
The almost constant period of stage B superhumps
is also consistent with this small $q$
(cf. \cite{kat15wzsge}).
The object is a WZ Sge-type dwarf nova likely in
relatively evolved state.

% SI

% SI

% SI

\begin{table*}
\caption{Superhump maxima of ASASSN-15uj (2015)}\label{tab:asassn15ujoc2015}
\begin{center}
\begin{tabular}{rp{55pt}p{40pt}r@{.}lrrp{55pt}p{40pt}r@{.}lr}
\hline
\multicolumn{1}{c}{$E$} & \multicolumn{1}{c}{max\commenta} & \multicolumn{1}{c}{error} & \multicolumn{2}{c}{$O-C$\commentb} & \multicolumn{1}{c}{$N$\commentc} & \multicolumn{1}{c}{$E$} & \multicolumn{1}{c}{max\commenta} & \multicolumn{1}{c}{error} & \multicolumn{2}{c}{$O-C$\commentb} & \multicolumn{1}{c}{$N$\commentc} \\
\hline
0 & 57386.5738 & 0.0038 & $-$0&0117 & 14 & 84 & 57391.2865 & 0.0013 & 0&0061 & 75 \\
1 & 57386.6264 & 0.0020 & $-$0&0150 & 14 & 85 & 57391.3379 & 0.0006 & 0&0016 & 128 \\
2 & 57386.6878 & 0.0031 & $-$0&0095 & 14 & 86 & 57391.3937 & 0.0004 & 0&0015 & 128 \\
18 & 57387.5930 & 0.0010 & 0&0015 & 12 & 87 & 57391.4475 & 0.0005 & $-$0&0006 & 128 \\
19 & 57387.6486 & 0.0012 & 0&0012 & 14 & 88 & 57391.5047 & 0.0006 & 0&0007 & 129 \\
20 & 57387.7049 & 0.0014 & 0&0016 & 14 & 89 & 57391.5596 & 0.0009 & $-$0&0003 & 140 \\
21 & 57387.7605 & 0.0008 & 0&0013 & 19 & 90 & 57391.6210 & 0.0028 & 0&0052 & 13 \\
22 & 57387.8179 & 0.0012 & 0&0028 & 20 & 91 & 57391.6753 & 0.0013 & 0&0036 & 12 \\
35 & 57388.5473 & 0.0015 & 0&0056 & 12 & 92 & 57391.7240 & 0.0014 & $-$0&0035 & 13 \\
36 & 57388.6023 & 0.0012 & 0&0047 & 12 & 93 & 57391.7842 & 0.0014 & 0&0008 & 19 \\
37 & 57388.6588 & 0.0017 & 0&0053 & 12 & 94 & 57391.8375 & 0.0018 & $-$0&0019 & 16 \\
38 & 57388.7128 & 0.0016 & 0&0034 & 11 & 107 & 57392.5657 & 0.0018 & $-$0&0003 & 19 \\
53 & 57389.5514 & 0.0009 & 0&0036 & 13 & 108 & 57392.6194 & 0.0022 & $-$0&0024 & 14 \\
54 & 57389.6064 & 0.0020 & 0&0027 & 13 & 109 & 57392.6725 & 0.0023 & $-$0&0053 & 14 \\
55 & 57389.6615 & 0.0020 & 0&0020 & 12 & 110 & 57392.7328 & 0.0017 & $-$0&0008 & 14 \\
56 & 57389.7199 & 0.0024 & 0&0045 & 13 & 111 & 57392.7862 & 0.0016 & $-$0&0033 & 18 \\
57 & 57389.7739 & 0.0019 & 0&0025 & 19 & 112 & 57392.8410 & 0.0033 & $-$0&0044 & 14 \\
58 & 57389.8270 & 0.0027 & $-$0&0002 & 19 & 125 & 57393.5711 & 0.0016 & $-$0&0009 & 15 \\
67 & 57390.3330 & 0.0004 & 0&0027 & 128 & 126 & 57393.6230 & 0.0026 & $-$0&0049 & 14 \\
68 & 57390.3873 & 0.0004 & 0&0012 & 128 & 127 & 57393.6800 & 0.0031 & $-$0&0038 & 13 \\
69 & 57390.4448 & 0.0004 & 0&0028 & 128 & 128 & 57393.7377 & 0.0020 & $-$0&0019 & 15 \\
70 & 57390.5014 & 0.0006 & 0&0035 & 126 & 129 & 57393.7938 & 0.0020 & $-$0&0018 & 19 \\
\hline
  \multicolumn{12}{l}{\commenta BJD$-$2400000.} \\
  \multicolumn{12}{l}{\commentb Against max $= 2457386.5855 + 0.055892 E$.} \\
  \multicolumn{12}{l}{\commentc Number of points used to determine the maximum.} \\
\end{tabular}
\end{center}
\end{table*}

\subsection{ASASSN-15ux}\label{obj:asassn15ux}

\begin{figure}
  \begin{center}
%    \FigureFile(85mm,110mm){asassn15uxeshpdm.eps}
    \FigureFile(85mm,110mm){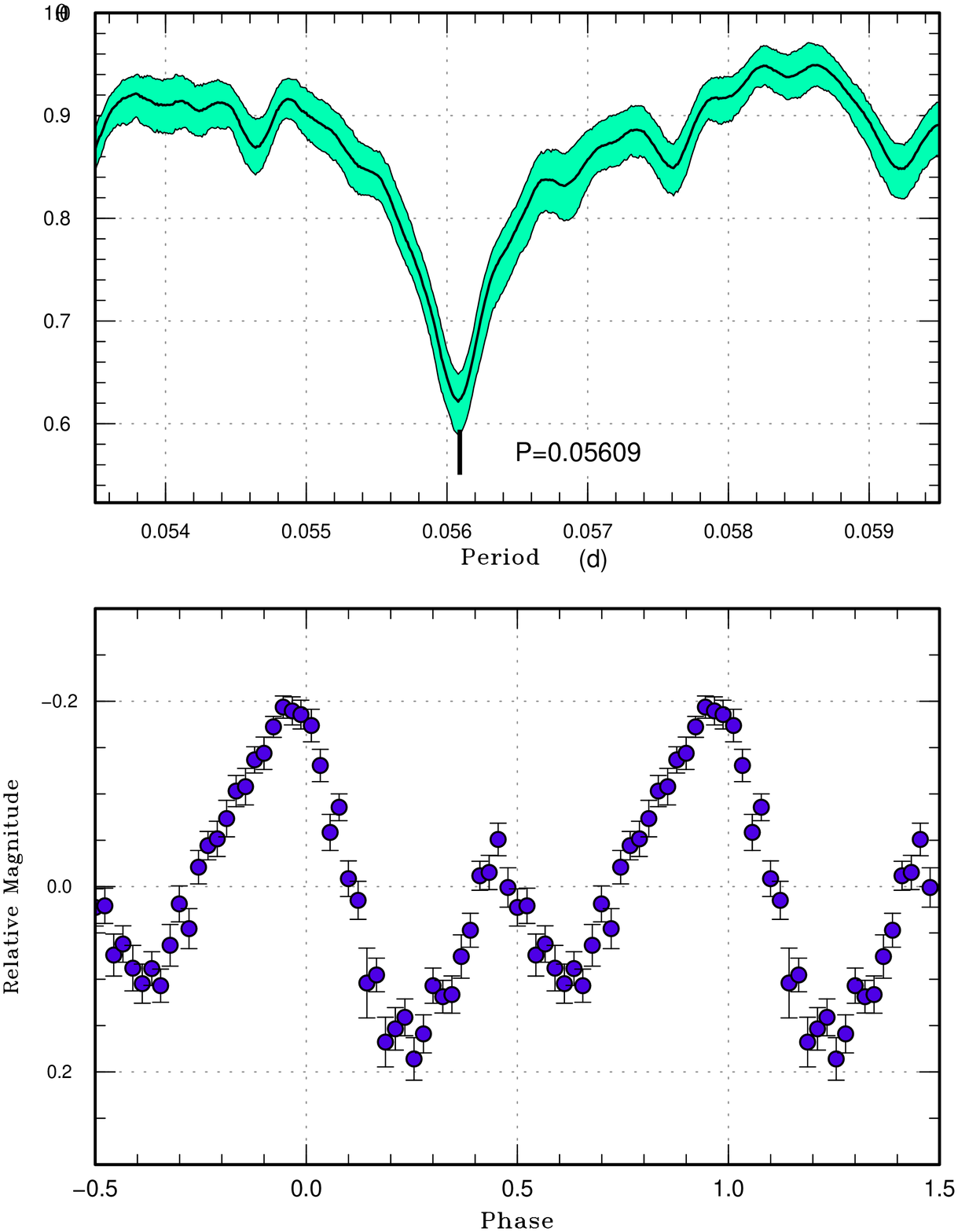}
  \end{center}
  \caption{Early superhumps in ASASSN-15ux (2015).
     The data before BJD 2457396 were used.
     (Upper): PDM analysis.
     (Lower): Phase-averaged profile.}
  \label{fig:asassn15uxeshpdm}
\end{figure}

\begin{figure}
  \begin{center}
%    \FigureFile(85mm,110mm){asassn15uxshpdm.eps}
    \FigureFile(85mm,110mm){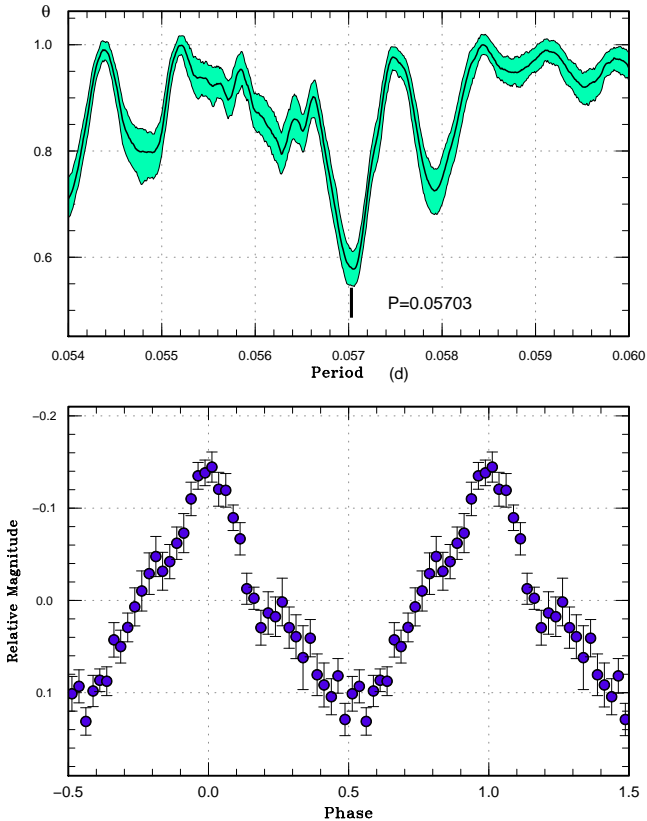}
  \end{center}
  \caption{Superhumps in ASASSN-15ux (2015).
     The data after BJD 2457400 were used.  The averaged
     profile outsides the eclipses is shown.  The mean period
     and profile were likely affected by the beat phenomenon
     with the orbital variation.
     (Upper): PDM analysis.
     (Lower): Phase-averaged profile.}
  \label{fig:asassn15uxshpdm}
\end{figure}

\begin{figure}
  \begin{center}
%    \FigureFile(85mm,120mm){asassn15uxlc.eps}
    \FigureFile(85mm,120mm){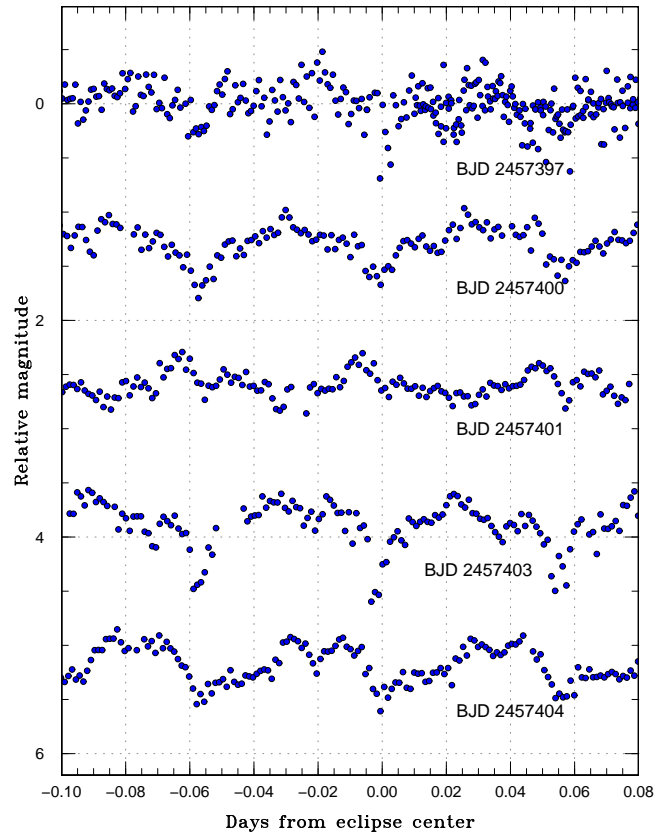}
  \end{center}
  \caption{Light curve of ASASSN-15ux after the appearance
  of the eclipses.  Shallow eclipses with variable profiles
  were recorded.}
  \label{fig:asassn15uxlc}
\end{figure}

   This object was detected as a large-amplitude transient
at $V$=14.4 on 2015 December 29 by the ASAS-SN team.
No quiescent counterpart is known.
Subsequent observations detected early superhumps
(these modulations were initially reported as
superhumps) (vsnet-alert 19390, 19391, 19396;
figure \ref{fig:asassn15uxeshpdm}).
The object started to show ordinary superhumps
on 2016 January 12 (vsnet-alert 19409, 19411;
figure \ref{fig:asassn15uxshpdm}).
The large amplitude of early superhumps and
the complex profile of individual superhumps suggested
the presence of eclipses (vsnet-alert 19391, 19409).
The eclipsing nature was confirmed by further
observations (vsnet-alert 19423; figure \ref{fig:asassn15uxlc}).

   The eclipse ephemeris was determined
by using MCMC analysis \citep{Pdot4}
of the observations after BJD 2457396.5
(when eclipses became apparent):
\begin{equation}
{\rm Min(BJD)} = 2457400.82908(10) + 0.056109(2) E .
\label{equ:asassn15uxecl}
\end{equation}
This ephemeris is not intended for long-term prediction
of eclipses.  The epoch refers to the center of
the observations used.
The period of early superhumps was 0.056091(4)~d,
0.03\% shorter than the orbital period.
This value is in very good agreement with the other
eclipsing WZ Sge-type dwarf novae \citep{kat15wzsge}.

   The times of superhump maxima determined from
observations outside the eclipses are listed in
table \ref{tab:asassn15uxoc2015}.  Although hump maxima
for $E \le 3$ looked like stage A superhumps,
they may have been a transition phase from early superhumps
to ordinary superhumps (e.g. WZ Sge, figure 126
in \cite{Pdot}), we have disregarded these maxima
in determining the period of stage A superhumps.
There also remains ambiguity in cycle counts between
$E$=3 and $E$=55.  The maxima for $E \ge$73 are clearly
stage B superhumps.

   Although the object is very faint, the object is
a rare WZ Sge-type dwarf nova with high amplitude of
early superhumps (the mean amplitude of 0.38 mag is
one of the largest, cf. \cite{kat15wzsge}) and eclipses.
The object will deserve further detailed observations.

% SI

% SI

% SI

\begin{table}
\caption{Superhump maxima of ASASSN-15ux (2015)}\label{tab:asassn15uxoc2015}
\begin{center}
\begin{tabular}{rp{50pt}p{30pt}r@{.}lcr}
\hline
$E$ & max\commenta & error & \multicolumn{2}{c}{$O-C$\commentb} & phase\commentc & $N$\commentd \\
\hline
0 & 57397.1599 & 0.0031 & $-$0&0010 & 0.35 & 54 \\
1 & 57397.2180 & 0.0022 & 0&0002 & 0.59 & 46 \\
2 & 57397.2716 & 0.0027 & $-$0&0033 & 0.41 & 101 \\
3 & 57397.3300 & 0.0028 & $-$0&0019 & 0.52 & 83 \\
55 & 57400.3020 & 0.0030 & 0&0053 & 0.79 & 17 \\
56 & 57400.3483 & 0.0017 & $-$0&0054 & 0.41 & 37 \\
57 & 57400.4049 & 0.0019 & $-$0&0058 & 0.46 & 35 \\
58 & 57400.4634 & 0.0010 & $-$0&0043 & 0.27 & 38 \\
59 & 57400.5249 & 0.0018 & 0&0002 & 0.37 & 33 \\
60 & 57400.5787 & 0.0008 & $-$0&0030 & 0.29 & 52 \\
73 & 57401.3286 & 0.0008 & 0&0056 & 0.23 & 31 \\
74 & 57401.3876 & 0.0012 & 0&0076 & 0.39 & 31 \\
75 & 57401.4419 & 0.0008 & 0&0050 & 0.37 & 28 \\
76 & 57401.4993 & 0.0011 & 0&0053 & 0.38 & 30 \\
77 & 57401.5561 & 0.0012 & 0&0051 & 0.53 & 27 \\
108 & 57403.3219 & 0.0023 & 0&0034 & 0.53 & 30 \\
109 & 57403.3762 & 0.0009 & 0&0007 & 0.43 & 31 \\
110 & 57403.4344 & 0.0023 & 0&0019 & 0.56 & 30 \\
111 & 57403.4898 & 0.0009 & 0&0003 & 0.38 & 35 \\
112 & 57403.5472 & 0.0013 & 0&0007 & 0.54 & 35 \\
113 & 57403.6052 & 0.0013 & 0&0016 & 0.33 & 36 \\
126 & 57404.3424 & 0.0009 & $-$0&0024 & 0.27 & 33 \\
127 & 57404.3987 & 0.0007 & $-$0&0031 & 0.31 & 33 \\
128 & 57404.4555 & 0.0014 & $-$0&0033 & 0.40 & 36 \\
129 & 57404.5128 & 0.0011 & $-$0&0030 & 0.32 & 32 \\
130 & 57404.5695 & 0.0011 & $-$0&0033 & 0.30 & 34 \\
131 & 57404.6271 & 0.0012 & $-$0&0027 & 0.35 & 25 \\
\hline
  \multicolumn{7}{l}{\commenta BJD$-$2400000.} \\
  \multicolumn{7}{l}{\commentb Against max $= 2457397.1609 + 0.057015 E$.} \\
  \multicolumn{7}{l}{\commentc Orbital phase.} \\
  \multicolumn{7}{l}{\commentd Number of points used to determine the maximum.} \\
\end{tabular}
\end{center}
\end{table}

\subsection{ASASSN-16af}\label{obj:asassn16af}

\begin{figure}
  \begin{center}
%    \FigureFile(85mm,110mm){asassn16afshpdm.eps}
    \FigureFile(85mm,110mm){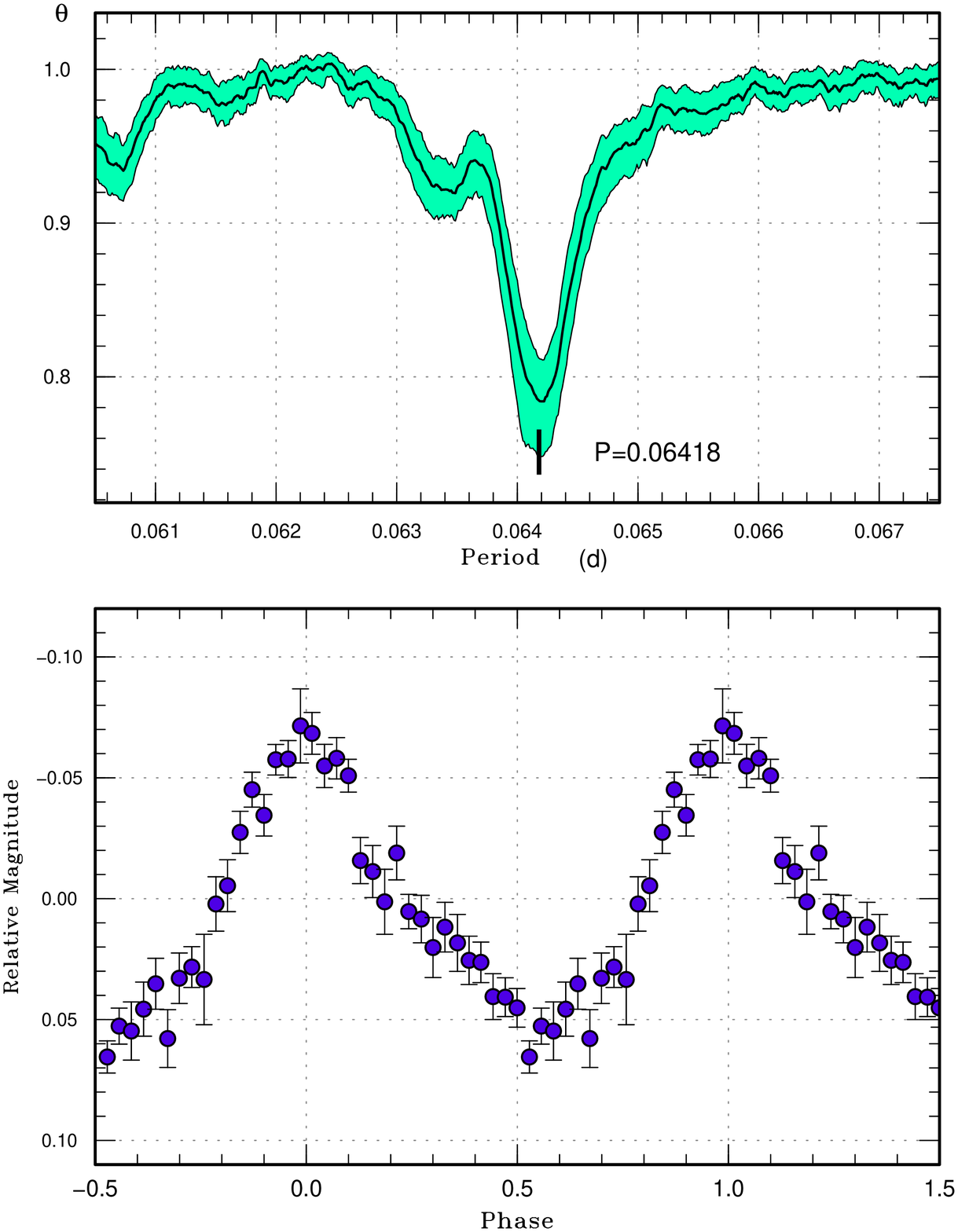}
  \end{center}
  \caption{Superhumps in ASASSN-16af (2016).
     (Upper): PDM analysis.
     (Lower): Phase-averaged profile.}
  \label{fig:asassn16afshpdm}
\end{figure}

   This object was detected as a transient
at $V$=15.5 on 2016 January 10 by the ASAS-SN team.
Subsequent observations detected superhumps
(vsnet-alert 19417, 19419, 19424;
figure \ref{fig:asassn16afshpdm}).
The times of superhump maxima are listed in
table \ref{tab:asassn16afoc2016}.  A positive
$P_{\rm dot}$ typical for this superhump period
was recorded.

% SI

% SI

\begin{table}
\caption{Superhump maxima of ASASSN-16af (2016)}\label{tab:asassn16afoc2016}
\begin{center}
\begin{tabular}{rp{55pt}p{40pt}r@{.}lr}
\hline
\multicolumn{1}{c}{$E$} & \multicolumn{1}{c}{max\commenta} & \multicolumn{1}{c}{error} & \multicolumn{2}{c}{$O-C$\commentb} & \multicolumn{1}{c}{$N$\commentc} \\
\hline
0 & 57401.4789 & 0.0013 & 0&0046 & 26 \\
1 & 57401.5401 & 0.0006 & 0&0016 & 58 \\
2 & 57401.6045 & 0.0009 & 0&0019 & 52 \\
10 & 57402.1163 & 0.0010 & $-$0&0000 & 51 \\
11 & 57402.1816 & 0.0005 & 0&0011 & 111 \\
12 & 57402.2446 & 0.0006 & $-$0&0001 & 125 \\
13 & 57402.3098 & 0.0007 & 0&0009 & 124 \\
20 & 57402.7560 & 0.0010 & $-$0&0023 & 22 \\
21 & 57402.8215 & 0.0028 & $-$0&0010 & 22 \\
26 & 57403.1429 & 0.0012 & $-$0&0006 & 104 \\
27 & 57403.2050 & 0.0005 & $-$0&0028 & 125 \\
28 & 57403.2703 & 0.0006 & $-$0&0017 & 97 \\
29 & 57403.3340 & 0.0009 & $-$0&0021 & 35 \\
35 & 57403.7202 & 0.0009 & $-$0&0012 & 19 \\
36 & 57403.7852 & 0.0021 & $-$0&0004 & 22 \\
37 & 57403.8502 & 0.0025 & 0&0004 & 14 \\
51 & 57404.7461 & 0.0019 & $-$0&0026 & 20 \\
52 & 57404.8113 & 0.0015 & $-$0&0016 & 20 \\
63 & 57405.5183 & 0.0017 & $-$0&0008 & 60 \\
64 & 57405.5788 & 0.0017 & $-$0&0045 & 55 \\
65 & 57405.6480 & 0.0014 & 0&0005 & 58 \\
66 & 57405.7120 & 0.0062 & 0&0003 & 16 \\
67 & 57405.7773 & 0.0035 & 0&0014 & 20 \\
75 & 57406.2988 & 0.0028 & 0&0093 & 45 \\
\hline
  \multicolumn{6}{l}{\commenta BJD$-$2400000.} \\
  \multicolumn{6}{l}{\commentb Against max $= 2457401.4743 + 0.064204 E$.} \\
  \multicolumn{6}{l}{\commentc Number of points used to determine the maximum.} \\
\end{tabular}
\end{center}
\end{table}

\subsection{ASASSN-16ag}\label{obj:asassn16ag}

\begin{figure}
  \begin{center}
%    \FigureFile(85mm,110mm){asassn16agshpdm.eps}
    \FigureFile(85mm,110mm){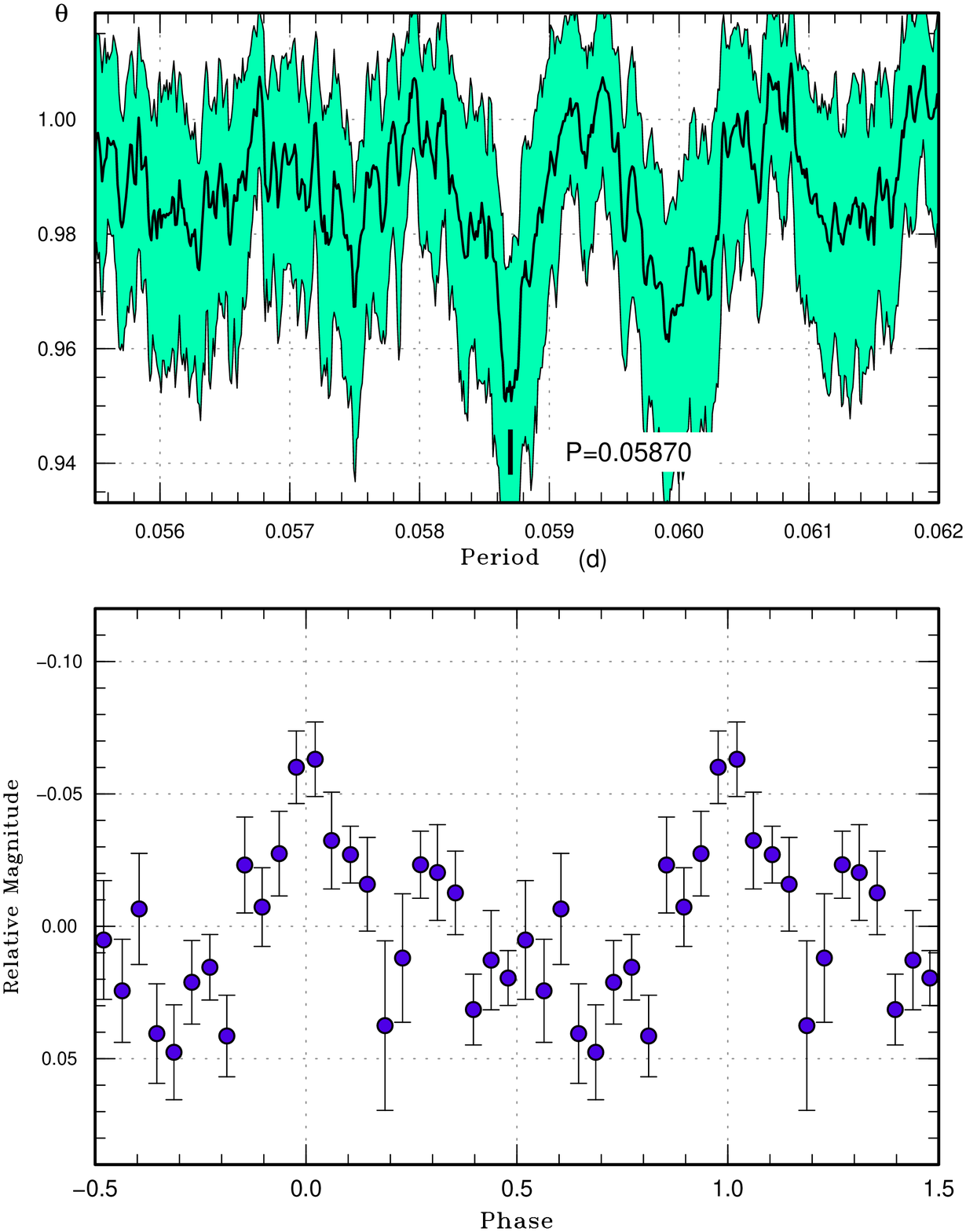}
  \end{center}
  \caption{Superhumps in ASASSN-16ag (2016).
     (Upper): PDM analysis.
     (Lower): Phase-averaged profile.}
  \label{fig:asassn16agshpdm}
\end{figure}

   This object was detected as a transient
at $V$=16.2 on 2016 January 11 by the ASAS-SN team.
Although the observations on January
12.4--12.6 UT did not detect superhumps, superhumps
were detected immediately after these observations
(vsnet-alert 19407).  Although these superhumps
were possibly stage A superhumps, we could not
determine the stage due to the lack of observations
and the low signal-to-noise ratios caused by faintness
of the object.  In table \ref{tab:asassn16agoc2016},
we list times of maxima using a period which
reasonably expresses all the observations
(see also figure \ref{fig:asassn16agshpdm}).
The object apparently has a low outburst amplitude
(3.5 mag) for an SU UMa-type dwarf nova.

% SI

% SI

\begin{table}
\caption{Superhump maxima of ASASSN-16ag (2015)}\label{tab:asassn16agoc2016}
\begin{center}
\begin{tabular}{rp{55pt}p{40pt}r@{.}lr}
\hline
\multicolumn{1}{c}{$E$} & \multicolumn{1}{c}{max\commenta} & \multicolumn{1}{c}{error} & \multicolumn{2}{c}{$O-C$\commentb} & \multicolumn{1}{c}{$N$\commentc} \\
\hline
0 & 57400.3447 & 0.0014 & $-$0&0009 & 51 \\
1 & 57400.4013 & 0.0019 & $-$0&0028 & 54 \\
2 & 57400.4598 & 0.0013 & $-$0&0027 & 62 \\
3 & 57400.5202 & 0.0012 & $-$0&0008 & 53 \\
17 & 57401.3369 & 0.0041 & $-$0&0028 & 52 \\
18 & 57401.4137 & 0.0024 & 0&0156 & 45 \\
19 & 57401.4521 & 0.0049 & $-$0&0045 & 31 \\
34 & 57402.3342 & 0.0055 & 0&0004 & 63 \\
44 & 57402.9099 & 0.0035 & $-$0&0087 & 21 \\
45 & 57402.9746 & 0.0036 & $-$0&0025 & 47 \\
46 & 57403.0445 & 0.0015 & 0&0090 & 58 \\
47 & 57403.0882 & 0.0107 & $-$0&0058 & 32 \\
53 & 57403.4544 & 0.0014 & 0&0095 & 49 \\
96 & 57405.9564 & 0.0044 & $-$0&0031 & 54 \\
\hline
  \multicolumn{6}{l}{\commenta BJD$-$2400000.} \\
  \multicolumn{6}{l}{\commentb Against max $= 2457400.3432 + 0.058596 E$.} \\
  \multicolumn{6}{l}{\commentc Number of points used to determine the maximum.} \\
\end{tabular}
\end{center}
\end{table}

\subsection{ASASSN-16ao}\label{obj:asassn16ao}

\begin{figure}
  \begin{center}
%    \FigureFile(85mm,70mm){asassn16aoshlc.eps}
    \FigureFile(85mm,70mm){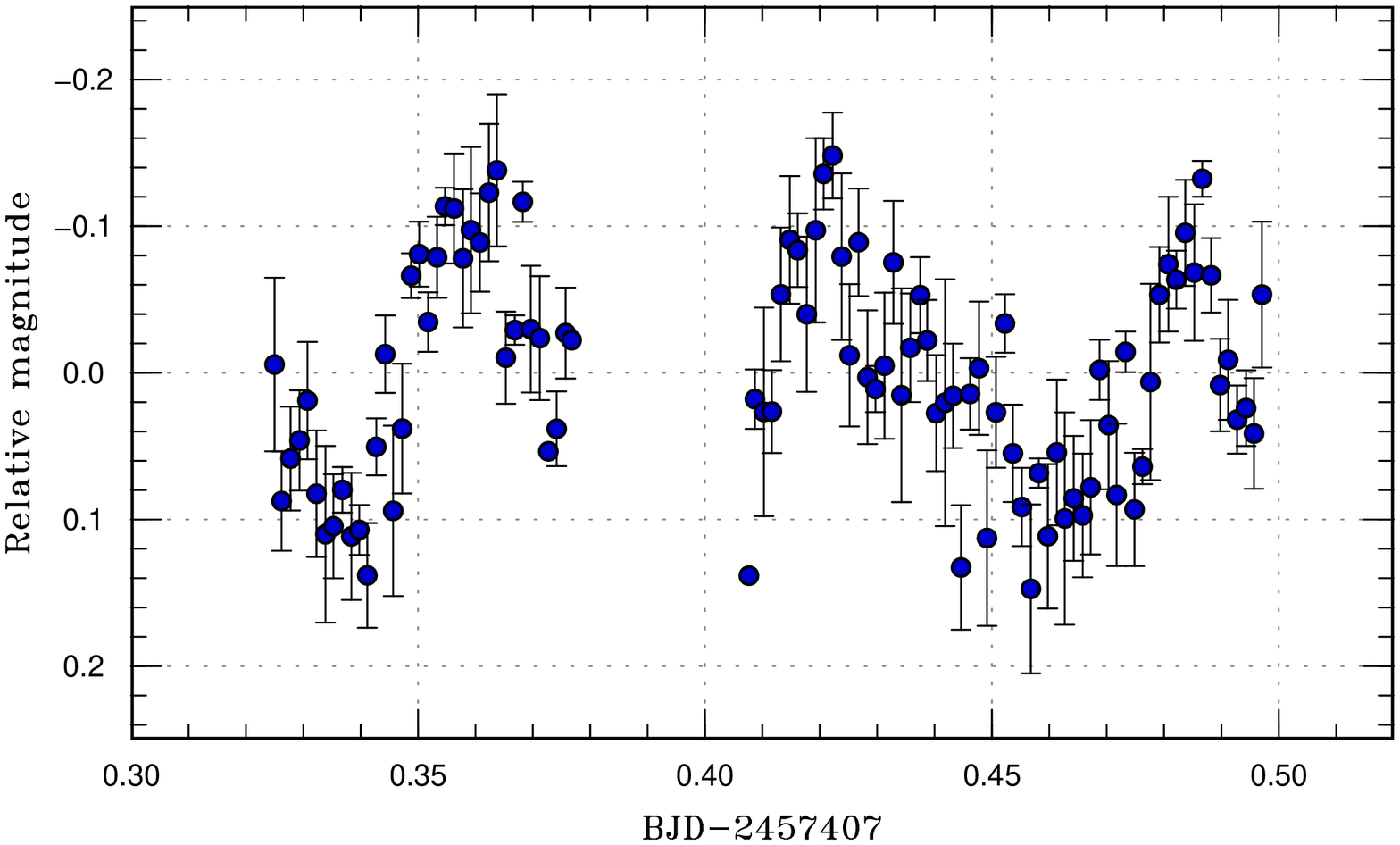}
  \end{center}
  \caption{Possible superhumps in ASASSN-16ao (2016).
  The data were binned to 0.0015~d.
  }
  \label{fig:asassn16aoshlc}
\end{figure}

   This object was detected as a transient
at $V$=16.1 on 2016 January 13 by the ASAS-SN team.
The outburst once faded below $V$=18.0 on January 16
(ASAS-SN data).  The object was observed bright by
B. Monard on January 19 when the object was at about
18.7 mag.  These observations detected modulations
attributable to superhumps (vsnet-alert 19433;
figure \ref{fig:asassn16aoshlc}).
The identification, however, is not secure since
the object was not in very bright state (the identification
of a precursor in vsnet-alert 19433 was probably
incorrect).  The best period from these observations
was 0.0639(5)~d.  Further observations are needed
to firmly characterize the nature of this object.

\subsection{ASASSN-16aq}\label{obj:asassn16aq}

\begin{figure}
  \begin{center}
%    \FigureFile(85mm,70mm){asassn16aqshlc.eps}
    \FigureFile(85mm,70mm){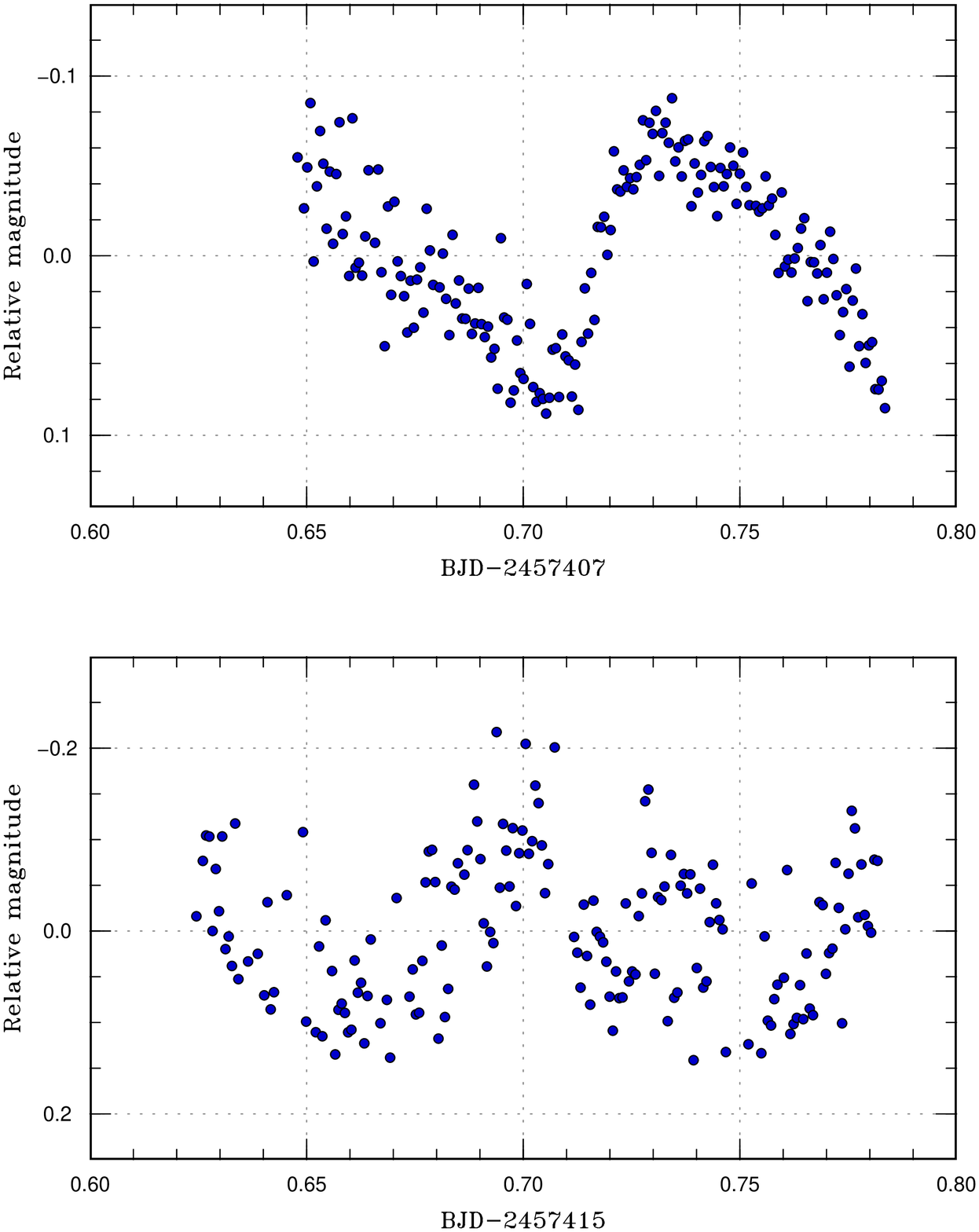}
  \end{center}
  \caption{Superhumps in ASASSN-16aq (2016).
  }
  \label{fig:asassn16aqshlc}
\end{figure}

   This object was detected as a transient
at $V$=15.1 on 2016 January 17 by the ASAS-SN team.
The large outburst amplitude suggested
a possible WZ Sge-type dwarf nova (cf. vsnet-alert 19428).
Superhump-type modulations were immediately recorded
(vsnet-alert 19432; figure \ref{fig:asassn16aqshlc}).
Two superhump maxima were detected:
BJD 2457407.7398(6) ($N$=97) and 2457415.7009(20) ($N$=83).
Since there was a long gap of observations between
these two maxima, we could not choose
an unambiguous period.  The light curve suggests
that the period is relatively longer (longer than
0.08~d).  This period is, however, too long for
a WZ Sge-type dwarf nova and the object may be
a kind of large-amplitude, long-$P_{\rm orb}$
SU UMa-type dwarf novae such as V1251 Cyg and
RZ Leo (see \cite{kat15wzsge}).

\subsection{ASASSN-16bh}\label{obj:asassn16bh}

\begin{figure}
  \begin{center}
%    \FigureFile(85mm,110mm){asassn16bheshpdm.eps}
    \FigureFile(85mm,110mm){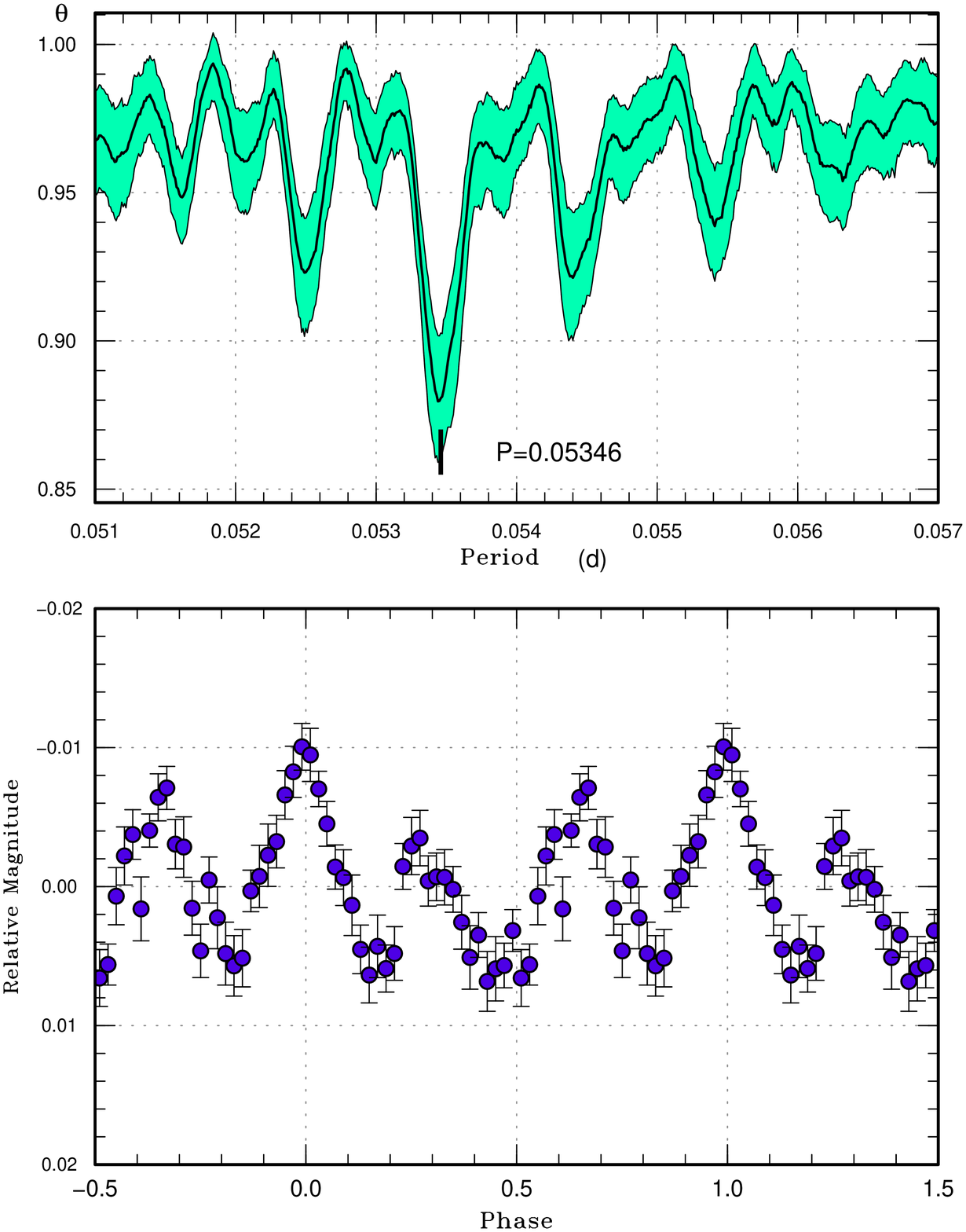}
  \end{center}
  \caption{Early superhumps in ASASSN-16bh (2016).
     The high-quality data by MLF before BJD 2457431 were used.
     (Upper): PDM analysis.
     (Lower): Phase-averaged profile.}
  \label{fig:asassn16bheshpdm}
\end{figure}

   This object was detected as a transient
at $V$=12.7 on 2016 February 6 by the ASAS-SN team
\citep{sim16asassn16bhatel8648}.
The object was suspected to be a WZ Sge-type dwarf nova
based on the large ($\sim$8 mag) outburst amplitude.
Early observations detected low-amplitude modulations,
which were suspected to be early superhumps
(vsnet-alert 19476, 19479).  After the development
of ordinary superhumps, the period of early superhumps
was established (vsnet-alert 19513).
In figure \ref{fig:asassn16bheshpdm}, we show the mean
profile of early superhumps using the high-quality data
from the southern hemisphere (MLF).
The best period with the PDM method was 0.05346(2)~d.
It is noteworthy that the profile has three peaks
(in contrast to two peaks in many WZ Sge-type dwarf
novae, cf. \cite{kat15wzsge}) in one cycle.

\begin{figure}
  \begin{center}
%    \FigureFile(85mm,110mm){asassn16bhshpdm.eps}
    \FigureFile(85mm,110mm){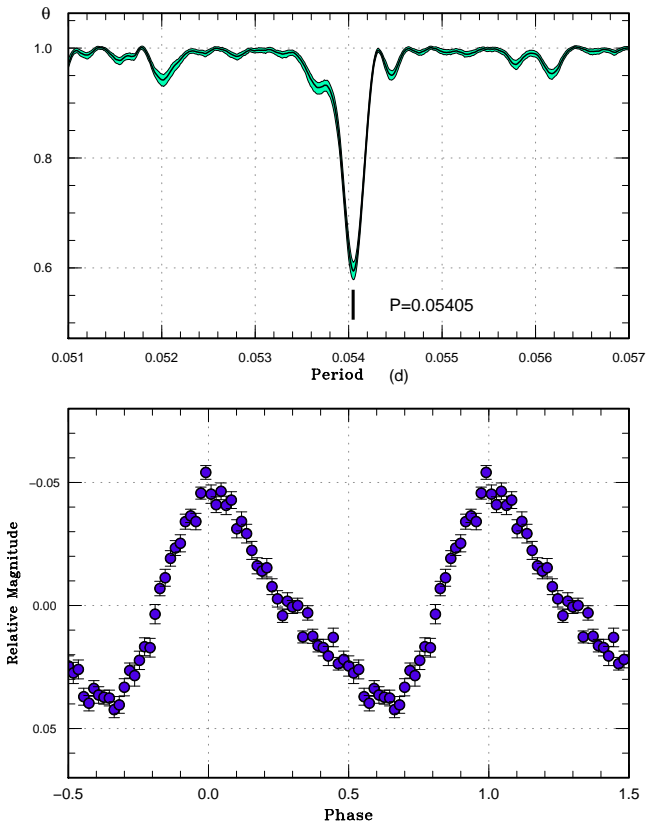}
  \end{center}
  \caption{Ordinary superhumps in ASASSN-16bh (2016).
     The data between BJD 2457432 and 2457444 were used.
     (Upper): PDM analysis.
     (Lower): Phase-averaged profile.}
  \label{fig:asassn16bhshpdm}
\end{figure}

   Ordinary superhumps then appeared (vsnet-alert 19484,
19513, 19520, 19521; figure \ref{fig:asassn16bhshpdm}).
The times of superhump maxima are listed in
table \ref{tab:asassn16bhoc2016}.
The maxima for $E \le$15 were very clearly
stage A superhumps.  After $E=$206 (rapid fading phase),
there was probably a phase jump.  Similar phenomena
were recorded in other WZ Sge-type dwarf novae
(e.g. GW Lib: figure 33 in \cite{Pdot}; FL Psc =
ASAS J002511$+$1217.2: figure 34 in \cite{Pdot};
V355 UMa: figure 43 in \cite{Pdot3}).

\begin{figure}
  \begin{center}
%    \FigureFile(85mm,100mm){asassn16bhhumpall.eps}
    \FigureFile(85mm,100mm){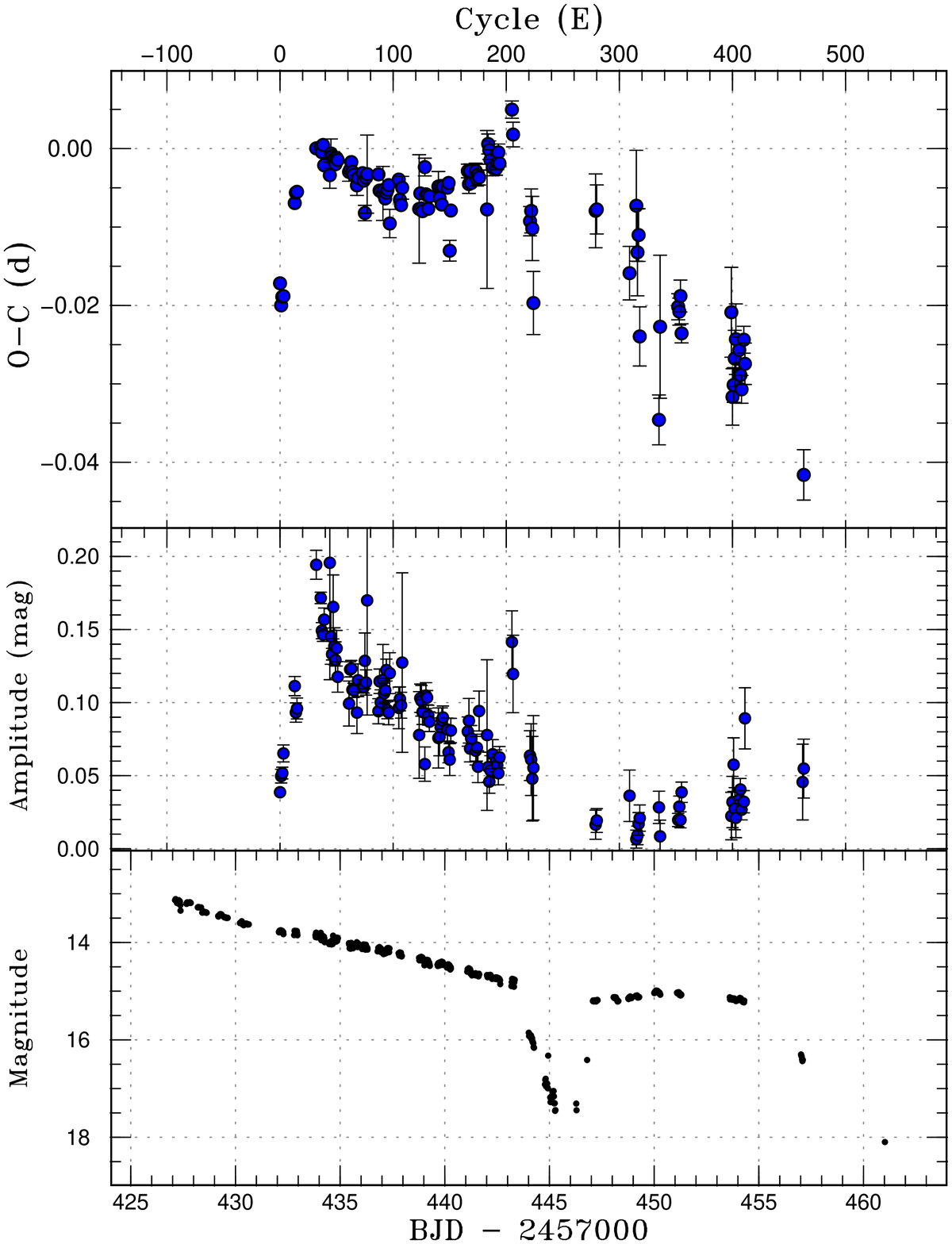}
  \end{center}
  \caption{$O-C$ diagram of superhumps in ASASSN-16bh (2016).
     (Upper:) $O-C$ diagram.
     We used a period of 0.05403~d for calculating the $O-C$ residuals.
     (Middle:) Amplitudes of superhumps.
     (Lower:) Light curve.  The data were binned to 0.017~d.
  }
  \label{fig:asassn16bhhumpall}
\end{figure}

\begin{figure}
  \begin{center}
%    \FigureFile(85mm,110mm){asassn16bhshrebpdm.eps}
    \FigureFile(85mm,110mm){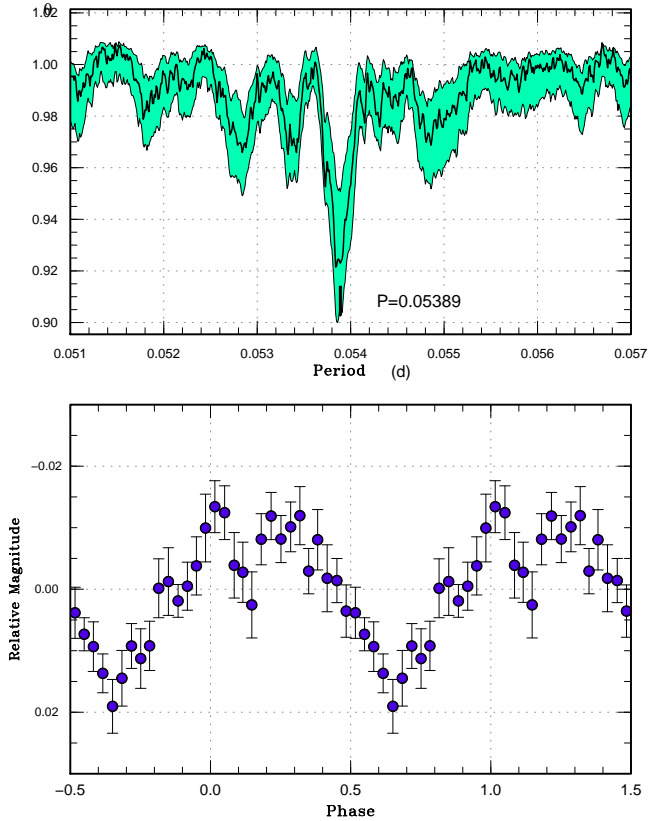}
  \end{center}
  \caption{Superhumps in ASASSN-16bh in the rebrightening phase (2016).
     (Upper): PDM analysis.
     (Lower): Phase-averaged profile.}
  \label{fig:asassn16bhshrebpdm}
\end{figure}

   The object rapidly faded from the superoutburst
plateau on February 25. This fading was actually
a ``dip'' seen in many WZ Sge-type dwarf novae
\citep{kat15wzsge} (see figure \ref{fig:asassn16bhhumpall}).
The object brightened again
on February 28 (vsnet-alert 19536) and this rebrightening
was a plateau-type one without major fluctuations
(vsnet-alert 19567).  On March 9, the object rapidly
faded from the rebrightening phase (vsnet-alert 19569).
During the rebrightening phase, superhumps were present
and grew in amplitudes (figure \ref{fig:asassn16bhhumpall}).
The times of superhump maxima during the rebrightening
phase are listed in table \ref{tab:asassn16bhoc2016reb},
although the data were rather noisy due to the faintness.
The superhumps, however, were very apparent
on March 6 (BJD 2457454), the final night before
the rapid fading.  The mean period of the superhumps
during the rebrightening phase was determined to be
0.05389(3)~d with the PDM method
(figure \ref{fig:asassn16bhshrebpdm}).

   The resultant $\epsilon^*$ for stage A superhumps
was 0.0283(3), corresponding to $q$=0.076(1).
Using the relation between $P_{\rm dot}$ and $q$
in equation (6) in \citet{kat15wzsge}, we can obtain
$q$=0.076(6), consistent with that from stage A
superhumps.

   It took 7~d for this object from the outburst detection
to the emergence of ordinary superhumps.  This value is
relatively short among WZ Sge-type dwarf novae,
particularly among objects with type-A rebrightenings
[see figure 18 in \citet{kat15wzsge}].
Since the gap in the ASAS-SN data before the outburst
detection was very short, this delay of appearance of
superhumps should not exceed 9~d.

   We can estimate the disk radius during the rebrightening
using the superhump period as shown in subsection 4.3
in \citet{kat13qfromstageA}.
The mean superhump period during the rebrightening phase
gives $\epsilon^*$ of 0.0080(7).  This value corresponds
to a radius of 0.26(2)$a$, where $a$ is the binary
separation, if we can ignore the pressure effect.
This value is small compared to the values (0.30$a$--0.38$a$)
in post-outburst state of WZ Sge-type dwarf novae
\citep{kat13qfromstageA}.  Although pressure effect
may have reduced $\epsilon^*$ and give a systematically
small disk radius, it is likely that the disk radius
during the plateau-type rebrightening was indeed small.

   The object resembles AL Com (cf. \cite{pat96alcom};
\cite{how96alcom}; \cite{nog97alcom}; \cite{kat96alcom};
\cite{ish02wzsgeletter}) in many respects.
The apparent brightness may even quality this object
to be the southern counterpart of AL Com.
Since the superoutbursts in AL Com appear to have some
degree of diversity (cf. \cite{uem08alcom}; \cite{kim16alcom}),
further monitoring for outbursts and time-resolved
photometry during each superoutburst will be rewarding.
Photometry and spectroscopy in quiescence to determine
the exact orbital period are also desired.

% SI

% SI

% SI

% SI

\begin{table*}
\caption{Superhump maxima of ASASSN-16bh (2016)}\label{tab:asassn16bhoc2016}
\begin{center}
\begin{tabular}{rp{55pt}p{40pt}r@{.}lrrp{55pt}p{40pt}r@{.}lr}
\hline
\multicolumn{1}{c}{$E$} & \multicolumn{1}{c}{max\commenta} & \multicolumn{1}{c}{error} & \multicolumn{2}{c}{$O-C$\commentb} & \multicolumn{1}{c}{$N$\commentc} & \multicolumn{1}{c}{$E$} & \multicolumn{1}{c}{max\commenta} & \multicolumn{1}{c}{error} & \multicolumn{2}{c}{$O-C$\commentb} & \multicolumn{1}{c}{$N$\commentc} \\
\hline
0 & 57432.1129 & 0.0007 & $-$0&0112 & 33 & 107 & 57437.9040 & 0.0006 & $-$0&0022 & 53 \\
1 & 57432.1641 & 0.0006 & $-$0&0140 & 34 & 108 & 57437.9603 & 0.0015 & $-$0&0000 & 18 \\
2 & 57432.2192 & 0.0005 & $-$0&0130 & 34 & 123 & 57438.7680 & 0.0069 & $-$0&0028 & 28 \\
3 & 57432.2733 & 0.0006 & $-$0&0129 & 31 & 124 & 57438.8241 & 0.0006 & $-$0&0009 & 54 \\
13 & 57432.8255 & 0.0004 & $-$0&0011 & 54 & 125 & 57438.8762 & 0.0007 & $-$0&0028 & 53 \\
14 & 57432.8808 & 0.0005 & 0&0002 & 54 & 126 & 57438.9298 & 0.0006 & $-$0&0031 & 49 \\
15 & 57432.9350 & 0.0005 & 0&0003 & 54 & 128 & 57439.0435 & 0.0011 & 0&0025 & 30 \\
32 & 57433.8590 & 0.0003 & 0&0057 & 24 & 129 & 57439.0941 & 0.0006 & $-$0&0010 & 34 \\
36 & 57434.0753 & 0.0003 & 0&0058 & 19 & 130 & 57439.1480 & 0.0003 & $-$0&0011 & 34 \\
37 & 57434.1287 & 0.0003 & 0&0052 & 33 & 131 & 57439.2003 & 0.0005 & $-$0&0029 & 34 \\
38 & 57434.1836 & 0.0002 & 0&0061 & 33 & 132 & 57439.2559 & 0.0005 & $-$0&0013 & 34 \\
39 & 57434.2351 & 0.0004 & 0&0035 & 28 & 140 & 57439.6894 & 0.0019 & $-$0&0001 & 25 \\
44 & 57434.5040 & 0.0017 & 0&0022 & 28 & 141 & 57439.7420 & 0.0011 & $-$0&0016 & 25 \\
45 & 57434.5608 & 0.0019 & 0&0049 & 52 & 142 & 57439.7975 & 0.0008 & $-$0&0001 & 79 \\
46 & 57434.6145 & 0.0002 & 0&0046 & 120 & 143 & 57439.8492 & 0.0007 & $-$0&0025 & 68 \\
47 & 57434.6680 & 0.0010 & 0&0041 & 11 & 144 & 57439.9055 & 0.0006 & $-$0&0002 & 66 \\
48 & 57434.7223 & 0.0006 & 0&0043 & 17 & 148 & 57440.1215 & 0.0006 & $-$0&0004 & 34 \\
49 & 57434.7755 & 0.0007 & 0&0035 & 17 & 149 & 57440.1761 & 0.0008 & 0&0002 & 34 \\
50 & 57434.8303 & 0.0006 & 0&0043 & 26 & 150 & 57440.2215 & 0.0013 & $-$0&0084 & 29 \\
51 & 57434.8841 & 0.0006 & 0&0040 & 22 & 151 & 57440.2807 & 0.0007 & $-$0&0033 & 28 \\
61 & 57435.4229 & 0.0012 & 0&0024 & 36 & 166 & 57441.0962 & 0.0009 & 0&0016 & 34 \\
62 & 57435.4772 & 0.0004 & 0&0027 & 77 & 167 & 57441.1485 & 0.0012 & $-$0&0001 & 31 \\
63 & 57435.5322 & 0.0003 & 0&0037 & 92 & 168 & 57441.2042 & 0.0009 & 0&0016 & 31 \\
64 & 57435.5850 & 0.0002 & 0&0024 & 124 & 169 & 57441.2567 & 0.0008 & 0&0001 & 34 \\
65 & 57435.6386 & 0.0006 & 0&0020 & 66 & 173 & 57441.4744 & 0.0010 & 0&0016 & 125 \\
68 & 57435.7994 & 0.0013 & 0&0007 & 40 & 174 & 57441.5274 & 0.0009 & 0&0005 & 125 \\
69 & 57435.8544 & 0.0007 & 0&0016 & 64 & 175 & 57441.5820 & 0.0013 & 0&0011 & 124 \\
73 & 57436.0711 & 0.0003 & 0&0022 & 33 & 176 & 57441.6356 & 0.0010 & 0&0007 & 93 \\
74 & 57436.1242 & 0.0002 & 0&0012 & 33 & 183 & 57442.0098 & 0.0101 & $-$0&0034 & 15 \\
75 & 57436.1741 & 0.0010 & $-$0&0029 & 67 & 184 & 57442.0721 & 0.0013 & 0&0049 & 34 \\
76 & 57436.2326 & 0.0006 & 0&0016 & 108 & 185 & 57442.1254 & 0.0012 & 0&0041 & 34 \\
77 & 57436.2871 & 0.0050 & 0&0020 & 29 & 186 & 57442.1782 & 0.0013 & 0&0028 & 34 \\
87 & 57436.8274 & 0.0006 & 0&0019 & 54 & 188 & 57442.2853 & 0.0010 & 0&0019 & 25 \\
88 & 57436.8793 & 0.0005 & $-$0&0002 & 54 & 191 & 57442.4472 & 0.0008 & 0&0017 & 124 \\
89 & 57436.9333 & 0.0006 & $-$0&0003 & 52 & 192 & 57442.5020 & 0.0009 & 0&0025 & 121 \\
91 & 57437.0411 & 0.0034 & $-$0&0006 & 16 & 193 & 57442.5574 & 0.0011 & 0&0038 & 123 \\
92 & 57437.0956 & 0.0004 & $-$0&0001 & 33 & 194 & 57442.6100 & 0.0008 & 0&0023 & 124 \\
93 & 57437.1485 & 0.0005 & $-$0&0012 & 34 & 205 & 57443.2112 & 0.0011 & 0&0091 & 52 \\
94 & 57437.2032 & 0.0003 & $-$0&0005 & 120 & 206 & 57443.2620 & 0.0016 & 0&0059 & 52 \\
95 & 57437.2576 & 0.0004 & $-$0&0001 & 103 & 221 & 57444.0614 & 0.0019 & $-$0&0053 & 34 \\
96 & 57437.3123 & 0.0007 & 0&0004 & 46 & 222 & 57444.1167 & 0.0028 & $-$0&0040 & 34 \\
97 & 57437.3614 & 0.0018 & $-$0&0045 & 17 & 223 & 57444.1685 & 0.0041 & $-$0&0062 & 34 \\
105 & 57437.7992 & 0.0008 & 0&0011 & 43 & 224 & 57444.2131 & 0.0040 & $-$0&0157 & 28 \\
106 & 57437.8507 & 0.0005 & $-$0&0015 & 52 & \multicolumn{1}{c}{--} & \multicolumn{1}{c}{--} & \multicolumn{1}{c}{--} & \multicolumn{2}{c}{--} & \multicolumn{1}{c}{--}\\
\hline
  \multicolumn{12}{l}{\commenta BJD$-$2400000.} \\
  \multicolumn{12}{l}{\commentb Against max $= 2457432.1241 + 0.054039 E$.} \\
  \multicolumn{12}{l}{\commentc Number of points used to determine the maximum.} \\
\end{tabular}
\end{center}
\end{table*}

% SI

\begin{table}
\caption{Superhump maxima of ASASSN-16bh in the rebrightening phase (2016)}\label{tab:asassn16bhoc2016reb}
\begin{center}
\begin{tabular}{rp{55pt}p{40pt}r@{.}lr}
\hline
\multicolumn{1}{c}{$E$} & \multicolumn{1}{c}{max\commenta} & \multicolumn{1}{c}{error} & \multicolumn{2}{c}{$O-C$\commentb} & \multicolumn{1}{c}{$N$\commentc} \\
\hline
0 & 57447.1965 & 0.0047 & 0&0000 & 22 \\
1 & 57447.2507 & 0.0031 & 0&0004 & 26 \\
30 & 57448.8094 & 0.0034 & $-$0&0025 & 59 \\
36 & 57449.1422 & 0.0071 & 0&0072 & 26 \\
37 & 57449.1903 & 0.0056 & 0&0015 & 22 \\
38 & 57449.2465 & 0.0034 & 0&0038 & 26 \\
39 & 57449.2876 & 0.0038 & $-$0&0089 & 18 \\
56 & 57450.1955 & 0.0032 & $-$0&0165 & 21 \\
57 & 57450.2614 & 0.0091 & $-$0&0044 & 26 \\
73 & 57451.1284 & 0.0017 & 0&0011 & 26 \\
74 & 57451.1818 & 0.0017 & 0&0006 & 22 \\
75 & 57451.2379 & 0.0020 & 0&0028 & 26 \\
76 & 57451.2871 & 0.0012 & $-$0&0018 & 22 \\
120 & 57453.6671 & 0.0057 & 0&0089 & 22 \\
121 & 57453.7104 & 0.0036 & $-$0&0017 & 22 \\
122 & 57453.7659 & 0.0022 & $-$0&0000 & 22 \\
123 & 57453.8233 & 0.0036 & 0&0035 & 28 \\
124 & 57453.8798 & 0.0045 & 0&0062 & 36 \\
127 & 57454.0405 & 0.0016 & 0&0053 & 27 \\
128 & 57454.0914 & 0.0013 & 0&0023 & 26 \\
129 & 57454.1436 & 0.0018 & 0&0007 & 25 \\
131 & 57454.2580 & 0.0017 & 0&0074 & 25 \\
132 & 57454.3089 & 0.0026 & 0&0045 & 11 \\
183 & 57457.0306 & 0.0045 & $-$0&0201 & 26 \\
184 & 57457.1043 & 0.0032 & $-$0&0002 & 15 \\
\hline
  \multicolumn{6}{l}{\commenta BJD$-$2400000.} \\
  \multicolumn{6}{l}{\commentb Against max $= 2457447.1964 + 0.053848 E$.} \\
  \multicolumn{6}{l}{\commentc Number of points used to determine the maximum.} \\
\end{tabular}
\end{center}
\end{table}

\subsection{ASASSN-16bi}\label{obj:asassn16bi}

\begin{figure}
  \begin{center}
%    \FigureFile(85mm,110mm){asassn16bieshpdm.eps}
    \FigureFile(85mm,110mm){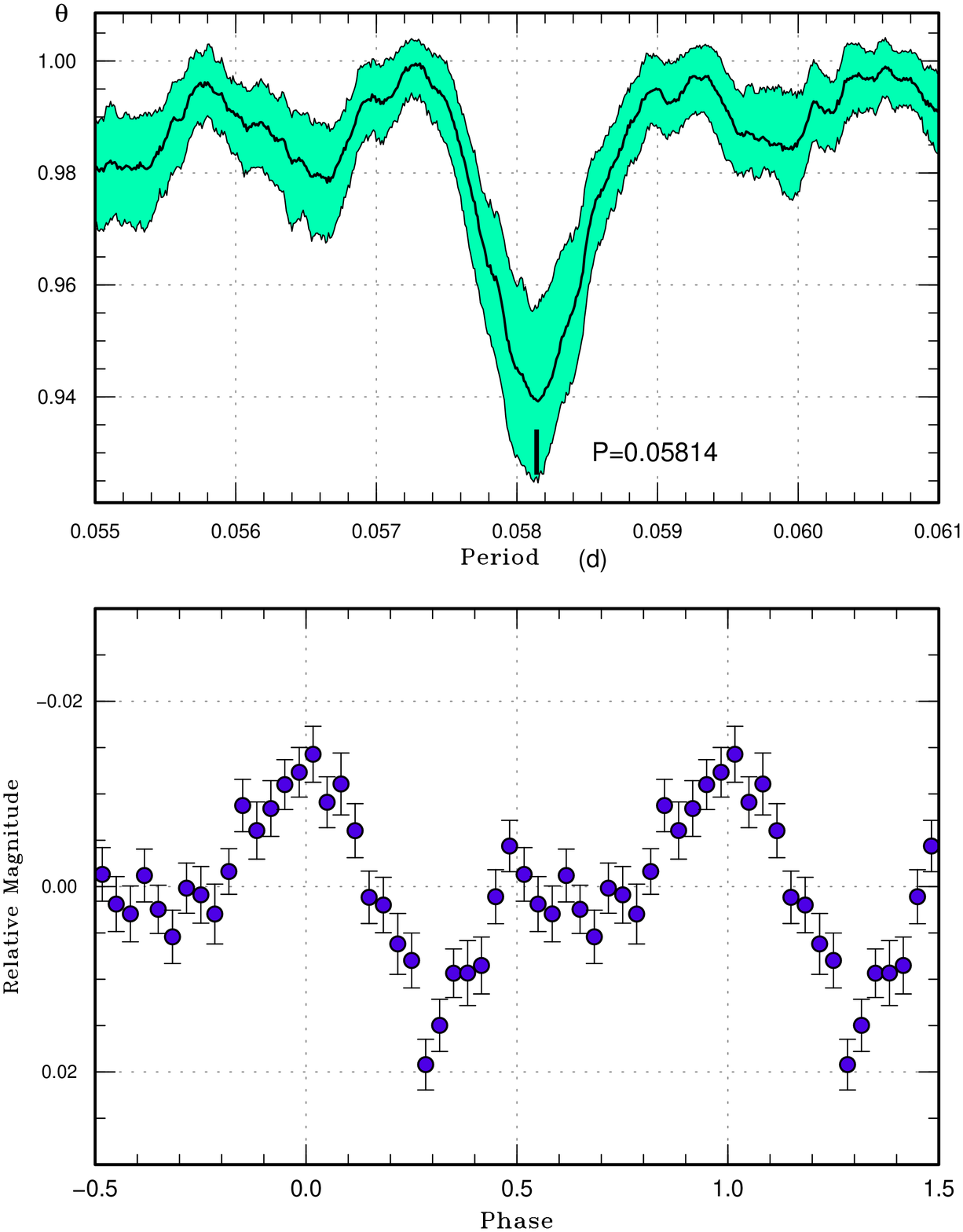}
  \end{center}
  \caption{Early superhumps in ASASSN-16bi (2016).
     The data before BJD 2457432 were used.
     (Upper): PDM analysis.
     (Lower): Phase-averaged profile.}
  \label{fig:asassn16bieshpdm}
\end{figure}

\begin{figure}
  \begin{center}
%    \FigureFile(85mm,110mm){asassn16bilc.eps}
    \FigureFile(85mm,110mm){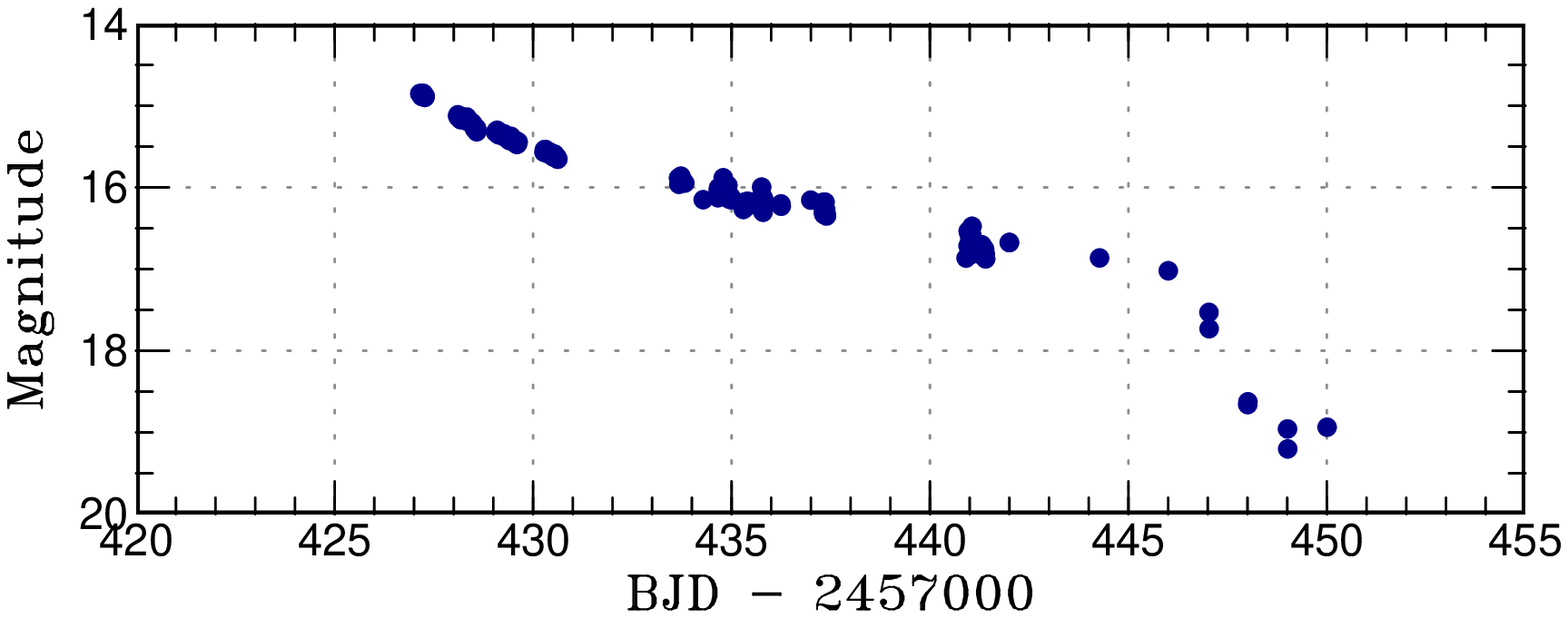}
  \end{center}
  \caption{Light curve of the superoutburst of ASASSN-16bi
     (2016).  The data were binned to 0.01~d.
     The ASAS-SN detection corresponds to BJD 2457424.53.}
  \label{fig:asassn16bilc}
\end{figure}

   This object was detected as a transient
on the rise from $V$=15.2 to $V$=14.3 on 2016 February 6
by the ASAS-SN team.  The object was also detected
by Gaia as Gaia16ads on February 13.\footnote{
  $<$http://gsaweb.ast.cam.ac.uk/alerts/alert/Gaia16ads/$>$.
}
Although an $R$=20.6 star was initially suggested to be
the quiescent counterpart, it is \timeform{6''} from
the Gaia position and is unlikely the counterpart.
The large outburst amplitude suggested
a possible WZ Sge-type dwarf nova.
Early observations detected double-wave early
superhumps (vsnet-alert 19514; figure \ref{fig:asassn16bieshpdm}).
The best period with the PDM method is 0.05814(5)~d.
Although emerging ordinary superhumps were detected
(vsnet-alert 19519), the period was not well determined
due to the faintness of the object.
The only available superhumps maxima are BJD 2457437.2963(33)
$N$=39 and 2457437.3571(8) $N$=61.
The outburst light curve is shown in figure \ref{fig:asassn16bilc}.
The phase of early superhumps lasted at least for 6~d
if it immediately started after the outburst detection.
The entire duration of the superoutburst was 23~d.

% SI

\subsection{ASASSN-16bu}\label{obj:asassn16bu}

\begin{figure}
  \begin{center}
%    \FigureFile(85mm,110mm){asassn16bushpdm.eps}
    \FigureFile(85mm,110mm){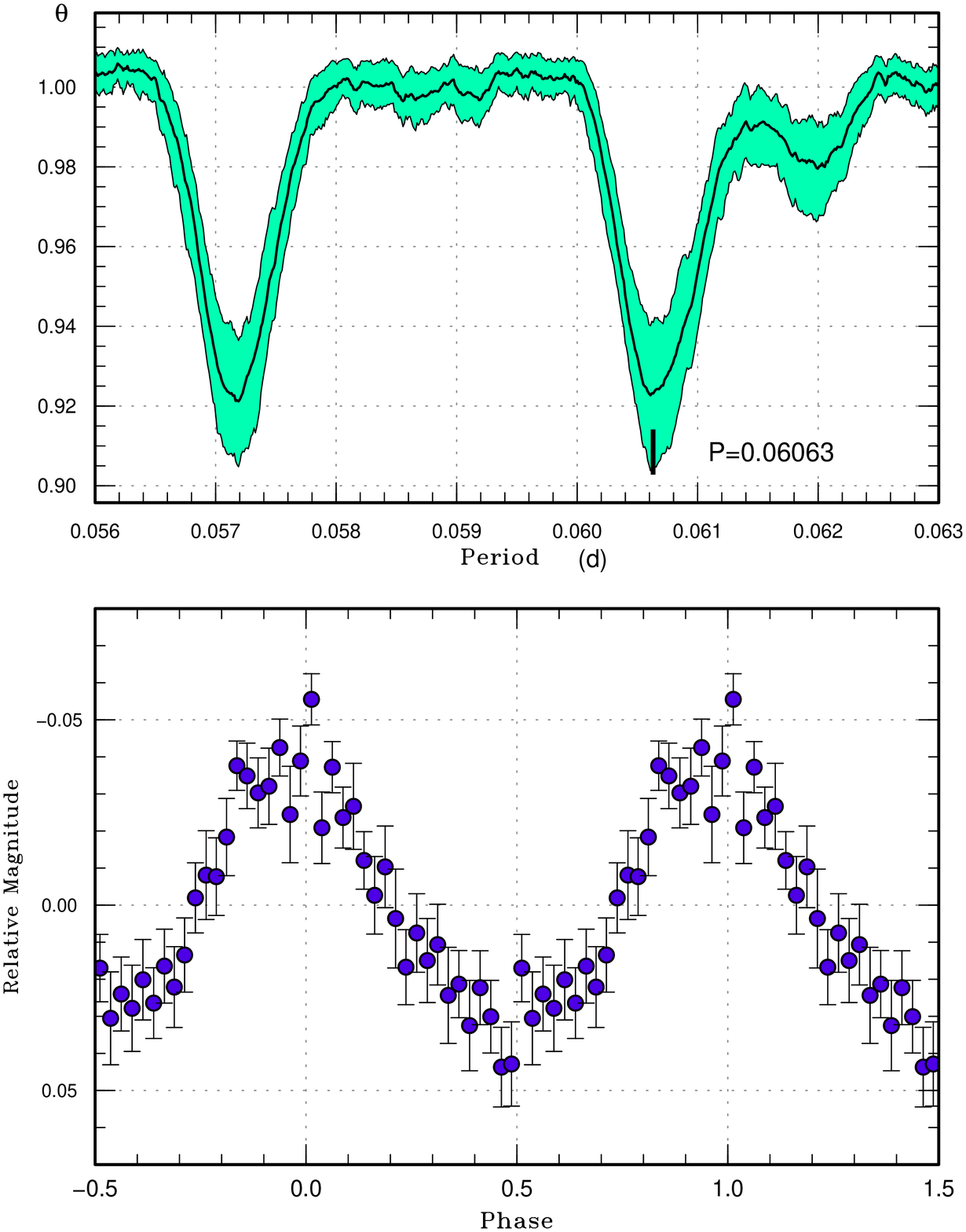}
  \end{center}
  \caption{Superhumps in ASASSN-16bu (2016).
     The data between BJD 2457441 and 2457450 were used.
     (Upper): PDM analysis.  The true period was selected
     by $O-C$ analysis.
     (Lower): Phase-averaged profile.}
  \label{fig:asassn16bushpdm}
\end{figure}

\begin{figure}
  \begin{center}
%    \FigureFile(85mm,110mm){asassn16bueshpdm.eps}
    \FigureFile(85mm,110mm){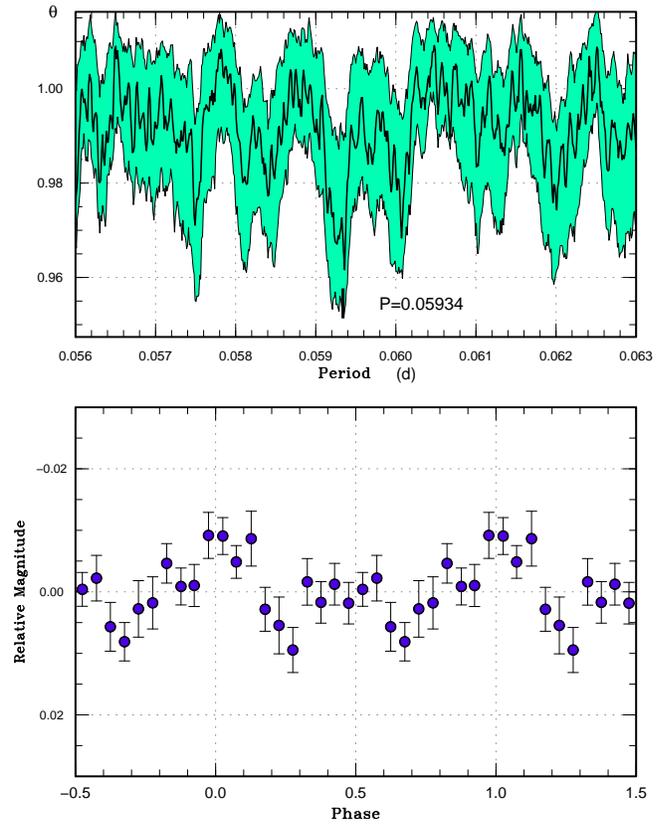}
  \end{center}
  \caption{Possible early superhumps in ASASSN-16bu (2016).
     The high-quality data before BJD 2457440 were used.
     (Upper): PDM analysis.
     (Lower): Phase-averaged profile.}
  \label{fig:asassn16bueshpdm}
\end{figure}

   This object was detected as a transient
at $V$=14.5 on 2016 February 15 by the ASAS-SN team
(vsnet-alert 19491).
The large outburst amplitude suggested
a possible WZ Sge-type dwarf nova (cf. vsnet-alert 19500).
After nine nights, ordinary superhumps emerged
(vsnet-alert 19525; figure \ref{fig:asassn16bushpdm}).
The times of superhump maxima are listed in
table \ref{tab:asassn16buoc2016}.
The maxima for $E \le$30 are clearly stage A
superhumps with growing amplitudes.
An analysis of the earlier part of the observation
yielded a weak signal of possible early superhumps
(figure \ref{fig:asassn16bueshpdm}).  The period
with the PDM method was 0.05934(13)~d.
By using this period and the period of stage A
superhumps, the value of $\epsilon^*$=0.037(4).
This value corresponds to $q$=0.10(1).
Since the periods of early superhumps and stage A
superhumps were not very well determined, this $q$
value needs to be treated with caution.
The other features of the behavior, including
the slow growth of ordinary superhumps and
small amplitudes of superhumps, likely suggest
a low $q$ comparable to a period bouncer
\citep{kat15wzsge}.

% SI

% SI

% SI

\begin{table}
\caption{Superhump maxima of ASASSN-16bu (2016)}\label{tab:asassn16buoc2016}
\begin{center}
\begin{tabular}{rp{55pt}p{40pt}r@{.}lr}
\hline
\multicolumn{1}{c}{$E$} & \multicolumn{1}{c}{max\commenta} & \multicolumn{1}{c}{error} & \multicolumn{2}{c}{$O-C$\commentb} & \multicolumn{1}{c}{$N$\commentc} \\
\hline
0 & 57442.3543 & 0.0035 & $-$0&0204 & 61 \\
15 & 57443.2903 & 0.0010 & 0&0023 & 76 \\
16 & 57443.3410 & 0.0008 & $-$0&0079 & 120 \\
17 & 57443.4056 & 0.0011 & $-$0&0041 & 95 \\
18 & 57443.4688 & 0.0016 & $-$0&0019 & 58 \\
27 & 57444.0286 & 0.0009 & 0&0099 & 136 \\
28 & 57444.0843 & 0.0005 & 0&0047 & 190 \\
29 & 57444.1454 & 0.0007 & 0&0049 & 111 \\
30 & 57444.2070 & 0.0010 & 0&0057 & 63 \\
42 & 57444.9445 & 0.0011 & 0&0124 & 65 \\
43 & 57444.9985 & 0.0005 & 0&0056 & 67 \\
44 & 57445.0632 & 0.0017 & 0&0094 & 125 \\
45 & 57445.1138 & 0.0014 & $-$0&0009 & 95 \\
46 & 57445.1781 & 0.0016 & 0&0025 & 60 \\
47 & 57445.2401 & 0.0018 & 0&0036 & 59 \\
59 & 57445.9678 & 0.0018 & 0&0007 & 59 \\
60 & 57446.0288 & 0.0015 & 0&0008 & 59 \\
61 & 57446.0879 & 0.0022 & $-$0&0010 & 60 \\
62 & 57446.1501 & 0.0015 & 0&0003 & 58 \\
63 & 57446.2038 & 0.0031 & $-$0&0069 & 34 \\
76 & 57446.9945 & 0.0009 & $-$0&0078 & 65 \\
77 & 57447.0606 & 0.0009 & $-$0&0026 & 66 \\
82 & 57447.3585 & 0.0006 & $-$0&0091 & 57 \\
\hline
  \multicolumn{6}{l}{\commenta BJD$-$2400000.} \\
  \multicolumn{6}{l}{\commentb Against max $= 2457442.3747 + 0.060889 E$.} \\
  \multicolumn{6}{l}{\commentc Number of points used to determine the maximum.} \\
\end{tabular}
\end{center}
\end{table}

\subsection{ASASSN-16de}\label{obj:asassn16de}

\begin{figure}
  \begin{center}
%    \FigureFile(85mm,70mm){asassn16deshlc.eps}
    \FigureFile(85mm,70mm){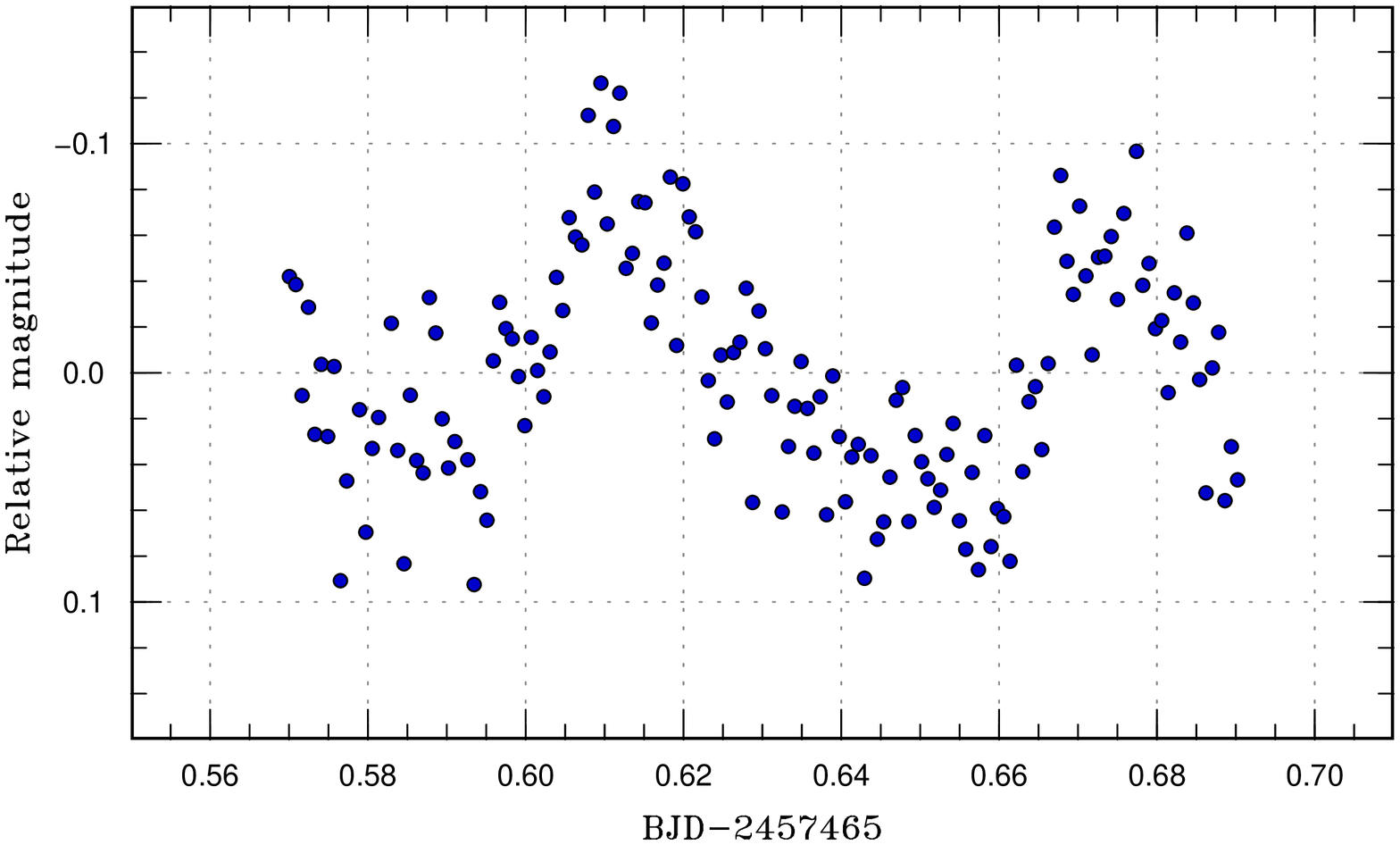}
  \end{center}
  \caption{Superhumps in ASASSN-16de (2016).
  }
  \label{fig:asassn16deshlc}
\end{figure}

   This object was detected as a transient
at $V$=14.5 on 2016 March 17 by the ASAS-SN team
(vsnet-alert 19491).
The object was probably
in outburst in USNO-A2.0 at $B$=16.1.  A blue object
is present in SDSS images, but it is not listed in \cite{SDSS7}.
Subsequent observations detected superhumps
(vsnet-alert 19610; figure \ref{fig:asassn16deshlc}).
This observation on the first night gave a period
of 0.063(1)~d (PDM method).
The times of superhump maxima were BJD 2457465.6155(8) ($N$=60)
and 2457465.6774(7) ($N$=46).
Although the object was observed on five nights
in the late stage of the superoutburst
(6~d after the initial observation of superhumps),
the superhump signal was too weak to determine
the period.  The object faded rapidly on March 28,
11~d after the outburst detection.

\subsection{CRTS J081936.1$+$191540}\label{obj:j0819}

   This object (hereafter CRTS J081936) was confirmed
to be an SU UMa-type dwarf nova on its outburst in
2013 \citep{Pdot7}.  Another superoutburst was detected
on 2015 March 10 by the ASAS-SN team (vsnet-alert 18411).
Two superhump maxima were measured:
BJD 2457093.4786(13) ($N$=85) and 2457093.5492(18) ($N$=79).

\subsection{CRTS J095926.4$-$160147}\label{obj:j0959}

   This object (=CSS110226:095926$-$160147, hereafter CRTS J095926)
was detected as a transient by CRTS on 2011 February 26.
Five outbursts were recorded in the CRTS data
up to 2013 July.

\begin{figure}
  \begin{center}
%    \FigureFile(85mm,110mm){j0959shpdm.eps}
    \FigureFile(85mm,110mm){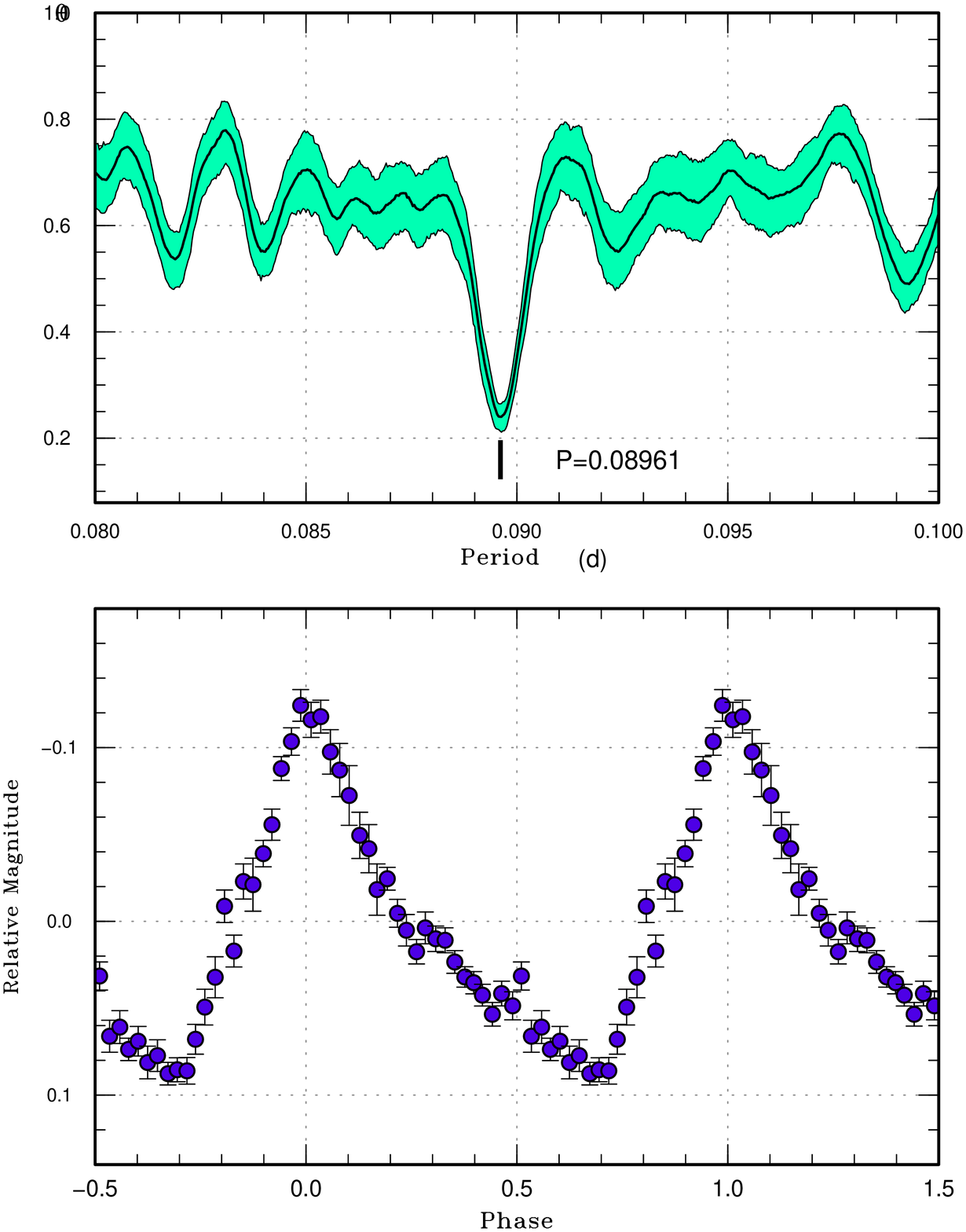}
  \end{center}
  \caption{Superhumps in CRTS J095926 (2015).
     (Upper): PDM analysis.
     (Lower): Phase-averaged profile.}
  \label{fig:j0959shpdm}
\end{figure}

   The 2015 outburst was detected on May 14 by the ASAS-SN
team (vsnet-alert 18622).
Our observations starting after two nights detected
superhumps (vsnet-alert 18628, 18633, 18648;
figure \ref{fig:j0959shpdm}).
The times of superhump maxima are listed in
table \ref{tab:j0959oc2015}.  Although it was initially
difficult to distinguish one-day aliases due to the short
observing time, observations at two different longitudes
established the alias selection (vsnet-alert 18648).
It is possible that maxima for $E \le$12 correspond to
stage A superhumps since there was a large decrease
in the period (by 1.5\%) around $E$=11.  If it is the case,
stage A in this system likely lasted more than 3~d.

% SI

% SI

\begin{table}
\caption{Superhump maxima of CRTS J095926 (2015)}\label{tab:j0959oc2015}
\begin{center}
\begin{tabular}{rp{55pt}p{40pt}r@{.}lr}
\hline
\multicolumn{1}{c}{$E$} & \multicolumn{1}{c}{max\commenta} & \multicolumn{1}{c}{error} & \multicolumn{2}{c}{$O-C$\commentb} & \multicolumn{1}{c}{$N$\commentc} \\
\hline
0 & 57159.4812 & 0.0099 & $-$0&0144 & 10 \\
1 & 57159.5829 & 0.0011 & $-$0&0024 & 11 \\
11 & 57160.4845 & 0.0011 & 0&0028 & 18 \\
12 & 57160.5800 & 0.0007 & 0&0087 & 12 \\
23 & 57161.5658 & 0.0014 & 0&0084 & 11 \\
34 & 57162.5467 & 0.0008 & 0&0032 & 16 \\
41 & 57163.1710 & 0.0041 & 0&0001 & 29 \\
42 & 57163.2616 & 0.0003 & 0&0010 & 207 \\
43 & 57163.3509 & 0.0009 & 0&0006 & 103 \\
45 & 57163.5295 & 0.0006 & $-$0&0000 & 20 \\
56 & 57164.5126 & 0.0010 & $-$0&0030 & 18 \\
67 & 57165.4966 & 0.0021 & $-$0&0050 & 25 \\
\hline
  \multicolumn{6}{l}{\commenta BJD$-$2400000.} \\
  \multicolumn{6}{l}{\commentb Against max $= 2457159.4956 + 0.089643 E$.} \\
  \multicolumn{6}{l}{\commentc Number of points used to determine the maximum.} \\
\end{tabular}
\end{center}
\end{table}

\subsection{CRTS J120052.9$-$152620}\label{obj:j1200}

   This object (=CSS110205:120053$-$152620, hereafter
CRTS J120052) was discovered by the CRTS on 2011 February 5. 
The 2011 superoutburst was observed only on two nights
and the analysis was reported in \citet{Pdot3}.
Due to the lack of observations, there remained
ambiguity in selecting the superhump period \citep{Pdot3}.

\begin{figure}
  \begin{center}
%    \FigureFile(85mm,110mm){j1200shpdm.eps}
    \FigureFile(85mm,110mm){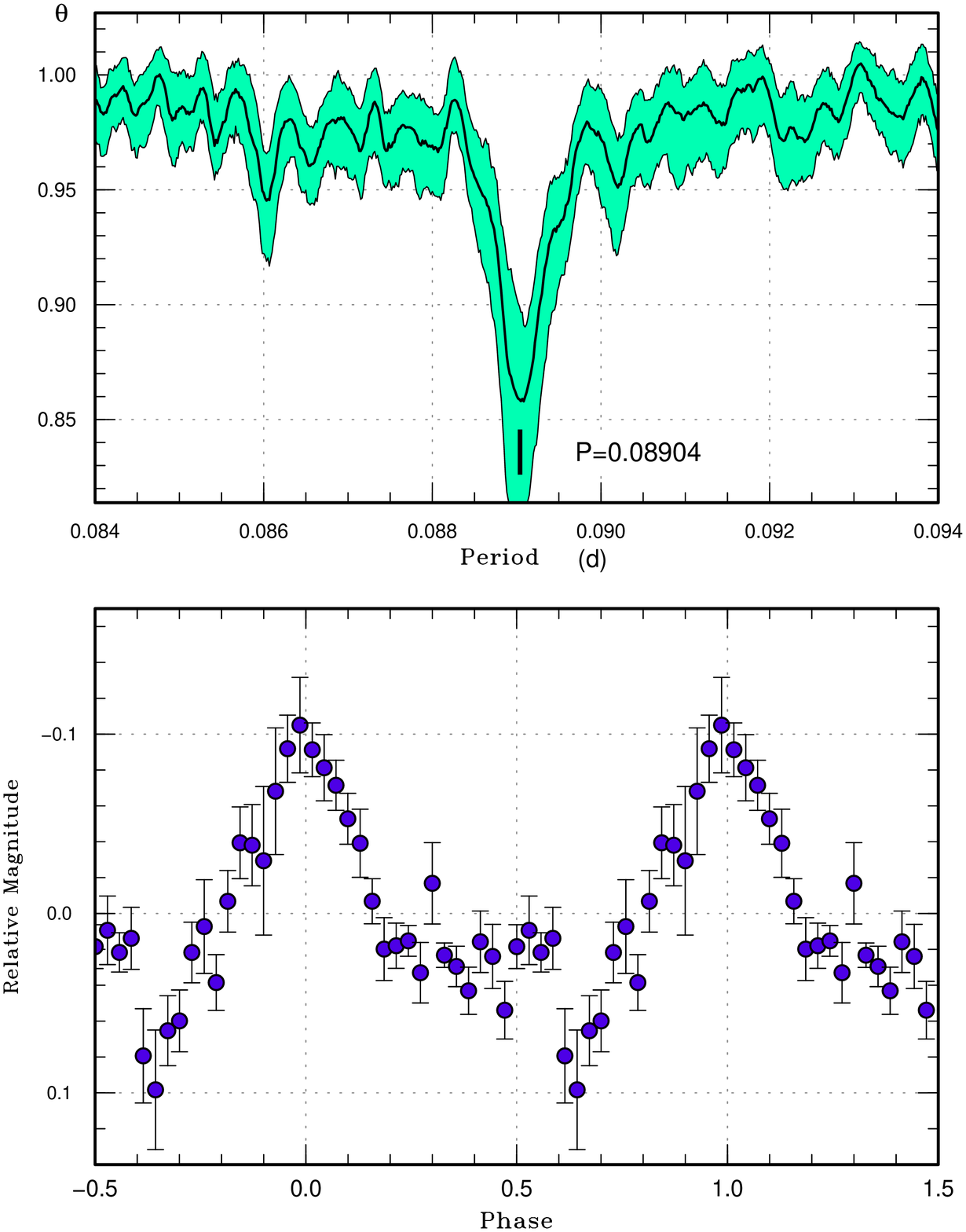}
  \end{center}
  \caption{Superhumps in CRTS J120052 (2016).
     (Upper): PDM analysis.
     (Lower): Phase-averaged profile.}
  \label{fig:j1200shpdm}
\end{figure}

   The 2016 superoutburst was detected by the ASAS-SN team
at $V$=13.78 on March 14 (cf. vsnet-alert 19590).
Subsequent observations detected superhumps
(vsnet-alert 19609, 19618; figure \ref{fig:j1200shpdm}).
The times of superhump maxima are listed in
table \ref{tab:j1200oc2016}.  These observations
recorded the relatively late phase of the superoutburst
and these superhumps may be of stage C.
In table \ref{tab:perlist}, we gave a mean period.

   Thanks to the new observation, it has become evident
that an alias different from that reported in
\citet{Pdot3} was the correct superhump period
during the 2011 superoutburst (vsnet-alert 19618).
The corrected period for the 2011
superoutburst based on this identification is 0.08882(3)~d.

% SI

% SI

\begin{table}
\caption{Superhump maxima of CRTS J120052 (2016)}\label{tab:j1200oc2016}
\begin{center}
\begin{tabular}{rp{55pt}p{40pt}r@{.}lr}
\hline
\multicolumn{1}{c}{$E$} & \multicolumn{1}{c}{max\commenta} & \multicolumn{1}{c}{error} & \multicolumn{2}{c}{$O-C$\commentb} & \multicolumn{1}{c}{$N$\commentc} \\
\hline
0 & 57464.1267 & 0.0004 & $-$0&0030 & 153 \\
1 & 57464.2172 & 0.0003 & $-$0&0015 & 190 \\
12 & 57465.1966 & 0.0006 & $-$0&0005 & 84 \\
28 & 57466.6205 & 0.0008 & 0&0002 & 31 \\
29 & 57466.7118 & 0.0006 & 0&0025 & 39 \\
40 & 57467.6882 & 0.0010 & 0&0005 & 40 \\
41 & 57467.7781 & 0.0031 & 0&0014 & 8 \\
51 & 57468.6679 & 0.0027 & 0&0017 & 42 \\
52 & 57468.7623 & 0.0045 & 0&0072 & 19 \\
62 & 57469.6489 & 0.0022 & 0&0043 & 42 \\
63 & 57469.7294 & 0.0056 & $-$0&0042 & 35 \\
73 & 57470.6199 & 0.0032 & $-$0&0031 & 38 \\
74 & 57470.7053 & 0.0047 & $-$0&0068 & 35 \\
85 & 57471.6919 & 0.0025 & 0&0014 & 35 \\
\hline
  \multicolumn{6}{l}{\commenta BJD$-$2400000.} \\
  \multicolumn{6}{l}{\commentb Against max $= 2457464.1298 + 0.088950 E$.} \\
  \multicolumn{6}{l}{\commentc Number of points used to determine the maximum.} \\
\end{tabular}
\end{center}
\end{table}

\subsection{CRTS J163120.9$+$103134}\label{obj:j1631}

   This object (=CSS080505:163121$+$103134, hereafter CRTS J163120)
was detected as a transient by CRTS on 2008 May 5.
Refer to \cite{Pdot} for the history.
The 2008 and 2010 superoutbursts were reported in
\citet{Pdot} and \citet{Pdot2}, respectively.

   The 2015 superoutburst was detected by the ASAS-SN
team (cf. vsnet-alert 18603) and superhumps were
observed on single night (table \ref{tab:j1631oc2015}).
The period by the PDM method is listed in table \ref{tab:perlist}.

% SI

\begin{table}
\caption{Superhump maxima of CRTS J163120 (2015)}\label{tab:j1631oc2015}
\begin{center}
\begin{tabular}{rp{55pt}p{40pt}r@{.}lr}
\hline
\multicolumn{1}{c}{$E$} & \multicolumn{1}{c}{max\commenta} & \multicolumn{1}{c}{error} & \multicolumn{2}{c}{$O-C$\commentb} & \multicolumn{1}{c}{$N$\commentc} \\
\hline
0 & 57149.6529 & 0.0023 & 0&0027 & 64 \\
1 & 57149.7092 & 0.0011 & $-$0&0047 & 58 \\
2 & 57149.7790 & 0.0026 & 0&0015 & 63 \\
3 & 57149.8417 & 0.0016 & 0&0006 & 62 \\
\hline
  \multicolumn{6}{l}{\commenta BJD$-$2400000.} \\
  \multicolumn{6}{l}{\commentb Against max $= 2457149.6503 + 0.063621 E$.} \\
  \multicolumn{6}{l}{\commentc Number of points used to determine the maximum.} \\
\end{tabular}
\end{center}
\end{table}

\subsection{CRTS J200331.3$-$284941}\label{obj:j2003}

   This object (=SSS100615:200331$-$284941, hereafter CRTS J200331)
was discovered by CRTS Siding Spring Survey (SSS)
on 2010 June 15.  Since the 2010 outburst had a fading tail
resembling those of WZ Sge-type dwarf novae
(cf. vsnet-alert 18763), the object received special
attention.

   The 2015 outburst, the second known outburst of this
object, was detected by the ASAS-SN team on June 20 at $V$=14.84.
The initial observation revealed that this object
is an eclipsing SU UMa-type (or WZ Sge-type)
dwarf nova (vsnet-alert 18788).  The eclipses became
clearer as the outburst proceeded and
we have obtained the eclipse ephemeris
by using MCMC analysis \citep{Pdot4}
of the present observations:
\begin{equation}
{\rm Min(BJD)} = 2457200.79900(6) + 0.0587048(3) E .
\label{equ:j2003ecl}
\end{equation}
This ephemeris is not intended for long-term prediction
of eclipses.

\begin{figure}
  \begin{center}
%    \FigureFile(85mm,120mm){j2003lc.eps}
    \FigureFile(85mm,120mm){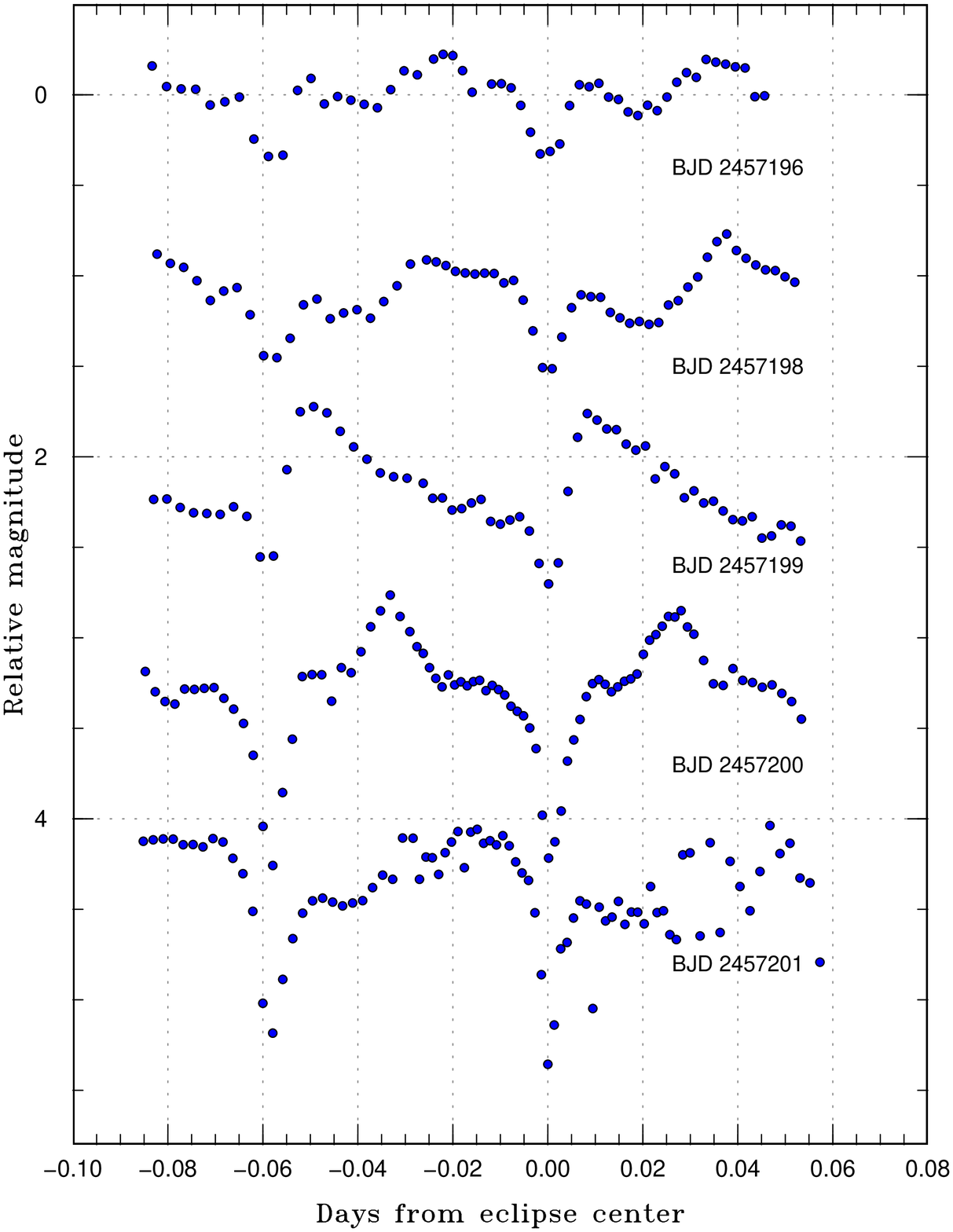}
  \end{center}
  \caption{Light curve of CRTS J200331 on the first five nights.
  Superposition of growing superhumps and eclipses are well visible.}
  \label{fig:j2003lc}
\end{figure}

   The object showed growing superhumps up to June 26
(figure \ref{fig:j2003lc})
and the object significantly brightened after this epoch.
Before June 26 superhumps with a long period of
0.06058(2)~d were observed (note that the period
reported in vsnet-alert 18798 referred to this period).
We identified them to be
stage A superhumps.  On the three subsequent nights,
stable superhumps were recorded, which we identified
to be stage B superhumps.  The times of superhump maxima
are listed in table \ref{tab:j2003oc2015}.
The cycle count between $E$=84 and $E$=250 was ambiguous.
The period of stage A superhumps corresponds to
$q$=0.084(1), which is close to those of WZ Sge-type
dwarf novae \citep{kat15wzsge}.  The duration of stage A
superhumps (more than 50 cycles) is also long, consistent
with the small $q$.
All the pieces of evidence suggest that this object
is located near the borderline of SU UMa-type and WZ Sge-type
objects.  Since our initial observation started 3~d after
the ASAS-SN detection and it was not clear whether there
were early superhumps before this observation.
We can, however, probably rule out the long-lasting
(more than 10~d) phase of early superhumps seen in
typical WZ Sge-type dwarf novae.

   This object is probably the first one in which
the period of stage A superhumps were sufficiently measured in
a deeply eclipsing system.  Determination of system parameters
in quiescence using eclipse modeling in such a system
would provide a direct test for our method od
$q$ determination using stage A superhumps
(cf. \cite{kat13qfromstageA}).

% SI

\begin{table}
\caption{Superhump maxima of CRTS J200331 (2015)}\label{tab:j2003oc2015}
\begin{center}
\begin{tabular}{rp{50pt}p{30pt}r@{.}lcr}
\hline
$E$ & max\commenta & error & \multicolumn{2}{c}{$O-C$\commentb} & phase\commentc & $N$\commentd \\
\hline
0 & 57196.8438 & 0.0011 & $-$0&0335 & 0.62 & 15 \\
1 & 57196.9019 & 0.0009 & $-$0&0349 & 0.62 & 15 \\
33 & 57198.8418 & 0.0008 & $-$0&0028 & 0.66 & 19 \\
34 & 57198.9026 & 0.0007 & $-$0&0016 & 0.70 & 21 \\
49 & 57199.8111 & 0.0004 & 0&0127 & 0.17 & 13 \\
50 & 57199.8717 & 0.0005 & 0&0137 & 0.20 & 18 \\
66 & 57200.8240 & 0.0009 & 0&0121 & 0.43 & 26 \\
67 & 57200.8846 & 0.0007 & 0&0131 & 0.46 & 29 \\
82 & 57201.7783 & 0.0019 & 0&0126 & 0.68 & 11 \\
83 & 57201.8396 & 0.0013 & 0&0143 & 0.73 & 26 \\
84 & 57201.9067 & 0.0060 & 0&0217 & 0.87 & 15 \\
250 & 57211.7723 & 0.0030 & $-$0&0089 & 0.92 & 18 \\
251 & 57211.8339 & 0.0020 & $-$0&0068 & 0.97 & 28 \\
252 & 57211.8889 & 0.0015 & $-$0&0115 & 0.91 & 15 \\
\hline
  \multicolumn{7}{l}{\commenta BJD$-$2400000.} \\
  \multicolumn{7}{l}{\commentb Against max $= 2457196.8773 + 0.059616 E$.} \\
  \multicolumn{7}{l}{\commentc Orbital phase.} \\
  \multicolumn{7}{l}{\commentd Number of points used to determine the maximum.} \\
\end{tabular}
\end{center}
\end{table}

\subsection{CRTS J212521.8$-$102627}\label{obj:j2125}

   This object (=CSS080927:212522$-$102627, hereafter CRTS J212521)
was discovered by CRTS on 2008 September 27.
There were at least 6 outbursts (up to 2013 September)
in the CRTS data.

\begin{figure}
  \begin{center}
%    \FigureFile(85mm,110mm){j2125shpdm.eps}
    \FigureFile(85mm,110mm){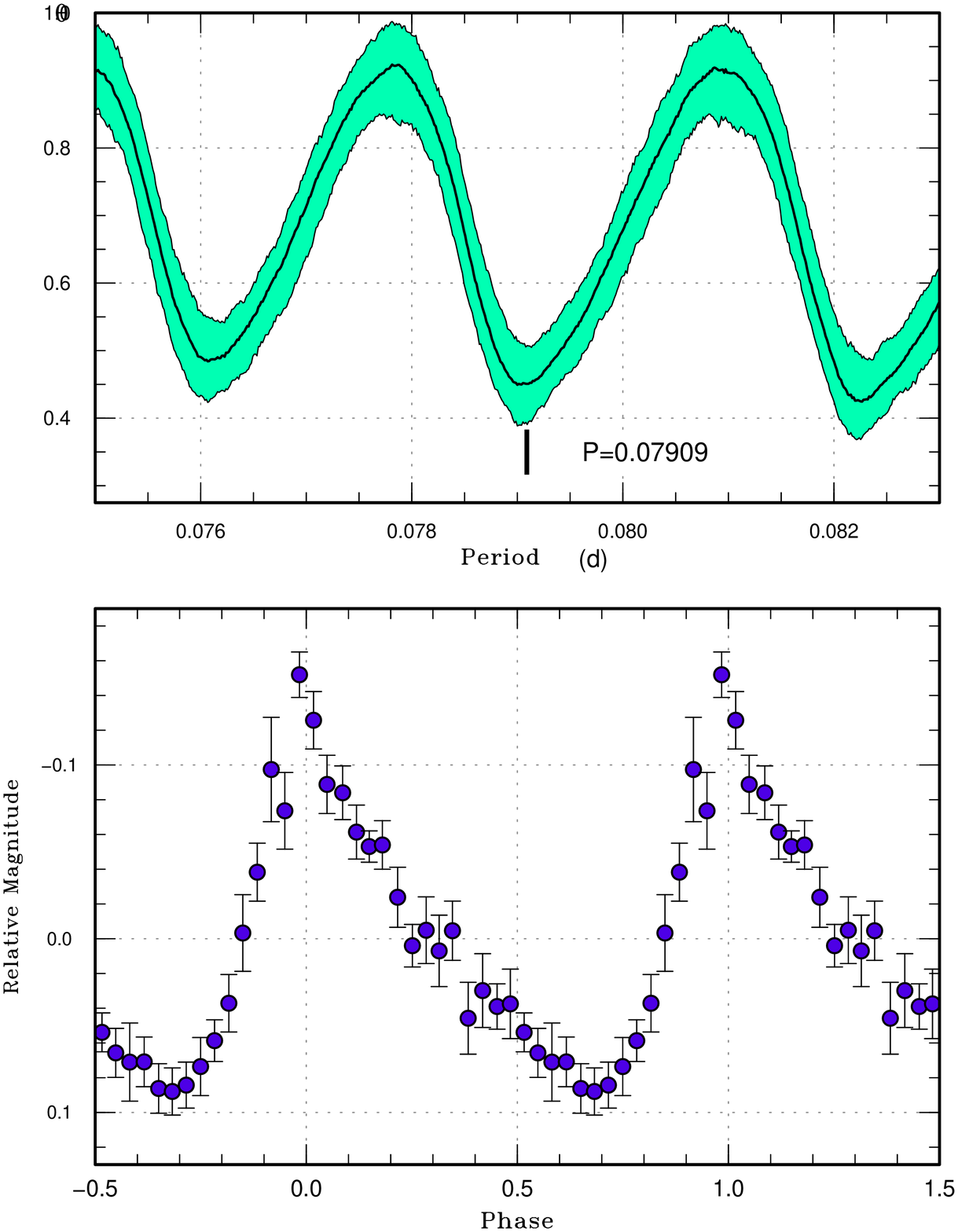}
  \end{center}
  \caption{Superhumps in CRTS J212521 (2015).
     (Upper): PDM analysis.  The data before BJD 2457262 were used.
     The alias selection was based on $O-C$ analysis.
     (Lower): Phase-averaged profile.}
  \label{fig:j2125shpdm}
\end{figure}

   The 2015 outburst was detected by the ASAS-SN team
on August 23 at a magnitude of $V$=15.0.  Since the past
outbursts recorded in the ASAS-SN data resembled superoutbursts,
an SU UMa-type dwarf nova was suspected (vsnet-alert 19000).
Observations soon recorded superhumps (vsnet-alert 19006,
19010; figure \ref{fig:j2125shpdm}).
The times of superhump maxima are listed in
table \ref{tab:j2125oc2015}.  Although the basic superhump
period was determined to be 0.0791(1)~d from
the observations on the first two nights,
the cycle count is ambiguous between $E=$26 and
$E=$100.  The maxima for $E \ge$100 may represent
stage C superhumps.
We listed a period only based on the initial
two nights in table \ref{tab:perlist}.

% SI

% SI

\begin{table}
\caption{Superhump maxima of CRTS J212521 (2015)}\label{tab:j2125oc2015}
\begin{center}
\begin{tabular}{rp{55pt}p{40pt}r@{.}lr}
\hline
\multicolumn{1}{c}{$E$} & \multicolumn{1}{c}{max\commenta} & \multicolumn{1}{c}{error} & \multicolumn{2}{c}{$O-C$\commentb} & \multicolumn{1}{c}{$N$\commentc} \\
\hline
0 & 57259.4157 & 0.0008 & 0&0045 & 43 \\
1 & 57259.4967 & 0.0007 & 0&0058 & 44 \\
25 & 57261.3968 & 0.0025 & $-$0&0038 & 107 \\
26 & 57261.4702 & 0.0009 & $-$0&0100 & 135 \\
100 & 57267.3726 & 0.0016 & 0&0038 & 57 \\
101 & 57267.4480 & 0.0011 & $-$0&0004 & 53 \\
\hline
  \multicolumn{6}{l}{\commenta BJD$-$2400000.} \\
  \multicolumn{6}{l}{\commentb Against max $= 2457259.4113 + 0.079575 E$.} \\
  \multicolumn{6}{l}{\commentc Number of points used to determine the maximum.} \\
\end{tabular}
\end{center}
\end{table}

\subsection{CRTS J214738.4$+$244554}\label{obj:j2147}

   This object (=CSS111004:214738+244554, hereafter CRTS J214738)
was discovered by CRTS on 2011 November 4 \citep{bre14CRTSCVs}.
The 2011 and 2014 superoutbursts were reported in
\citet{Pdot4} and \citet{Pdot7}, respectively.

   The 2015 superoutburst was visually detected by
C. Chiselbrook on December 15 (cf. vsnet-alert 19351).
Only one superhump maximum was recorded:
BJD 2457376.8752(4) ($N$=84).

\subsection{CSS J221822.9$+$344511}\label{obj:j2218}

   This object (hereafter CSS J221822) was originally
discovered by CRTS (CSS120812:221823$+$344509) as a suspected
dwarf nova on 2012 August 12 at an unfiltered CCD magnitude
of 15.85.  There is an X-ray counterpart (1RXS J221823.7$+$344507).

\begin{figure}
  \begin{center}
%    \FigureFile(85mm,110mm){j2218shpdm.eps}
    \FigureFile(85mm,110mm){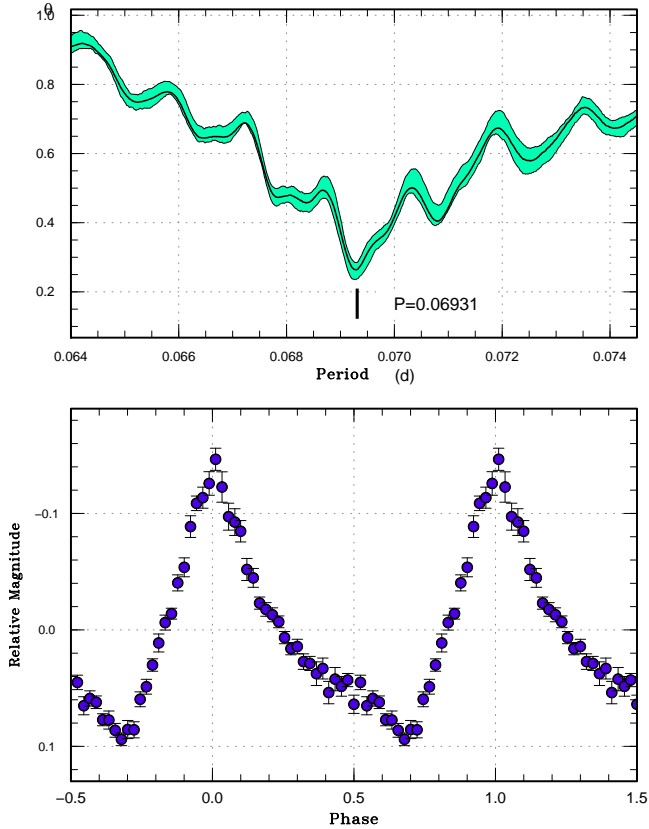}
  \end{center}
  \caption{Superhumps in CSS J221822 (2015).
     (Upper): PDM analysis.
     (Lower): Phase-averaged profile.}
  \label{fig:j2218shpdm}
\end{figure}

   The 2015 outburst was detected by the ASAS-SN team
on October 1 at $V$=15.42--15.24 (two measurements).
Subsequent observations detected superhumps
(vsnet-alert 19109, 19110, 19118, 19188;
figure \ref{fig:j2218shpdm}).
The times of superhump maxima are listed in
table \ref{tab:j2218oc2015}.

% SI

% SI

\begin{table}
\caption{Superhump maxima of CSS J221822 (2015)}\label{tab:j2218oc2015}
\begin{center}
\begin{tabular}{rp{55pt}p{40pt}r@{.}lr}
\hline
\multicolumn{1}{c}{$E$} & \multicolumn{1}{c}{max\commenta} & \multicolumn{1}{c}{error} & \multicolumn{2}{c}{$O-C$\commentb} & \multicolumn{1}{c}{$N$\commentc} \\
\hline
0 & 57298.0261 & 0.0030 & 0&0015 & 59 \\
1 & 57298.0923 & 0.0003 & $-$0&0016 & 142 \\
2 & 57298.1628 & 0.0003 & $-$0&0004 & 143 \\
3 & 57298.2300 & 0.0004 & $-$0&0025 & 137 \\
4 & 57298.3020 & 0.0003 & 0&0002 & 63 \\
5 & 57298.3711 & 0.0003 & 0&0000 & 68 \\
6 & 57298.4394 & 0.0003 & $-$0&0009 & 65 \\
7 & 57298.5096 & 0.0004 & $-$0&0001 & 48 \\
8 & 57298.5840 & 0.0011 & 0&0051 & 28 \\
19 & 57299.3403 & 0.0011 & $-$0&0009 & 49 \\
47 & 57301.2789 & 0.0042 & $-$0&0026 & 38 \\
48 & 57301.3491 & 0.0014 & $-$0&0016 & 47 \\
50 & 57301.4931 & 0.0030 & 0&0038 & 45 \\
\hline
  \multicolumn{6}{l}{\commenta BJD$-$2400000.} \\
  \multicolumn{6}{l}{\commentb Against max $= 2457298.0246 + 0.069294 E$.} \\
  \multicolumn{6}{l}{\commentc Number of points used to determine the maximum.} \\
\end{tabular}
\end{center}
\end{table}

\subsection{DDE 26}\label{obj:dde26}

\begin{figure}
  \begin{center}
%    \FigureFile(85mm,70mm){dde26comp.eps}
    \FigureFile(85mm,70mm){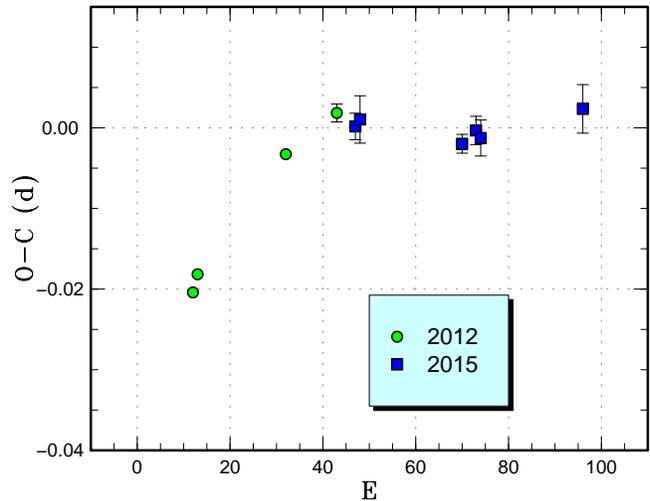}
  \end{center}
  \caption{Comparison of $O-C$ diagrams of DDE 26 between different
  superoutbursts.  A period of 0.08860~d was used to draw this figure.
  Approximate cycle counts ($E$) after the outburst detections
  were used.  The actual starts of the outbursts were unknown.
  }
  \label{fig:dde26comp}
\end{figure}

   DDE 26 is a dwarf nova discovered by \citep{den12USNOCVs}.
See \citet{Pdot5} for more information.
The 2015 outburst was detected by the ASAS-SN team
at $V$=15.7 on December 1 (cf. vsnet-alert 19311, 19320).
Subsequent observations detected superhumps
(vsnet-alert 19324, 19331).
The times of superhump maxima are listed in
table \ref{tab:dde26oc2015}.
A comparison of $O-C$ diagrams (figure \ref{fig:dde26comp})
indicates that the superhump period was longer in 2012.
It may be possible that the 2012 observations recorded
stage A superhumps, since some of long-$P_{\rm orb}$
systems are known to show long-lasting stage A
(e.g. \cite{kat16v1006cyg}; subsection \ref{sec:longstagea}).
This interpretation, however, does not agree with
the large amplitudes of superhumps during the 2012 observation.
More observations are needed to clarify the superhump
variation in this system.

% SI

\begin{table}
\caption{Superhump maxima of DDE 26 (2015)}\label{tab:dde26oc2015}
\begin{center}
\begin{tabular}{rp{55pt}p{40pt}r@{.}lr}
\hline
\multicolumn{1}{c}{$E$} & \multicolumn{1}{c}{max\commenta} & \multicolumn{1}{c}{error} & \multicolumn{2}{c}{$O-C$\commentb} & \multicolumn{1}{c}{$N$\commentc} \\
\hline
0 & 57361.9325 & 0.0016 & 0&0005 & 108 \\
1 & 57362.0220 & 0.0029 & 0&0014 & 89 \\
23 & 57363.9681 & 0.0012 & $-$0&0020 & 133 \\
26 & 57364.2356 & 0.0018 & $-$0&0004 & 74 \\
27 & 57364.3233 & 0.0022 & $-$0&0014 & 96 \\
49 & 57366.2761 & 0.0030 & 0&0019 & 88 \\
\hline
  \multicolumn{6}{l}{\commenta BJD$-$2400000.} \\
  \multicolumn{6}{l}{\commentb Against max $= 2457361.9320 + 0.088617 E$.} \\
  \multicolumn{6}{l}{\commentc Number of points used to determine the maximum.} \\
\end{tabular}
\end{center}
\end{table}

\subsection{IPHAS J230538.39$+$652158.7}\label{obj:j2305}

\begin{figure}
  \begin{center}
%    \FigureFile(85mm,110mm){j2305shpdm.eps}
    \FigureFile(85mm,110mm){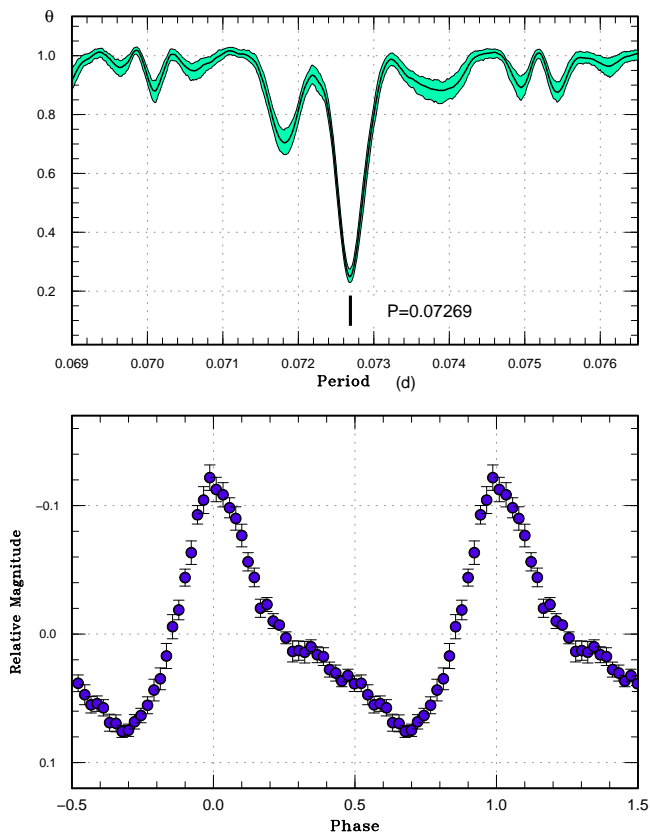}
  \end{center}
  \caption{Superhumps in IPHAS J230538 (2015).
     (Upper): PDM analysis.  The data after BJD 2457182 were used.
     (Lower): Phase-averaged profile.}
  \label{fig:j2305shpdm}
\end{figure}

   This object (hereafter IPHAS J230538) was detected
as an H$\alpha$ emission line object in INT/WFC Photometric
H$\alpha$ Survey (IPHAS: \cite{wit08IPHAS}).
The first known outburst was detected on 2015 June 4
by the ASAS-SN team (cf. vsnet-alert 18690).  The object
was still rising (vsnet-alert 18695) and growing
superhumps were detected (vsnet-alert 18702, 18709).
Further evolution of superhumps were observed
(vsnet-alert 18715, 18730, 18789; figure \ref{fig:j2305shpdm}).
The times of superhump maxima are listed in
table \ref{tab:j2305oc2015}.  Although observations were
rather sparse, we could identify stages A--C.

% SI

% SI

\begin{table}
\caption{Superhump maxima of IPHAS J230538 (2015)}\label{tab:j2305oc2015}
\begin{center}
\begin{tabular}{rp{55pt}p{40pt}r@{.}lr}
\hline
\multicolumn{1}{c}{$E$} & \multicolumn{1}{c}{max\commenta} & \multicolumn{1}{c}{error} & \multicolumn{2}{c}{$O-C$\commentb} & \multicolumn{1}{c}{$N$\commentc} \\
\hline
0 & 57181.4144 & 0.0028 & $-$0&0092 & 35 \\
1 & 57181.4980 & 0.0017 & 0&0016 & 65 \\
2 & 57181.5634 & 0.0023 & $-$0&0056 & 24 \\
15 & 57182.5175 & 0.0004 & 0&0031 & 46 \\
28 & 57183.4623 & 0.0003 & 0&0024 & 76 \\
29 & 57183.5358 & 0.0003 & 0&0033 & 77 \\
43 & 57184.5529 & 0.0004 & 0&0022 & 69 \\
82 & 57187.3918 & 0.0004 & 0&0050 & 74 \\
83 & 57187.4664 & 0.0005 & 0&0068 & 74 \\
96 & 57188.4074 & 0.0005 & 0&0024 & 74 \\
97 & 57188.4788 & 0.0004 & 0&0012 & 74 \\
110 & 57189.4218 & 0.0007 & $-$0&0012 & 39 \\
111 & 57189.4945 & 0.0008 & $-$0&0013 & 33 \\
123 & 57190.3708 & 0.0110 & 0&0024 & 9 \\
124 & 57190.4362 & 0.0006 & $-$0&0049 & 51 \\
125 & 57190.5056 & 0.0011 & $-$0&0082 & 49 \\
\hline
  \multicolumn{6}{l}{\commenta BJD$-$2400000.} \\
  \multicolumn{6}{l}{\commentb Against max $= 2457181.4236 + 0.072722 E$.} \\
  \multicolumn{6}{l}{\commentc Number of points used to determine the maximum.} \\
\end{tabular}
\end{center}
\end{table}

\subsection{MASTER OT J003831.10$-$640313.7}\label{obj:j0038}

\begin{figure}
  \begin{center}
%    \FigureFile(85mm,110mm){j0038shpdm.eps}
    \FigureFile(85mm,110mm){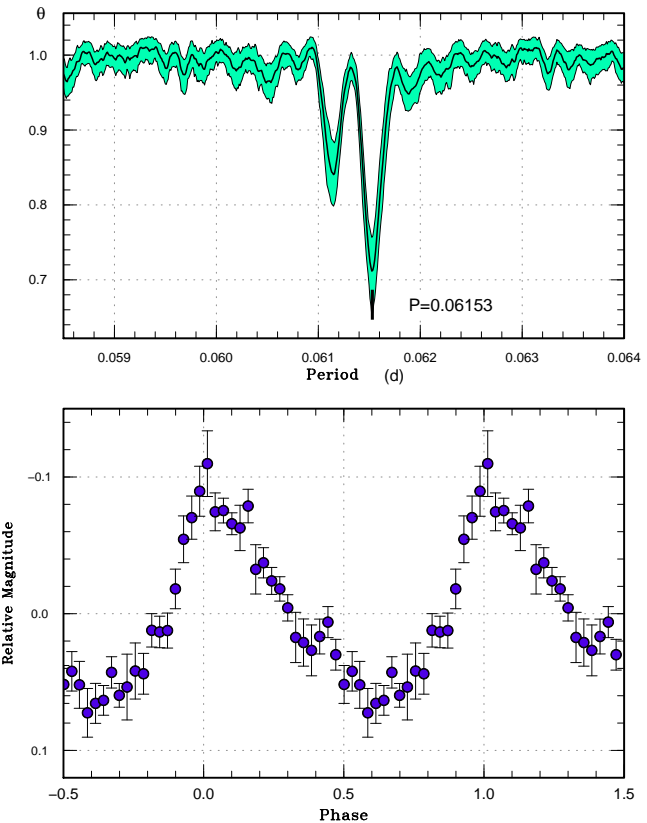}
  \end{center}
  \caption{Superhumps in MASTER J003831 during the plateau phase (2016).
     (Upper): PDM analysis.
     (Lower): Phase-averaged profile.}
  \label{fig:j0038shpdm}
\end{figure}

   This object (hereafter MASTER J003831)
was detected as a transient at an unfiltered
CCD magnitude of 12.7 on 2016 January 26 by the MASTER network
\citep{gre16j0038atel8596}.  Although \citet{gre16j0038atel8596}
suggested either a CV or a BL Lac-type object,
the presence of several past outbursts in the ASAS-3
data confirmed the dwarf nova-type classification
(vsnet-alert 19443).  
There is a GALEX counterpart with an NUV magnitude of
18.9(1).
Subsequent observations detected superhumps
(vsnet-alert 19449, 19454, 19467; figure \ref{fig:j0038shpdm}).
The times of superhump maxima are listed in
table \ref{tab:j0038oc2016}.  There are clear stages
B and C, with a strongly positive $P_{\rm dot}$
characteristic to this $P_{\rm SH}$.
The object showed a post-superoutburst rebrightening
on February 15 (vsnet-alert 19508).

% SI

% SI

\begin{table}
\caption{Superhump maxima of MASTER J003831 (2016)}\label{tab:j0038oc2016}
\begin{center}
\begin{tabular}{rp{55pt}p{40pt}r@{.}lr}
\hline
\multicolumn{1}{c}{$E$} & \multicolumn{1}{c}{max\commenta} & \multicolumn{1}{c}{error} & \multicolumn{2}{c}{$O-C$\commentb} & \multicolumn{1}{c}{$N$\commentc} \\
\hline
0 & 57416.5648 & 0.0004 & 0&0042 & 25 \\
1 & 57416.6248 & 0.0006 & 0&0026 & 25 \\
16 & 57417.5471 & 0.0005 & 0&0013 & 38 \\
17 & 57417.6078 & 0.0004 & 0&0005 & 40 \\
32 & 57418.5269 & 0.0014 & $-$0&0040 & 23 \\
33 & 57418.5910 & 0.0006 & $-$0&0014 & 40 \\
49 & 57419.5725 & 0.0005 & $-$0&0051 & 40 \\
50 & 57419.6339 & 0.0008 & $-$0&0052 & 28 \\
65 & 57420.5594 & 0.0009 & $-$0&0034 & 39 \\
66 & 57420.6209 & 0.0009 & $-$0&0034 & 36 \\
81 & 57421.5486 & 0.0025 & 0&0007 & 39 \\
98 & 57422.5985 & 0.0011 & 0&0039 & 39 \\
114 & 57423.5909 & 0.0008 & 0&0111 & 40 \\
130 & 57424.5676 & 0.0007 & 0&0027 & 40 \\
131 & 57424.6297 & 0.0021 & 0&0032 & 18 \\
146 & 57425.5489 & 0.0008 & $-$0&0012 & 39 \\
147 & 57425.6124 & 0.0009 & 0&0007 & 31 \\
179 & 57427.5747 & 0.0018 & $-$0&0073 & 37 \\
\hline
  \multicolumn{6}{l}{\commenta BJD$-$2400000.} \\
  \multicolumn{6}{l}{\commentb Against max $= 2457416.5606 + 0.061572 E$.} \\
  \multicolumn{6}{l}{\commentc Number of points used to determine the maximum.} \\
\end{tabular}
\end{center}
\end{table}

\subsection{MASTER OT J073325.52$+$373744.9}\label{obj:j0733}

\begin{figure}
  \begin{center}
%    \FigureFile(85mm,110mm){j0733shpdm.eps}
    \FigureFile(85mm,110mm){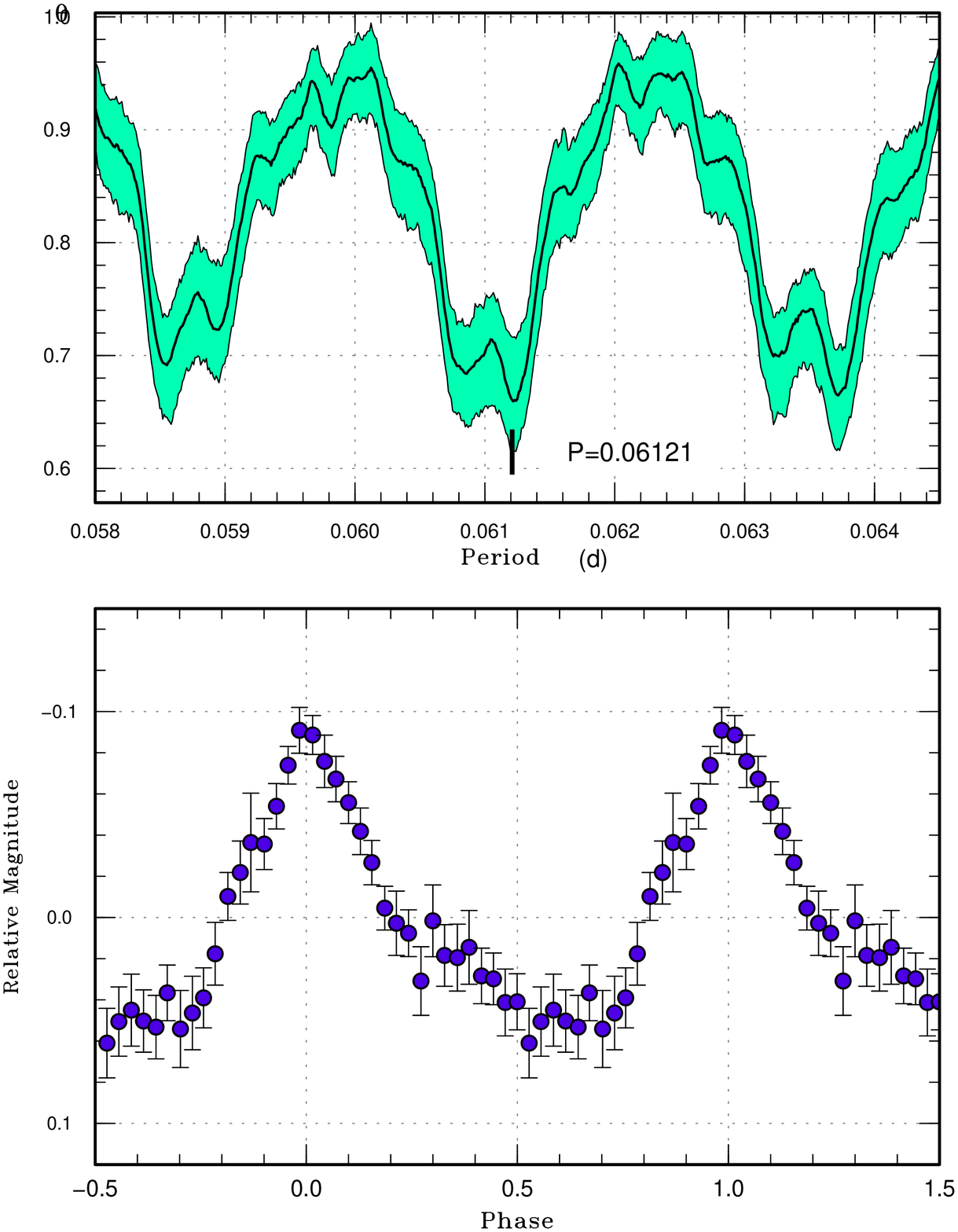}
  \end{center}
  \caption{Superhumps in MASTER J073325 (2016).
     The data after BJD 2457447 (stage B) were used.
     (Upper): PDM analysis.  The alias selection was based
     on $O-C$ analysis.
     (Lower): Phase-averaged profile.}
  \label{fig:j0733shpdm}
\end{figure}

   This object (hereafter MASTER J073325)
was detected as a transient at an unfiltered
CCD magnitude of 15.1 on 2016 February 24 by the MASTER network
\citep{gre16j0733atel8730}.  There was at least
one outburst in the CRTS data (15.9 mag on 2008 January 31,
vsnet-alert 19528).  Subsequent observations
immediately detected growing superhumps
(vsnet-alert 19532, 19537; figure \ref{fig:j0733shpdm}).
The times of superhump maxima are listed in
table \ref{tab:j0733oc2016}.
The maxima for $E \le$19 were undoubtedly stage A
superhumps as judged from the $O-C$ values and
growing amplitudes.  The likely positive $P_{\rm dot}$
for stage B is typical for this $P_{\rm SH}$ and
the early appearance of (ordinary) superhumps indicates
that this object is an ordinary SU UMa-type dwarf nova
rather than a WZ Sge-type dwarf nova as suspected
from the large outburst amplitude \citep{gre16j0733atel8730}.
The presence of a past outburst is consistent with
this identification.

% SI

% SI

\begin{table}
\caption{Superhump maxima of MASTER J073325 (2016)}\label{tab:j0733oc2016}
\begin{center}
\begin{tabular}{rp{55pt}p{40pt}r@{.}lr}
\hline
\multicolumn{1}{c}{$E$} & \multicolumn{1}{c}{max\commenta} & \multicolumn{1}{c}{error} & \multicolumn{2}{c}{$O-C$\commentb} & \multicolumn{1}{c}{$N$\commentc} \\
\hline
0 & 57445.4247 & 0.0009 & $-$0&0114 & 63 \\
9 & 57445.9832 & 0.0026 & $-$0&0042 & 33 \\
10 & 57446.0493 & 0.0018 & 0&0007 & 63 \\
11 & 57446.1046 & 0.0013 & $-$0&0052 & 65 \\
14 & 57446.2989 & 0.0020 & 0&0052 & 17 \\
17 & 57446.4804 & 0.0008 & 0&0030 & 61 \\
18 & 57446.5409 & 0.0011 & 0&0023 & 59 \\
19 & 57446.6046 & 0.0019 & 0&0047 & 28 \\
32 & 57447.4013 & 0.0006 & 0&0051 & 64 \\
33 & 57447.4602 & 0.0005 & 0&0027 & 99 \\
34 & 57447.5226 & 0.0005 & 0&0039 & 80 \\
35 & 57447.5833 & 0.0005 & 0&0034 & 50 \\
58 & 57448.9855 & 0.0008 & $-$0&0032 & 61 \\
59 & 57449.0465 & 0.0006 & $-$0&0035 & 60 \\
60 & 57449.1103 & 0.0017 & $-$0&0009 & 34 \\
80 & 57450.3340 & 0.0008 & $-$0&0023 & 29 \\
162 & 57455.3588 & 0.0012 & $-$0&0002 & 58 \\
\hline
  \multicolumn{6}{l}{\commenta BJD$-$2400000.} \\
  \multicolumn{6}{l}{\commentb Against max $= 2457445.4361 + 0.061253 E$.} \\
  \multicolumn{6}{l}{\commentc Number of points used to determine the maximum.} \\
\end{tabular}
\end{center}
\end{table}

\subsection{MASTER OT J120251.56$-$454116.7}\label{obj:j1202}

\begin{figure}
  \begin{center}
%    \FigureFile(85mm,110mm){j1202shpdm.eps}
    \FigureFile(85mm,110mm){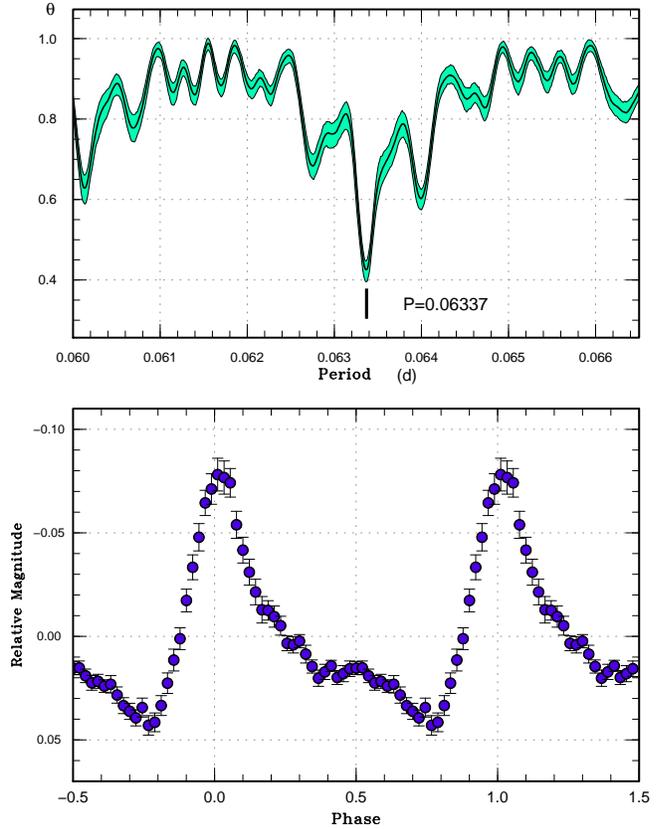}
  \end{center}
  \caption{Superhumps in MASTER J120251 during the plateau phase (2015).
     (Upper): PDM analysis.
     (Lower): Phase-averaged profile.}
  \label{fig:j1202shpdm}
\end{figure}

   This object (hereafter MASTER J120251)
was detected as a transient at an unfiltered
CCD magnitude of 16.0 on 2015 March 15 by the MASTER network
\citep{gre15j1202atel7237}.
This object was also detected at $V$=16.7
on 2015 March 15 by the ASAS-SN team (=ASASSN-15fp,
\cite{dan15j1202atel7260}).
The object faded to fainter than $V$=17.5 on March 18 and 
then brightened to $V$=14.4 on March 20
\citep{dan15j1202atel7260}.
Superhumps were immediately detected (vsnet-alert 18483,
18486, 18499; figure \ref{fig:j1202shpdm}).
The times of superhump maxima are listed in
table \ref{tab:j1202oc2015}.
Due to the 4-d gap in the observation, we were not
able to identify the stage classification and gave
a global value in table \ref{tab:perlist}.
The initial part ($E \le 27$) probably recorded
stage B superhumps.

% SI

% SI

\begin{table}
\caption{Superhump maxima of MASTER J120251 (2015)}\label{tab:j1202oc2015}
\begin{center}
\begin{tabular}{rp{55pt}p{40pt}r@{.}lr}
\hline
\multicolumn{1}{c}{$E$} & \multicolumn{1}{c}{max\commenta} & \multicolumn{1}{c}{error} & \multicolumn{2}{c}{$O-C$\commentb} & \multicolumn{1}{c}{$N$\commentc} \\
\hline
0 & 57103.3175 & 0.0002 & $-$0&0004 & 146 \\
1 & 57103.3806 & 0.0003 & $-$0&0006 & 146 \\
16 & 57104.3310 & 0.0003 & $-$0&0009 & 145 \\
17 & 57104.3953 & 0.0004 & 0&0001 & 138 \\
18 & 57104.4520 & 0.0023 & $-$0&0066 & 32 \\
19 & 57104.5220 & 0.0003 & 0&0000 & 145 \\
20 & 57104.5925 & 0.0007 & 0&0072 & 47 \\
27 & 57105.0303 & 0.0006 & 0&0014 & 49 \\
99 & 57109.5939 & 0.0026 & 0&0022 & 28 \\
111 & 57110.3510 & 0.0005 & $-$0&0012 & 146 \\
112 & 57110.4143 & 0.0006 & $-$0&0013 & 146 \\
113 & 57110.4828 & 0.0038 & 0&0038 & 18 \\
115 & 57110.6017 & 0.0018 & $-$0&0040 & 34 \\
131 & 57111.6199 & 0.0063 & 0&0003 & 28 \\
\hline
  \multicolumn{6}{l}{\commenta BJD$-$2400000.} \\
  \multicolumn{6}{l}{\commentb Against max $= 2457103.3179 + 0.063372 E$.} \\
  \multicolumn{6}{l}{\commentc Number of points used to determine the maximum.} \\
\end{tabular}
\end{center}
\end{table}

\subsection{MASTER OT J131320.24$+$692649.1}\label{obj:j1313}

\begin{figure}
  \begin{center}
%    \FigureFile(85mm,110mm){j1313shpdm.eps}
    \FigureFile(85mm,110mm){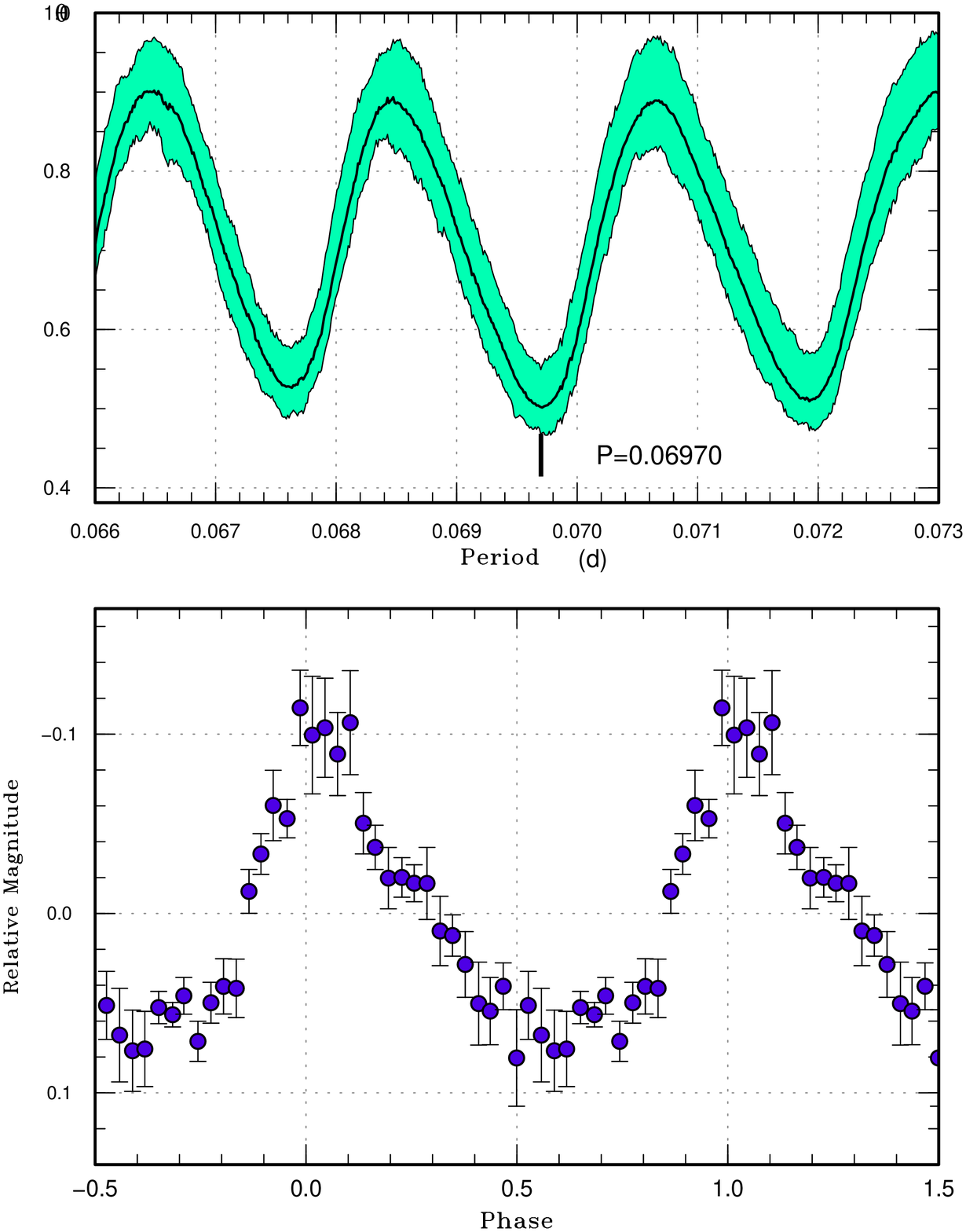}
  \end{center}
  \caption{Superhumps in MASTER J131320 (2016).
     (Upper): PDM analysis.  The alias selection was based on
     $O-C$ analysis.
     (Lower): Phase-averaged profile.}
  \label{fig:j1313shpdm}
\end{figure}

   This object (hereafter MASTER J131320)
was detected as a transient at an unfiltered
CCD magnitude of 14.7 on 2013 May 14 by the MASTER network
\citep{den13j1313atel5065}.
There is a GALEX counterpart with an NUV magnitude of
20.6(2).  Two more outbursts were recorded between 2014
and 2015 in the ASAS-SN data.  
The 2016 outburst was detected on February 15 at $V$=15.01
by the ASAS-SN team (cf. vsnet-alert 19505).
Subsequent observations detected superhumps
(vsnet-alert 19511; figure \ref{fig:j1313shpdm}).
The times of superhump maxima are listed in
table \ref{tab:j1313oc2016}.
Although we selected the alias period to minimize
absolute $O-C$ residuals, the superhump amplitudes
were significantly smaller on the first night.
It was possible that the stage A superhumps were
recorded on the initial night.  If there was
a strong variation in the period between two nights,
our method may have failed to select the correct
period.  Although there was no indication of
such a strong variation in the $O-C$ values,
the period should be treated with caution.

% SI

% SI

\begin{table}
\caption{Superhump maxima of MASTER J131320 (2016)}\label{tab:j1313oc2016}
\begin{center}
\begin{tabular}{rp{55pt}p{40pt}r@{.}lr}
\hline
\multicolumn{1}{c}{$E$} & \multicolumn{1}{c}{max\commenta} & \multicolumn{1}{c}{error} & \multicolumn{2}{c}{$O-C$\commentb} & \multicolumn{1}{c}{$N$\commentc} \\
\hline
0 & 57435.4849 & 0.0006 & $-$0&0004 & 65 \\
1 & 57435.5554 & 0.0006 & 0&0004 & 70 \\
32 & 57437.7164 & 0.0009 & 0&0004 & 35 \\
33 & 57437.7836 & 0.0009 & $-$0&0021 & 47 \\
34 & 57437.8569 & 0.0008 & 0&0016 & 46 \\
\hline
  \multicolumn{6}{l}{\commenta BJD$-$2400000.} \\
  \multicolumn{6}{l}{\commentb Against max $= 2457435.4853 + 0.069709 E$.} \\
  \multicolumn{6}{l}{\commentc Number of points used to determine the maximum.} \\
\end{tabular}
\end{center}
\end{table}

\subsection{MASTER OT J181523.78$+$692037.4}\label{obj:j1815}

\begin{figure}
  \begin{center}
%    \FigureFile(85mm,110mm){j1815shpdm.eps}
    \FigureFile(85mm,110mm){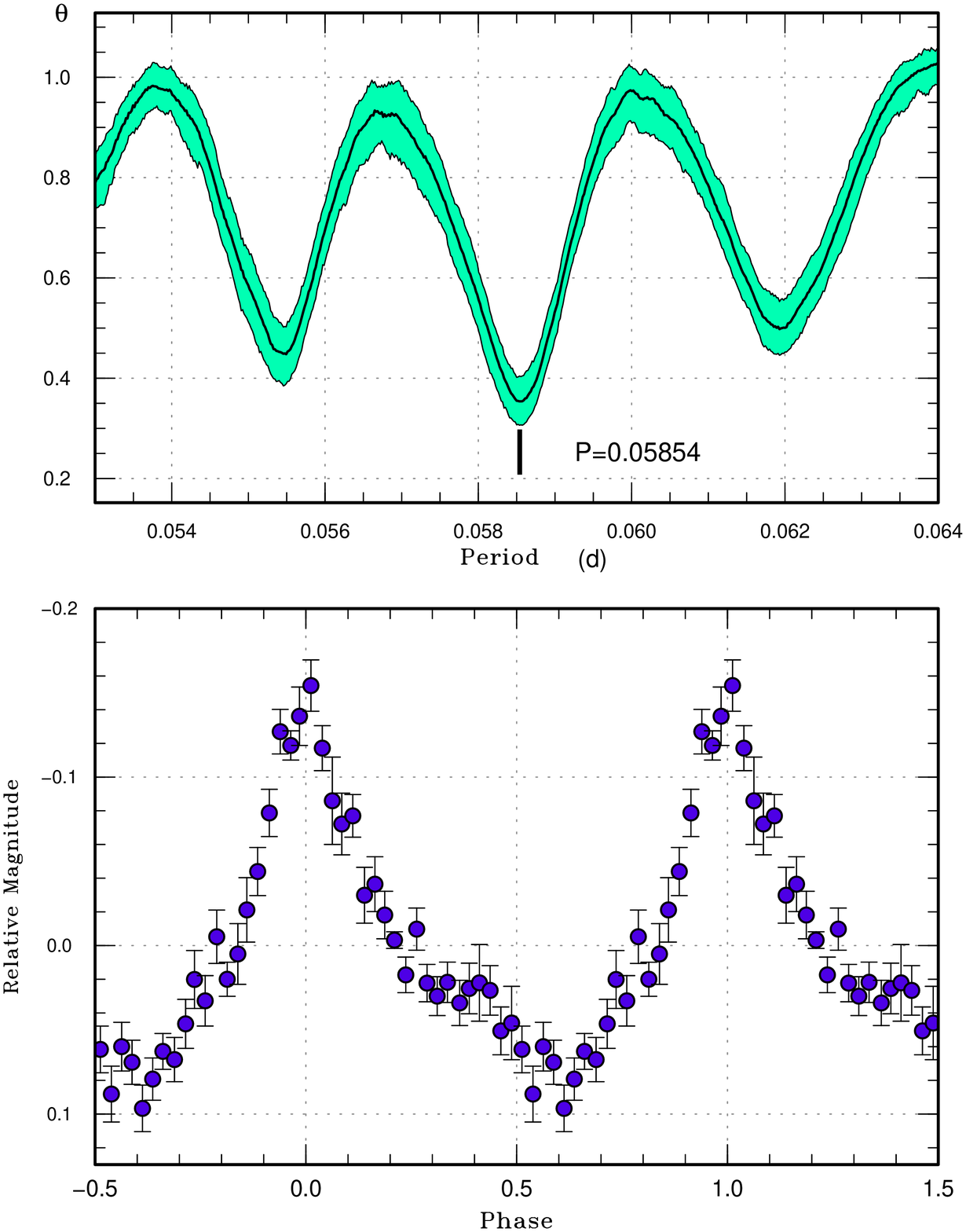}
  \end{center}
  \caption{Superhumps in MASTER J181523 (2015).
     (Upper): PDM analysis.  The alias selection was based on
     $O-C$ analysis.
     (Lower): Phase-averaged profile.}
  \label{fig:j1815shpdm}
\end{figure}

   This object (hereafter MASTER J181523)
was detected as a transient at an unfiltered
CCD magnitude of 15.7 on 2015 August 28 by the MASTER network
\citep{gre15j1815atel7972}.
There is a GALEX counterpart with an NUV magnitude of
22.8(5).  Superhumps were soon detected
(vsnet-alert 19026, 19041; figure \ref{fig:j1815shpdm}).
The times of superhump maxima are listed in
table \ref{tab:j1815oc2015}.

% SI

% SI

\begin{table}
\caption{Superhump maxima of MASTER J181523 (2015)}\label{tab:j1815oc2015}
\begin{center}
\begin{tabular}{rp{55pt}p{40pt}r@{.}lr}
\hline
\multicolumn{1}{c}{$E$} & \multicolumn{1}{c}{max\commenta} & \multicolumn{1}{c}{error} & \multicolumn{2}{c}{$O-C$\commentb} & \multicolumn{1}{c}{$N$\commentc} \\
\hline
0 & 57267.4270 & 0.0005 & 0&0006 & 52 \\
1 & 57267.4830 & 0.0015 & $-$0&0019 & 24 \\
2 & 57267.5448 & 0.0008 & 0&0014 & 38 \\
3 & 57267.6016 & 0.0008 & $-$0&0003 & 51 \\
18 & 57268.4805 & 0.0008 & 0&0009 & 49 \\
19 & 57268.5385 & 0.0006 & 0&0004 & 55 \\
20 & 57268.5954 & 0.0010 & $-$0&0012 & 53 \\
\hline
  \multicolumn{6}{l}{\commenta BJD$-$2400000.} \\
  \multicolumn{6}{l}{\commentb Against max $= 2457267.4264 + 0.058512 E$.} \\
  \multicolumn{6}{l}{\commentc Number of points used to determine the maximum.} \\
\end{tabular}
\end{center}
\end{table}

\subsection{MASTER OT J212624.16$+$253827.2}\label{obj:j2126}

   This object (hereafter MASTER J212624)
was detected as a transient at an unfiltered
CCD magnitude of 14.1 on 2013 June 26 by the MASTER network
\citep{den13j2126atel5111}.  The 2013 superoutburst
was well observed and a large positive $P_{\rm dot}$
despite the long $P_{\rm SH}$ was detected \citep{Pdot5}.
For more information, see \citet{Pdot5}.

   The 2015 superoutburst was detected by the ASAS-SN team
at $V$=14.24 on August 27 (cf. vsnet-alert 19012).
Our observation on September 1 recorded superhumps
(vsnet-alert 19031).  The times of superhump maxima
are listed in table \ref{tab:j2126oc2015}.
Since the observation was 5~d after the outburst detection,
our observation did not cover the early part
of the superoutburst.  Although the 2013 observation
started 2~d after the outburst detection, it may have
not been detected sufficiently early.
Our present observations were insufficient to
verify the large positive $P_{\rm dot}$.
Further observations, particularly in the early phase,
are still needed.

% SI

\begin{table}
\caption{Superhump maxima of MASTER J212624 (2015)}\label{tab:j2126oc2015}
\begin{center}
\begin{tabular}{rp{55pt}p{40pt}r@{.}lr}
\hline
\multicolumn{1}{c}{$E$} & \multicolumn{1}{c}{max\commenta} & \multicolumn{1}{c}{error} & \multicolumn{2}{c}{$O-C$\commentb} & \multicolumn{1}{c}{$N$\commentc} \\
\hline
0 & 57267.3587 & 0.0003 & $-$0&0013 & 97 \\
1 & 57267.4486 & 0.0004 & $-$0&0026 & 104 \\
2 & 57267.5407 & 0.0004 & $-$0&0017 & 98 \\
3 & 57267.6397 & 0.0021 & 0&0061 & 41 \\
30 & 57270.0905 & 0.0038 & $-$0&0053 & 57 \\
31 & 57270.1917 & 0.0075 & 0&0047 & 31 \\
\hline
  \multicolumn{6}{l}{\commenta BJD$-$2400000.} \\
  \multicolumn{6}{l}{\commentb Against max $= 2457267.3600 + 0.091193 E$.} \\
  \multicolumn{6}{l}{\commentc Number of points used to determine the maximum.} \\
\end{tabular}
\end{center}
\end{table}

\subsection{N080829A}\label{obj:n080829a}

   This object was originally reported as a transient
by H. Mikuz on 2008 August 29 at $R$=15.98(4)
(vsnet-alert 10485).  The object was observed at
around $R$=16.3 on August 31.

\begin{figure}
  \begin{center}
%    \FigureFile(85mm,110mm){n080829ashpdm.eps}
    \FigureFile(85mm,110mm){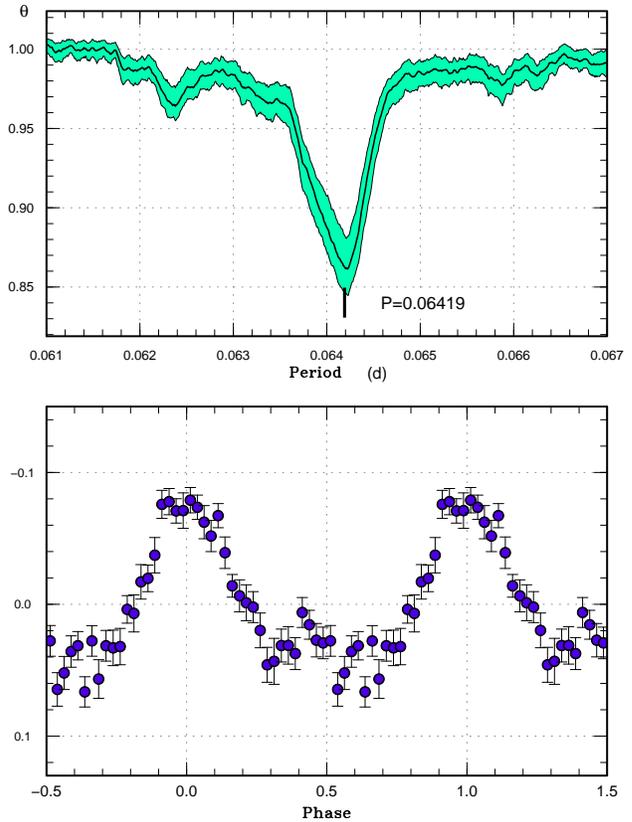}
  \end{center}
  \caption{Superhumps in N080829A (2015).
     (Upper): PDM analysis.  The alias selection was based on
     $O-C$ analysis.
     (Lower): Phase-averaged profile.}
  \label{fig:n080829ashpdm}
\end{figure}

   The 2015 outburst was detected by CRTS on October 12
at an unfiltered CCD magnitude of 15.98.
Subsequent observations detected superhumps
(vsnet-alert 19162, 19164, 19167, 19169;
figure \ref{fig:n080829ashpdm}).
The times of superhump maxima are listed in
table \ref{tab:n080829aoc2015}.
The strongly positive $P_{\rm dot}$ is typical
for stage B superhumps with this superhump period.

% SI

% SI

\begin{table}
\caption{Superhump maxima of N080829A (2015)}\label{tab:n080829aoc2015}
\begin{center}
\begin{tabular}{rp{55pt}p{40pt}r@{.}lr}
\hline
\multicolumn{1}{c}{$E$} & \multicolumn{1}{c}{max\commenta} & \multicolumn{1}{c}{error} & \multicolumn{2}{c}{$O-C$\commentb} & \multicolumn{1}{c}{$N$\commentc} \\
\hline
0 & 57308.9554 & 0.0012 & 0&0077 & 116 \\
2 & 57309.0800 & 0.0011 & 0&0037 & 160 \\
3 & 57309.1480 & 0.0013 & 0&0075 & 179 \\
11 & 57309.6584 & 0.0004 & 0&0036 & 67 \\
12 & 57309.7199 & 0.0004 & 0&0008 & 66 \\
13 & 57309.7840 & 0.0004 & 0&0006 & 67 \\
15 & 57309.9033 & 0.0087 & $-$0&0086 & 44 \\
16 & 57309.9739 & 0.0015 & $-$0&0023 & 136 \\
17 & 57310.0413 & 0.0011 & 0&0008 & 182 \\
18 & 57310.1046 & 0.0006 & $-$0&0002 & 72 \\
19 & 57310.1658 & 0.0013 & $-$0&0033 & 88 \\
32 & 57311.0006 & 0.0007 & $-$0&0041 & 62 \\
33 & 57311.0686 & 0.0008 & $-$0&0004 & 99 \\
34 & 57311.1323 & 0.0037 & $-$0&0010 & 41 \\
47 & 57311.9633 & 0.0010 & $-$0&0056 & 131 \\
48 & 57312.0358 & 0.0028 & 0&0025 & 138 \\
49 & 57312.0886 & 0.0014 & $-$0&0089 & 74 \\
69 & 57313.3800 & 0.0017 & $-$0&0031 & 67 \\
70 & 57313.4490 & 0.0012 & 0&0016 & 33 \\
78 & 57313.9565 & 0.0035 & $-$0&0052 & 71 \\
79 & 57314.0242 & 0.0034 & $-$0&0017 & 113 \\
80 & 57314.0898 & 0.0043 & $-$0&0004 & 119 \\
97 & 57315.1828 & 0.0116 & $-$0&0002 & 19 \\
98 & 57315.2523 & 0.0042 & 0&0049 & 36 \\
99 & 57315.3162 & 0.0021 & 0&0046 & 33 \\
100 & 57315.3823 & 0.0024 & 0&0065 & 20 \\
\hline
  \multicolumn{6}{l}{\commenta BJD$-$2400000.} \\
  \multicolumn{6}{l}{\commentb Against max $= 2457308.9477 + 0.064282 E$.} \\
  \multicolumn{6}{l}{\commentc Number of points used to determine the maximum.} \\
\end{tabular}
\end{center}
\end{table}

\subsection{OT J191443.6$+$605214}\label{obj:j1914}

   This object (hereafter OT J191443) was discovered by
K. Itagaki \citep{yam08j1914cbet1535}.  The 2008 superoutburst
was studied in \citet{boy09j1914} and \citet{Pdot}.
See \citet{Pdot5} for more history.

\begin{figure}
  \begin{center}
%    \FigureFile(85mm,70mm){j1914comp.eps}
    \FigureFile(85mm,70mm){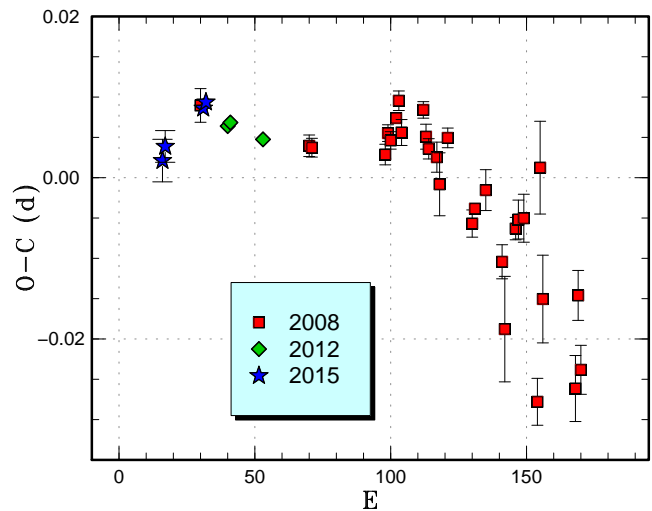}
  \end{center}
  \caption{Comparison of $O-C$ diagrams of OT J191443 between different
  superoutbursts.  A period of 0.07138~d was used to draw this figure.
  Approximate cycle counts ($E$) after the start of the superoutburst
  were used.  Since the starts of the 2008 and 2012 superoutbursts
  were not well constrained, we shifted the $O-C$ diagrams
  to fit the 2015 one, whose cycle counts are considered to be
  best determined.}
  \label{fig:j1914comp}
\end{figure}

    The 2015 outburst was detected by the ASAS-SN team on July 24.
The outburst detection was probably early enough and
initial observations detected low-amplitude stage A
superhumps (vsnet-alert 18887).
The times of superhump maxima are listed in
table \ref{tab:j1914oc2015}.  Although the maxima
for $E \le$ 1 were stage A superhumps, we could not
determine the period of stage A superhumps
(see also figure \ref{fig:j1914comp} for the comparison
of $O-C$ diagrams).

% SI

\begin{table}
\caption{Superhump maxima of OT J191443 (2015)}\label{tab:j1914oc2015}
\begin{center}
\begin{tabular}{rp{55pt}p{40pt}r@{.}lr}
\hline
\multicolumn{1}{c}{$E$} & \multicolumn{1}{c}{max\commenta} & \multicolumn{1}{c}{error} & \multicolumn{2}{c}{$O-C$\commentb} & \multicolumn{1}{c}{$N$\commentc} \\
\hline
0 & 57229.0977 & 0.0026 & $-$0&0006 & 137 \\
1 & 57229.1709 & 0.0020 & 0&0007 & 146 \\
15 & 57230.1749 & 0.0003 & $-$0&0002 & 68 \\
16 & 57230.2471 & 0.0002 & 0&0001 & 77 \\
\hline
  \multicolumn{6}{l}{\commenta BJD$-$2400000.} \\
  \multicolumn{6}{l}{\commentb Against max $= 2457229.0984 + 0.071784 E$.} \\
  \multicolumn{6}{l}{\commentc Number of points used to determine the maximum.} \\
\end{tabular}
\end{center}
\end{table}

\subsection{SDSS J074859.55$+$312512.6}\label{obj:j0748}

\begin{figure}
  \begin{center}
%    \FigureFile(85mm,70mm){j0748porb.eps}
    \FigureFile(85mm,70mm){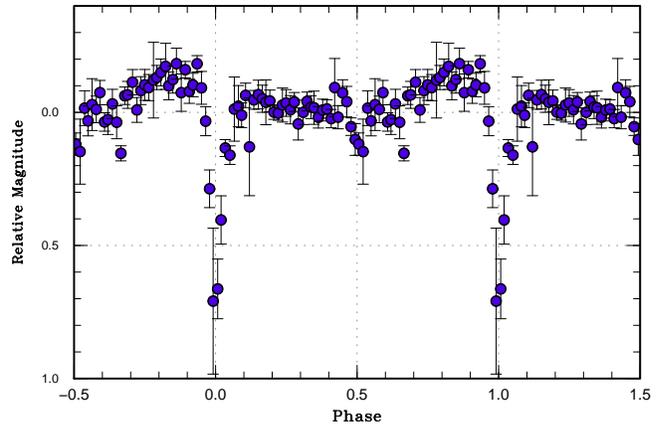}
  \end{center}
  \caption{Mean orbital light curve SDSS J074859.
      The CRTS data and our observations are used.
      The ephemeris of equation (\ref{equ:j0748ecl}) is used.}
  \label{fig:j0748porb}
\end{figure}

\begin{figure}
  \begin{center}
%    \FigureFile(85mm,110mm){j0748shpdm.eps}
    \FigureFile(85mm,110mm){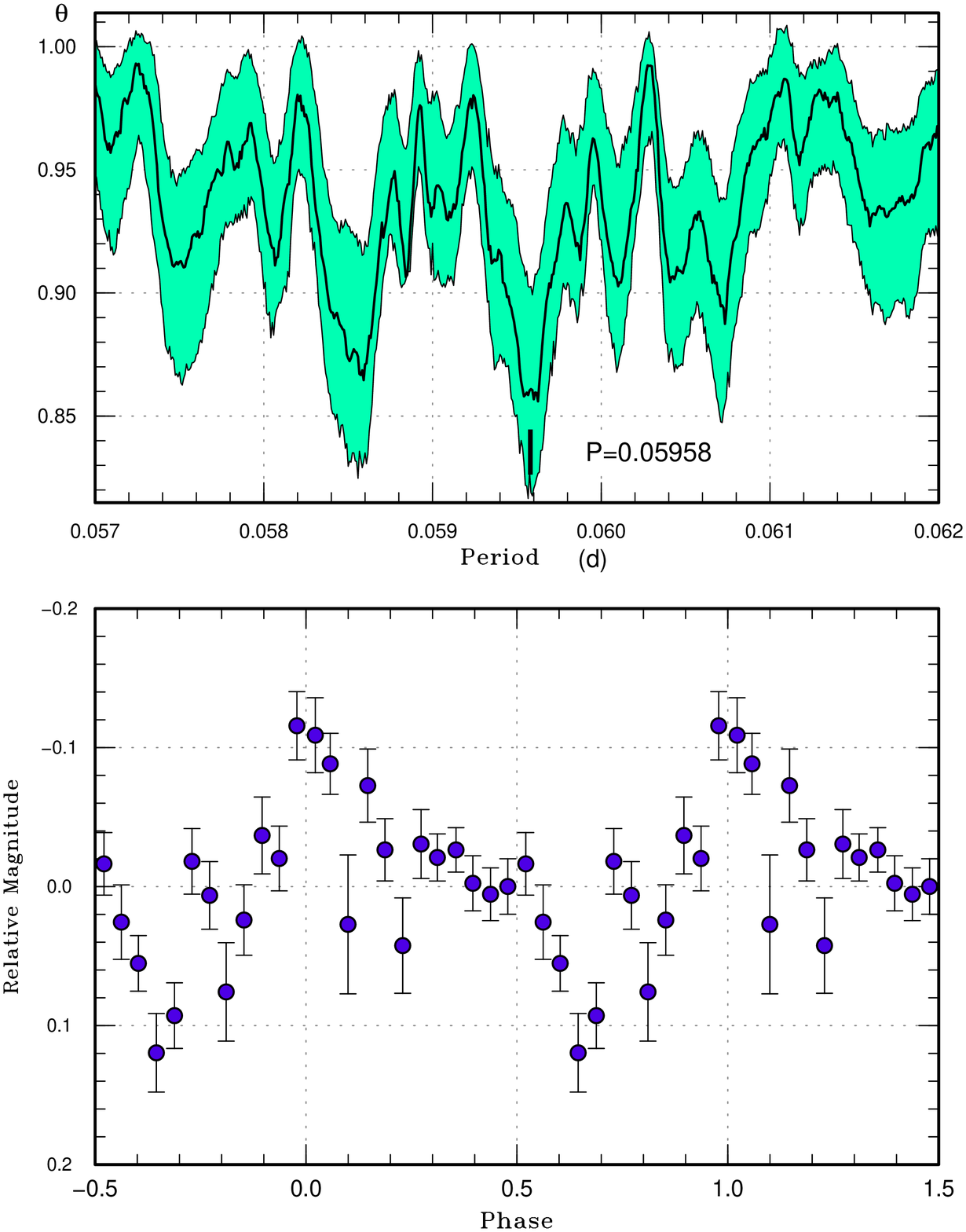}
  \end{center}
  \caption{Superhumps in SDSS J074859 (2015).
     (Upper): PDM analysis outside the eclipses.
     (Lower): Phase-averaged profile.}
  \label{fig:j0748shpdm}
\end{figure}

   This object (hereafter SDSS J074859) is a dwarf nova
selected by \citet{wil10newCVs}.  The 2015 outburst
was detected by CRTS on November 20 at an unfiltered CCD
magnitude of 15.86 (cf. vsnet-alert 19292).
There were frequent outbursts in the past CRTS data.
Observations on Nov. 24 detected eclipses and
an analysis of the CRTS data combined with the new data
yielded the following ephemeris (vsnet-alert 19297):
\begin{equation}
{\rm Min(BJD)} = 2457351.21283(2) + 0.0583110901(7) E .
\label{equ:j0748ecl}
\end{equation}
The orbital light curve (figure \ref{fig:j0748porb}) indicates
deep eclipses and pre-eclipse orbital humps.
These features, combined with the short orbital period,
suggest that this object is an eclipsing SU UMa-type
dwarf nova.  Although we could not determine individual
superhump maxima, a PDM analysis yielded a superhump
period of 0.05958(3)~d, 2.1\% longer than the orbital period
(figure \ref{fig:j0748shpdm}).  This period is listed in
table \ref{tab:perlist}.

% SI

\subsection{SDSS J145758.21$+$514807.9}\label{obj:j1457}

   This object (hereafter SDSS J145758) is a CV selected
by \citet{szk05SDSSCV4}.  The spectrum in \citet{szk05SDSSCV4}
strongly suggested a dwarf nova with a very low
mass-accretion rate.
The object was found to contain a pulsating white dwarf
as in GW Lib \citep{uth12j1457bwscl}.
The object has a photometric orbital period of
0.054087(5)~d \citep{uth11CVthesis}.
No previous outburst was known.

\begin{figure}
  \begin{center}
%    \FigureFile(85mm,110mm){j1457eshpdm.eps}
    \FigureFile(85mm,110mm){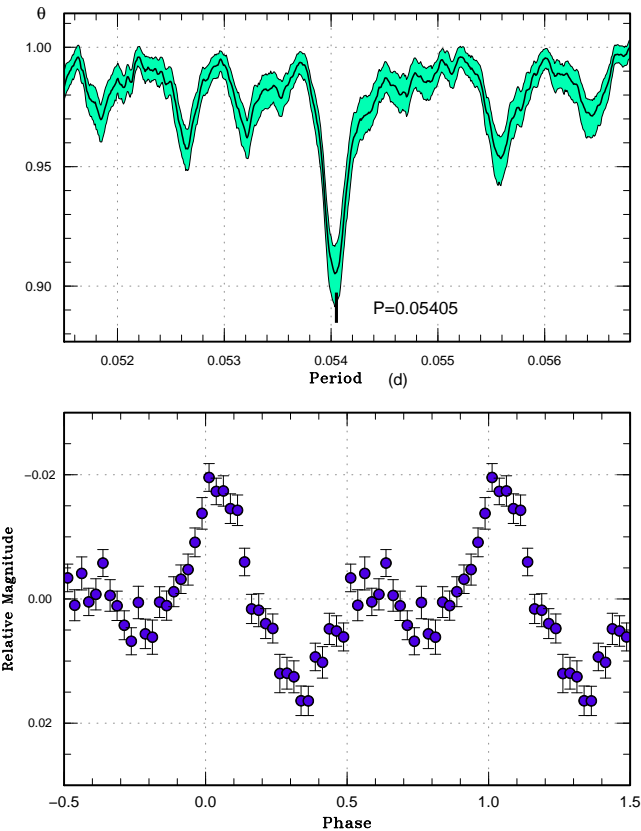}
  \end{center}
  \caption{Early superhumps in SDSS J145758 (2015).
     (Upper): PDM analysis.  The data for
     BJD 2457296--2457305 were used.
     (Lower): Phase-averaged profile.}
  \label{fig:j1457eshpdm}
\end{figure}

   J. Shears detected the first known outburst on 2015
September 29 at an unfiltered CCD magnitude of 15.3
(vsnet-outburst 18727).  The object further rose to
11.9 (visual magnitude) on September 30 (vsnet-alert 19097, 19098).
Subsequent observations detected double-wave
early superhumps (vsnet-alert 19100, 19103;
figure \ref{fig:j1457eshpdm}).
The period of early superhumps was shorter than
the orbital period by 0.07(2)\%, confirming
the relation reported in other WZ Sge-type dwarf novae
\citep{kat15wzsge}.

\begin{figure}
  \begin{center}
%    \FigureFile(85mm,110mm){j1457shpdm.eps}
    \FigureFile(85mm,110mm){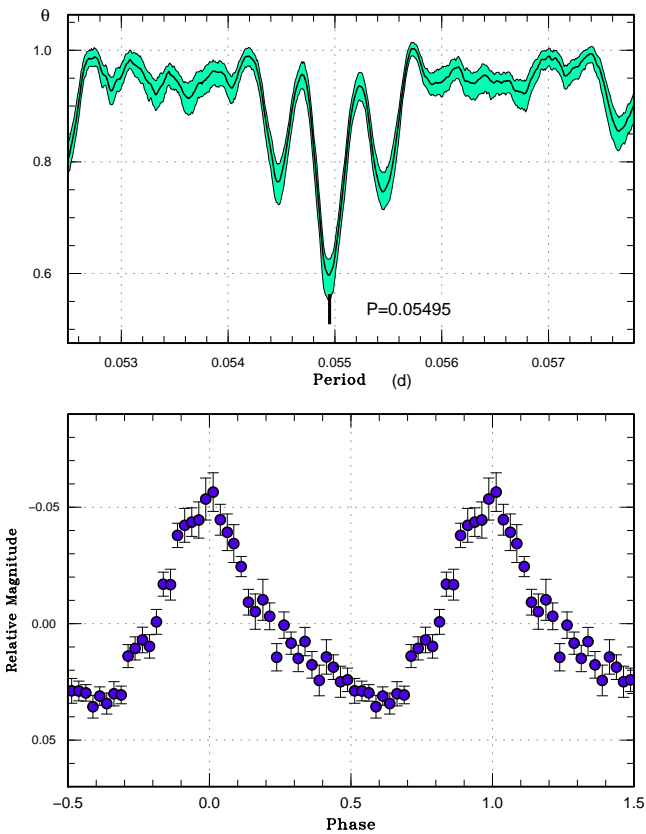}
  \end{center}
  \caption{Ordinary superhumps in SDSS J145758 (2015).
     (Upper): PDM analysis.  The data for
     BJD 2457305--2457313 were used.
     (Lower): Phase-averaged profile.}
  \label{fig:j1457shpdm}
\end{figure}

   Due to the short visibility in the evening sky,
the development of ordinary superhumps was not
well observed.  Ordinary superhumps were confidently
detected only during the final stage of the plateau
phase (vsnet-alert 19187; figure \ref{fig:j1457shpdm}).
The object started fading rapidly 3~d after these
observations of ordinary superhumps.
The times of maxima of ordinary superhumps are listed in
table \ref{tab:j1457oc2015}.  The maxima for
$E \ge$181 refer to superhumps during the rapid fading
and they were not used in determining the superhump
period in table \ref{tab:perlist}.
No post-superoutburst observations were available
and it was not known whether this fading was a ``dip''
as in other WZ Sge-type dwarf novae or not.

% SI

% SI

% SI

\begin{table}
\caption{Superhump maxima of SDSS J145758 (2015)}\label{tab:j1457oc2015}
\begin{center}
\begin{tabular}{rp{55pt}p{40pt}r@{.}lr}
\hline
\multicolumn{1}{c}{$E$} & \multicolumn{1}{c}{max\commenta} & \multicolumn{1}{c}{error} & \multicolumn{2}{c}{$O-C$\commentb} & \multicolumn{1}{c}{$N$\commentc} \\
\hline
0 & 57305.2918 & 0.0006 & $-$0&0075 & 185 \\
1 & 57305.3592 & 0.0032 & 0&0050 & 87 \\
6 & 57305.6266 & 0.0007 & $-$0&0020 & 32 \\
7 & 57305.6832 & 0.0016 & $-$0&0002 & 18 \\
91 & 57310.2907 & 0.0004 & $-$0&0028 & 124 \\
92 & 57310.3503 & 0.0010 & 0&0020 & 61 \\
108 & 57311.2274 & 0.0004 & 0&0010 & 112 \\
109 & 57311.2847 & 0.0009 & 0&0034 & 82 \\
110 & 57311.3373 & 0.0024 & 0&0011 & 40 \\
126 & 57312.2173 & 0.0006 & 0&0030 & 104 \\
127 & 57312.2724 & 0.0005 & 0&0032 & 101 \\
128 & 57312.3273 & 0.0006 & 0&0032 & 82 \\
181 & 57315.2264 & 0.0020 & $-$0&0064 & 59 \\
182 & 57315.2846 & 0.0044 & $-$0&0030 & 57 \\
\hline
  \multicolumn{6}{l}{\commenta BJD$-$2400000.} \\
  \multicolumn{6}{l}{\commentb Against max $= 2457305.2993 + 0.054881 E$.} \\
  \multicolumn{6}{l}{\commentc Number of points used to determine the maximum.} \\
\end{tabular}
\end{center}
\end{table}

\subsection{SDSS J164248.52$+$134751.4}\label{obj:j1642}

   This object (hereafter SDSS J164248) is a CV selected
by \citet{szk09SDSSCV7}.  \citet{szk09SDSSCV7}
suggested an orbital period of 1.3~hr.  \citet{sou08CVperiod}
obtained a spectroscopic orbital period of 0.07889(11)~d
and also reported Doppler tomography with an unusual brightness
distribution in the accretion disk.  Despite one well-recorded
outburst detection at an unfiltered CCD magnitude of 14.7
in the CRTS data, there had not been outbursts until
2012 September 6, when E. Muyllaert recorded an outburst
at an unfiltered CCD magnitude of 15.9 (cvnet-outburst 4910).
The 2012 outburst quickly faded.

\begin{figure}
  \begin{center}
%    \FigureFile(85mm,110mm){j1642shpdm.eps}
    \FigureFile(85mm,110mm){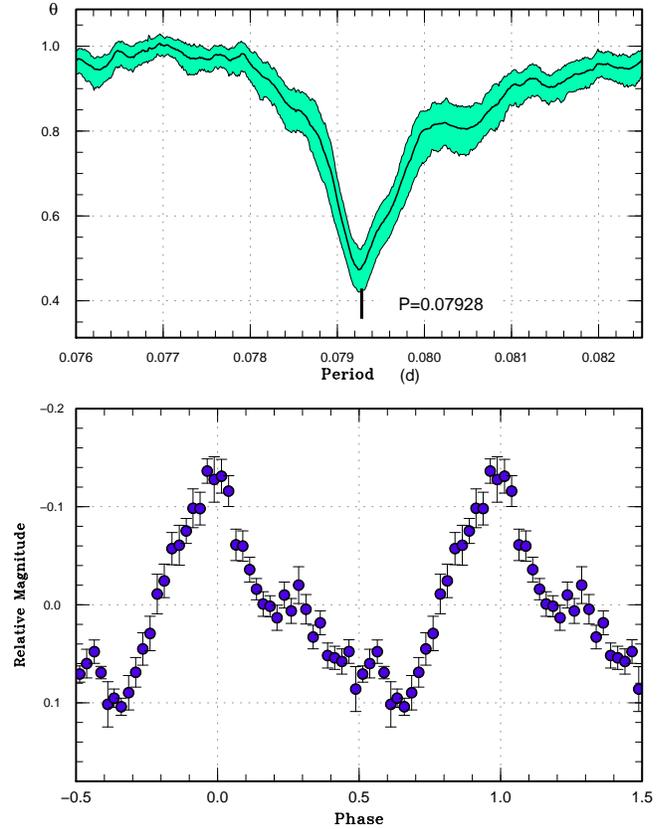}
  \end{center}
  \caption{Ordinary superhumps in SDSS J164248 (2016).
     (Upper): PDM analysis.
     (Lower): Phase-averaged profile.}
  \label{fig:j1642shpdm}
\end{figure}

   The 2016 outburst was detected at $V$=15.46 by
the ASAS-SN team (cf. vsnet-alert 19575).  Subsequent
observations detected superhumps (vsnet-alert 19582,
19593; figure \ref{fig:j1642shpdm}).
The times of superhump maxima are listed in
table \ref{tab:j1642oc2016}.
Since the superhump period of 0.07928(2)~d is too close
to the suggested orbital period, this orbital period
does not seem to have been well determined.
It is likely that the baseline for the spectroscopic
observations was not sufficient to obtain
an accurate orbital period.
For this reason, we did not include this orbital period
in table \ref{tab:perlist}.

% SI

% SI

\begin{table}
\caption{Superhump maxima of SDSS J164248 (2016)}\label{tab:j1642oc2016}
\begin{center}
\begin{tabular}{rp{55pt}p{40pt}r@{.}lr}
\hline
\multicolumn{1}{c}{$E$} & \multicolumn{1}{c}{max\commenta} & \multicolumn{1}{c}{error} & \multicolumn{2}{c}{$O-C$\commentb} & \multicolumn{1}{c}{$N$\commentc} \\
\hline
0 & 57461.5908 & 0.0004 & $-$0&0005 & 81 \\
1 & 57461.6717 & 0.0004 & 0&0011 & 74 \\
12 & 57462.5327 & 0.0038 & $-$0&0105 & 32 \\
13 & 57462.6265 & 0.0005 & 0&0039 & 82 \\
14 & 57462.7048 & 0.0014 & 0&0029 & 20 \\
15 & 57462.7748 & 0.0073 & $-$0&0064 & 13 \\
16 & 57462.8617 & 0.0010 & 0&0012 & 31 \\
28 & 57463.8154 & 0.0013 & 0&0030 & 28 \\
29 & 57463.9027 & 0.0036 & 0&0109 & 13 \\
41 & 57464.8425 & 0.0011 & $-$0&0012 & 25 \\
53 & 57465.7935 & 0.0013 & $-$0&0021 & 18 \\
54 & 57465.8725 & 0.0018 & $-$0&0024 & 25 \\
\hline
  \multicolumn{6}{l}{\commenta BJD$-$2400000.} \\
  \multicolumn{6}{l}{\commentb Against max $= 2457461.5913 + 0.079327 E$.} \\
  \multicolumn{6}{l}{\commentc Number of points used to determine the maximum.} \\
\end{tabular}
\end{center}
\end{table}

\section{Discussion}\label{sec:discuss}

\subsection{Statistics of objects}\label{sec:stat}

   In \citet{Pdot7}, we introduced the statistics of
the sources of the objects studied in our surveys
and noticed that the rapid increase of the objects
registered as ASAS-SN CVs.
Several dwarf novae received new variable star designations 
in the latest updates of the General Catalog of
Variable Stars (\cite{NameList81a}; \cite{NameList81b})
since \citet{Pdot7}, and these newly named objects
are included in the GCVS category.
The tendency pointed out in \citet{Pdot7}
became more prominent and roughly two thirds of the objects
studied in this survey are now ASAS-SN CVs.
The present GCVS
names appear to be almost complete discoveries 
up to 2008 in the literature (dwarf novae reported in
IAUCs and CBETs appear to be designated more quickly
than other literature) and let's hope that the GCVS team
could give more final designations to newly discovered
dwarf novae.

\begin{figure}
  \begin{center}
%    \FigureFile(80mm,70mm){objsource.eps}
    \FigureFile(80mm,70mm){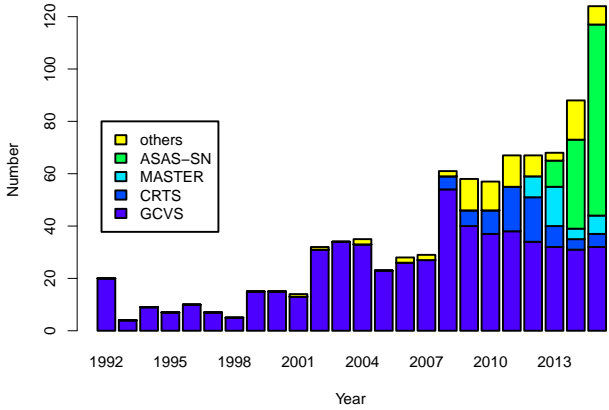}
  \end{center}
  \caption{Object categories in our survey.  Superoutbursts
  with measured superhump periods are included.
  The year represents the year of outburst.
  The year 1992 represents outbursts up to 1992 and the year
  2015 includes the outbursts in 2016, respectively.
  The category GCVS includes the objects named in the General
  Catalog of Variable Stars \citet{GCVS} in the latest version
  and objects named in New Catalog of Suspected Variable Stars
  (NSV: \cite{NSV}).  The categories CRTS, MASTER, ASAS-SN represent
  objects which were discovered in respective surveys.
  A small fraction of objects discovered by these surveys
  are already named in GCVS and are included in the category GCVS.
  }
  \label{fig:objsource}
\end{figure}

\subsection{Period distribution}\label{sec:perdist}

\begin{figure}
  \begin{center}
%    \FigureFile(80mm,135mm){phist.eps}
    \FigureFile(80mm,135mm){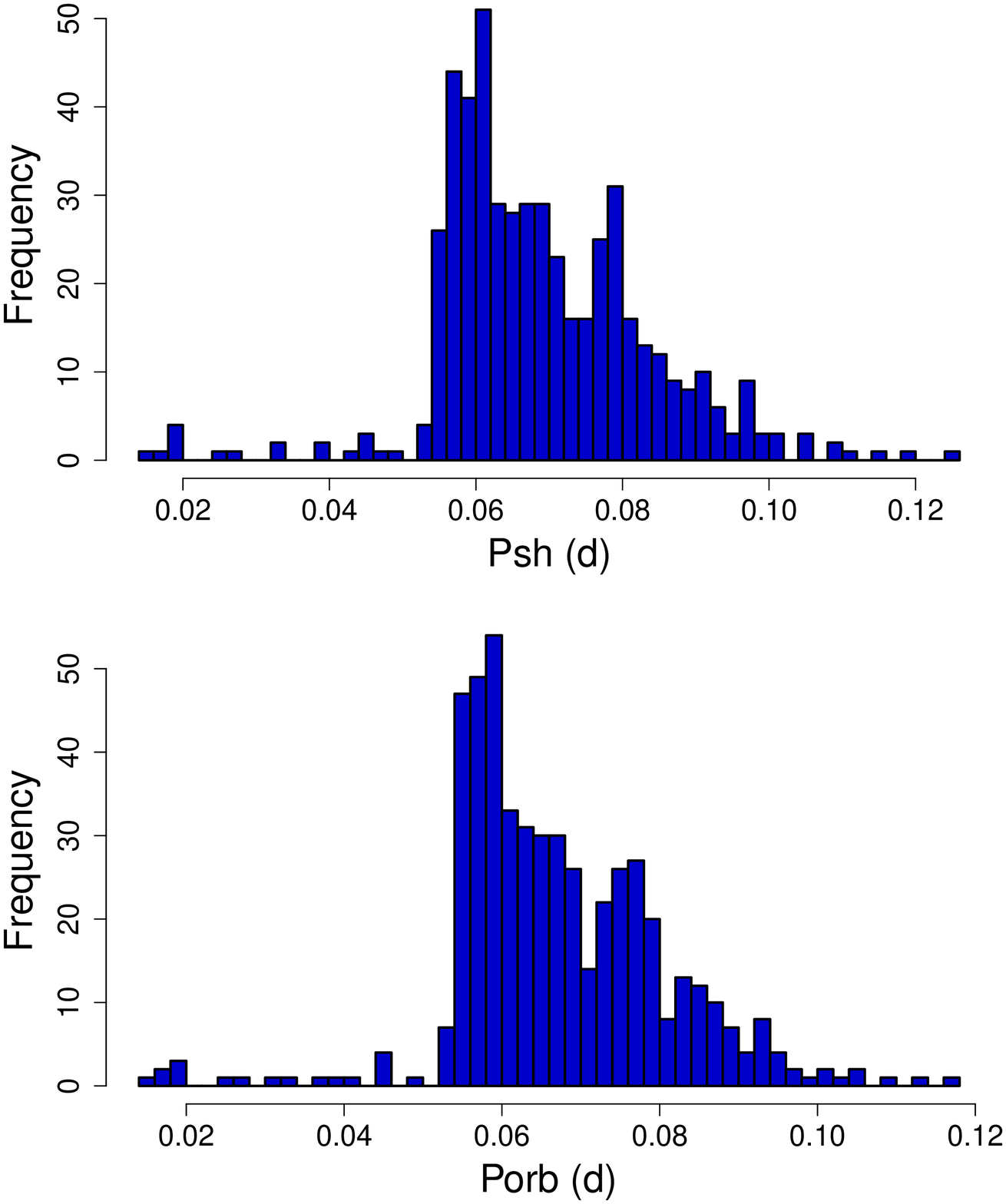}
  \end{center}
  \caption{Distribution of superhump periods in this survey.
  The data are from \citet{Pdot}, \citet{Pdot2}, \citet{Pdot3},
  \citet{Pdot4}, \citet{Pdot5}, \citet{Pdot6}, \citet{Pdot7}
  and this paper.
  The mean values are used when multiple superoutbursts
  were observed.  The number of objects is 511.
  (Upper) distribution of superhump periods.
  (Lower) distribution of orbital periods.  For objects with
  superhump periods shorter than 0.053~d, the orbital periods
  were assumed to be 1\% shorter than superhump periods.
  For objects with superhump periods longer than 0.053~d,
  we used the calibration in \citet{Pdot3} to estimate
  orbital periods.
  }
  \label{fig:phist}
\end{figure}

\begin{figure}
  \begin{center}
%    \FigureFile(80mm,135mm){phistRK.eps}
    \FigureFile(80mm,135mm){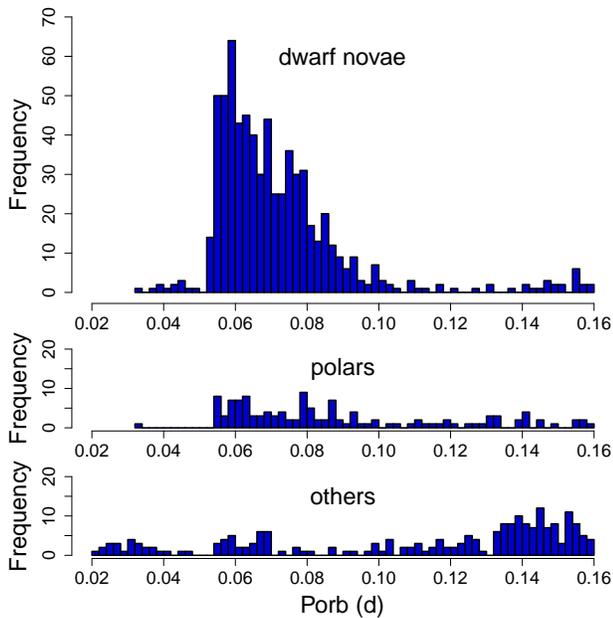}
  \end{center}
  \caption{Distribution of orbital periods in the latest
  version of RKCat (\cite{RKCat}; Edition 7.23, 2015 June 30).
  The three histograms represent distributions of
  dwarf novae (DN in RKCat), polars (AM in RKCat) and
  others.  The number of dwarf novae in the region of
  0.02--0.12~d (corresponding to figure \ref{fig:phist})
  is 676.
  }
  \label{fig:phistRK}
\end{figure}

   In figure \ref{fig:phist}, we give distributions of
superhump and estimated orbital periods
(see the caption for details) since \citet{Pdot}.
For readers' convenience, we also listed new ephemerides of
eclipsing systems newly determined or updated in this study
in table \ref{tab:eclipsing}.
When there are multiple observations of superoutbursts
of the same object, we adopted an average of
the measurements.
Since most of non-magnetic CVs below
the period gap are considered to be SU UMa-type
dwarf novae, this distribution reflects the distribution
of non-magnetic CVs below the period gap.
As already pointed out in \citet{Pdot7}, the sharp
cut-off at a period of 0.053~d
(the objects below this period are either AM CVn-type
systems and EI Psc-type objects) and the apparent
absence of the strong signature of the lower edge of
the period gap are even more apparent.  The updated
statistics confirms the findings in \citet{Pdot7}.
The same statistics using the latest
version of RKCat (\cite{RKCat}; Edition 7.23, 2015 June 30)
is shown in figure \ref{fig:phistRK}.  The result
confirms the general trend seen in our sample,
although it is not surprising since more than half
of dwarf novae in this $P_{\rm orb}$ region
in RKCat are from our surveys.
The disrupted magnetic braking may be weaker
or more CVs may be formed in the period gap
than had been supposed.

\begin{table}
\caption{Ephemerides of eclipsing systems.}\label{tab:eclipsing}
\begin{center}
\begin{tabular}{cccccc}
\hline
Object & Epoch (BJD) & Period (d) \\
\hline
V2051 Oph    & 2453189.48679(1) & 0.0624278552(2) \\
ASASSN-15sl  & 2457341.23671(7) & 0.0870484(7) \\
ASASSN-15ux  & 2457400.82908(10) & 0.056109(2) \\
CRTS J200331 & 2457200.79900(6) & 0.0587048(3) \\
SDSS J074859 & 2457351.21283(2) & 0.0583110901(7) \\
\hline
\end{tabular}
\end{center}
\end{table}

\subsection{Period derivatives during stage B}\label{sec:stagebpdot}

\begin{figure*}
  \begin{center}
%    \FigureFile(110mm,88mm){pdotporb8.eps}
    \FigureFile(110mm,88mm){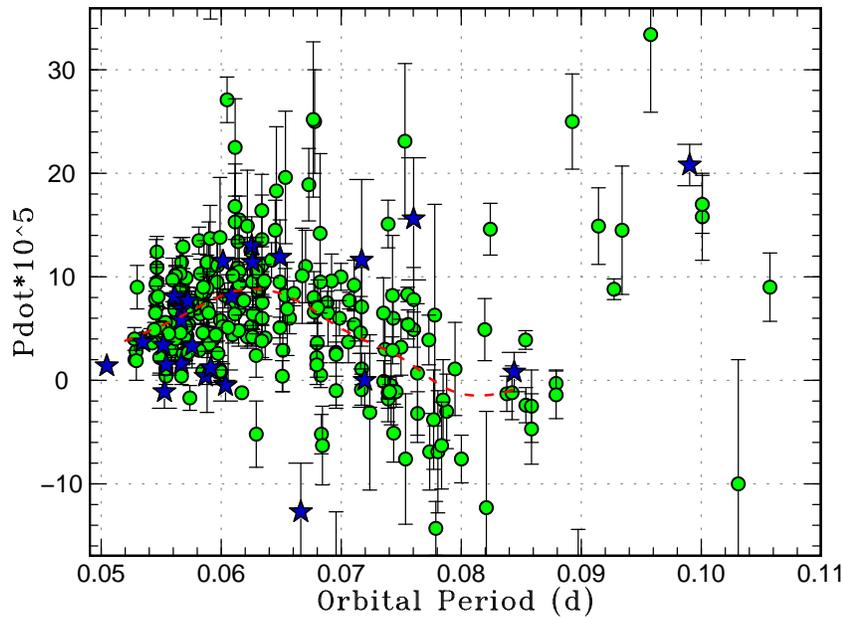}
  \end{center}
  \caption{$P_{\rm dot}$ for stage B versus $P_{\rm orb}$.
  Filled circles and filled stars represent samples in 
  \citet{Pdot}--\citet{Pdot7} and this paper, respectively.
  The curve represents the spline-smoothed global trend.
  }
  \label{fig:pdotporb8}
\end{figure*}

   Figure \ref{fig:pdotporb8} represents updated relation
between $P_{\rm dot}$ for stage B versus $P_{\rm orb}$.
Although this is essentially an updated version of
the corresponding figures in the earlier series of
papers, we have omitted poor quality observation
(quality C) and simplified the symbols.
The object listed in this paper with large negative
$P_{\rm dot}$ is PM J03338, which had a separate precursor
and a long stage A \citep{kat16j0333}.  Other objects 
with large negative $P_{\rm dot}$ in earlier papers
are UV Gem (2003), PU UMa (2012) and CY UMa (2014).
UV Gem is famous for the large negative $P_{\rm dot}$
and it is possibly interpreted as a stage A-B transition
rather than period variation during stage B
(cf. \cite{kat16v1006cyg}).  PU UMa is an eclipsing
object and the $P_{\rm dot}$ determination may have
suffered from the beat phenomenon.  In CY UMa (2014),
the stage transition was rather smooth and it was
difficult to define the border of stage B.
These outliers have their own reasons to be outside
the distribution of the majority of objects,
and the main trends in this figure seem to apply
to most of ordinary superoutbursts.

\subsection{Long-period objects with long-lasting stage A}\label{sec:longstagea}

   It had been known that some long-$P_{\rm orb}$ systems
show a strong decrease of the superhump
periods [cf. MN Dra and UV Gem, see subsection 4.10
in \citet{Pdot}].  Although the origin of this strong
period variation had remained a mystery, \citet{Pdot6}
proposed a working hypothesis that these strong period
variations are a result of the combination of
stages A and B.  This interpretation requires that
stage A in these systems is unusually long.
In \citet{Pdot6}, the case of MN Dra was studied,
which lacked the spectroscopically determined
orbital period.  \citet{kat16v1006cyg} presented
a more convincing example of V1006 Cyg, whose
orbital period had been determined spectroscopically.
It appears to have been established that at least
some long-$P_{\rm orb}$ systems show long-lasting
stage A, which implies that the 3:1 resonance grows
slowly in these systems.  \citet{Pdot6} and \citet{kat16v1006cyg}
suggested that the mass ratios close to the borderline
of the 3:1 resonance is responsible for this phenomenon.

   An updated list of long-$P_{\rm orb}$ SU UMa-type objects
with long phase of stage A superhumps is given
in table \ref{tab:longPsuuma}.  For V452 Cas, we used
the best observed superoutburst (2007) in \citet{she09v452cas}
and modified the superhump stages listed in \citet{Pdot}
according to the modern interpretation
(see also subsection \ref{obj:v452cas} and figure
\ref{fig:v452cascomp3}).  The duration of stage A in
KK Tel is from the combined $O-C$ diagram
(see also subsection \ref{obj:kktel} and figure
\ref{fig:kktelcomp2}).
ASASSN-15rs (subsection \ref{obj:asassn15rs}) and
DDE 26 (subsection \ref{obj:dde26})
may belong to this category.

\begin{table*}
\caption{Comparison of SU UMa-type objects with long phase of stage A superhumps}\label{tab:longPsuuma}
\begin{center}
\begin{tabular}{cccccccc}
\hline
Object & $P_{\rm orb}$\commenta & $P_{\rm A}$\commentb
       & $P_{\rm B}$\commentc & $P_{\rm C}$\commentd
       & dur\commente & $q$\commentf & References \\
\hline
V1006 Cyg (2015) & 0.09903(9) & 0.1093(3) & 0.10541(4) & 0.10444(5) & $\ge$32 & 0.34(2) & \citet{kat16v1006cyg} \\
MN Dra (2012) & 0.0998(2) & 0.10993(9) & 0.10530(6) & -- & $\ge$39 & 0.327(5) & \citet{Pdot6} \\
MN Dra (2013) & 0.0998(2) & 0.1082(1) & 0.10504(7) & -- & $\ge$18 & 0.258(5) & \citet{Pdot6}
 \\
CRTS J214738.4$+$244554 (2011) & 0.09273(3) & 0.0992(3) & 0.09715(2) & -- & $\ge$21 & 0.204(11) & \citet{Pdot7} \\
OT J064833.4$+$065624 (2014) & -- & 0.1052(4) & 0.10033(3) & -- & $\ge$38 & -- & \citet{Pdot7} \\
V452 Cas (2007) & -- & 0.08943(7) & 0.08869(2) & -- & 20--35 & -- & this work \\
KK Tel (2015) & -- & 0.09005(12) & 0.08761(2) & -- & $\ge$25 & -- & this work \\
ASASSN-15cl (2016) & -- & 0.0961(3) & 0.09463(10) & 0.09391(7) & $\ge$22 & -- & this work \\
\hline
  \multicolumn{8}{l}{\commenta Orbital period (d).} \\
  \multicolumn{8}{l}{\commentb Period of stage A superhumps (d).} \\
  \multicolumn{8}{l}{\commentc Period of stage B superhumps (d).} \\
  \multicolumn{8}{l}{\commentd Period of stage C superhumps (d).} \\
  \multicolumn{8}{l}{\commente Duration of stage A (cycles).} \\
  \multicolumn{8}{l}{\commentf Determined from stage A superhumps.} \\
\end{tabular}
\end{center}
\end{table*}

\subsection{Mass ratios from stage A superhumps}\label{sec:stagea}

   Since the new interpretation of stage A
as representing the dynamical precession rate
at the 3:1 resonance in \citet{kat13qfromstageA},
the application of this method produced a steady stream
of $q$ measurements.  We list new estimates for $q$ 
from stage A superhumps in table \ref{tab:newqstageA}.
This table also includes new objects that were studied
in detail in other papers.  The appropriate
references are listed in table \ref{tab:perlist}.

   In table \ref{tab:pera}, we list all stage A superhumps
recorded in the present study.

\begin{figure*}
  \begin{center}
%    \FigureFile(110mm,88mm){qall5.eps}
    \FigureFile(110mm,88mm){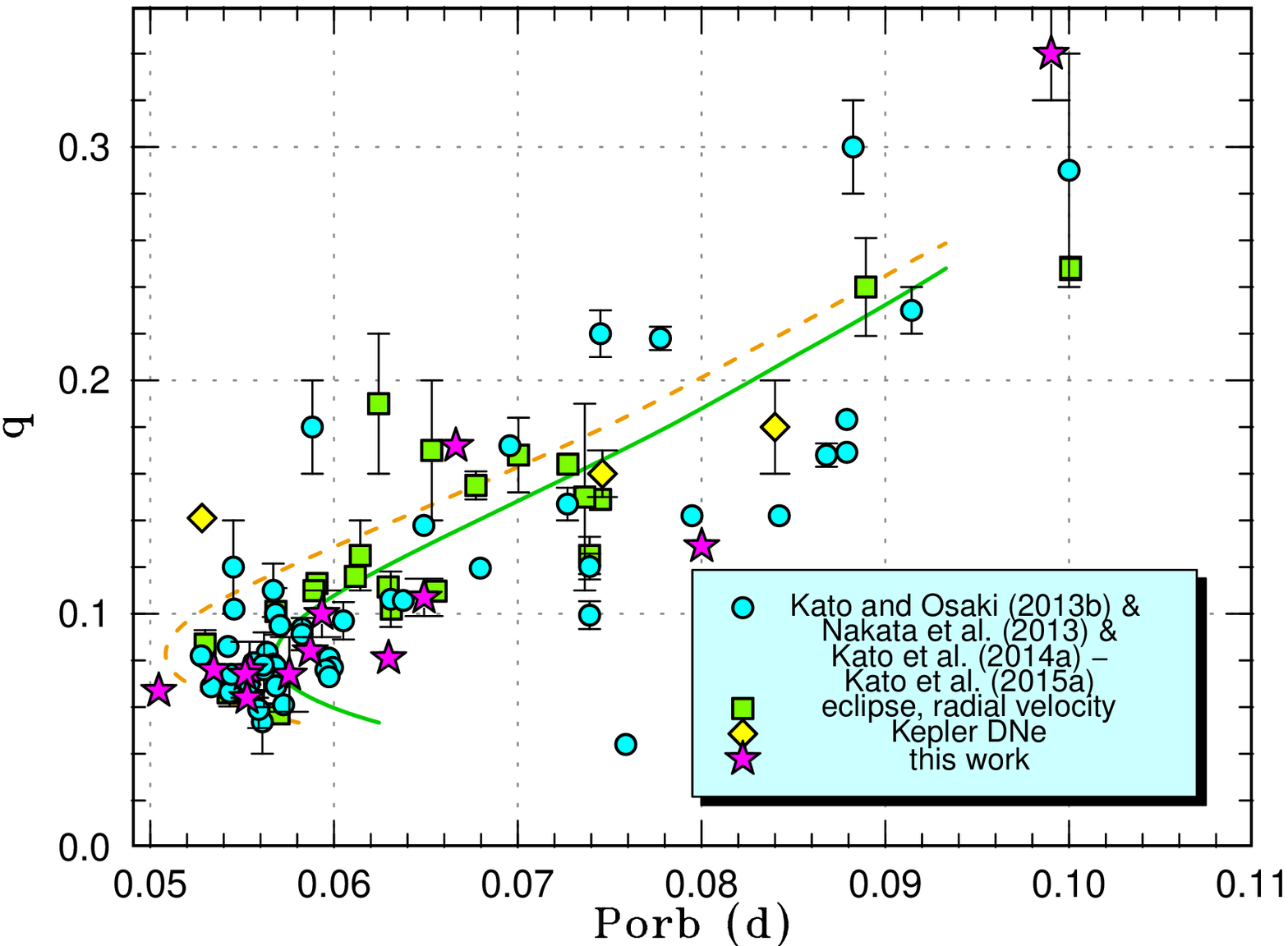}
  \end{center}
  \caption{Mass ratio versus orbital period.
  The dashed and solid curves represent the standard and optimal
  evolutionary tracks in \citet{kni11CVdonor}, respectively.
  The filled circles, filled squares, filled stars, filled diamonds
  represent $q$ values from a combination of the estimates
  from stage A superhumps published in four preceding
  sources (\cite{kat13qfromstageA}; \cite{nak13j2112j2037};
  \cite{Pdot5}; \cite{Pdot6}; \cite{Pdot7}),
  known $q$ values from quiescent eclipses or 
  radial-velocity study (see \cite{kat13qfromstageA} for
  the data source), $q$ estimated in this work and dwarf novae
  in the Kepler data (see text for the complete reference),
  respectively.  The objects in ``this work'' includes
  objects studied in other papers but listed in table \ref{tab:outobs}}
  \label{fig:qall5}
\end{figure*}

   A updated distribution of mass ratios is shown in
figure \ref{fig:qall5} [for the list of objects, see
\citet{kat13qfromstageA} and \citet{Pdot7}].
We have newly added PHL 1445 with $P_{\rm orb}$=0.052985~d
and $q$=0.087(6) (\cite{mca15phl1445}, eclipse observation).
It would be worth mentioning that \citet{har16wzsge}
derived $q \le$0.071 for WZ Sge ($P_{\rm orb}$=0.056688~d)
by infrared spectroscopy of the secondary (not plotted
in this figure).
The present study has strengthened the concentration
of WZ Sge-type dwarf novae around $q=0.07$
just above the period minimum, as reported in \citet{Pdot7}.

\begin{table}
\caption{New estimates for the binary mass ratio from stage A superhumps}\label{tab:newqstageA}
\begin{center}
\begin{tabular}{ccc}
\hline
Object         & $\epsilon^*$ (stage A) & $q$ from stage A \\
\hline
V1006 Cyg      & 0.094(3)   & 0.34(2) \\
V493 Ser       & 0.0449(13) & 0.129(5) \\
ASASSN-15gq    & 0.038(2)   & 0.107(8) \\
ASASSN-15hd    & 0.028(4)   & 0.076(12) \\
ASASSN-15na    & 0.030(2)   & 0.081(5) \\
ASASSN-15ni    & 0.0027(2)  & 0.074(2) \\
ASASSN-15po    & 0.0251(5)  & 0.067(2) \\
ASASSN-15pu    & 0.028(5)   & 0.074(16) \\
ASASSN-15uj    & 0.0243(13) & 0.064(4) \\
ASASSN-16bh    & 0.0283(3)  & 0.076(1) \\
ASASSN-16bu    & 0.037(4)   & 0.10(1) \\
CRTS J200331   & 0.0310(2)  & 0.084(1) \\
PM J03338      & 0.0604(13) & 0.172(4) \\
\hline
\end{tabular}
\end{center}
\end{table}

\begin{table}
\caption{Superhump Periods during Stage A}\label{tab:pera}
\begin{center}
\begin{tabular}{cccc}
\hline
Object & Year & period (d) & err \\
\hline
EG Aqr & 2015 & 0.08109 & 0.00022 \\
V1006 Cyg & 2015 & 0.10930 & 0.00030 \\
V844 Her & 2015 & 0.05703 & 0.00019 \\
V493 Ser & 2015 & 0.08377 & 0.00011 \\
KK Tel & 2015 & 0.09005 & 0.00012 \\
ASASSN-15cl & 2016 & 0.09613 & 0.00027 \\
ASASSN-15dp & 2015 & 0.06145 & 0.00013 \\
ASASSN-15ee & 2015 & 0.05794 & 0.00009 \\
ASASSN-15gn & 2015 & 0.06453 & 0.00003 \\
ASASSN-15gq & 2015 & 0.06748 & 0.00018 \\
ASASSN-15hd & 2015 & 0.05703 & 0.00024 \\
ASASSN-15hm & 2015 & 0.05662 & 0.00010 \\
ASASSN-15hn & 2015 & 0.06322 & 0.00016 \\
ASASSN-15kh & 2015 & 0.06155 & 0.00003 \\
ASASSN-15lt & 2015 & 0.06213 & 0.00024 \\
ASASSN-15na & 2015 & 0.06491 & 0.00012 \\
ASASSN-15ni & 2015 & 0.05673 & 0.00017 \\
ASASSN-15po & 2015 & 0.05178 & 0.00001 \\
ASASSN-15pu & 2015 & 0.05920 & 0.00030 \\
ASASSN-15sc & 2015 & 0.05867 & 0.00009 \\
ASASSN-15uj & 2015 & 0.05664 & 0.00008 \\
ASASSN-15ux & 2015 & 0.05743 & 0.00031 \\
ASASSN-16bh & 2016 & 0.05502 & 0.00010 \\
ASASSN-16bu & 2016 & 0.06159 & 0.00023 \\
CRTS J095926 & 2015 & 0.09079 & 0.00090 \\
CRTS J200331 & 2015 & 0.06058 & 0.00002 \\
MASTER J073325 & 2016 & 0.06209 & 0.00017 \\
PM J03338 & 2015 & 0.07067 & 0.00005 \\
\hline
\end{tabular}
\end{center}
\end{table}

\subsection{WZ Sge-type objects}\label{sec:wzsgetype}

   In table \ref{tab:wztab}, we list the parameters of
WZ Sge-type dwarf novae (including likely ones).

\begin{table*}
\caption{Parameters of WZ Sge-type superoutbursts.}\label{tab:wztab}
\begin{center}
\begin{tabular}{cccccccccccc}
\hline
Object & Year & $P_{\rm SH}$ & $P_{\rm orb}$ & $P_{\rm dot}$\commenta & err\commenta & $\epsilon$ & Type\commentb & $N_{\rm reb}$\commentc & delay\commentd & Max & Min \\
\hline
RZ Leo & 2016 & 0.078675 & 0.076030 & 15.6 & 5.9 & 0.035 & C & 1 & -- & ]13.0 & 18.5 \\
V2051 Oph & 2015 & 0.064708 & 0.062428 & -- & -- & 0.037 & -- \\
ASASSN-15dp & 2015 & 0.060005 & -- & 0.4 & 1.1 & -- & -- & -- & -- & ]14.1 & 19.4: \\
ASASSN-15ee & 2015 & 0.057136 & -- & 8.1 & 1.2 & -- & -- & -- & 6 & 12.6 & 19.9: \\
ASASSN-15gq & 2015 & 0.066726 & 0.06490 & 11.9 & 0.8 & 0.028 & -- & -- & $\ge$5 & ]15.4 & [21.6 \\
ASASSN-15hd & 2015 & 0.056105 & 0.05541 & 1.5 & 0.3 & 0.013 & C? & $\ge$1 & 11 & 14.0 & 21.7 \\
ASASSN-15hn & 2015 & 0.061831 & -- & $-$0.5 & 1.5 & -- & -- & -- & 13--14 & 12.9 & 21.9: \\
ASASSN-15kh & 2015 & 0.060480 & -- & 1.2 & 1.6 & -- & -- & -- & 13 & 13.2 & [21.0 \\
ASASSN-15na & 2015 & 0.063720 & 0.06297 & 3.1 & 2.6 & 0.012 & -- & -- & $\ge$9 & ]14.8 & 21.5: \\
ASASSN-15ni & 2015 & 0.055854 & 0.05517 & 3.4 & 0.6 & 0.012 & -- & -- & 10 & 12.9 & 21.0: \\
ASASSN-15po & 2015 & 0.050916 & 0.050457 & 1.1 & 0.1 & 0.009 & A/B & $\ge$5 & 11 & 13.7 & 21.6 \\
ASASSN-15pu & 2015 & 0.058254 & 0.05757 & 3.3 & 2.1 & 0.012 & -- & -- & 10 & 13.7 & 22.1: \\
ASASSN-15se & 2015 & 0.063312 & -- & -- & -- & -- & A/B or B & $\ge$2 & $\ge$5 & ]13.0 & 20.6 \\
ASASSN-15sl & 2015 & 0.091065 & 0.087048 & 9.1 & 2.6 & 0.046 & -- \\
ASASSN-15uj & 2015 & 0.055805 & 0.055266 & $-$1.1 & 1.6 & 0.010 & -- & -- & 10 & 14.3 & 21.0: \\
ASASSN-15ux & 2015 & 0.056857 & 0.056109 & -- & -- & 0.013 & -- & -- & 14 & 14.4 & [21.0 \\
ASASSN-16bh & 2016 & 0.054027 & 0.05346 & 3.7 & 0.3 & 0.011 & A & 1 & 7 & 12.7 & 20.3: \\
ASASSN-16bi & 2016 & -- & 0.05814 & -- & -- & -- & -- & -- & 12: & 14.3 & [20.6 \\
ASASSN-16bu & 2016 & 0.060513 & 0.05934 & -- & -- & 0.020 & -- & -- & 9 & 14.5 & 22.1 \\
CRTS J200331 & 2015 & 0.059720 & 0.058705 & -- & -- & 0.017 & -- \\
SDSS J074859 & 2015 & 0.05958 & 0.058311 & -- & -- & 0.022 & -- \\
SDSS J145758 & 2015 & 0.054912 & 0.054087 & 2.2 & 2.9 & 0.015 & -- & -- & -- & 11.9 & 29.5 \\
\hline
  \multicolumn{12}{l}{\commenta Unit $10^{-5}$.} \\
  \multicolumn{12}{l}{\commentb A: long-lasting rebrightening; B: multiple rebegitehnings; C: single rebrightening; D: no rebrightening.} \\
  \multicolumn{12}{l}{\commentc Number of rebrightenings.} \\
  \multicolumn{12}{l}{\commentd Days before ordinary superhumps appeared.} \\
\end{tabular}
\end{center}
\end{table*}

\begin{figure*}
  \begin{center}
%    \FigureFile(110mm,88mm){wzpdottype8.eps}
    \FigureFile(110mm,88mm){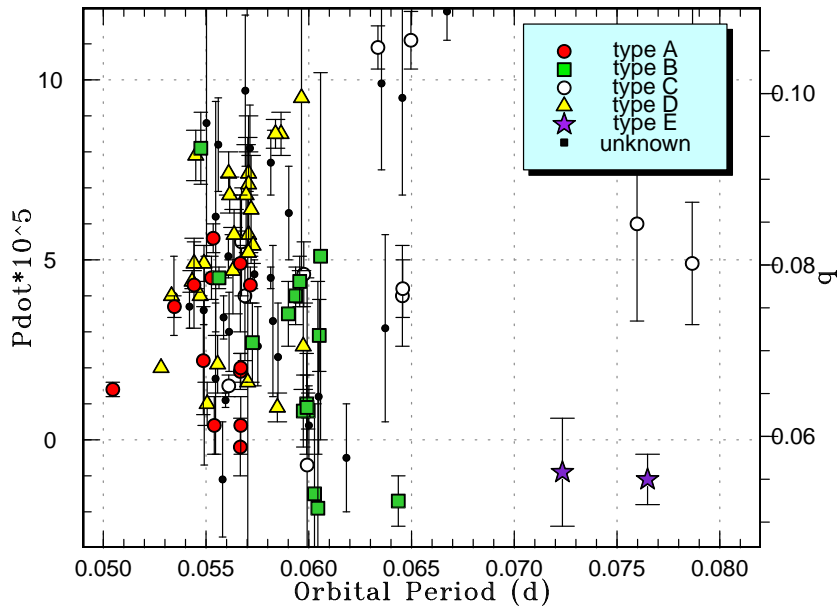}
  \end{center}
  \caption{$P_{\rm dot}$ versus $P_{\rm orb}$ for WZ Sge-type
  dwarf novae.  Symbols represent the type of outburst:
  type-A (filled circles), type-B (filled squares),
  type-C (filled triangles), type-D (open circles)
  and type-E (filled stars) (see text for details).
  On the right side, we show mass ratios estimated
  using equation equation (6) in \citet{kat15wzsge}.
  We can regard this figure as to represent
  an evolutionary diagram.
  }
  \label{fig:wzpdottype8}
\end{figure*}

   It has been known that $P_{\rm dot}$
and $P_{\rm orb}$ are correlated with the rebrightening type
[starting with figure 36 in \cite{Pdot} and refined in
\citet{Pdot}--\citet{Pdot7} and \citet{kat15wzsge}].
The five types of outbursts based on rebrightenings are:
type-A outbursts (long-duration rebrightening),
type-B outbursts (multiple rebrightenings),
type-C outbursts (single rebrightening),
type-D outbursts (no rebrightening) and
type-E outbursts (double superoutburst, with ordinary superhumps
only during the second one).
In figure \ref{fig:wzpdottype8}, we show the updated
result up to this paper.
In this figure, we also added objects without known
rebrightening types.  These objects have been confirmed
to follow the same trend, which we consider
the evolutionary track.

   ASASSN-16bh, the very noteworthy and well-observed
WZ Sge-type dwarf nova in this study is located
at the minimum period [$P_{\rm orb}$=0.05346~d,
$P_{\rm dot}$=$+$3.7(3)].  This object showed
a typical type-A rebrightening, in agreement with
the trend in other WZ Sge-type dwarf novae.

   ASASSN-15po is an outlier ($P_{\rm orb}$=0.05092~d)
in the figure below the period minimum of most of the objects.
This object may be similar to OV Boo = SDSS J150722.33$+$523039.8,
which has an orbital period of 0.046258~d
(\cite{lit07j1507}; \cite{pat08j1507}; \cite{uth11j1507}).
Although the spectroscopic features strongly suggest
the dwarf nova-type (probably WZ Sge-type) nature,
OV Boo has not been yet recorded in major outburst.
ASASSN-15po is the first object below the period
minimum which has undergone a typical WZ Sge-type
superoutburst.  The details of the outburst are discussed
in \Namekataprep .

\subsection{Lessons from recent observations}\label{sec:lessons}

\begin{figure}
  \begin{center}
%    \FigureFile(80mm,70mm){obsqual.eps}
    \FigureFile(80mm,70mm){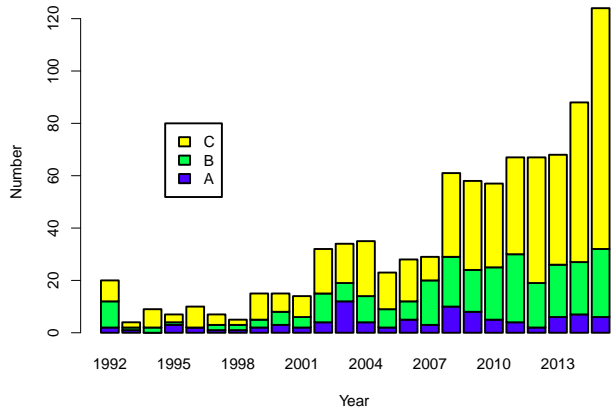}
  \end{center}
  \caption{Quality of observations (A: excellent, B: partial coverage
  or slightly low quality, C: insufficient coverage or observations
  with large scatter).
  The year represents the year of outburst.
  The year 1992 represents outbursts up to 1992 and the year
  2015 includes the outbursts in 2016, respectively.
  }
  \label{fig:obsqual}
\end{figure}

   Thanks to the increase of discoveries of new
dwarf novae and detections of outbursts
by modern surveys (such as ASAS-SN),
the number of studied objects has dramatically
increased in recent years.
This increase has indeed improved our knowledge
in the distribution of CVs below the period gap
(e.g. subsection \ref{sec:stat}).
The fraction of well-observed superoutbursts,
however, largely decreased.  For example, the number
of ``A''-class (well-observed) observations
decreased from 58 (out of 363 outbursts) in
\citet{Pdot}, 6 (out of 65 outbursts) in \citet{Pdot2}
to 5 (out of 107 outbursts).
Although such qualification of observations
are subjective and the criteria may have not necessarily been
the same, the increase of ``underobserved'' outbursts
is apparent despite the increase of observations
(figure \ref{fig:obsqual}).

   The same trend is even more apparent in
WZ Sge-type outbursts.  In \citet{Pdot}--\citet{Pdot4},
55 WZ Sge-type outbursts (out of 66) had observations
to classify the rebrightening pattern.
In the present study, only 5 WZ Sge-type outbursts
(out of 18) have rebrightening classifications.
Such a trend is fatal since rebrightenings are one of
key elements in the study of WZ Sge-type dwarf novae
(cf. \cite{kat15wzsge}). 

   These trends in observations probably reflect
the increase of freshly discovered objects
or outbursts, which would easily divert observers'
attention.  In order that the observations will be
more astrophysically beneficial and rewarding
to observers, we propose the following lessons from
recent observations.  Although some of the lessons
may be evident, we list them since they will be
useful for those who wish to start contributing
to this field, and they are not usually written
in practical textbooks (such as \cite{hel01book}).

\begin{itemize}

\item Single-night observations have very limited
value (except classification of the object
and the initial detection of superhumps).
If there are more than observations on two nights
(hopefully consecutive nights), we can determine
the superhump period better than to 0.2\% (1$\sigma$ error),
necessary to make comparison with the orbital one.
Periods from single-night observations have large errors
typically 1--3\%, which is entirely insufficient
to make comparison with the orbital one.

\item Once the object is observed, do not lightly
change the target.  In general, fresh outbursts tend to be
``overobserved'' (observations are sometimes redundant)
while they become underobserved as the progress
of the outbursts.  There may not be many observations in
the later phase and observations such a phase can be
relatively more important.  We should note, however,
that objects may become too faint or the amplitudes of
superhumps became to small to make useful observations.
In such cases, we recommend nightly snapshots.

\item Even after the superoutburst ends, regularly
visit the target and obtain snapshot observations.
This is particularly true for WZ Sge-type dwarf novae.
If there is a major rebrightening, restart
time-resolved photometry.

\item For detecting stage A superhumps, which is very
important to estimate mass ratios, early observations
are very important.  Even a 1-d gap in the observation
could be fatal.  In WZ Sge-type dwarf novae,
there is usually a long waiting time ($\sim$10~d)
before stage A superhumps appear.  Although observations
in this phase may not appear so appealing since
early superhumps may become less apparent and amplitudes
of variations became smaller, this phase is astrophysically
more important (compared to the phase after full growth
of superhumps) and it is a waste to miss this phase.

\end{itemize}

\section{Summary}\label{sec:summary}

   In addition to the updated statistics of
the period distribution, $P_{\rm orb}$--$P_{\rm dot}$
relation, the updated evolutionary track using
stage A superhumps and refined relationship between
$P_{\rm orb}$--$P_{\rm dot}$ versus the rebrightening
type in WZ Sge-type dwarf novae, the objects of
special interest in this paper can be summarized as follows.

\begin{itemize}

\item V452 Cas ($P_{\rm SH} \sim$0.0888~d),
KK Tel ($P_{\rm SH} \sim$0.0876~d) 
and ASASSN-15cl ($P_{\rm SH} \sim$0.0946~d) appear
to have a long-lasting stage A.
They would be members of growing group of
long-$P_{\rm orb}$ objects with slowly growing superhumps.
A slow growth of the 3:1 resonance near
the stability border has been proposed
(\cite{kat16v1006cyg}; also \cite{Pdot6}).
If the mass ratios for these objects are determined
by measuring $P_{\rm orb}$, they would provide
an excellent test for this interpretation.

\item The WZ Sge-type object RZ Leo underwent
a well-observed superoutburst in 2016.
No clear evidence of early superhumps was detected.
This object showed a strong beat phenomenon
between the superhump and orbital periods.

\item ASASSN-15cy is an object below the period
minimum ($P_{\rm SH} \sim$0.0500~d).
This object showed a superoutburst resembling
the EI Psc-type object CSS J174033.5$+$414756.

\item ASASSN-15hd is a WZ Sge-type dwarf nova
showing large-amplitude early superhumps with 
a ``saw-tooth''-like profile.

\item ASASSN-15gn ($P_{\rm SH} \sim$0.0636~d),
ASASSN-15hn ($P_{\rm SH} \sim$0.0618~d),
ASASSN-15kh ($P_{\rm SH} \sim$0.0605~d) and 
ASASSN-16bu ($P_{\rm SH} \sim$0.0609~d)
are possibly period bouncers as judged from the slow growth
of ordinary superhumps and small amplitudes of
superhumps.

\item ASASSN-15na is a WZ Sge-type dwarf with
a relatively long orbital period (0.06297~d).
The object, however, appears to have a larger
$q$ than expected for a period bouncer.

\item ASASSN-15ni is a WZ Sge-type dwarf nova
showing a superoutburst typical for this class.

\item ASASSN-15sl and SDSS J074859 are eclipsing
systems and we have also determined the orbital
periods using eclipse observations.

\item ASASSN-15uj is a WZ Sge-type dwarf nova
with a low $q$=0.064(4), indicating the relatively
evolved state.

\item ASASSN-15ux is a rare eclipsing WZ Sge-type
dwarf nova.

\item ASASSN-16bh is a WZ Sge-type dwarf nova
showing a relatively rare plateau-type, long
rebrightening (without small rebrightenings in it).
This object also showed early superhumps with
three maxima in one cycle.

\item CRTS J200331 is an eclipsing SU UMa-type
or WZ Sge-type dwarf nova probably near the border
of SU UMa-type and WZ Sge-type objects.

\end{itemize}

\section*{Acknowledgements}

This work was supported by the Grant-in-Aid
``Initiative for High-Dimensional Data-Driven Science through Deepening
of Sparse Modeling'' (25120007) 
from the Ministry of Education, Culture, Sports, 
Science and Technology (MEXT) of Japan.
The authors are grateful to observers of VSNET Collaboration and
VSOLJ observers who supplied vital data.
We acknowledge with thanks the variable star
observations from the AAVSO International Database contributed by
observers worldwide and used in this research.
We are also grateful to the VSOLJ database.
This work is deeply indebted to outburst detections and announcement
by a number of variable star observers worldwide, including participants of
CVNET and BAA VSS alert.
The CCD operation of the Bronberg Observatory is partly sponsored by
the Center for Backyard Astrophysics.
We are grateful to the Catalina Real-time Transient Survey
team for making their real-time
detection of transient objects available to the public.
This research has made use of the SIMBAD database,
operated at CDS, Strasbourg, France.
This research has made use of the International Variable Star Index 
(VSX) database, operated at AAVSO, Cambridge, Massachusetts, USA.

\end{document}